\documentclass[a4paper,openright,10pt,twoside]{book} 
\usepackage{dcolumn}
\usepackage{bm}
\usepackage{amssymb}
\usepackage{multirow}
\usepackage{rotating}
\usepackage{cite}
\usepackage{hyperref}
\hypersetup{colorlinks, linkcolor={blue}}
\usepackage{fancyhdr}
\usepackage{amsmath}
\usepackage{geometry}
\usepackage{textcomp}
\usepackage[printwatermark]{xwatermark}
\usepackage{graphicx}
\usepackage{lipsum}

\RequirePackage[calcwidth]{titlesec}
\titleformat{\section}[hang]{\sffamily\bfseries}
{\Large\thesection}{12pt}{\Large}[{\titlerule[0.5pt]}]
\def\rpv{\not\!\!{R_p}}
\def\L{\not\!\!{L}}
\def\D{\not\!\!{D}}
\def\k{\not\!\!{k}}

\def\ptmiss{\not\!\!{p_T}}
\def\bea{\begin{eqnarray}}
\def\eea{\end{eqnarray}}
\def\beq{\begin{equation}}
\def\eeq{\end{equation}}
\def\lam{\lambda}
\def\kp{\kappa}
\def\ovl{\overline}
\def\wt{\widetilde}
\def\al{\alpha}
\def\del{\delta}
\def\ep{\epsilon}
\def\pl{\partial}
\def\rt2{\sqrt{2}}
\def\nn{\nonumber}
\def\vp{\varepsilon}
\def\bN{\mathbf N}
\def\bU{\mathbf U}
\def\bV{\mathbf V}
\def\bRs{\mathbf R^{S^0}}
\def\bRp{\mathbf R^{P^0}}
\def\bRc{\mathbf R^{S^{\pm}}}

\def\Rsq{\mathbf R^{\widetilde q}}
\def\Rsu{\mathbf R^{\widetilde u}}
\def\Rsd{\mathbf R^{\widetilde d}}
\def\Ru{\mathbf R^{u}}
\def\Rd{\mathbf R^{d}}
\def\B{\mathbf B}
\def\q{\widetilde q}
\def\u{\widetilde u}
\def\d{\widetilde d}
\def\ntrli{\widetilde{\chi}^0_i}
\def\ntrlj{\widetilde{\chi}^0_j}
\def\n{{\widetilde \chi}^0}
\def\ps{\displaystyle{\not} p}

\def\ntrli{\widetilde{\chi}^0_i}
\def\ntrlj{\widetilde{\chi}^0_j}
\def\tg{\widetilde g}

\newcommand{\bsq}[1]{\widetilde{b}_#1}
\newcommand{\tsq}[1]{\widetilde{t}_#1}
\newcommand{\ns}[1]{S^0_#1}
\newcommand{\nps}[1]{P^0_#1}
\newcommand{\cs}[1]{S^\pm_#1}
\newcommand{\ntrl}[1]{\widetilde{\chi}^0_#1}
\newcommand{\chpm}[1]{\widetilde{\chi}^{\pm}_#1}
\def\gsim{\lower0.5ex\hbox{$\:\buildrel >\over\sim\:$}}
\def\lsim{\lower0.5ex\hbox{$\:\buildrel <\over\sim\:$}}

\begin{document}
\begin{figure}
\centering
\includegraphics[width=1\textwidth]{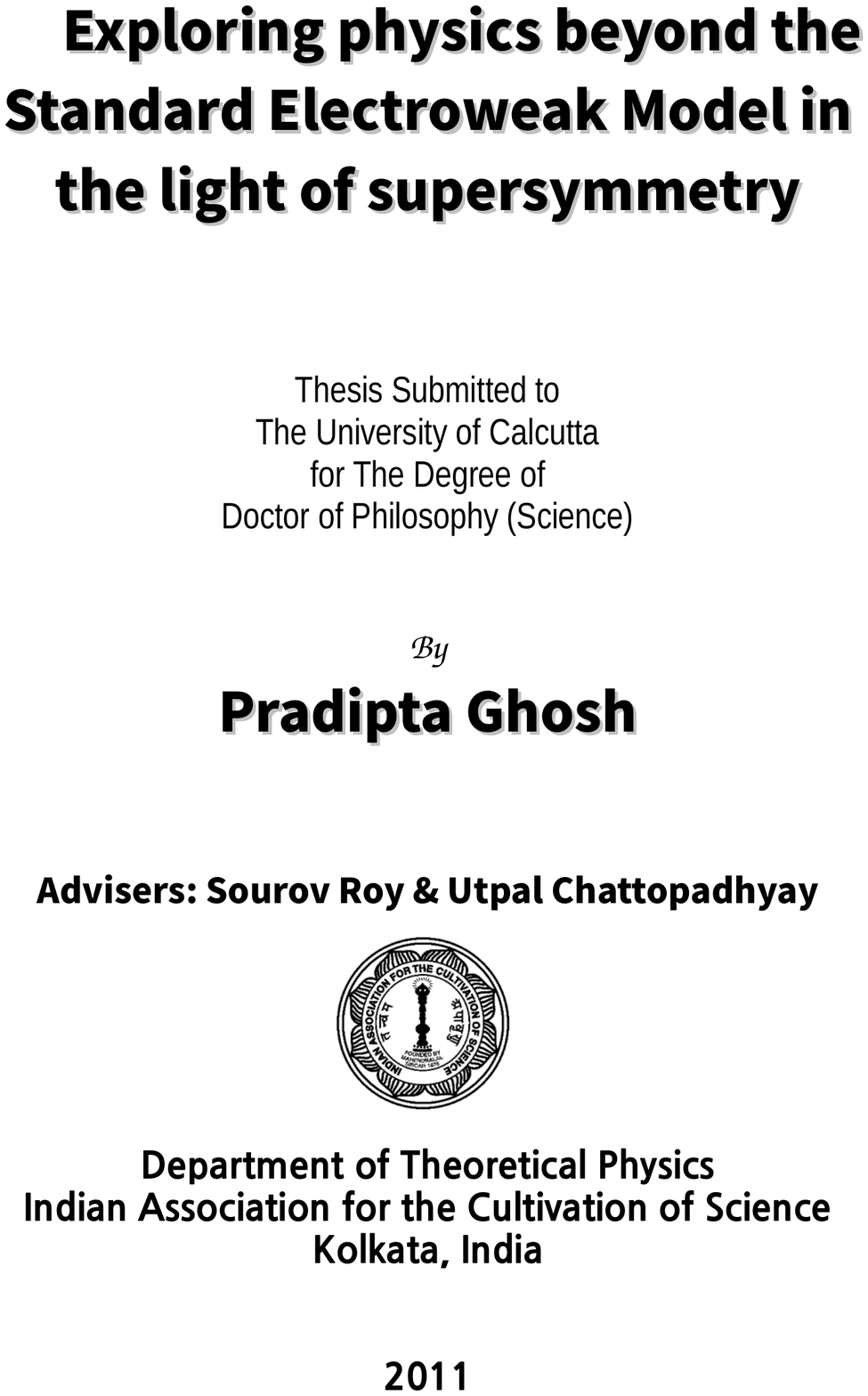}
\label{title-page}
\end{figure}
\pagenumbering{roman}
\chapter*{\sffamily{{\bf A}bstract}}

Weak scale supersymmetry has perhaps become the most popular choice for 
explaining new physics beyond the standard model. An extension beyond the
standard model was essential to explain issues like gauge-hierarchy
problem or non-vanishing neutrino mass. With the initiation 
of the large hadron collider era at CERN,  
discovery of weak-scale supersymmetric particles and, of course, Higgs boson
are envisaged. In this thesis we try to discuss certain phenomenological aspects
of an $R_p$-violating non-minimal supersymmetric model, called $\mu\nu$SSM. We show that
$\mu\nu$SSM can provide a solution
to the $\mu$-problem of supersymmetry and can simultaneously accommodate the 
existing three flavour global data from neutrino experiments even at the tree 
level with the simple choice of flavour diagonal neutrino Yukawa couplings. 
We show that it is also possible to achieve different mass hierarchies
for light neutrinos at the tree level itself. In $\mu\nu$SSM, the effect of 
$R$-parity violation together with a seesaw mechanism with TeV scale right-handed
neutrinos are instrumental for light neutrino mass generation. We also
analyze the stability of tree level neutrino masses and mixing with
the inclusion of one-loop radiative corrections. In addition, we investigate
the sensitivity of the one-loop corrections to different light neutrino mass orderings.
Decays of the lightest supersymmetric particle were also computed and ratio of certain
decay branching ratios was observed to correlate with certain neutrino mixing 
angle. We extend our analysis for different natures of the lightest
supersymmetric particle as well as with various light neutrino mass hierarchies.
We present estimation for the length of
associated displaced vertices for various natures of the lightest
supersymmetric particle which can act as a discriminating feature
at a collider experiment.
We also present an unconventional signal of Higgs boson in supersymmetry
which can lead to a discovery, even at the initial stage 
of the large hadron collider running. Besides, we show that a signal of this
kind can also act as a probe to the seesaw scale. Certain other phenomenological
issues have also been addressed.

\chapter*{\sffamily{{\bf }}}
\thispagestyle{empty}
\vspace{7cm}
\begin{center}
\sffamily
{\bf{ {{\it To a teacher who is like the pole star to me and many others}}}}\\
\vspace*{2cm}
\begin{flushright}
{\it{ {\Huge {Dr. Ranjan Ray}}}} \\
\vspace*{0.5cm}
{\it{ {{1949  - 2001}}}}
\end{flushright}

\end{center}


\chapter*{\sffamily{{\bf A}cknowledgment}}

I am grateful to the Council of Scientific and Industrial Research,
Government of India for providing me the financial assistance for the completion
of my thesis work (Award No. 09/080(0539)/2007-EMR-I (Date 12.03.2007)). I am
also thankful to my home institute for a junior research fellowship that I
had enjoyed from August, 2006 to January, 2007.

\vspace{0.1cm}
I have no words to express my gratitude to the members of theoretical
high energy physics group of the Department of Theoretical Physics 
of my home institute, particularly Dr. Utpal Chattopadhyay
and Dr. Sourov Roy. Their gratuitous infinite patience for my academic and personal problems,
unconditional affection to me, heart-rending analysis of my performance, 
cordial pray for my success, spontaneous and constant motivations for
facing new challenges were beyond the conventional teacher-student relation.
I am also grateful to Dr. Dilip Kumar Ghosh, Dr. Pushan Majumdar, 
Dr. Koushik Ray, Prof. Siddhartha Sen and Prof. Soumitra SenGupta of the same group for their
encouragement, spontaneous affection, crucial guidance and of course criticism,
in academics and life beyond it. It is also my pleasure to thank 
Dr. Shudhanshu Sekhar Mandal and Dr. Krishnendu Sengupta of the 
condensed matter group. It is an honour for me to express my respect
to Prof. Jayanta Kumar Bhattacharjee not only for his marvelous teaching,
but also for explaining me a different meaning of academics.

\vspace{0.1cm}
I sincerely acknowledge the hard efforts and sincere 
commitments of my collaborators Dr. Priyotosh Bandopadhyay and Dr. Paramita Dey
to the research projects. I am thankful to them for their level of tolerance
to my infinite curiosity in spite of their extreme busy schedules. I 
learned several new techniques and some rare insights
of the subjects from them. In the course of scientific collaboration I
have been privileged to work with Prof. Biswarup Mukhopadhyaya,
who never allowed me to realize the two decades of age difference between us.
Apart from his precious scientific guidance (I was also fortunate enough to attend
his teaching), his affection and inspiration for me has earned an eternal mark
in my memory just like his signature smile.

\vspace{0.1cm}
It is my duty to express my sincere gratitude to all of my teachers, starting
from the very basic level till date. It was their kind and hard efforts which help
me to reach here. 
I am especially grateful to Ms. Anuradha SenSarma and Mr. Malay Ghosh
for their enthusiastic efforts and selfless sacrifices during my school days.
I have no words to express my respect to Prof. Anirban Kundu and Prof. Amitava
Raychaudhuri for their precious guidance and unconventional teaching during
my post-graduate studies. I am really fortunate enough to receive their
affection and guidance till date.
In this connection I express my modest gratitude to some
of the renowned experts 
of the community for their valuable advise and encouragement.
They were always very generous to answer even some of my stupid questions,
in spite of their extremely busy professional schedules. I am particularly
grateful to Dr. Satyaki Bhattacharya, Prof. Debajyoti Choudhury, Dr. Anindya Datta,
Dr. Aseshkrishna Datta, Dr. Manas Maity, Prof. Bruce Mellado,  
Dr. Sujoy Poddar, Dr. Subhendu Rakshit and Prof. Sreerup
Raychaudhuri for many useful suggestions and very helpful discussions.

\vspace{0.1cm}
I also express my humble thanks to my home institute, 
Indian Association for the Cultivation of Science, for providing all the facilities
like high-performance personal desktop, constant and affluent access to high-speed
internet, a homely atmosphere and definitely a world class library. I am also
thankful to all the non-teaching staff members of my department 
(Mr. Subrata Balti, Mr. Bikash Darji, Mr. Bhudeb Ghosh, Mr. Tapan Moulik and Mr. Suresh Mondal)
who were always there to assist us. It is my honour to thank the Director of my home institute,
Prof. Kankan Bhattacharyya for the encouragement I received from him.

\vspace{0.1cm}
It is a pleasure to express my thanks to my colleagues and friends who were
always there to cheer me up when things were not so smooth either in
academics or in personal life. 
My cordial and special thanks to
Dr. Naba Kumar Bera, Dr. Debottam Das, Sudipto Paul Chowdhury, Dwipesh Majumder
and Joydip Mitra who were not just my colleagues but were, are and always
will be my brothers. I am really thankful to them and also to
Dr. Shyamal Biswas, Amit Chakraborty, Dr. Dipanjan Chakrabarti, Manimala Chakraborty, 
Sabyasachi Chakraborty, Anirban Datta, Ashmita Das, Sanjib Ghosh,
Dr. R. S. Hundi, Dr. Ratna Koley, Dr. Debaprasad Maity, 
Sourav Mondal, Subhadeep Mondal, Sanhita Modak, Shreyoshi Mondal,
Dr. Soumya Prasad Mukherjee, Sutirtha Mukherjee, Tapan Naskar, Dr. Himadri Sekhar Samanta,
Kush Saha, Ipsita Saha and Ankur Sensharma for making my office my second home.

\vspace{0.1cm}
It is definitely the worst injustice to acknowledge the support of my family as without 
them I believe it is just like getting lost in crowd.   

\vspace{0.1cm}
I cannot resist myself to show my humble tribute to three personalities, who by the 
philosophy of their lives and works have influenced diverse aspects of my life.
The scientist who was born long before his time, Richard P. Feynman, the writer
who showed that field of expertise is not really a constraint, Narayan Sanyal
and my old friend, Mark.

\begin{flushright}
\vspace{1.5cm}
Pradipta Ghosh 
\end{flushright}


\clearpage
\chapter*{\sffamily{ {\bf L}ist of Publications}}
\vspace*{0.02cm}
\begin{center}
{\bf{\sf{\underline{In refereed journals}}}}. 
\end{center}
\vspace*{0.02cm}
\begin{itemize}
\item
{\textbf{Radiative 
contribution to neutrino masses and mixing in $\mu\nu$SSM.}}\\
{{\em Pradipta Ghosh}},~{\em Paramita Dey},~{\em Biswarup Mukhopadhyaya},~{\em Sourov Roy.}\\
{\bf \href{http://dx.doi.org/10.1007/JHEP05(2010)087}{\textit{Journal 
of High Energy Physics} 05 (2010) 087}, 
\href{http://arxiv.org/abs/arXiv:1002.2705}{arXiv:1002.2705 [hep-ph]}.}

\vspace*{1cm}
\item
{\textbf{Neutrino masses and mixing, lightest neutralino decays 
and a solution to the $\mu$ problem in supersymmetry.}}\\
{{\em Pradipta Ghosh}},~{\em Sourov Roy.}\\
{\bf \href{http://dx.doi.org/10.1088/1126-6708/2009/04/069}{\textit{Journal 
of High Energy Physics} 04 (2009) 069}, 
\href{http://arxiv.org/abs/arXiv:0812.0084}{arXiv:0812.0084 [hep-ph]}.}
\end{itemize}
\vspace*{0.02cm}
\begin{center}
{\bf{\sf{\underline{Preprints}}}}. 
\end{center}
\vspace*{0.02cm}
\begin{itemize}
\item
{\textbf{An unusual signal of Higgs boson in supersymmetry at the LHC.}}\\
{{\em Priyotosh Bandyopadhyay}},~{{\em Pradipta Ghosh}},~{\em Sourov Roy.}\\
{\bf \href{http://dx.doi.org/10.1103/PhysRevD.84.115022}{\textit{Phys. Rev. D} 84 (2011) 115022}, 
\href{http://arxiv.org/abs/arXiv:1012.5762}{arXiv:1012.5762 [hep-ph]}\footnote{Now published
as ``Unusual Higgs boson signal in R-parity violating nonminimal 
supersymmetric models at the LHC'' in {\bf \href{http://dx.doi.org/10.1103/PhysRevD.84.115022}{\textit{Phys. Rev. D} 84 (2011) 115022}.} }.}
%
\end{itemize}
\vspace*{0.02cm}
\begin{center}
{\bf{\sf{\underline{In proceedings}}}}. 
\end{center}
\vspace*{0.02cm}
\begin{itemize}
\item
{\textbf{Neutrino masses and mixing in $\mu\nu$SSM.}}\\
{\href{http://dx.doi.org/10.1088/1742-6596/259/1/012063}{\textit{2010 
J. Phys.: Conf. Ser.} 259 012063}, 
\href{http://arxiv.org/abs/arXiv:1010.2578}{arXiv:1010.2578 [hep-ph]},}
~{PASCOS 2010}

\item
{\textbf{Neutrino masses and mixing in a supersymmetric model 
and characteristic signatures at the LHC.}}\\
{Proceedings of the XVIII DAE-BRNS symposium, Vol.~18, 2008, 140-143.}
\end{itemize}

\chapter*{\sffamily{{\bf M}otivation and plan of the thesis}}\label{motiv}

The standard model of the particle physics is extremely successful in
explaining the elementary particle interactions, as has been firmly
established by a host of experiments. However, unfortunately
there exist certain issues where the standard model is an apparent
failure, like unnatural fine tuning associated with the mass of the 
hitherto unseen Higgs boson or explaining massive neutrinos, as
confirmed by neutrino oscillation experiments. A collective approach to
address these shortcomings requires extension beyond the standard model 
framework. The weak scale supersymmetry has been a very favourite
choice to explain physics beyond the standard model where by virtue
of the construction, the mass of Higgs boson is apparently free
from fine-tuning problem. On the other hand, violation
of a discrete symmetry called $R$-parity is an intrinsically
supersymmetric way of accommodating massive neutrinos.
But, in spite of all these successes supersymmetric theories
are also not free from drawbacks and that results in a wide variety
of models. Besides, not a single supersymmetric particle has
been experimentally discovered yet. Nevertheless, possibility 
of discovering weak scale supersymmetric particles
as well as Higgs boson are highly envisaged with the initiation of
the large hadron collider experiment at CERN. 

In this thesis we plan to study a few phenomenological aspects
of a particular variant of $R$-parity violating supersymmetric model, popularly known as
the $\mu\nu$SSM. This model offers a solution for the $\mu$-problem
of the minimal supersymmetric standard model and simultaneously accommodate massive 
neutrinos with the use of a common set of right-handed neutrino superfields.
Initially, we aimed to accommodate massive neutrinos in this model 
consistent with the three flavour global neutrino data with tree level
analysis for different schemes of light neutrino masses. Besides, 
as the lightest supersymmetric particle is unstable
due to $R$-parity violation, we also tried to explore the possible correlations
between light neutrino mixing angles with the branching ratios of the 
decay modes of the lightest supersymmetric particle 
(which is usually the lightest neutralino for an appreciable region of the parameter space)
as a possible check of this model 
in a collider experiment. Later on we looked forward to re-investigate
the tree level analysis with the inclusion of one-loop radiative corrections.
We were also keen to study the sensitivity of our one-loop corrected results to the 
light neutrino mass hierarchy. Finally, we proposed an unconventional background free 
signal for Higgs boson in $\mu\nu$SSM which can concurrently act as a probe to the 
seesaw scale. A signal of this kind not only can lead to an early discovery,
but also act as an unique collider signature of $\mu\nu$SSM.

This thesis is organized as follows, we start with a brief introduction of the 
standard model in chapter \ref{SM}, discuss the very basics of mathematical
formulations and address the apparent successes and shortcomings. We start our
discussion in chapter \ref{susy} by studying how the quadratic
divergences in the standard model Higgs boson mass can be handled in a supersymmetric theory.
We also discuss the relevant mathematical formulations,
address the successes and drawbacks of the minimal supersymmetric standard model
with special attentions on the $\mu$-problem and the $R$-parity. A small discussion
on the next-to-minimal supersymmetric standard model has also been addressed. We
devote chapter \ref{neut} for neutrinos. The issues of neutrino mass generation
both in supersymmetric and non-supersymmetric models have been addressed for
tree level as well as for one-loop level analysis. Besides, implications of neutrino
physics in a collider analysis has been discussed. Light neutrino masses
and mixing in $\mu\nu$SSM both for tree level and one-loop level analysis are given in 
chapter \ref{munuSSM-neut}. The $\mu\nu$SSM model has been discussed more extensively
in this chapter. We present the results of correlation study between the neutrino 
mixing angles and the branching ratios of the decay modes of the lightest neutralino
in $\mu\nu$SSM in chapter \ref{munuSSM-LSP}. Our results are given for different natures of the 
lightest neutralino with different hierarchies in light neutrino masses. Finally,
in chapter \ref{munuSSM-Higgs} we present an unusual background free signal for 
Higgs boson in $\mu\nu$SSM, which can lead to early discovery. We list our conclusions in 
chapter \ref{con-sum}. Various technical details, like different mass matrices, couplings,
matrix element squares of the three-body decays of the lightest supersymmetric particle, 
Feynman diagrams etc. are relegated to the appendices.


%
\tableofcontents
%
\newpage
\pagenumbering{arabic}
\chapter{ \sffamily{{\bf T}he Standard Model and beyond...}}\label{SM}

\section{{\bf T}he Standard Model}\label{SM-intro}

The quest for explaining diverse physical phenomena with a single ``supreme'' 
theory is perhaps deeply embedded in the human mind. The journey was started
long ago with {\it{Michael Faraday}} and later with {\it{James Clerk Maxwell}}
with the unification of the electric and the magnetic forces
as the electromagnetic force. The inspiring successful past
has finally led us to the Standard Model (SM) 
(see reviews \cite{c1Abers-Lee,c1Beg-Sirlin} and 
\cite{c1Halzen-Martin,c1Cheng-Li,c1Griffiths,c1Burgess-Moore}) 
of elementary Particle Physics. In the SM three of the four fundamental 
interactions, namely electromagnetic, weak and strong interactions
are framed together. The first stride towards 
the SM was taken by {\it{Sheldon Glashow}} 
\cite{c1Glashow-1961} by unifying the theories of 
electromagnetic and weak interactions as the electroweak theory. 
Finally, with pioneering contributions from {\it{Steven Weinberg}} 
\cite{c1Weinberg-1967} and {\it{Abdus Salam}} \cite{c1Salam-1968}
and including the third fundamental interaction of nature, namely
the strong interaction the Standard Model of particle physics
emerged in its modern form. Ever since, the SM has successfully 
explained host of experimental results and precisely predicted a 
wide variety of phenomena. Over time and through many experiments by 
many physicists, the Standard Model has become established as a well-tested 
physics theory.

\begin{flushleft}
{\it{$\maltese$ The quarks and leptons}}
\end{flushleft}

The SM contains elementary particles which are the basic ingredients 
of all the matter surrounding us. These particles are divided
into two broad classes, namely, {\it{quarks}} and {\it{leptons}}.
These particles are called {\it{fermions}} since they are spin $\frac{1}{2}$
particles. Each group of quarks and leptons consists of six members, which
are ``paired up'' or appear in {\it{generations}}. The lightest 
and most stable particles make up the first generation, whereas 
the heavier and less stable particles belong to the second and 
third generations. The six quarks are paired in the three 
generations, namely the `up quark $(u)$' and the `down quark $(d)$' 
form the first generation, followed by the second generation containing the 
`charm quark $(c)$' and `strange quark $(s)$', and finally the 
`top quark $(t)$' and `bottom quark $(b)$' of the third generation. 
The leptons are similarly arranged in three generations, namely the 
`electron $(e)$' and the `electron-neutrino $(\nu_e)$', 
the `muon $(\mu)$' and the `muon-neutrino $(\nu_\mu)$', 
and the `tau $(\tau)$' and the `tau-neutrino $(\nu_\tau)$'.

\begin{flushleft}
{\it{$\maltese$ There are gauge bosons too}}
\end{flushleft}

Apart from the quarks and leptons the SM also contains
different types of {\it{spin-1}}  bosons, responsible
for mediation of the electromagnetic, weak and 
the strong interaction. These force mediators essentially 
emerge as a natural consequence
of the theoretical fabrication of the SM, which relies on the principle 
of local gauge invariance with the gauge group ${\rm SU}(3)_C \times 
{\rm SU}(2)_L \times {\rm U}(1)_Y$. The force mediator gauge bosons
are $\bf{n^2-1}$ in number for an ${\rm SU}(n)$ group and 
belong to the adjoint representation of the group.

The group ${\rm SU}(3)_C$ is
associated with the {\it{colour}} symmetry in the quark sector and
under this group one obtains the so-called colour triplets. 
Each quark $(q)$ can carry a {\it{colour charge}} under the ${\rm SU}(3)_C$
group\footnote{The colour quantum number was introduced
for quarks \cite{c1Greenberg:1964pe} to save the {\it{Fermi}} statistics. 
These are some hypothetical charges having no connection with the 
real life colour of light.} (very similar to electric 
charges under ${\rm U}(1)_{em}$ 
symmetry). Each quark carries one of the three 
fundamental colours (${\bf{3}}$ representation), namely,
red $(R)$, green $(G)$ and blue $(B)$. In a similar fashion 
an anti-quark $(\bar{q})$ has the complementary colours 
(${\bf{\bar 3}}$ representation), cyan $(\ovl R)$, 
magenta $(\ovl G)$ and yellow $(\ovl B)$. The accompanying eight 
force mediators are known as gluons $(G^a_\mu)$. 
The gluons belong to the adjoint representation of ${\rm SU}(3)_C$. 
However, all of the hadrons (bound states of quarks) are colour singlet.
Three weak bosons $(W^a_\mu)$ are the force mediators 
for ${\rm SU}(2)_L$ group, under which left-chiral quark and lepton
fields transform as doublets. The remaining gauge group ${\rm U}(1)_Y$
provides hypercharge quantum number $(Y)$ to all the SM particles
and the corresponding gauge boson is denoted by $B_\mu$. In describing
different gauge bosons the index `$\mu$' ($=1,..,4$) has been used to 
denote {\it{Lorentz}} index. The index `$a$' appears for the 
non-Abelian gauge groups\footnote{Yang and Mills \cite{c1Yang-Mills}.}
and they take values $1,..,8$ for ${\rm SU}(3)_C$
and $1,2,3$ for ${\rm SU}(2)_L$.

Different transformations for the SM fermions and gauge bosons under
the gauge group ${\rm SU}(3)_C \times {\rm SU}(2)_L \times {\rm U}(1)_Y$
are shown below\footnote{We 
choose $Q=T_3 + \frac{Y}{2}$, where $Q$ is the electric charge,
$T_3$ is the third component of the weak isospin ($\pm\frac{1}{2}$ 
for an ${\rm SU(2)}$ doublet) and $Y$ is the weak hypercharge.}

\bea
&&L_{i_L} = \left(\begin{array}{c}
\nu_{\ell_i} \\
\ell_i 
\end{array}\right)_L \sim ({\bf{1,2,-1}}),
~~\ell_{i_R} \sim ({\bf{1,1,-2}})
,\nonumber \\
&&Q_{i_L} = \left(\begin{array}{c}
u_i \\
d_i 
\end{array}\right)_L \sim ({\bf{3,2,{\frac{1}{3}}}}),~~u_{i_R} \sim 
({\bf{3,1,{\frac{4}{3}}}}),~~d_{i_R} \sim ({\bf{3,1,-{\frac{2}{3}}}}),\nonumber \\
&& G^a_\mu \sim ({\bf{8,0,0}}),~~W^a_\mu \sim ({\bf{1,3,0}}),
~~B_\mu \sim ({\bf{1,1,0}}),
\label{SM-gauge-group}
\eea
where $\ell_i=e,\mu,\tau$, $u_i=u,c,t$ and $d_i=d,s,b$. The 
singlet representation is given by ${\bf{1}}$.

\begin{flushleft}
{\it{$\maltese$ Massive particles in the SM~?}}
\end{flushleft}

Principle of gauge invariance demands for massless
gauge bosons which act as the force mediators. In addition,
all of the SM fermions (quarks and leptons) are supposed 
to be exactly massless, as a consequence of the gauge invariance.
But these are in clear contradiction to observational facts. In
reality one encounters with massive fermions. Also, 
the short range nature of the weak interaction indicates towards 
some massive mediators.  This apparent contradiction
between gauge invariance and massive gauge boson 
was resolved by the celebrated method
of {\it{spontaneous breaking of gauge symmetry
}} \cite{c1Anderson-1958,c1Englert-1964,c1Higgs1-1964,c1Higgs-1964,
c1Guralnik-1964}. The initial SM gauge group 
after spontaneous symmetry breaking (SSB) reduces to ${\rm SU}(3)_C  
\times {\rm U}(1)_{em}$, leaving the colour and electric charges
to be conserved in nature. Consequently, the corresponding gauge
bosons, gluons and photon, respectively remain massless ensuing gauge invariance,
whereas the weak force mediators ($W^{\pm}$ and $Z$ bosons) become massive.
Symbolically,
\bea
{\rm SU}(3)_C \times {\rm SU}(2)_L \times {\rm U}(1)_Y
~~~\underrightarrow{SSB}~~~
{\rm SU}(3)_C \times {\rm U}(1)_{em}.
\label{SSB}
\eea
Since ${\rm SU}(3)_C$ is unbroken in nature, all the particles
existing freely in nature are forced to be colour neutral. In a similar
fashion unbroken ${\rm U}(1)_{em}$ implies that any charged particles
having free existence in nature must have their charges as integral
multiple of that of a electron or its antiparticle. It is interesting
to note that quarks have fractional charges but they are not free 
in nature since ${\rm SU}(3)_C$ is unbroken.

\begin{list}{}{}
\item
\begin{flushleft}
{{$\blacklozenge$ Spontaneous symmetry breaking}}
\end{flushleft} 
Let us consider a Hamiltonian $H_0$ which is invariant under some 
symmetry transformation. If this symmetry of $H_0$ is
not realized by the particle spectrum, the symmetry is
spontaneously broken. A more illustrative example
is shown in figure \ref{SSB-pic}. Here the minima
of the potential lie on a circle (white dashed)
rather than being a specific point. Each of these points are
equally eligible for being the minimum and whenever the 
red ball chooses a specific minimum, the symmetry of the 
ground state (the state of minimum energy) is spontaneously broken. 
In other words,
when the symmetry of $H_0$ is not respected by the ground
state, the symmetry is spontaneously broken. It turns
out that the degeneracy in the ground state is essential
for spontaneous symmetry breaking.
\begin{figure}[ht]
\centering
\includegraphics[width=5.45cm,keepaspectratio]{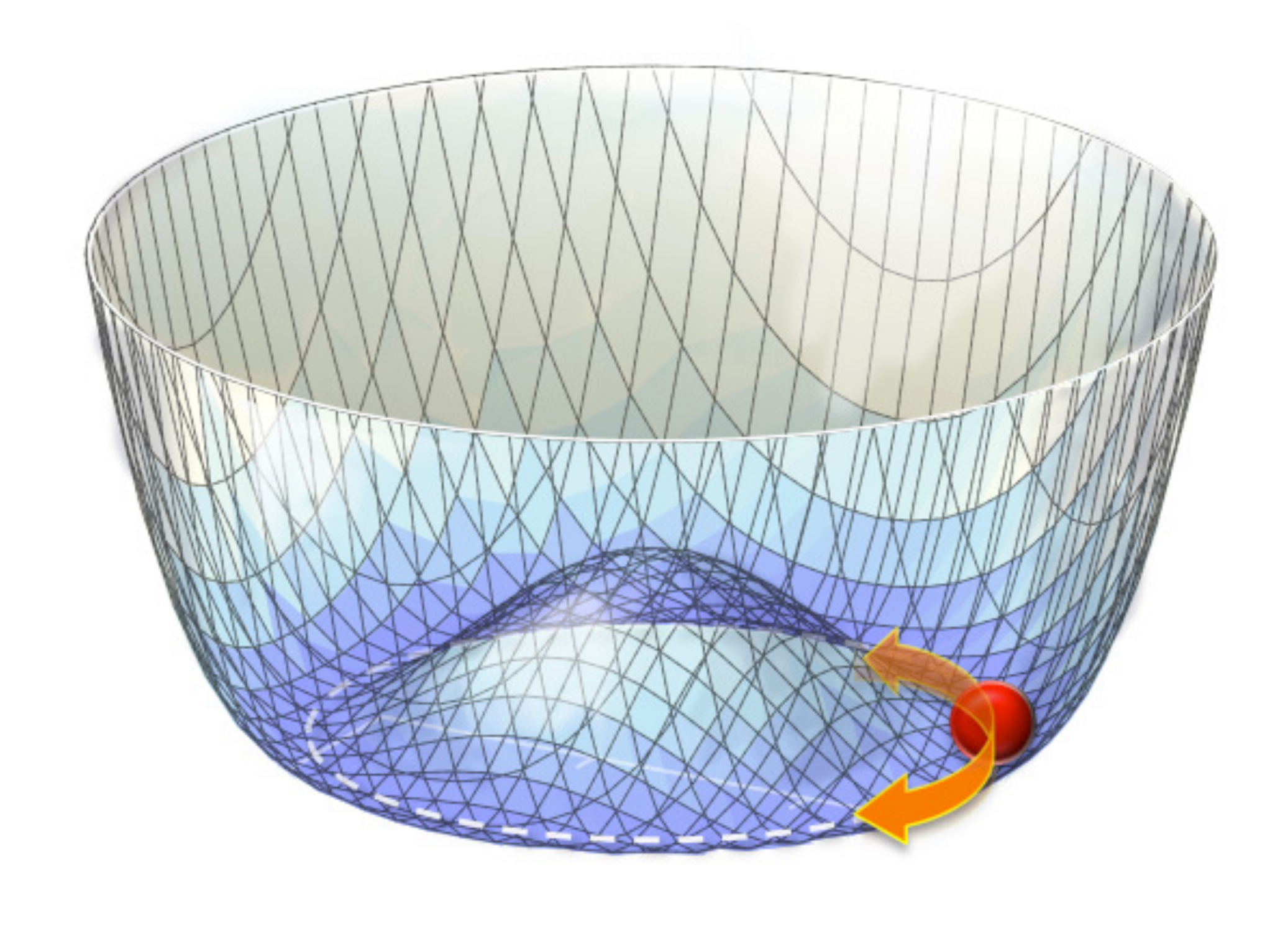}
\caption{Spontaneous breaking of symmetry through
the choice of a specific degenerate ground state.}
\label{SSB-pic}
\end{figure}
\end{list}

Everything seems to work fine with the massive gauge bosons.
But the demon lies within the method of spontaneous symmetry
breaking itself. The {\it{spontaneous breakdown}} of a {\it{continuous symmetry}}
implies the existence of {\it{massless, spinless particles}} as
suggested by {\it{Goldstone}} theorem.\footnote{Initially by Nambu 
\cite{c1Nambu-1960}, Nambu and Jona-Lasino. \cite{c1Nambu1-1961,c1Nambu2-1961}.
General proof by Goldstone \cite{c1Goldstone1-1961,c1Goldstone2-1962}.}
They are known as {\it{Nambu-Goldstone}} or simply {\it{Goldstone}}
bosons. So the SSB apart from generating gauge boson masses also
produces massless scalars which are not yet experimentally detected.
This is the crisis point when the celebrated 
``Higgs-mechanism''\footnote{The actual name
should read as Brout-Englert-Higgs-Guralnik-Hagen-Kibble mechanism 
after all the contributors. Brout and Englert \cite{c1Englert-1964}, 
Higgs \cite{c1Higgs1-1964,c1Higgs-1964}, Guralnik, Hagen and Kibble \cite{c1Guralnik-1964}.} 
resolves the crisis situation. 
The unwanted
massless scalars are now eaten up by the gauge boson fields
and they turn out to be the badly needed  longitudinal polarization
mode for the ``massive'' gauge bosons. 
So this is essentially the reappearance of three degrees of freedom
associated with three massless scalars in the form of three 
longitudinal polarization modes for the massive gauge bosons.
This entire mechanism happens
without breaking the gauge invariance of the theory explicitly.
This mechanism for generating gauge boson masses is also consistent
with the renormalizability of a theory with massive gauge 
bosons.\footnote{Veltman and 't Hooft, 
\cite{c1'tHooft-1971,c1'tHooft-1972}.} The fermion masses also
emerge as a consequence of Higgs mechanism.

\begin{flushleft}
{\it{$\maltese$ Higgs sector of the SM and mass generation}}
\end{flushleft}

So the only scalar (spin-$0$) in the SM is the Higgs boson.
Higgs mechanism is incorporated in the SM through a complex scalar doublet 
$\Phi$ with the following transformation properties under the SM
gauge group.

\bea
\Phi = \left(\begin{array}{c}
\phi^+ \\
\phi^0
\end{array}\right) \sim ({\bf{1,2,1}}).
\label{SM-Higgs}
\eea

The potential for $\Phi$ is written as
\bea
V(\Phi) =  \mu^2 \Phi^{\dagger} \Phi + \lambda (\Phi^{\dagger} \Phi)^2,
\label{SM-Higgs-pot}
\eea
with $\mu^2 < 0$ and $\lambda > 0$ (so that the potential is bounded
from below). Only a colour and charge (electric) neutral component 
can acquire a vacuum expectation value (VEV), since 
even after SSB the theory remains invariant under 
${\rm SU(3)}_C \times {\rm U(1)}_{em}$
(see eqn.(\ref{SSB})). Now with a suitable choice of gauge
(``unitary gauge''), so that the Goldstone bosons disappear
one ends up with 
\bea
\Phi = \frac{1}{\sqrt{2}}\left(\begin{array}{c}
0 \\
v + h^0
\end{array}\right),
\label{SM-Higgs-VEV}
\eea
where $h^0$ is the physical Higgs field and `$v$' is the 
VEV for $\mathcal{R}e(\phi^0)$ (all other fields
acquire zero VEVs) with 
$v^2 = -\frac{\mu^2}{\lambda}$. At this moment it is apparent 
that eqn.(\ref{SSB}) can be recasted as
\bea
{\rm SU}(2)_L \times {\rm U}(1)_Y
~~~\underrightarrow{SSB}~~~
{\rm U}(1)_{em},
\label{EWSB}
\eea
which is essentially the breaking of the electroweak symmetry 
since the ${\rm{SU(3)}}_C$ sector remains unaffected.
Thus the phenomena of SSB in the context of the SM is identical
with the electroweak symmetry breaking (EWSB). The weak bosons, 
$W^a_\mu$ and $\rm U(1)_Y$ gauge boson $B_\mu$ now mix together and
finally yield three massive vector bosons $(W^\pm_\mu, Z^0_\mu)$
and one massless photon $(A^0_\mu)$:
\bea
&&W^\pm_\mu = \frac{W^1_\mu \mp i W^2_\mu}{\sqrt{2}},\nonumber\\
&&Z^0_\mu = \rm{cos}\theta_W W^3_\mu - \rm{sin}\theta_W B_\mu,\nonumber\\
&&A^0_\mu = \rm{sin}\theta_W W^3_\mu + \rm{cos}\theta_W B_\mu,
\label{WZboson-Photon}
\eea
where $\theta_W$ is the {\it{Weinberg}} angle or weak mixing angle.\footnote{
At present $\rm{sin}^2\theta_W = 0.231$ (evaluated at $M_Z$ with
renormalization scheme $\ovl{MS}$) \cite{c1Nakamura-c2}.} In 
terms of the $\rm SU(2)_L$ and $\rm U(1)_Y$ gauge couplings $(g_2, g_1)$ 
one can write
\bea
g_2 ~{\rm{sin}}\theta_W = g_1 ~{\rm{cos}}\theta_W.
\label{Weak-angle}
\eea
The $W^\pm_\mu, Z^0_\mu$ boson masses are given by
\bea
M_{W} = \frac{g_2 v}{2}, 
~~~~M_{Z} = \frac{v}{2} \sqrt{g^2_1 + g^2_2},
\label{WZ-mass}
\eea
with $v^2 = -\frac{\mu^2}{\lambda}$. The mass of physical
Higgs boson $(h^0)$ is given by $m^2_{h^0} = 2 v^2 \lam$.
Note that $m_{h^0} > 0$ since $\mu^2 < 0$. 
Interestingly, ratio of the quantities $M^2_W$ and
$M^2_Z \cos^2\theta_W$ is equal to one at the tree level
(see eqns. (\ref{Weak-angle}) and (\ref{WZ-mass})). This
ratio is defined as the $\rho$-parameter, which is an
important parameter for electroweak precision test:
\bea
\rho = \frac{M^2_W}{M^2_Z {\rm cos}^2\theta_W} = 1.
\label{rho-param}
\eea
There exists an alternative realization of the $\rho$-parameter. The
$\rho$-parameter specifies the relative strength of the neutral current
(mediated through $Z$-bosons) to the charged current (mediated through $W^\pm$-boson)
weak interactions.

For the purpose of fermion mass generation consider the 
Lagrangian containing interactions between Higgs field
and matter fermions.

\bea
-\mathcal{L}_{\text{Yukawa}} = y_{\ell_i} \ovl{L_i} \Phi e_i
+ y_{d_i} \ovl{Q_i} \Phi d_i +  y_{u_i} \ovl{Q_i} \tilde{\Phi} u_i
+ \text{Hermitian conjugate},
\label{fermion-mass}
\eea
where $y_{\ell_i,u_i,d_i}$ are the {\it{Yukawa}} couplings for the 
charged leptons, up-type quarks and down-type quarks, 
respectively. The $\rm SU(2)_L$ doublet and singlet
quark and lepton fields are shown in eqn.(\ref{SM-gauge-group}).
The field $\tilde{\Phi}$ is used to generate masses for 
the up-type quarks and it is given by
\bea
\tilde\Phi = - i \sigma_2 \Phi^* = 
i \left(\begin{array}{cc}
0 & -i \\
i & 0
\end{array}\right)\left(\begin{array}{c}
\phi^- \\
{\phi^0}^*
\end{array}\right) = 
\left(\begin{array}{c}
-{\phi^0}^* \\
\phi^-
\end{array}\right).
\label{conju-Higgs}
\eea
The fermion masses and their interactions with Higgs field
emerge after the EWSB using eqn.(\ref{fermion-mass}). For example
considering the {\it{electron}} these terms are as follows
\bea
\mathcal{L}^{\text{electron}}_{\text{Yukawa}} = - \frac{y_e (v + h^0)}{\sqrt{2}}
(\ovl {e_L} e_R + \ovl {e_R} e_L),
\label{fermion-mass-Higgs}
\eea
where $e_{L} = P_L L_e $ (see eqn.(\ref{SM-gauge-group})).

So with four component spinor $e$ as 
$\left(\begin{array}{c}
e_L \\
e_R
\end{array}\right)$, eqn.(\ref{fermion-mass-Higgs})
can be rewritten as
\bea
\mathcal{L}^{\text{electron}}_{\text{Yukawa}} = - m_e \ovl{e} e
- \frac{m_e}{v} \ovl{e} e h^0,
\label{fermion-mass-Higgs-2}
\eea
with $m_e = \frac{Y_e v}{\sqrt{2}}$ as mass of the 
electron. The particle spectrum of the SM can be 
written in a tabular form as shown in table \ref{SM-spectrum}.
\begin{table}[t]
\footnotesize
\centering
\begin{tabular}{ c  c  c  c  c}
\hline \hline 
Particle  & mass in GeV & Spin & Electric Charge & Colour charge \\ \hline \hline
electron $(e)$ & 5.109$\times10^{-4}$ & $\frac{1}{2}$ & -1 & 0 \\
\vspace*{0.1cm}
muon $(\mu)$ & 0.105 & $\frac{1}{2}$ & -1 & 0 \\ 
\vspace*{0.1cm}
tau $(\tau)$ & 1.776 & $\frac{1}{2}$ & -1 & 0 \\
\vspace*{0.1cm}
neutrinos $(\nu_{e,\mu,\tau})$ & 0 & $\frac{1}{2}$ & 0 & 0   \\ 
\vspace*{0.1cm}
up-quark $(u)$ & 2.49$\times10^{-3}$ & $\frac{1}{2}$ & $\frac{2}{3}$ & yes   \\ 
\vspace*{0.1cm}
down-quark $(d)$ & 5.05$\times10^{-3}$ & $\frac{1}{2}$ & $-\frac{1}{3}$ & yes  \\ 
\vspace*{0.1cm}
charm-quark $(c)$ & 1.27 & $\frac{1}{2}$ & $\frac{2}{3}$ &  yes  \\ 
\vspace*{0.1cm}
strange-quark $(s)$ & 0.101 & $\frac{1}{2}$ & $-\frac{1}{3}$ &  yes  \\ 
\vspace*{0.1cm}
top-quark $(t)$ & 172.0 & $\frac{1}{2}$ & $\frac{2}{3}$ & yes   \\ 
\vspace*{0.1cm}
bottom-quark $(b)$ & 4.19 & $\frac{1}{2}$ & $-\frac{1}{3}$ &  yes  \\ 
\vspace*{0.1cm}
W-boson $(W^\pm)$ & 80.399 & 1 & $\pm 1$ & 0  \\ 
\vspace*{0.1cm}
Z-boson $(Z^0)$ & 91.187 & 1 & 0 & 0   \\ 
\vspace*{0.1cm}
photon $(\gamma)$ & 0 & 1 & 0 & 0  \\ 
\vspace*{0.1cm}
gluon $(g)$ & 0 & 1 & 0 & yes \\
\vspace*{0.1cm}
Higgs $(h^0)$ & {\bf{?}} & 0 & 0 & 0 \\ \hline \hline
\end{tabular}
\caption{\label{SM-spectrum}
The particle spectrum of the SM \cite{c1Nakamura-c2}. 
Each of the charged particles are accompanied 
by charge conjugate states of equal mass. 
The charge neutral particles act as their own 
antiparticles with all charge like quantum numbers
as opposite to that of the corresponding particles.
Evidence for Higgs boson is yet experimentally 
missing and thus Higgs mass is denoted as `$?$'.
The neutrinos are presented with zero masses since
we are considering the SM only (see section \ref{suc-prob}).  
}
\end{table}
\begin{flushleft}
{\it{$\maltese$ SM interactions}}
\end{flushleft}
Based on the discussion above, the complete Lagrangian for the SM 
can be written as
\bea
\mathcal{L}_{SM} = \mathcal{L}_{1} + \mathcal{L}_{2}
+ \mathcal{L}_{3} + \mathcal{L}_{4},
\label{SM-Lagrangian}
\eea
where 
\begin{enumerate}
 \item 
$\mathcal{L}_{1}$ is the part of the Lagrangian which
contains kinetic energy terms and self-interaction terms
for the gauge bosons. After the EWSB these gauge bosons are 
known as $W^\pm, Z^0$, gluons and photon. So we have
\bea
\mathcal{L}_{1} = \sum_{\mu,\nu=1}^{4} \left[
-\frac{1}{4} \sum_{a=1}^{8} G^a_{\mu\nu} G^{\mu\nu}_a
- \frac{1}{4} \sum_{i=1}^{3} W^i_{\mu\nu} W^{\mu\nu}_i
- \frac{1}{4} B_{\mu\nu} B^{\mu\nu}\right],
\label{Lagrangian-gauge}
\eea
where
\bea
&&G^a_{\mu\nu} = \partial_\mu G^a_\nu - \partial_\nu G^a_\mu
-g_3 f_{abc} G^b_\mu G^c_\nu,\nonumber\\
&&W^i_{\mu\nu} = \partial_\mu W^i_\nu - \partial_\nu W^i_\mu
-g_2 \epsilon_{ijk} W^j_\mu W^k_\nu,\nonumber\\
&&B_{\mu\nu} = \partial_\mu B_\nu - \partial_\nu B_\mu,
\label{non-Abelian-gauge}
\eea
with $f_{abc}$ and $\epsilon_{ijk}$ as the structure constants of the 
respective non-Abelian groups. $g_3$ is the coupling constant for 
$\rm {SU(3)_C}$ group.

\item
Kinetic energy terms for quarks and leptons belong to 
$\mathcal{L}_2$. This part of the Lagrangian also contains
the interaction terms between the elementary fermions and gauge
bosons. Symbolically,
\bea
\mathcal{L}_{2} = i \ovl{\chi_L} \D \chi_L + 
i \ovl{\chi_R} \D \chi_R,
\label{Lagrangian-fermion}
\eea
where $\D = \gamma^\mu D_\mu$ with $D_\mu$ as the covariant
derivative.\footnote{Replacement of ordinary derivative $(\partial_\mu)$
by $D_\mu$ is essential for a gauge transformation, so that $D_\mu \psi$
transforms covariantly under gauge transformation, similar to
the matter field, $\psi$.} The quantity $\chi_L$ stands for lepton and
quark $\rm{SU(2)_L}$ doublets whereas $\chi_R$ denotes $\rm{SU(2)_L}$ 
singlet fields (see eqn.(\ref{SM-gauge-group})). The 
covariant derivative $D_\mu$ for different fermion
fields are written as (using eqn.(\ref{SM-gauge-group}))
\bea
D_\mu Q_i &=& \left[\partial_\mu + i g_1 \frac{1}{6} B_\mu
+ i \sum_{i=1}^{3} g_2 \frac{1}{2} \sigma_i.W^i_\mu \right] Q_i,\nonumber\\
D_\mu u_i &=& \left[\partial_\mu + i g_1 \frac{2}{3} B_\mu \right] u_i,\nonumber\\
D_\mu d_i &=& \left[\partial_\mu - i g_1 \frac{1}{3} B_\mu \right] d_i,\nonumber\\
D_\mu L_i &=& \left[\partial_\mu - i g_1 \frac{1}{2} B_\mu
+ i \sum_{i=1}^{3} g_2 \frac{1}{2} \sigma_i.W^i_\mu \right] L_i,\nonumber\\
D_\mu e_i &=& \left[\partial_\mu - i g_1  B_\mu \right] e_i.\nonumber\\
\label{Dmu-fermions}
\eea
But these are the information for $\rm{SU(2)_L}\times\rm{U(1)_Y}$
only. What happens to the $\rm{SU(3)_C}$ part$?$ Obviously, for the 
leptons there will be no problem since they are $\rm{SU(3)_C}$ singlet 
after all (see eqn.(\ref{SM-gauge-group})). For the quarks the 
$\rm{SU(3)_C}$ part can be taken care of in the following fashion, 
\bea
D_\mu 
\left(\begin{array}{c}
{q_i}_R \\
{q_i}_G \\
{q_i}_B
\end{array}\right) 
&=& \left[\partial_\mu  
+ i \sum_{a=1}^{8} g_3 \frac{1}{2} \lam_a.G^a_\mu \right] 
\left(\begin{array}{c}
{q_i}_R \\
{q_i}_G \\
{q_i}_B
\end{array}\right),
\label{Dmu-QCD-quarks}
\eea
where $R,G$ and $B$ are the three types of colour charge and 
$\lam_a$'s are eight Gell-Mann matrices. $q_i$ is
triplet under $\rm{SU(3)_C}$, where `$i$' stands for different
types of left handed or right handed 
(under ${\rm{SU(2)_L}}$) quark flavours, namely 
$u,d,c,s,t$ and $b$. 

\item
The terms representing physical Higgs mass and Higgs self-interactions
along with interaction terms between Higgs and the gauge bosons
are inhoused in $\mathcal{L}_3$
\bea
\mathcal{L}_{3} = (D^\mu \Phi)^{\dagger} (D_\mu \Phi)
- V(\Phi).
\label{Lagrangian-Higgs}
\eea
The expressions for $\Phi$ and $V(\Phi)$ are given in
eqns.(\ref{SM-Higgs}) and (\ref{SM-Higgs-pot}), respectively.
For $\Phi$ the covariant derivative $D_\mu$ is given by
\bea
D_\mu \Phi &=& \left[\partial_\mu + i g_1 \frac{1}{2} B_\mu
+ i \sum_{i=1}^{3} g_2 \frac{1}{2} \sigma_i.W^i_\mu \right] \Phi.
\label{Dmu-Higgs}
\eea

\item
The remaining Lagrangian $\mathcal{L}_4$ contains lepton
and quark mass terms and their interaction terms with Higgs
field $(h^0)$ (after EWSB). The expression for $\mathcal{L}_4$
is shown in eqn.(\ref{fermion-mass}). The elementary fermions get
their masses through respective Yukawa couplings, which are 
{\it{free parameters}} of the theory. It turns out that in the SM the flavour
states are not necessarily the mass eigenstates, and it is
possible to relate them through an {\it{unitary transformation}}.
In case of the quarks this matrix is known as the 
CKM (Cabibbo-Kobayashi-Maskawa) \cite{c1Cabibbo-1963,c1Kobayashi-1973} 
matrix. This $3\times3$
unitary matrix contains three mixing angles and one phase.
The massless neutrinos in the SM make the corresponding
leptonic mixing matrix a trivial one (Identity matrix).
All possible interactions of the SM are shown in figure \ref{SM-int}. 
The loops represent self-interactions like $h^0h^0h^0$, $h^0h^0h^0h^0$
(from the choice of potential, see eqn. (\ref{SM-Higgs-pot}))
$W^\pm W^\pm W^\mp W^\mp$, $ggg$ or $gggg$ (due to non-Abelian interactions)
and also interactions like $W^\pm W^\mp Z Z$, $W^\pm W^\mp \gamma \gamma$,
$W^\pm W^\mp Z$, $W^\pm W^\mp \gamma$ etc. 
\begin{figure}[ht]
\centering
\includegraphics[width=8.95cm,keepaspectratio]{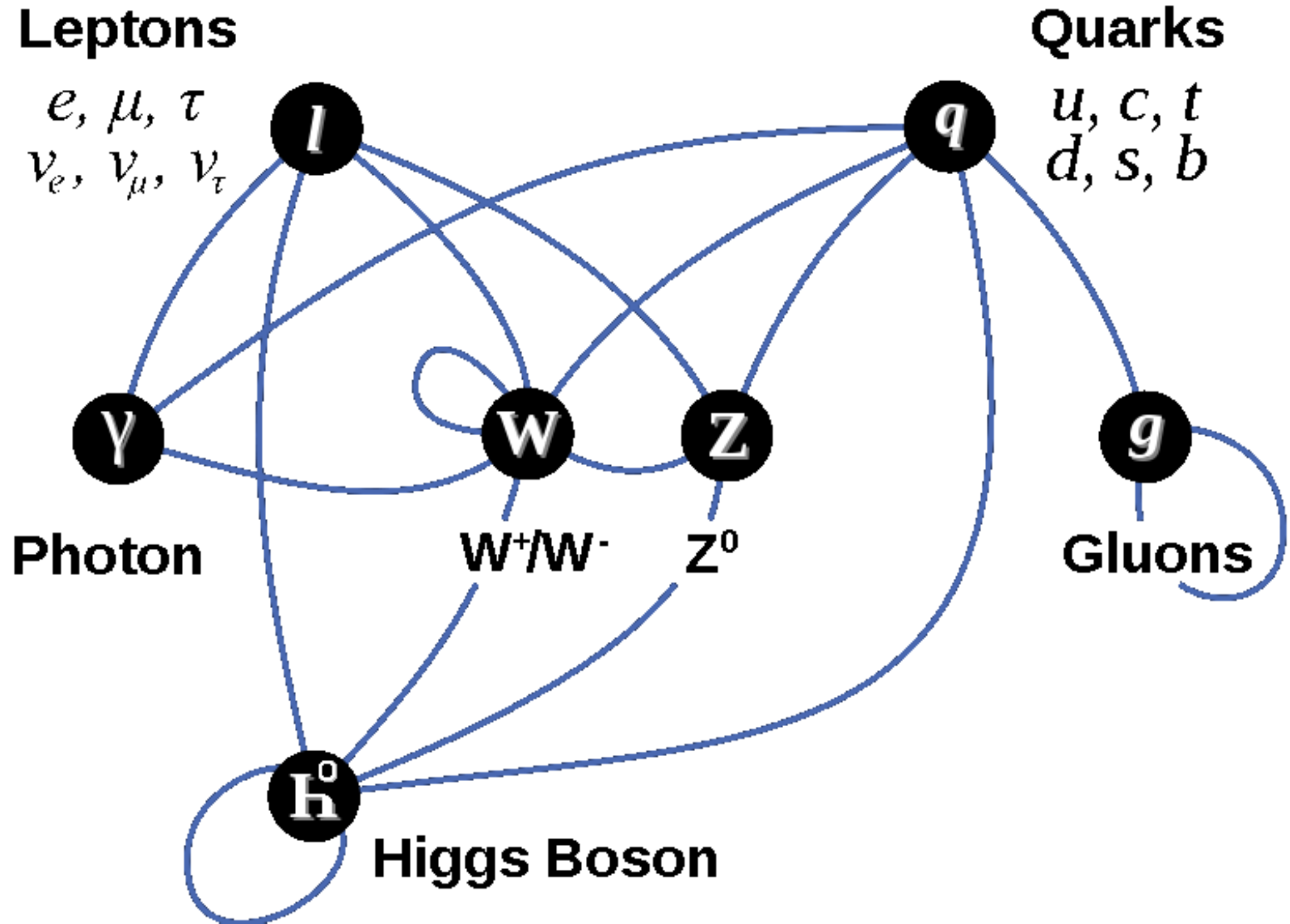}
\caption{Interactions of the Standard Model. See text for more details.}
\label{SM-int}
\end{figure}
\end{enumerate}

\section{{\bf A}pparent successes and the dark sides}\label{suc-prob}

The SM is an extremely successful theory to explain a host of
elementary particle interactions. Masses of the $W^\pm$ and $Z$
bosons as predicted by the SM theory are very close to their
experimentally measured values. The SM also predicted the existence
of the charm quark from the requirement to suppress flavour 
changing neutral current (FCNC)\footnote{Glashow, 
Iliopoulos and Maiani \cite{c1Glashow-1970}.} before it was actually
discovered in 1974. In a similar fashion the SM also predicted
the mass of the heavy top quark in the right region
before its discovery. Besides, all of the SM particles except Higgs boson
have been discovered already and their masses are also measured very precisely 
\cite{c1Nakamura-c2}. Indeed, apart from Higgs sector, rest of the SM has
been already analysed for higher order processes and their spectacular
accuracy as revealed by a host of experiments has firmly established the 
success of the SM.

Unfortunately, the so-called glorious success of the SM suffers serious 
threat from various theoretical and experimental perspective. One of
the main stumbling blocks is definitely the Higgs boson, yet to
be observed in an experiment and its mass. Some of these shortcomings 
are listed below.

\begin{enumerate}
 \item 
The SM has a large number of free parameters ({\bf{19}}).
The parameters are {\bf{9}} Yukawa couplings (or elementary 
fermion masses) + {\bf{3}} angles and {\bf{one}} phase
of CKM matrix + {\bf {3}} gauge couplings $g_1,g_2,g_3$\footnote{An 
alternate set could be $g_3$, $e$ (the unit of electric charge) and 
the Weinberg angle $\theta_W$.} + {\bf {2}} parameters
$(\mu, \lam)$ from scalar potential (see eqn.(\ref{SM-Higgs-pot}))
+ {\bf{one}} vacuum angle for quantum chromodynamics (QCD).
The number of free parameters is rather large for a fundamental theory.

\item
There are no theoretical explanation why there exist only
three generations of quarks and leptons. Also the huge mass
hierarchy between different generations (from first to third), 
that is to say why mass of the top quark $(m_t) \gg$ mass of
the up-quark $(m_u)$ (see table \ref{SM-spectrum}), is unexplained.

\item
The single phase of CKM matrix accounts for
many Charge-Parity (CP) violating processes. However, one needs 
additional source of CP-violation to account for 
the large matter-anti matter asymmetry of the universe.

 \item
The most familiar force in our everyday lives, gravity, is not a part of 
the SM. Since the effect of gravity dominates near the ``{\it{Planck Scale $(M_P)$}}'',
$(\sim 10^{19}$ GeV) the SM still works fine despite its reluctant exclusion of the 
gravitational interaction. In conclusion, the Standard Model cannot be a theory which is 
valid for all energy scales.

\item
There is no room for a cold Dark Matter candidate inside
the SM framework, which has been firmly established by now from the
observed astrophysical and cosmological evidences.

\item 
Neutrinos are exactly massless in the Standard Model as a 
consequence of the particle content, gauge invariance, renormalizability
and Lorentz invariance. However, the 
experimental results from atmospheric, solar and reactor neutrino experiments 
suggest that the neutrinos do have non-zero masses with non-trivial mixing 
among different neutrino flavours \cite{c1Schwetz:2008er,c1GonzalezGarcia:2010er}. 
In order to generate masses and mixing for 
the neutrinos, one must extend the SM framework by introducing additional 
symmetries or particles or both.

But in reality the consequence of a massive neutrino is far serious
than asking for an extension of the SM. As written earlier, the 
massive neutrinos trigger a non-trivial mixing in the charged lepton
sector just like the CKM matrix\footnote{Known as the PMNS matrix, will
be addressed in chapter \ref{neut} in more details.}, but with large
off-diagonal entries. It remains to explain why the structure of the 
mixing matrix for the leptons are so different from the quarks?

\item 
Perhaps the severe most of all the drawbacks is associated
with Higgs boson mass. In the Standard Model, 
Higgs boson mass is totally unprotected by any 
symmetry argument. In other words putting $m_{h^0}=0$, does not
enhance any symmetry of the theory.\footnote{Note 
that putting zero for fermion or
gauge boson mass however enhances the symmetry of the Lagrangian. 
In this case the chiral and gauge symmetry, respectively.}
Higgs mass can be as large as the ``Grand Unified Theory (GUT)" 
scale ($10^{16} ~\rm{GeV}$) or the ``{\it{Planck Scale}}" ($10^{19} ~\rm{GeV}$)
when radiative corrections are included. This is the so called 
{\it{gauge hierarchy problem}}. However, from several theoretical 
arguments \cite{c1Cornwall-1973,c1Cornwall-1974,c1Smith-1973,c1Lee-1977,
c1Dashen:1983ts,c1Lindner:1985uk,c1Stevenson:1985zy,c1Hasenfratz:1987tk,
c1Luscher:1987ay,c1Hasenfratz:1987eh,c1Luscher:1987ek,c1Kuti:1987nr,c1Sher:1988mj,
c1Lindner:1988ww,c1Callaway:1988ya,c1Ford:1992mv,c1Sher:1993mf,
c1Choudhury:1997bn} and various experimental searches 
\cite{c1Abbiendi:1998rd,c1Barate:2003sz,c1Nakamura-c2} Higgs boson mass is 
expected to be in the  range of a few hundreds of GeV, 
which requires unnatural fine tuning of parameters 
($\sim$ one part in $10^{38}$) for all orders in perturbation 
theory. Different one-loop
diagrams contributing to the radiative correction to 
Higgs boson mass are shown in figure \ref{Higgs-loop}.

\begin{figure}[ht]
\centering
\includegraphics[width=10.95cm]{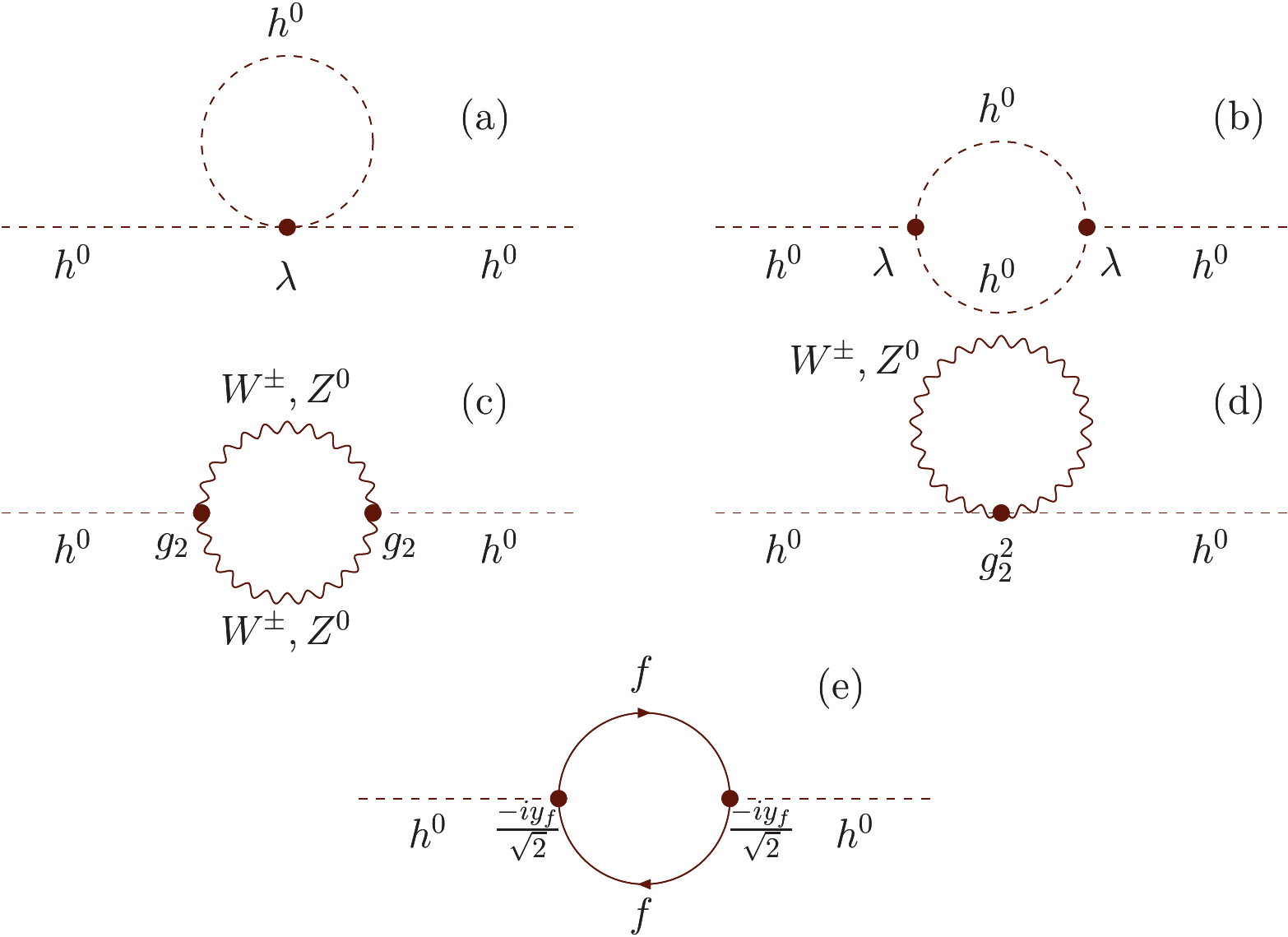}
\caption{One-loop radiative corrections to the Higgs boson mass from (a) and (b) 
self-interactions (c) and (d) interactions with gauge bosons
and (e) interactions with fermions (f).}
\label{Higgs-loop}
\end{figure}
It is clear from figure \ref{Higgs-loop}, the contribution from the 
fermion loop is proportional to the  squared Yukawa couplings $(y^2_f)$. 
As a corollary these contributions are negligible
except when heavy quarks are running in the loop. Contributions from the
diagrams (b) and (c) are {\it{logarithmically divergent}} which
is well under control due to the behaviour of {\it{log}} function. The contributions
from diagrams (a), (d) and (e) are {\it{quadratically divergent}},
which are the sources of the hierarchy problem.
\begin{list}{}{}
\item
\begin{flushleft}
{{$\blacklozenge$ Loop correction and divergences}}
\end{flushleft} 
Consider the diagram (e) of figure \ref{Higgs-loop}, which
represents the fermionic loop contribution to the scalar
two point function. Assuming the loop momentum to be `$k$' 
and the momentum for the external leg to be `$p$'
this contribution can be written as

\bea
\Pi^f_{h^0h^0}(p^2=0) &=& (-1) \int \frac{d^4k}{(2\pi)^4}
(\frac{-i y_f}{\sqrt{2}})^2{\rm{Tr}}\left[ \frac{i}{\k-m_f} 
 \frac{i}{\k-m_f} \right],\nonumber\\
&=& (-\frac{y^2_f}{2}) \int \frac{d^4k}{(2\pi)^4}
{\rm{Tr}}\left[\frac{(\k+m_f)(\k+m_f)}{(k^2-m^2_f)^2}\right],\nonumber\\
&=& (-2 y^2_f) \int \frac{d^4k}{(2\pi)^4}
\left[\frac{(k^2 + m^2_f)}{(k^2-m^2_f)^2}\right],\nonumber\\
&=& {-2 y^2_f} \int \frac{d^4k}{(2\pi)^4}
\left[\frac{1}{(k^2-m^2_f)}
+ \frac{2 m^2_f}{(k^2-m^2_f)^2} \right],\nonumber\\
\label{Higgs-fermion-loop}
\eea
where the $(-1)$ factor appears for closed fermion loop and 
`$i$' comes from the {\it{Feynman}} rules (see eqn.(\ref{fermion-mass})).
Fermion propagator is written as $i/(\k-m_f)$. Here some of 
the properties of Dirac Gamma matrices have been used. 

Now in eqn.(\ref{Higgs-fermion-loop}) Higgs mass appears
nowhere which justifies the fact that setting $m_{h^0}=0$ does
not increase any symmetry of the Lagrangian. From naive
power counting argument the second term of 
eqn.(\ref{Higgs-fermion-loop}) is logarithmically divergent 
whereas the first term is quadratically divergent. Suppose the 
theory of the SM is valid upto Planck scale and the cut off
scale $\Lambda$ (scale upto which a certain theory is valid) 
lies there, then the correction to the 
Higgs boson mass goes as (using eqn.(\ref{Higgs-fermion-loop})),
\bea
\delta m^2_{h^0} \approx - \frac{y^2_f}{8 \pi^2} \Lambda^2 + 
{\text{logarithmic terms}}.
\label{Higgs-fermion-approx}
\eea
The renormalized Higgs mass squared is then given by
\bea
\tilde{m}^2_{h^0} = m^2_{h^0,{\text{bare}}} + \delta m^2_{h^0},
\label{Higgs-masssq-renorm}
\eea
and looking at eqn.(\ref{Higgs-fermion-approx}) the requirement of
fine tuning for a TeV scale Higgs mass is apparent.
Note that mass generation for all of the SM particles
solely depend on Higgs. So in a sense the entire mass
spectrum of the SM will be driven towards a high scale
with the radiative correction in Higgs boson mass. 

\end{list}
\end{enumerate}

The list of drawbacks keep on increasing with 
issues like unification of gauge couplings at a high scale and 
a few more. To summarize, all of these unanswered questions 
have opened up an entire new area of physics, popularly known as 
``Beyond the Standard Model (BSM)'' physics. Some of the well-known candidates are 
{\it{supersymmetry}}\footnote{First proposed in the context of hadronic physics, 
by Hironari Miyazawa (1966).} \cite{c1Gervais:1971ji,c1Golfand-1971,c1Volkov-1973,
c1Wess:1974tw,c1Wess:1973kz}, {\it{theories with extra spatial dimensions}} 
\cite{c1ArkaniHamed1-1998,c1ArkaniHamed2-1998,c1Randall-1999} and many others. 
In this proposed thesis we plan to study some of the problems mentioned 
earlier in the context of a supersymmetric theory and look for signatures 
of such a theory at the ongoing Large Hadron Collider (LHC) experiment.



\chapter{ \sffamily{{\bf S}upersymmetry
 }}\label{susy}

\section{{\bf W}aking up to the idea}\label{intro}
The effect of radiative correction drives the 
``natural'' Higgs mass, and therefore the entire SM particle spectra
to some ultimate cutoff of the theory, namely, the Planck scale. 
A solution to this hierarchy problem could
be that, either the Higgs boson is some sort of composite
particle rather than being a fundamental particle
or the SM is an effective theory valid upto a certain energy
scale so that the cutoff scale to the theory lies far below the Planck scale.
It is also a viable alternative that there exists no Higgs boson at
all and we need some alternative mechanism
to generate masses for the SM particles\footnote{These issues are well 
studied in the literature and beyond the theme of this thesis.}.
However, it is also possible that even in the presence of
quadratic divergences the Higgs boson mass can be in the 
range of a few hundreds of GeV to a TeV provided different sources of 
radiative corrections cancel the quadratic divergent pieces.
It is indeed possible to cancel the total one-loop
quadratic divergences (shown in chapter \ref{SM}, section \ref{suc-prob}) by
explicitly canceling contributions between bosonic
and fermionic loop with some postulated relation 
between their masses. However, this cancellation is not motivated
by any symmetry argument and thus a rather accidental
cancellation of this kind fails for higher order loops. 

Driven by this simple argument let us assume that there are
two additional complex scalar fields $\widetilde f_L$ and
$\widetilde f_R$ corresponding to a fermion $f$ which couples
to field $\Phi$ (see eqn.(\ref{SM-Higgs})) in the following manner
\bea
\mathcal{L}_{\widetilde f \widetilde f h^0}
&=& \widetilde {\lam_f} |\Phi|^2 
(|\widetilde f_L|^2 + |\widetilde f_R|^2), \nonumber \\
&\underrightarrow{EWSB}& \frac{1}{2} \widetilde {\lam_f} h^{0^2} 
(|\widetilde f_L|^2 + |\widetilde f_R|^2) + 
v \widetilde {\lam_f} h^{0} 
(|\widetilde f_L|^2 + |\widetilde f_R|^2) +..,
\label{c2-susy-motivation}
\eea
where $h^0$ is the physical Higgs field (see eqn.(\ref{SM-Higgs-VEV})).
A Lagrangian of the form of eqn.(\ref{c2-susy-motivation}) will yield
additional one-loop contributions to Higgs mass. Note that in order
to get a potential bounded from below, $\widetilde {\lam_f}<0$.
\begin{figure}[ht]
\centering
\includegraphics[width=10.95cm]{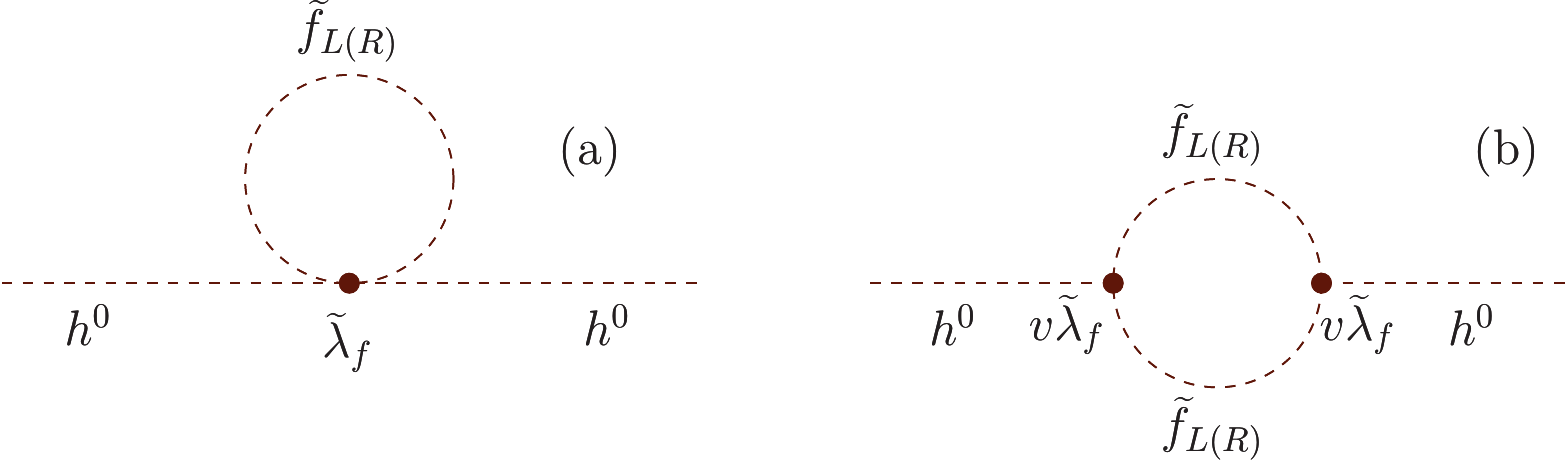}
\caption{New diagrams contributing to Higgs mass correction
from Lagrangian $\mathcal{L}_{\widetilde f \widetilde f h^0}$
(eqn.(\ref{c2-susy-motivation})).}
\label{Higgs-loop-c2}
\end{figure}
The additional contributions to the two point function for Higgs mass 
via the loops (figure \ref{Higgs-loop-c2}) can be written as

\bea
\Pi^{\widetilde f}_{h^0h^0}(p^2=0) &=&
-\widetilde {\lam_f} \int \frac{d^4k}{(2\pi)^4}
\left(\frac{1}{k^2-m^2_{\widetilde f_L}}
+ \frac{1}{k^2-m^2_{\widetilde f_R}} \right) \nonumber \\
&+& (v \widetilde {\lam_f})^2 \int \frac{d^4k}{(2\pi)^4}
\left(\frac{1}{(k^2-m^2_{\widetilde f_L})^2}
+ \frac{1}{(k^2-m^2_{\widetilde f_R})^2} \right).
\label{c2-susy-loop1}
\eea
Eqn.(\ref{c2-susy-loop1}) contains two types of divergences, 
(a) the first line which is quadratically divergent and (b) 
second line, which is logarithmically divergent. Following
similar procedure to that of deriving eqn.(\ref{Higgs-fermion-approx}), 
one can see that the total two point function $\Pi^{\widetilde f}_{h^0h^0}(p^2=0)$
$ +~\Pi^{f}_{h^0h^0}(p^2=0)$ (see eqn.(\ref{Higgs-fermion-loop}))
is {\it{completely}} free from quadratic divergences, provided
\bea
\widetilde \lam_f = - y^2_f.
\label{coup-equality}
\eea
It is extremely important to note that eqn.(\ref{coup-equality})
is {\it{independent}} of mass of $f,\widetilde f_L$ and $\widetilde f_R$,
namely $m_f, m_{\widetilde f_L}$ and $m_{\widetilde f_R}$ respectively.
The remaining part of $\Pi^{\widetilde f}_{h^0h^0}(p^2=0)$
$ +~\Pi^{f}_{h^0h^0}(p^2=0)$, containing logarithmic divergences
can be explicitly written as (using eqn.(\ref{coup-equality}) 
and dropping $p^2$)
\bea
\Pi^{\widetilde f}_{h^0h^0}(0) +  
\Pi^{f}_{h^0h^0}(0) &=&
\frac{i y^2_f}{16 \pi^2}
\left[-2m^2_f(1-ln \frac{m^2_f}{\mu^2_R}) 
+ 4 m^2_f ln \frac{m^2_f}{\mu^2_R}\right] \nonumber \\
&+& \frac{i y^2_f}{16 \pi^2}
\left[+2m^2_{\wt f}(1-ln \frac{m^2_{\wt f}}{\mu^2_R}) 
- 4 m^2_{\wt f} ln \frac{m^2_{\wt f}}{\mu^2_R}\right],
\label{c2-susy-loop-log}
\eea
with $m_{\wt f_L} = m_{\wt f_R} = m_{\wt f}$. $\mu_R$
is the scale of renormalization. If further one considers
{\bf{$m_{\wt f} = m_{f}$}} then from eqn.(\ref{c2-susy-loop-log}),
$\Pi^{\widetilde f}_{h^0h^0}(0) + \Pi^{f}_{h^0h^0}(0) = 0$, i.e. sum of the 
two point functions via the loop vanishes! This theory
is absolutely free from hierarchy problem. However, in order
to achieve a theory free from quadratic divergences, such cancellation between
fermionic and bosonic contributions must persists for all higher orders also.
This is indeed a unavoidable feature of a theory, if there exists a symmetry
relating fermion and boson masses and couplings\footnote{The hint of such a 
symmetry is evident from $m_{\wt f} = m_{f}$.}.

\section{{\bf B}asics of supersymmetry algebra}\label{susy-basics}

A symmetry which transforms a fermionic state into a bosonic one
is known as supersymmetry (SUSY) 
\cite{c2Gervais:1971ji,c2Neveu:1971rx,c2Ramond:1971gb,c2Aharonov:1971kf,c2Wess:1974tw,c2Wess:1973kz,
c2Ferrara:1974ac,c2Iwasaki:1973pw,c2Fayet:1976cr,c2Nilles-1983,c2Gates-1983,c2Haber:1984rc,
c2Sohnius-1985,c2Drees-1996,c2Lykken-1996,c2Dine-1996,c2Martin-1997,c2Bilal-2001,c2Wess-1992,
c2Bailin:1994qt,
c2Weinberg-2000,c2Drees-2004,c2Baer-2006} (also see references of \cite{c2Martin-1997}). 
The generator $(Q)$ of SUSY thus satisfies
\bea
Q|Boson\rangle = |Fermion\rangle,
~~Q|Fermion\rangle = |Boson\rangle.
\label{susy-generator}
\eea
In eqn.(\ref{susy-generator}) spin of the left and right hand 
side differs by {\it{half-integral}} number and thus
$Q$ must be a {\it{spinorial}} object in nature and hence follows anti-commutation
relation. Corresponding Hermitian conjugate $(\ovl{Q})$ is also 
another viable generator since spinors are complex objects.
It is absolutely important to study the space-time property
of $Q$, because they change the spin (and hence statistics also)
of a particle and spin is related to the behaviour under {\it{spatial
rotations}}. 

Let us think about an unitary operator $\mathcal{U}$, representing
a rotation by $360^\circ$ about some axis in configuration space,
then
\bea
&&\mathcal{U}Q|Boson\rangle = 
\mathcal{U}Q\mathcal{U}^{-1}\mathcal{U}|Boson\rangle 
= \mathcal{U}|Fermion\rangle,\nonumber \\
&&\mathcal{U}Q|Fermion\rangle = 
\mathcal{U}Q\mathcal{U}^{-1}\mathcal{U}|Fermion\rangle 
= \mathcal{U}|Boson\rangle.
\label{susy-gen-prop}
\eea
However, under a rotation by $360^\circ$ (see ref.\cite{c2Feynman-1987})
\bea
\mathcal{U}|Boson\rangle = |Boson\rangle,
~~\mathcal{U}|Fermion\rangle = -|Fermion\rangle.
\label{susy-gen-prop-2}
\eea
Combining eqns.(\ref{susy-gen-prop}),
(\ref{susy-gen-prop-2}) one ends up with
\bea
\mathcal{U}Q\mathcal{U}^{-1} = -Q,
~~\Rrightarrow \{Q,\mathcal{U}\}=0.
\label{susy-gen-prop-3}
\eea
Extending this analysis for any Lorentz transformations it is possible
to show that $Q$ does not commute with the generators of Lorentz transformation.
On the contrary, under space-time translation,
\bea
P_\mu|Boson\rangle = |Boson\rangle,
~~P_\mu|Fermion\rangle = |Fermion\rangle.
\label{susy-gen-prop-3a}
\eea
Eqns.(\ref{susy-gen-prop-3a}) and (\ref{susy-generator}) together
imply that $Q$ (also $\bar{Q}$) is invariant under space-time translations. 
that is 
\beq
[Q,P^\mu ]=[\bar{Q},P^\mu ]=0.
\label{susy-gen-prop-4}
\eeq
It is obvious from eqns.(\ref{susy-gen-prop-3}) and (\ref{susy-gen-prop-4}), 
that supersymmetry is indeed a space-time symmetry. In fact now the 
largest possible space-time symmetry is no longer {\it{Poincar\'{e}}}
symmetry but the supersymmetry itself with larger number of generators,\footnote{This
statement is consistent with the statement of {\it{Coleman-Mandula}}
theorem \cite{c2Coleman:1967ad} and {\it{Haag-Lopuszanski-Sohnius}} theorem \cite{c2Haag:1974qh}.}
$M^{\mu\nu}$ (Lorentz transformation $\Rrightarrow$ spatial rotations and boosts), 
$P^{\mu}$ (Poincar\'{e} transformation $\Rrightarrow$ translations) and 
$Q,\bar{Q}$ (SUSY transformations). It has been argued earlier that the SUSY 
generators ${Q,\bar{Q}}$ are {\it{anti-commuting}} rather than
being commutative. So what is $\{Q,\bar{Q}\}$? Since $Q,\bar{Q}$ 
are spinorial in nature, then expression for $\{Q,\bar{Q}\}$ must 
be bosonic in nature and definitely has to be another symmetry 
generator of the larger group.
In general, one can expect that $\{Q,\bar{Q}\}$ should be a combination of $P^\mu$ 
and $M^{\mu\nu}$ (with appropriate index contraction), However, after
a brief calculation one gets
\beq
\{Q,\bar{Q}\}\propto P^\mu.
\label{susy-gen-prop-5}
\eeq
Eqn.(\ref{susy-gen-prop-5}) is the basic of the 
SUSY algebra which contains generators of the SUSY transformations
$(Q,\bar{Q})$ on the left hand side and generator for space-time
translations, $P^\mu$ on the other side. This suggests that
successive operation of two finite SUSY transformations will
induce a space-time translation on the states under operation.
The quantity $\{Q,\bar{Q}\}$ is a Hermitian operator with
positive definite eigenvalue, that is
\beq
\langle...|\{Q,\bar{Q}\}|...\rangle
=|Q|...\rangle|^2 + |\bar{Q}|...\rangle|^2
\ge 0.
\label{susy-gen-prop-6}
\eeq
Summing over all the SUSY generators and using eqns.(\ref{susy-gen-prop-5})
and (\ref{susy-gen-prop-6}) one gets
\beq
\sum_{Q}\{Q,\bar{Q}\}\propto P^0,
\label{susy-gen-prop-7}
\eeq
where $P^0$ is the total energy of the system or the eigenvalue
of the Hamiltonian, thus 
Hamiltonian of supersymmetric theory contains no negative eigenvalues.

If $|0\rangle$ denotes the {\it{vacuum}} or the lowest energy state 
of any supersymmetric theory then
following eqns.(\ref{susy-gen-prop-6}) and (\ref{susy-gen-prop-7}) one 
obtains $P^0|0\rangle = 0$. This is again true if $Q|0\rangle = 0$ and 
$\bar{Q}|0\rangle = 0$ for all $Q,\bar{Q}$. This implies that any one-particle
state with non-zero energy cannot be invariant under SUSY transformations.
So there must be one or more supersymmetric partners ({\it{superpartners}}) 
$Q|1\rangle$ or $\bar{Q}|1\rangle$ for every one-particle state $|1\rangle$.
Spin of superpartner state differs by $\frac{1}{2}$ unit from that of $|1\rangle$.
The state $|1\rangle$ together with its superpartner state said to form a {\it{supermultiplet}}.
In a supermultiplet different states are connected in between through one or more
SUSY transformations. Inside a supermultiplet the number of fermionic
degrees of freedom $(n_F)$ must be equal to that for bosonic one 
$(n_B)$. A supermultiplet must contain at least one boson and
one fermion state. This simple most supermultiplet is known as the {\it{chiral}} 
supermultiplet which contains a Weyl spinor (two degrees of freedom) and one
complex scalar (two degrees of freedom). It is important to note that the translational 
invariance of SUSY generators (see eqn.(\ref{susy-gen-prop-4})) imply 
{\it{All states in a supermultiplet must have same mass}}\footnote{It 
is interesting to note that supercharge $Q$ satisfies 
$[Q,P^2] = 0$ but $[Q,W^2] \neq 0$, where $W^\mu( = \frac{1}{2}\ep^{\mu\nu\rho\sigma}
M_{\nu\rho}P_\sigma)$ is the {\it{Pauli-Lubanski}} vector. Note that eigenvalue
of $W^2\propto s(s+1)$ where $s$ is spin of a particle. Thus in general
members of a supermultiplet should have same mass but different spins, which is
the virtue of supersymmetry.}. 
%
It must be emphasized here that throughout the calculation
indices for $Q$ and $\bar{Q}$ have been suppressed. In reality
$Q\equiv Q^i_a$ where `$i=1,2,...N$' is the number of supercharges
and `$a$' is the spinor index. To be specific one should explicitly
write (for $i=1$), $Q_\al,\bar{Q}_{\dot{\al}}$, where $\al,\dot{\al}$
are spinorial indices belonging to two different representations 
of the Lorentz group. We stick to $i=1$ for this thesis. 
Details of SUSY algebra is given in 
refs.\cite{c2MullerKirsten-1986,c2Simonsen-1995}.

\section{{\bf C}onstructing a supersymmetric Lagrangian} \label{susy-Lagrangian}

Consider a supersymmetric Lagrangian with a single Weyl fermion,
$\psi$ (contains two helicity states, $\Rrightarrow n_F=2$) and a 
complex scalar, $\phi$ ($\Rrightarrow n_B=2$) without any
interaction terms. This two component Weyl spinor 
and the associated complex scalar are said to form a chiral supermultiplet. 
The free Lagrangian, which contains only kinetic terms is written as
\beq
\mathcal{L}^{susy} = -\partial_\mu \phi^* \partial^\mu \phi 
+ i \psi^\dagger \ovl{\sigma}^\mu \partial_\mu \psi,
\label{susy-Lag-1}
\eeq
where $\ovl{\sigma}^\mu = \bf{1},-\sigma_i$.
Eqn.(\ref{susy-Lag-1}) represents a massless, non-interacting supersymmetric
model known as {\it{Wess-Zumino}} model \cite{c2Wess:1974tw}. The action
$\mathcal{S}^{susy} (=\int d^4x \mathcal{L}^{susy})$ 
is invariant under the set of transformations, given as
\bea
&& \del \phi = \epsilon_\al \psi^\al \equiv \ep \psi, 
~~\Rrightarrow \del \phi^* = \epsilon^\dagger \psi^\dagger,\nn \\
&& \del \psi_\al = -i(\sigma^\mu \epsilon^\dagger)_\al \partial \phi, 
~~\Rrightarrow \del \psi^\dagger_{\dot{\al}} 
= i(\epsilon \sigma^\mu)_{\dot{\al}} \partial \phi^*,
\label{susy-Lag-2}
\eea
where $\epsilon^\al$ parametrizes infinitesimal SUSY transformation.
It is clear from eqn.(\ref{susy-Lag-2}), on the basis of dimensional 
argument that $\ep^\al$ must be spinorial object
and hence anti-commuting in nature. They have mass dimension 
$[M]^{-\frac{1}{2}}$. It is important to note that 
$\partial_\mu \ep^\al = 0$ for global SUSY transformation. 

\begin{flushleft}
{\it{$\maltese$ Is supersymmetry algebra closed?}}
\end{flushleft}

It has already been stated that $\mathcal{S}^{susy}$ is
invariant under SUSY transformations (eqn.(\ref{susy-Lag-2})).
But does it also indicate that the SUSY algebra is closed?
In other words, is it true that two successive SUSY transformations 
(parametrized by $\ep_1,\ep_2$) is indeed another symmetry of the 
theory? In reality one finds 
\bea
[\del_{\ep2},\del_{\ep1}] X
&=& -i(\ep_1\sigma_\mu\ep^\dagger_2 - \ep_2\sigma_\mu\ep^\dagger_1)
\partial^\mu X,
\label{susy-Lag-3}
\eea
where $X=\phi,\psi_\al$, which means that
commutator of two successive supersymmetry transformations 
is equivalent to the space-time translation of the respective fields.

This is absolutely consistent with our realization of 
eqn.(\ref{susy-gen-prop-5}). But there is a {\it{flaw}}
in the above statement. In order to obtain eqn.(\ref{susy-Lag-3})
one has to use the equation of motion for the massless fermions
and therefore the SUSY algebra closes only in 
on-shell limit. So how to close SUSY algebra even
in off-shell. A more elucidate statement for this problem
should read as how to match the bosonic degrees of freedom
to that of a fermionic one in off-shell? The remedy of this
problem can come from adding some auxiliary field, $F$
(with mass dimension $2$) in the theory which can provide 
the required extra bosonic degrees of freedom.
Being auxiliary, $F$ cannot posses a kinetic term ($\mathcal{L}_{auxiliary} = F^*F$,
Euler-Lagrange equation is $F = F^* = 0$). So the modified set of
transformations read as
\bea
&& \del \phi = \ep \psi, 
~~\Rrightarrow \del \phi^* = \epsilon^\dagger \psi^\dagger,\nn \\
&& \del \psi_\al = -i(\sigma^\mu \epsilon^\dagger)_\al \partial \phi
+ \ep_\al F, 
~~\Rrightarrow \del \psi^\dagger_{\dot{\al}} 
= i(\epsilon \sigma^\mu)_{\dot{\al}} \partial \phi^*
+ \ep^\dagger_{\dot{\al}} F^*\nn \\
&& \del F = -i \ep^\dagger \bar{\sigma}^\mu \partial_\mu \psi
~~\Rrightarrow \del F^* 
= i \partial_\mu \psi^\dagger \bar{\sigma}^\mu \ep.
\label{susy-Lag-4}
\eea
Eqn.(\ref{susy-Lag-1}) also receives modification and for `$i$' number
of chiral supermultiplets is given by
\beq
\mathcal{L}^{chiral} = 
-\underbrace{\partial_\mu \phi^{i^*} \partial^\mu \phi_i}_{\mathcal{L}_{scalar}} 
+\underbrace{i \psi^{i^\dagger} \ovl{\sigma}^\mu \partial_\mu \psi_i}_{\mathcal{L}_{fermion}}
+\underbrace{F^{i^*} F_i}_{\mathcal{L}_{auxiliary}}.
\label{susy-Lag-5}
\eeq
\begin{flushleft}
{\it{$\maltese$ Gauge bosons}}
\end{flushleft}
Theory of the SM also contains different types of gauge bosons. So in
order to supersymmetrize the SM one must consider some ``fermionic counterparts'' 
also to complete the set. The massless spin one gauge boson $(A^a_\mu)$ 
and the accompanying spin $\frac{1}{2}$ supersymmetric partner 
(two component Weyl spinor, called {\it{gauginos ($\lam^a$)}}) also belong 
to the same multiplet, known as the gauge supermultiplet. The index `$a$' 
runs over adjoint representation of the associated $\rm{SU(N)}$ group. It is 
interesting to note that since gauge bosons belong to the adjoint representation, 
hence a gauge supermultiplet is a real representation. Just like the case
of chiral supermultiplet one has to rely on some auxiliary fields
$D^a$ to close off-shell SUSY algebra. The corresponding Lagrangian
is written as 
\beq
\mathcal{L}^{gauge} = 
-\overbrace{\frac{1}{4}F^a_{\mu\nu}F^{\mu\nu}_a}^{F^a_{\mu\nu}
= \pl_\mu A^a_\nu -\pl_\nu A^a_\mu + g f^{abc} A^b_{\mu} A^c_{\nu}}
+\underbrace{i \lam^{a\dagger} \bar{\sigma}^\mu D_\mu \lam^a}_{D_\mu
\lam^a = \pl_\mu \lam^a + g f^{abc} A^b_\mu \lam^c} 
+~ \frac{1}{2} D^a D^a,
\label{susy-Lag-6}
\eeq
where $F^a_{\mu\nu}$ is the Yang-Mills field strength
and $D_\mu \lam^a$ is the covariant derivative for gaugino
field, $\lam^a$. The set of SUSY transformations which leave the
action $\mathcal{S}^{gauge} (=\int d^4x \mathcal{L}^{gauge})$ 
invariant are written as
\bea
&& \del A^a_\mu = -\frac{1}{\rt2} (\ep^\dagger \bar{\sigma}_\mu \lam^a
+ \lam^{a^\dagger} \bar{\sigma}_\mu \ep),\nn \\
&& \del \lam^a_\al = \frac{i}{2\rt2} (\sigma^\mu \bar{\sigma}^\nu \ep)_\al
F^a_{\mu\nu} + \frac{1}{\rt2} \ep_\al D^a,\nn \\
&& \del D^a = -\frac{i}{\rt2} (\ep^\dagger \bar{\sigma}_\mu D_\mu \lam^a
- D_\mu \lam^{a^\dagger} \bar{\sigma}_\mu \ep).
\label{susy-Lag-7}
\eea
\begin{flushleft}
{\it{$\maltese$ Interactions in a supersymmetric theory}}
\end{flushleft}
A supersymmetrize version of the SM should include an interaction
Lagrangian invariant under SUSY transformations. From the 
argument of renormalizability and naive power counting
the most general interaction Lagrangian (without gauge interaction)
appears to be
\beq
\mathcal{L}^{int} = \underbrace{-\frac{1}{2}W^{ij}\psi_i\psi_j}^{[W^{ij}] = [mass]^1}
+ \overbrace{W^{i}F_i}^{[W^{i}] = [mass]^2} 
+ \overbrace{x^{ij} F_i F_j}^{[x^{ij}] = [mass]^0}
+~ c.c - \underbrace{U}_{[U] = [mass]^4},
\label{susy-Lag-8}
\eeq
where $x^{ij},W^{ij},W^i,U$ all are polynomials of
$\phi,\phi^*$ (scalar fields) with degrees $0,1,2,4$. However, invariance under 
SUSY transformations restricts the form of eqn. (\ref{susy-Lag-8})
as
\beq
\mathcal{L}^{int} = (-\frac{1}{2} W^{ij}\psi_i\psi_j
+ W^{i}F_i) + c.c.
\label{susy-Lag-9}
\eeq
It turns out that in order to maintain the interaction Lagrangian
invariant under supersymmetry transformations,
the quantity $W^{ij}$ must to be analytic function of $\phi_i$
and thus cannot contain a $\phi^*_i$. It is convenient to define
a quantity $W$ such that $W^{ij} = \frac{\pl W}{\pl \phi_i \pl \phi_j}$
and $W^{i} = \frac{\pl W}{\pl \phi_i}$. The entity $W$ in 
most general form looks like
\beq
W = h_i \phi_i + \frac{1}{2} M^{ij} \phi_i \phi_j
+ \frac{1}{3!} f^{ijk} \phi_i \phi_j \phi_k.
\label{susy-Lag-10}
\eeq
First term of eqn.(\ref{susy-Lag-10}) vanishes for the supersymmetric
version of the SM as $h^i = 0$ in the absence of a
gauge singlet scalar field. It is important to note that in an equivalent language,
the quantity $W$ is said to be a function of the chiral superfields 
\cite{c2Ferrara:1974ac,c2Salam:1974yz}. A superfield is a single object
that contains as components all of the bosonic, fermionic, and auxiliary 
fields within the corresponding supermultiplet. That is
\bea
&&{\bf \Phi} \supset (\phi,\psi,F),~{\rm{or}},\nn\\
&&{\bf \Phi}(y^\mu,\theta) = \phi(y^\mu) + \theta \psi(y^\mu) 
+ \theta\theta F(y^\mu)~{\rm{and}},\nn\\
&&{\bf \Phi}^\dagger(\bar{y}^\mu,\bar{\theta}) = \phi^*(\bar{y}^\mu) 
+ \bar{\theta} \bar{\psi}(\bar{y}^\mu) 
+ \bar{\theta}\bar{\theta} F(\bar{y}^\mu),
\label{superfield-defn}
\eea
where $y^\mu~(=x^\mu-i\theta \sigma^\mu \bar{\theta})$ and 
$\bar{y}^\mu~(=\bar{x}^\mu+i\theta \sigma^\mu \bar{\theta})$ represent left and
right chiral superspace coordinates, respectively. It is important
to note that in case of the $(3+1)$ dimensional field theory $x^\mu$ represents
the set of coordinates. However, for implementation
of SUSY with $(3+1)$ dimensional field theory one needs to consider 
{\it{superspace}} with supercoordinate $(x^\mu,\theta^\al,\bar{\theta}_{\dot{\al}})$.
$\theta^\al,\bar{\theta}_{\dot{\al}}$ are spinorial coordinates spanning the 
fermionic subspace of the superspace. Any superfield, which is
a function of $y$ and $\theta$ ($\bar{y}$ and $\bar{\theta}$) only,
would be known as a left(right) chiral superfield. Alternatively,
if one defines chiral covariant derivatives $\bf{\mathcal{D}}_A$ and
$\bf{\mathcal{\bar{D}}}_{\bar A}$ as 
\bea
{\bf{\mathcal{D}}_A} \bar{y}^\mu = 0,~{\bf{\mathcal{\bar{D}}}_{\bar A}} y^\mu=0,
\label{chiral-covariant-defn}
\eea
then a left and a right chiral superfield is defined as
\bea
\bf{\mathcal{D}}_A {\bf \Phi} = 0
~{\rm{and}}
~\bf{\mathcal{\bar{D}}}_{\bar A} {\bf \Phi^\dagger}=0,
\label{chiral-field-defn2}
\eea

The gauge quantum numbers and the mass dimension of a chiral superfield are the 
same as that of its scalar component, thus in the superfield
formulation, eqn.(\ref{susy-Lag-10}) can be recasted as
\beq
W = h_i {\bf \Phi}_i + \frac{1}{2} M^{ij} {\bf \Phi}_i {\bf \Phi}_j
+ \frac{1}{3!} f^{ijk} {\bf \Phi}_i {\bf \Phi}_j {\bf \Phi}_k.
\label{susy-Lag-10-superfield}
\eeq
The quantity $W$ is now called a superpotential. The superpotential $W$
now not only determines the scalar interactions of the theory,
but also determines fermion masses as well as different Yukawa
couplings. Note that $W(W^\dagger)$ is an
analytical function of the left(right) chiral superfield.

Coming back to interaction Lagrangian, using the equation of motion for $F$ and $F^*$ 
finally one ends up with
\beq
\mathcal{L}^{int} = -\frac{1}{2}(W^{ij}\psi_i\psi_j 
+ W^*_{ij}\psi^{\dagger^i}\psi^{\dagger^j}) - 2 W^i W^*_i.
\label{susy-Lag-11}
\eeq

The last and remaining interactions are coming from
the interaction between gauge and chiral supermultiplets.
In presence of the gauge interactions SUSY transformations
of eqn.(\ref{susy-Lag-4}) suffer the following modification,
$\Rrightarrow \pl_\mu \rightarrow D_\mu$. It is also interesting
to know that in presence of interactions, Euler-Lagrange equations
for $D^a$ modify as $D^a = -g(\phi^{i^*} T^a \phi^i)$ with $T^a$
as the generator of the group.

So finally with the help of eqns.(\ref{susy-Lag-5}), (\ref{susy-Lag-6}),
(\ref{susy-Lag-11}) and including the effect of gauge interactions
the complete supersymmetric Lagrangian looks like

\bea
\mathcal{L}^{total} &=&  - \partial_\mu \phi^{i^*} \partial^\mu \phi_i
+ i \psi^{i^\dagger} \ovl{\sigma}^\mu \partial_\mu \psi_i
-\frac{1}{4}F^a_{\mu\nu}F^{\mu\nu}_a 
+i \lam^{a\dagger} \bar{\sigma}^\mu D_\mu \lam^a \nn \\
-&&\left[\{\frac{1}{2}(W^{ij}\psi_i\psi_j 
+\rt2 g (\phi^*_iT^a_{ij}\psi_j)\lam^a \} + h.c\right]
\nn \\
-&&V(\phi,\phi^*) \left\{\Rrightarrow {
\underbrace{W^*_i W^i}_{F^*_i F^i} +
\overbrace{\frac{1}{2}\sum_{a} g^2_a 
(\phi^{i^*} T^a \phi_i)^2}^{\sum \frac{1}{2}D^aD^a}}\right\}.
\label{susy-Lag-12}
\eea
In eqn.(\ref{susy-Lag-12}) index `$a$' runs over three of the SM
gauge group, ${\rm SU}(3)_C \times {\rm SU}(2)_L \times {\rm U}(1)_Y$. 
Potential $V(\phi,\phi^*)$, by definition (see eqn.(\ref{susy-Lag-12})) 
is bounded from below with minima at the origin.

\section{{\bf S}USY breaking} \label{susy-breaking}
In a supersymmetric theory fermion and boson belonging to the 
same supermultiplet must have equal mass. 
This statement can be re-framed in 
a different way. Consider the supersymmetric partner 
of electron (called selectron, $\wt e$), then SUSY invariance
demands, $m_e = m_{\wt e} = 5.109\times10^{-4} {\rm{GeV}}$ (see table \ref{SM-spectrum}),
where $m_{\wt e}$ is mass of the selectron. But till date there exists no experimental 
evidence (see ref.\cite{c2Nakamura-c2}) for a selectron. That simply indicates that
supersymmetry is a broken symmetry in nature.
The immediate question arises then what is the pattern of SUSY
breaking? Is it a spontaneous or an explicit breaking? With the 
successful implementation of massive gauge bosons in the SM, it
is naturally tempting to consider a spontaneous SUSY breaking first.

\begin{figure}[ht]
\centering
\includegraphics[width=8.95cm,keepaspectratio]{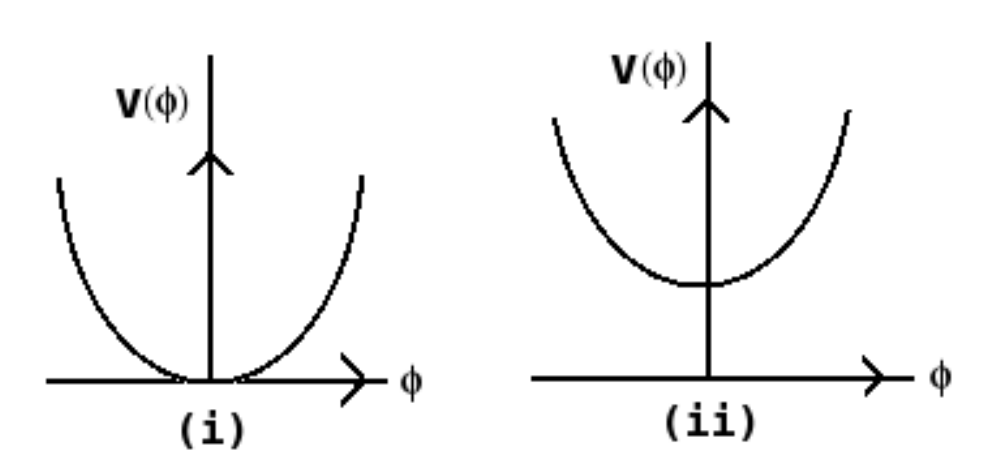}
\caption{Vacua of a supersymmetric theory. (i) exactly supersymmetric
and (ii) SUSY is spontaneously broken.}
\label{SSB-susy}
\end{figure}

\begin{flushleft}
{\it{$\blacklozenge$ Spontaneous breaking of SUSY}}
\end{flushleft}
In the case of spontaneous SUSY breaking the supersymmetric Lagrangian
remains unchanged, however, vacuum of the theory is no longer symmetric
under SUSY transformations. This will in turn cause
splitting in masses between fermionic and bosonic states within the 
same multiplet connected by supersymmetry transformation. 
From the argument given in section \ref{susy-basics} 
it is evident that the spontaneous breaking of supersymmetry 
occurs when the supercharges $Q,\bar{Q}$ (the SUSY generators)
fail to annihilate the vacuum of the theory. In other words if supersymmetry
is broken spontaneously (see figure \ref{SSB-susy}), the vacuum must have 
positive energy, i.e. $\langle0|\mathcal{H}^{susy}|0\rangle \equiv 
\langle\mathcal{H}^{susy}\rangle> 0$ (see eqn.(\ref{susy-gen-prop-6})).
$\mathcal{H}^{susy}$ is the SUSY Hamiltonian. Neglecting the space-time
effects one gets

\beq
\langle0|\mathcal{H}^{susy}|0\rangle = \langle0|\mathcal{V}^{susy}|0\rangle,
\label{susy-Lag-13}
\eeq
where $\mathcal{V}^{susy}$ is given by $V(\phi,\phi^*)$ (see eqn.(\ref{susy-Lag-12})).
Therefore spontaneous breaking of SUSY implies
\beq
\underbrace{\langle F \rangle \neq 0}_{\tt{F-type~breaking}} 
~~{\rm{or}}~~
\overbrace{\langle D \rangle \neq 0}^{\tt{D-type~breaking}}.
\label{susy-break-1}
\eeq
It is interesting to note that eqn.(\ref{susy-break-1}) does not
contain $D^a$ because if the theory is gauge invariant
then $\langle D \rangle = 0$ holds for Abelian vector superfield
only. It is informative to note that the spontaneous breaking of
a supersymmetric theory through $F$-term is known as
{\it{O'raifeartaigh}} mechanism \cite{c2O'Raifeartaigh:1975pr} and 
the one from $D$-term as {\it{Fayet-Iliopoulos}} 
mechanism \cite{c2Fayet:1974jb,c2Fayet:1975yh}. In the case of
global \footnote{The infinitesimal SUSY transformation parameter 
$\ep_\al$ is a space-time independent quantity.} SUSY breaking, 
the broken generator is $Q$, and hence the Nambu-Goldstone particle must 
be a massless neutral spin $\frac{1}{2}$ Weyl fermion (known as {\it{goldstino}}).
The goldstino in not the supersymmetric partner of Goldstone boson,
but a Goldstone fermion itself.

But there are drawbacks with this simple approach. 
The supersymmetric particle spectrum is known to follow 
certain sum rules, known as the supertrace
sum rules which must vanish. 
The supertrace of the tree-level squared-mass eigenvalues is
defined with a weighted sum over all particles with spin $j$ as
STr$(m^2) \equiv \sum (-1)^j (2j+1) Tr(m^2_j) = 0$ 
\cite{c2Ferrara:1979iz,c2Ferrara:1979wa}. This theorem holds
for sets of states having same quantum numbers.

But, a vanishing supertrace indicates 
that some of the supersymmetric particles must be lighter compared 
to that of the SM, which is of course not observed experimentally so far.
However, this relation holds true at the tree level and for renormalizable
theories. So supersymmetry can be spontaneously broken in some
``hidden sector'' which only couples to the ``visible'' or ``observable''
SM sector through loop mediated or through non-renormalizable interactions.
These intermediate states which appear in loops or are integrated out
to produce non-renormalizable interactions are known as the ``messengers''
or ``mediators''. 
Some of the well-motivated  
communication schemes are {\it{supergravity, anomaly mediation, gauge mediation, 
gaugino mediation}} and many others 
(see review \cite{Kolda:1997wt,c2Intriligator:2007cp}). 
In all of these scenario SUSY is
spontaneously broken at some hidden or secluded
sector, containing fields singlet under the SM gauge group at
some distinct energy scale and the information of breaking
is communicated to the observable minimal sector via
some messenger interaction. A discussion on these issues
is beyond the scope of this thesis. 

\begin{flushleft}
{\it{$\blacklozenge$ Explicit SUSY breaking and soft-terms}}
\end{flushleft}
It is now well understood that with the minimal field content
SUSY has to be broken explicitly.
But what happens to Higgs mass hierarchy if SUSY is broken in nature?
It turns out that in order to have a theory free from quadratic divergence
as well as to have the desired convergent behaviour of supersymmetric
theories at high energies along with the nonrenormalization of its
superpotential couplings, the explicit SUSY breaking terms must be 
soft \cite{c2Girardello:1981wz,c2Dimopoulos:1981zb,c2Sakai:1981gr,c2Hall:1990ac}.
The word soft essentially implies that all field operators occurring
in explicit SUSY breaking Lagrangian must have a mass dimension 
less than four. 

The possible most general \cite{c2Fayet:1976cr,c2Girardello:1981wz} 
soft supersymmetry breaking terms inhoused in
$\mathcal{L}_{soft}$ are\footnote{It is interesting to note that 
terms like $-\frac{1}{2}c^{jk}_i \phi^{i^*} \phi_j \phi_k + c.c$ 
are also viable candidates for $\mathcal{L}_{soft}$, however they can 
generate quadratic divergence from the loop in the presence of gauge singlet
chiral superfields. A term like this becomes soft \cite{c2Hall:1990ac} 
in the absence of singlet superfields. 
One more important lesson is that the mass dimension of any coupling
in $\mathcal{L}_{soft}$ has to be less than four is a necessary but not sufficient
condition for the softness of any operator.}
\bea
\mathcal{L}_{soft} &&= -\left(\frac{1}{2}M_a \lam^a \lam^a
+ \frac{1}{3!} a^{ijk} \phi_i \phi_j \phi_k + \frac{1}{2} b^{ij} \phi_i \phi_j
+ t^i \phi_i\right) + c.c \nonumber \\
&&- (m^2)^i_j \phi^{j^*} \phi_i.
\label{susy-soft-1}
\eea
In eqn.(\ref{susy-soft-1}) terms like $t^i \phi_i$ are possible only if
there exist gauge singlet superfields and thus these terms are
absent from the minimal supersymmetric version of the SM. $M_a$'s are
the gaugino soft mass terms, $(m^2)^j_i$ are the coefficients for  
scalar squared mass terms and $b^{ij},a^{ijk}$ are the couplings for 
quadratic and cubic scalar interactions. 

\begin{flushleft}
{\it{$\blacklozenge$ Higgs mass hierarchy and $\mathcal{L}_{soft}$}}
\end{flushleft}
\begin{list}{}{}
\item
The form of eqn.(\ref{susy-soft-1}) indicates modification of
Lagrangian shown in eqn.(\ref{c2-susy-motivation}). Adding a possible 
interaction term of the form 
$\frac{\lam_f A_{\wt f}}{\rt2} \wt{f_L}\wt{f^*_R}h^0 + h.c$ 
(scalar cubic interaction) in eqn.(\ref{c2-susy-motivation})
in turn modifies the two-point function via the loop 
(see eqn.(\ref{c2-susy-loop-log})) as
\bea
\Pi^{\widetilde f}_{h^0h^0}(0) +  
\Pi^{f}_{h^0h^0}(0) &=&
-\frac{i y^2_f}{16 \pi^2}
\left[4 \del^2 + (2 \del^2 + |A_{\wt f}|^2) ln \frac{m^2_f}{\mu^2_R}\right]\nn\\
&+& {\text{higher orders}},
\label{c2-susy-loop-soft-mod}
\eea
where $\del^2 = m^2_{\wt f} - m^2_f$ and we 
assume $|\del|,|A_{\wt f}|\ll m_f$. The most important observation about 
eqn.(\ref{c2-susy-loop-soft-mod}) is that, in the exact supersymmetric
limit
\beq
m^2_{\wt f} = m^2_f,~~A_{\wt f} = 0,
\label{c2-susy-limit}
\eeq
that is, entire one loop renormalization of the Higgs self 
energy vanishes \footnote{Actually this condition is true for
all orders of perturbation theory and is a consequence of the 
nonrenormalization theorem \cite{c2Wess:1973kz,c2Iliopoulos:1974zv,
c2Ferrara:1974fv,c2Zumino:1974bg,c2Grisaru:1979wc}.}. It is also clear
from eqn.(\ref{c2-susy-loop-soft-mod}) that Higgs self energy 
is linearly proportional to the SUSY breaking parameters 
$(\del^2,|A_{\wt f}|^2)$. Thus supersymmetric theories are
free from quadratic divergences, unless $m^2_{\wt f} \gg m^2_f$. 
This is an extremely important relation, which indicates that
in order to have a TeV scale Higgs boson mass (theoretical limit)
the soft terms $(A_{\wt f})$ and the sparticle masses $(m_{\wt f})$
must lie in the same energy scale (reason why we are dreaming to
discover SUSY at the large hadron collider experiment). 
\end{list}

\section{{\bf M}inimal Supersymmetric Standard Model} \label{susy-MSSM}
\begin{figure}[ht]
\centering
\includegraphics[width=10.75cm,keepaspectratio]{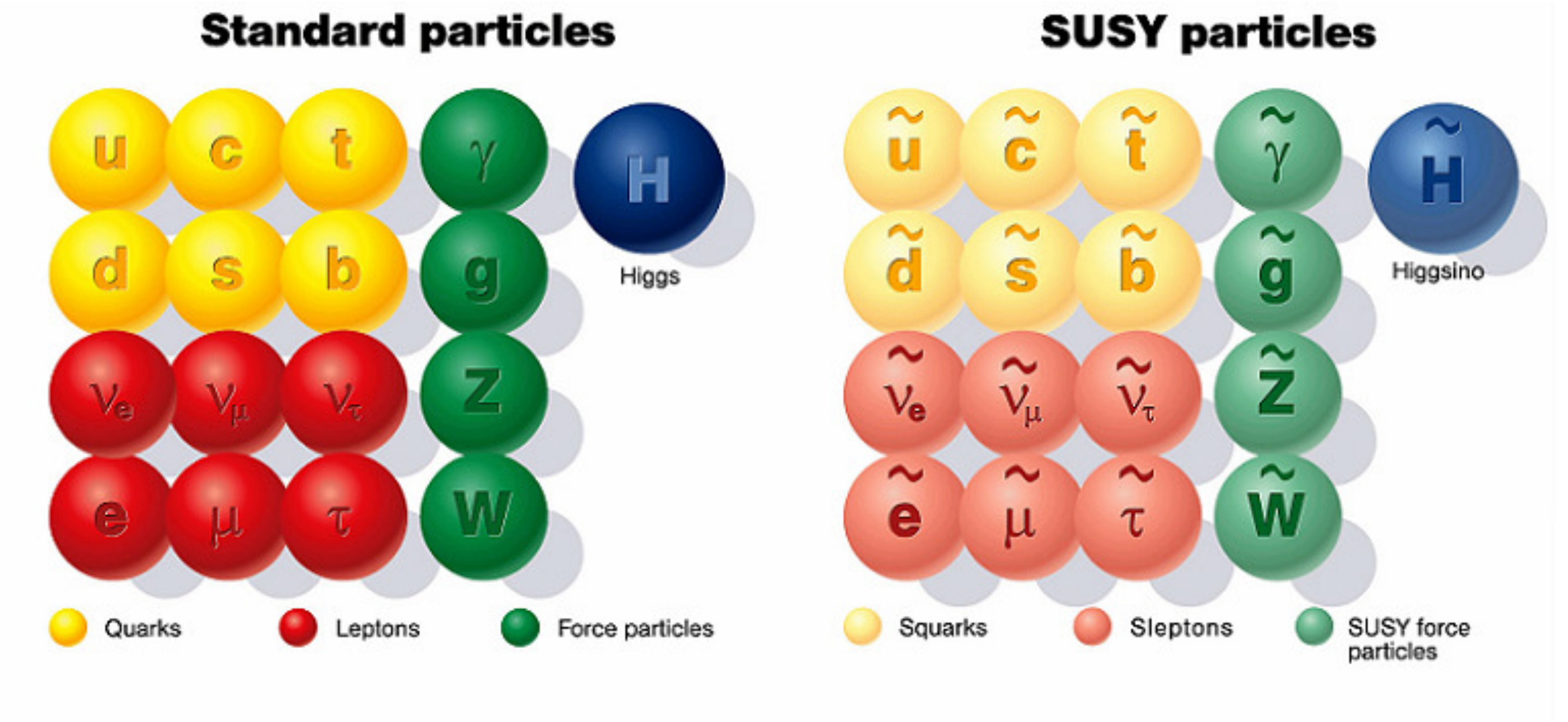}
\caption{particle content of the MSSM.}
\label{MSSM-particle}
\end{figure}

We are now well equipped to study the Minimal Supersymmetric Standard
Model or MSSM (see reviews \cite{c2Nilles-1983,c2Sohnius-1985,c2Martin-1997}). 
It is always illuminating to start with a description of
the particle content. Each of the SM fermions have their bosonic counterparts,
known as sfermions. Fermionic counterpart for a gauge boson is known as
a gaugino. Higgsino is the fermionic counter part for a Higgs boson. 
It is important to re-emphasize that since a superpotential 
is invariant under supersymmetry transformation it cannot involve an chiral and a anti-chiral
superfield at the same time. In other words a superpotential $(W)$ is
an analytical functions of chiral superfields only ($W^\dagger$
contains anti-chiral superfields only) and thus two Higgs doublets are essential for MSSM.
In addition, the condition for
anomaly cancellation in the higgsino sector, which is a requirement of renormalizability
also asks for two Higgs doublets, $H_u$ and $H_d$. It must be remembered that
each of the supersymmetric particle ({\it{sparticle}}) has same set of gauge 
quantum numbers under the SM gauge group as their SM counterpart, 
as shown in eqn.(\ref{SM-gauge-group}). The Higgs doublet $H_u$ behaves like 
eqn.(\ref{SM-Higgs}), whereas the other doublet
$H_d$ under ${\rm SU}(3)_C \times {\rm SU}(2)_L \times {\rm U}(1)_Y$ transforms as,
\bea
H_d = \left(\begin{array}{c}
H^0_d \\
H^-_d
\end{array}\right) \sim ({\bf{1,2,-1}}).
\label{Hd-Higgs}
\eea
The particle content of the MSSM is shown in figure \ref{MSSM-particle}.
Every lepton $(\ell)$ and quark $(q)$ of the SM (spin $\frac{1}{2}$)
is accompanied by a {\it{slepton $(\wt{\ell})$}} and 
{\it{squark $(\wt{q})$}} (spin $0$). Corresponding to two Higgs fields
$H_u$ and $H_d$ (denoted as $H$ in figure \ref{MSSM-particle}) 
there exist two {\it{Higgsino}} fields ($\wt{H_U},\wt{H_d}$) as well 
(denoted as $\wt H$ in figure \ref{MSSM-particle}). The electroweak gauge bosons
$W,Z$, gluons $(g)$ and photon $(\gamma)$ are associated with their
superpartner states, namely, {\it{wino $(\wt W)$}}, {\it{zino $(\wt Z)$}},
{\it{gluino $(\wt g)$}} and {\it{photino $(\wt \gamma)$}}\footnote{Another 
alternative set in lieu of $Z,\gamma$ 
could be $B,W^3$, where $B$ and $W^3$ are the $\rm{U(1)_Y}$ and $\rm{SU(2)_L}$ 
gauge bosons, respectively. Correspondingly on the right hand side
one should have $\wt{W}_3,\wt{B}\Longleftrightarrow \wt \gamma, \wt Z$.}.
Without further clarification we will concentrate first on the MSSM superpotential
and then on the soft terms. We will not talk about the 
kinetic terms i.e, the free Lagrangian and the gauge interactions
(see ref.\cite{c2Drees-2004} for an extensive discussions).

\begin{flushleft}
{\it{$\maltese$ MSSM superpotential and soft terms}} 
\end{flushleft}
The superpotential for the MSSM is written as
\bea 
W^{MSSM} &=& \epsilon_{ab}(Y^{ij}_u\hat H^b_u\hat Q^a_i\hat u^c_j +
Y^{ij}_d\hat H^a_d \hat Q^b_i\hat d^c_j + Y^{ij}_e\hat H^a_d\hat
L^b_i\hat e^c_j - \mu \hat H^a_d\hat H^b_u ),\nonumber \\
\label{MSSM-superpotential-1}
\eea
where $\hat H_d$ and $\hat H_u$ are the down-type and up-type Higgs 
superfields, respectively. The $\hat Q_i$ are ${SU(2)}_L$ doublet quark superfields, 
${\hat u}^c_j$ [${\hat d}^c_j$] are $\rm{SU(2)}_L$ singlet up-type [down-type] quark 
superfields. The $\hat L_i$ are the doublet lepton superfields, and the 
${\hat e}^c_j$ are the singlet charged lepton superfields. Here $a,b$ are SU(2) 
indices, and $\epsilon_{12}$ = --$\epsilon_{21}$ = 1. Note that
$u^c_i,d^c_i,e^c_i \equiv u^*_{i_R},d^*_{i_R},\ell^*_{i_R}$ 
(see eqn.(\ref{SM-gauge-group})). The only coupling of the superpotential $W$,
that has a positive mass dimension is the $\mu$-parameter.

The corresponding soft SUSY breaking Lagrangian can be
written as
\bea
-\mathcal{L}^{MSSM}_{\text{soft}} &=&
(m_{\tilde{Q}}^2)^{ij} {\tilde Q^{a^*}_i} \tilde{Q^a_j}
+(m_{\tilde u^c}^{2})^{ij}
{\tilde u^{c^*}_i} \tilde u^c_j
+(m_{\tilde d^c}^2)^{ij}{\tilde d^{c^*}_i}\tilde d^c_j
- \epsilon_{ab} B_\mu \hat H^a_d\hat H^b_u \nonumber \\
&+&(m_{\tilde{L}}^2)^{ij} {\tilde L^{a^*}_i}\tilde{L^a_j}
+ (m_{\tilde e^c}^2)^{ij}{\tilde e^{c^*}_i}\tilde e^c_j 
+ m_{H_d}^2 {H^{a^*}_d} H^a_d + m_{H_u}^2 {H^{a^*}_u} H^a_u \nonumber \\
&+&  \left[ \epsilon_{ab} \left\{ 
(A_uY_u)^{ij} H_u^b\tilde Q^a_i \tilde u_j^c +
(A_dY_d)^{ij} H_d^a \tilde Q^b_i \tilde d_j^c \right. \right.\nonumber \\
&+& \left. \left.
(A_eY_e)^{ij} H_d^a \tilde L^b_i \tilde e_j^c  
\right\} - \frac{1}{2} \sum_{i=1}^3  M_i \wt {\lam_i}+ \text{h.c.} \right]. 
\label{Lsoft-MSSM}
\eea
In eqn.(\ref{Lsoft-MSSM}), the first two lines consist of squared-mass terms
of squarks, sleptons and Higgses along with a bilinear 
term $(\ep_{ab} B_\mu \hat H^a_d\hat H^b_u)$ in two Higgs superfields. 
The next line contains the trilinear scalar 
couplings. Finally, in the last line $M_3, M_2$, and $M_1$ are 
Majorana masses corresponding to $\rm{SU(3)}_c$, $\rm{SU(2)}_L$ 
and $\rm{U(1)}_Y$  gauginos $\wt{\lambda}_3, \wt{\lambda}_2$, 
and $\wt{\lambda}_1$,  respectively. 

The tree level scalar potential is given by (see eqn.(\ref{susy-Lag-12}))
\bea 
V^{MSSM}_{scalar} &=& V^{MSSM}_{soft} + \frac{1}{2} D^a D^a 
+ \left|\frac{\partial W^{MSSM}}{\partial \phi^{MSSM}}
\right|^2,
\label{MSSM-scalar-pot}
\eea
where $V^{MSSM}_{soft}$ contains only the scalar couplings of
eqn.(\ref{Lsoft-MSSM}) and $\Phi^{MSSM}$ represents scalar
component of any of the MSSM chiral superfields. Only the neutral scalar 
fields develop vacuum expectation values while minimizing 
the scalar potential $V^{MSSM}_{scalar}$ as follows 

\bea 
\langle H_d^0 \rangle = v_1,~\langle H_u^0 \rangle = v_2.
\label{MSSM-VEV}
\eea

It is evident from eqns.(\ref{MSSM-superpotential-1}) and
(\ref{Lsoft-MSSM}) that the MSSM has a very rich particle
spectra. Note that the matrices associated with bilinear terms in fields
(particles or sparticles) are often appear with off-diagonal
entries after EWSB (see chapter \ref{SM}). Clearly entries of these matrices cannot  
represent any physical masses. So in general these off-diagonal matrices
of the gauge or flavour basis can be rotated in a diagonal basis 
using suitable unitary or bi-unitary transformations.
All the scalar mass squared matrices are inhoused in $V^{MSSM}_{scalar}$.

\begin{flushleft}
{\it{$\maltese$ Gauge versus Mass eigen-basis}} 
\end{flushleft}
\begin{itemize}
\item
The squarks $(\wt q)$ and sleptons $(\wt l)$

The squark and slepton mass square matrices in the flavour basis
are bilinears in $\wt {f_L}^* \wt {f_L}$, $\wt {f_R}^* \wt {f_R}$
and $\wt {f_L}^* \wt {f_R} + \text{c.c}$ where $\wt f \equiv \wt l
/ \wt q$. It is always possible to rotate them into another basis $\wt f_1,\wt f_2$
where only combination like $\wt {f_1}^* \wt {f_1}, \wt {f_2}^* \wt {f_2}$
exists. The basis $\wt f_{1,2}$ is known as the mass basis for squarks
and sleptons. The orthogonal mixing matrix relating $\wt f_{L,R}$ and 
$\wt f_{1,2}$ contains an angle `$\theta$'
which depends on the ratio of the off-diagonal entry in $\wt f_{L,R}$ basis
and the difference in diagonal entries in the same basis. It can be shown 
(see for example ref. \cite{c2Drees-2004}) that
for the first two generations of squark and charged slepton the effect of
off diagonal mixing is negligible and to a very good approximation
$\wt f_{L,R}$ can be treated as the mass basis. So we conclude that
\begin{table}[ht]
\footnotesize
\centering
\begin{tabular}{c  c}
\hline \hline 
Gauge or flavour basis  & Mass basis  \\ \hline \hline
$\wt e_L,\wt e_R,\wt \mu_L,\wt \mu_R$   & 
 $\wt e_L,\wt e_R,\wt \mu_L,\wt \mu_R$ \\
\vspace*{0.1cm}
$\wt u_L,\wt u_R,\wt d_L,\wt d_R$ 
& $\wt u_L,\wt u_R,\wt d_L,\wt d_R$ \\
\vspace*{0.1cm}
$\wt c_L,\wt c_R,\wt s_L,\wt s_R$ 
& $\wt c_L,\wt c_R,\wt s_L,\wt s_R$ \\ \hline \hline
\end{tabular}
\end{table}

%
However, this simple minded approach fails for the 
third family of slepton and squark due to relatively large Yukawa 
coupling. This is because, it is the effect of Yukawa coupling
which controls the size of the off-diagonal term. Summing up, 
for the third family
\begin{table}[ht]
\footnotesize
\centering
\begin{tabular}{c  c}
\hline \hline 
Gauge or flavour basis  & Mass basis  \\ \hline \hline
$\wt \tau_L,\wt \tau_R$   & 
$\wt \tau_1,\wt \tau_2$ \\
\vspace*{0.1cm}
$\wt b_L,\wt b_R,\wt t_L,\wt t_R$ 
& $\wt b_1,\wt b_2,\wt t_1,\wt t_2$ \\ \hline \hline
\end{tabular}
\end{table}

%
It remains to talk about the left sneutrinos which do not
have any right handed counter part. The degenerate squared mass for 
all three generations of left sneutrino is given by
\bea 
M^2_{\wt \nu} = m^2_{\wt L} + \frac{1}{2} M^2_Z {\rm cos} 2\beta,
\label{left-sneutrinos}
\eea
where $\rm{tan}\beta=\frac{v_2}{v_1}$ is the ratio of two Higgs VEVs and
$M_Z$ is the $Z$ boson mass given by $M^2_Z = \frac{1}{2}(g^2_1
+g^2_2)(v^2_1+v^2_2)$. Mass for the $W^\pm$-bosons are given
by $M^2_W = \frac{g^2_2}{2}(v^2_1+v^2_2)$.

\item
The neutral and charged supersymmetric fermions

The neutral supersymmetric fermions $(-i\wt B^0,-i\wt W^0_3,\wt H^0_d,
\wt H^0_u)$ are known to form a $4\times4$ symmetric matrix
in the flavour basis. This symmetric matrix is diagonalizable
using a single unitary matrix $N$ and the corresponding four mass
eigenstates are known as neutralinos, $\chi^0_i$ (two-component spinor). 
Mathematically,

\beq 
\chi^0_i = N_{i1} \wt B^0 + N_{i2} \wt W^0_3 + N_{i3} \wt H^0_d
+ N_{i4} \wt H^0_u,
\label{MSSM-neutralinos}
\eeq
with $N_{ij}$ as the elements of the matrix $N$.

The charged fermions $\psi^+=-i\wt W^+,\wt H^+_u$ and 
$\psi^-=-i\wt W^-,\wt H^-_d$ on the other hand form a 
$4\times4$ mass matrix in the Lagrangian as follows 
\bea 
\mathcal{L}^{chargino}_{MSSM} = 
-\frac{1}{2}
\left(\begin{array}{c c}
\psi^+ & \psi^-
\end{array}\right)
\left(\begin{array}{c c}
0 & (M^{chargino}_{MSSM})^T_{2\times2}\\
(M^{chargino}_{MSSM})_{2\times2} & 0
\end{array}\right)
\left(\begin{array}{c}
\psi^+ \\
\psi^-
\end{array}\right) + h.c.\nn\\
\label{MSSM-charginos-Lagrangian}
\eea

The $2\times2$ non-symmetric matrix $(M^{chargino}_{MSSM})_{2\times2}$
(see appendix \ref{appenA}) requires a bi-unitary transformation to 
go the two-component physical charged fermion 
eigenstates, known as charginos, $\chi^\pm_i$. If $U,V$ are the two required 
transformation matrices, then
\bea 
&&\chi^+_i = V_{i1} \wt W^+ + V_{i2} \wt H^+_u, \nn \\
&&\chi^-_i = U_{i1} \wt W^- + U_{i2} \wt H^-_d.
\label{MSSM-charginos}
\eea
It is important to re-emphasize that all the charged and neutral
spinors considered here are two-component Weyl spinors. They
can be used further to construct the corresponding four-component
spinors. The neutralino and chargino mass matrices for MSSM are given
in appendix \ref{appenA}.
\item
The neutral and the charged leptons and the
quarks are treated in MSSM similar to that of the SM.
\item
The gauge bosons are also treated in similar fashion.
\item
Higgs bosons in MSSM

Let us write down two Higgs doublet of the MSSM 
in the real $(\Re)$ and imaginary $(\Im)$ parts as follows
\bea
&&H_d = \left(\begin{array}{c}
H^0_d \\
H^-_d
\end{array}\right) = 
\left(\begin{array}{c}
\Re{H^0_d} + i \Im{H^0_d} \\
\Re{H^-_d} + i \Im{H^-_d}
\end{array}\right) =
\left(\begin{array}{c}
h_1 + i h_2 \\
h_3 + i h_4
\end{array}\right),\nonumber\\
&&H_u = \left(\begin{array}{c}
H^+_u \\
H^0_u
\end{array}\right) = 
\left(\begin{array}{c}
\Re{H^+_u} + i \Im{H^+_u} \\
\Re{H^0_u} + i \Im{H^0_u}
\end{array}\right) =
\left(\begin{array}{c}
h_5 + i h_6 \\
h_7 + i h_8
\end{array}\right),
\label{MSSM-Higgs-doublets}
\eea
Out of this eight Higgs field $(h_i)$, only the neutral real
fields can develope a {\it{non-zero}} VEV which are 
(recasting eqn.(\ref{MSSM-VEV}))
\bea 
\langle \Re{H_d^0} \rangle = v_1,~\langle \Re{H_u^0} \rangle = v_2.
\label{MSSM-VEV-2}
\eea

These eight Higgs fields are further classifiable into three groups,
namely (1) CP-even $(h_1,h_7)$, (2) CP-odd $(h_2,h_8)$
and (3) charged $(h_{3-6})$. In the mass basis these
give five physical Higgs states, $h^0,H^0,A^0,H^\pm$ and 
three Goldstone bosons $(G^0,G^\pm)$. In terms of mathematical relations,
\bea
&&H^0 =  \rt2\left((\Re{H^0_d}-v_1){\rm cos}\al+(\Re{H^0_u}-v_2){\rm sin}\al\right),\nn \\
&&h^0 =  \rt2\left(-(\Re{H^0_d}-v_1){\rm sin}\al+(\Re{H^0_u}-v_2){\rm cos}\al\right),\nn \\
&&H^- =  \left((\Re{H^-_d} + i \Im{H^-_d}){\rm sin}\beta 
+ (\Re{H^+_u} + i \Im{H^+_u})^\dagger {\rm cos}\beta\right),\nn\\
&&A^0 = \rt2 \left(-\Im{H^0_d}{\rm sin}\beta + \Im{H^0_u}{\rm cos}\beta\right),\nn\\
&&G^0 = \rt2 \left(\Im{H^0_d}{\rm cos}\beta - \Im{H^0_u}{\rm sin}\beta\right),\nn\\
&&G^- = \left((\Re{H^-_d} + i \Im{H^-_d}){\rm cos}\beta 
- (\Re{H^+_u} + i \Im{H^+_u})^\dagger {\rm sin}\beta\right),\nn\\
&& H^+ = (H^-)^\dagger,~~G^+ = (G^-)^\dagger,
\label{Higgs-mass-gauge}
\eea
where $\al$ is a mixing angle relating the gauge and mass basis of
CP-even Higgs fields. Scalar (CP-even), pseudoscalar (CP-odd)
and charged scalar mass squared matrices in the flavour basis for 
MSSM Higgs fields are given in appendix \ref{appenA}.
Physical Higgs boson squared masses are given by
(using eqns.(\ref{MSSM-superpotential-1}),(\ref{Lsoft-MSSM}))
\bea
&&m^2_{A^0} =  \frac{2 B_\mu} {{\rm sin}{2\beta}} ,\nn \\
&&m^2_{H^0} =  \frac{1}{2}\left[m^2_{A^0} + M^2_Z
+ \sqrt{(m^2_{A^0} + M^2_Z)^2-4 m^2_{A^0} M^2_Z {\rm cos}^22\beta}\right],\nn \\
&&m^2_{h^0} =  \frac{1}{2}\left[m^2_{A^0} + M^2_Z
- \sqrt{(m^2_{A^0} + M^2_Z)^2-4 m^2_{A^0} M^2_Z {\rm cos}^22\beta}\right],\nn \\
&&m^2_{H^\pm} =  m^2_{A^0} + M^2_W.
\label{Higgs-mass-MSSM}
\eea

From eqn.(\ref{Higgs-mass-MSSM}) one can achieve a theoretical upper
limit of the lightest Higgs boson mass \cite{c2Casas:1994us,c2Casas:1994qy}, 
$(m_{h^0})$ at the tree level after a bit of algebraic
exercise as \cite{c2Gunion:1989we,c2Gunion:1992hs,c2Gunion:1984yn},
\beq
m_{h^0} \le M_Z |{\rm cos}2\beta|.
\label{Higgs-mass-MSSM-lightest}
\eeq
The lightest Higgs mass can however, receives significant radiative
corrections from higher order processes, which are capable of
altering the lightest Higgs mass bound drastically. Note that
the value for angle $\beta$ is between $0$ to $\frac{\pi}{2}$. Thus
it is easy to conclude that $m_{h^0}$ at the tree level can be
at most of the order of the $Z$-boson mass. But this is already 
ruled out by the LEP experiment 
\cite{c2Barate:2003sz,c2Nakamura-c2}. So it is evident that inclusion of 
loop correction \cite{c2Haber:1990aw,c2Okada:1990vk,c2Okada:1990gg,
c2Ellis:1990nz,c2Ellis:1991zd,c2Espinosa:1991fc} (see also ref.\cite{c2Chankowski:1992er} and 
references therein) to lightest Higgs boson mass in MSSM is 
extremely important. The dominant contribution arises from top-stop
loop and assuming masses for sparticles below $1$ TeV we get $m_{h^0} \le 135$ 
GeV\footnote{This limit can be further relaxed to $m_{h^0} \le 150$
GeV, assuming all couplings in the theory remain perturbative up
to the unification scale \cite{c2Kane:1992kq,c2Espinosa:1992hp}.}.

The conditions for the tree level
Higgs potential to be bounded from below (in the direction $v_1 = v_2$) 
as well as the condition for EWSB are
\bea
&&m_{H_d}^2 + m_{H_u}^2 + 2 |\mu|^2 \ge 2 |B_\mu|,\nn \\
&&(m_{H_d}^2 + |\mu|^2)(m_{H_u}^2 + |\mu|^2)< B^2_\mu.
\label{Higgs-pot-cond}
\eea
It is extremely important to note that if $B_\mu,m_{H_d}^2,m_{H_u}^2$
all are zero, i.e. there exist no soft SUSY breaking terms, the EWSB
turns out to be impossible. So in a sense SUSY breaking is somehow
related to the EWSB.

We conclude the description of the MSSM with a note on the corresponding set
of Feynman rules. The number of vertices are extremely large for 
a supersymmetric theory even in the minimal version, and consequently
there exist a huge number of Feynman rules. The rules are far more
complicated compared to the SM because of the presence of Majorana
particles (particles, that are antiparticles of their own, neutralinos for 
example). For a complete set of Feynman rules 
for the MSSM see references \cite{c2Haber:1984rc,c2Gunion:1984yn,c2Gunion:1989we,
c2Rosiek:1989rs,c2Rosiek:1995kg}. A detailed analysis for the Higgs boson in supersymmetry
and related phenomenology are addressed in a series of references 
\cite{c2Gunion:1984yn,c2Gunion:1986nh,c2Gunion:1988yc,c2Gunion:1992tq}.
\end{itemize}


\section{{\bf T}he $R$-parity} \label{R-parity}

The superpotential for MSSM was shown in eqn.(\ref{MSSM-superpotential-1}).
This superpotential is gauge (the SM gauge group) invariant, Lorentz invariant
and maintains renormalizability. However, it is natural to ask that
what is preventing the following terms to appear in $W^{MSSM}$, which
are also gauge and Lorentz invariant and definitely renormalizable:

\bea 
W^{extra} &=& \epsilon_{ab}(- \varepsilon^i \hat L^a_i\hat H^b_u  
+ \frac{1}{2} \lambda_{ijk}\hat L^a_i\hat L^b_j\hat e^c_k  
+ \lambda^{'}_{ijk} \hat L^a_i \hat Q^b_j\hat d^c_k
+ \frac{1}{2} \lambda^{''}_{ijk} \hat u^c_i \hat d^c_j \hat d^c_k).\nn\\
\label{MSSM-superpotential-2}
\eea
Of course, all of these terms violate either lepton $(L)$ \cite{c2Weinberg:1981wj,
c2Hall:1983id} or baryon $(B)$ \cite{c2Weinberg:1981wj,
c2Zwirner:1984is} number by odd units. The second and the third terms of
eqn.(\ref{MSSM-superpotential-2}) violate lepton number by one unit whereas
the fourth term violates baryon number by one unit.

\begin{figure}[ht]
\centering
\includegraphics[width=7.45cm]{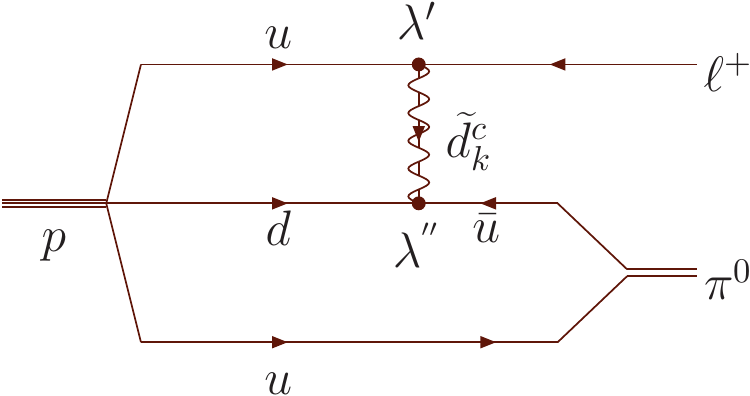}
\caption{Feynman diagrams for the process $p \to \ell^+ \pi^0$ with 
$\ell^+=e^+,\mu^+$.}
\label{proton-decay}
\end{figure}

Now it is well known that in the SM, lepton and baryon numbers
are conserved at the perturbative level. In the SM, $L$ and $B$ are the accidental
symmetry of the Lagrangian, that is to say that these are not symmetries
imposed on the Lagrangian, rather they are consequence of the gauge
and Lorentz invariance, renormalizability and, of course, particle content
of the SM. Moreover, these numbers are no way related to any fundamental 
symmetries of nature, since they are known to be violated by non-perturbative
electroweak effects \cite{c2'tHooft:1976up}. So it is rather difficult to drop these terms
from a general MSSM superpotential unless one assumes $B,L$ conservation
as a postulate for the MSSM. However, in the presence of these terms
there exists new contribution to the proton decay process 
$(p\to\ell^+\pi^0$ with $\ell^+=e^+,\mu^+$) as shown
in figure \ref{proton-decay}.
%
This process (see figure \ref{proton-decay}) will yield a 
proton life time $\approx 10^{-9}$ sec,
assuming $\lam',\lam^{''}\sim \cal {O}$ $(10^{-1})$ and TeV scale
squark masses. However, the known experimental bound for proton
lifetime is $>10^{32}$ years \cite{c2NH:2009gd,c2Nakamura-c2}. So in order to 
explain proton stability either these new couplings $(\lam,\lam^{'},\lam^{''})$
are extremely small (which again requires explanation) or their products
(appear in the decay width for the process $p\to\ell^+\pi^0$) are
very small or these
terms are somehow forbidden from the MSSM superpotential. In fact,
to avoid very fast proton decay mediated through squarks of masses 
of the order of the electroweak scale, simultaneous presence of
$\lam',\lam''$ type couplings must be forbidden unless the product
$\lam'\lam''$ is severely constrained (see figure \ref{proton-decay}).
The $\lam$ type of operators are not so stringently suppressed, and therefore 
still a lot of freedom remains (see ref.\cite{c2Dreiner:2006gu} and references therein).

\begin{figure}[ht]
\centering
\includegraphics[height=3.10cm]{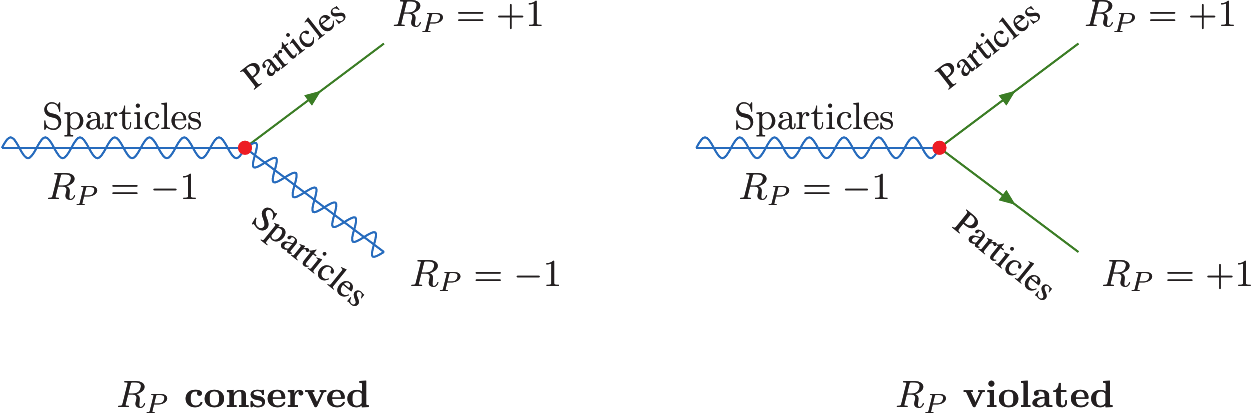}
\caption{With $R_p$ conservation the LSP is
forced to be stable due to unavailability of any lighter sparticle 
states ({\it{left}}), whereas for the $R_p$-violating scenario the LSP can 
decay into SM particles ({\it{right}}).}
\label{Rpc-Rpv}
\end{figure}

It turns out that since these new terms (see eqn.(\ref{MSSM-superpotential-2}))
violate either lepton or baryon number by odd units it is possible
to restrict them from appearing in $W^{MSSM}$ by imposing a discrete 
symmetry called $R$-parity $(R_p)$,\footnote{See also matter parity 
\cite{c2Dimopoulos:1981zb,c2Weinberg:1981wj,c2Sakai:1981pk,c2Dimopoulos:1981dw}.}
\cite{c2Farrar:1978xj,c2Weinberg:1981wj,c2Sakai:1981pk,c2Dimopoulos:1981dw} defined as,
\beq
R_p = (-1)^{3(B-L)+2s},
\label{R-parity-defn}
\eeq
where $s$ is the spin of the particle. Since $L$ is an integer, an
alternative expression for $R_p$ is also given by
\beq
R_p = (-1)^{3B+L+2s}.
\label{R-parity-defn-2}
\eeq
It is interesting to
note that since different states within a supermultiplet have
different spins, they must have different $R_p$. It turns out
that by construction all the SM particles have $R_p = +1$
and for all superpartners, $R_p = -1$. This is a discrete $Z_2$
symmetry and multiplicative in nature. It is important to note that
$R_p$ conservation would require (1) even number of 
sparticles at each interaction vertex, and (2) the lightest
supersymmetric particle (LSP) has no lighter $R_p=-1$ states to decay
and thus it is absolutely stable (see figure \ref{Rpc-Rpv}). 
Thus the LSP for a supersymmetric model with conserved $R_p$ can 
act as a natural dark matter candidate. It 
must be remembered that the soft supersymmetry breaking Lagrangian
will also contain $R_p$ violating terms \cite{c2Allanach:2003eb,c2de-Carlos:1996du}.

\begin{figure}[ht]
\centering
\includegraphics[height=3.10cm]{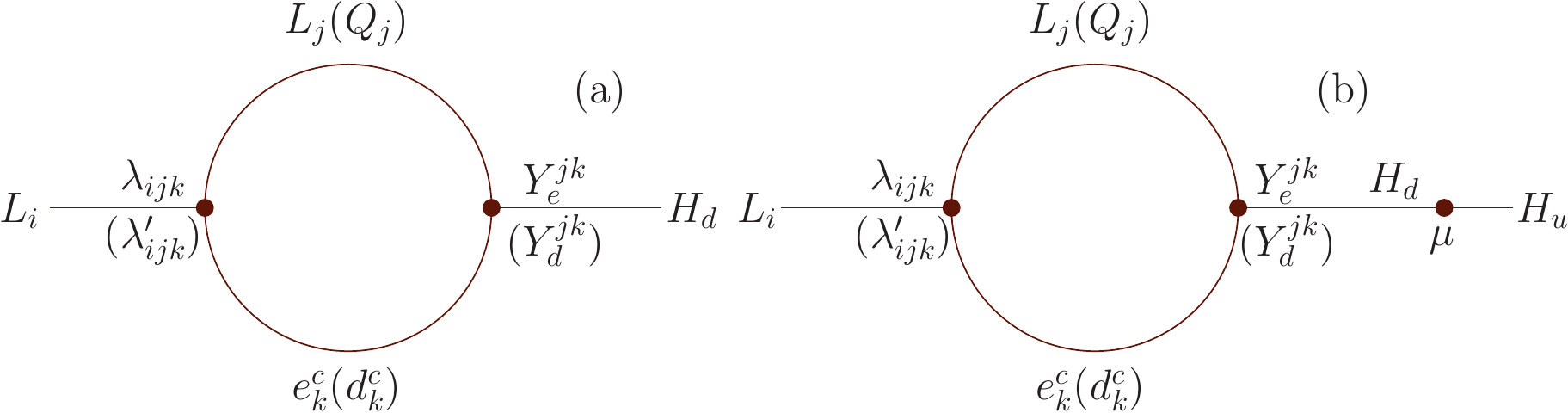}
\caption{One loop diagrams contributing to bilinear terms like $L_iH_u,L_iH_d$ 
using the trilinear couplings $\lam,\lam'$.}
\label{t-b}
\end{figure}

Looking at eqn.(\ref{MSSM-superpotential-2}) it is clear that
sources for $R_p$ violation $(\rpv)$ (see references 
\cite{c2Ellis:1984gi,c2Ross:1984yg,c2Dawson:1985vr,c2Barbieri:1985ty,c2Barger:1989rk,
c2Ibanez:1991pr,c2Bhattacharyya:1996nj,c2Dreiner:1997uz,c2Bhattacharyya:1997vv,
c2Barbier:1998fe,
c2Allanach:1999bf,c2Barbier:2004ez,c2Chemtob:2004xr}) are either bilinear $(\ep)$ 
\cite{c2Joshipura:1994wm,c2Joshipura:1994ib,c2de-Campos:1995av,c2Smirnov:1995ey,
c2Nowakowski:1995dx,c2Hempfling:1995wj,c2Nilles:1996ij,c2deCarlos:1996yh,c2Roy:1996bua,
c2Diaz:1997xc,c2Akeroyd:1997iq,c2Hirsch:2000ef,c2RestrepoQuintero:2001pk} or trilinear 
$(\lam,\lam^{'},\lam^{''})$ \cite{c2Barger:1989rk,
c2Enqvist:1992ef,c2de-Carlos:1996du,c2deCarlos:1996yh,c2Dreiner:1997uz,
c2Choudhury:1997dt,c2Rakshit:1998kd,c2Raychaudhuri:1999zg} in nature. The simple
most example of $\rpv$ turns out to be bilinear. It is interesting to
note that these bilinear terms are removable from superpotential by using field 
redefinitions, however they reappear as trilinear couplings both in superpotential
and in soft SUSY breaking Lagrangian \cite{c2Hall:1983id,c2Lee:1984kr,c2Lee:1984tn}
along with the original bilinear $R_p$-violating terms, that were in
the soft SUSY breaking Lagrangian to start with.
The effect of rotating away $L_iH_u$ term from the superpotential
by a redefinition of the lepton and Higgs superfields are bound to
show up via the scalar potential \cite{c2de-Campos:1995av}.
Also even if bilinear terms are rotated away at one energy scale, they reappear
in some other energy scale as the couplings evolve radiatively \cite{c2Barger:1995qe}.
The trilinear couplings can also give rise to bilinear terms in one-loops 
(see figure \ref{t-b}) \cite{c2de-Carlos:1996du}. 
Note that $\rpv$ can be either explicit 
(like eqn.(\ref{MSSM-superpotential-2})) \cite{c2Hall:1983id,c2Lee:1984kr,
c2Lee:1984tn,c2Ellis:1984gi} or spontaneous \cite{c2Aulakh:1982yn,c2Ellis:1984gi,c2Ross:1984yg,
c2Santamaria:1987uq,c2Santamaria:1988zm,c2Santamaria:1988ic,
c2Masiero:1990uj,c2Nogueira:1990wz,c2Romao:1992vu}.

Here as a digression it should be mentioned that $R_p$ can
be embedded into a larger continuous group (see, for example, ref.\cite{c2Joshipura:2000sn} 
and references therein) which is finally abandoned
for phenomenological reasons\footnote{A continuous symmetry
would prefer massless gauginos, which is already ruled out 
by experiments.}. However, its $Z_2$ subgroup could still be retained,
which is the $R_p$.

To summarize, it seems that $R_p$ violation is a natural
feature for supersymmetric theories, since $R_p$-violating terms
(see eqn.(\ref{MSSM-superpotential-2})) are not forbidden
to appear in the MSSM superpotential by the arguments of 
gauge and Lorentz invariance or renormalizability. 
On the contrary, assumption of $R_p$-conservation to prevent proton decay 
appears to be an ad hoc one. Besides,
models with $R_p$-violation are also phenomenologically very rich. Of course, 
it is natural to ask about the fate of the proton. But considering
either lepton or baryon number violation at a time proton stability can be achieved. 

It is true that with$~\rpv$ the LSP is no longer stable and can decay
into the SM particles. The stable LSP (in case it is colour and charge neutral) can be a 
natural candidate for the Dark matter \cite{c2Goldberg:1983nd,c2Ellis:1983ew}. 
However, their exist other viable dark matter candidates even for a theory with$~\rpv$,
namely, gravitino \cite{c2Borgani:1996ag,c2Takayama:2000uz,c2Hirsch:2005ag}, 
axion \cite{c2Kim:1986ax,c2Raffelt:1996wa} and 
axino \cite{c2Chun:1999cq,c2Chun:2006ss} (supersymmetric partner of axion). 

It is important to note that a decaying LSP has very different
and enriched implications in a collider study. Unlike models
with $R_p$ conservation, which yield large missing energy signature
at the end of any supersymmetric process, effect of $\rpv$ can often
produce interesting visible final states detectable in a collider
experiments. Models with bilinear $\rpv$ are especially interesting
concerning collider studies \cite{Guchait:1996xw,Huitu:1997qu,
BarShalom:1998av,Choudhury:1998ta,c2Mukhopadhyaya:1998xj,c2Choi:1999tq,
Ghosh:1999ix,Hikasa:1999wy,Chiappetta:1999cd,Lebedev:1999ze,Moreau:2000bs,
Ghosh:2000gx,Choudhury:2002av,c2Hirsch:2005ag,Li:2006he}, as they admit direct 
mixing between neutrino and neutralinos.

Finally, it remains to be mentioned the most important
aspect of $R_p$ violation, namely, generation of the neutrino mass.
It is impossible to generate neutrino masses in a supersymmetric
model with $R_p$ conservation along with minimal field content 
(see eqn.(\ref{MSSM-superpotential-1})). It is
rather important to clarify the importance of $\rpv$ in neutrino
mass generation. There are other ways to generate
light neutrino masses, both in supersymmetric or non-supersymmetric
models like adding extra particles or enhancing the gauge group 
(left-right symmetric models \cite{c2Pati:1974yy} for example) and many others. 
But generating massive neutrinos with $\rpv$ is a pure 
supersymmetric phenomenon without any SM analog. 
More on the issue of light neutrino mass generation 
and $\rpv$ will be addressed in the next chapter.

To complete the discussion, it is important to mention that 
these $\rpv$ couplings are highly constrained by experimental
limits on different physical processes, like neutron-anti neutron 
scattering \cite{c2BasteroGil:1996ps,c2BasteroGil:1996vc,c2Brahmachari:1994wd,
c2Goity:1994dq}, neutrinoless double beta decay \cite{c2Enqvist:1992ef,c2Hirsch:1995zi,
c2Hirsch:1995ek,c2Faessler:1996ph,c2Faessler:1997db,c2Hirsch:1998kc}, 
precision measurements of $Z$ decay \cite{c2Bhattacharyya:1995bw,
c2Lebedev:1999ze,c2Takeuchi:2000dc}, proton decay \cite{c2Smirnov:1996bg,
c2Bhattacharyya:1998dt,c2Bhattacharyya:1998bx}, Majorana masses for
neutrinos \cite{c2Dimopoulos:1988jw,c2Godbole:1992fb,c2Drees:1997id,
c2Rakshit:1998kd,c2Borzumati:2002bf,c2Bhattacharyya:1999tv} etc.
Discussion on different supersymmetric models with and without $R_p$
conservation, proposed 
in the literature is given in a recent review \cite{c2Munoz:2007fa}.

\section{{\bf S}uccesses of supersymmetry} \label{MSSM-success}
So far, we tried to formulate the theory of MSSM step
by step starting from the very basics. It is perhaps the appropriate
place to discuss the success of the supersymmetric 
theories over most of the shortcomings of the SM (see section \ref{suc-prob}). 
We are about to discuss all the seven points made in section \ref{suc-prob}
but in reverse order.

\begin{enumerate}
 \item 
The last point deals with Higgs mass hierarchy in the SM. It has 
been shown earlier that how a supersymmetric theory can predict 
a finite Higgs mass without any {\it{quadratic divergences}} even 
though SUSY is broken in nature. 
\item
It is true that MSSM with $R_p$ conservation predicts massless neutrinos
similar to the SM. However, as argued in the earlier section,
supersymmetric theories are capable of accommodating massive neutrinos
if $R_p$ is broken. Just for the sake of completeness, let us mention that there 
exist also certain non-minimal supersymmetric models, which can account for the
neutrino masses with seesaw mechanism. Such models include e.g. right-
handed neutrinos or other very heavy particles. In the next chapter
these possibilities will be explored in detail.
\item
The SM hardly offers any room for a suitable dark matter candidate. 
But as described in
section \ref{R-parity} the lightest supersymmetric particle is a
good candidate for the dark matter in a supersymmetric model with
$R_p$ conservation. Nevertheless, as stated in section \ref{R-parity},
there exist other viable dark matter candidates (gravitino, axion etc.)
even for an $R_p$-violating supersymmetric theory.
\item
The apparent exclusion of gravitational interaction from the SM 
is still maintained in supersymmetric theories, so long one considers 
global supersymmetry.
A locally supersymmetric theory together with the theory of general relativity
can incorporate gravitational interaction in SUSY. This theory is popularly 
known as supergravity theory.
\item
Concerning point no.(3) of section \ref{suc-prob}, there are other
sources of CP-violation in the MSSM itself, which can account for 
the large matter-anti matter asymmetry of the universe. In general,
apart from one CKM phase there exist many different phases in the MSSM,
particularly in the soft supersymmetry breaking sector. However, some of 
these are subjected to strong phenomenological constraints.
\item
It is true that the number of free parameters in a general
MSSM theory is larger $(> 100)$ \cite{c2Dimopoulos:1995ju,c2Giudice:1998bp} 
compared to that of the SM. However, there are models where most of these parameters can 
be achieved through evolution of a fewer number of parameters at a higher
scale. For example in minimal supergravity \cite{c2Nath:1983fp,c2Nilles-1983} 
theory the number of free parameters is just five.
\end{enumerate}

It has to be emphasized here, that this will be a rather incorrect
statement that supersymmetric theories are free from any drawbacks. It is
definitely true that they provide explanations to
some of the shortcomings of the SM in a few
occasions, but not always. As an example supersymmetric
theories are more prone to FCNC through the sparticle mediated
processes \cite{c2Hall:1985dx,c2Dine:1993np,c2Dine:1993su,c2Hagelin:1992tc,
c2Dimopoulos:1995ju,c2Sutter:1995kp,c2Haber:1997if}. This 
problem can be removed using clever tricks, but
a related discussion is beyond the scope of this thesis. Another,
well known problem of MSSM, the $\mu$-problem will be addressed
in the following section.

The main stumbling block for any supersymmetric theory is that
there are no experimental evidence for supersymmetry till date.
All the experimental bounds on different phenomenological processes
with supersymmetric effects are basically exclusion limit.

\section{{\bf T}he $\mu$-problem} \label{mu-prob}

The $\mu$-parameter, associated with the bilinear term in Higgs 
superfields (see eqn.(\ref{MSSM-superpotential-1})) is the only
coupling in the MSSM superpotential having a non-zero positive mass
dimension. The problem appears when
one consider the EWSB condition, which is given by
\beq
{\frac{1}{2}} M^2_Z = 
\frac{m^2_{H_d}-m^2_{H_u} \tan^2{\beta}}{{\tan^2{\beta}}-1}-|\mu^2|,
\label{mu-problem}
\eeq 
where $m^2_{H_d},m^2_{H_u}$ are given by eqn.(\ref{Lsoft-MSSM}), 
$\rm{tan}\beta = \frac{v_2}{v_1}$ and $M_Z$ is the $Z$ boson mass.
The $Z$ boson mass is very preciously measured to be $91.187$ GeV
(see table \ref{SM-spectrum}). So it is expected that all the entries
of the right hand side of eqn.(\ref{mu-problem}) (without any fine cancellation) 
should have the same order of magnitudes. But how could this happen?
$m^2_{H_d},m^2_{H_u}$ are coming from the soft supersymmetric breaking
sector with entries at the TeV scale. On the other hand, $\mu$ belongs to SUSY invariant 
$W^{MSSM}$ (eqn.(\ref{MSSM-superpotential-1})), which naturally can be as large as the Planck 
scale.
So why these two scales appear to be of the same order of magnitude
without having any a priori connection in between? This defines
the $\mu$-problem \cite{c2Kim:1983dt}. An alternative statement could be 
why $\mu^2\sim m^2_{soft}$ and not $\sim M^2_{Planck}$.

It seems easy to solve this problem by starting with $\mu=0$ at $W^{MSSM}$
and then use the favour of radiative corrections to generate a non-zero $\mu$
term. But there are some phenomenological problems of this approach and
moreover $\mu=0$ will give zero VEV for $H_d$ along with the presence
of unwanted Weinberg-Wilczek axion \cite{c2Weinberg:1977ma,c2Wilczek:1977pj}. 
So it is apparent that one needs to consider either $\mu\neq0$ or 
require extra fields. 
The requirement of additional fields often lead to other problems
and consequently do not predict satisfactory models \cite{c2Polchinski:1982an,
c2Nilles:1982mp,c2Lahanas:1982bk,c2Ferrara:1982ke,c2AlvarezGaume:1983gj}. 
There exist indeed various solutions to the $\mu$-problem where in most 
of the occasions the $\mu$-term is absent at the tree level 
and a TeV or electroweak scale $\mu$-term arises from
the VEV(s) of new fields. These VEVs are obtained by minimizing the
potential which also involves soft SUSY breaking terms. Thus, the fact
$\mu^2\sim m^2_{soft}$ turns out to be rather natural. Different
solutions to the $\mu$-problem have been addressed in references
\cite{c2Giudice:1988yz,c2Kim:1991mv,c2Chun:1991xm,c2Casas:1992mk,
c2Giudice:1993rc,c2Kim:1994eu,c2Yanagida:1997yf,c2Dimopoulos:1997je,
c2Kim:1999fg,c2Langacker:1999hs,c2Choi:1999yaa,c2Mafi:2000kg}. 
Some of these mechanisms are operational at very high energies and 
thus are hardly testable experimentally. 

Perhaps the simple most dynamical solution to the $\mu$-problem is
offered by next-to minimal supersymmetric standard model or 
NMSSM (see review \cite{c2Ellwanger:2009dp} and references therein).
In NMSSM the bilinear term $\epsilon_{ab} \hat H^a_d\hat H^b_u$
gets replaced by $\epsilon_{ab} \lambda\hat S \hat H^a_d\hat H^b_u$.
The superfield $\hat S$ is singlet \cite{c2Fayet-NMSSM,c2Kaul-NMSSM,c2Barbieri-NMSSM,
c2Nilles-NMSSM,c2Frere-NMSSM,c2Derendinger-NMSSM} under the SM gauge group. After
the EWSB an effective $\mu$ term is given by
\beq
\mu = \lam v_s,
\label{mu-gen-NMSSM}
\eeq
where $v_s= \langle S \rangle$, is the VEV acquired by the 
scalar component of the superfield $\hat{S}$. The VEV calculation
invokes the soft SUSY breaking terms and hence in general the VEVs
are at the TeV scale. It is now clear that the $\mu$-term
of eqn.(\ref{mu-gen-NMSSM}) is of the right order of magnitude and
it is indeed connected to $m^2_{soft}$. The NMSSM superpotential
assumes a $Z_3$ symmetry which forbids any bilinear term in superpotential.

It is important to note that any term in the superpotential
with a non-zero positive mass dimension suffers the similar fate.
In fact the bilinear $\rpv$ terms (see eqn.(\ref{MSSM-superpotential-2}))
are also associated with similar kind of problem known as the 
$\ep$-problem \cite{c2Nilles:1996ij}. A common origin for the $\vp_i$ (to 
account for the neutrino oscillation data), and the $\mu$-term
can be achieved using a horizontal family symmetry as suggested in
ref.\cite{c2Mira:2000gg}.

\section{{\bf N}ext-to-Minimal Supersymmetric Standard Model} \label{NMSSM}
It is perhaps logical and consistent with the theme of this thesis 
to give a brief introduction of the NMSSM. The NMSSM superpotential,
is given by (see review \cite{c2Maniatis:2009re,c2Ellwanger:2009dp})
\beq
W^{NMSSM} = W^{'^{MSSM}} - \epsilon_{ab} \lambda\hat S \hat H^a_d\hat H^b_u
+ \frac{1}{3} \kappa \hat{S}^3,
\label{NMSSM-superpotential}
\eeq
where $W^{'^{MSSM}}$ is the MSSM superpotential (eqn.(\ref{MSSM-superpotential-1}))
without the $\mu$-term. In a similar fashion if $\mathcal{L}^{'^{MSSM}}_{soft}$
denotes $\mathcal{L}^{MSSM}_{soft}$ without the $B_\mu$ term (see eqn.(\ref{Lsoft-MSSM})),
then
\beq
-\mathcal{L}^{NMSSM}_{soft} = -\mathcal{L}^{'^{MSSM}}_{soft} 
+ (m^2_{\wt S}) \wt S \wt S
- \ep_{ab} (A_{\lambda}\lambda) \tilde S H_d^a  H_u^b 
+ \frac{1}{3} (A_{\kappa}\kappa) \tilde S^3 + h.c.
\label{Lsoft-NMSSM}
\eeq
However, even in NMSSM, if $R_p$ is conserved, light neutrinos
are exactly massless. NMSSM models of neutrino mass generation
will be discussed in the next chapter.

Particle spectrum for NMSSM will be enlarged over that of the MSSM
due to extra particle content. However, $\hat S$ being SM gauge
singlet only the neutralino sector and the neutral Higgs sector receives
modifications. The neutralino mass matrix is now a $5\times5$
symmetric matrix and there will be one more CP-odd and CP-even neutral
scalar states, compared to that of the MSSM. The phenomenology of NMSSM is definitely much
enriched compared to the MSSM. This is essentially due to the admixture of new singlet 
states with MSSM fields. For example, theoretical lower bound on the lightest Higgs mass is 
now given by \cite{c2Drees:1988fc} (For Higgs sector in NMSSM also see references
\cite{c2Ellis:1988er,c2Haber:1990aw,c2Binetruy:1991mk,c2Espinosa:1991gr,c2Pietroni:1992in,
c2Moroi:1992zk,c2Elliott:1993uc,c2Elliott:1993bs,c2Pandita:1993tg,c2Pandita:1993hx,c2Wu:1994eb,
c2Franke:1995xn,c2King:1995ys,c2Franke:1995tc,c2Choudhury:1995bx,c2Ham:1996mi,c2Daikoku:2000eq,c2Ellwanger:2000fh,
c2Panagiotakopoulos:2000wp,c2Davies:2001uv,c2Ellwanger:2001iw,c2Ellwanger:2004gz,c2Ham:2004nv,
c2Ellwanger:2004xm,c2Ellwanger:2005fh,c2Ellwanger:2006rm,c2Ellwanger:2006rn,c2Cavicchia:2008fn,
c2Degrassi:2009yq})
\beq
m^{'^2}_{h^0} \le M^2_Z \left[{\rm cos}^22\beta
+ \frac{2\lam^2{\rm cos}^2_{\theta_W}}{g^2_2}~{\rm sin}^22\beta\right].
\label{Higgs-mass-NMSSM-lightest}
\eeq
which is clearly different from eqn.(\ref{Higgs-mass-MSSM-lightest}).
It is interesting to note that the lower limit of tree level
lightest Higgs boson mass in NMSSM depends on $\lam$ and hence
it is in general difficult to put some upper bound on $m'_{h^0}$
without constraining $\lam$.

The last term in eqn.(\ref{NMSSM-superpotential}) with coefficient $\kp$ is 
included in order to avoid an unacceptable axion associated to the breaking of a 
global ${\rm{U(1)}}$ symmetry \cite{c2Ellis:1988er}. This term is perfectly allowed 
by all symmetry. However, the discrete $Z_3$ symmetry of the NMSSM superpotential 
(see section \ref{mu-prob}) when spontaneously broken leads to three degenerate 
vacua. Casually disconnected parts of the Universe then would have randomly chosen 
one of these three equivalent minima leading to the formation of the dangerous 
cosmological domain wall \cite{c2Ellis:1986mq,c2Rai:1992xw,c2Abel:1995wk}. However, 
solutions to this problem exist 
\cite{c2Abel:1996cr,c2Panagiotakopoulos:1998yw,c2Panagiotakopoulos:1999ah}, 
but these issues are beyond the scope of this thesis\footnote{One 
solution of this problem is to put $\kp=0$ in the NMSSM superpotential by some 
symmetry argument. This simplified version is known as Minimally NMSSM or MNMSSM.}. 
Another major problem of the NMSSM 
theories are associated with the stability of gauge hierarchy arising from the 
{\it{tadpole}} contribution of the singlet field. 

Diverse phenomenological aspects of NMSSM models are discussed in 
references \cite{c2Franke:1994hj,c2Ananthanarayan:1995xq,c2King:1995vk,c2Franke:1995tf,c2Ananthanarayan:1995zr,
c2Franke:1995tc,
c2Ananthanarayan:1996zv,c2Kim:1996jh,c2Asatrian:1996dg,c2Ellwanger:1998vi,c2Ellwanger:1998jk,c2Demir:1999mm,
c2BasteroGil:2000bw,c2Nevzorov:2000uv,c2Nevzorov:2001vj,c2Menon:2004wv,c2Hiller:2004ii,c2Cerdeno:2004xw,
c2Ellwanger:2005uu,c2Funakubo:2005pu,c2Ellwanger:2005uu,c2Ellwanger:2005fh,c2Belanger:2005kh,c2Kraml:2005nx,
c2Ellwanger:2005dv,c2Gunion:2005rw,c2Dermisek:2005gg,c2Schuster:2005py,c2Arhrib:2006sx,c2Huber:2006wf,
c2Moretti:2006hq,c2Ferrer:2006hy,c2Dermisek:2006wr,c2Ma:2006te,c2Dermisek:2006py,c2Cerdeno:2007sn,
c2Mangano:2007gi,c2Balazs:2007pf,c2Dermisek:2007yt,c2Hugonie:2007vd,c2Dermisek:2007ah,c2Domingo:2007dx,
c2Barbieri:2007tu,c2Forshaw:2007ra,c2Akeroyd:2007yj,c2Heng:2008rc,c2Djouadi:2008uw,c2Djouadi:2008yj,
c2Chun:2008pg,c2Ellwanger:2008py,c2He:2008zw,c2Belyaev:2008gj,c2Domingo:2008bb,c2Chang:2008np,
c2Gogoladze:2008wz,c2Cao:2008un,c2Domingo:2008rr,c2Kraml:2008zr,c2Djouadi:2008uj,c2Belanger:2008nt,
c2Dermisek:2008uu,c2Cao:2009ad,c2Cerdeno:2009dv,c2Dermisek:2010mg,c2Das:2010ww,c2Staub:2010ty,c2Abada:2010ym}.

\vspace{1cm}
The prime focus of this thesis remains the issues of neutrino masses and mixing
in supersymmetric theories. It has been already argued that massive neutrinos
can be accommodated in supersymmetric theories either through $\rpv$ or using
seesaw mechanism with non-minimal field contents. 
Besides, mass generation is possible both with the tree level and loop 
level analysis. However, even before it is important to
note the evidences as well as the basics of the neutrino oscillation.
It is also interesting to note the implications of the massive neutrinos
in an accelerator experiment. We aim to discuss these issues in details
in the next chapter along with other phenomenological implications.


\chapter{ \sffamily{{\bf N}eutrinos
 }}\label{neut}

Long back in 1930, a new particle was suggested by {\it{Pauli}}
to preserve the conservation of energy, conservation of momentum, and 
conservation of angular momentum in beta decay \cite{c3Fermi:1934sk,
c3Fermi:1934hr}\footnote{To be specific, this was an electron neutrino.
$\nu_\mu$ and $\nu_\tau$ were hypothesized later in 1940 and 1970, respectively.}. 
The name {\it{neutrino}} was coined by {\it{Fermi}} in 1934. The much
desired experimental evidence for neutrinos (actually $\nu_e$) was finally
achieved in 1956 \cite{c3Cowan:1992xc}. In 1962, muon neutrino was discovered
\cite{c3Danby:1962nd}. However, it took a long time till 2000 to discover  
$\nu_\tau$ \cite{c3Kodama:2000mp}.

Neutrino sources are everywhere, however, they are broadly classifiable in
two major classes, namely, (1) natural sources and (2) man made neutrinos.
Natural neutrino sources are nuclear $\beta$ decay ($\nu_e$), solar neutrinos ($\nu_e$),
atmospheric neutrinos ($\nu_e,\nu_\mu$ and their anti-neutrinos) and supernovae neutrinos 
(all flavours) mainly. Man made neutrinos are produced by the particle accelerators and  
neutrinos coming out of nuclear reactors.

Neutrino physics has been seeking huge attention for the last few decades. 
Different aspects of neutrino physics have been discussed in references
\cite{c3Frampton:1982qi,c3Boehm:1987fc,c3Bilenky:1987ty,c3Bahcall:1989ks,c3Kayser:1989iu,
c3Mohapatra:1991ng,c3Gelmini:1994az,c3Barbieri:1998mq,c3Bilenky:1998dt,c3Fisher:1999fb,
c3Kayser:2000pe,c3Langacker:2000fp,c3Caldwell:2001pc,c3GonzalezGarcia:2002dz,c3Pakvasa:2003zv,
c3Barger:2003qi,c3Grossman:2003eb,c3King:2003jb,c3Altarelli:2003ph,c3Altarelli:2004za,c3Giunti:2004yg,
c3Kayser:2005cd,c3Mohapatra:2005wg,c3Mohapatra:2006gs,c3Strumia:2006db,c3Valle:2006vb,
c3GonzalezGarcia:2007ib}. 

\section{{\bf N}eutrinos in the Standard Model}\label{neut-SM}

The neutrinos as discussed in chapter \ref{SM}, appear to be a part of the SM. 
Confining our attention within the SM, it is worth listing the information about 
neutrinos, that ``lies within the SM'' 

\begin{enumerate}
\item
They are spin $\frac{1}{2}$ objects and thus follow Fermi-Dirac statistics 
\cite{c3Fermi,c3Dirac:1926jz}.
\item 
Neutrinos are electrically neutral fundamental particles, belonging to
the lepton family. The SM contains three neutrinos, corresponding
to three charged leptons.
\item
They are a part of the weak isospin ($\rm{SU(2)_L}$) doublet.
Being charge and colour neutral neutrinos are sensitive to weak interaction only.
\item
There exist two kinds of neutrino interactions in nature,
(1) neutral and (2) charge current interactions (see figure \ref{CC-NC}).
\begin{figure}[ht]
\centering
\includegraphics[width=7.45cm]{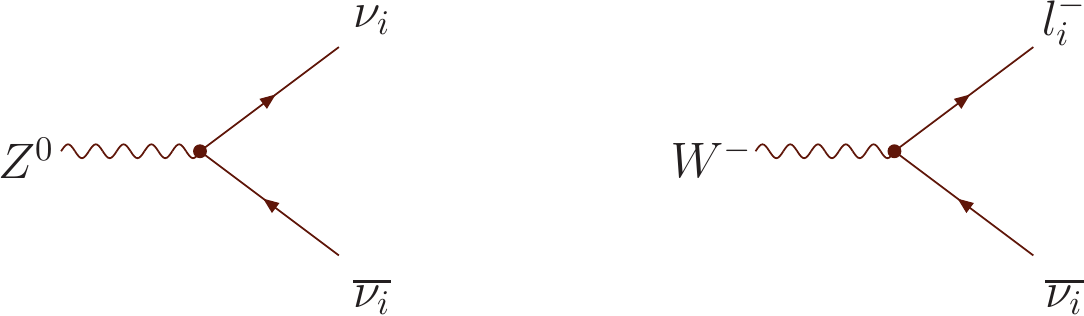}
\caption{Feynman diagram for the neutral and charged
current interactions. $\nu_i$ stand for different neutrino flavours
like $\nu_e,\nu_\mu,\nu_\tau$. The charged leptons $(e,\mu,\tau)$ 
are represented by $l_i$s.}
\label{CC-NC}
\end{figure}
\item 
There are only left-chiral \cite{c3Lee:1956qn,c3Wu:1957my}
(spin anti-parallel to the momentum direction) 
neutrinos in nature, without any right-handed counter part. But there
exists anti-neutrinos of right chirality (spin parallel to momentum).
\item 
Neutrinos are exactly mass less in the SM.
\item 
Since the neutrinos are massless within the framework of
the SM, the mass basis and the weak interaction basis are same for the charged leptons.
In other words, there exists no leptonic analogue
of the CKM matrix (see ref.\cite{c3Cabibbo-1963,c3Kobayashi-1973})
with vanishing neutrino mass.
\end{enumerate}

The massless neutrinos seem to work fine with the SM until the 
first hint of neutrino oscillation appeared in 1969 \cite{c3Davis:1968cp}, 
which requires massive neutrinos!\footnote{The first idea of neutrino
oscillation was given by Bruno Pontecorvo  
\cite{c3Pontecorvo:1957cp,c3Pontecorvo:1957qd}.}. However, maintaining
the gauge invariance, Lorentz invariance and renormalizability, there is
absolutely no room for massive neutrinos in the SM (see reviews
\cite{c3Akhmedov:1999uz,c3Mohapatra:2005wg}). It is then apparent
that to explain neutrino oscillation the SM framework requires extension.
We leave these modifications for time being until section \ref{neut-mass}.
It is rather more important to know the phenomenon of neutrino oscillation. 
Besides, it is also important to know if neutrinos posses non-zero mass,
what will be the possible experimental impressions?

\section{{\bf N}eutrino oscillation}\label{neut-osc}

\begin{flushleft}
{\it{$\maltese$ Evidences of neutrino oscillation}}
\end{flushleft}

\noindent
{\bf{I. Atmospheric neutrino problem $\blacktriangleright$}}
Consider the atmospheric neutrinos, which are coming from the interaction
of cosmic rays with the earth's atmosphere. The charged pion $(\pi^\pm)$
produced in the interaction, has the following decay chain
\bea
\pi^\pm \to \mu^\pm + \nu_\mu(\ovl{\nu_\mu}),
\label{neut-atm1}
\eea
followed by
\bea
\mu^\pm \to e^\pm + \nu_e(\ovl{\nu_e}) + \ovl{\nu_\mu}({\nu_\mu}).
\label{neut-atm2}
\eea
These neutrinos(anti-neutrinos) take part in charge current interaction 
(see figure \ref{CC-NC}) and produce detectable charged leptons. Looking
at eqns. (\ref{neut-atm1}, \ref{neut-atm2}) one would naively expect 
number wise\footnote{This number is actually not exactly $2$, because of
various uncertainties like, geometry of cosmic ray flux and neutrino flux, 
solar activities,uncertainty in cross section measurements, etc.},
\beq
R_{\frac{\mu}{e}}=\frac{\nu_\mu(\ovl{\nu_\mu})}{\nu_e(\ovl{\nu_e})} 
= 2.
\label{neut-atm3}
\eeq
However, in reality $R_{\frac{\mu}{e}}$ is much smaller $(\sim 0.6)$, as observed by
experiments like Kamiokande \cite{c3Hirata:1992ku,
c3Fukuda:1994mc}, NUSEX \cite{c3Aglietta:1988be}, IMB \cite{c3Casper:1990ac,c3BeckerSzendy:1992hq},
Soudan-2 \cite{c3Allison:1996yb}, MACRO \cite{c3Ambrosio:1998wu,c3Ronga:1998ei},
Super-Kamoikande \cite{c3Fukuda:1998mi,c3Fukuda:1998ah}. The diminution in $R_{\frac{\mu}{e}}$ 
as observed by a host of experiments indicates a deficit of muon (anti)neutrino flux. 
This apparent discrepancy between predicted and observed neutrino flux defines the 
atmospheric neutrino problem.


\vspace{0.2cm}
\noindent
{\bf{II. Solar neutrino problem $\blacktriangleright$}}
The Sun gets huge energy by fusing hydrogen $(^1_1{\rm H})$ to helium $(^4_2{\rm He})$
in thermonuclear reactions. There exist a few viable candidates for this reaction
chain, like proton-proton $(pp)$ cycle, carbon-nitrogen-oxygen (CNO) cycle 
\cite{c3Bethe:1939bt,c3Bahcall:2000xc} etc, 
although the $pp$ cycle appears to be the dominant one. The sun is a major
source of electron neutrinos (see also ref.\cite{c3Bahcall:1995gw,c3Bahcall:2000kh}) 
following the process
\beq
4p \to {^4_2{\rm He}} +  2 e^+ + 2 \nu_e,
\label{neut-sol1}
\eeq
where $e^+$ is a positron. There exist a host of literature concerning
the standard solar model \cite{c3Bahcall:1987jc,c3Bahcall:1989ks,c3TurckChieze:1988tj,
c3TurckChieze:1993dw,c3Bahcall:1992hn,c3Bahcall:1998wm}, which account for the 
number of solar neutrinos expected to be detected in an earth based detector. 
However, only one-third of the expected solar neutrino flux has been detected
by experiments like Homestake \cite{c3Davis:1968cp,c3Davis:1994jw,c3Cleveland:1998nv},
SAGE \cite{c3Abdurashitov:1996dp,c3Abdurashitov:1999zd,
c3Abdurashitov:2009tn}, GALLEX \cite{c3Anselmann:1992um,
c3Hampel:1998xg}, GNO \cite{c3Altmann:2000ft},
Kamiokande\cite{c3Hirata:1990xa}, Super-Kamiokande \cite{c3Fukuda:2001nj,
c3Cravens:2008zn}, SNO \cite{c3Boger:1999bb,c3Ahmad:2002jz,
c3Aharmim:2005gt} etc. The disappearance of a large fraction of solar neutrinos
defines the solar neutrino problem.

There were numerous attempts to explain the discrepancy between the measured
and the predicted neutrino flux for the solar and the atmospheric neutrinos. 
In fact, these neutrino deficits lead to the proposal of various theoretical 
models\footnote{A discussion of these models is beyond the scope of this thesis.
See ref.\cite{c3Vicente:2011pf} for further discussions.}.
However, with the idea of Bruno Pontecorvo \cite{c3Pontecorvo:1957cp,
c3Pontecorvo:1957qd,c3Pontecorvo:1967fh}, it seems more logical to think
about some sort of {\it{conversion}} among neutrino flavours while they
propagate through vacuum or matter, which can lead to diminution of
a specific type of flavor as reported by experiments.

\begin{flushleft}
{\it{$\maltese$ Theory of neutrino oscillation}}
\end{flushleft}

In order to explain the Solar and atmospheric neutrino 
deficits, as discussed earlier it is expected that a neutrino of a 
specific flavour, say $a$, during propagation can alter its flavour to 
some other one, say $b$, at a later stage of time. Now from our knowledge of
quantum mechanics it is evident that, 

\begin{enumerate}
 \item 
The set of linearly independent mass eigenstates form a complete
basis.
\item
Any arbitrary state can be expressed as a linear combination of
the linearly independent mass eigenstates.
\end{enumerate}

So, if neutrinos oscillate, 
\cite{c3Kayser:1981ye,c3Giunti:1991ca,c3Rich:1993wu,c3Grossman:1996eh}
the flavour eigenstates, $\nu_e,\nu_\mu,\nu_\tau$ must
differ from the physical or mass eigenstates and it is possible to express 
them as a linear combination of the neutrino mass eigenstates, 
$\nu_1,\nu_2,\nu_3$\footnote{Assuming three active light neutrino flavour 
\cite{c3Nakamura-c2}. There are controversies
concerning more than three light neutrino flavours 
\cite{c3Athanassopoulos:1996jb,c3Athanassopoulos:1997pv,
c3AguilarArevalo:2007it,c3AguilarArevalo:2010wv}. 
This may be the so-called sterile neutrino which
mixes with three light neutrinos, but is phobic to weak interactions, so that
invisible decay width of $Z$-boson remains sacred. Nevertheless, there
exists literature \cite{c3Ma:1995gf,c3Goswami:1995yq,c3Sarkar:1995dd,c3Gaur:1998uk} 
which deals with more than three neutrino species.}. Thus, we define
\bea 
|\nu'_a\rangle = U^*_{a i} |\nu_i\rangle,
\label{neut-osc-1}
\eea
where $\nu'_a,\nu_i$ are flavour and mass eigenstates for neutrinos, 
respectively and $U^*_{a i}$ are the coefficients, carrying information of 
``conversion''. So if at time, $t=0$ we have a flavour state $\nu_a$,
then the probability for transforming to another flavour state $\nu_b$
at a later time $t$ is given by (using eqn.(\ref{neut-osc-1})),
\bea 
P(\nu_a\to \nu_b;t) =  \sum_{j} |U_{bj}e^{-i E_j t} U^*_{aj}|^2.
\label{neut-osc-2}
\eea
Eqn.(\ref{neut-osc-2}) is the key equation for neutrino oscillation
and the underlying physics can be explained in three pieces,

\vspace{0.1cm}
\noindent
I. $U^*_{aj}$ is the amplitude of transformation of a flavour state $\nu_a$
into some mass eigenstate $\nu_j$.

\vspace{0.1cm}
\noindent
II. Immediately after that, the factor $e^{-i E_j t}$ governs the evolution
of mass eigenstate $\nu_j$ with time.

\vspace{0.1cm}
\noindent
III. Finally, $U_{bj}$ is the amplitude of transformation of a time evolved
mass eigenstate $\nu_j$ into some other flavor state $\nu_b$.

A bit of algebraic trick for relativistic neutrinos of momentum $p$, 
($E_j \simeq p + \frac{m^2_j}{2E}$) yields
\bea 
P(\nu_a\to \nu_b;t) =  \sum_{j,k} U^*_{bk} U_{ak} U_{bj} U^*_{aj}
e^{-i \Delta m^2_{jk} \frac{L}{2E}},
\label{neut-osc-3}
\eea
where $\Delta m^2_{jk} = m^2_{j} - m^2_{k}$. $m_i$ is the mass of $\nu_i$ 
state and $L\simeq t$ (L is the distance traversed by a neutrino in time $t$ 
to change its flavour) using natural unit system for relativistic neutrinos.
It is clear from eqn.(\ref{neut-osc-3}) that oscillation probability
depends on the squared mass differences rather than individual masses,
thus it is impossible to probe the absolute mass scale for neutrinos
with oscillation data. 

It is important to note from eqn.(\ref{neut-osc-3}), one can define
the survival probability for a flavour $\nu_a$ as
\bea 
P(\nu_a\to \nu_a;t) =  1- \sum_{j,k} U^*_{ak} U_{ak} U_{aj} U^*_{aj}
e^{-i \Delta m^2_{jk} \frac{L}{2E}}.
\label{neut-osc-4}
\eea
With the aid of eqn.(\ref{neut-osc-4}), deficit of a particular flavour
in the solar and the atmospheric neutrino flux can be explained.
However, even using eqn.(\ref{neut-osc-4}) it is hardly possible to 
account for the solar neutrino problem. This
was an apparent puzzle until the matter effects in the enhancement of neutrino 
oscillation were understood. Eqn.(\ref{neut-osc-4}) works only for 
oscillations in vacuum\cite{c3Gribov:1968kq}. The much desired modification
for explaining matter effect induced enhanced oscillations to accommodate
the solar neutrino deficit was given by Mikheyev, Smirnov and Wolfenstein 
\cite{c3Wolfenstein:1977ue,c3Mikheev:1986gs,c3Mikheev:1986wj}. 
This is popularly known as the MSW effect.

\begin{flushleft}
{\it{$\maltese$ What do we know about oscillations?}}
\end{flushleft}

It has been argued already that the theory of neutrino oscillation
is sensitive to squared mass differences. It is also confirmed by this 
time that, it is indeed possible to explain oscillation phenomena
with two massive neutrinos, consequently, two squared mass differences
are enough. We define them as $\Delta m^2_{solar}$ and $\Delta m^2_{atm}$,
where the word atmospheric is abbreviated as {\it{atm}}. From the observational 
fact $\Delta m^2_{atm}$($\sim$ $10^{-3} {\rm eV}^2$) $\gg$ 
$\Delta m^2_{solar}$($\sim$ $10^{-5} {\rm eV}^2$). The sign of $\Delta m^2_{solar}$
has been affirmed experimentally to be positive, but $\Delta m^2_{atm}$
can be either positive or negative. With this sign ambiguity, two types
of light neutrino mass spectrum are possible, namely {\it{normal}} and
{\it{inverted}}. Mathematically, (i) normal hierarchy: 
$m_1 < m_2 \sim \sqrt{\Delta m^2_{solar}}$ ,~$m_3 \sim
\sqrt{|\Delta m^2_{atm}|}$, (ii) inverted hierarchy: $m_1 \approx m_2 \sim
\sqrt{|\Delta m^2_{atm}|}$, $m_3 \ll \sqrt{|\Delta m^2_{atm}|}$, where
$m_1,m_2,m_3$ are light neutrino masses\footnote{It is useful to note that
$m_2>m_1$ irrespective of mass hierarchy, since $\Delta m^2_{solar}>0$ always.
However, $\Delta m^2_{atm}>0$ for normal hierarchy whereas $<0$ for the 
inverted one.}. There exists a third possibility
of light neutrino mass ordering, where $m_1 \approx m_2 \approx m_3 \gg 
\sqrt{|\Delta m^2_{atm}|}$ with finely splitted $m_i$s in
order to satisfy oscillation data. This is known as the quasi-degenerate
spectrum. Note that, it is impossible to
accommodate quasi-degenerate spectrum unless all three neutrinos are massive
whereas for the normal or inverted hierarchical scheme of light neutrino mass
at least two neutrinos must be massive \cite{c3Fukuda:1998mi,c3Ahmad:2002jz,c3Eguchi:2002dm}.

Probability of flavour oscillation also contains the elements of
conversion matrix, $U$ (eqn.(\ref{neut-osc-3})). The matrix $U$
acts as the bridge between flavour and mass eigenstates, having
one index from both the basis. This matrix is the leptonic analogue
of the CKM matrix (see chapter \ref{SM}) and is known as the 
Pontecorvo-Maki-Nakagawa-Sakata or the PMNS matrix 
\cite{c3Pontecorvo:1957cp,c3Pontecorvo:1957qd,c3Maki:1962mu,c3Pontecorvo:1967fh}. 
In three flavour model this can be parametrized as 
\cite{c3Schechter:1980gr,c3Chau:1984fp,c3Eidelman:2004wy}
\beq
U =
\left(\begin{array}{ccc}
{c_{12}}{c_{13}} & {s_{12}}{c_{13}} & {s_{13}} e^{-i\del} \\ \\
-{s_{12}}{c_{23}}-{c_{12}}{s_{23}}{s_{13}}e^{i\del} & {c_{12}}{c_{23}}
-{s_{12}}{s_{23}}{s_{13}}e^{i\del}  & {s_{23}}{c_{13}}\\ \\
{s_{12}}{s_{23}}-{c_{12}}{c_{23}}{s_{13}}e^{i\del} & -{c_{12}}{s_{23}}
-{s_{12}}{c_{23}}{s_{13}}e^{-i\del}  & {c_{23}}{c_{13}}
\end{array}\right).U'(\al),
\label{PMNS1}
\eeq 
where $c_{ij} = \cos{\theta_{ij}}$, $s_{ij} = \sin{\theta_{ij}}$ and
$U'(\al) = diag(e^{-i\al_1/2},1,e^{-i\al_2/2})$.
Here $\al_1,\al_2,\del$ are complex phases. The phases $\al_1,\al_2$
can be non-zero only if neutrinos are Majorana particle in nature
(will be addressed later). Neutrino oscillation is insensitive to Majorana phases.
The phase $\del$ is a Dirac CP-violating phase and can appear in oscillation
dynamics. We stick to Charge-Parity(CP)-preserving case (zero phases) 
throughout this thesis.

It is interesting to note that unlike the CKM matrix of quark sector,
the PMNS matrix has a rather non-trivial structure. Present
experimental evidence favours a tri-bimaximal mixing in
light neutrino sector \cite{c3Harrison:2002er}, though there exist other alternatives 
\cite{c3Harrison:1994iv,c3Altarelli:1998sr,c3Barger:1998ta,c3Baltz:1998ey,c3Barger:1998ed
,c3Bjorken:2005rm} (see also \cite{c3Mohapatra:1998ka} and references 
therein)\footnote{Some of these proposals are now experimentally ruled out.}.
The atmospheric mixing angle $(\theta_{23})$ is close to maximal $(\sim \pi/4)$ and the 
solar mixing angle $(\theta_{12})$ is also large $(\sim 35^\circ)$. The third
and the remaining mixing angle, the reactor angle $(\theta_{13})$ is 
the most difficult one to measure. There exist a host of literature, 
both on theoretical prediction and experimental observation
for the value and the measurement of the angle $\theta_{13}$.
(see ref. \cite{c3Ardellier:2006mn,c3Guo:2007ug,c3Fogli:2008jx,c3Schwetz:2008er,
c3Maltoni:2008ka,c3GonzalezGarcia:2010er} for recent updates. Also see ref.
\cite{c3Mezzetto:2009cr}). At the zeroth-order approximation 
\beq
\theta_{23} = \frac{\pi}{4},~\theta_{12} \simeq 35^\circ,~\theta_{13} =0.
\label{PMNS2}
\eeq 

Recently, non-zero value for the angle $\theta_{13}$ has been
reported by the T2K collaboration \cite{Abe:2011sj} both
for the normal and inverted hierarchy in the light neutrino masses.
For normal (inverted) hierarchy and at $90\%$ C.L. this value
is reported to be
\beq
0.03(0.04)< {\rm sin}^2{2\theta_{13}} < 0.28(0.34) .
\label{T2K-13}
\eeq 

The oscillation parameters $(\Delta m^2_{sol},\Delta m^2_{atm},\theta_{23},
\theta_{12},\theta_{13})$ are highly constrained by experiments.
In table \ref{osc-para} best-fit values of these parameters for the global 
three-flavor neutrino oscillation data are given \cite{c3Schwetz:2008er}.
\begin{table}[ht]
\centering
\begin{tabular}{c || c || c}
\hline \hline 
Parameter  & Best fit  & 3$\sigma$ limit\\ \hline \hline
$\Delta m^2_{sol} \times 10^5$ ${\rm eV}^2$   & 
$7.65^{+0.23}_{-0.20}$ & $7.05 - 8.34$\\
$|\Delta m^2_{atm}| \times 10^3$ ${\rm eV}^2$ 
& $2.40^{+0.12}_{-0.11}$ & $2.07 - 2.75$\\
${\rm sin}^2\theta_{23}$
& $0.50^{+0.07}_{-0.06}$ & $0.25 - 0.37$ \\ 
${\rm sin}^2\theta_{12}$
& $0.304^{+0.022}_{-0.016}$ & $0.36 - 0.67$ \\ 
${\rm sin}^2\theta_{13}$
& $0.01^{+0.016}_{-0.011}$ & $\leqslant 0.056$ \\  \hline \hline
\end{tabular}
\caption{\label{osc-para}
Best fit values and $3\sigma$ ranges of oscillation parameters from
three flavour global data \cite{c3Schwetz:2008er}.
}
\end{table}
The experiments like Borexino \cite{c3Arpesella:2008mt},
CHOOZ \cite{c3Apollonio:1999ae,c3Apollonio:2002gd}, 
Double Chooz \cite{c3Ardellier:2004ui,c3Ardellier:2006mn},
KamLAND \cite{c3:2008ee,c3Gando:2010aa}, Kamiokande \cite{c3Wendell:2010md},
Super-Kamiokande \cite{c3Cravens:2008zn,c3Abe:2010hy},
K2K \cite{c3Ahn:2006zza}, MINOS \cite{c3Adamson:2008zt,c3Adamson:2011ig},
GNO \cite{c3Kirsten:2003ev,c3Altmann:2005ix}, SNO \cite{c3Aharmim:2009gd} 
and others are now in the era of precision measurements. More, sophisticated experiments
like RENO \cite{c3Ahn:2010vy}, OPERA \cite{c3Acquafredda:2006ki,c3Acquafredda:2009zz} etc. 
have already been initiated and an extremely precise global fit
is well anticipated in near future. One can go through ref. \cite{c3Schwetz:2011qt}
for a recent analysis of the precision results. 
 
\vspace*{0.5cm}
\begin{flushleft}
{\it{$\maltese$ Searching for a neutrino mass}}
\end{flushleft}

Theory of neutrino oscillation depends on squared mass differences, which
are shown in table \ref{osc-para}. It is then, natural to ask that what is
the absolute scale for a neutrino mass. Is it small, $\sim$ a few eV so that
small squared mass differences (see table \ref{osc-para}) seem natural or
the absolute masses are much larger and have unnatural fine splittings
among them. 

Possible evidences of absolute neutrino mass scale can come from different
experiments which are discussed below.

\begin{figure}[ht]
\centering
\includegraphics[width=7.05cm]{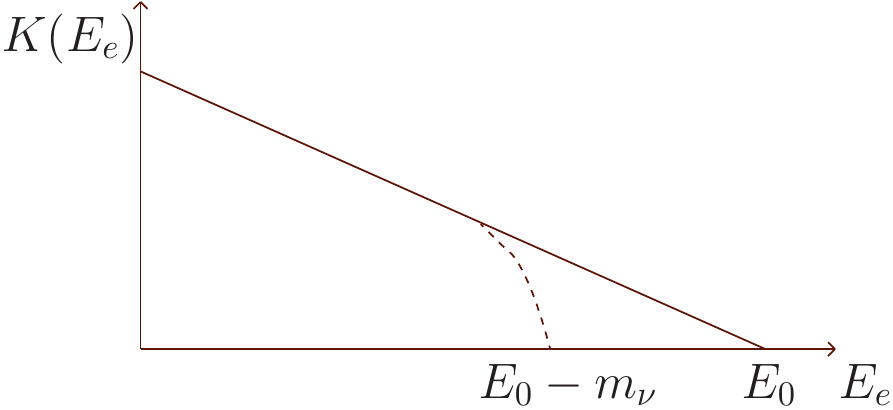}
\caption{Kurie function, $K(E_e)$ versus energy $(E_e)$ of $\beta$-particle $(e^-)$ plot 
for neutrino mass, $m_\nu=0$ (solid line) and $m_\nu \neq 0$ (dashed line). $E_0$ is the
energy release.}
\label{Kurie}
\end{figure}

\vspace{0.1cm}
\noindent
I. {\bf{Tritrium beta decay}} $\blacktriangleright$ There are a host of experimental
collaboration (Mainz \cite{c3Bonn:2001tw,c3Kraus:2004zw}, 
Troitsk \cite{c3Belesev:1995sb,c3Lobashev:2001uu}, 
KARTIN \cite{c3Osipowicz:2001sq,c3Lobashev:2003kt}) looking for the 
modification in the beta spectrum in the process $^3{\rm H}\to {^3{\rm {He}}}+e^-+\ovl{\nu_e}$ 
in the presence of non-zero $m_i$. Indeed, the Kurie plot \cite{c3PhysRev.49.368} 
shows deviation near the endpoint with $m_i\neq0$ (see figure \ref{Kurie}). 
The experiments, however in reality
measure an effective neutrino mass $m_\beta = \sqrt{\sum |U_{ei}|^2 m^2_i}$ 
($U$ is the PMNS matrix). The present bound is \cite{c3Nakamura-c2}
\bea 
m_\beta < 2.0 ~{\rm eV}.
\label{neut-nonosc-1}
\eea

\begin{figure}[ht]
\centering
\includegraphics[width=3.15cm]{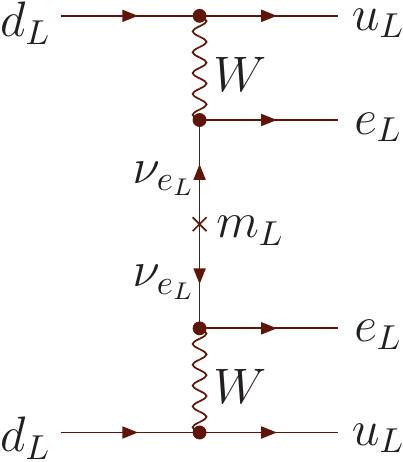}
\caption{Diagram for neutrinoless double beta decay in the SM. 
Subscript $L$ signifies the left handed nature of weak interaction.}
\label{0nubb1}
\end{figure}

\vspace{0.1cm}
\noindent
II. {\bf{Neutrinoless double beta decay}} $\blacktriangleright$ Consider two beta decays, 
($n\to p^++e^-+\nu_e$ or $d\to u+e^-+\nu_e$, in the quark level of proton (neutron) 
simultaneously, such that (anti)neutrino emitted in one decay is
somehow absorbed in the other. This leads to the neutrinoless double beta decay, 
$0\nu\beta\beta$ (figure \ref{0nubb1}).
However, it is clear from figure \ref{0nubb1}, this breaks lepton
number conservation. The quantity $m_L$ represents Majorana mass 
(will be addressed soon) of a left-handed neutrino, which is responsible 
for this lepton number violation $(\L)$.
Not only the lepton number is broken in this interaction, but
through Majorana mass term $m_L$ process like this also breaks
chirality conservation \cite{c3Beshtoev:2009zz}. The measurable quantity is defined as
$m_{\beta\beta} = \sum U_{ei}^2 m_i$. Since
$m_{\beta\beta} \propto U_{ei}^2$, rather than $|U_{ei}|^2$, information
about the $\del$-phase is not lost until one asks for CP-preservation
in the lepton sector. Experimental reporting of neutrinoless double beta decay
is controversial. The result obtained
by Heidelberg-Moscow experiment \cite{c3Baudis:1999xd,
c3KlapdorKleingrothaus:2004wj,
c3KlapdorKleingrothaus:2006ff}, CUORICINO \cite{c3Arnaboldi:2008ds} suggests
\bea 
m_{\beta\beta} < 0.2-0.6 ~{\rm eV}.
\label{neut-nonosc-2}
\eea
However, there are experiments like CUORE \cite{c3Arnaboldi:2008ds}, 
EXO \cite{c3Gornea:2010zz}, GERDA \cite{c3Smolnikov:2008fu}, MAJORANA and
a few others, which are expected to shed light on this 
claim in near future. One important point about $0\nu\beta\beta$ is 
that unless a neutrino possesses a Majorana mass term,
$m_{\beta\beta}=0$. This is true for different models 
and has been confirmed by model independent analysis \cite{c3Schechter:1981bd}. 
\begin{figure}[ht]
\centering
\includegraphics[width=3.15cm]{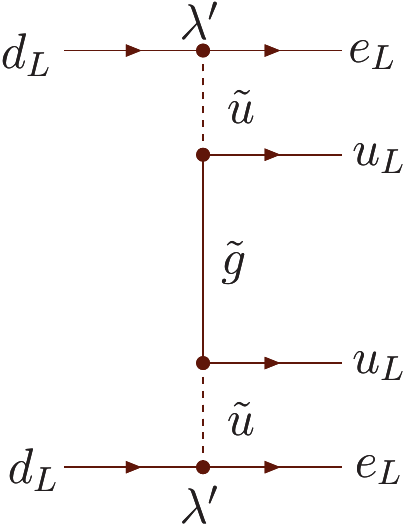}
\caption{Diagram for neutrinoless double beta decay in $R_p$-violating 
supersymmetric models. See text for further details.}
\label{0nubb2}
\end{figure}
It must be emphasized here that, actually the presence
of Majorana mass term is a sufficient condition for non-zero $m_{\beta\beta}$,
but not necessary, for example $0\nu\beta\beta$ in $R_p$-violating  
(section \ref{R-parity}) supersymmetric model (see figure \ref{0nubb2})
can occur without a neutrino Majorana mass term (see ref. \cite{c3Hirsch:1995ek}). 
In figure \ref{0nubb2},
$\tilde g$ represents a gluino, superpartner of a gluon and $\tilde u$ is a 
up-type squark (see figure \ref{MSSM-particle}).

\vspace{0.1cm}
\noindent
III. {\bf{Cosmology}} $\blacktriangleright$ Neutrino masses are also constrained by 
the standard big-bang cosmology. However, in this case the bound exists
on sum of neutrino masses. There were earlier works \cite{c3Gershtein:1966gg,
c3Cowsik:1972gh,c3Szalay:1976ef} in this connection, where, a bound on the sum 
of neutrino mass was obtained from the bound on the 
total density of the universe. However, the present bound as obtained from
sophisticated experiment like WMAP \cite{c3Komatsu:2008hk,c3Dunkley:2008ie,
c3Hinshaw:2008kr,c3Komatsu:2010fb,c3Larson:2010gs,c3Jarosik:2010iu} is much 
stringent and is given by
\bea 
\sum_{i=1}^{3} m_{i} \leq 0.58 ~{\rm eV}(95\%~{\rm C.L.}).
\label{neut-nonosc-3}
\eea
Note that only for the case of quasi-degenerate
light neutrino masses individual masses can be much larger
compared to the squared mass differences (see table \ref{osc-para}).
Thus quasi-degenerate
neutrinos masses are highly constrained by eqn.(\ref{neut-nonosc-3}). 

So far we have addressed the features of oscillation and non-oscillation
experiments to constrain the neutrino physics. It is the time to
demonstrate the origin of neutrino mass. But even before that, it
is important to discuss the nature of neutrino masses, that is
whether they are Dirac or Majorana particles \cite{c3Fiorini:2006cg,
c3Kayser:2009zz}.

\section{{\bf M}odels of neutrino mass}\label{neut-mass}

\begin{flushleft}
{\it{$\maltese$ Nature of neutrino mass, Dirac versus Majorana}}
\end{flushleft}

It is well-known that the charge conjugation operator
$\hat C$ is defined as
\bea 
\hat C: \psi \to \psi^c = C {\bar{\psi}}^T,
\label{neut-mass-a1}
\eea
where $C$ is the matrix representation for $\hat C$, $T$ denotes transpose
and $\psi$ is a four component spinor. It is then apparent that for
a charge neutral fermion
\bea 
\psi^c = \psi.
\label{neut-mass-a2}
\eea
Any $\psi$ which obeys eqn.(\ref{neut-mass-a2}) is known as a 
Majorana fermion\footnote{A free Majorana field is an eigenstate of charge conjugation operator.}. 
On the contrary, the so-called Dirac fermions are
known to follow $\psi\neq\psi^c$. Now, since the neutrinos are
the only charge neutral particle in the SM there is a possibility, 
that a neutrino is a Majorana particle, 
\cite{c3Majorana:1937vz} rather than a Dirac particle \cite{c3Dirac:1928hu}.

\vspace{0.1cm}
\noindent
I. {\bf{Dirac Mass}} $\blacktriangleright$ If there were right-handed neutrinos 
$(\nu_R)$ in the SM, a non-zero Dirac mass $(m_D)$ is well expected. The 
mass term using eqn.(\ref{fermion-mass}) can be written as
\bea 
y_\nu \ovl{L} \tilde{\Phi} \nu_R~+~h.c = \frac{y_\nu .v}{\rt2} \ovl{\nu_L} \nu_R
~+~h.c= m_D \ovl{\nu_L} \nu_R~+~h.c,
\label{neut-mass-a3}
\eea 
where $\nu_{L(R)}$ is a left(right) handed neutrino field and $y_\nu$
is the neutrino Yukawa coupling. Demanding a neutrino mass $\sim$
1 eV one gets $y_\nu$ $\sim$ $10^{-11}$. But immediately then, it
is legitimate to ask why $m_D$ is extremely small compared to other 
masses as shown in table \ref{SM-spectrum} or alternatively why
$y_\nu$ is much smaller, compared to say electron Yukawa coupling, $Y_e
$ $\sim$ $10^{-6}$. 
The Dirac mass terms respect the total lepton number $L$, but not the 
family lepton number, $L_e,L_\mu,L_\tau$. If $m_D\neq0$, a neutrino
is different from an anti-neutrino.

\vspace{0.1cm}
\noindent
II. {\bf{Majorana Mass}} $\blacktriangleright$ A Majorana mass term not only violate
the family lepton number, but also the total lepton number. In
general they are given by
\bea 
m_{ii} \ovl{\nu_i} \nu^c_i,
\label{neut-mass-a4}
\eea 
where, $i=L,R$. $\nu^c_{L(R)}$ represents a CP conjugated state.
A Majorana spinor has only two degrees of freedom because of 
eqn.(\ref{neut-mass-a2}), whereas a Dirac spinor has four degrees of freedom.
Thus, two degenerate Majorana neutrinos of opposite CP, when maximally
mixed form a Dirac neutrino. A Majorana neutrino is believed to
be one of the main reasons for non-vanishing amplitude in $0\nu\beta\beta$
(see section \ref{neut-osc}). However, just like the Dirac case
it is also important to explain how a neutrino gets a tiny Majorana
mass? The answer will be given in the subsequent paragraph.

In the most general occasion, a neutrino can posses a ``Dirac + Majorana'' mass
term. A term of this kind can lead to extremely interesting 
neutrino-anti neutrino oscillation which violates total lepton 
number (see ref. \cite{c3Bilenky:1998dt} for detailed discussion).
It is also important to mention that since neutrino oscillation
does not violate total lepton number, it is hardly possible to
differentiate between a Dirac and a Majorana nature from the 
theory of neutrino oscillation. The $0\nu\beta\beta$ is definitely
an evidence for Majorana nature. Besides, one can use the favour
of electric and magnetic dipole moment measurement, to discriminate
these scenarios \cite{c3Bernstein:1963qh,c3Okun:1986hi,c3Schechter:1981hw,
c3Nieves:1981zt,c3Kayser:1982br,c3Shrock:1982sc,c3Babu:1987be,c3Raffelt:1999gv}. 

\subsection{{\bf M}ass models I}\label{neut-mass-I}

It is apparent by now that we need to extend either the particle
content of the SM or need to enlarge the gauge group to accommodate neutrino
mass. In this subsection we discuss the models for neutrino
mass generation without introducing supersymmetry. Some of these models
generate masses at the tree level and the remaining through loop
processes. 

\vspace*{0.02cm}
\begin{center}
{\bf{1. Renormalizable interaction with triplet Higgs}}. 
\end{center}
\vspace*{0.02cm}
Consider a term in the Lagrangian as
\bea 
f_{\Delta} (\ell^T C i \sigma_2 {\bm{\sigma}} \ell) {\Delta} ~+h.c,
\label{neut-mass-I1}
\eea 
where $C$ is the charge conjugation matrix, ${\bm{\sigma}}$'s are Pauli 
matrices and $\Delta$ is a {\it{triplet}} Higgs field with hypercharge,
$Y=2$. If further we assume that ${\Delta}$ has $L=-2$, then Lagrangian
given by eqn.(\ref{neut-mass-I1}) conserves lepton number. The mass term
for neutrinos will be then
\bea 
m_\nu \approx {f_\Delta v_\Delta},
\label{neut-mass-I2}
\eea 
where $v_\Delta$ is the VEV of ${\Delta}$ field. But this will also 
produce massless triplet Majoron \cite{c3Gelmini:1980re,c3Chikashige:1980ui,
c3Georgi:1981pg} since lepton number is spontaneously
broken by the VEV of $\rm{SU(2)_L}$ triplet ${\Delta}$ field. Missing 
experimental evidence has ruled out this model. 
One alternative could be to put $L=0$ for ${\Delta}$, which breaks the lepton number
explicitly. This situation, though free of triplet Majoron are highly
constrained from $\rho$ parameter measurement (eqn.(\ref{rho-param})), which requires 
$v_\Delta<8$ GeV. Once again for $m_\nu \sim$ 1 eV, $f_\Delta \sim$
$10^{-10}$.

\vspace*{0.05cm}
\begin{center}
{\bf{2. Non-renormalizable interactions}}. 
\end{center}
If we want to build a model for tiny neutrino mass with the SM
degrees of freedom, the immediate possibility that emerges is the
so-called {\it{dimension five}} Weinberg operator \cite{c3Weinberg:1979sa,c3Weinberg:1980bf}
\bea 
Y_{ij}\frac{(\ell_i \Phi)(\Phi \ell_j)}{M},
\label{neut-mass-I3}
\eea 
where $\ell$ are the SM ${\rm{SU(2)}_L}$ lepton doublets (eqn.(\ref{SM-gauge-group})) 
and $\Phi$ is the SM Higgs doublet (eqn.(\ref{SM-Higgs})) with VEV $\frac{v}{\rt2}$.
$Y_{ij}$ stands for some dimensionless coupling. $M$ is some high scale in the theory, 
and is the messenger of some new physics. Thus the small Majorana neutrino mass 
coming from this $\Delta L = 2$ operator is
\bea 
(m_\nu)_{ij}=\frac{Y_{ij} v^2}{2 M}.
\label{neut-mass-I4}
\eea 
Note that if $M$ is large enough ($\sim$ $10^{14}$ GeV) the coupling, 
$Y_{ij} \sim$ 1 (close to perturbative cutoff), for right order of 
magnitude in the neutrino mass $(m_\nu)$ $\sim 0.1$ eV. However,
this is a rather challenging scenario, since it is hardly possible
to probe $M$ ($\sim$ $10^{14}$ GeV) in a collider experiment. 
One viable alternative is a TeV scale $M$, 
which is possible to explore in a collider experiment.
Note that for such a choice of $M$, $Y_{ij}$ is much smaller. 

Maintaining the gauge invariance and renormalizability, the effective
operator can arise from three possible sources.

\begin{figure}[ht]
\centering
\includegraphics[width=9.95cm]{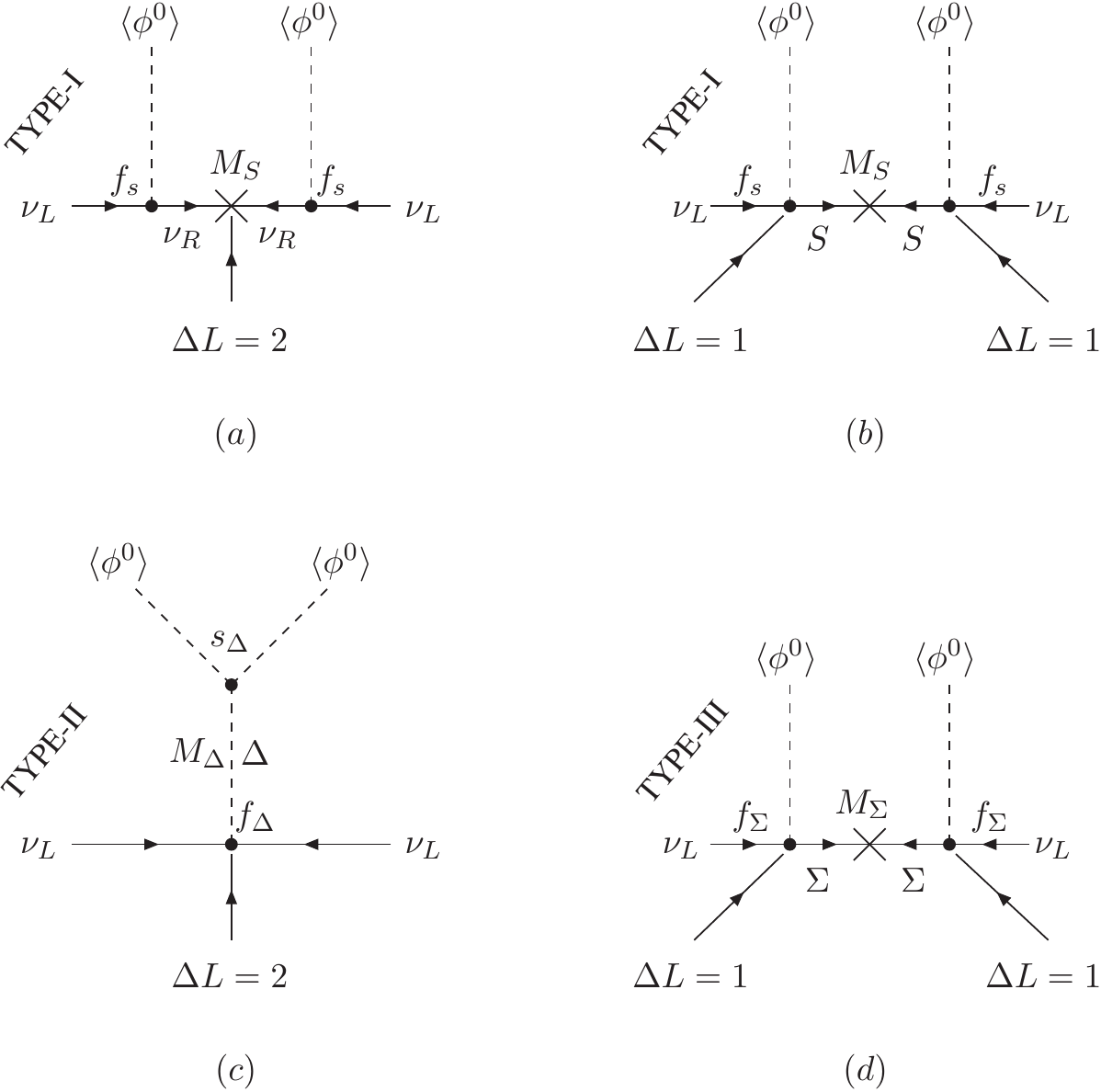}
\caption{Different types of seesaw mechanism. The cross on the fermionic propagator 
signifies a Majorana mass term for the corresponding fermion.}
\label{seesaw}
\end{figure}

\vspace{0.1cm}
\noindent
I. {\bf{Fermion singlet}} $\blacktriangleright$ The intermediate massive particle is
a SM gauge singlet particle, $(S)$. This is the example of Type-I seesaw
mechanism \cite{c3Minkowski:1977sc,c3Yanagida:1979as,c3Glashow:1979nm,c3Mohapatra:1979ia,
c3GellMann:1980vs,c3Schechter:1980gr,c3Schechter:1981cv} (seesaw mechanism will be 
discussed later in more details). 
The light neutrino mass is given by  
\bea 
m_\nu=\frac{f^2_s v^2}{2 {M_S}},
\label{neut-mass-I5}
\eea 
where $M_S$ is the mass of the singlet fermion and $f_s$ is the 
$\ell \Phi S$ coupling. (see figure \ref{seesaw} (a) and (b)). It
is important to note that the $\Delta L=2$ effect can arise either 
using a singlet with non-zero lepton number (right handed neutrino, 
$\nu_R$) (figure \ref{seesaw} (a)) or through a singlet, $S$ without 
lepton number (figure \ref{seesaw} (b)).

\vspace{0.1cm}
\noindent
II. {\bf{Scalar triplet}} $\blacktriangleright$ The intermediate massive particle is
a scalar triplet $(\Delta)$ under the group ${\rm{SU(2)}_L}$. It is singlet under 
${\rm{SU(3)}_C}$ and has hypercharge, $Y = 2$. This is the so-called Type-II 
seesaw mechanism \cite{c3Konetschny:1977bn,c3Magg:1980ut,c3Lazarides:1980nt,c3Cheng:1980qt,
c3Mohapatra:1980yp}. The light neutrino mass is given by  
\bea 
m_\nu=\frac{f_\Delta s_\Delta v^2}{2 {M^2_{\Delta}}},
\label{neut-mass-I6}
\eea 
where $M_\Delta$ is the mass of the scalar triplet. $f_\Delta$
and $s_\Delta$ are the $LL\Delta$ and $\Phi\Phi\Delta$ coupling, 
respectively (see figure \ref{seesaw} (c)).

\vspace{0.1cm}
\noindent
III. {\bf{Fermion triplet}} $\blacktriangleright$ A triplet fermion $(\Sigma)$ acts
as the mediator and this is an illustration of the Type-III seesaw 
mechanism \cite{c3Foot:1988aq,c3Ma:1998dn}.
This $\Sigma$ is a triplet under ${\rm{SU(2)}_L}$ but singlet under 
${\rm{SU(3)}_C}$. Hypercharge for $\Sigma$ is zero. 
The light neutrino mass is given by  
\bea 
m_\nu=\frac{f^2_\Sigma v^2}{2 {M_{\Sigma}}},
\label{neut-mass-I7}
\eea 
where $M_\Sigma$ is the mass of the fermion triplet. $f_\Sigma$
is the $L\Phi\Sigma$ coupling (see figure \ref{seesaw} (d)).

A very important aspect of these seesaw models and the associated 
Majorana nature is that they can produce same sign dilepton at a 
collider experiment \cite{c3Keung:1983uu}
apart from a non-zero amplitude for the $0\nu\beta\beta$ process. 
The collider phenomenology for Type-II or III seesaw models
are more attractive compared to the Type-I scenario, due to
the involvement of a SM gauge singlet fermion in the latter case. Also a seesaw generated
neutrino mass can have implications in flavour violating processes, 
\cite{c3Hall:1983id,c3Lee:1984kr,c3Lee:1984tn,c3Dawson:1985vr,c3Dimopoulos:1988jw}
leptogenesis \cite{c3Fukugita:1986hr,c3Luty:1992un,c3Buchmuller:2004nz,
c3Davidson:2008bu}. However, none of these issues are 
addressed here.

There exist other interesting seesaw models like (a) Inverse seesaw 
\cite{c3Mohapatra:1986bd} (requires an extra SM singlet $S$ apart from $\nu_R$), 
(b) Linear seesaw \cite{c3Malinsky:2005bi}, (c) Double seesaw 
\cite{c3Mohapatra:1986aw,c3Mohapatra:1986bd}, (d) Hybrid seesaw 
\cite{c3Wetterich:1981bx,c3Wetterich:1998vh,c3Antusch:2004xd,c3Chen:2005jm} 
etc. Some of these models have
definite group theoretical motivation. Also neutrino masses can arise
in the left-right symmetric model \cite{c3Pati:1974yy,c3Mohapatra:1974gc,
c3Senjanovic:1975rk,c3Senjanovic:1978ev}. It is important to
note that the Weinberg operator can also give rise to neutrino
mass via loop effects \cite{c3Zee:1980ai,c3Zee:1985id,c3Babu:1988ki,c3Branco:1988ex,
c3Ma:2006km,c3Ma:2008uza}. Some of the very early attempts
in this connection have been addressed in references \cite{c3Georgi:1972hy,c3Cheng:1977ir}. 
However, any more involved discussion of these 
topics are beyond the scope of this thesis. 
A comprehensive information about various neutrino mass models is given 
in ref. \cite{c3Vicente:2011pf}.

\begin{flushleft}
{\it{$\maltese$ The seesaw mechanism}}
\end{flushleft}

It has been already argued that the seesaw mechanism (Type-I,II,III and others)
is perhaps the most convenient way to generate small Majorana masses for neutrinos.
But what is actually a seesaw mechanism and how does it lead to small Majorana mass?
It is true that a Majorana mass term violates lepton number by {\it{two units}},
but this could happen either through a pair of $\Delta L=1$ effects 
(see figure \ref{seesaw} (b),(d)) or by a $\Delta L=2$ vertex 
(see figure \ref{seesaw} (a),(c)). We will discuss the canonical seesaw mechanism
(Type-I, however this analysis is also applicable for Type-III) using a simple model 
containing left-handed neutrino, $\nu_L$ and some fermion $f$, either
a SM gauge singlet (Type-I seesaw) or an ${\rm{SU(2)}_L}$ triplet (Type-III seesaw).
Further we assume that to start with Majorana mass for $\nu_L$ is absent.
Majorana mass for $f$ is given by $M_f$ and the co-efficient of the 
mixing term $(\nu_Lf)$ is written as $m_m$. The mass matrix in the $\nu_L,f$
basis is given by
\bea
\mathcal{M} = \left(\begin{array}{cc}
0 & m_m \\
m_m & M_f
\end{array}\right).
\label{seesaw-mat}
\eea
If $M_f\gg m_m$, the eigenvalues are given as
\bea
m_{light} \simeq -\frac{m^2_m}{M_f},{\rm{~and}}~m_{heavy} \simeq M_f.
\label{seesaw-eigval1}
\eea
If $\chi_1,\chi_2$ form a new basis where 
$\mathcal{M}\to~diag(m_{light},m_{heavy})$, then mixing between
$\chi_1,\chi_2$ and $\nu_L,f$ basis is parametrized by an angle
$\theta$ with 
\bea
{\rm tan}2\theta = \frac{2 m_m}{M_f}.
\label{seesaw-mixing}
\eea
Eqn.(\ref{seesaw-eigval1}) is the celebrated seesaw formula 
for neutrino mass generation. Now the left neutrino possesses
a non-zero Majorana mass term which was {\it{zero}} to start with. Also
the mass $m_{light}$, is suppressed by a factor $\frac{m_m}{M_f}$,
and thus always is small as long as $M_f\gg m_m$. 
Considering three generations of light neutrinos,
the unitary matrix (orthogonal in the absence of complex phases) $U'$ which rotates 
the off-diagonal basis ($m_{light}$ is a $3\times3$ matrix for 
the three generation case) to the
diagonal one is known as the PMNS matrix (eqn.(\ref{PMNS1})). 
Mathematically,
\bea
U^{'^T} m_{light} U' = diag(m_{\nu_i}),~~i=1,2,3.
\label{seesaw-PMNS}
\eea
In the Type-I and Type-III seesaw process, the 
effective leptonic mixing matrix or PMNS matrix looses its unitarity 
\cite{c3Langacker:1988up,c3Antusch:2006vwa,c3Ma:2009du} $\sim \frac{m_m}{M_f}$.
The unitary nature is restored when $M_f\to\infty$. This feature is however
absent in Type-II seesaw mechanism. A discussion on the phenomenological implications
of this non-unitarity is beyond the scope of this thesis.

It is essential to note that when $f$ is a right handed neutrino, $\nu_R$,
then $m_m\equiv m_D$, the Dirac mass term. Further replacing $M_f$ by
$M_R$, in the limit $m_D\gg M_R$, we get from eqn.(\ref{seesaw-eigval1})
\bea
m_{light} \simeq m_D-\frac{M_R}{2},~{\rm{and}}~|m_{heavy}| \simeq m_D+\frac{M_R}{2}.
\label{seesaw-pseudo-Dirac}
\eea
This pair is known to behave as Dirac neutrino in various
aspects and is named as {\it{quasi-Dirac neutrinos}} 
\cite{c3Wolfenstein:1981kw,c3Bilenky:1987ty}.

\subsection{{\bf M}ass models II}\label{neut-mass-II}

In this subsection we try to address the issues of neutrino
mass generation in a supersymmetric theory \cite{c3Hirsch:2004he}, which is 
one of the prime themes of this thesis. $\rpv$ through bilinear terms ($\vp_i$, 
see eqn.(\ref{MSSM-superpotential-2})) is the simplest extension of the MSSM 
\cite{c3Diaz:1997xc}, which provides a framework for neutrino masses and mixing angles 
consistent with experiments. It is important to clarify
that there are various sources for light neutrino mass
generation in supersymmetry without $\rpv$ (section \ref{R-parity}), 
for example see refs. \cite{c3Borzumati:1986qx,c3Hisano:1995cp,c3Hisano:1995nq,
c3Hirsch:1997vz,c3Grossman:1997is,c3Hirsch:1997dm,c3Davidson:1998bi,
c3Aulakh:1999cd,c3Casas:1999ac,c3ArkaniHamed:2000bq,c3Das:2010wp,c3Aoki:2010ib,c3Abada:2010ym}. 
But we stick to a very special case where the origin of neutrino mass 
generation is entirely supersymmetric, namely through $R_p$-violation. 
An introduction to $\rpv$ was given in section \ref{R-parity} and
here we will concentrate only on the effect of $\rpv$ in neutrino mass 
generation.

The effect of $\rpv$ and neutrino masses in a supersymmetric theory
has received immense interest for a long time and there exist a host
of analyses to shed light on different phenomenological
aspects of broken $R_p$ (see section \ref{R-parity} and references therein). 
We quote a few of these references having connections with the theme of this 
thesis, namely (a) neutrino mass generation either through explicit
$\rpv$ \cite{c3Joshipura:1994ib,c3de-Campos:1995av,c3Banks:1995by,
c3Nowakowski:1995dx,c3Diaz:1997xc,c3Akeroyd:1997iq,c3Faessler:1997db,c3Hirsch:1998kc,
c3Chang:1999nu,c3Datta:1999yd,c3Diaz:1999is,c3Aulakh:2000sn,c3Restrepo:2001me,
c3Abada:2006qn,c3Abada:2006gh,c3Hall:1983id,
c3Lee:1984kr,c3Lee:1984tn,c3Ellis:1984gi,c3Dimopoulos:1988jw,
c3Barbieri:1990qj,c3Enqvist:1992ef,c3Hempfling:1995wj,c3de-Carlos:1996du,
c3Borzumati:1996hd,c3Nilles:1996ij,c3Nardi:1996iy,c3Roy:1996bua,c3Drees:1997id,
c3Chun:1998gp,c3Bednyakov:1998cx,c3Ferrandis:1998ii,c3Grossman:1998py,c3Chun:1998ub,
c3Rakshit:1998kd,c3Kaplan:1999ds,c3Ma:1999ni,c3Joshipura:1999hr,c3Choi:1999tq,
c3Grossman:1999hc,c3Romao:1999up,c3Abada:1999ai,c3Haug:1999kr,c3Chun:1999bq,
c3Takayama:1999pc,c3Davidson:1999mc,c3Adhikari:1999pa,c3Hirsch:2000jt,c3Hirsch:2000ef,
c3Davidson:2000uc,c3Grossman:2000ex,
c3Abada:2000xr,c3Mira:2000gg,c3Davidson:2000ne,c3Porod:2000hv,c3Joshipura:2001mq,c3Barger:2001xe,
c3Abada:2001zh,c3Joshipura:2002fc,c3Chun:2002rh,c3Borzumati:2002bf,c3Abada:2002ju,c3Chun:2002vp,c3Diaz:2003as,
c3Grossman:2003gq,c3Rakshit:2004rj,c3Jung:2004rd,c3Chun:2004mu,c3Chemtob:2006ur,
c3Dedes:2006ni,c3Mukhopadhyaya:2006is,c3Chun:2006ss,c3Dedes:2007ef,c3Dey:2008ht,c3Hundi:2009yf,c3JeanLouis:2009du} 
or through spontaneous $\rpv$ \cite{c3Romao:1991ex,
c3Nogueira:1990wz,c3Romao:1991tp,c3Giudice:1992jg,c3Kitano:1999qb,c3Frank:2007un,c3Hirsch:2008ur,
c3Mitra:2009jj,c3Ross:1984yg,c3Dawson:1985vr,
c3Santamaria:1987uq,c3Umemura:1993wc} (tree and(or) loop corrections)
and (b) neutrino mass generation and(or) collider phenomenology 
\cite{c3Ellis:1984gi,c3Barger:1989rk,c3Dawson:1985vr,c3Dimopoulos:1988jw,
c3Barbieri:1989vb,c3Nogueira:1990wz,c3GonzalezGarcia:1991ap,c3Romao:1991tp,
c3Dreiner:1991pe,c3Enqvist:1992ef,c3Godbole:1992fb,c3Butterworth:1992tc,
c3de-Campos:1995av,c3Bhattacharyya:1995pr,c3Bhattacharyya:1995bw,c3Nowakowski:1995dx,
c3Adhikari:1996bm,c3Choudhury:1996ia,c3Nilles:1996ij,c3Roy:1996bua,c3Kim:1997rr,c3Choudhury:1997dt,
c3Akeroyd:1997iq,c3Mukhopadhyaya:1998xj,c3Diaz:1998wq,c3Chun:1998ub,c3Bisset:1998hy,
c3Mukhopadhyaya:1999gy,c3Choi:1999tq,c3Ghosh:1999ix,c3Allanach:1999bf,c3Romao:1999up,
c3Takayama:1999pc,c3Davidson:1999mc,c3Adhikari:1999pa,c3Porod:2000hv,c3Datta:2000ci,
c3Restrepo:2001me,c3Barger:2001xe,c3Saha:2002kt,c3Chun:2002rh,c3Hirsch:2003fe,c3Jung:2004rd,
c3Datta:2006ak,c3Hirsch:2008ur,c3Arhrib:2009mz,c3Bandyopadhyay:2009xa,c3Perez:2009mu,c3Nath:2010zj,
c3DeCampos:2010yu}.

We start with a brief discussion of spontaneous $\rpv$ and later
we will address the issues of neutrino mass generation with explicit
breaking of $R_p$.

\vspace*{0.02cm}
\begin{center}
{\bf{I. Spontaneously broken $R$-parity}}. 
\end{center}
\vspace*{0.02cm}
The idea of spontaneous $\rpv$ was first implemented in ref.\cite{c3Aulakh:1982yn}
through spontaneous violation of the lepton number. The lepton number
violation occurs through the left sneutrino VEVs. It was revealed in ref. 
\cite{c3Nieves:1983kn} that if supersymmetry breaking terms include
trilinear scalar couplings and gaugino Majorana masses, only one
neutrino mass would be generated at the tree level. Remaining two small
masses are generated at the one-loop level \cite{c3Aulakh:1982yn}. Different
phenomenological implications for such a model 
were addressed in references \cite{c3Ellis:1984gi,c3Ross:1984yg,
c3Santamaria:1987uq,c3Santamaria:1988ic,c3Santamaria:1988zm}. A consequence of 
spontaneous $\L$ appears in the form of a massless Nambu-Goldstone boson
called Majoron \cite{c3Chikashige:1980ui,c3Gelmini:1980re}. Unfortunately,
a Majoron, arising from the breaking of gauge non-singlet fields
(in this case a doublet Majoron from the left sneutrino VEVs which is a 
member of ${\rm{SU(2)}_L}$ family) is strongly disfavored by
electroweak precision measurements ($Z$-boson decay width) 
\cite{c3Adeva:1989mn,c3Decamp:1989tu,c3Akrawy:1989pi,c3Aarnio:1989tv}
and astrophysical constraints \cite{c3Georgi:1981pg,c3Fukugita:1982ep,c3Raffelt:1996wa,
c3Kachelriess:2000qc}. Thus this doublet-Majoron model is ruled out 
\cite{c3GonzalezGarcia:1989zh,c3Romao:1989yh,c3Nogueira:1990wz}.

The possible shortcomings of a doublet Majoron model are removable by
using the VEV of a gauge-singlet field as suggested in ref.\cite{c3Masiero:1990uj}.
Most of these models break the lepton number spontaneously by giving VEV to a 
singlet field carrying one unit of lepton number \cite{c3Masiero:1990uj,c3Romao:1991ex,
c3Romao:1992vu}. However, there exists model where the singlet field carries
two unit of lepton number \cite{c3Giudice:1992jg}. This singlet Majoron model
\cite{c3Chikashige:1980ui} is not ruled out by LEP data. More phenomenological
implications of this class of models are addressed in refs. \cite{c3Romao:1991tp,
c3Hirsch:2004rw,c3Hirsch:2005wd,c3Hirsch:2006di,c3Hirsch:2008ur,c3Hirsch:2009ee,
c3Bhattacharyya:2010kr}.

We just briefly mentioned the basics of spontaneous $\rpv$ for the sake
of completeness. These issues are not a part of this thesis work and hence we
do not elaborate further. A dedicated discussion on the spontaneous violation of 
$R_p$ has been given in ref.\cite{c3Vicente:2011pf}.

\vspace*{0.02cm}
\begin{center}
{\bf{II. Explicit breaking of $R$-parity}}. 
\end{center}
\vspace*{0.02cm}
 
The MSSM superpotential with $R_p$ violating terms was given by
eqns. (\ref{MSSM-superpotential-1}) and (\ref{MSSM-superpotential-2}).
Since we aim to generate Majorana masses for the light neutrinos,
we consider violation of the lepton number only and thus the 
baryon number violating terms 
($\frac{1}{2} \lambda^{''}_{ijk} \hat u^c_i \hat d^c_j \hat d^c_k$)
are dropped for the rest of the discussions. It is perhaps, best
to start with the simple most example of $\rpv$, namely bilinear
$R_p$-violation ($bR_pV$) and continue the discussion with the trilinear 
terms ($tR_pV$) later.

\begin{flushleft}
{\it{$\maltese$ Bilinear $R$-parity violation}} 
\end{flushleft}

The superpotential and soft terms are given by 
(see eqns. (\ref{MSSM-superpotential-1}), (\ref{Lsoft-MSSM}) 
and (\ref{MSSM-superpotential-2}))

\bea 
W^{bR_pV} &=& W^{MSSM} - \epsilon_{ab}  \varepsilon^i \hat L^a_i\hat H^b_u, \nn \\
-\mathcal{L}^{bR_pV}_{\text{soft}} 
&=& -\mathcal{L}^{MSSM}_{\text{soft}} 
- \epsilon_{ab} B_{\varepsilon_i} \hat L^a_i\hat H^b_u.
\label{MSSM-brpv}
\eea

Now what are the implications of eqn.(\ref{MSSM-brpv})?

\vspace{0.1cm}
\noindent
1. $R_p$ is violated through lepton number violation by odd unit,
$\Delta L = 1$. This is an explicit breaking and so 
there is no possibility for 
an experimentally disfavored doublet Majoron emission.

\vspace{0.3cm}
\noindent
2. Similar to eqn.(\ref{MSSM-scalar-pot}) one can construct the 
neutral scalar potential, $V^{bR_pV}_{neutral~scalar}$. Interestingly
now one get non-zero VEVs for the left sneutrino fields using the 
suitable minimization condition
\beq
\sum_{j} (m^2_{\widetilde{L}})^{ji} {v'_j}- B_{\varepsilon_i} v_2 
+{\gamma_g}{\xi_{\upsilon}}{v'_i}
+{\varepsilon_i} {\eta} =0,
\label{Minim-Lsneut}
\eeq
where
\bea
\eta &=&\sum_{i} {\varepsilon^i}{v'_i}  + {\mu} v_1 ,
~\gamma_{g} = \frac{1}{4}({g_1^2 + g_2^2}), 
~\xi_{\upsilon} ={\sum_{i} {v'^2_i} + v_1^2 -v_2^2}. 
\label{Abbrevations-1}
\eea
$v_1,v_2$ are VEVs for down and up-type Higgs fields, respectively.
$v'_i$ is the VEV acquired by `$i$'-th sneutrino
field. The soft masses $(m^2_{\widetilde{L}})^{ji}$ are assumed to
be symmetric in `$i$' and `$j$' indices. 

The masses for $W,Z$ bosons now should be given by
\bea
M_W = \frac{g_2 v_{new}}{\rt2} ,
~~M_Z = \frac{v_{new}}{\rt2} \sqrt{g^2_1 + g^2_2},
\label{wz-mod-mass}
\eea
where $v^2_{new} = {\sum {v'^2_i} + v_1^2 + v_2^2}$.
It is apparent from eqn.(\ref{wz-mod-mass}) that to
maintain the electroweak precision, $\sum {v'^2_i} \ll v^2_1,v^2_2$,
so that ${\sum {v'^2_i} + v_1^2 + v_2^2} \simeq v_1^2 + v_2^2$ to
a very good approximation.

\vspace{0.3cm}
\noindent
3. Significance of the lepton number is lost, indeed without
a designated lepton number there is no difference between a
lepton superfield $(\hat L_i)$ and a down-type Higgs superfield, 
$\hat H_d$. As a consequence now the neutral sleptons 
(left sneutrinos $(\wt{\nu})$ in this
case) can mix with CP-odd (pseudoscalar) and even (scalar) 
neutral Higgs bosons. Similar mixing is allowed between charged Higgs 
and the charged sleptons. These enlarged scalar and 
pseudoscalar mass squared matrices in the 
basis $(\Re{H^0_d},\Re{H^0_u},\Re{\wt{\nu}_\al})$ 
and $(\Im{H^0_d},\Im{H^0_u},\Im{\wt{\nu}_\al})$ 
respectively, are given by
\bea
(a)~(M^2_{scalar})_{5\times5} = \left(\begin{array}{c c}
(\mathcal{M}^2_{MSSM-scalar})_{2\times2} &
(\mathcal{S}^2_{\wt{\nu}_\al H^0_i})_{2\times3}\\
(\mathcal{S}^2_{\wt{\nu}_\al H^0_i})^T 
& (\mathcal{S}^2_{\wt{\nu}_\al \wt{\nu}_\beta})_{3\times3}
\end{array}\right),
\label{brpMSSM-cp-even-1}
\eea
where $i=\left(\begin{array}{c}
d\\
u
\end{array}\right)$ with $\al,\beta=1,2,3$ or $e,\mu,\tau$ and 
\bea
(\mathcal{S}^2_{\wt{\nu}_\al H^0_d}) &=&
(\mu \varepsilon_\al + 2\gamma_{g} v'_\al v_1), 
~(\mathcal{S}^2_{\wt{\nu}_\al H^0_u}) =
(- B_{\varepsilon_\al} - 2\gamma_{g} v'_\al v_2), \nn \\
(\mathcal{S}^2_{\wt{\nu}_\al \wt{\nu}_\beta}) &=&
{\varepsilon_\al}{\varepsilon_\beta}
+{\gamma_g}{\xi_{\upsilon}}{\delta_{\al\beta}}
+2{\gamma_g}{v'_\al}{v'_\beta}+{(m^2_{\tilde{L}})_{\al\beta}},
\label{brpMSSM-cp-even-2}
\eea
and $(b)$
\bea
(M^2_{pseudoscalar})_{5\times5} = \left(\begin{array}{c c}
(\mathcal{M}^2_{MSSM-pseudoscalar})_{2\times2} &
(\mathcal{P}^2_{\wt{\nu}_\al H^0_i})_{2\times3}\\
(\mathcal{P}^2_{\wt{\nu}_\al H^0_i})^T 
& (\mathcal{P}^2_{\wt{\nu}_\al \wt{\nu}_\beta})_{3\times3}
\end{array}\right),
\label{brpMSSM-cp-odd-1}
\eea
with
\bea
(\mathcal{P}^2_{\wt{\nu}_\al H^0_d}) &=&
(-\mu \varepsilon_\al), 
~(\mathcal{P}^2_{\wt{\nu}_\al H^0_u}) =
(B_{\varepsilon_\al}), \nn \\
(\mathcal{P}^2_{\wt{\nu}_\al \wt{\nu}_\beta}) &=&
{\varepsilon_\al}{\varepsilon_\beta}
+{\gamma_g}{\xi_{\upsilon}}{\delta_{\al\beta}}
+{(m^2_{\tilde{L}})_{\al\beta}}.
\label{brpMSSM-cp-odd-2}
\eea
Here `$\Re$' and `$\Im$' correspond to the real and imaginary part
of a neutral scalar field.

The charged scalar mass squared matrix with the basis choice 
$({H^+_d},{H^+_u},{\wt{\ell}^+_{\al_R}},
{\wt{\ell}^+_{\al_L}})$ looks like
\bea
(M^2_{charged~scalar})_{8\times8} = \left(\begin{array}{c c}
(\mathcal{M}^2_{MSSM-charged})_{2\times2} &
(\mathcal{C}^2_{\wt{\ell}_{\al_X} H_i})_{2\times6}\\
(\mathcal{C}^2_{\wt{\ell}_{\al_X} H_i})^T 
& (\mathcal{C}^2_{\wt{\ell}_{\al_X} \wt{\ell}_{\beta_Y}})_{6\times6}
\end{array}\right),
\label{brpMSSM-charged-1}
\eea
where $X,Y=L,R$ and 
\bea
(\mathcal{C}^2_{\wt{\ell}_{\al_R} H_d})_{1\times3} &=&
(Y^{\al\beta}_e \varepsilon_\beta  v_2 - (A_e Y_e )^{\al\beta} v'_\beta),\nn \\
(\mathcal{C}^2_{\wt{\ell}_{\al_L} H_d})_{1\times3} &=&
(\mu \vp_\al - Y^{\al a}_e Y^{\beta a}_e 
v'_\beta v_1 + \frac{g_2^2}{2} v'_\al v_1),\nn \\
(\mathcal{C}^2_{\wt{\ell}_{\al_R} H_u})_{1\times3} &=&
(- \mu Y^{\beta \al}_e v'_\beta + Y^{\beta \al}_e \vp_\beta v_1),
~(\mathcal{C}^2_{\wt{\ell}_{\al_L} H_u})_{1\times3} =
(\frac{g_2^2}{2} v'_\al v_2 + B_{\vp_\al}),\nn \\
(\mathcal{C}^2_{\wt{\ell}_{\al_L} \wt{\ell}_{\beta_L}})_{3\times3} &=&
(\vp_\al \vp_\beta + Y^{\al \rho}_e Y^{\beta \rho}_e 
v^2_1 + \gamma_g {\xi_\upsilon} \delta_{\al\beta} - 
\frac{g_2^2}{2} \mathcal{D}_{\al\beta} + (m^2_{\tilde{L}})^{\al\beta}), \nn \\
(\mathcal{C}^2_{\wt{\ell}_{\al_R} \wt{\ell}_{\beta_R}})_{3\times3} &=&
(Y^{\rho\al}_e Y^{\sigma\beta}_e v'_\rho v'_\sigma 
+ Y^{\rho\al}_e Y^{\rho\beta}_e v^2_1 + (m^2_{\tilde{e}^c})^{\al\beta} - 
\frac{g_1^2}{2} {\xi_\upsilon} \delta_{\al\beta}), \nn \\
(\mathcal{C}^2_{\wt{\ell}_{\al_L} \wt{\ell}_{\beta_R}})_{3\times3} &=&
(- \mu Y^{\al\beta}_e v_2 + (A_e Y_e)^{\al\beta} v_1),
\label{brpMSSM-charged-2}
\eea
where $\mathcal{D}_{\al\beta} = \{{\xi_\upsilon} 
\delta_{\al\beta} -  v'_\al v'_\beta\}$. 
The soft-terms are assumed to be symmetric. The ${2\times2}$ MSSM scalar,
pseudoscalar and charged scalar mass squared matrices are given in 
appendix \ref{appenA}. 

\vspace{0.3cm}
\noindent
4. In a similar fashion charged leptons $(\ell_\al \equiv e, \mu, \tau)$ 
mix with charged gauginos 
as well as with charged higgsinos
and yield an enhanced chargino mass matrix.
In the basis, $-i \tilde {\lambda}^{+}_{2}, \tilde{H}^+_u, \ell^+_{\al_R}$ (column)
and $-i \tilde {\lambda}^{-}_{2}, \tilde{H}^-_d, \ell^-_{\beta_L}$ (row)
\bea
(M_{chargino})_{5\times5} =
\left(\begin{array}{c c}
(M^{chargino}_{MSSM})_{2\times2} & 
\left(\begin{array}{c}
0\\
Y^{\rho\al}v'_{\rho}
\end{array}\right)_{2\times3}\\
\left(\begin{array}{c c}
g_2 v'_\al & \vp_\al
\end{array}\right)_{3\times2}
& (Y^{\beta\al}v_1)_{3\times3}
\end{array}\right).
\label{brpMSSM-chargino}
\eea
With this enhancement eqn.(\ref{MSSM-charginos}) looks like
\bea 
&&\chi^+_i = V_{i1} \wt W^+ + V_{i2} \wt H^+_u + V_{i,\al+2} \ell^+_{\al_R}, \nn \\
&&\chi^-_i = U_{i1} \wt W^- + U_{i2} \wt H^-_d + U_{i,\al+2} \ell^-_{\al_L}.
\label{brpMSSM-charginos-reln}
\eea

The neutral fermions also behave in a similar manner. The neutralino mass
matrix now can accommodate three light neutrinos $(\nu \equiv \nu_L)$ apart from 
the four MSSM neutralinos. The extended neutralino mass matrix in the basis 
$\tilde B^0, \tilde W_3^0, \tilde H_d^0,\tilde H_u^0,{\nu_\al}$ is written as
\bea
(M_{neutralino})_{7\times7} =
\left(\begin{array}{c c}
(M^{neutralino}_{MSSM})_{4\times4} & 
((m)_{3\times4})^T\\
(m)_{3\times4}
& (0)_{3\times3}
\end{array}\right),
\label{brpMSSM-neutralino}
\eea
with 
\bea
(m)_{3\times4} =
\left(\begin{array}{c c c c}
-\frac{g_1}{\sqrt{2}}{v'_\al} & \frac{g_2}{\sqrt{2}}{v'_\al} 
& 0 & -{\ep_\al}
\end{array}\right).
\label{brpMSSM-neutralino2}
\eea
Just like the charginos, for the neutralinos one can rewrite
eqn.(\ref{MSSM-neutralinos}) in modified form as
\beq 
\chi^0_i = N_{i1} \wt B^0 + N_{i2} \wt W^0_3 + N_{i3} \wt H^0_d
+ N_{i4} \wt H^0_u + N_{i,\al+4} \nu_\al.
\label{MSSM-neutralinos-reln}
\eeq
Chargino and neutralino mass matrices for MSSM are given 
in appendix \ref{appenA}.

\begin{figure}[ht]
\centering
\includegraphics[width=9.95cm]{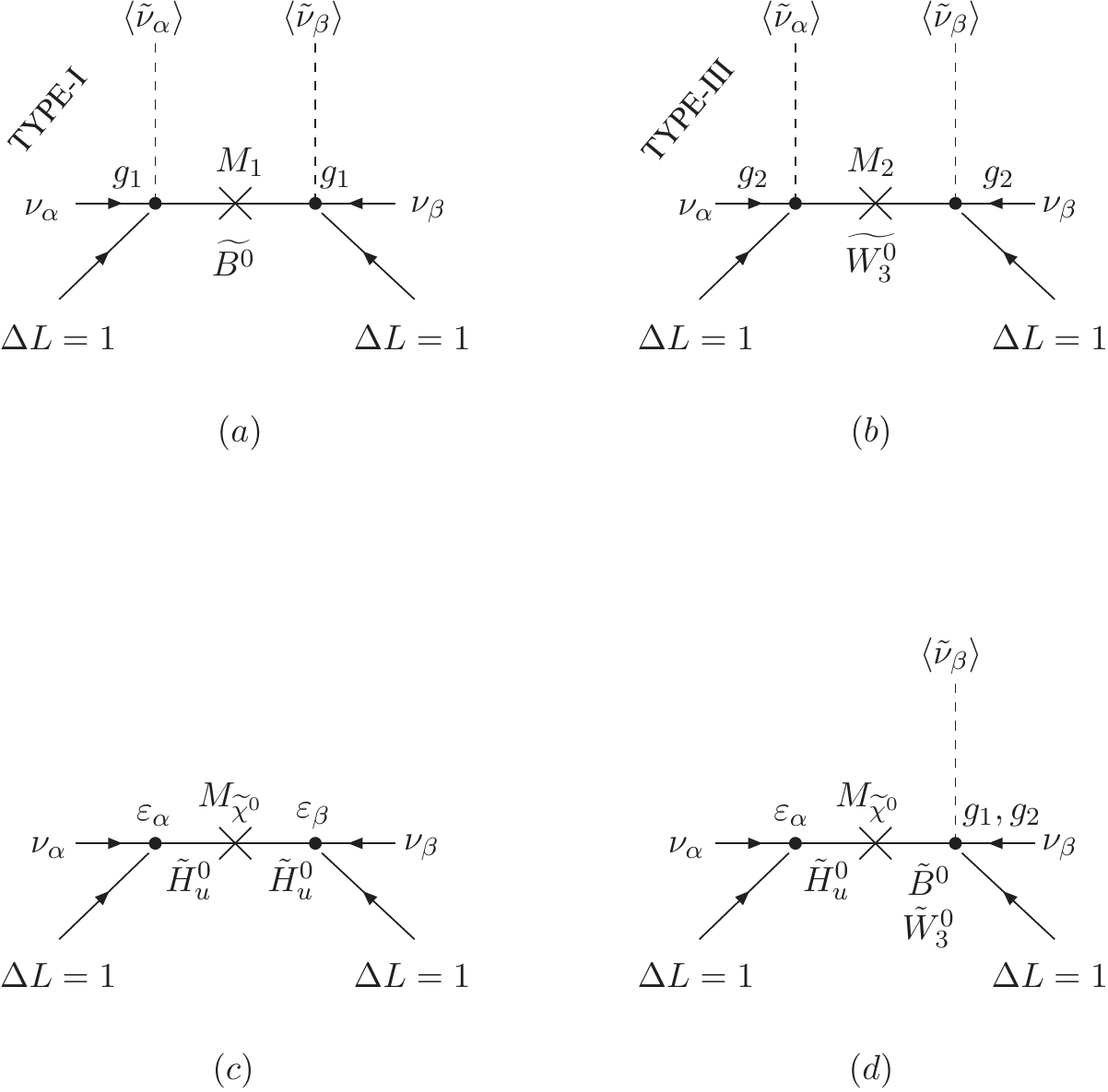}
\caption{Different types of tree level contributions to the neutrino mass in a 
$bR_pV$ supersymmetric model. The cross on the neutralino propagator signifies
a Majorana mass term for the neutralino.
}
\label{brpv-seesaw}
\end{figure}

\vspace{0.3cm}
\noindent
5. In eqn.(\ref{brpMSSM-neutralino}) entries of the $4\times4$ MSSM
block are $\sim$ TeV scale, which are $\gg$ entries of $(m)_{3\times4}$.
Besides, the $3\times3$ null matrix $(0)_{3\times3}$ signifies the 
absence of Majorana mass terms for the left-handed neutrinos. This matrix has
a form similar to that of eqn.(\ref{seesaw-mat}), thus the effective
light neutrino mass matrix is given by (using eqn.(\ref{seesaw-eigval1}))
\beq 
m_{seesaw} = -(m)_{3\times4}\{(M^{neutralino}_{MSSM})_{4\times4}\}^{-1}
\{(m)_{3\times4}\}^T,
\label{seesaw-brpv}
\eeq
or in component form
\beq 
(m_{seesaw})_{\al\beta} = \frac{g^2_1M_2+g^2_2M_1}
{2 ~Det[(M^{neutralino}_{MSSM})_{4\times4}]}
(\mu v'_\al - \vp_\al v_1)(\mu v'_\beta- \vp_\beta v_1).
\label{seesaw-brpv-component}
\eeq
Assuming $M_1,M_2,\mu,v_1,v_2$ are $\sim \wt{m}$, a generic mass scale
(say EWSB scale or the scale of the soft supersymmetry breaking terms) 
and $g_1,g_2\sim$ $\cal{O}$ $(1)$ we get from
eqn.(\ref{seesaw-brpv-component})
%
\beq 
(m_{seesaw})_{\al\beta} \approx \underbrace{\frac{v'_\al v'_\beta}{\wt{m}}}
_{I} + \overbrace{\frac{\vp_\al \vp_\beta}{\wt{m}}}^{II}
-\underbrace{\frac{(\vp_\al v'_\beta + \al\leftrightarrow\beta)}{\wt{m}}}_{III}.
\label{seesaw-brpv-component-2}
\eeq
The first term of eqn.(\ref{seesaw-brpv-component-2}) is coming from the 
{\it{gaugino seesaw}} effect, which is
originating though the mixing of light neutrinos with either a bino $(\wt B^0)$
or a neutral wino $(\wt W^0_3)$. This is also another example for a 
Type-I (bino) $+$ Type-III (wino) seesaw (see figure \ref{brpv-seesaw} (a),(b)).
The second and third contributions are represented by $(c)$ and $(d)$
of figure \ref{brpv-seesaw}. There is one extremely important point to note
about this analysis, that is if $\vp_\al=0$ but $B_{\vp_\al}$ are not,
even then $v'_\al\neq0$ (see eqn.(\ref{Minim-Lsneut})). Thus even if
$R_p$ violation is rotated away from the superpotential, effect of $\rpv$
in the soft terms can still trigger non-zero neutrino mass as shown by $(a,b)$
of figure \ref{brpv-seesaw}. However, this analysis is strictly valid
if $B_{\vp_\al} \not\propto \vp_\al$.

If we define $\mu'_\al = (\mu v'_\al - \vp_\al v_1)$, then using the following
set of relations, namely, $g^2_1/(g^2_1+g^2_2) = {\rm sin}^2\theta_W$, 
$g^2_2/(g^2_1+g^2_2) = {\rm cos}^2\theta_W$, $M^2_Z = (1/2)(g^2_1+g^2_2)(v^2_1+v^2_2)$
and the fact $Det[(M^{neutralino}_{MSSM})_{4\times4}] =
(g^2_1M_2+g^2_2M_1)v_1v_2\mu - M_1M_2\mu^2$, we get an alternative expression
of eqn.(\ref{seesaw-brpv-component-2}) 
\beq 
(m_{seesaw})_{\al\beta} \approx 
\frac{\mu'_\al \mu'_\beta}{\wt{m}}{\rm cos}^2\beta,
\label{seesaw-brpv-component-3}
\eeq
where $tan\beta = v_2/v_1$ holds good with $v'_\al \ll v_1,v_2$.
The problem with this tree level effective light neutrino mass matrix 
is that, it gives only one {\it{non-zero}} eigenvalue, given by
\beq 
m_{neut} = 
\frac{|\mu'_\al|^2}{\wt{m}}{\rm cos}^2\beta.
\label{seesaw-brpv-component-4}
\eeq
The {\it{only non-zero}} neutrino mass at the tree level of a $bR_pV$
model is suppressed by squared $R_p$-violating parameter and also
by ${\rm tan}^{-2}\beta$ for ${\rm tan}\beta\gg1$. With $\vp_\al\sim$ $10^{-4}$ 
GeV and $\wt m\sim$ $1$ TeV one gets $m_{neut}\sim$ $10^{-{11}}$ GeV, 
which is the scale for the atmospheric neutrinos\footnote{Assuming normal hierarchy 
in light neutrino masses.}. But to accommodate three flavour global data 
\cite{c3Schwetz:2008er,c3GonzalezGarcia:2010er}
one requires at least two massive neutrinos!

\begin{figure}[ht]
\centering
\includegraphics[width=12.05cm]{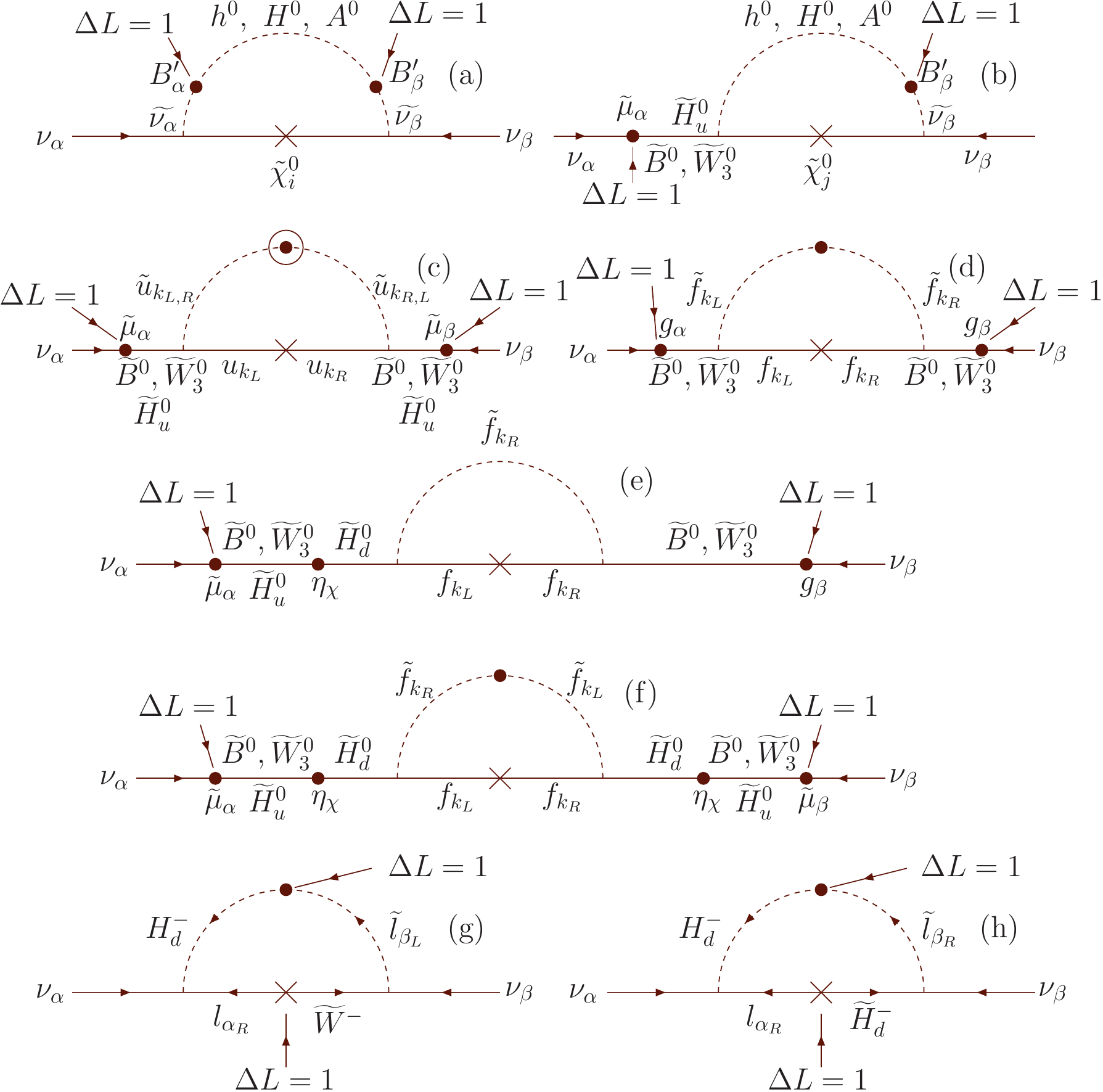}
\caption{Neutrino mass generation through loops in a
model with $bR_pV$. For details of used symbols see text.}
\label{brpv-loops}
\end{figure}

\begin{flushleft}
{\it{$\blacklozenge$ Loop corrections in bilinear $R$-parity violation}} 
\end{flushleft}
The remedy to this problem can come from the one-loop
contributions to the light neutrino masses. The dominant diagrams are 
shown in figure \ref{brpv-loops}. Before discussing these diagrams 
and their contributions further it is
worthy to explain the meaning of symbols used in figure \ref{brpv-loops}.
The quantity $B'_\al$ denotes mixing between a left handed
sneutrino $\wt \nu_\al$ (see eqns.(\ref{brpMSSM-cp-even-2}), (\ref{brpMSSM-cp-odd-2})) 
and physical MSSM Higgs bosons (eqn.(\ref{Higgs-mass-gauge})). 
$\wt{\mu}_\al$ is either $\vp_\al$ ($\nu_\al \wt H_u$ 
mixing, see eqn.(\ref{brpMSSM-neutralino2})) or $g_1 v'_\al,g_2 v'_\al$ 
($\nu_\al \wt B^0,\nu_\al \wt W^0_3$ mixing, see eqn.(\ref{brpMSSM-neutralino2}))
(figure $(a)$ and $(b)$). In figure $(c)$ a {\it{blob}} on the scalar line
indicates a mixing between left and right handed up-type squarks, which exists
if one has either gauginos $(\wt B^0,\wt W^0_3)$ or up-type higgsino $(\wt H^0_u)$ 
on both the sides. However, if one puts gauginos on one side and higgsino 
on the other, then this left-right mixing is absent. This situation is represented by
a void circle on the scalar line around the blob. In figure $(d)$ $g_\al,g_\beta$ 
represents neutrino-gaugino mixing (eqn.(\ref{brpMSSM-neutralino2})). $f$ denotes a 
{\it{down-type}} fermion, that is either a charged lepton, $(\ell_k=e,\mu,\tau)$
or a down quark, $(d_k=d,s,b)$. There also exist more complicated diagrams
for down-type fermion loops as shown in figure $(e,f)$. $\eta_\chi$ represents
mixing of a down-type higgsino with neutral gauginos and up-type higgsino 
(see eqn.(\ref{MSSM-neutralino})). The last two diagrams $(g,h)$ arise from
chargino-charged scalar contribution to neutrino mass. A {\it{cross}} on the 
fermion line represents a mass insertion, responsible for a chirality flip.
In all of these diagrams $\Delta L=2$ effect is coming from a pair of $\Delta L=1$ 
contributions. For diagrams $(g,h)$ the blobs and the cross represent mixing only 
without any chirality flip (see eqns.(\ref{brpMSSM-charged-2}),(\ref{brpMSSM-chargino})).

These diagrams are shown for a general
basis where both of the bilinear $R_p$-violating parameters $(\vp_\al)$
and sneutrino VEVs $(v'_\al)$ are non-vanishing. When $v'_\al=0$, using the 
minimization condition for left sneutrinos (eqn. (\ref{Minim-Lsneut})), 
diagram $(a)$ of figure \ref{brpv-loops} reduces to the well-known 
$BB$-loop \cite{c3Grossman:1997is,c3Davidson:2000uc,c3Davidson:2000ne,c3Grossman:2003gq}.

%
This $BB$ loop can either give mass to one more neutrino state 
(not to that one which was already massive at the tree level 
so long $B_{\vp_\al} \not\propto \vp_\al$) when sneutrino 
masses are degenerate or can contribute to all three
light neutrino masses with non-degenerate sneutrinos. Assuming all
the scalar and neutralino masses $\sim \wt m$, an approximate
expression for this loop contribution to light neutrino masses
with degenerate sneutrinos is given by \cite{c3Grossman:2003gq,c3Rakshit:2004rj}.
\beq 
m^{BB}_{\al\beta} \simeq 
\frac{g^2_2}{64 \pi^2 \cos^2\beta}
\frac{B'_\al B'_\beta}{\wt m^3}.
\label{BB-loop}
\eeq
It is important to mention that in order to generate
solar neutrino mass square difference
using loop corrections one
should naively expect $B'\sim$ $(0.1-1)$ ${\rm {GeV^2}}$, with
the assumption of normal hierarchical structure in light neutrino masses.

In a similar fashion the loop shown by diagram $(b)$ of figure \ref{brpv-loops}
is an example of the $\mu B$-type loop at $v'_\al=0$ \cite{c3Davidson:2000uc,
c3Davidson:2000ne,c3Grossman:2003gq,c3Rakshit:2004rj}. This loop involves 
neutrino-gaugino or neutrino-higgsino mixing (collectively labeled as $\wt \mu_\al$, 
see eqn.(\ref{brpMSSM-neutralino2})) together with sneutrino-Higgs mixing ($B'_\beta$, 
see eqns.(\ref{brpMSSM-cp-even-2}), (\ref{brpMSSM-cp-odd-2}), (\ref{Higgs-mass-gauge})). 
Assuming all the masses (Higgs, sneutrino, 
neutralino) are at the weak scale $\wt m$, an approximate contribution is
given by \cite{c3Davidson:2000uc,c3Davidson:2000ne,c3Grossman:2003gq,c3Rakshit:2004rj}
\beq 
m^{\mu B}_{\al\beta} \simeq 
\frac{g^2_2}{64 \pi^2 \cos\beta}
\frac{\wt \mu_\al B'_\beta + \wt \mu_\beta B'_\al}{\wt m^2}.
\label{muB-loop}
\eeq
It is evident from the structure of right hand side of eqn.(\ref{muB-loop}) that 
the $\mu B$ loop contributes to more than one neutrino masses. However, presence
of $\wt\mu_\al$ makes this loop contribution sub-leading to neutrino
masses compared to the $BB$ loop \cite{c3Chun:2002vp,c3Grossman:2003gq,c3Rakshit:2004rj}.
For large values of $\tan\beta$
$(\tan\beta \gg 1)$ the $BB$-loop and the $\mu B$-loop are enhanced by
$\tan^2\beta$ and $\tan\beta$, respectively.

Contributions to neutrino masses from quark-squark loops are given
by diagrams $(c,d,e,f)$ of figure \ref{brpv-loops}. Diagram $(c)$ represents
an up-type quark-squark loops. This diagram can yield large contribution
to neutrino mass particularly when it is a top-stop $(t-\tilde {t})$ loop,
because of the large top Yukawa
coupling, $Y_t$. This loop contribution is proportional
to $\wt \mu_\al \wt \mu_\beta$, which is exactly same as the tree level
one (see eqn.(\ref{seesaw-brpv-component-3})), thus this entire
effect eventually gives a correction to a neutrino mass which is already
massive at the tree level \cite{c3Mukhopadhyaya:2006is}. An approximate expression 
for this loop is given by

\beq 
m^{u_k {\wt {u}_k}}_{\al\beta}(no~blob) \simeq 
\frac{N_c f^2_{u \wt u}}{16 \pi^2}
\frac{m_{u_k}\wt \mu_\al \wt \mu_\beta}{\wt m^2},
~~m^{u_k {\wt {u}_k}}_{\al\beta}(blob) \simeq 
\frac{N_c f^2_{u \wt u}}{16 \pi^2}
\frac{m^2_{u_k}\wt \mu_\al \wt \mu_\beta}{\wt m^3},
\label{upsup-loop}
\eeq
where $m_{u_k}$ is the mass of up-quark of type $k$. The coupling factor
$f^2_{u \wt u}$ is either $g_ig_j$ or $g_iY_{u_k}$\footnote{$u_k
\equiv u,c,t$.} with $i=1,2$. $N_c$ is the colour factor which is $3$ for 
quarks. For the case of left right sfermion mixing we use the relation 
\beq 
m^{2^{LR}}_{\wt f_k} \approx m_{f_k} \wt m.                                                         
\label{fermion-sfermion-relation}
\eeq

In a similar fashion for a down type fermion-sfermion, 
$f_k-\wt f_k$ (charged lepton-slepton or down quark-squark) (see diagram 
$(d)$ of figure \ref{brpv-loops}) an approximate expression is given by 
(using eqn.(\ref{fermion-sfermion-relation}))
%
\beq 
m^{f_k {\wt {f}_k}}_{\al\beta} \simeq 
\frac{N_c f^2_{f \wt f}}{16 \pi^2}
\frac{m^2_{f_k}g_\al g_\beta}{\wt m^3},
\label{fdsfd-loop}
\eeq
where $m_{f_k}$ is the mass of down-type fermion of type $k$\footnote{
$f=d_k \equiv d,s,b$ or $f=\ell_k\equiv e,\mu,\tau$.}. $N_c=3$ 
for quarks but $=1$ for leptons. The coupling factor
$f^2_{f \wt f}$ is $g_ig_j$  with $i=1,2$. The quantity $g_\al$ denotes mixing 
between a neutrino and a gaugino. However,
for down-type fermion-sfermion there exist other complicated loop diagrams 
like $(e,f)$ \cite{c3Mukhopadhyaya:2006is} of figure 
\ref{brpv-loops}. These loops give contribution of the approximate form
\beq 
m'^{f_k {\wt {f}_k}}_{\al\beta} \simeq 
\frac{N_c f'^2_{f \wt f} \eta_\chi}{16 \pi^2}
\frac{m_{f_k}(\wt \mu_\al g_\beta + \al\leftrightarrow\beta)}{\wt m^3},
\label{fdsfd-loop-2}
\eeq
for diagram $(e)$ and
\beq 
m''^{f_k {\wt {f}_k}}_{\al\beta} \simeq 
\frac{N_c f''^2_{f \wt f} \eta^2_\chi}{16 \pi^2}
\frac{m^2_{f_k}\wt \mu_\al \wt \mu_\beta}{\wt m^5},
\label{fdsfd-loop-3}
\eeq
for diagram $(f)$, respectively. The quantity $f'^2_{f \wt f}$ is $g_i Y_{f_k}$ 
whereas $f''^2_{f \wt f}$ represents $Y^2_{f_k}$ with $i=1,2$
and $Y_{f_k}$ being either charged lepton or down quark Yukawa couplings.
It is apparent that eqns.(\ref{fdsfd-loop}), (\ref{fdsfd-loop-3})
once again contribute to the ``{\it{same neutrino}}'' which is already massive
at the tree level. However, eqn.(\ref{fdsfd-loop-2}) will contribute
to more than one neutrino masses. Note that since 
contributions of these set of diagrams are proportional to the fermion mass, 
$m_{f_k}$, they are important only for bottom quark and tau-lepton along with the corresponding
scalar states running in the loop.

Diagrams $(g,h)$ are the chargino-charged scalar loop contributions
to light neutrino mass \cite{c3Dedes:2006ni}. An approximate form for these loops 
are given by
\beq 
m^{(g)}_{\al\beta} \simeq 
\frac{g^2_2 Y_{\ell_k}}{16 \pi^2}
\frac{v'_\al B''_\beta}{\wt m},
~~
m^{(h)}_{\al\beta} \simeq 
\frac{Y^3_{\ell_k}}{16 \pi^2}
\frac{v'_\al B''_\beta}{\wt m},
\label{csc-loop}
\eeq
where $Y_{\ell_k}$ are the charged lepton Yukawa couplings and
$B''_\beta$ $(\sim B')$ represents a generic charged slepton-charged Higgs mixing 
(see eqn.(\ref{brpMSSM-charged-2})). These contributions vanishes identically 
when $v'_\al=0$. These contributions
being proportional to small parameters like $v',Y_\ell$, are {\it{much smaller}}
compared to the other types of loops. Various couplings needed here
can be found in references like \cite{c3Haber:1984rc,c3Gunion:1984yn,c3Rosiek:1989rs,c3Rosiek:1995kg,
c3Hempfling:1995wj,c3Hirsch:2000ef,c3Diaz:2003as}.

\begin{figure}[ht]
\centering
\includegraphics[width=12.35cm]{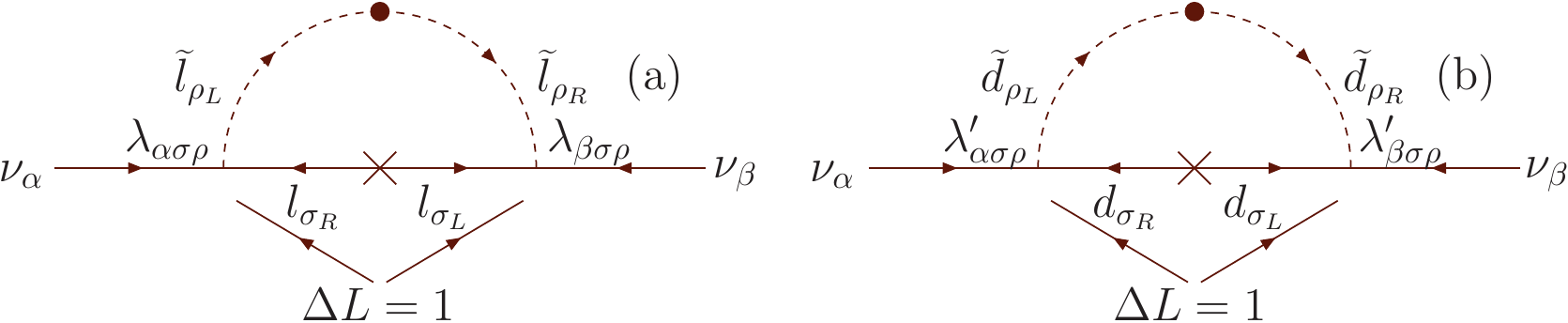}
\caption{Neutrino mass generation through loops in a
model with $tR_pV$.}
\label{trpv-loops}
\end{figure}

\begin{flushleft}
{\it{$\blacklozenge$ Trilinear $R$-parity violation and loop corrections}} 
\end{flushleft}
The so-called trilinear couplings, 
contribute to light neutrino mass through loops only \cite{c3Grossman:1997is,
c3Grossman:1998py,c3Grossman:2003gq,c3Davidson:2000uc}. Possible
diagrams  are shown in figure \ref{trpv-loops}. 

Using eqn.(\ref{fermion-sfermion-relation})
these contributions can be written as
\beq 
m^{\lam\lam}_{\al\beta} \simeq 
\frac{N_c\lam_{\al\sigma\rho}\lam_{\beta\sigma\rho}}{8 \pi^2}
\frac{m_{\ell_\sigma} m_{\ell_\rho}}{\wt m},
~~
m^{\lam'\lam'}_{\al\beta} \simeq 
\frac{N_c\lam'_{\al\sigma\rho}\lam'_{\beta\sigma\rho}}{8 \pi^2}
\frac{m_{d_\sigma} m_{d_\rho}}{\wt m},
\label{ll-lplp-loop}
\eeq
where $N_c$ is $1(3)$ for $\lam\lam(\lam'\lam')$ loop. Contributions
of these diagrams are suppressed by squared $R_p$ violating couplings 
$\lam^2,\lam'^2$ and squared charged lepton, down-type quark masses
apart from usual loop suppression factor. Thus usually these loop
contributions are quiet small \cite{c3Grossman:1998py}. 

\begin{figure}[ht]
\centering
\includegraphics[width=12.35cm]{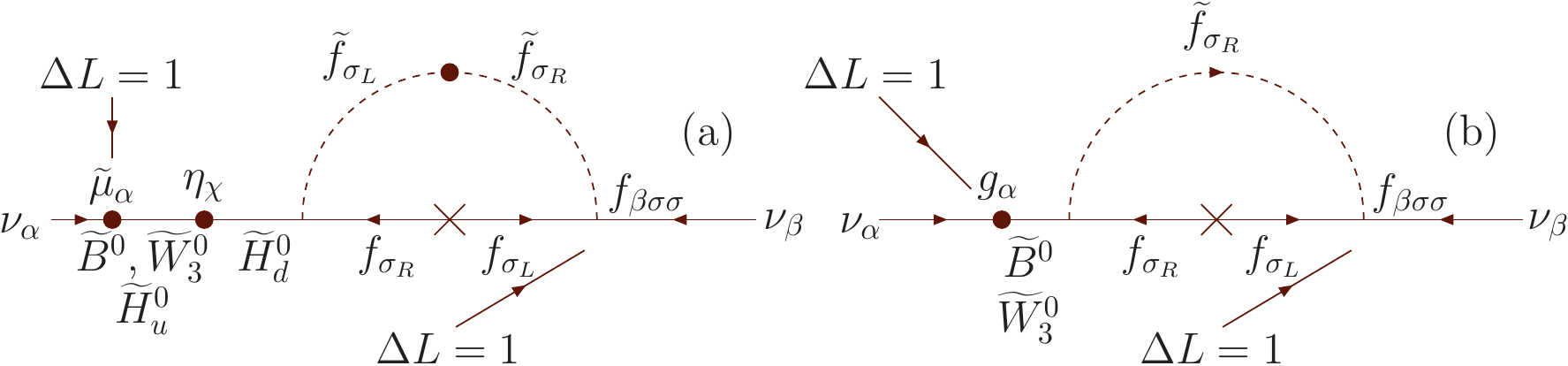}
\caption{Neutrino mass generation through loops in a
model with both $bR_pV$ and $tR_pV$. $\wt{f}_\sigma$ is either
a charged slepton with $f_{\beta\sigma\sigma}=\lam_{\beta\sigma\sigma}$ 
or a down-type squark with $f_{\beta\sigma\sigma}=\lam'_{\beta\sigma\sigma}$.
$\wt\mu_\al,\eta_\chi,g_\al$ are same as explained in figure \ref{trpv-loops}.
The cross have similar explanation as discussed
in figure \ref{brpv-loops}.}
\label{brpv-trpv-loops}
\end{figure}
\begin{flushleft}
{\it{$\blacklozenge$ Loop corrections in $bR_pV+tR_pV$}} 
\end{flushleft}
There also exist a class of one-loop diagrams which involve
both bilinear and trilinear $R_p$ violating couplings, as
shown figure \ref{brpv-trpv-loops} \cite{c3Davidson:2000uc,c3Davidson:2000ne
,c3Grossman:2003gq,c3Rakshit:2004rj}. One can write down
these loop contributions approximately as
\beq 
(i)~~~m^{\mu f}_{\al\beta} \simeq 
\frac{N_c \wt \mu_\al \eta_\chi Y_{f_\sigma} f_{\beta\sigma\sigma}}{16 \pi^2}
\frac{m^2_{f_\sigma} }{\wt m^3} + \al \leftrightarrow \beta,
\label{mu-ll-lplp-loop-1}
\eeq
for diagram $(a)$ where $N_c=1(3)$ for charged lepton (down-type quark),
$Y_{f_\sigma}$ is either a charged lepton or a down-type Yukawa coupling and
\beq 
(ii)~~~m^{\mu f}_{\al\beta} \simeq 
\frac{N_c g_\al g_i f_{\beta\sigma\sigma}}{16 \pi^2}
\frac{m_{f_\sigma} }{\wt m} + \al \leftrightarrow \beta,
\label{mu-ll-lplp-loop-2}
\eeq
for diagram $(b)$.
The quantity $f_{\al\sigma\sigma}$ represents either $\lam_{\al\sigma\sigma}$
or $\lam'_{\al\sigma\sigma}$ couplings. $g_\al$ 
represents a neutrino-gaugino mixing (see eqn.(\ref{brpMSSM-neutralino2})).
$i=1,2$. These contributions are suppressed by a pair of $R_p$-violating
couplings $(\mu\lam/\mu\lam')$ or product of sneutrino VEVs and trilinear
$R_p$-violating couplings $(v'\lam/v'\lam')$, a loop factor and at least by a 
fermion mass ($\propto$ Yukawa coupling) \cite{c3Grossman:2003gq,c3Rakshit:2004rj}. 
Also contributions of diagram $(a)$ is negligible
compared to that of $(b)$ by a factor of squared Yukawa coupling. Contributions
of these loops are second order in the above mentioned suppression factors
(similar to that of $\mu B$ loop) once the tree level effect is taken into account.

There are literature 
where these loop contributions are analysed in a basis independent formalism 
\cite{c3Davidson:2000uc,c3Davidson:2000ne,
c3Rakshit:2004rj} (also see refs.\cite{c3Davidson:1996cc,
c3Davidson:1997mc,c3Davidson:1998yy,c3Ferrandis:1998ii,c3Grossman:2000ex} for basis 
independent parameterizations of $\rpv$). For this discussion we
stick to the ``mass insertion approximation'' but alternatively it is also 
possible to perform these entire analysis in physical or mass basis 
\cite{c3Hempfling:1995wj,c3Hirsch:2000ef,c3Dedes:2006ni}. The mass
insertion approximation works well since the effect of $R_p$-violating parameters
are expected to be small in order to account for neutrino data. 
All of these calculations are performed assuming no
flavour mixing for the sfermions. 
                   
\vspace{0.2cm}
\begin{flushleft}
{\it{$\blacklozenge$ A comparative study of different loop contribution}} 
\end{flushleft}
Usually the trilinear loops $(\lam\lam, \lam'\lam')$ are doubly Yukawa 
suppressed (through fermion masses) and they yield rather small contributions. 
The $\mu B$-type, $\mu \lam, \mu \lam'$ loop contributions to the light neutrino
masses are second order in suppression factors. The $\mu \lam, \mu \lam'$
loop contributions are also suppressed by single Yukawa coupling. The Yukawa
couplings (either double or single) are also present in the quark-squark or
charged lepton-slepton loops. However, in most of the occasions they give
corrections to the tree level neutrino mass, though other contributions
can also exist (see eqn.(\ref{fdsfd-loop-2})). These loops are sometimes 
dominant \cite{c3Kaplan:1999ds,c3Hirsch:2000ef,c3Diaz:2003as} provided the 
$BB$-type loop suffers large cancellation among 
different Higgs contributions. In general the second neutrino receives
major contribution from the $BB$ loop. 

In the situation when $\tan\beta$ is large, the tree 
level contribution (see eqn.(\ref{seesaw-brpv-component-3})) can be smaller
compared to the loop contributions. In this situation, the tree level result
usually account for the solar neutrino mass scale whereas the loop corrections 
generate the atmospheric mass scale. In conventional scenario when tree level
effect is leading, it is easy to fit the normal hierarchical spectrum of
neutrino mass in an $R_p$-violating theory.

\section{{\bf T}esting neutrino oscillation at Collider}\label{corre}

We have already spent enough time to discuss the issue of light
neutrino mass generation. It is then legitimate to ask what
are the possible experimental implications of a massive neutrino?
It was first advocated in ref.\cite{c3Mukhopadhyaya:1998xj} that
in a simple supersymmetric model with only $bR_pV$ it is possible
to get some kind of relation between the neutrino sector and the decays
of the LSP.
This kind of model predicts comparable numbers of muons and taus, produced
together with the $W$-boson, in decays of the lightest neutralino. Usually 
for an appreciable region of parameter space the lightest neutralino is the LSP.
Additionally, the appearance of a measurable ``displaced vertex'' 
was also addressed in ref.\cite{c3Mukhopadhyaya:1998xj} which is
extremely useful for a collider related study to efface undesired backgrounds.
This novel feature also has been addressed in refs.\cite{c3Chun:1998ub,c3Choi:1999tq,
c3Porod:2000hv,c3DeCampos:2010yu}. See also refs.\cite{c3Bartl:2000yh,c3Restrepo:2001me,
c3Hirsch:2002ys,c3Bartl:2003uq,c3Hirsch:2003fe,c3Bartl:2003sr} for tests of neutrino
properties at accelerator experiments.

The correlation between a LSP decay and neutrino physics is apparent
for supersymmetric models with bilinear $\rpv$, since the same parameter $\vp_\al$ 
is involved in both the analysis. For example, if the neutralino LSP, $\ntrl1$
decays into a charged lepton and $W$-boson \cite{c3Mukhopadhyaya:1998xj} then
following \cite{c3Romao:1999up,c3Porod:2000hv} one can get approximately

\beq 
\frac{(\ntrl1\to\mu^\pm W^\mp)}{(\ntrl1\to\tau^\pm W^\mp)}
\simeq \left(\frac{\mu'_\mu}{\mu'_\tau}\right)^2= \tan^2\theta_{23},
\label{mu-tau-23}
\eeq
where $\mu_\al = \mu v'_\al - \vp_\al v_1$ with $\al=e,\mu,\tau$
and $\tan^2\theta_{23}$ is the atmospheric mixing angle. 
Similar correlations with trilinear $\rpv$ parameters are lost \cite{c3Choi:1999tq}
since the model became less predictive with a larger set of parameters.
A rigorous discussion of these correlations has
been given in ref.\cite{c3Nath:2010zj}.

We note in passing that when the LSP is no longer stable (due to $\rpv$) it is
not necessary for them to be charge or colour neutral \cite{c3Wolfram:1978gp,c3Dover:1979sn,
c3Ellis:1983ew}. With broken $R_p$ any sparticle (charginos \cite{c3Feng:1999fu}, 
squarks, gluinos \cite{c3Raby:1997bpa,c3Baer:1998pg,c3Raby:1998xr}, sneutrinos 
\cite{c3Hagelin:1984wv}, (see also ref.\cite{c3Ellis:1983ew})) 
can be the LSP. It was pointed out in ref.\cite{c3Hirsch:2003fe} that whatever be the LSP, 
measurements of branching ratios at future accelerators will provide a definite test of 
bilinear $R_p$ breaking as the model of neutrino mass.




\chapter{ \sffamily{{\bf $\mu$}$\nu$SSM: neutrino masses and mixing
 }}\label{munuSSM-neut}

\section{{\bf I}ntroducing $\mu\nu$SSM}
As discussed earlier, the minimal supersymmetric
standard model (MSSM) is not free from drawbacks. We have addressed
these issues in the context of the $\mu$-problem \cite{c4Kim:1983dt}
and light neutrino mass generation. Supersymmetric theories
can accommodate massive neutrinos either through $\rpv$ or using 
seesaw mechanism. Regarding the $\mu$-problem, as discussed in section \ref{NMSSM},
a simple solution is given by the NMSSM. There exist a host of NMSSM models
where the superpotential contains either explicit $R_p$-violating 
couplings \cite{c4Abada:2006qn,c4Abada:2006gh,c4Hundi:2009yf,c4JeanLouis:2009du,
c4Pandita:1999jd,c4Pandita:2001cv,c4Chemtob:2006ur}  
or use spontaneous violation of $R_p$ \cite{c4Kitano:1999qb} to accommodate 
light neutrino masses apart from offering a solution to the $\mu$-problem. 
Unfortunately, NMSSM models of neutrino mass generation with $bR_pV$ suffer from the 
$\epsilon$-problem \cite{c4Nilles:1996ij}. Besides, with bilinear $\rpv$ not 
all the neutrino masses are generated at 
the tree level. Thus loop corrections are unavoidable to account for the three flavour 
oscillation data. Loop effects are compulsory 
for models with $tR_pV$ where all of the neutrino masses appear at loop level. 
Certainly, larger number
of trilinear couplings reduce the predictability of these models. An elegant alternative
is given by NMSSM models with spontaneous $\rpv$ where apart from a singlet superfield, $\hat S$
(to solve the $\mu$-problem) one requires a right-handed neutrino superfield,
$\hat \nu^c$ to accommodate massive neutrinos. The
issues of light neutrino mass generation together with a solution to the $\mu$-problem in 
$R_p$-conserving NMSSM models have been addressed in references 
\cite{c4Frank:2007un,c4Das:2010wp,c4Abada:2010ym}.
 
So, in a nutshell, the well-known NMSSM models of neutrino mass generation either suffer 
from the naturalness problem or are less predictive due to the presence of 
either large number of couplings 
or additional superfields. Now following the structure of the SM it seems rather
natural to add right-handed neutrino superfields with the MSSM superfields in order
to generate neutrino masses. Also being a SM gauge singlet, a right-handed neutrino superfield,
$\hat \nu^c$ can act as a viable alternative for the singlet field $(\hat S)$ of NMSSM
used to solve the $\mu$-problem.  

The novel idea of solving the $\mu$-problem and light neutrino mass generation
simultaneously in a supersymmetric model using {\it{only}} right-handed neutrino 
superfields, $\hat \nu^c_i$ was advocated in ref.\cite{c4LopezFogliani:2005yw}. 
This model is popularly known as the {\it{``$\mu$ from $\nu$'' supersymmetric standard
model}} or $\mu\nu$SSM \cite{c4LopezFogliani:2005yw}. Details of this model
will be provided in the next sub-section.

In this chapter we plan to discuss the $\mu\nu$SSM model first with necessary details
like neutral scalar potential, minimization conditions, scalar sector, fermionic sector etc.
and later we aim to discuss the issues of light neutrino masses and mixing in the $\mu\nu$SSM
at the tree level as well as with one-loop radiative corrections.

\section{{\bf T}he model} \label{munuSSM-model}

In this section we introduce the model along the lines of ref.\cite{c4LopezFogliani:2005yw}, 
discuss its basic features and set our notations. Throughout this thesis we consider 
three generations of right-handed neutrino superfield $(\hat \nu^c_i)$ apart from the 
MSSM superfields as proposed in ref.\cite{c4LopezFogliani:2005yw}. We start with 
the model superpotential and the soft terms and continue our discussion with the minimization
conditions later.

\begin{flushleft}
{\it{$\maltese$ Superpotential}} 
\end{flushleft}
The $\mu\nu$SSM superpotential is given by
\beq
W^{\mu\nu SSM} = W^{'^{MSSM}} + \epsilon_{ab} Y^{ij}_\nu \hat H^b_u\hat L^a_i\hat \nu^c_j
-\underbrace{\epsilon_{ab} \lambda^i\hat \nu^c_i\hat H^a_d\hat H^b_u}_{\Delta L=1}
+\overbrace{\frac{1}{3}\kappa^{ijk}\hat \nu^c_i\hat \nu^c_j\hat \nu^c_k}^{\Delta L=3},
\label{munuSSM-superpotential}
\eeq
where $W^{'^{MSSM}}$ is the MSSM superpotential (see eqns.(\ref{MSSM-superpotential-1}),
(\ref{NMSSM-superpotential})) 
but without the $\epsilon_{ab}\mu \hat H^a_d\hat H^b_u$-term. The superfields
$\hat H_d,\hat H_u,\hat L_i$ are usual MSSM down-type Higgs, up-type Higgs
and $\rm{SU(2)_L}$ doublet lepton superfields. Since right-handed neutrinos carry
a non-zero lepton number, the third and fourth terms of eqn.(\ref{munuSSM-superpotential})
violate lepton number by odd unit(s) ({\it{one}} and {\it{three}}, respectively).
Violation of lepton number by odd units is the source of $\rpv$ (eqn.(\ref{R-parity-defn-2}))
is $\mu\nu$SSM.

It is important to mention the implications of different terms of 
eqn.(\ref{munuSSM-superpotential}) at this stage.

\vspace*{0.1cm}
\noindent
$\bullet$ The second term $\epsilon_{ab} Y^{ij}_\nu \hat H^b_u\hat L^a_i\hat \nu^c_j$
respects lepton number conservation to start with. However, after EWSB these
terms give rise to effective bilinear $R_p$-violating terms as $\vp^i L_i H_u$
with $\vp^i=Y^{ij}_\nu v^c_j$. $v^c_j$ denotes the VEV acquired
by $j$-th right-handed sneutrino. Besides, a term of this kind also give rise
to Dirac neutrino mass matrix with entries as $m_{D_{ij}} = Y^{ij}_\nu v_2$.

\vspace*{0.1cm}
\noindent
$\bullet$ The third term $\epsilon_{ab} \lambda^i\hat \nu^c_i\hat H^a_d\hat H^b_u$
after EWSB generates an effective $\mu$-term as $\mu=\sum \lam_i v^c_i$. This term
violates lepton number by one unit.

\vspace*{0.1cm}
\noindent
$\bullet$ The last term $\frac{1}{3}\kappa^{ijk}\hat \nu^c_i\hat \nu^c_j\hat \nu^c_k$
violates lepton number by three units. Note that this term is allowed
by all possible symmetry arguments. Now if $\kappa^{ijk}=0$ to start with then
the Lagrangian has a global $\rm{U(1)}$
symmetry which is broken spontaneously by the VEVs of the scalar
fields and leads to unacceptable massless axion.
In order to avoid axions non-zero values for $\kappa^{ijk}$
are essential \cite{c4Ellis:1988er}. Besides, after EWSB the last term of 
eqn.(\ref{munuSSM-superpotential}) produces entries for the right-handed neutrino 
Majorana mass matrix as $m_{\nu^c_{ij}}=2\kappa^{ijk} v^c_k$.

\vspace*{0.1cm}
\noindent
$\bullet$ As already mentioned, a $Z_3$ symmetry is imposed on the $\mu\nu$SSM superpotential
(eqn.(\ref{munuSSM-superpotential})) to forbid appearance of any bilinear term.
This feature is similar to the NMSSM models as stated in section \ref{NMSSM}. Thus
similar to the NMSSM, the $\mu\nu$SSM also suffers from the problem of 
cosmological domain wall formation \cite{c4Ellis:1986mq,c4Rai:1992xw,c4Abel:1995wk}.
However, the problem can be ameliorated through well known methods \cite{c4Abel:1996cr,
c4Panagiotakopoulos:1998yw,c4Panagiotakopoulos:1999ah}.

\begin{figure}[ht]
\centering
\includegraphics[width=7.45cm]{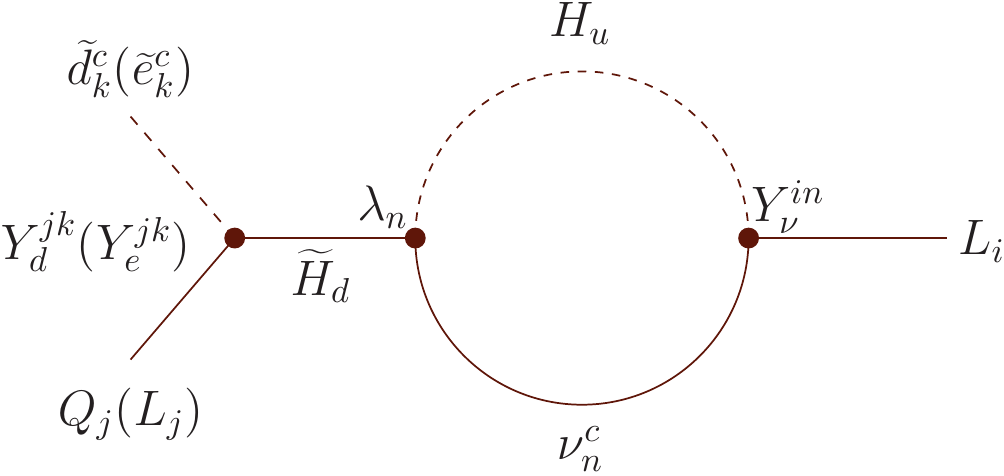}
\caption{One-loop generation of the $\lam^{ijk},\lam'^{ijk}$ terms in the superpotential. These terms are 
proportional to product of two Yukawa couplings and $\lam$. Product of two Yukawa couplings
assures smallness of the $\lam^{ijk},\lam'^{ijk}$ couplings.}
\label{munuSSM-llp}
\end{figure}

\vspace*{0.1cm}
\noindent
$\bullet$ The conventional trilinear couplings $\lam^{ijk},\lam'^{ijk}$ (see eqn.(\ref{MSSM-superpotential-2}))
can be generated in $\mu\nu$SSM at one-loop level as shown in figure \ref{munuSSM-llp} 
\cite{c4Escudero:2008jg}. 

A term of the type 
$Y_\nu L H_u \nu^c$ has been considered also in ref. 
\cite{c4Mukhopadhyaya:2006is} in the context of light
neutrino mass generation, but without offering any attempts to solve the $\mu$-problem.
In ref.\cite{c4Farzan:2005ez} couplings 
of the form $Y_\nu L H_u \nu^c$ and $\lam H_d H_u \nu^c$ were considered
along with Majorana mass terms $\frac{1}{2}M^{ij}\nu^c_i\nu^c_j$ for right-handed neutrinos. 
However, in this case 
the contribution of $\lam H_d H_u \nu^c$ for generating the $\mu$-term is negligible 
because the right-handed sneutrinos $(\wt \nu^c_i)$ , being super heavy, do not acquire
sizable vacuum expectation values. In reference \cite{c4Farzan:2005ez} the term 
$\lam H_d H_u \nu^c$ has been utilized for the purpose of thermal seesaw leptogenesis.

\begin{flushleft}
{\it{$\maltese$ Soft terms}} 
\end{flushleft}
Confining ourselves in the framework of supergravity mediated supersymmetry breaking, the
Lagrangian $\mathcal{L}^{\mu\nu SSM}_{soft}$, containing the soft-supersymmetry-breaking 
terms is given by
\bea
-\mathcal{L}^{\mu\nu SSM}_{soft} &=& -\mathcal{L}^{'^{MSSM}}_{soft} 
+ (m^2_{\wt \nu^c})^{ij} \wt \nu^{c^*}_i \wt \nu^c_j \nn \\
&+& \{ \ep_{ab} (A_{\nu}Y_\nu)^{ij} H_u^b \wt L^a_i \wt \nu^c_j
- \ep_{ab} (A_{\lambda}\lambda)^i \wt \nu^c_i H_d^a  H_u^b \nn \\
&+& \frac{1}{3} (A_{\kappa}\kappa)^{ijk} \wt \nu^c_i \wt \nu^c_j \wt \nu^c_k  + h.c\},
\label{munuSSM-soft}
\eea
where $\mathcal{L}^{'^{MSSM}}_{soft}$ denotes $\mathcal{L}^{MSSM}_{soft}$ 
without the $B_\mu$ term (see eqn.(\ref{Lsoft-MSSM})). $(m^2_{\wt \nu^c})^{ij}$
denote soft square masses for right-handed sneutrinos.

\vspace{1cm}
\begin{flushleft}
{\it{$\maltese$ Scalar potential and minimization}} 
\end{flushleft}
The tree-level scalar potential receives the usual D and F term 
(see eqn.(\ref{MSSM-scalar-pot}), where $\left|\frac{\partial W^{\mu\nu SSM}}
{\partial \phi^{\mu\nu SSM}} \right|^2 \equiv F^*F$ 
with $\phi$ as any superfields of the $\mu\nu$SSM)
contributions, in addition to the terms from $\mathcal{L}^{\mu\nu SSM}_{soft}$.
We adhere to the $CP$-preserving case, so that only the real parts of
the neutral scalar fields develop, in general, the following VEVs,
\bea
\langle H_d^0 \rangle = v_1 \, , \quad \langle H_u^0 \rangle = v_2 \,
, \quad \langle \widetilde \nu_i \rangle = v'_i \, , \quad
\langle \widetilde \nu_i^c \rangle = v^c_i.
\label{munuSSM-vevs}
\eea
In eqn.(\ref{munuSSM-vevs}) $i=1,2,3 \equiv e, \mu, \tau$.
The tree level neutral scalar potential looks like \cite{c4LopezFogliani:2005yw,
c4Escudero:2008jg,c4Ghosh:2008yh,c4Bartl:2009an,c4Ghosh:2010zi}
\bea 
\langle V_{\text{neutral}}\rangle &=&
\left|\sum_{i,j}Y^{ij}_{\nu}{v^{\prime}_i}{v^c_j}
-  \sum_{i}\lambda^i{v^c_i} v_1\right|^2\ \nonumber \\
&+& \sum_{j}\left|\sum_{i}  Y^{ij}_{\nu}{v^{\prime}_i}v_2
  -\lambda^jv_1v_2+\sum_{i,k} \kappa^{ijk}{v^c_i}{v^c_k}\right|^2 
+ \left|\sum_{i}\lambda^i{v^c_i}v_2\right|^2 \nonumber \\
&+& \sum_{i}\left|\sum_{j}Y^{ij}_{\nu}v_2 {v^c_j}\right|^2
+(\frac{g_1^2+g_2^2}{8}) \left[\sum_{i}|v'_i|^2+|v_1|^2 -
  |v_2|^2\right]^2 \nonumber \\
&+& \left [\sum_{i,j}(A_\nu Y_\nu )^{ij} {v'_i} {v^c_j}v_2 -\sum_{i}
  (A_\lambda \lambda )^{i} {v^c_i}v_1v_2 + {\rm H.c.} \right]\nonumber \\
&+& \left [\sum_{i,j,k}\frac{1}{3}(A_\kappa
  \kappa )^{ijk}v^c_iv^c_jv^c_k + {\rm H.c.} \right]
+ \sum_{i,j} (m_{\widetilde{L}}^2)^{ij}{v'_i}^* {v'_j}\nonumber \\
&+& \sum_{i,j}(m_{\widetilde{\nu}^c}^2)^{ij} {v^{c^*}_i}{v^c_j} +
m_{H_u}^2|v_2|^2 + m_{H_d}^2|v_1|^2 .
\label{Vneut-munuSSM}
\eea
One important thing is to notice  that the potential is bounded from below 
because the coefficient of the fourth power of all the eight superfields are 
positive (see eqn.(\ref{Vneut-munuSSM})). We shall further assume that all the parameters 
present in the scalar potential are real. From eqn.(\ref{Vneut-munuSSM}), the minimization 
conditions with respect to $v^c_i,~v'_i,v_2,~v_1$ are 
\bea
&&2{\sum_{j}} {{u^{ij}_c}} {\zeta^{j}} + \sum_{k} Y^{ki}_{\nu} {r^{k}_c} {v_2^2}
+\sum_{j} (m^2_{\widetilde{\nu}^c})^{ji} 
{v^c_j}+ {\rho^i \eta} + {\mu} {\lambda^i v_2^2}+(A_x x)^{i} =0,\nn\\
&&{\sum_{j}} {Y_{\nu}}^{ij} {v_2} {\zeta^{j}}  +\sum_{j} (m^2_{\widetilde{L}})^{ji} {v'_j}
+\sum_{j} (A_{\nu} Y_{\nu})^{ij} {v^c_j} v_2 +{\gamma_g}{\xi_{\upsilon}}{v'_i}
+{r^i_c} {\eta} =0,\nn\\
&&{\sum_{j}}{\rho^{j}}{\zeta^{j}} + {{\sum_{i}} {r^i_c}^2 v_2}+ {\sum_{i}}({A_{\nu}Y_{\nu}})^{ij} 
{v'_i} {v^c_j}-\sum_{i}({A_{\lambda} {\lambda}})^i {v^c_i} v_1 + X^u v_2 = 0,\nn \\
&&-{\sum_{j}}{\lambda^j}v_2 {\zeta^{j}}- {\mu} {\sum_{j}} {r^j_c} {v'_j} -\sum_{i}({A_{\lambda} 
{\lambda}})^i {v^c_i} v_2 + X^d {v_1} = 0,
\label{Minim-munuSSM}
\eea
with $X^u = {m^2_{H_u}} + {\mu^2} -{\gamma_g}{\xi_{\upsilon}}$, 
$X^d = {m^2_{H_d}} + {\mu^2} +{\gamma_g}{\xi_{\upsilon}}$ and
\bea
(A_x x)^{i} &=& \sum_{j} (A_{\nu} Y_{\nu})^{ji} {v'_j} v_2 + \sum_{j,k} 
({A_\kappa} {\kappa})^{ijk} {v^c_j} {v^c_k} 
- (A_{\lambda} {\lambda})^i v_1 v_2,  \nonumber \\
{r^{i}_c} &=& \varepsilon^i = {\sum_{j}Y^{ij}_{\nu} {v^c_{j}}}, 
~~{r^{i}} = {\sum_{j}Y^{ij}_{\nu} {v'_{j}}}, 
~~{u^{ij}_c} = {\sum_{k}}\kappa^{ijk}{v^c_k}, \nonumber \\
\zeta^{j} &=& \sum_{i} 
{u^{ij}_c} {v^c_i} + {r^j} v_2 - 
{\lambda}^j v_1 v_2 ,  ~~\mu = \sum_{i}\lambda^{i} {v^c_i},
~~\rho^{i} = {r^i} - \lambda^i v_1,\nonumber \\ 
\eta &=&\sum_{i} {r^i_c}{v'_i}  - {\mu} v_1 , 
~~\gamma_{g} = \frac{1}{4}({g_1^2 + g_2^2}), 
~~\xi_{\upsilon} ={\sum_{i} {v'^2_i} + v_1^2 -v_2^2}.  \nonumber \\
\label{Abbrevations}
\eea
In deriving the above equations, it has been assumed that 
${\kappa}^{ijk}$, $({{A_\kappa} {\kappa}})^{ijk}$,  $Y^{ij}_{\nu}$,
$(A_{\nu} Y_{\nu})^{ij}$, $(m^2_{\widetilde{\nu}^c})^{ij}$, 
$(m^2_{\widetilde{L}})^{ij}$ are all symmetric in their indices.

It is important to know that now the Majorana masses for right-handed
neutrinos $(2\kappa^{ijk}v^c_k)$ are at the TeV scale with $\kappa\sim$
$\cal{O}$ $(1)$ and TeV scale $v^c_i$ (see first one of eqn.(\ref{Minim-munuSSM})).
For neutrino Dirac masses $(Y_{\nu}^{ij} v_2)$ $\sim$ $10^{-4}$ GeV
the neutrino Yukawa couplings $(Y^{ij}_\nu)$ must also be very small 
$\sim \cal{O}$ $(10^{-7})$, in order to get 
correct neutrino mass scale using a seesaw mechanism involving TeV scale 
right-handed neutrino. This immediately 
tells us that in the limit $Y_{\nu}^{ij} \rightarrow 0$,  
(see second one of eqn.(\ref{Minim-munuSSM}))
$v'_{i} \rightarrow 0$. So in order to get appropriate neutrino 
mass scale both $Y_{\nu}^{ij}$ and $v'_{i}$ have to be small.

Ignoring the terms of the second order in $Y_{\nu}^{ij}$ 
and considering $({v'^2_i}+v^2_1-v^2_2) \approx (v^2_1-v^2_2)$ 
(which is a good approximation), and
\textbf{$(m^2_{\tilde{L}})^{ij} = (m^2_{\tilde{L}}) \delta^{ij}$}, we can 
easily solve second one of eqn.(\ref{Minim-munuSSM}) as (using eqn. (\ref{Abbrevations}))
\beq
{v'_i}\approx - \left\{{\frac{{Y_{\nu}}^{ik}{u^{kj}_c} {v_2} - {\mu}{v_1} 
{Y^{ij}_{\nu}}  +(A_{\nu} Y_{\nu})^{ij} v_2}{{\gamma_{g}} ({v_1^2 -v_2^2})+ 
(m^2_{\tilde{L}})}}\right\}{v^c_{j}} + 
\left\{{\frac{{Y_{\nu}}^{ij}{\lambda}^j v_1 v^2_2}{{\gamma_{g}} 
({v_1^2 -v_2^2})+ (m^2_{\tilde{L}})}}\right\}.
\label{sneutrino_VEV_simplified}
\eeq
Note from eqn.(\ref{sneutrino_VEV_simplified}), that the left-handed sneutrinos
can acquire, in general, non-vanishing, non-degenerate VEVs even in the limit 
of zero vacuum expectation values of the gauge singlet sneutrinos \cite{c4Ghosh:2008yh}.
However, zero VEVs of all the three gauge singlet sneutrinos is not an acceptable 
solution since in that case no $\mu$-term $(\sum \lam_i v^c_i)$ will be generated. 
It is essential to ensure that the extremum value of the potential corresponds to the 
minimum of the potential, by studying the second derivatives. 

The neutral scalar potential and the minimization conditions
in $\mu\nu$SSM but for complex VEVs, have been discussed in ref.\cite{c4Fidalgo:2009dm} in the 
context of spontaneous $CP$ violation and its implications in neutrino physics.

\section{{\bf S}calar sector of $\mu\nu$SSM}\label{munuSSM-scalar}

It is evident from eqns.(\ref{munuSSM-superpotential}) and (\ref{munuSSM-soft})
that lepton number $(L)$ is no longer conserved in $\mu\nu$SSM. 
In this situation states having zero lepton number can mix with states having $L\neq0$. These
lepton number violating mixings in turn result in larger $(8\times8)$ mass squared matrices  
for $CP$-even neutral scalar, $CP$-odd neutral pseudoscalar 
and charged scalar states. This is a consequence of the fact that
in $\mu\nu$SSM three generations of left and right-handed sneutrinos
can mix with neutral Higgs bosons. In a similar fashion charged sleptons
mix with the charged Higgs bosons. The enhancement
over the $2\times2$ MSSM structure (see appendix \ref{appenA}) is phenomenologically
very rich. Detailed structures for neutral scalar, pseudoscalar and the charged scalar mass 
squared matrices are given in appendix \ref{appenB}. In our numerical analysis we
confirm the existence of two charged and one neutral Goldstone boson(s) in 
the charged scalar and pseudoscalar sector. In addition, we have checked 
that all the eigenvalues of the scalar, pseudoscalar, and charged scalar mass-squared 
matrices (apart from the Goldstone bosons) appear to be positive ({\it{non-tachyonic}}) 
for a minima. These matrices are addressed in refs.\cite{c4Escudero:2008jg,c4Ghosh:2008yh,
c4Bartl:2009an}. In appendix \ref{appenB} squark mass squared matrices are also addressed 
\cite{c4Escudero:2008jg,c4Ghosh:2008yh}.

Before discussing the scalar sector of this model further, it is important
to point out the approximation and simplification used for involved numerical
analysis. 
For numerical calculations we assume all soft-masses, $\lam^i,\kappa^{ijk}$ and the 
corresponding soft parameters $(A_\lam\lam)^i,(A_\kappa\kappa)^{ijk}$ to be flavour
diagonal as well as flavour blind. However the neutrino Yukawa couplings $(Y^{ij}_\nu)$
and the respective soft parameters $(A_\nu Y_\nu)^{ij}$ are chosen to be flavour
diagonal. For simplicity all three right sneutrino VEVs are assumed to be 
degenerate $(v^c)$. Mathematically,
%
\bea
&&\kappa^{ijk} = \kappa \delta^{ij} \delta^{jk},
~~(A_\kappa\kappa)^{ijk} = (A_\kappa\kappa) \delta^{ij} \delta^{jk},\nn\\
&& Y^{ij}_\nu = Y^{ii}_\nu \delta^{ij},~~(A_\nu Y_\nu)^{ij} = (A_\nu Y_\nu)^{ii} \delta^{ij},\nn\\
&& \lam^i = \lam,~~(A_\lam\lam)^i=(A_\lam\lam),~~v^c_i=v^c,\nn\\
&& (m^2_{\tilde{L}})^{ij}=(m^2_{\tilde{L}})\delta^{ij},
~~(m^2_{\tilde{\nu}^c})^{ij}=(m^2_{\tilde{\nu}^c})\delta^{ij}.
\label{assumption1}
\eea

Coming back to the scalar sector of the $\mu\nu$SSM, apart from excluding 
the corner of parameter space responsible for tachyons,
additional constraints on the parameter space can come from the existence 
of false minima as well as from the perturbativity of the model parameters 
(free from Landau pole). 
A detailed discussion on this issue has been presented in 
ref. \cite{c4Escudero:2008jg} and the regions excluded by the existence of false 
minima have been shown. One can check from these figures that mostly the lower 
part of the region allowed by the absence of tachyons, are excluded by the 
existence of false minima. In our analysis, we have chosen the parameter 
points in such a way that they should be well above the regions disallowed by 
the existence of false minima. Nevertheless, in the case of gauge-singlet 
neutrino ($\nu^c$) dominated lightest neutralino (to be discussed in the next chapter), 
the value of $\kappa$ that we have chosen is $0.07$ with two different values of 
$\lambda$, namely, $0.1$ and $0.29$. In this case, there is a possibility that 
these points might fall into the regions disallowed by the existence of false 
minima. However, we have checked that even if we take the value of $\kappa$ 
to be higher ($0.2$ or so), with appropriately chosen $\lambda$, our conclusions 
do not change much. For such a point in the parameter space, it is likely that 
the existence of false minima can be avoided.  

Let us also mention here that the sign of the $\mu$-term is controlled by 
the sign of the VEV $v^c$ (assuming a positive $\lambda$), which is 
controlled by the signs of $A_\lambda \lambda$ and $A_\kappa \kappa$. 
If $A_\lambda \lambda$ is negative and $A_\kappa \kappa$ is positive then the 
sign of the $\mu$ parameter is negative whereas for opposite signs of the 
above quantities, we get a positive sign for the $\mu$ parameter. 

\begin{figure}[ht]
\centering
\includegraphics[width=6.45cm]{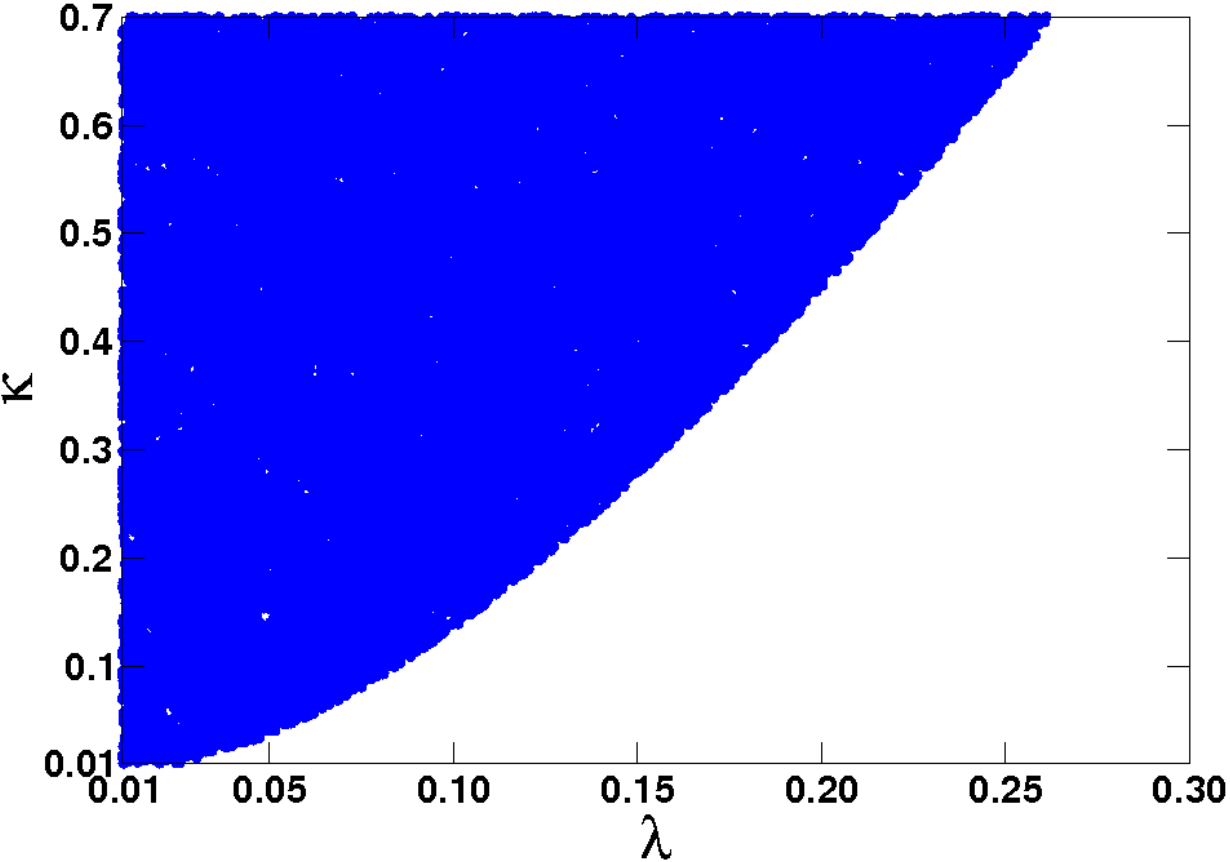}
\caption{Allowed regions in ($\lambda$--$\kappa$) plane which satisfy  
various constraints on the scalar sector, for tan$\beta =10$. $\lambda$ and 
$\kappa$ were allowed to vary from $0.005$ to $0.50$ and $0.005$ to $0.70$, respectively. 
Corresponding set of other parameters are given in table \ref{lkplot-parameters}.}
\label{scalar_sector_for_tan_beta_values}
\end{figure}

The eigenvalues of the scalar mass-squared matrices and the right-handed sneutrino 
VEVs $(v^c)$ are not very sensitive to the change in neutrino 
Yukawa couplings ($Y_\nu$ $\sim \cal {O}$ $(10^{-7})$) and the corresponding 
soft parameter $A_\nu Y_\nu$ ($\sim \cal {O}$ $(10^{-4})$ GeV). On the other 
hand, the values of $\tan\beta$ and the coefficients $\lambda$ and $\kappa$ are 
very important in order to satisfy various constraints on the scalar sector 
mentioned earlier. In figure \ref{scalar_sector_for_tan_beta_values}, we have 
plotted the allowed regions in the ($\lambda$--$\kappa$) plane for 
$\tan\beta=10$ \cite{c4Ghosh:2008yh}. Relevant parameters are given in 
table \ref{lkplot-parameters}.

\begin{table}[ht]
\centering
\begin{tabular}{c | c || c | c}
\hline \hline 
Parameter  & Chosen Value &  Parameter  & Chosen Value  \\ \hline \hline
$(A_\lam \lam)$   & $1000\times\lam$ GeV& $(A_\kappa \kappa)$ & $1000\times\kappa$ GeV\\
$Y^{11}_\nu$   & $5.0\times10^{-7}$ & $(A_\nu Y_\nu)^{11}$ & $5.0\times10^{-4}$ GeV\\
$Y^{22}_\nu$   & $4.0\times10^{-7}$ & $(A_\nu Y_\nu)^{22}$ & $4.0\times10^{-4}$ GeV\\
$Y^{33}_\nu$   & $3.0\times10^{-7}$ & $(A_\nu Y_\nu)^{33}$ & $3.0\times10^{-4}$ GeV\\
$m^2_{\wt L}$   & $400^2$ GeV$^2$ & $m^2_{\wt \nu^c}$ & $300^2$ GeV$^2$\\
\hline \hline
\end{tabular}
\caption{\label{lkplot-parameters}
Relevant parameter choices for figure \ref{scalar_sector_for_tan_beta_values} 
consistent with the EWSB conditions and non-tachyonic nature for squared scalar masses. 
Eqn.(\ref{assumption1}) has been used and we choose $\tan\beta=10$.}
\end{table}
The upper limit of the value
of $\kappa$ is taken to be $\sim$ 0.7 because of the constraints coming from 
the existence of Landau pole \cite{c4Escudero:2008jg}.  
With the values of different parameters satisfying the constraints in the 
scalar sector (see figure \ref{scalar_sector_for_tan_beta_values}), we will go on to 
calculate the neutrino masses and the mixing patterns in the next few sections.

It is also important to discuss the bounds on the lightest Higgs boson mass
in $\mu\nu$SSM. Neglecting small neutrino Yukawa couplings $Y^{ij}_\nu$,
the tree level upper bound on the lightest neutral
Higgs mass \cite{c4Drees:1988fc,c4Ellis:1988er,c4Binetruy:1991mk,c4Espinosa:1991gr,
c4Espinosa:1992hp} is given by (see eqn.(\ref{Higgs-mass-NMSSM-lightest}))
\beq
m^{'^2}_{h^0} \lesssim M^2_Z \left[\cos^22\beta
+ 3.62~\lam^i\lam_i~\sin^22\beta\right].
\label{Higgs-mass-NMSSM-lightest-2}
\eeq
Apparently, one can optimize this bound by choosing small $\tan\beta$ and 
large $\lam^i\lam_i$ values simultaneously.
Similar to the NMSSM \cite{c4Espinosa:1998re,c4Daikoku:2000eq,c4Ellwanger:2006rm} 
the upper bound for the lightest ${\rm{SU(2)_L}}$ doublet-like
Higgs boson mass in the $\mu\nu$SSM is $\sim 140$ GeV 
for $\tan\beta \sim2$ \cite{c4Escudero:2008jg}. Such a conclusion strictly demands small mixing 
among the MSSM Higgs and the right-handed sneutrinos $\wt \nu^c_i$ (see eqns. 
(\ref{element_of_scalar_mass_matrix}), (\ref{element_of_pseudoscalar_mass_matrix})).

It should be mentioned at this point that the radiative corrections to the 
lightest Higgs boson mass, can be significant in some regions of the parameter space
as discussed in ref.\cite{c4Escudero:2008jg}. It has been shown that the light 
Higgs mass larger than the LEP lower limit of 114 GeV can be obtained with the
value of $A_t$ (trilinear coupling in the scalar sector for the stop) within
1-2.4 TeV and when the mixing of the light Higgs with the right-handed 
sneutrino is small. The latter requirement is fulfilled in most of the cases
that we have considered and in some cases the mixing is slightly larger.
However, there is always the freedom of choosing the value of $A_t$
appropriately. Hence, it would be fair to say that the experimental limits on
the light Higgs boson mass can be satisfied in our analysis. 

Before starting the next section we want to emphasize that the 
parameters chosen for our numerical analysis are just for illustrative purpose.
These are not some particular and specific choices in some sacred corner of the 
model space. Since we have a large parameter space, it is always possible
to choose a different parameter point with the same characteristic features
satisfying all the experimental constraints.

\section{{\bf F}ermions in $\mu\nu$SSM}\label{munuSSM-fermion}

Effect of $\rpv$ in the superpotential and in the soft terms (eqns.(\ref{munuSSM-superpotential}),
(\ref{munuSSM-soft})) is responsible for enrichment in the scalar sector. In an identical
fashion, the neutral and the charged fermion mass matrices also receive enhancement
through lepton number violating couplings. 

\begin{flushleft}
{\it{$\maltese$ Neutralino mass matrix}}
\end{flushleft}

The neutral fermions of the MSSM $(\wt B^0,\wt W^0_3,\wt H^0_d,\wt H^0_u)$,
through second, third and fourth terms of $\mu\nu$SSM superpotential 
(eqn.(\ref{munuSSM-superpotential})), can mix with
three generations of left and right-handed neutrinos, $\nu_i$ and $\nu^c_i$ 
respectively. The neutralino mass matrix for $\mu\nu$SSM is thus a $10\times10$ symmetric
matrix \cite{c4LopezFogliani:2005yw,c4Escudero:2008jg,c4Ghosh:2008yh,c4Bartl:2009an,c4Ghosh:2010zi}.

In the weak interaction basis defined by
\bea
{\Psi^0}^T = \left(\tilde B^0, \tilde W_3^0, \tilde H_d^0, 
\tilde H_u^0,{\nu^c_\al},{\nu_{\al}} \right),
\label{neutralino_basis}
\eea
where $\al=1,2,3\equiv e, \mu, \tau$.
The neutral fermion mass term in the Lagrangian is of the form
\bea
{\mathcal{L}_{neutral}^{mass}} = -\frac{1}{2}{{\Psi^0}^T} \mathcal{M}_n 
{\Psi^0} + \text{H.c.},
\label{weak-basis-Lagrangian-neutralino}
\eea
The massless neutrinos now can acquire masses due to their mixing with the MSSM neutralinos 
and the gauge singlet right-handed
neutrinos. The three lightest eigenvalues of this $10\times10$ neutralino mass 
matrix correspond to the three light physical neutrinos, which are expected 
to be very small in order to satisfy the experimental data on massive 
neutrinos (see table \ref{osc-para}). The matrix $\mathcal{M}_n$ can be written 
in the following fashion 
\beq
\mathcal{M}_n =
\left(\begin{array}{cc}
M_{7\times 7} & m_{3\times 7}^T \\
m_{3\times 7} & 0_{3\times 3}
\end{array}\right),
\label{neutralino-seesaw}
\eeq
where using eqn.(\ref{Abbrevations})
\beq
M_{7\times7} =
\left(\begin{array}{ccccccc}
M_1 & 0 & -\frac{g_1}{\sqrt{2}}v_1 & \frac{g_1}{\sqrt{2}}v_2 & 0 & 0 & 0 \\ \\
0 & M_2 & \frac{g_2}{\sqrt{2}}v_1 & -\frac{g_2}{\sqrt{2}}v_2 & 0 & 0 & 0 \\ \\
-\frac{g_1}{\sqrt{2}}v_1 & \frac{g_2}{\sqrt{2}}v_1 & 0 & -{\mu} & 
-{\lambda^e}v_2 & -{\lambda^{\mu}}v_2 & -{\lambda^{\tau}}v_2 \\ \\
\frac{g_1}{\sqrt{2}}v_2 & -\frac{g_2}{\sqrt{2}}v_2 & -{\mu} & 0 & {\rho^e} 
& {\rho^{\mu}} & {\rho^{\tau}}\\ \\
0 & 0 & -{\lambda^e}v_2 & {\rho^e} & 2 {u^{ee}_c} & 2 {u^{e{\mu}}_c} & 
2 {u^{e{\tau}}_c}\\ \\
0 & 0 & -{\lambda^{\mu}}v_2 & {\rho^{\mu}} & 2{u^{{\mu}e}_c} & 
2 {u^{{\mu}{\mu}}_c} & 2 {u^{{\mu}{\tau}}_c}\\ \\
0 & 0 & -{\lambda^{\tau}}v_2 & {\rho^{\tau}} & 2 {u^{{\tau}e}_c} & 
2{u^{{\tau}{\mu}}_c} & 2 {u^{{\tau}{\tau}}_c}
\end{array}\right),
\label{neutralino_7x7}
\eeq
and
\beq
m_{3\times7} =
\left(\begin{array}{ccccccc}
-\frac{g_1}{\sqrt{2}}{v'_e} & \frac{g_2}{\sqrt{2}}{v'_e} & 0 & {r^e_c} & 
Y_{\nu}^{ee} v_2 & Y_{\nu}^{e{\mu}} v_2 & Y_{\nu}^{e{\tau}} v_2\\ \\
-\frac{g_1}{\sqrt{2}}{v'_{\mu}} & \frac{g_2}{\sqrt{2}}{v'_{\mu}} & 0 & 
{r^{\mu}_c} & Y_{\nu}^{{\mu}e} v_2 & Y_{\nu}^{{\mu}{\mu}} v_2 & 
Y_{\nu}^{{\mu}{\tau}} v_2\\ \\
-\frac{g_1}{\sqrt{2}}{v'_{\tau}} & \frac{g_2}{\sqrt{2}}{v'_{\tau}} & 0 &
{r^{\tau}_c}  & Y_{\nu}^{{\tau}e} v_2 & Y_{\nu}^{{\tau}{\mu}} v_2 & 
Y_{\nu}^{{\tau}{\tau}} v_2
\end{array}\right).
\label{neutralino_3x7}
\eeq

Note that the top-left $4\times4$ block of the matrix $M_{7\times7}$ is the 
usual neutralino mass matrix of the MSSM (see eqn.(\ref{MSSM-neutralino})). 
The bottom right $3\times3$ block represents
the Majorana mass matrix for gauge singlet neutrinos, which will be taken as 
diagonal (see eqn.(\ref{assumption1})) in the subsequent analysis. 
The entries of $M_{7\times7}$ are in
general of the order of the electroweak scale whereas the entries of 
$m_{3\times7}$ are much smaller $\sim$ $\cal O$ $(10^{-5})$ GeV. 
Hence, the matrix (\ref{neutralino-seesaw}) has a seesaw structure, 
which will give 
rise to three very light eigenvalues corresponding to three light neutrinos. 
The correct neutrino mass scale of $\sim$ $10^{-2}$ eV can easily be obtained
with such a structure of the $10\times10$ neutralino mass matrix. It has
been shown in ref.\cite{c4Ghosh:2008yh} that one can obtain the correct mass-squared
differences and the mixing pattern for the light neutrinos even with the choice
of flavour diagonal neutrino Yukawa couplings in eqn.(\ref{neutralino_3x7}). 
Besides, the choice of  flavour diagonal neutrino Yukawa couplings
(eqn.(\ref{assumption1})) makes the analysis simpler 
with a reduced number of parameters and makes the model more predictive. As 
we will show later, it is possible to find out the correct mixing pattern and 
the mass hierarchies (both normal and inverted) among the light neutrinos in 
such a situation, even at the tree level \cite{c4Ghosh:2008yh}. 

In order to obtain the physical neutralino states, one needs to diagonalize 
the $10\times10$ matrix $\mathcal{M}_n$. As in the case of MSSM, the symmetric 
mass matrix $\mathcal{M}_n$ can be diagonalized with one unitary matrix {\bf{N}}. 
The mass eigenstates $\chi^0_i$ are related to flavour eigenstates 
$\Psi^0_j$ (eqn.(\ref{neutralino_basis})) as 
\beq 
 \chi^0_i = \bN_{i1} \wt B^0 + \bN_{i2} \wt W^0_3 + \bN_{i3} \wt H^0_d
+ \bN_{i4} \wt H^0_u + \bN_{i,\al+4} \nu^c_\al + \bN_{i,\al+7} \nu_\al.
\label{munuSSM-neutralinos-reln}
\eeq
where the $10\times10$ unitary matrix {\bf{N}} satisfies
\beq\label{neutralino_mass_eigenstate_matrixform}
{\bN}^* \mathcal{M}_n {\bN}^{-1} = \mathcal{M}^0_D = 
diag(m_{\wt \chi^0_i}, m_{\nu_j}),
\eeq
with the diagonal neutralino mass matrix denoted as $\mathcal{M}^0_D$.
$i$ and $j$ runs from $1$ to $7$ and $1$ to $3$, respectively. The quantity
$m_{\wt \chi^0_i}$ represent neutralino masses. Physical neutrino
masses are being represented by $m_{\nu_j}$.
It is, in general, very difficult to predict the nature of the lightest neutralino 
(out of seven $\chi^0_i$) state since that depends on several unknown 
parameters. Neutralino mass matrix for $\mu\nu$SSM with complex VEVs
is given in ref.\cite{c4Fidalgo:2009dm}.

\begin{flushleft}
{\it{$\maltese$ Chargino mass matrix}}
\end{flushleft}

Similar augmentation in the charged lepton sector result in a $5\times5$
chargino mass matrix where the charged electroweak gauginos $(-i \wt {\lambda}^{\pm}_{2})$
and higgsinos $(\wt{H}^+_u,\wt{H}^-_d)$ mix with charged leptons through
$R_p$ violating couplings. These mixings are coming from the second term of
eqn.(\ref{munuSSM-superpotential}) and as well as from non-zero left-handed
sneutrino VEVs. 

In the weak interaction basis defined by
\beq
{\Psi^{+T}} = (-i \tilde {\lambda}^{+}_{2}, \wt{H}_u^{+}, \ell_{R}^{+}), 
~~{\Psi^{-T}} = (-i \tilde {\lambda}_{2}^{-}, \wt{H}_d^{-}, \ell_{L}^{-}), \nonumber \\
\label{chargino_basis}
\eeq
where $\ell=e,\mu,\tau$.
The charged fermion mass term in the Lagrangian is of the form
\beq\label{chargino_mass_Lagrangian}
{\mathcal{L}_{charged}^{mass}} = -\frac{1}{2} 
\left(\begin{array}{cc}
\Psi^{+^T} & \Psi^{-^T}
\end{array}\right)
\left(\begin{array}{cc}
0_{5\times5} & m_{5\times5}^T \\ \\
m_{5\times5} & 0_{5\times5}
\end{array}\right)
\left(\begin{array}{c}
\Psi^+ \\ \\
\Psi^-
\end{array}\right).
\eeq
Here we have included all three generations of charged leptons and assumed 
that the charged lepton Yukawa couplings are in the diagonal form. The matrix 
$m_{5\times5}$ using eqn.(\ref{Abbrevations}) is given by 
\cite{c4Escudero:2008jg,c4Ghosh:2008yh,c4Ghosh:2010zi}
\beq\label{chargino_mass_matrix}
m_{5\times5} =
\left(\begin{array}{ccccc}
M_2 & {g_2}{v_2} & 0 & 0 & 0 \\ \\
{g_2}{v_1} & {\mu} & -{Y_{e}^{ee}}{v'_e} & -{Y_{e}^{{\mu}{\mu}}}{v'_{\mu}} &
-{Y_{e}^{{\tau}{\tau}}}{v'_{\tau}} \\ \\
{g_2}{v'_e} & -{r^e_c} & {Y_{e}^{ee}}{v_1} & 0 & 0 \\ \\
{g_2}{v'_{\mu}} & -{r^{\mu}_c} & 0 & {Y_{e}^{{\mu}{\mu}}}{v_1} & 0 \\ \\
{g_2}{v'_{\tau}} & -{r^{\tau}_c} & 0 & 0 & {Y_{e}^{{\tau}{\tau}}}{v_1}
\end{array}\right).
\eeq
The charged fermion masses are obtained by applying a bi-unitary 
transformation like 
\beq\label{chargino_mass_eigenstate_matrixform}
\bU^* m_{5\times5} \bV^{-1} = \mathcal{M}^{\pm}_D,
\eeq
where $\bU$ and $\bV$ are two unitary matrices and $\mathcal{M}^{\pm}_D$ is the
diagonal matrix. Relations between the mass $\chi^\pm_i$ and flavour eigenstates
for charginos are same as eqn.(\ref{brpMSSM-charginos-reln}), namely
\bea 
&& \chi^+_i = \bV_{i1} \wt W^+ + \bV_{i2} \wt H^+_u + \bV_{i,\al+2} \ell^+_{\al_R}, \nn \\
&& \chi^-_i = \bU_{i1} \wt W^- + \bU_{i2} \wt H^-_d + \bU_{i,\al+2} \ell^-_{\al_L},
\label{munuMSSM-charginos-reln}
\eea
where $ \wt W^\pm \equiv {-i\wt {\lambda}}^{\pm}_{2}$.

It is important to note that the off-diagonal elements (except for 12 and 21 elements) of the chargino mass 
matrix (eqn. (\ref{chargino_mass_matrix})) either contain $Y^{ij}_\nu$ $({r^{i}_c} = {\sum Y^{ij}_{\nu} 
{v^c_{j}}})$ or left-handed sneutrino VEVs $v'_i$, both of which are very small $\sim$ $\cal {O}$ $(10^{-4}$ 
GeV). This indicates that the physical charged lepton eigenstates will 
have very small admixture of charged higgsino and charged gaugino states. So it is safe to assume 
(also verified numerically) that these lepton number violating mixing have very little effect on 
the mass eigenstates of the charged leptons. Thus, while writing down the 
PMNS matrix \cite{c4Pontecorvo:1957cp,c4Pontecorvo:1957qd,c4Maki:1962mu,c4Pontecorvo:1967fh} 
(eqn.(\ref{PMNS1})), it is justified to assume that one is  working in 
the basis where the charged lepton mass matrix is already in the diagonal form \cite{c4Ghosh:2008yh}. 

So far all of the neutralinos and charginos are considered in two-component form. Corresponding 
four component neutralino, chargino and charge conjugated chargino spinors are respectively defined as
\bea
& &{\widetilde \chi}^0_i =
\left(\begin{array}{c}
\chi^0_i \\
\ovl{\chi^0_i}\\
\end{array}\right),\quad
{\widetilde \chi}_i =
\left(\begin{array}{c}
\chi^+_i \\
\ovl{\chi^-_i}\\
\end{array}\right),\quad
{\widetilde \chi}^c_i =
\left(\begin{array}{c}
\chi^-_i \\
\ovl{\chi^+_i}\\
\end{array}\right),
\label{neutralino-chargino}
\eea
where $\chi^0_i$ and $\chi^{\pm}_i$ are two component neutral and charged spinors, respectively.
In our analysis the charged leptons are represented by their charged conjugate fields 
\cite{c4Nowakowski:1995dx}, which are positively charged.

Unlike the scalar mass squared matrices, eigenvalues of the neutralino or chargino 
mass matrix can be either positive or negative. 
It is always possible to remove the wrong signs via appropriate rotations.
However, then one should be very 
careful about the corresponding Feynman rules. A viable
alternative is to live with the signs of fermion masses ($\eta_i$ for neutralinos and
$\epsilon_i$ for charginos) and incorporate them properly in the respective Feynman rules
\cite{c4Gunion:1984yn}.

For the sake of completeness we also write down the quark mass matrices in $\mu\nu$SSM
in appendix \ref{appenB}.

\section{{\bf N}eutrinos at the tree level}\label{tree-neut}

It has been already emphasized that the $10\times10$ neutralino mass matrix
$\mathcal{M}_n$ possesses a seesaw like structure.
The effective light neutrino mass matrix ${M^{seesaw}_{\nu}}$, arising
via the seesaw mechanism in the presence of explicit lepton number violation, 
is in general given by
\beq 
{M^{seesaw}_{\nu}} = -{m_{3\times7}} {M_{7\times7}^{-1}}
{m_{3\times7}^T}.
\label{seesaw_formula}
\eeq

With small $\rpv$, it is possible to carry out a
perturbative diagonalization of the $10\times10$ neutralino mass
matrix (see \cite{c4Schechter:1981cv}), by defining \cite{c4Hirsch:1998kc,
c4Hirsch:2000jt} a matrix $\xi$ as
%
\beq
\label{expansion-parameter}
\xi=m_{3\times7}.M^{-1}_{7\times7}.
\eeq
%
If the elements of $\xi$ satisfy $\xi_{ij} \ll 1$, then this can be
used as an expansion parameter to get an approximate analytical
solution for the matrix $\bN$ (see
eqn.(\ref{neutralino_mass_eigenstate_matrixform})). A general expression for the 
elements of $\xi$ with simplified assumptions can be written in the form 
$\mathcal{A'}a_i + \mathcal{B'}b_i + \mathcal{C'}c_i$, where
\beq 
a_i = Y_{\nu}^{ii} v_2, ~c_i = {v'_i}, ~b_i = (Y_{\nu}^{ii} v_1 + 3 {\lambda}
{v'_i}) = (a_i\cot\beta + 3 \lambda c_i),
\label{specifications}
\eeq
with ${i} = {e,\mu,\tau} ~\equiv{1,2,3}$, $\tan\beta = \frac{v_2}{v_1}$ and
$\mathcal{A'},\mathcal{B'},\mathcal{C'}$ are complicated functions of various
parameters of the model \cite{c4Ghosh:2010zi}. The complete expressions for the elements of 
$\xi$ \cite{c4Ghosh:2010zi} are given in appendix \ref{appenC}. In deriving 
detailed expression for $\xi$'s we neglect the sub-dominant terms $\sim$ $\cal{O}$  
($\frac{v'^3}{{\tilde m}^3}$, $\frac{Y_\nu v'^2}{{\tilde m}^2}$,
$\frac{Y_\nu^2 v'}{{\tilde m}}$), where $\tilde m$ is the electroweak (or 
supersymmetry breaking) scale. 

With the help of eqn.(\ref{expansion-parameter}), eqn.(\ref{seesaw_formula})
reduces to 
\beq 
{M^{seesaw}_{\nu}} = -\xi{m_{3\times7}^T}.
\label{seesaw_formula-2}
\eeq
Using the favour of eqn.(\ref{assumption1}) in eqn.(\ref{seesaw_formula-2}), 
together with the expressions for $\xi^{ij}$ given 
in appendix \ref{appenC}, entries for the $3\times3$ matrix ${M^{seesaw}_{\nu}}$
are approximately given as (neglecting terms $\propto$ fourth power in $Y^{ij}_\nu,v'_i$ 
(separately or in a product) \cite{c4Ghosh:2008yh,c4Fidalgo:2009dm})
\bea
({M^{seesaw}_{\nu}})_{ij}& \approx& {\frac{v^2_2}{6 \kappa {v^c}}}{Y^{ii}_{\nu}} 
{Y^{jj}_{\nu}}(1-3 {\delta{ij}}) \nonumber \\
&-& {\frac{1}{2 M_{eff}}}\left[{v'_i}{v'_j} +{\frac{v_1 {v^c} 
(Y^{ii}_{\nu}{v'_j}+Y^{jj}_{\nu}{v'_i})}{\mu}
+{\frac{Y^{ii}_{\nu}Y^{jj}_{\nu}v^2_1 {v^c}^2}{\mu^2}}}\right].\nonumber \\
\label{mnuij-compact2}
\eea
Here we have used 
\bea
& & M_{eff}= M \left[1-{\frac{v^2}{2 M A {\mu}}}\left({\kappa}{v^c}^2 
\sin{2\beta} +{\frac{\lambda {v^2}}{2}}\right)\right],\nonumber \\ 
& &v_2 = v {\rm{sin}{\beta}},~v_1 = v {\rm{cos}}{\beta},~\mu = 3 \lam v^c,\nonumber \\
& & A = ({\kappa}{v^c}^2 + {\lambda} v_1 v_2),~\frac{1}{M} = \frac{g^2_1}{M_1} 
+\frac{g^2_2}{M_2}.
\label{mnuij-compact2-details}
\eea
Before proceeding further it is important to discuss eqn.(\ref{mnuij-compact2}) in
more details \cite{c4Ghosh:2008yh,c4Ghosh:2010zi}. 

\vspace*{0.1cm}
\noindent
I.~First consider the limit ${v^c}\rightarrow \infty$ and $v \rightarrow 0$ 
($\Rrightarrow v_1,v_2 \to 0$). 
Immediately eqn. (\ref{mnuij-compact2}) reduces to
\beq
\label{gauginoseesaw}
({M^{seesaw}_{\nu}})_{ij} \approx -{\frac{{v'_i}{v'_j}}{2 M}}
~~\Rrightarrow m_\nu \sim \frac{(g_1  c_i)^2}{M_1} + \frac{(g_2 c_i)^2}{M_2},
\eeq
which is the first part of the second term of eqn.(\ref{mnuij-compact2}). In 
this case the elements of the neutrino mass matrix are bilinear in the 
left-handed sneutrino VEVs and they appear due to a seesaw effect involving 
the gauginos. This is known as the ``gaugino seesaw'' effect and neutrino mass 
generation through this effect is a characteristic feature of the bilinear 
$R_p$ violating model. This effect is present in this model because we 
have seen earlier that the effective bilinear $R_p$ violating terms are 
generated in the scalar potential as well as in the superpotential through the 
vacuum expectation values of the gauge singlet sneutrinos $(\vp^i=Y^{ij}v^c_j)$. 
In gaugino seesaw the role of the {\it{Dirac}} mass terms are played by $g_1 v'_i$ and $g_2 v'_i$, 
where $g_1,~g_2$ are the $U(1)$ and the $SU(2)$ gauge couplings respectively
and $v'_i$ ($\equiv c_i$ (eqn.(\ref{specifications}))) stand for the left-handed sneutrino 
VEVs. The role of the {\it{Majorana}} masses are played by the gaugino soft
masses $M_1,~M_2$. The gaugino seesaw effect is closely analogous to the
TYPE-I \cite{c4Minkowski:1977sc,c4Yanagida:1979as,c4Glashow:1979nm,c4Mohapatra:1979ia,
c4GellMann:1980vs,c4Schechter:1980gr,c4Schechter:1981cv} + Type-III seesaw mechanism 
\cite{c4Foot:1988aq,c4Ma:1998dn} due to simultaneous involvement of a  
singlet $(\wt  B^0)$ and triplet fermion $(\wt  W^0_3)$ 
(see section \ref{neut-mass-II}, figure \ref{brpv-seesaw}, diagrams $(a,b)$).
This analogy has been pointed out in ref.\cite{c4Ghosh:2010zi}.
Note that the gaugino seesaw effect can generate mass for only one doublet neutrino,
as shown in eqn.(\ref{gauginoseesaw}).

\vspace*{0.1cm}
\noindent
II.~In the limit $M\rightarrow \infty$, eqn.(\ref{mnuij-compact2}) reduces to
\beq
\label{ordinaryseesaw}
({M^{seesaw}_{\nu}})_{ij} \approx{\frac{v^2_2}{6 \kappa {v^c}}}{Y^{ii}_{\nu}}
{Y^{jj}_{\nu}}(1-3 {\delta{ij}})\equiv {\frac{a_ia_j}{3 m_{\nu^c}}}(1-3 {\delta{ij}}),
\eeq
which corresponds to the ``ordinary seesaw'' effect between the left-handed and
gauge singlet right-handed neutrinos. Remember that the effective Majorana masses for 
the gauge singlet neutrinos are given by $m_\nu^c = 2 \kappa v^c$  
and the usual Dirac masses are given by $a_i=Y^{ii}_\nu v_2$. 
The ordinary seesaw effect can generate, in general, masses for more than one neutrinos. 
Thus depending on the magnitudes and the hierarchies of various diagonal
neutrino Yukawa couplings $Y_\nu^{ii}$, one can generate normal or inverted
hierarchy of neutrino masses (combining with the ``gaugino seesaw" effect) 
corresponding to atmospheric and solar mass squared differences \cite{c4Ghosh:2008yh}.

It is also
interesting to note that a conventional ordinary seesaw (generated only
through the mixing between left-handed and right-handed neutrinos) in contrast to
eqn.(\ref{ordinaryseesaw}) would give rise to a mass matrix of the form \cite{c4Fidalgo:2009dm}
\beq
\label{ordinaryseesaw-2}
({M^{seesaw}_{\nu}})_{ij} \approx - {\frac{v^2_2}{2 \kappa {v^c}}}{Y^{{ii}^2}_{\nu}}.
\eeq
The off-diagonal contributions as shown in eqn.(\ref{ordinaryseesaw}) are arising from
an effective mixing between the right-handed neutrinos and Higgsinos.
Hence, when right-handed neutrinos are also decoupled $(v^c\rightarrow \infty)$, 
the neutrino masses are zero as corresponds to the case
of a seesaw with only Higgsinos \cite{c4Fidalgo:2009dm}.


\subsection{{\bf N}eutrino masses at the tree level}\label{tree-neut-mass}
Eqn.(\ref{mnuij-compact2}) can be re-casted in a compact form 
using eqns.(\ref{specifications}) (\ref{mnuij-compact2-details}) as
\beq
({M^{seesaw}_{\nu}})_{ij} = {\frac{1}{6 \kappa {v^c}}} {a_{i}}
{a_{j}}(1-3\delta_{ij}) + {\frac{2 A {v^c}}{3 \Delta}} {b_{i}}
{b_{j}},
\label{mnuij-compact1}
\eeq
or alternatively using eqn.(\ref{specifications}) in a more elucidate form as
\beq
({M^{seesaw}_{\nu}})_{ij} = f_1 a_i a_j + f_2 c_i c_j + f_3 (a_i c_j + a_j c_i),
\label{mnuij-compact-recasted}
\eeq
with
\bea
f_1 = \frac{1}{6 \kappa v^c} (1-3\delta_{ij}) + \frac{2 A v^c {\rm{cot}}^2\beta}{3 \Delta}, 
~~f_2 = \frac{2 A \lambda \mu}{\Delta},~~f_3 = \frac{2 A \mu {\rm{cot}}\beta}{3 \Delta},
\label{specifications2}
\eea
and $\Delta = \lambda^2 (v^2_1 + v^2_2)^2 + 4 \lambda \kappa v_1 v_2 {v^c}^2 - 4 \lambda A \mu M$.
It is apparent from eqn.(\ref{mnuij-compact1}) that the second term $(\propto b_ib_j)$ can
contribute to only one neutrino mass, $\propto b^2_i$. However, presence of $(1-3\delta^{ij})$ factor
in the first term assures non-zero masses for other neutrinos. If we concentrate on the normal hierarchical
scheme of light neutrino masses, then with suitable choice of model ingredients it is possible
to generate the atmospheric neutrino mass scale $(\sim$ $\cal O$ $(10^{-11}$ GeV)) from the second term, 
whereas relatively small solar scale $(\sim$ $\cal O$ $(10^{-12}$ GeV)) emerges from the first term 
of eqn.(\ref{mnuij-compact1}). The imposed order of magnitude difference between the first and
the second term of eqn.(\ref{mnuij-compact1}) through certain choices of model parameters can be used to extract 
the eigenvalues of eqn.(\ref{mnuij-compact1}) analytically. Choosing the dominant terms to be $\propto b_ib_j,$ 
which contribute to only one neutrino mass, it is possible to apply the techniques of degenerate perturbation
theory to extract the effect of the perturbed term $(\propto a_ia_j)$ over the unperturbed one 
$(\propto b_ib_j)$ \cite{c4Ghosh:2008yh}. It has to be clarified here that actually in $\mu\nu$SSM
for a novel region of the parameter space $b_i \sim a_i$ \cite{c4Ghosh:2008yh}, however, with a clever 
choice of the $\lam$ and $\kappa$ parameter it is possible to vary the order of magnitude of the 
co-efficients in front $(\frac{1}{6\kappa v^c},\frac{2 A v^c}{3 \Delta}$, see eqn.(\ref{mnuij-compact1})).
For the chosen set of parameters (see table \ref{table-tree-parameters}) co-efficients of the $a_ia_j$
term is an order of magnitude smaller compared to that of $b_ib_j$ \cite{c4Ghosh:2008yh}. So the perturbative
approach is well justified. As shown in ref.\cite{c4Ghosh:2008yh} it is possible to extract 
simple analytical form for light neutrino masses in this approach. Detailed expressions for 
the eigenvectors and eigenvalues of eqn.(\ref{mnuij-compact1})
obtained through perturbative calculations are given in appendix \ref{appenC}. It is interesting to
see from eqn.(\ref{eigenvalues}) that the correction to unperturbed eigenvalues are proportional 
to the effect of ordinary seesaw \cite{c4Ghosh:2008yh}. 

The numerical values of the solar and atmospheric mass squared differences 
$\Delta{m^2_{solar}}$ $(\equiv \Delta{m^2_{21}})$ and $\Delta{m^2_{atm}}$ $(\equiv \Delta{m^2_{31}})$ 
as obtained from full numerical calculations (Using eqn.(\ref{seesaw_formula}))
and from appropriate analytical formulae (Using eqn.(\ref{eigenvalues}))
have been shown in table \ref{table-tree}\footnote{A typo has been corrected
compared to ref.\cite{c4Ghosh:2008yh}. Also to denote individual neutrino masses, $m_{\nu_i}$
are used instead of $m_i$ (ref.\cite{c4Ghosh:2008yh}).} and the results show good agreement \cite{c4Ghosh:2008yh}. 
The numerical calculations have been performed with the help of a code developed by us using 
Mathematica \cite{c4math-wolfram}. 
In our numerical analysis for the normal hierarchical pattern in
light neutrino masses, we choose ${m_2}|_{max} < 1.0 \times 10^{-11}$ GeV  \cite{c4Ghosh:2008yh}.
Results of table \ref{table-tree} are consistent with the three flavour global neutrino data
\cite{c4Schwetz:2008er,c4GonzalezGarcia:2010er} as shown in table \ref{osc-para} in the $3\sigma$ limit.
It is interesting to observe that unlike conventional bilinear $R_p$ violating models, in 
$\mu\nu$SSM all three neutrinos are massive itself at the tree level. Consequently, it is
possible to accommodate the three flavour global neutrino data (table \ref{osc-para}) at the tree 
level even with the choice of diagonal neutrino Yukawa couplings (see table 
\ref{table-tree-parameters}) \cite{c4Ghosh:2008yh}.

\begin{table}[ht]
\centering
\begin{tabular}{c | c || c | c}
\hline \hline 
Parameter  & Chosen Value &  Parameter  & Chosen Value  \\ \hline \hline
$\lam$   & $0.06$ & $(A_\lam \lam)$ & $-60$ GeV\\
$\kappa$   & $0.65$ & $(A_\kappa \kappa)$ & $650$ GeV\\
$Y^{11}_\nu$   & $4.57\times10^{-7}$ & $(A_\nu Y_\nu)^{11}$ & $1.57\times10^{-4}$ GeV\\
$Y^{22}_\nu$   & $6.37\times10^{-7}$ & $(A_\nu Y_\nu)^{22}$ & $4.70\times10^{-4}$ GeV\\
$Y^{33}_\nu$   & $1.80\times10^{-7}$ & $(A_\nu Y_\nu)^{33}$ & $3.95\times10^{-4}$ GeV\\
$M_1$   & $325$ GeV & $M_2$ & $650$ GeV\\
$m^2_{\wt L}$   & $400^2$ GeV$^2$ & $m^2_{\wt \nu^c}$ & $300^2$ GeV$^2$\\
\hline \hline
\end{tabular}
\caption{\label{table-tree-parameters}
Parameter choices (consistent with figure \ref{scalar_sector_for_tan_beta_values}) for result 
presented in table \ref{table-tree}. Eqn.(\ref{assumption1}) has been used here and we choose 
$\tan\beta=10$.}
\end{table}

Both left and right sneutrino VEVs ($v'_i,v^c_i$, respectively) are derived using
the set of parameters given in table \ref{table-tree-parameters}. The relation between the 
gaugino soft masses $M_1$ and $M_2$ are assumed to be GUT (grand unified theory) motivated, 
so that, at the electroweak scale, we have $M_1 : M_2~=~1:2$.

\begin{table}[ht]
\footnotesize
\centering
\begin{tabular}{c|c|c|c|c|c}
\hline\hline
 & \multicolumn{3}{c|}{$m_\nu$ (eV) ($\times 10^3$)} 
& $\Delta{m^2_{21}}$(eV$^2$) &  $\Delta{m^2_{31}}$(eV$^2$) \\
\cline{2-4}
 & $m_{\nu_1}$ & $m_{\nu_2}$ & $m_{\nu_3}$ & $(\times 10^5)$ 
& $(\times 10^3)$ \\
\hline
\text{Using eqn.(\ref{seesaw_formula})} & 4.169 & 9.970 & 48.23 & 8.203 & 2.307 \\
\hline
\text{Using eqn.(\ref{eigenvalues})} & 4.168  & 9.468 & 47.71 & 7.228  & 2.187 \\
\hline\hline
\end{tabular}
\caption{\label{table-tree}
Absolute values of the neutrino masses and the mass-squared differences for a 
sample point of the parameter space \cite{c4Ghosh:2008yh}. Results for full numerical analysis 
have been obtained using eqn.(\ref{seesaw_formula}). Approximate analytical expressions of 
eqn.(\ref{eigenvalues}) have been used for comparison. Parameter choices are given in 
table \ref{table-tree-parameters}.}
\end{table}
\subsection{{\bf N}eutrino mixing at the tree level}\label{tree-neut-mixing}

The expansion parameter $\xi$ (see eqns.(\ref{expansion-parameter-terms})) has been introduced in 
eqn.(\ref{expansion-parameter}) to perform perturbative diagonalization of the $10\times10$ neutralino 
mass matrix $\mathcal{M}_n$. It is possible to express the neutralino mixing matrix $\bN$
(see eqn.(\ref{neutralino_mass_eigenstate_matrixform})) in leading order in $\xi$ as
\beq 
\bN^* = \left(\begin{array}{cc}
    \mathcal{N}^* & 0 \\
    0 & U^T
\end{array}\right)
\left(\begin{array}{cc}
1-\frac{1}{2} \xi^{\dagger} \xi & \xi^{\dagger} \\
-\xi & 1-\frac{1}{2} \xi \xi^{\dagger}
\end{array}\right).
\label{neutralino-mixing-matrix}
\eeq
The $10\times10$ neutralino mass matrix $\mathcal{M}_n$ can approximately be
block diagonalized to the form {\it{diag($M_{7\times7},{M^{seesaw}_{\nu}}$)}},
by the matrix defined in eqn.(\ref{neutralino-mixing-matrix}). The matrices
$\mathcal{N}$ and $U$, defined in eqn.(\ref{neutralino-mixing-matrix}),
are used to diagonalize ${M_{7\times7}}$ and ${M^{seesaw}_{\nu}}$ in the following 
manner (using eqn.(\ref{neutralino_mass_eigenstate_matrixform})),
\bea
\label{diag-matrix}
& &\mathcal{N}^* M_{7\times7} \mathcal{N}^{\dagger} = 
{\rm{diag}} (m_{{\widetilde \chi}^0_i}), \nonumber \\ 
& &{U}^T {M^{seesaw}_{\nu}} {U} = 
{\rm{diag}} (m_{\nu_1},m_{\nu_2},m_{\nu_3}).
\eea
Where $U$ is the non-trivial leptonic mixing matrix, known as PMNS matrix
\cite{c4Pontecorvo:1957cp,c4Pontecorvo:1957qd,c4Maki:1962mu,c4Pontecorvo:1967fh}.
As already stated in section \ref{neut-osc}, a non-trivial neutrino mixing is 
a consequence of massive neutrinos. If we adhere to a scenario where $CP$ is preserved, 
the PMNS matrix following eqn.(\ref{PMNS1}) can be written as
\beq
U =
\left(\begin{array}{ccc}
{c_{12}}{c_{13}} & {s_{12}}{c_{13}} & {s_{13}} \\ \\
-{s_{12}}{c_{23}}-{c_{12}}{s_{23}}{s_{13}} & {c_{12}}{c_{23}}
-{s_{12}}{s_{23}}{s_{13}}  & {s_{23}}{c_{13}}\\ \\
{s_{12}}{s_{23}}-{c_{12}}{c_{23}}{s_{13}} & -{c_{12}}{s_{23}}
-{s_{12}}{c_{23}}{s_{13}}  & {c_{23}}{c_{13}}
\end{array}\right),
\label{PMNS-CPC}
\eeq 
where $c_{ij} = \cos{\theta_{ij}}$, $s_{ij} = \sin{\theta_{ij}}$.

It is definitely possible to extract the mixing angles from $U$
in a full numerical analysis. However, it is always useful to
do the same with a simplified approximate analytical analysis (if at all possible)
to get an idea about the relative importance of the different parameters.
An analysis of this kind for light neutrino mixing angles using degenerate 
perturbation theory has been addressed in ref. \cite{c4Ghosh:2008yh}. We showed that it is 
possible to write down the PMNS matrix $U$ as (eqn.(\ref{mneutrino_mixing_numerical}))
\bea
{U}&=&
\left(\begin{array}{ccc}
{\mathcal{Y}_1} & {\mathcal{Y}_2} & {\mathcal{Y}_3}
\end{array}\right)_{3\times3},
\label{mneutrino_mixing_numerical2}
\eea
where ${\mathcal{Y}_i}$'s are defined in appendix \ref{tree-perturbed}. Using 
eqn.(\ref{mneutrino_mixing_numerical2}) it is possible to derive appropriate expressions
for the light neutrino mixing angles $\theta_{13},\theta_{23},\theta_{12}$ as \cite{c4Ghosh:2008yh}
\bea
\sin^2\theta_{13} = \frac{b^2_e}{b^2_e + b^2_\mu + b^2_\tau}.
\label{reactor_analytical}
\eea
\bea
\sin^2\theta_{23} = \frac{b^2_\mu}{b^2_\mu + b^2_\tau}.
\label{atmos_analytical}
\eea
\bea
\sin^2\theta_{12} \approx 1 - {(\alpha^\prime_1 
+ \alpha^\prime_2 {\frac{b_e}{b_\tau}})}^2,
\label{solar_analytical}
\eea
where $b_i$'s are given by eqn.(\ref{specifications}). The quantities 
$\alpha^\prime_1,\alpha^\prime_2$ are given by eqn.(\ref{alphas}).

It is apparent from eqn.(\ref{reactor_analytical}) that if we want the (13) mixing angle to be 
small (which is supported by data, see table \ref{osc-para}) then one must have 
$b^2_e \ll (b^2_\mu + b^2_\tau)$. On the other hand, since the (23) mixing angle
$\theta_{23}$ is maximal by nature $(\sim 45^\circ$, see table \ref{osc-para}), it is natural to 
expect $b^2_\mu = b^2_\tau$. The formula for solar mixing angle $\theta_{12}$ is a bit complicated. 
Nevertheless, in order to have $\theta_{12} \sim 35^\circ$, the square root of the second 
term on the right hand side of eqn.(\ref{solar_analytical}) should 
be approximately $0.8$. So these approximate analytical formulae clearly help
us to choose suitable corner of parameter space rather than performing a 
blind search, which is the power of the analytical approach.

We compare three light neutrino mixing angles as obtained from eqn.(\ref{mneutrino_mixing_numerical2})
to that obtained in full numerical analysis using eqn.(\ref{seesaw_formula}) in
table \ref{table-tree-mixing}. Neutrino masses are taken to be normal hierarchical.
\begin{table}[ht]
\footnotesize
\centering
\begin{tabular}{c|c|c}
\hline\hline
\text{mixing angles in degree}  & \text{Using (\ref{seesaw_formula})} & 
\text{Using (\ref{mneutrino_mixing_numerical})}\\
\hline
$\theta_{12}$ & 36.438 & 37.287 \\
\hline
$\theta_{13}$ & 9.424 & 6.428 \\
\hline
 $\theta_{23}$& 38.217 & 42.675 \\
\hline\hline
\end{tabular}
\caption{\label{table-tree-mixing}
Neutrino mixing angles using eqn.(\ref{seesaw_formula}) and
eqn.(\ref{mneutrino_mixing_numerical}) Parameter choices are given in 
table \ref{table-tree-parameters}. These values are consistent with entries of
table \ref{osc-para} in the $3\sigma$ limit \cite{c4Schwetz:2008er}.}
\end{table}

\begin{figure}[ht]
\vspace{0.5cm}
\centering
\includegraphics[width=10.00cm]{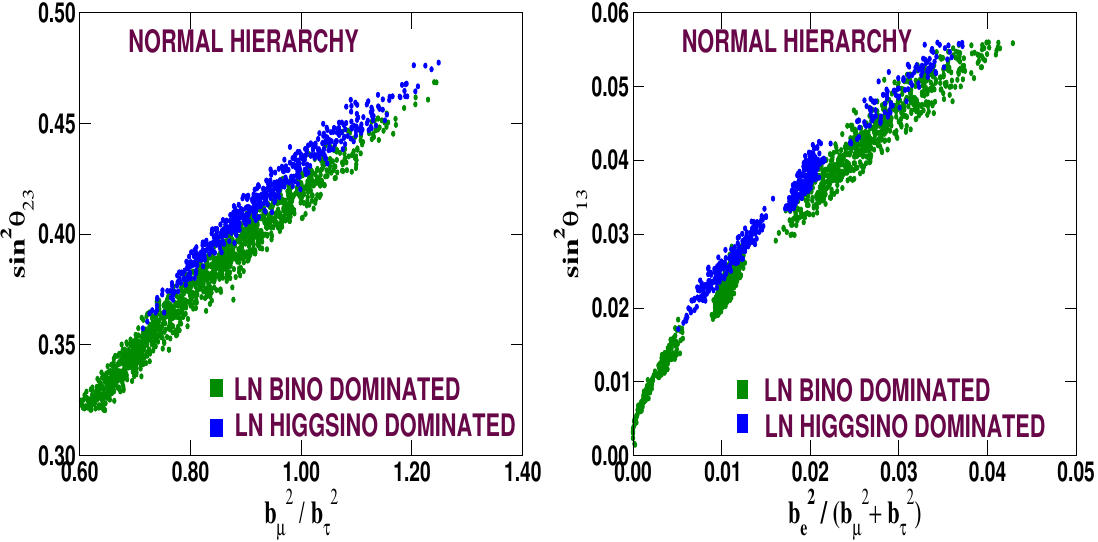}
\caption{Scatter plot of the neutrino mixing angle $\sin^2\theta_{23}$ (left) and  $\sin^2\theta_{13}$ (right) 
as a function of the ratio $\frac{b^2_\mu}{b^2_\tau}$ and $\frac{b^2_e}{b^2_\mu + b^2_\tau}$. Values of 
model parameters are given in table \ref{neut-tree-mix-param}. The lightest neutralino (LN) is either 
a bino $(\wt B^0)$ or a higgsino $(\wt H^0_u,\wt H^0_d)$ dominated. Light neutrino mass ordering is normal
hierarchical.}
\label{neut2313-mixing-tree}
\end{figure}

\begin{figure}[ht]
\vspace{0.5cm}
\centering
\includegraphics[width=5.00cm]{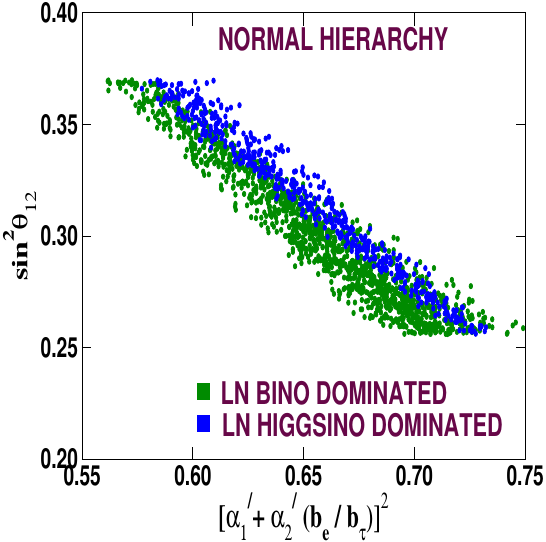}
\caption{$\sin^2\theta_{12}$ versus $(\alpha_1^{\prime}+\alpha_2^{\prime}\frac{b_e}{b_\tau})^2$ scatter
plot. Parameter choice and mass hierarchy is same as figure \ref{neut2313-mixing-tree}.}
\label{neut12-mixing-tree}
\end{figure}

We can see that for this set of chosen parameters (table \ref{table-tree-parameters}), numerical and 
approximate analytical results give quite good agreement. Naturally, one would 
be interested to check the predictions made in 
eqns. (\ref{reactor_analytical}), (\ref{atmos_analytical}), 
and (\ref{solar_analytical}) over a wide region in the parameter space and see
the deviations from the full numerical calculations. These are shown in 
figures.\ref{neut2313-mixing-tree}, \ref{neut12-mixing-tree} \cite{c4Ghosh:2008yh}.


It is apparent from the left diagram of figure \ref{neut2313-mixing-tree} that for $b^2_\mu = b^2_\tau$, 
the value of $\sin^2\theta_{23}$ varies in the range $0.41 -- 0.44$, which corresponds to 
$\theta_{23}$ between 40$^\circ$ and 42$^\circ$. On the other hand, 
eqn.(\ref{atmos_analytical}) tells that for $b^2_\mu = b^2_\tau$, $\sin^2\theta_{23} = 0.5$.
So for a wide region of parameter space result from the 
numerical calculation is reasonably close to the prediction from the 
approximate analytical formula.

Also from figure \ref{neut12-mixing-tree} as 
$(\alpha_1^{\prime}+\alpha_2^{\prime}\frac{b_e}{b_\tau})^2\rightarrow0.50$, 
$\sin^2\theta_{12}$ tends to be maximal, that is $\theta_{12}=45^\circ$,
which is well expected.

\begin{table}[ht]
\centering
\begin{tabular}{c | c || c | c}
\hline \hline 
Parameter  & Chosen Value &  Parameter  & Chosen Value  \\ \hline \hline
$Y^{11}_\nu$   & $3.55-5.45\times10^{-7}$ & $(A_\nu Y_\nu)^{11}$ & $1.25-1.95\times10^{-4}$ GeV\\
$Y^{22}_\nu$   & $5.55-6.65\times10^{-7}$ & $(A_\nu Y_\nu)^{22}$ & $3.45-4.95\times10^{-4}$ GeV\\
$Y^{33}_\nu$   & $1.45-3.35\times10^{-7}$ & $(A_\nu Y_\nu)^{33}$ & $2.35-4.20\times10^{-4}$ GeV\\
$m^2_{\wt L}$   & $400^2$ GeV$^2$ & $m^2_{\wt \nu^c}$ & $300^2$ GeV$^2$\\
$\lam$   & $0.06(0.13)$ & $(A_\lam \lam)$ & $-1000\times\lam$ GeV\\
$\kappa$   & $0.65$ & $(A_\kappa \kappa)$ & $1000\times\kappa$ GeV\\
$M_1$   & $110(325)$ GeV & $M_2$ & $2\times M_1$ GeV\\
\hline \hline
\end{tabular}
\caption{\label{neut-tree-mix-param}
Parameter choices (consistent with figure \ref{scalar_sector_for_tan_beta_values}) 
for figures \ref{neut2313-mixing-tree}, \ref{neut12-mixing-tree}. 
$\lam=0.06(0.13)$ for a bino(higgsino) dominated lightest neutralino. 
Similarly, $M_1=110~(325)$ GeV for a bino (higgsino) dominated lightest neutralino.
Eqn.(\ref{assumption1}) has been used here and we choose $\tan\beta=10$. The set of 
chosen parameters are consistent with the constraints of the scalar sector.}
\end{table}

Concerning table \ref{neut-tree-mix-param} it has to be emphasized here that
the allowed regions in the $\lambda -~\kappa$ plane (see figure 
\ref{scalar_sector_for_tan_beta_values}) are not very sensitive to the values of 
$Y_\nu$ and $A_\nu Y_\nu$ due to their smallness. Hence we choose to vary them randomly
(see table \ref{neut-tree-mix-param}), in order to accommodate the three flavour global neutrino data.

\begin{figure}[ht]
\vspace{0.5cm}
\centering
\includegraphics[width=6.00cm]{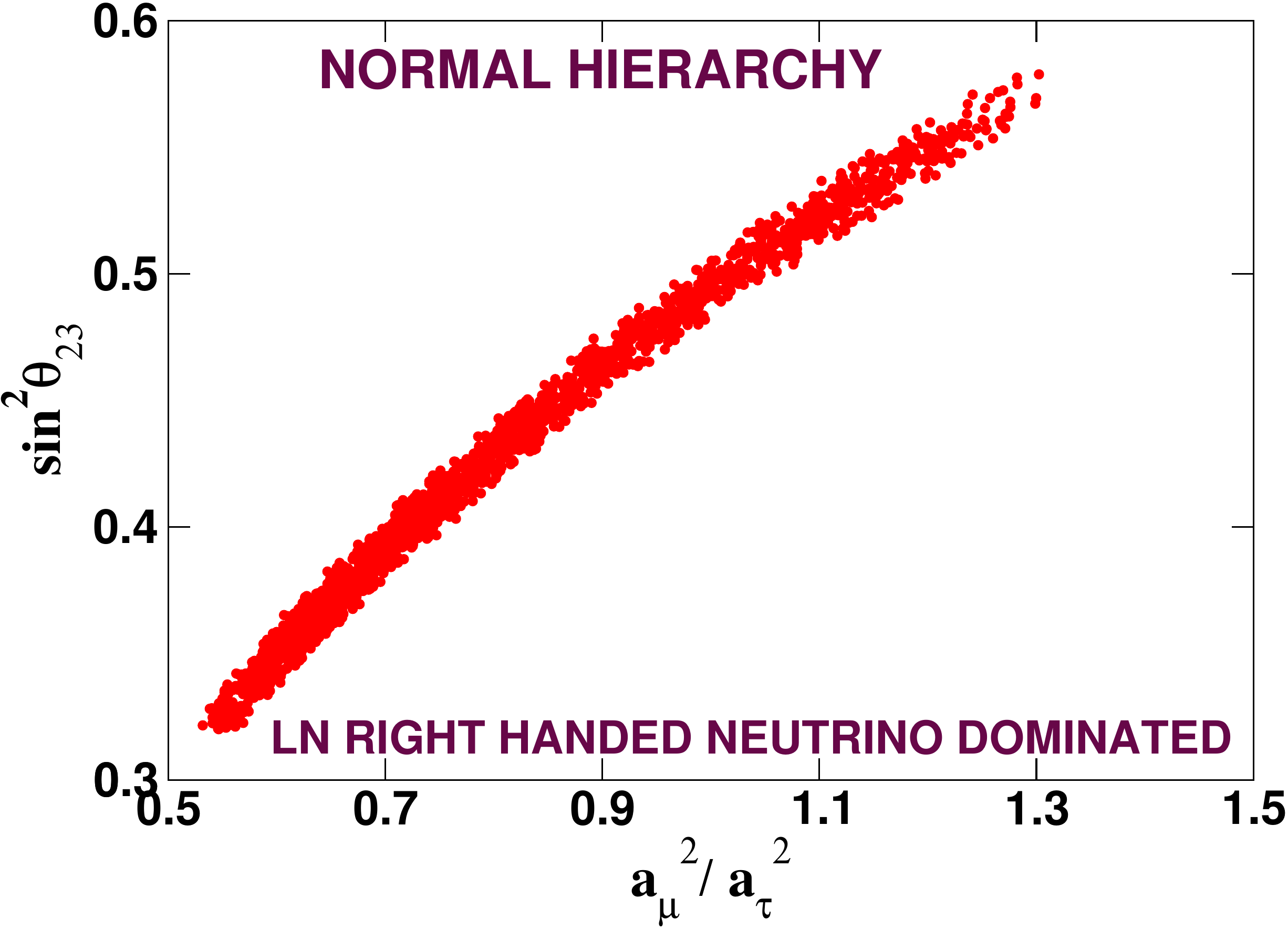}
\caption{Scatter plot of the neutrino mixing angle $\sin^2\theta_{23}$ as 
a function of the ratio $\frac{a^2_\mu}{a^2_\tau}$. The lightest neutralino (LN) is 
right-handed neutrino $(\nu^c)$ dominated.}
\label{neut23a-mixing-tree}
\end{figure}

So far we considered eqn.(\ref{mnuij-compact1}) in the limit when with suitable choice of
model parameters the terms $\propto a_ia_j$ can act as perturbation over the second term.
However, the huge parameter space for $\mu\nu$SSM always leaves room for the inverse 
situation. In other words there exists suitable corner of parameter space where the 
first term of eqn.(\ref{mnuij-compact1}) is the dominant one and then eqn.(\ref{atmos_analytical})
can be expressed as
\bea
\sin^2\theta_{23} = \frac{a^2_\mu}{a^2_\mu + a^2_\tau}.
\label{atmos_analyticala}
\eea
This is exactly what is shown by figure \ref{neut23a-mixing-tree}. 
Note that for $a^2_\mu = a^2_\tau$, the atmospheric mixing angle becomes maximal. 

\begin{figure}[ht]
\vspace{0.5cm}
\centering
\includegraphics[width=13.50cm]{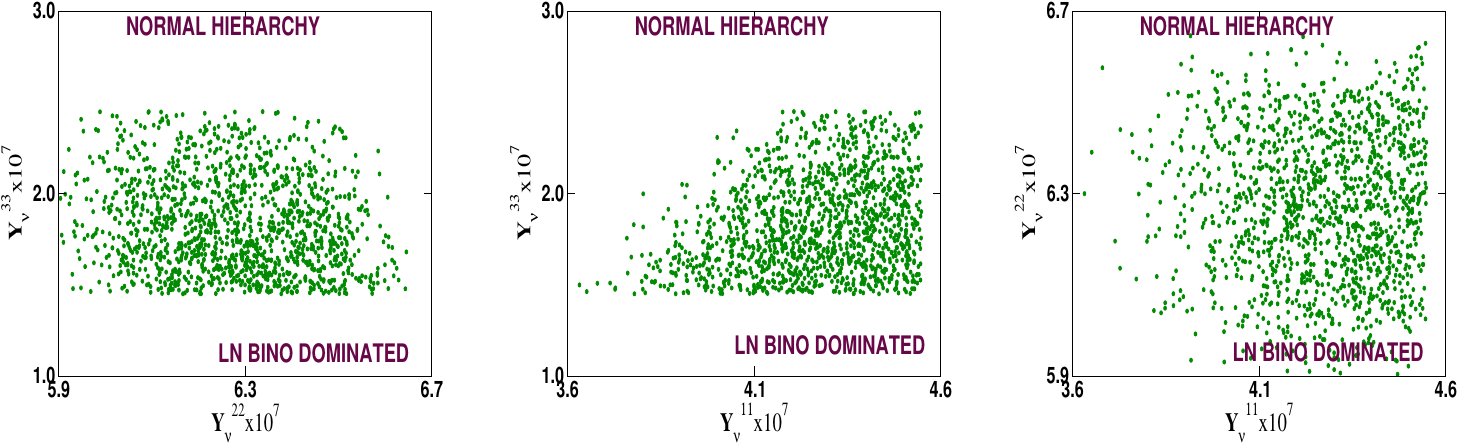}
\caption{Plots for normal hierarchical scheme of neutrino mass in 
$Y^{22}_\nu ~-~Y^{33}_\nu$,~$Y^{11}_\nu ~-~Y^{33}_\nu$ and  $Y^{11}_\nu ~-
~Y^{22}_\nu$ plane when the lightest neutralino (LN) is bino dominated.}
\label{normal_hierarchical_scheme_for_bino-dominated_case}
\end{figure}

In figure \ref{normal_hierarchical_scheme_for_bino-dominated_case}, we have shown
the regions in the various $Y_\nu$ planes satisfying the three flavour global 
neutrino data. The values of other parameters are as shown in 
table \ref{neut-tree-mix-param} for the case where the lightest neutralino $(\ntrl1)$ is 
bino dominated. We can see from these figures that the allowed values of 
$Y_\nu$s show a mild hierarchy such that $Y_\nu^{22} > Y_\nu^{11} > 
Y_\nu^{33}$ \cite{c4Ghosh:2008yh}. 

Similar studies have been performed for the inverted 
hierarchical case and the allowed region shows that the magnitudes of the 
neutrino Yukawa couplings are larger compared to the case of normal 
hierarchical scheme of the neutrino masses with a different hierarchy among
the $Y_\nu$'s themselves ($Y_\nu^{11} > Y_\nu^{22} > Y_\nu^{33}$). In this case
$\sin^2\theta_{12}$ shows an increasing behaviour with the ratio 
$b^2_e/b^2_\mu$, similar to the one shown by $\sin^2\theta_{23}$ with 
$b^2_\mu/b^2_\tau$ in the normal hierarchical scenario (see figure \ref{neut2313-mixing-tree}). 
On the other hand, $\sin^2\theta_{23}$ shows a decreasing behaviour with $b^2_\mu/b^2_\tau$. 
In all these cases, the solar and atmospheric mass-squared differences are within 
the 3$\sigma$ limits (table \ref{osc-para}).

\section{{\bf N}eutrinos at the loop level}\label{loop-neut}
\begin{figure}[ht]
\vspace{0.5cm}
\centering
\includegraphics[width=3.80cm]{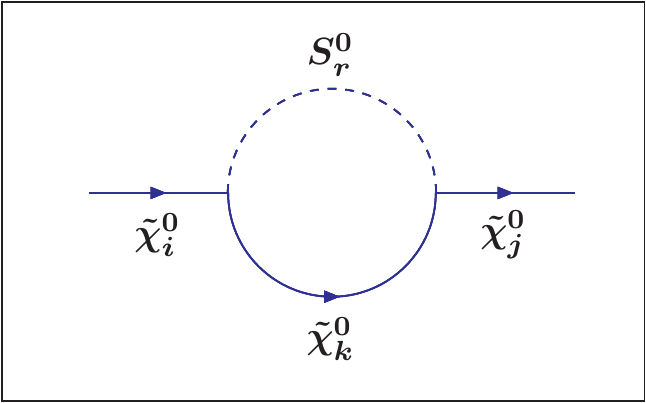}
\includegraphics[width=3.80cm]{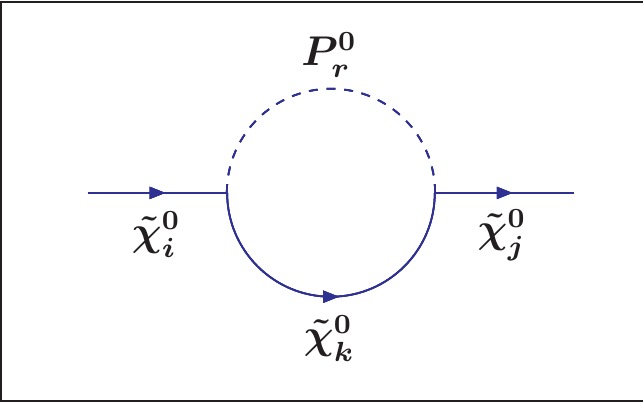}
\includegraphics[width=3.80cm]{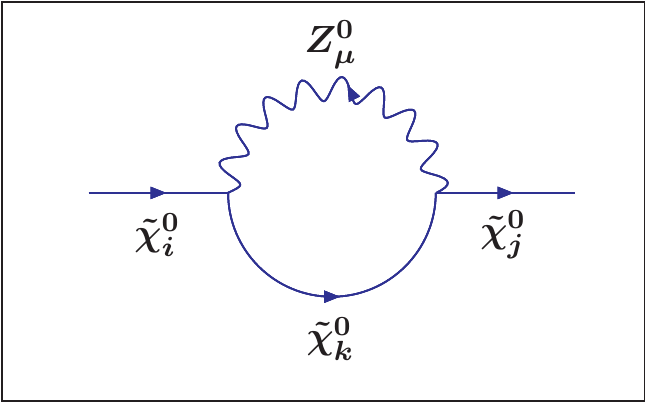}
\vspace{1cm}
\includegraphics[width=3.80cm]{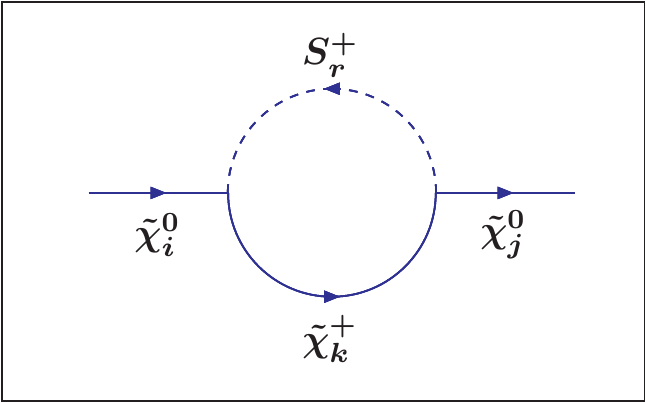}
\includegraphics[width=3.80cm]{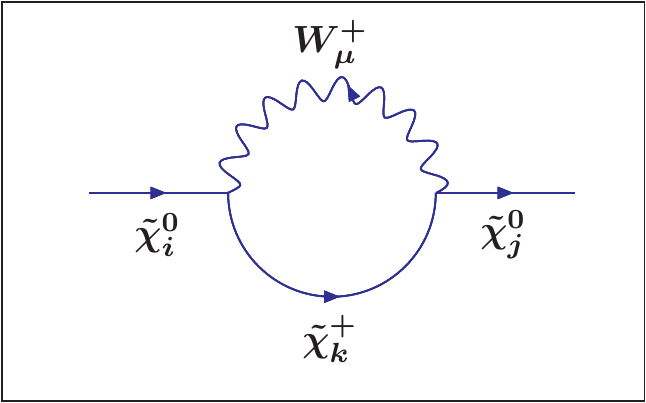}
\includegraphics[width=3.80cm]{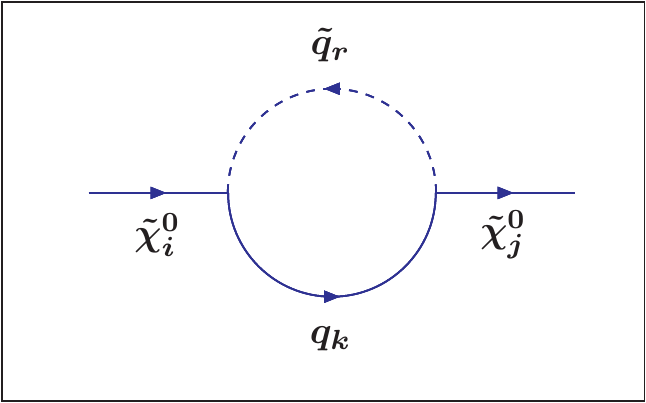}
\caption{One-loop diagrams contributing to the neutralino masses. The various
  contributions are arising from (clockwise from top left) 
  (a) neutralino-neutralino-neutral scalar loop, 
  (b) neutralino-neutralino-neutral pseudoscalar loop, 
  (c) neutralino-neutralino-$Z^0_{\mu}$ loop, 
  (d) neutralino-chargino-charged scalar loop, 
  (e) neutralino-chargino-$W^{\pm}_{\mu}$ loop, 
  (f) neutralino-quark-squark loop.}
\label{one-loop-diagrams}
\end{figure}

It is legitimate to ask that what is the motivation for performing loop calculations
in $\mu\nu$SSM when all three neutrinos can acquire masses at the tree level \cite{c4Ghosh:2008yh}?
In fact this is a feature where the $\mu\nu$SSM model is apparently successful over most
of the other models of light neutrino mass generation where loop corrections are unavoidable
in order to account for oscillation data. However, in the regime of 
renormalizable quantum field theories, stability of any tree
level analysis must be re-examined in the light of radiative
corrections. Following this prescription, the results of neutrino masses and
mixing will be more robust, once tree level analysis is further improved by
incorporating radiative corrections. The radiative corrections may have
sizable effect on the neutrino data at one-loop level. Thus, although all
three SM neutrinos acquire non-zero masses in the $\mu \nu$SSM even at the
tree level \cite{c4Ghosh:2008yh}, it is interesting to investigate the fate of
those tree level masses and mixing when exposed to one-loop corrections. With
this in view, in the following subsections we perform a systematic study of the neutrino
masses and mixing with all possible one-loop corrections both analytically and
numerically. In the subsequent subsections, while showing the results of one-loop
corrections, we try to explain the deviations (which may or may not be
prominent) from the tree level analysis. The complete set of one-loop diagrams are shown 
in figure \ref{one-loop-diagrams}.
Before going into the details, let us
discuss certain relevant issues of one-loop correction and renormalization for
the neutralino-neutrino sector. The most general one-loop 
contribution to the unrenormalized neutralino-neutrino two-point function can be expressed as
\beq 
i {\bf \Sigma}^{ij}_{\n \n}(p) = i\{\ps \left[P_L
  {\Sigma^L_{ij}}(p^2) + P_R {\Sigma^R_{ij}}(p^2)\right] -\left[P_L
  {\Pi^L_{ij}}(p^2) + P_R {\Pi^R_{ij}}(p^2)\right]\},
\label{loopgen-unrenorm}
\eeq
where $P_L$ and $P_R$ are defined as $\frac{1-\gamma_5}{2}$ and $\frac{1+\gamma_5}{2}$, respectively. 
$i,~j~=~1,...,10$ and $p$ is the external momentum. The unrenormalized self-energies $\Sigma^{L,R}$ 
and $\Pi^{L,R}$ depend on the squared external momentum $(p^2)$. The generic self
energies $\Sigma^{L(R)}_{ij},~\Pi^{L(R)}_{ij}$ of the Majorana neutralinos and neutrinos must
be symmetric in its indices, $i$ and $j$. $\ovl{DR}$ scheme \cite{c4Siegel:1979wq,c4Capper:1979ns,
c4Avdeev:1981vf,c4Avdeev:1982xy,c4Jack:1997sr} has been used to regularize one-loop contributions.
In the $\ovl{DR}$ scheme\footnote{In $\ovl{DR}$ scheme the subtraction
procedure is same as $\ovl{MS}$ \cite{c4Passarino:1978jh} scheme and the momentum integrals are also 
evaluated with $D$ dimensions. However, the {\it{Dirac algebras}} are done strictly in four dimensions
since only in four dimensions the numbers of fermions and bosons match in the case of
a supersymmetric system.}, the counter-terms cancel only the divergent pieces of the self-energies. 
Thus the self energies become finite but depend on the arbitrary scale
of renormalization. To resolve this scale dependency, the tree level masses are promoted to 
running masses in which they cancel the explicit scale dependence of the self energies 
$\Sigma,\Pi$ \cite{c4Hirsch:2000ef}. The resulting one-loop corrected mass matrix using dimensional 
reduction ($\ovl{DR}$) scheme is given by
\bea
\label{one-loop-corrected-mass}
(\mathcal{M}^{\rm{tree + 1-loop}}_{\chi^0})^{ij} &=& {m}_{\n}(\mu_R)\delta^{ij} + 
{\frac{1}{2}}\left(\widetilde \Pi^V_{ij}({{m^2_i}}) + \widetilde
  \Pi^V_{ij}({{m^2_j}})  \right.\nn\\
&-& {m_{\n_i}}{\widetilde
    \Sigma^V_{ij}}({{m^2_i}})
- \left. {m_{\n_j}}{\widetilde
    \Sigma^V_{ij}}({{m^2_j}})\right), \nn\\
\eea
with
\bea
\label{Sigma-Pi-renomalized}
\widetilde \Sigma^V_{ij} = \frac{1}{2} (\widetilde \Sigma^L_{ij} + \widetilde
\Sigma^R_{ij}),
~~\widetilde \Pi^V_{ij} = \frac{1}{2} (\widetilde \Pi^L_{ij} + \widetilde
\Pi^R_{ij}), 
 \eea
where the tree level neutralino mass $({m}_{\n})$ is defined at the
renormalization scale $\mu_R$, set at the electroweak scale. Here, the 
word {\it{neutralino mass}} stands for all the {\it{ten}} eigenvalues 
of the $10\times10$ neutralino mass matrix. The self-energies $\Sigma,~\Pi$ 
are also renormalized in the $\ovl{DR}$ scheme and denoted
by $\widetilde \Sigma$ and $\widetilde \Pi$ respectively. The detailed
expressions of $\widetilde \Sigma^V_{ij}$ and $\widetilde \Pi^V_{ij}$
depend on corresponding Feynman rules and the Passarino-Veltman functions 
\cite{c4'tHooft:1978xw,c4Passarino:1978jh}. 
%
In the next section we will describe our calculational approach.
\section{{\bf A}nalysis of neutrino masses and mixing at one-loop}\label{loop-neut-calc}
In this section we consider the effect of radiative corrections to the light
neutrino masses and mixing. Just for the sake of completeness it is always better to
recapitulate some of the earlier works regarding
one-loop corrections to the neutralino-neutrino sector. The complete set of
radiative corrections to the neutralino mass matrix in the $R_p$ conserving
MSSM was discussed in ref.\cite{c4Pierce:1993gj,c4Pierce:1994ew}, and the leading order
neutrino masses has been derived in ref.\cite{c4Hall:1983id}. One-loop radiative
corrections to the neutrino-neutralino mass matrix in the context of a
$R_P$-violating model were calculated in ref.\cite{c4Hempfling:1995wj} using
't-Hooft-Feynman gauge. In ref.\cite{c4Hirsch:2000ef}, $R_{\xi}$ gauge has been
used to compute the corrections to the neutrino-neutralino mass matrix at
one-loop level in an $R_p$-violating scenario. For our one-loop calculations
we choose to work with 't-Hooft-Feynman gauge, i.e. $\xi=1$.
Neutrino mass generation at the one-loop level in other variants of $R_P$-violating MSSM has
been widely addressed in literature, which are already given in the beginning of 
subsection \ref{neut-mass-II}. We note in passing that in a recent reference 
\cite{c4Liebler:2011tp} on-shell renormalization 
of neutralino and chargino mass matrices in $R_p$ violating models has been addressed, 
which also includes the $\mu\nu$SSM.

We begin by outlining the strategy of our analysis. We
start with a general $10\times10$ neutralino matrix, with off-diagonal entries
as well, which has a seesaw structure in the flavour-basis (see
eqn.(\ref{neutralino-seesaw})). Schematically, we can rewrite
eqn.(\ref{neutralino-seesaw}) as,
\beq
\mathcal{M}_n = \left(\begin{array}{cc}
    M_f & m^T_{D_f} \\
    m_{D_f} & 0
\end{array}\right),
\label{neutralino-seesaw-schematic}
\eeq
where the orders of the block matrices are same as those indicated in
eqn. (\ref{neutralino-seesaw}), and the subscript `$f$'denotes the flavour
basis. Here $M_f$ stands for the $7\times7$ Majorana mass matrix of the heavy
states, while $m_{D_f}$ contains the $3\times7$ Dirac type masses for the left-handed 
neutrinos. In the next step, instead of utilizing the seesaw structure
of this matrix to generate the effective light neutrino 
mass matrix for the three active light neutrino species, we {\it{diagonalize}} the entire
$10\times10$ matrix $\mathcal{M}_n$. The diagonal $10\times10$ matrix
$\mathcal{M}_D^{0}$ (eqn.(\ref{neutralino_mass_eigenstate_matrixform})) thus contains 
tree level neutralino masses, which we symbolically write as \cite{c4Ghosh:2010zi}
\beq
\mathcal{M}_D^{0} = \left(\begin{array}{cc}
    M_m & 0 \\
    0 & m_m
\end{array}\right),
\label{neutralino-tree-level-schematic}
\eeq
where $M_m~(m_m)$ are the masses of the heavy states (left-handed
neutrinos). Following eqn.(\ref{diag-matrix}) one can write 
\bea
M_m &=& diag (m_{\wt \chi^0_1},m_{\wt \chi^0_2},m_{\wt \chi^0_3},m_{\wt \chi^0_4}
,m_{\wt \chi^0_5},m_{\wt \chi^0_6},m_{\wt \chi^0_7}),\nn\\
m_m &=& diag (m_{\nu_1},m_{\nu_2},m_{\nu_3}).
\label{mass-mass-p1-p2}
\eea
At this stage we turn on all possible one-loop interactions as shown in
figure \ref{one-loop-diagrams}, so
that the $10\times10$ matrix $\mathcal{M}_D^{0}$ picks up radiatively
generated entries, both diagonal and off-diagonal. The resulting one-loop
corrected Lagrangian for the neutralino mass terms in the $\wt \chi^0$ basis, 
following eqn.(\ref{weak-basis-Lagrangian-neutralino}), can be written as
\beq
\label{loop-correcetd-Lagrangian-neutralino-a} {\mathcal{L}^{\prime}}
= -\frac{1}{2}{{\chi^{0^T}}} \left(\mathcal{M}^0_D +
  \mathcal{M}^1 \right) {\chi^0} + \text{H.c.}, 
\eeq
where $\mathcal{M}^1$ contains the effect of one-loop corrections. The
$10\times10$ matrix $\mathcal{M}^0_D$ is diagonal, but the matrix
$\mathcal{M}^1$ is a general symmetric matrix with off diagonal
entries.

One can rewrite the above equation, using
eqns.(\ref{munuSSM-neutralinos-reln}) and
(\ref{neutralino_mass_eigenstate_matrixform}), as
\beq
\label{loop-correcetd-Lagrangian-neutralino-b}
{\mathcal{L}^{\prime}} = -\frac{1}{2}{{\Psi^0}^T} \left(\mathcal{M}_n
  + \bN^T \mathcal{M}^1 \bN\right) {\Psi^0} + \text{H.c.}.  
\eeq
This is nothing but the {\it{one-loop corrected}} neutralino mass term in the Lagrangian in the
flavour basis. Symbolically \cite{c4Ghosh:2010zi},
\beq
\label{loop-correcetd-Lagrangian-neutralino-c}
{\mathcal{L}^{\prime}} = -\frac{1}{2}{{\Psi^0}^T} 
\mathcal{M}^{\prime} {\Psi^0} + \text{H.c.},
\eeq
with the $10\times10$ matrix $\mathcal{M}^{\prime}$ having the form \cite{c4Ghosh:2010zi}
\beq
\mathcal{M}^{\prime} = 
\left(\begin{array}{cc}
M_f + \Delta{M}_f & (m_{D_f} + \Delta{m_{D_f}})^T \\
m_{D_f} + \Delta{m_{D_f}} & \Delta{m}_f
\end{array}\right).
\label{neutralino-tree-plus-one-loop-level-schematic}
\eeq
The quantities $\Delta{M}_f$ and $\Delta{m}_f$ stand for one-loop corrections
to the heavy neutralino states and light neutrino states respectively, in the
flavour basis $\Psi^0$. The entity $\Delta{m_{D_f}}$ arises because of the off
diagonal interactions, i.e. between the heavy neutralinos and the light
neutrinos, in the same basis $(\Psi^0)$. Note that all of $\Delta M_f$, $\Delta m_{D_f}$,
$\Delta m_f$ in the $\chi_0$ basis are given by the second term on the right
hand side of eqn.(\ref{one-loop-corrected-mass}). We suitably transform them
into the basis $\Psi^0$ with the help of neutralino mixing matrix
$\bN$. From the order of magnitude approximations\footnote{The 
loop corrections are at least suppressed
by a loop factor $\frac{1}{16\pi^2}$ and thus tree level order of magnitude
approximations are still valid.}
the matrix $\mathcal{M}^{\prime}$ once again possesses a
seesaw structure, and one can therefore write down the 
one-loop corrected effective light neutrino mass matrix as
\beq ({M}^{\nu^{\prime}})_{\rm{eff}} \approx
\Delta{m}_f - (m_{D_f} + \Delta{m_{D_f}})(M_f +
\Delta{M}_f)^{-1}((m_{D_f} + \Delta{m_{D_f}})^T).
\label{mass-basis-seesaw-schematic}
\eeq
Let us now present an approximate form of
eqn.(\ref{mass-basis-seesaw-schematic}). For simplicity, let us begin by
assuming the quantities present in eqn.(\ref{mass-basis-seesaw-schematic}) to
be c-numbers (not matrices). In addition, assume $M_f \gg \Delta{M}_f$
(justified later), so that eqn.(\ref{mass-basis-seesaw-schematic}) may be
written as,
\beq 
({M}^{\nu^{\prime}})_{\rm{eff}} \approx \Delta{m}_f - \delta \times
{M_f}\left\{\left(\frac{m_{D_f}}{M_f}\right)^2 + 2
  \left(\frac{m_{D_f}}{M_f}\right) \left(\frac{\Delta
      m_{D_f}}{M_f}\right) + \left(\frac{\Delta m_{D_f}}{M_f}\right)^2
\right\},
\label{mass-basis-seesaw-schematic-2}
\eeq
with $\delta = \left(1-\frac{\Delta M_f}{M_f}\right)$. Now, even when $\Delta
m_{D_f}$ $\sim$ $\frac{1}{16 \pi^2}$ $m_{D_f}$ and $\Delta
M_f$ $\sim$ $\frac{1}{16 \pi^2}$ $M_f$, 
eqn.(\ref{mass-basis-seesaw-schematic-2}) looks like

\bea 
({M}^{\nu^{\prime}})_{\rm{eff}} &\approx& \Delta{m}_f - {M_f}
\left(1-\frac{1}{16 \pi^2}\right)\left\{\left(\frac{m_{D_f}}{M_f}\right)^2 +
\frac{2}{16 \pi^2} \left(\frac{m_{D_f}}{M_f}\right)^2 \right.\nn\\
&+&\left. \frac{1}{256 \pi^4}
\left(\frac{m_{D_f}}{M_f}\right)^2 \right\}.
\label{mass-basis-seesaw-schematic-3}
\eea
Thus, up to a very good approximation one can rewrite
eqn.(\ref{mass-basis-seesaw-schematic-3}) as
\beq
({M}^{\nu^{\prime}})_{\rm{eff}} \approx
\Delta{m}_f - {M_f} \left(\frac{m_{D_f}}{M_f}\right)^2.
\label{mass-basis-seesaw-schematic-4}
\eeq
Reimposing the matrix structure and using eqn.(\ref{seesaw_formula}),
eqn.(\ref{mass-basis-seesaw-schematic-4}) can be modified as,
\beq
({M}^{\nu^{\prime}})_{\rm{eff}} \approx
\Delta{m}_f + {M^{seesaw}_{\nu}}.
\label{mass-basis-seesaw-schematic-5}
\eeq
The eigenvalues of the $3\times3$ one-loop corrected neutrino mass matrix
$({M}^{\nu^{\prime}})_{\rm{eff}}$ thus correspond to one-loop corrected light
neutrino masses. In conclusion, it is legitimate to calculate one-loop
corrections to the $3\times3$ light neutrino mass matrix only 
(see eqn.(\ref{mass-basis-seesaw-schematic-5})), and diagonalize
it to get the corresponding one-loop corrected mass eigenvalues \cite{c4Ghosh:2010zi}. 

Let us denote the one-loop corrections to the masses of heavy neutralinos and
light neutrinos in the basis $\chi^0$ by $\Delta M$ and $\Delta m$
respectively. The one-loop corrections arising from neutralino-neutrino
interactions is denoted by $\Delta m_D$ in the same basis. The tree level
neutralino mixing matrix $\bN$, in the leading power of expansion matrix $\xi$ 
(eqn.(\ref{expansion-parameter})), using eqn.(\ref{neutralino-mixing-matrix}) 
can be written as,
\beq
\bN =
\left(\begin{array}{cc}
\mathcal{N} & \mathcal{N} \xi^T \\
-U^\dagger \xi^* & U^\dagger
\end{array}\right)=
\left(\begin{array}{cc}
\widetilde N_{7\times 7} & \widetilde N_{7\times 3} \\
\widetilde N_{3\times 7} & \widetilde N_{3\times 3}
\end{array}\right).
\label{neutralino-mixing-matrix-block-form}
\eeq
Now from the order of magnitude approximation of $\xi$ (eqn.(\ref{expansion-parameter}))
we get approximately $\xi \sim$ $({m_D^{\nu}}/{M_{\widetilde \chi^0}})$, where $m_D^{\nu}$
represents a generic entry of $m_{3\times7}$ matrix and ${M_{\widetilde \chi^0}}$ that
of $M_{7\times7}$ (see eqn.(\ref{neutralino-seesaw})). So apparently
the entries of the matrices $\widetilde N_{7\times
  3},~\widetilde N_{3\times 7}$ suffers a suppression $\sim \cal{O}$
$({m_D^{\nu}}/{M_{\widetilde \chi^0}})$, due to very small
neutrino-neutralino mixing \cite{c4Atre:2009rg}. The quantities $m_D^{\nu}$ $\sim$ $\cal{O}$ 
($10^{-4}$ GeV) and $M_{\widetilde \chi^0}$ $\sim$ $\cal{O}$ ($10^2$ GeV) represent the Dirac mass 
of a left-handed neutrino $(\nu_i)$ and the Majorana mass of a neutralino $(\chi^0_i)$,
respectively. From
eqns.(\ref{loop-correcetd-Lagrangian-neutralino-b}), (\ref{neutralino-mixing-matrix-block-form}) 
it is easy to figure out the relation between $\Delta m$ and $\Delta m_f$ as,
\beq 
\Delta m_{f} = {\widetilde N_{7\times 3}^T} {\Delta M} {\widetilde
  N_{7\times 3}} + {\widetilde N_{7\times 3}^T} {\Delta m_D^T} {\widetilde
  N_{3\times 3}} + {\widetilde N_{3\times 3}^T} {\Delta m_D} {\widetilde
  N_{7\times 3}} + {\widetilde N_{3\times 3}^T} {\Delta m} {\widetilde
  N_{3\times 3}}.
\label{flavour-mass-relation}
\eeq
Now as argued earlier, for a Dirac neutrino, the mass is $\lesssim$ $\cal{O}$
$(10^{-4}~\rm{GeV})$, while for a neutralino, the mass is 
$\sim$ $\cal{O}$ $(10^{2}~\rm{GeV})$. This means that the 
entries of the off-diagonal blocks in
eqn.(\ref{neutralino-mixing-matrix-block-form}) are $\lesssim$ $\cal{O}$
$(10^{-6})$. Therefore, for all practical purpose, one can neglect
the first three terms in comparison to the fourth term on the right
hand side of eqn.(\ref{flavour-mass-relation}). Thus,
\beq
\Delta m_{f} \approx {\widetilde N_{3\times 3}^T} {\Delta m} {\widetilde N_{3\times 3}}.
\label{flavour-mass-relation-2}
\eeq
up to a very good approximation. With this in view, our strategy is to compute
the one-loop corrections in the $\chi^0$ basis first, and then use
eqn.(\ref{flavour-mass-relation-2}) to obtain the corresponding corrections in
the flavour basis. Finally, adding tree level contribution $M^{seesaw}_\nu$ (eqn.(\ref{seesaw_formula})) 
to $\Delta m_f$ (eqn.(\ref{flavour-mass-relation-2})), we diagonalize
eqn.(\ref{mass-basis-seesaw-schematic-5}) to obtain the one-loop corrected
neutrino masses. We have performed all calculations in the 't-Hooft-Feynman
gauge. Let us also note in passing that the form of eqn.(\ref{one-loop-corrected-mass}) 
predicts off-diagonal entries ($i\neq j$). The off-diagonal elements are
responsible for the admixtures between diagonal entries, which become dominant
only when $\left({m}_{\n_i}-{m}_{\n_j}\right)~\lesssim(\frac{\alpha}{4 \pi})
\times {\rm{some ~electroweak ~scale ~mass}}$, (using the essence of eqn.(\ref{seesaw-mixing})) 
and then, one can choose $p^2 =\ovl{m^2} = ({{m^2}_{\n_i} + {m^2}_{\n_j}})/2$ for external 
momentum\cite{c4Hempfling:1995wj}. Thus, one can conclude that unless the tree level masses are highly 
degenerate, the off-diagonal radiative corrections can be neglected for all practical
purposes, when at least one indices $i$ or $j$ refers to a heavy states.

The self-energy corrections contain entries of the neutralino mixing matrix
$\bN$ through the couplings $O^{ff^{\prime}b}$ appearing in Feynman rules 
(see, appendix \ref{appenD}) \cite{c4Ghosh:2010zi}. This is because, the 
self energies ${\it{\widetilde \Sigma}_{ij}}$ and ${\it{\widetilde \Pi}_{ij}}$ in
general contain products of couplings of the form $O^{ff^{\prime}b}_{i..}O^{ff^{\prime}b}_{j..}$ 
(see, appendix \ref{appenE}\cite{c4Ghosh:2010zi} for detailed expressions of 
$\widetilde \Sigma^V_{ij}$ and $\widetilde \Pi^V_{ij}$). The matrix $\bN$, on the other
hand, contains the expansion parameter $\xi$ in the leading order (see
eqn.(\ref{neutralino-mixing-matrix})). This observation, together with the help
of eqn.(\ref{expansion-parameter-terms}), help us to express the
effective structure of the one-loop corrected neutrino mass matrix as \cite{c4Ghosh:2010zi},
\beq
[(\mathcal{M}^{\nu^{\prime}})_{eff}]_{ij} = A_1 a_i a_j + A_2 c_i c_j + A_3 (a_i c_j + a_j c_i),
\label{one-loop corrected structure of neutralino mass matrix}
\eeq
where $a_i$ and $c_i$ are given by eqn.(\ref{specifications}) and
$A_i$'s are functions of our model parameters and the Passarino-Veltman
functions $(B_0,B_1)$ \cite{c4'tHooft:1978xw,c4Passarino:1978jh,c4Hahn:1998yk} defined in 
appendix \ref{appenF}. The form of the loop corrected mass matrix thus obtained is identical 
to the tree level one (see, eqn.(\ref{mnuij-compact-recasted}))
with different coefficients $A_1$, $A_2$ and $A_3$ arising due to one-loop 
corrections.  

Note that the one-loop diagrams in figure \ref{one-loop-diagrams}, contributing to the neutrino mass matrix 
are very similar to those obtained in bilinear R-parity violating scenario \cite{c4Hirsch:2000ef,c4Davidson:2000uc,
c4Davidson:2000ne,c4Diaz:2003as,c4Grossman:2003gq,c4Dedes:2006ni}. However, it has been pointed
out in ref.\cite{c4Bartl:2009an}, that there is a new significant contribution coming from the loops containing the
neutral scalar and pseudoscalar with dominant singlet component. This contribution is proportional to the 
mass-splitting between the singlet scalar and pseudoscalar states 
\cite{c4Hirsch:1997vz,c4Grossman:1997is,c4Dedes:2007ef}.  
The corresponding mass splittings for the doublet sneutrinos are much smaller \cite{c4Bartl:2009an}.
In fact the sum of contributions of the singlet scalar $({\wt \nu}^c_{n{\mathcal R}})$ and pseudoscalar 
states $({\wt \nu}^c_{n{\mathcal I}})$ (see diagrams one and two of the top row of figure \ref{one-loop-diagrams}) 
is $\propto$ $\kappa^2 v^{c^2}$, squared mass difference between
the singlet scalar and pseudoscalar mass eigenstates \cite{c4Bartl:2009an}.
The effect of one-loop correction to light neutrino
masses and mixing has been considered in ref.\cite{c4Bartl:2009an} for one and two generations
of right-handed neutrinos. 

To conclude this section we finally concentrate on the one-loop contributions to
light neutrino mixing. The tree level
$3\times3$ orthogonal matrix $U$ diagonalizes the tree level seesaw matrix $M^{seesaw}_\nu$
as shown in eqn.(\ref{diag-matrix}). In a similar fashion the $3\times3$ orthogonal matrix 
(in the limit of all phases equal to zero) that 
diagonalizes the one-loop corrected neutrino mass matrix
$({M}^{\nu^{\prime}})_{\rm{eff}}$ (eqn.(\ref{mass-basis-seesaw-schematic-5})), can be denoted as
$U'$. Mathematically
\beq
\label{diag-one-loop-corrected-neutrino-mass}
{{U}^{{\prime}^{T}}} ({M}^{\nu^{\prime}})_{\rm{eff}}
{{U}^{\prime}} ={\rm{diag}} (m'_1,~m'_2,~m'_3),
\eeq
with $m'_1$, $m'_2$, $m'_3$ as the three one-loop corrected light neutrino
masses. The matrix $U'$ now can be used (see eqn.(\ref{PMNS-CPC})) to extract
the one loop corrected light neutrino mixing angles, $\theta'_{23},\theta'_{12},
\theta'_{13}$.

In the next section we will discuss the effect of one-loop corrections to the 
light neutrino masses and mixing in $\mu\nu$SSM for different light neutrino mass
hierarchy.
\section{{\bf O}ne-loop corrections and mass hierarchies}\label{loop-mass-hierarchy}

Analytical forms for the tree level and the one-loop corrected light neutrino mass matrices 
are given by eqn.(\ref{seesaw_formula}) and 
eqn.(\ref{one-loop corrected structure of neutralino mass matrix}), respectively.
Note that in both of the equations the first two terms $(\propto a_ia_j,
~\propto~c_ic_j)$ individually can generate only one neutrino mass, $\propto \sum a^2_i$
and $\propto \sum c^2_i$, respectively. These terms are the effect of the {\it{ordinary}}
and the {\it{gaugino}} seesaw, as already discussed in section \ref{tree-neut}. 
Together, they can generate two neutrino masses which is sufficient to
satisfy the neutrino oscillation data without the cross term $(a_ic_j+a_jc_i)$.
However, it is the effect of the mixing terms $(a_ic_j+a_jc_i)$ which together
with the first two terms along with different co-efficients for each term 
give masses to all three light neutrinos \cite{c4Ghosh:2008yh,c4Ghosh:2010zi}. 

In the following three consecutive subsections we will analyze the effect of one-loop radiative corrections
on the light neutrino masses and mixing when the mass orderings are (1) normal, (2) inverted
and (3) quasi-degenerate in nature. The choice of model parameters are given in table
\ref{loop-param} \cite{c4Ghosh:2008yh,c4Ghosh:2010zi}.
\begin{table}[ht]
\centering
\begin{tabular}{c | c || c | c}
\hline \hline 
Parameter  & Chosen Value &  Parameter  & Chosen Value  \\ \hline \hline
$\lam$   & $0.10$ & $(A_\lam \lam)$ & $-100$ GeV\\
$\kappa$   & $0.45$ & $(A_\kappa \kappa)$ & $450$ GeV\\
$m^2_{\wt e^c}$   & $300^2$ GeV$^2$ & $v^c$  & $-895$ to $-565$ GeV\\
$(A_\nu Y_\nu)^{ii}$   & $Y^{ii}_\nu\times1$ TeV & $\tan\beta$ & $10$\\
$M_1$   & $110$ GeV & $M_2$ & $220$ GeV\\
\hline \hline
\end{tabular}
\caption{\label{loop-param}
Choice of parameters for numerical analysis consistent with the EWSB conditions. These
choices are according to the eqn.(\ref{assumption1}). The gaugino soft masses $M_1$ and $M_2$ are
assumed to be GUT (grand unified theory) motivated, so that, at the
electroweak scale, we have $M_1 : M_2~=~1:2$.}
\end{table}
Apart from the right-handed sneutrino VEVs other variables are chosen to be
the left sneutrino VEVs $(v'_i)$ and the flavour diagonal neutrino Yukawa couplings $(Y^{ii}_\nu)$. 
These are given in table \ref{loop-param-2} \cite{c4Ghosh:2008yh,c4Ghosh:2010zi}.
\begin{table}[ht]
 \centering
\begin{tabular}{c||c|c|c||c|c|c}
\hline
 & \multicolumn{3}{c||}{$Y^{ii}_\nu \times 10^7$} 
& \multicolumn{3}{c}{$v'_i \times 10^5 (\rm{GeV})$ } \\
\cline{2-7}
 & $Y^{11}_\nu$ & $Y^{22}_\nu$ & $Y^{33}_\nu$ & $v'_1$ 
& $v'_2$ & $v'_3$ \\
\hline\hline
\text{Normal hierarchy} & 3.550 & 5.400 & 1.650 & 0.730 & 10.100 & 12.450\\
\text{Inverted hierarchy} & 12.800  & 3.300 & 4.450 & 8.350  & 8.680 & 6.400\\
\text{Quasi-degenerate-I} & 19.60 & 19.94 & 19.99 & 9.75 & 10.60 & 11.83\\
\text{Quasi-degenerate-II} & 18.50  & 18.00 & 18.00 & 9.85  & 10.50 & 10.10\\
\hline\hline
\end{tabular}
\caption{Values of the neutrino Yukawa couplings and the left-handed sneutrino VEVs,
  used as sample parameter points for numerical calculations. These are the
  values around which the corresponding parameters were varied. 
Other parameter choices are given in table \ref{loop-param}.\label{loop-param-2}}
\end{table}
To fit the three flavour global data we consider not only the oscillation
constraints (see table \ref{osc-para}) but also constraints from various
non-oscillation experiments like Tritrium beta decay, neutrinoless double 
beta decay and cosmology both for the tree level and the one-loop combined analysis.

\subsection{{\bf N}ormal hierarchy}\label{loop-normal}
In the normal hierarchical pattern of the three light neutrino masses 
(individual masses are denoted by $m_i$, $i=1,2,3$), 
the atmospheric and the solar mass squared differences, given by 
$\Delta m^2_{atm}= m^2_3 - m^2_2$ and $\Delta m^2_{solar}= m^2_2 - m^2_1$, 
are largely governed by the higher mass squared in each case, namely, 
$m_3^2$ and $m_2^2$, respectively. Before going into the discussion of 
the variation of the mass-squared values with the model parameter, some 
general remarks are in order. First of all, note that in eqn.(\ref{specifications}), 
if we choose $v^\prime_i$ such that $v'_i \gg \frac{Y^{ii}_\nu v_1}{3 \lambda}$, 
then $b_i \approx c_i$\cite{c4Fidalgo:2009dm}. 
Second, both the tree level and the one-loop corrected light neutrino mass matrix
have similar structure as shown in eqn.(\ref{mnuij-compact-recasted}) and 
eqn.(\ref{one-loop corrected structure of neutralino mass matrix}). Due to this 
structural similarity we expect both the tree and the one-loop corrected masses and mixing 
to show similar type of variations with certain relevant quantities, however
with some modifications, because of the inclusion of the one-loop corrections.
This similarity also indicates that the light neutrino masses and mixing are
entirely controlled by $a_i$ and $c_i$. 

\begin{figure}[ht]
\centering
\includegraphics[width=6.00cm]{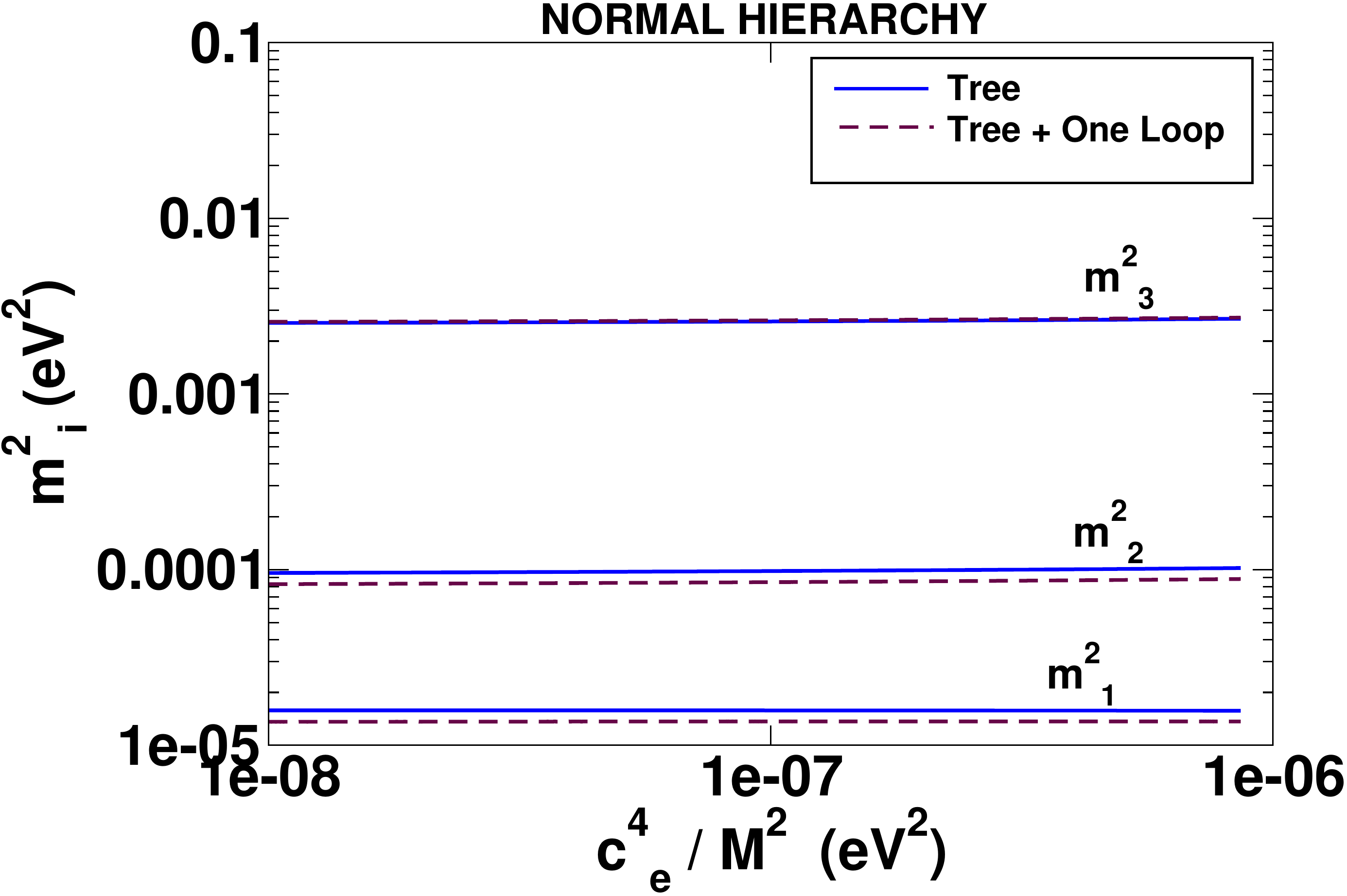}
\includegraphics[width=6.00cm]{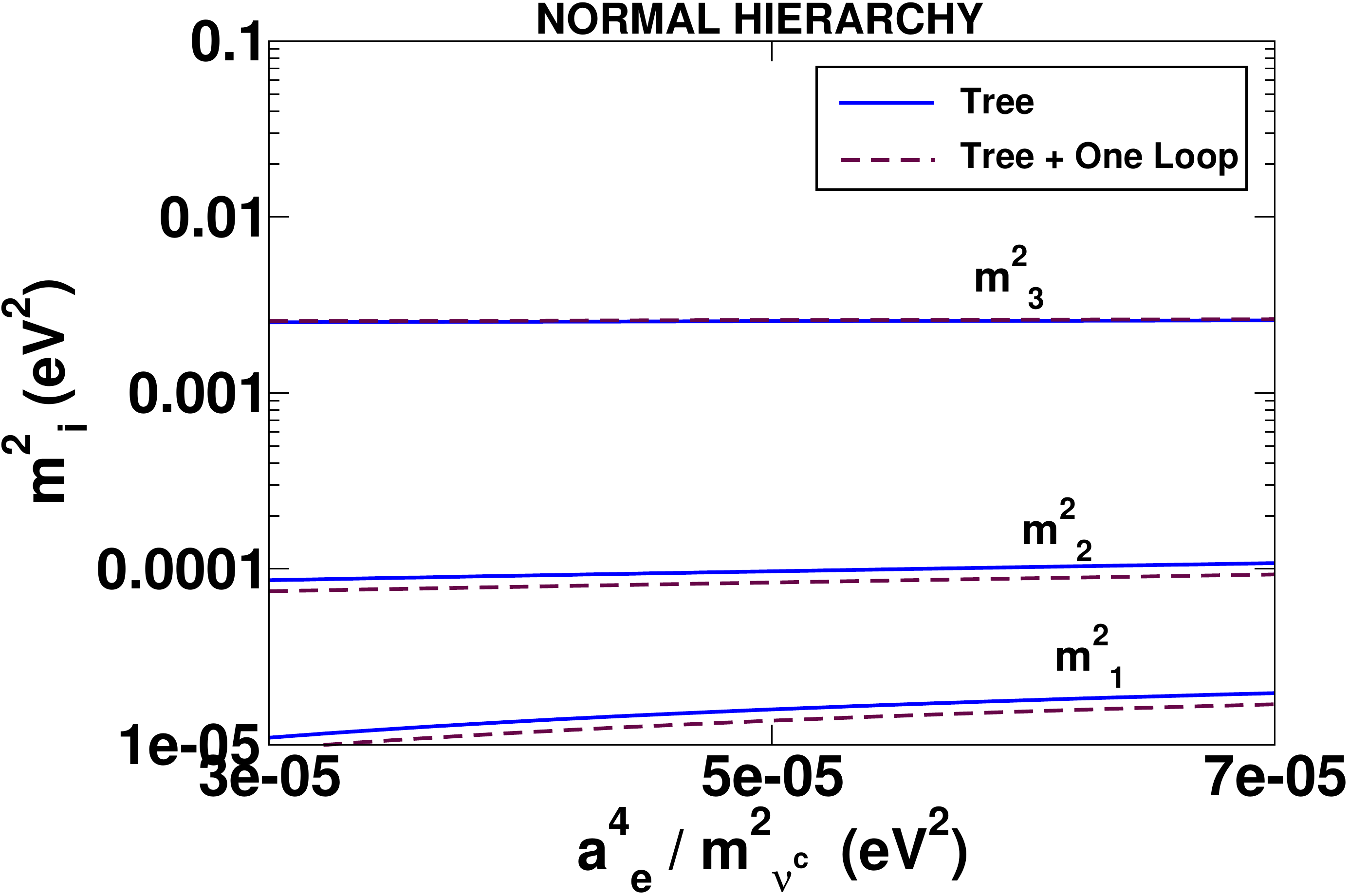}
\vspace{0.2cm}
\includegraphics[width=6.00cm]{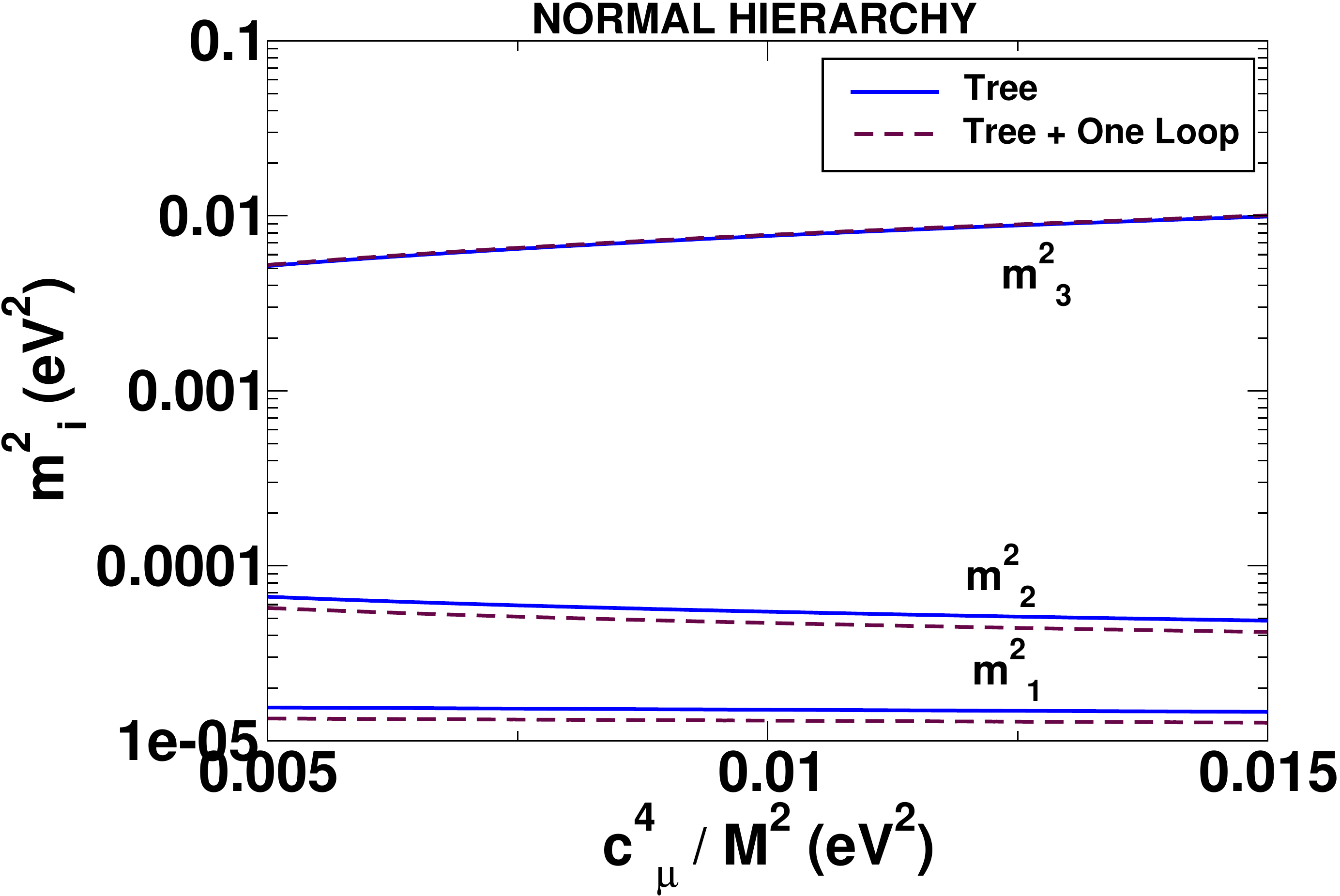}
\includegraphics[width=6.00cm]{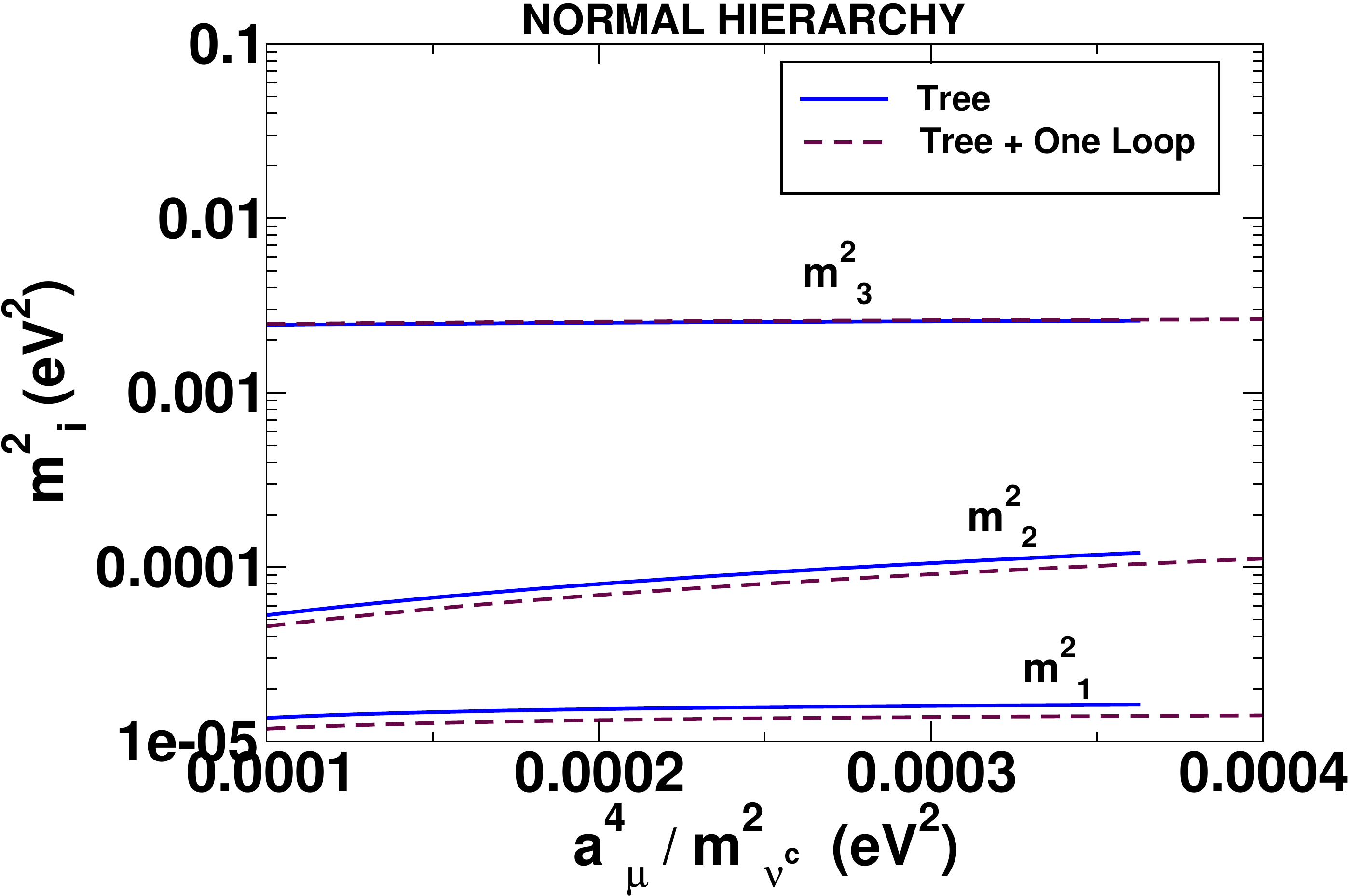}
\vspace{0.2cm}
\includegraphics[width=6.00cm]{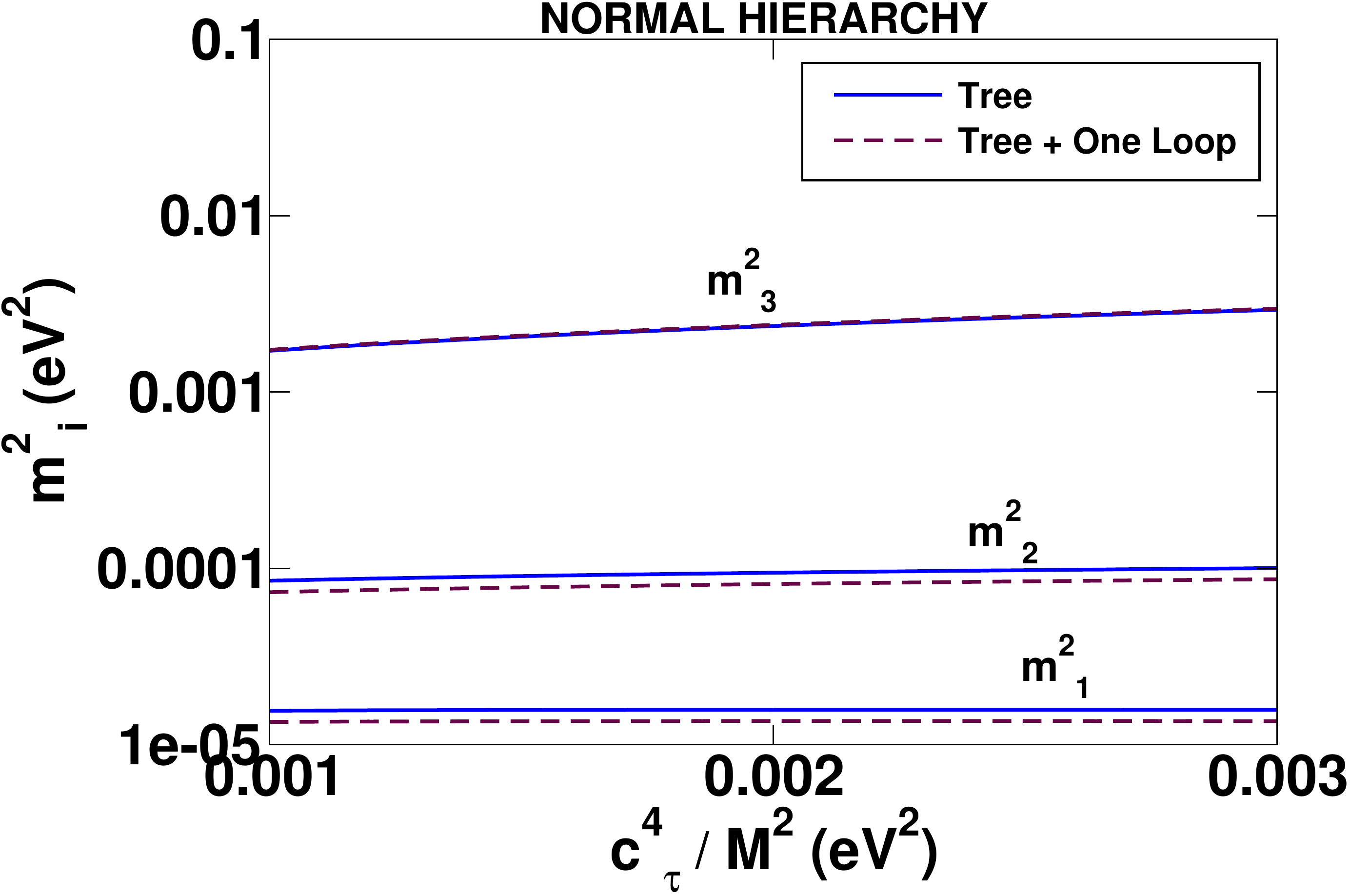}
\includegraphics[width=6.00cm]{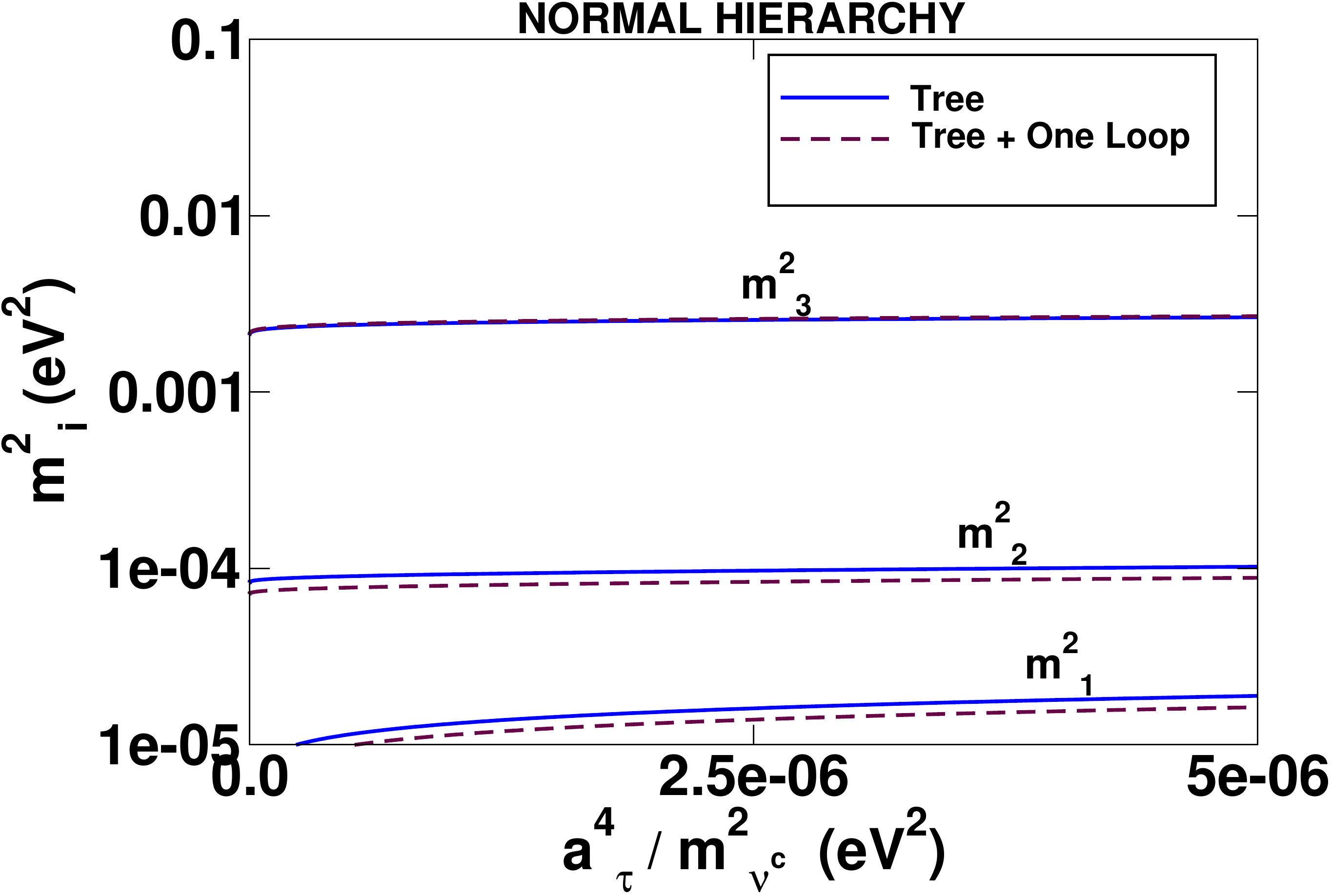}
\caption{Neutrino mass squared values ($m^2_i$) versus $\frac{c^4_i}{M^2}$ 
(left panel) and versus $\frac{a^4_i}{m^2_{\nu^c}}$ (right panel) plots for 
the {\it{normal hierarchical}} pattern of light neutrino masses, $i = e,
 \mu, \tau$. Parameter choices are shown in tables \ref{loop-param} 
  and \ref{loop-param-2}.}
\label{numsqNH}
\end{figure}
In this subsection, we show the variation of the neutrino squared 
masses ($m^2_i$) and the atmospheric and solar mass squared differences 
with the square of the seesaw parameters $\frac{c^2_i}{M}$ and 
$\frac{a^2_{i}}{m_{\nu^c}}$ for normal ordering in light neutrino masses. 
Results are shown for the tree level 
as well as the one-loop corrected neutrino masses. These plots also 
demonstrate the importance of one-loop corrections to neutrino masses 
compared to the tree level results \cite{c4Ghosh:2010zi}.  

Typical mass spectra are shown in figure \ref{numsqNH}. Note that a 
particular model parameter has been varied while the others are fixed
at values given in tables \ref{loop-param} and \ref{loop-param-2}. The effective 
light neutrino mass matrix given in eqn.(\ref{mnuij-compact1}) 
suggests that as long as $v'_i \gg \frac{Y^{ii}_\nu v_1}{3 \lambda}$ and 
$\kappa \gg \lambda$, the second term on the right hand side of 
eqn.(\ref{mnuij-compact1}) dominates over the first term and as a result 
the heaviest neutrino mass scale ($m_3$) is controlled mainly by the gaugino 
seesaw effect. This is because in this limit $b_i \approx c_i$, and, as discussed 
earlier, a neutrino mass matrix with a structure $\sim \frac{c_ic_j}{M}$ 
can produce only one non-zero neutrino mass. This feature is evident in 
figure \ref{numsqNH}, where we see that $m^2_3$ increases as a function of 
$c^4_i/M^2$. The other two masses are almost insensitive to $c^2_i/M$. A mild variation 
to $m^2_2$ comes from the combined effect of gaugino and ordinary seesaw (see 
the $(a_ic_j+a_jc_i)$ terms in eqns.(\ref{mnuij-compact-recasted}), 
(\ref{one-loop corrected structure of neutralino mass matrix})).
On the other hand, the two lighter neutrino mass scales ($m^2_2$ and
$m^2_1$) are controlled predominantly by the ordinary seesaw parameters
$a^2_i/{m_{\nu^c}}$. This behaviour is observed in the right panel figures of
figure \ref{numsqNH}. The heaviest neutrino mass scale is not much affected
by the quantities $a^2_i/{m_{\nu^c}}$.    

One can also see from these plots that the inclusion of one-loop 
corrections, for the chosen values of the soft SUSY breaking parameters,
reduces the values of $m^2_2$ and $m^2_1$, while increasing the 
value of $m^2_3$ only mildly. This is because, with such a choice,  the 
one-loop corrections cause partial cancellation in the generation of $m_1$ 
and $m_2$. For the heaviest state, it is just the opposite, since the 
diagonalization of the tree-level mass matrix already yields 
a negative mass eigenvalue, on which the loop correction has an additive effect. 
If, with all other parameters fixed, the signs of $\lambda$ and $A_\lambda$ are 
reversed (leading to a positive $\mu$ in the place of a negative one), $m_1$, 
$m_2$ and $m_3$ are all found to decrease through loop corrections. A flip in the
sign of $\kappa$ and the corresponding soft breaking terms, on the other hand, causes
a rise in all the mass eigenvalues, notably for $m_1$ and $m_2$.
 
\begin{figure}[ht]
\centering
\includegraphics[width=6.00cm]{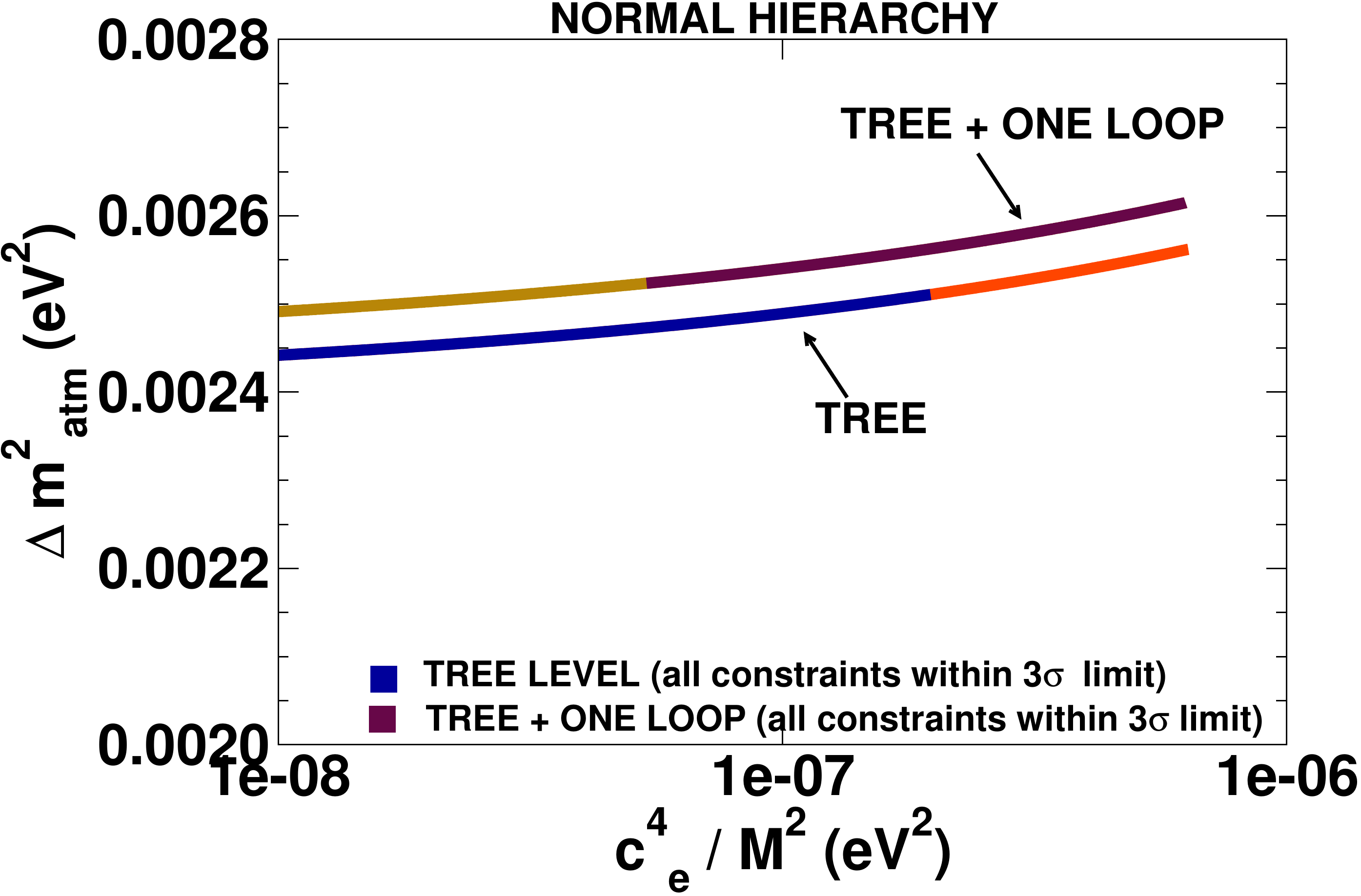}
\includegraphics[width=6.00cm]{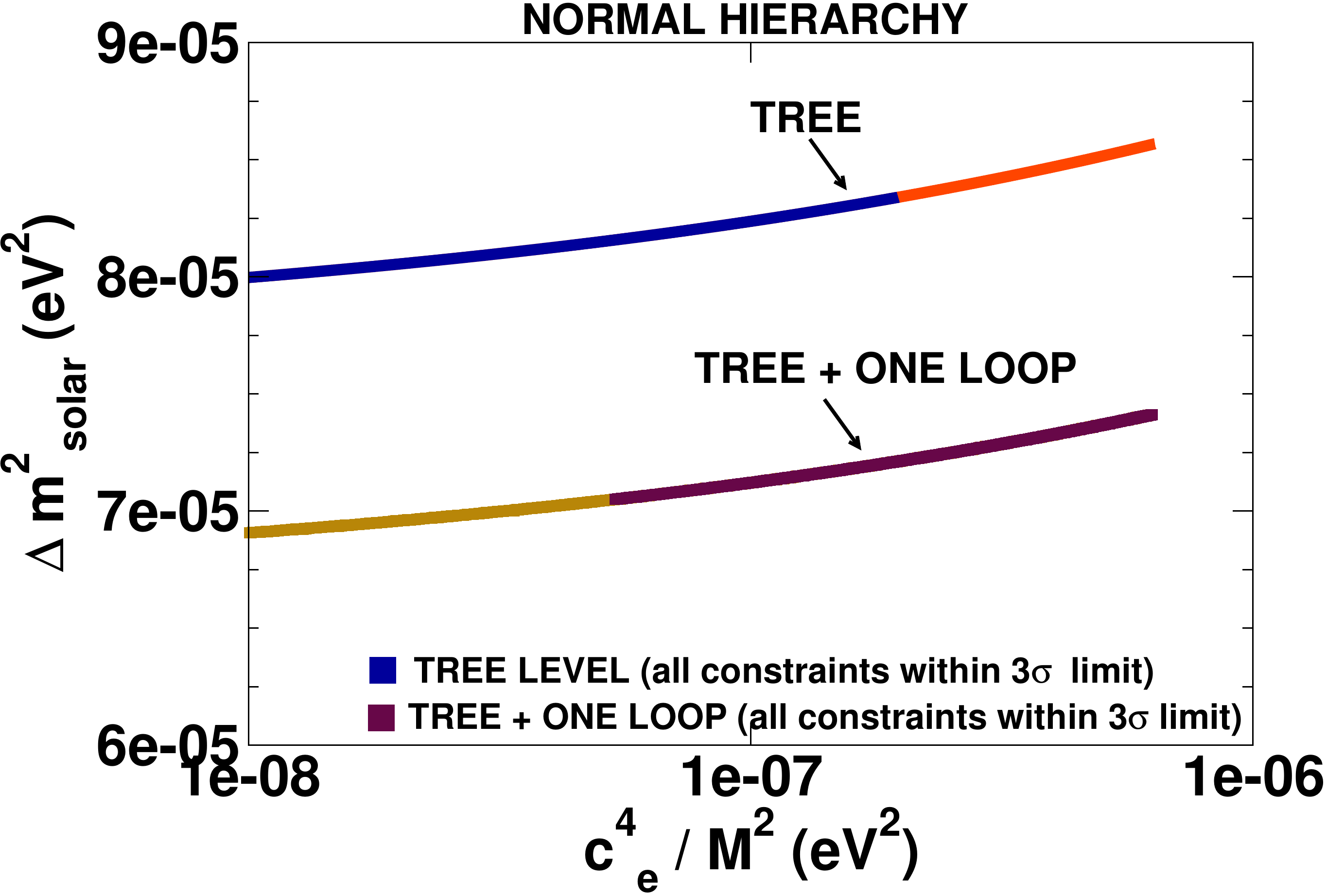}
\vspace{0.2cm}
\includegraphics[width=6.00cm]{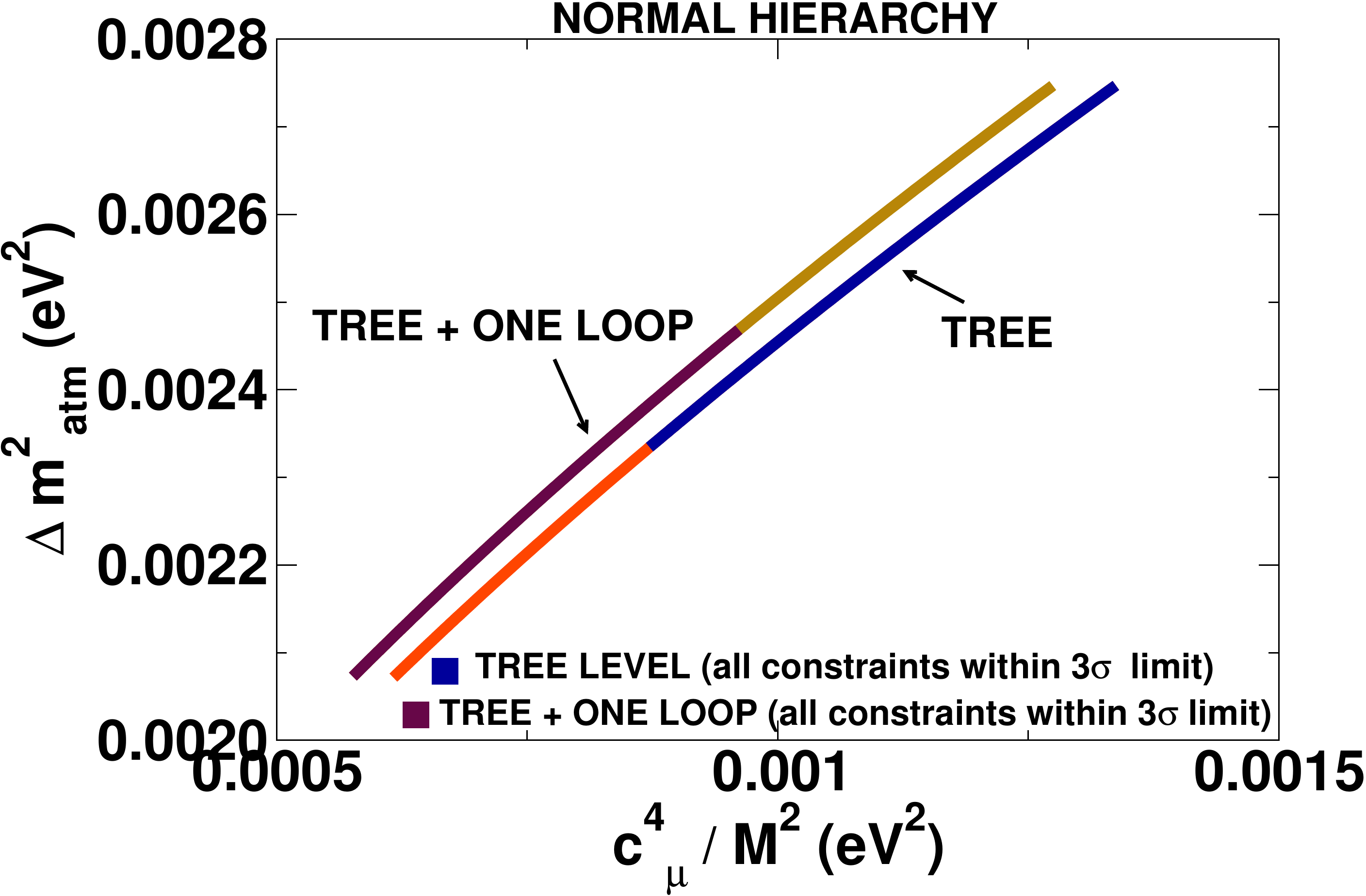}
\includegraphics[width=6.00cm]{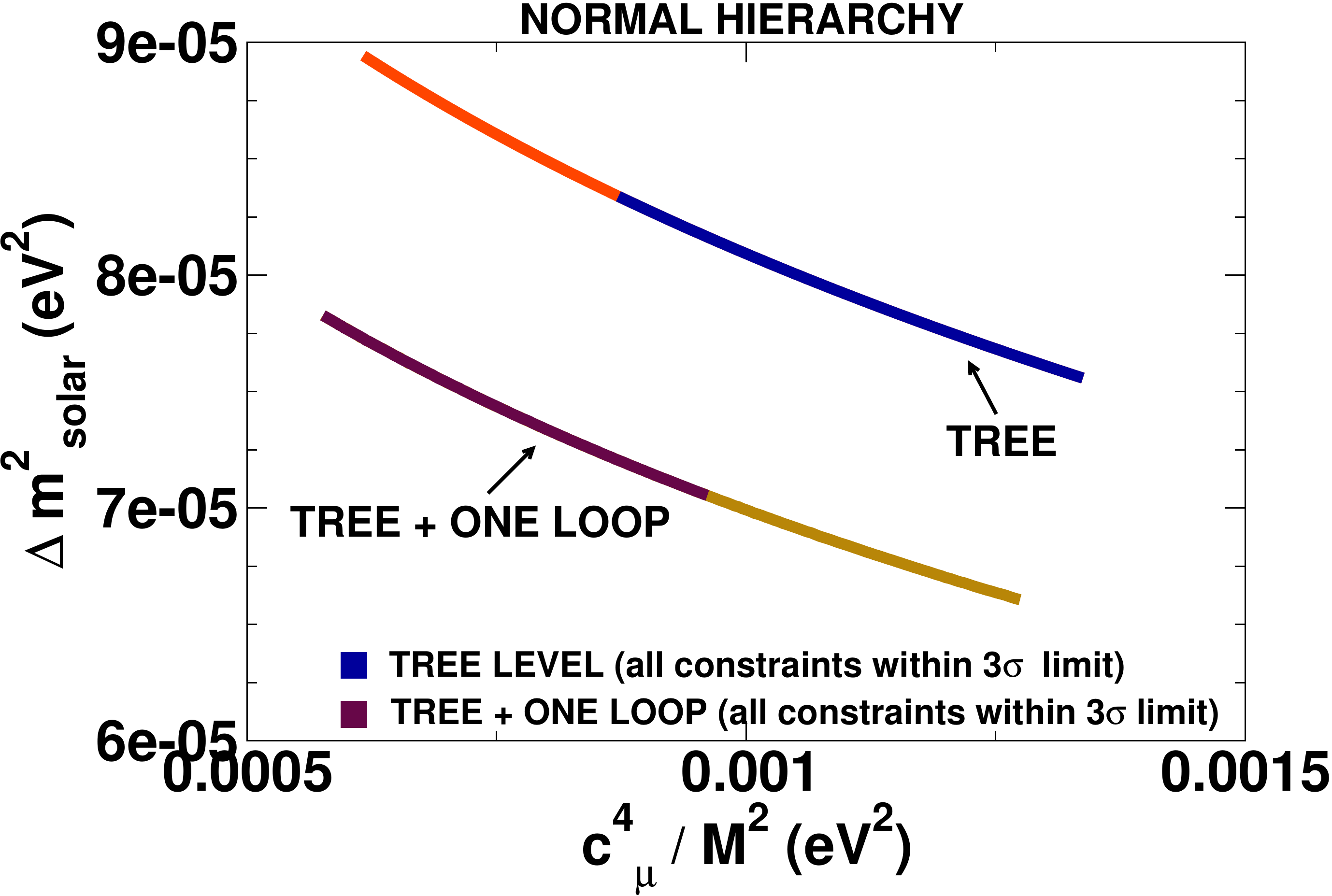}
\vspace{0.2cm}
\includegraphics[width=6.00cm]{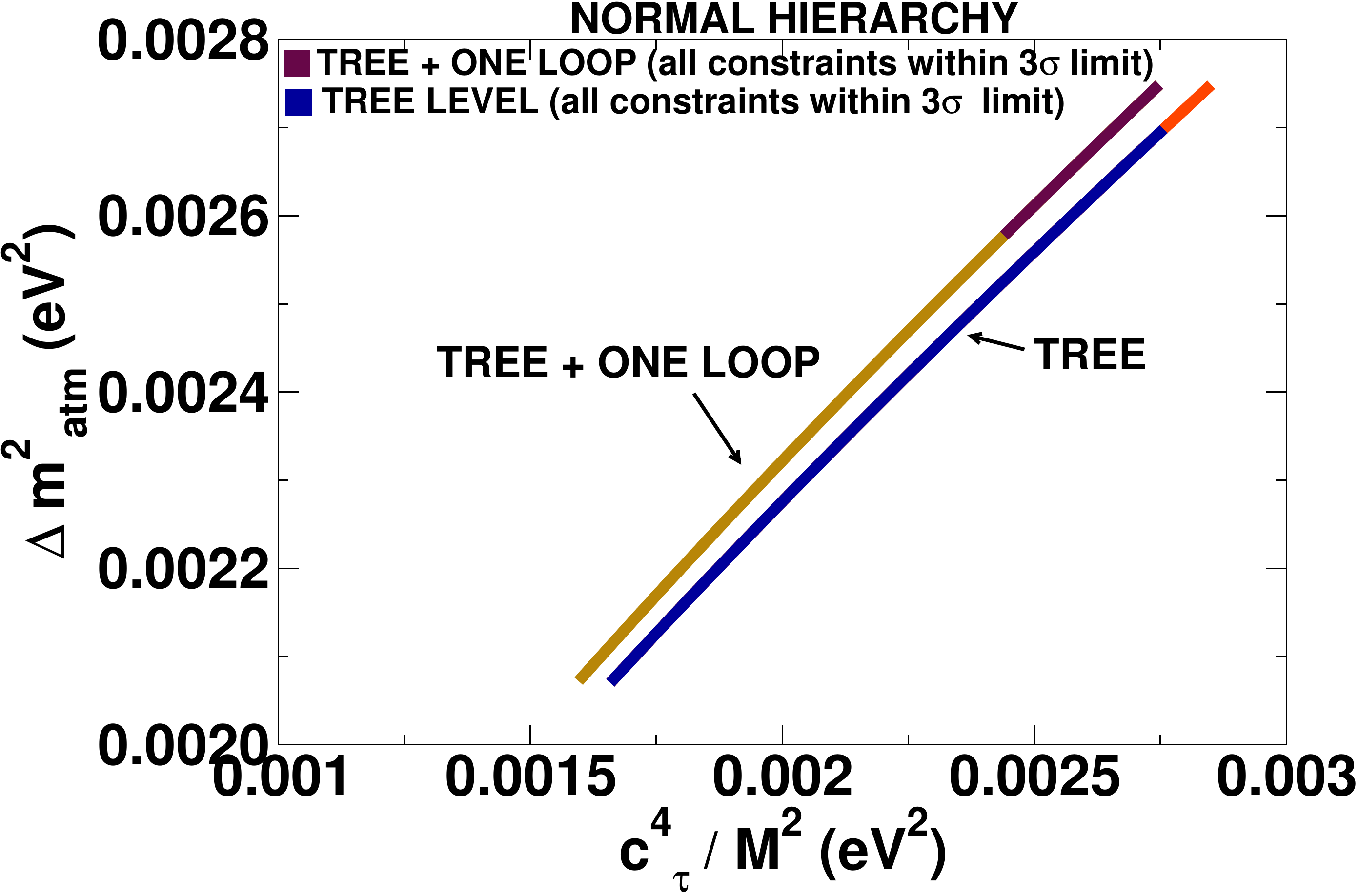}
\includegraphics[width=6.00cm]{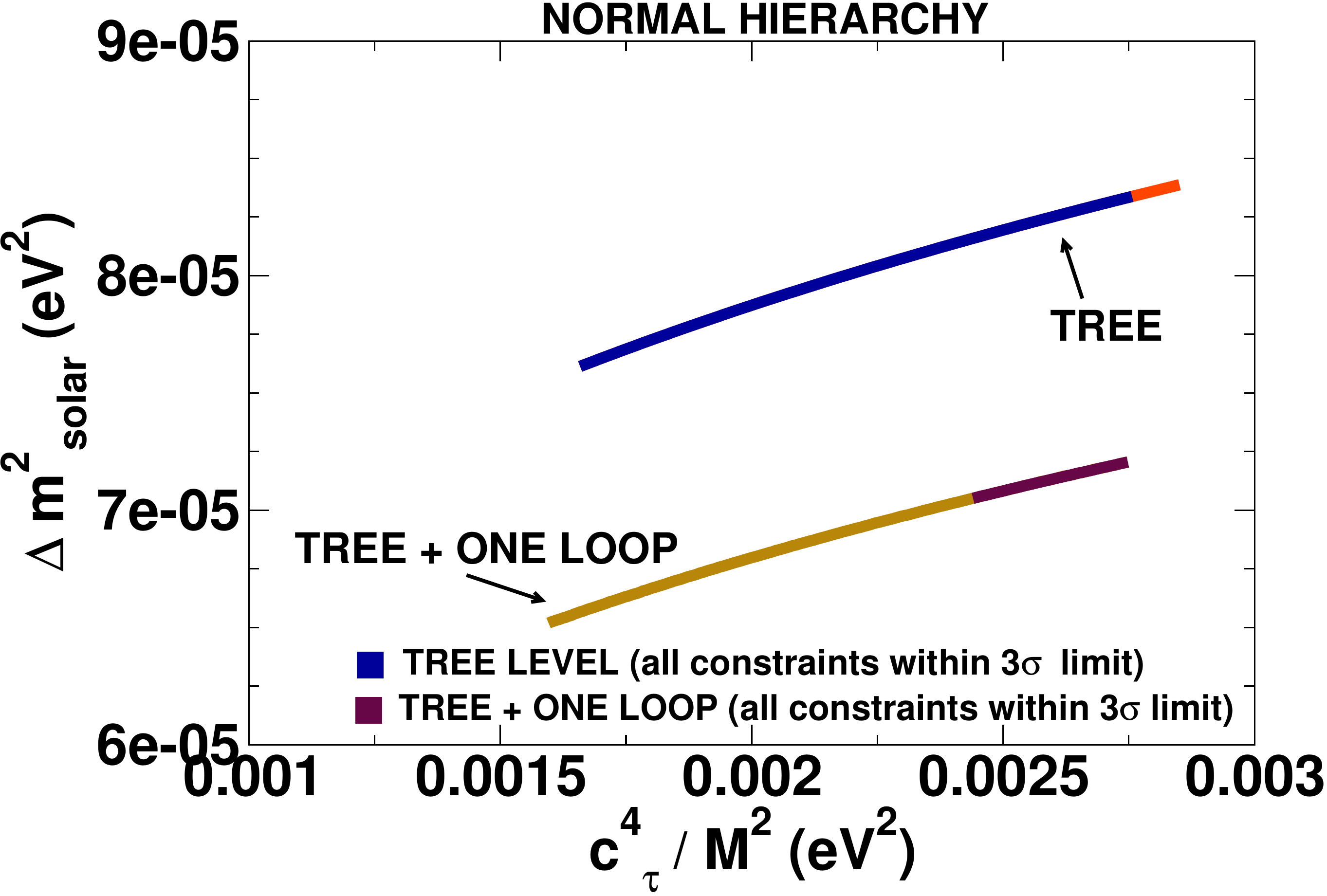}
\caption{Atmospheric and solar mass squared differences $(\Delta
  m^2_{atm},~\Delta m^2_{solar})$ vs $\frac{c^4_i}{M^2}$ plots for the
  {\it{normal hierarchical}} pattern of light neutrino masses, $i = e,
  \mu, \tau$. The full lines are shown for which only the constraints on 
  $\Delta m^2_{solar}$ is not within the 3$\sigma$ limit (see table \ref{osc-para}). 
  The dark coloured portions on these lines are the values of parameters 
  for which all the neutrino constraints are within the 3$\sigma$ limit. The 
  red (yellow) coloured lines in the plots correspond to the tree (one-loop corrected) 
  regions where all the constraints except $\Delta m^2_{solar}$ are within 3$\sigma$ 
  allowed region. Parameter choices are shown in tables \ref{loop-param} 
  and \ref{loop-param-2}.}
\label{gsNH}
\end{figure}

\begin{figure}[ht]
\centering
\includegraphics[width=6.00cm]{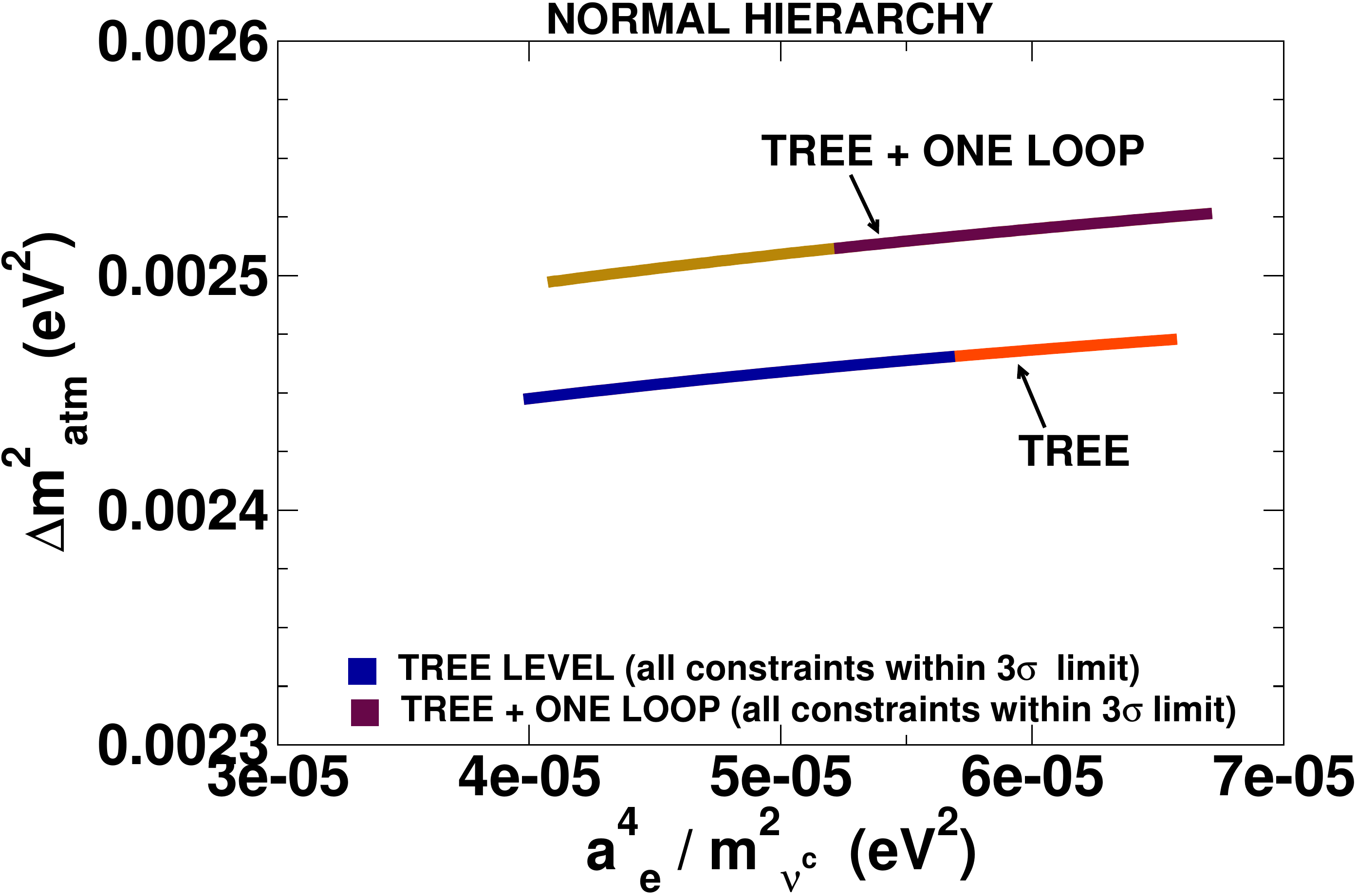}
\includegraphics[width=6.00cm]{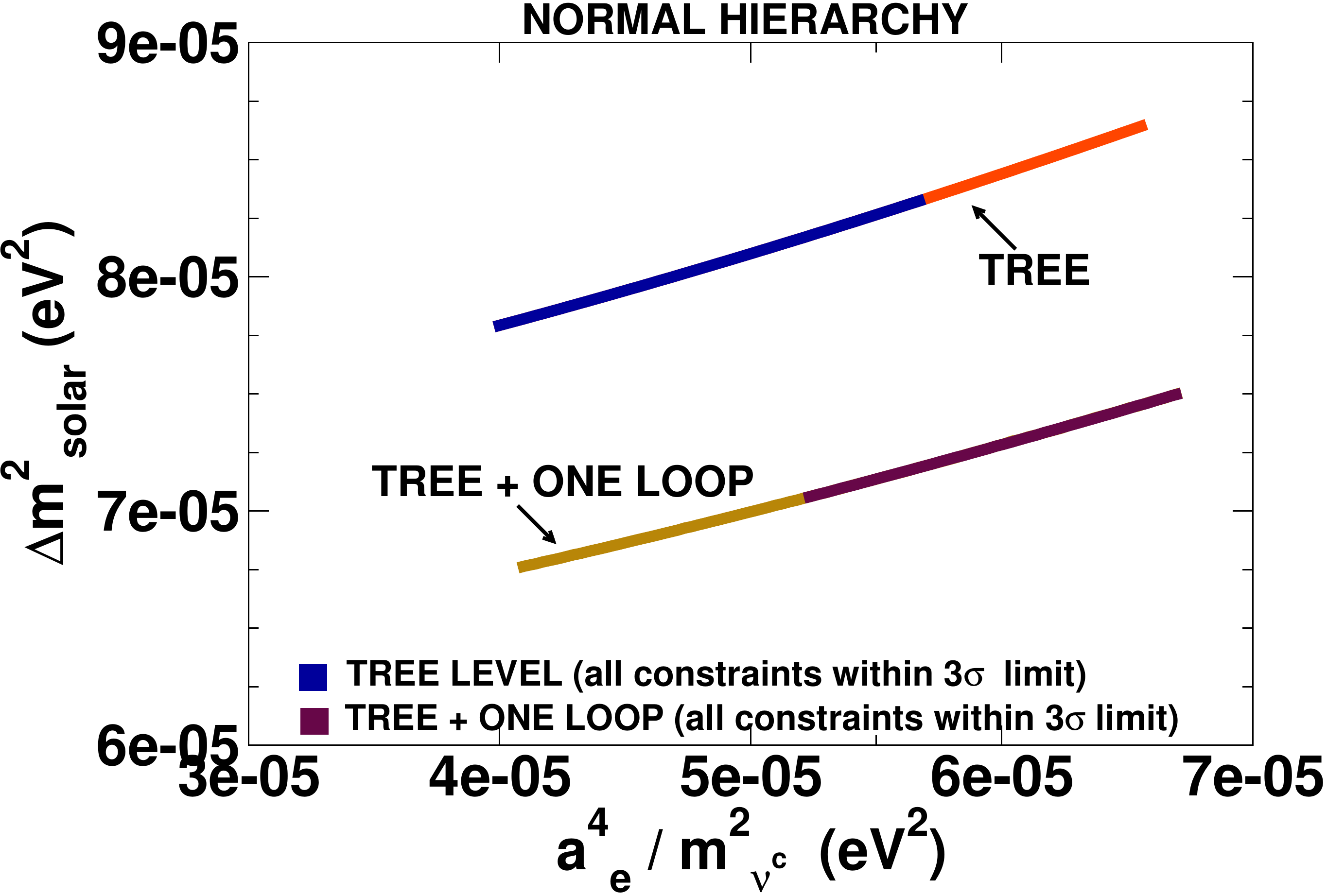}
\vspace{0.2cm}
\includegraphics[width=6.00cm]{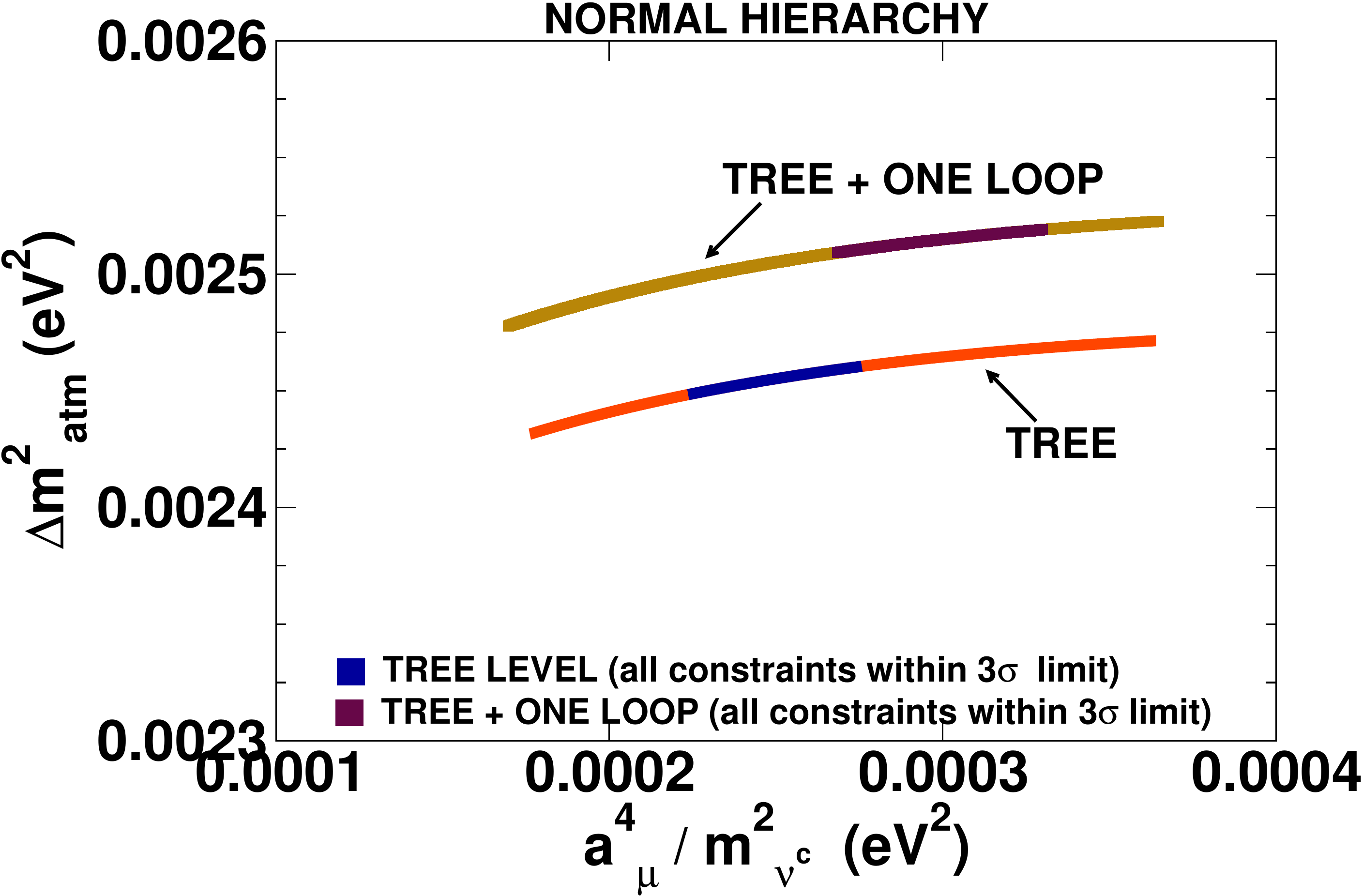}
\includegraphics[width=6.00cm]{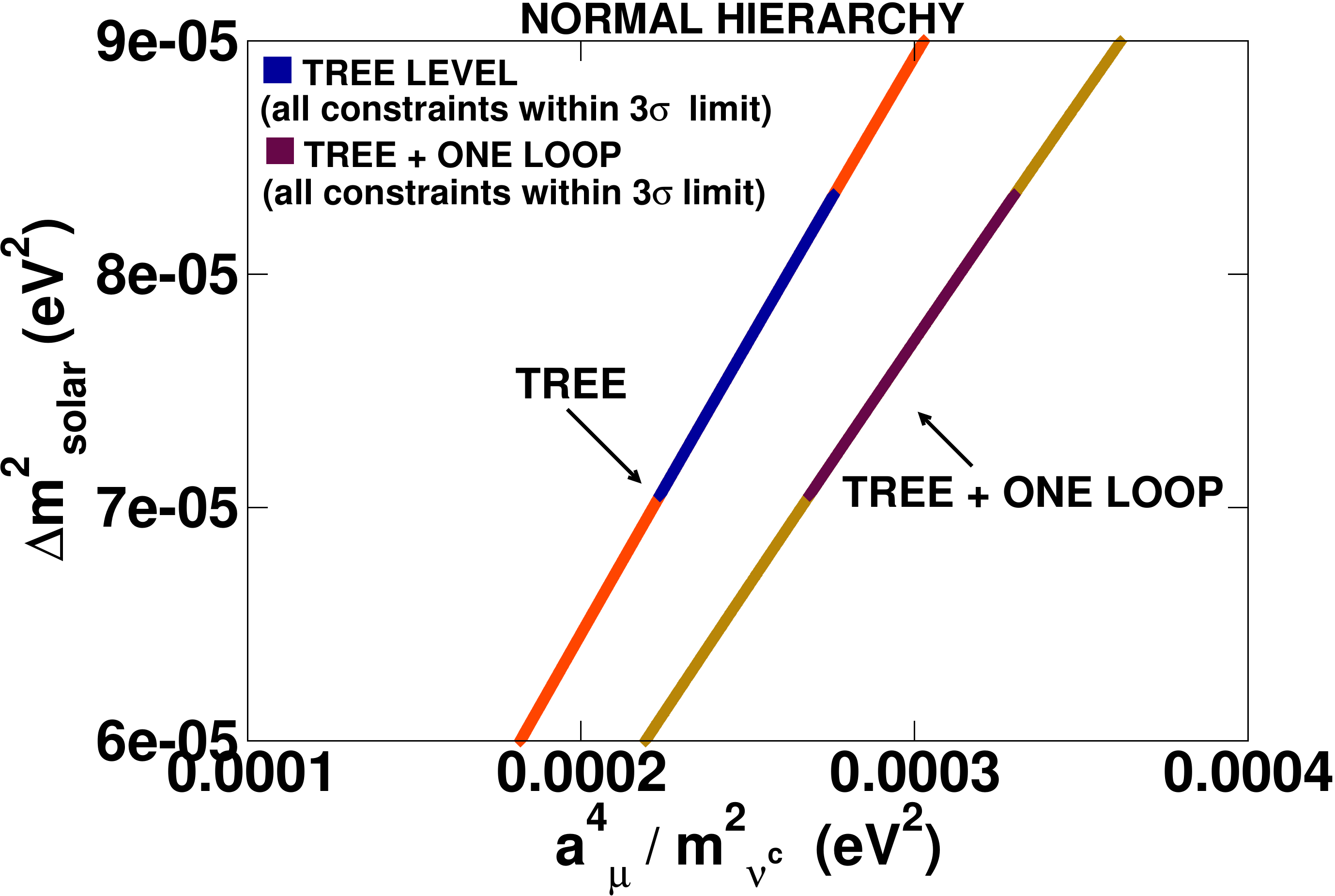}
\vspace{0.2cm}
\includegraphics[width=6.00cm]{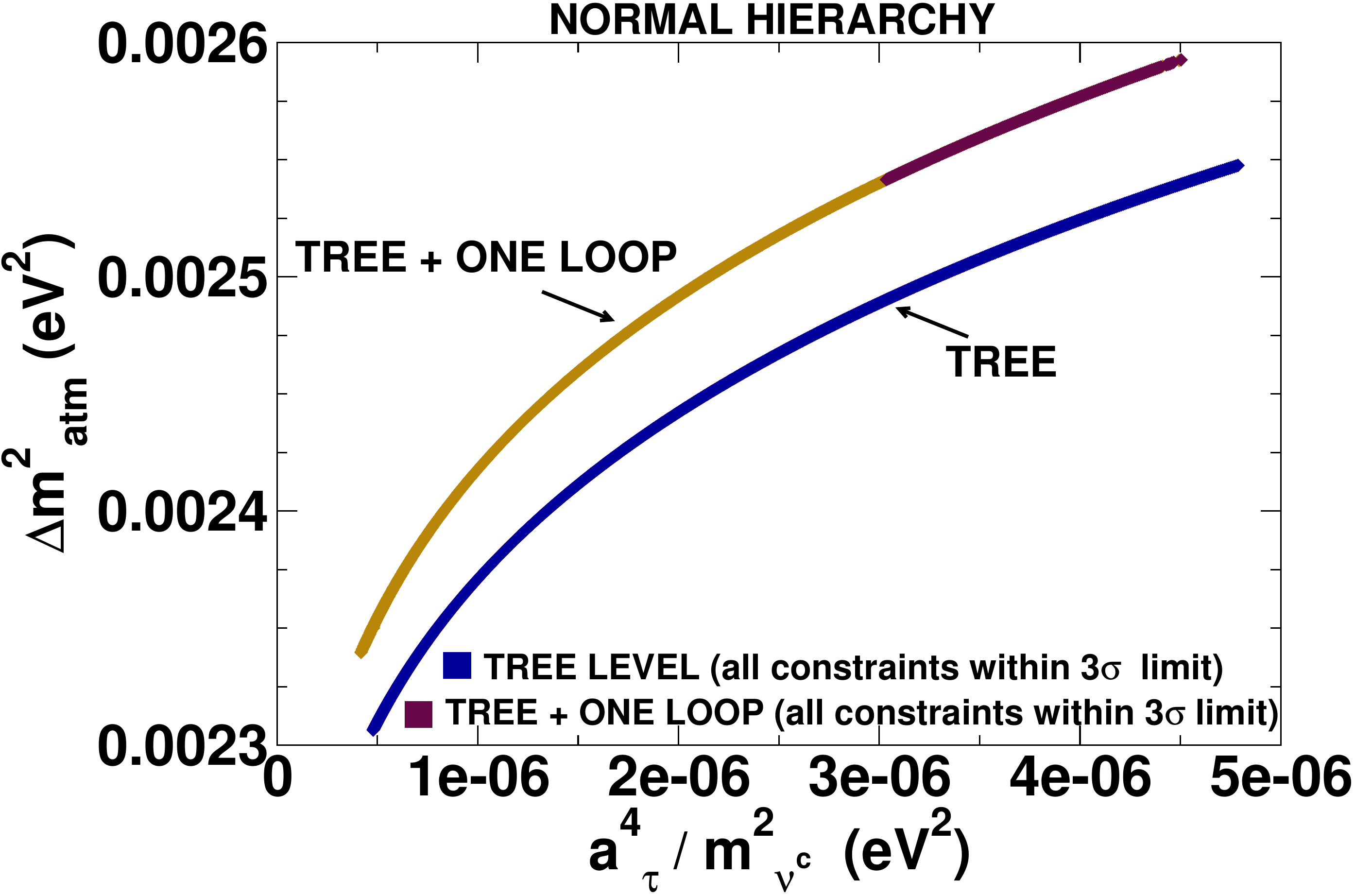}
\includegraphics[width=6.00cm]{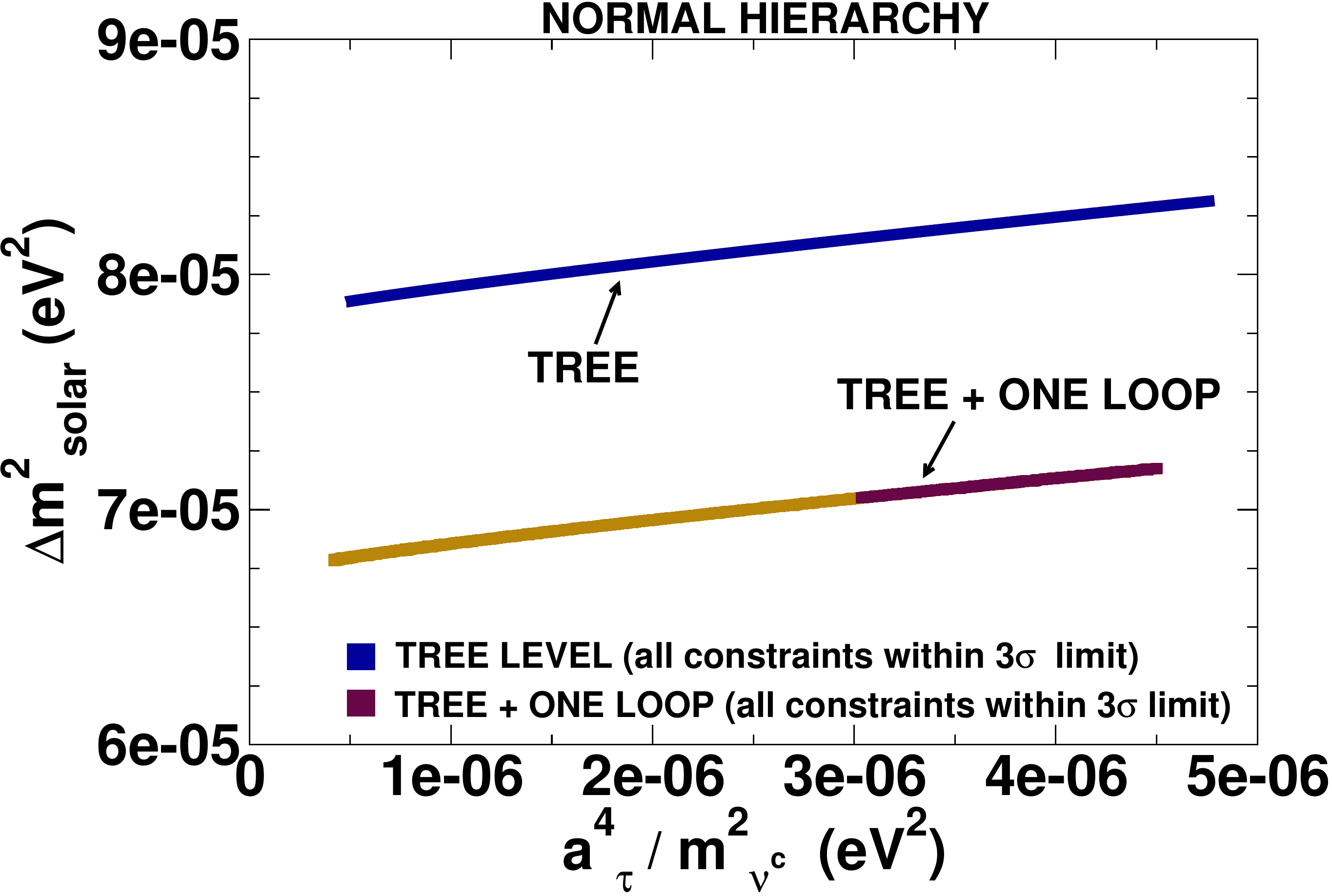}
\caption{Atmospheric and solar mass squared differences $(\Delta
  m^2_{atm},~\Delta m^2_{solar})$ vs $a^4_i/m^2_{\nu^c}$ plots for the
  {\it{normal hierarchical}} pattern of light neutrino masses with $i
  = e, \mu, \tau$. Colour specification is same as described 
  in the context of figure \ref{gsNH}.
  Parameter choices are shown in tables \ref{loop-param} 
  and \ref{loop-param-2}.}
\label{osNH}
\end{figure}

In the light of the discussion above, we now turn to explain the 
variation of $\Delta m^2_{atm}$ and $\Delta m^2_{solar}$
with $c^4_i/M^2$ and $a^4_i/{m^2_{\nu^c}}$ shown in figure \ref{gsNH} 
and figure \ref{osNH}. For our numerical analysis, in order to 
set the scale of the normal hierarchical spectrum, we choose 
$m_2|_{max}<0.011~\rm{eV}$. The left panel in 
figure \ref{gsNH} shows that $\Delta m^2_{atm}$ increases more 
rapidly with $c^4_{\mu,\tau}/M^2$, whereas the variation with 
$c^4_e/M^2$ is much slower as expected from figure \ref{numsqNH}. 
Similar behaviour is shown for the one-loop corrected 
$\Delta m^2_{atm}$. The small increase in the one-loop 
corrected result compared to the tree level one is essentially due 
to the splitting in $m^2_2$ value as shown earlier. The variation 
of $\Delta m^2_{solar}$ with $c^4_i/M^2$ 
can be explained in a similar manner. Obviously, in this case
the one-loop corrected result is smaller compared to the tree 
level one (see, figure \ref{numsqNH}). However, one should note
that $\Delta m^2_{solar}$ falls off with $c^4_{\mu}/M^2$ as 
opposed to the variation with respect to the other two 
gaugino seesaw parameters. This is due to the fact that $m^2_2$ 
slightly decreases with $c^4_{\mu}/M^2$ but show a slow increase 
with respect to $c^4_{e}/M^2$ and $c^4_{\tau}/M^2$. The dark solid 
lines in all these figures show the allowed values of various parameters
where all the neutrino mass and mixing constraints are satisfied.

The variation of $\Delta m^2_{atm}$ and $\Delta m^2_{solar}$ with 
$a^4_i/m^2_{\nu^c}$ in figure \ref{osNH} can be understood in a 
similar way by looking at the right panel plots of 
figure \ref{numsqNH}. $\Delta m^2_{atm}$ shows a very little increase 
with $a^4_{e,\mu}/m^2_{\nu^c}$ as expected, whereas the change is more 
rapid with $a^4_\tau/m^2_{\nu^c}$ for the range of values considered
along the x-axis. As in the case of figure \ref{gsNH}, the solid dark lines
correspond to the allowed values of parameters where all the neutrino 
mass and mixing constraints are satisfied. 

For higher values of $a^4_{e,\tau}/m^2_{\nu^c}$, $m^2_2$ increases 
very slowly with these parameters (see, figure \ref{numsqNH}) and this is
reflected in the right panel plots of figure \ref{osNH}, where 
$\Delta m^2_{solar}$ shows a very slow variation with 
$a^4_{e,\tau}/m^2_{\nu^c}$. On the other hand, $m^2_2$ increases 
more rapidly with $a^4_\mu/m^2_{\nu^c}$, giving rise to a faster
variation of $\Delta m^2_{solar}$. The plots of figure \ref{osNH} show 
that larger values of Yukawa couplings are required in order to satisfy
the global three flavour neutrino data, when one considers one-loop
corrected neutrino mass matrix. However, there are allowed ranges of
the parameters $a^4_i/m^2_{\nu^c}$, where the neutrino data can be 
satisfied with both tree and one-loop corrected analysis.

We have also considered the variation of light neutrino mass 
squared differences with the effective bilinear $R_P$
violating parameter, $\varepsilon_i = Y^{ij} v^c_j$. 
For this particular numerical study we vary both
$Y^{ii}_\nu$ and the right-handed sneutrino VEVs $(v^c_i)$ 
simultaneously, in the suitable ranges around the
values given in tables \ref{loop-param} and \ref{loop-param-2}. 
$\Delta m^2_{atm}$ is found to increase
with $\varepsilon_i$, whereas the solar mass squared
difference decreases with increasing $\varepsilon_i$. 
The $3\sigma$ allowed region for the solar and atmospheric mass 
squared differences were obtained for the lower values of 
$\varepsilon_i$s. In addition, we have noticed that the correlations 
of $\Delta m^2_{atm}$ with $\varepsilon_i$ is sharper compared to 
the correlations seen in the case of $\Delta m^2_{solar}$.

\begin{figure}[ht]
\centering
\includegraphics[width=6.00cm]{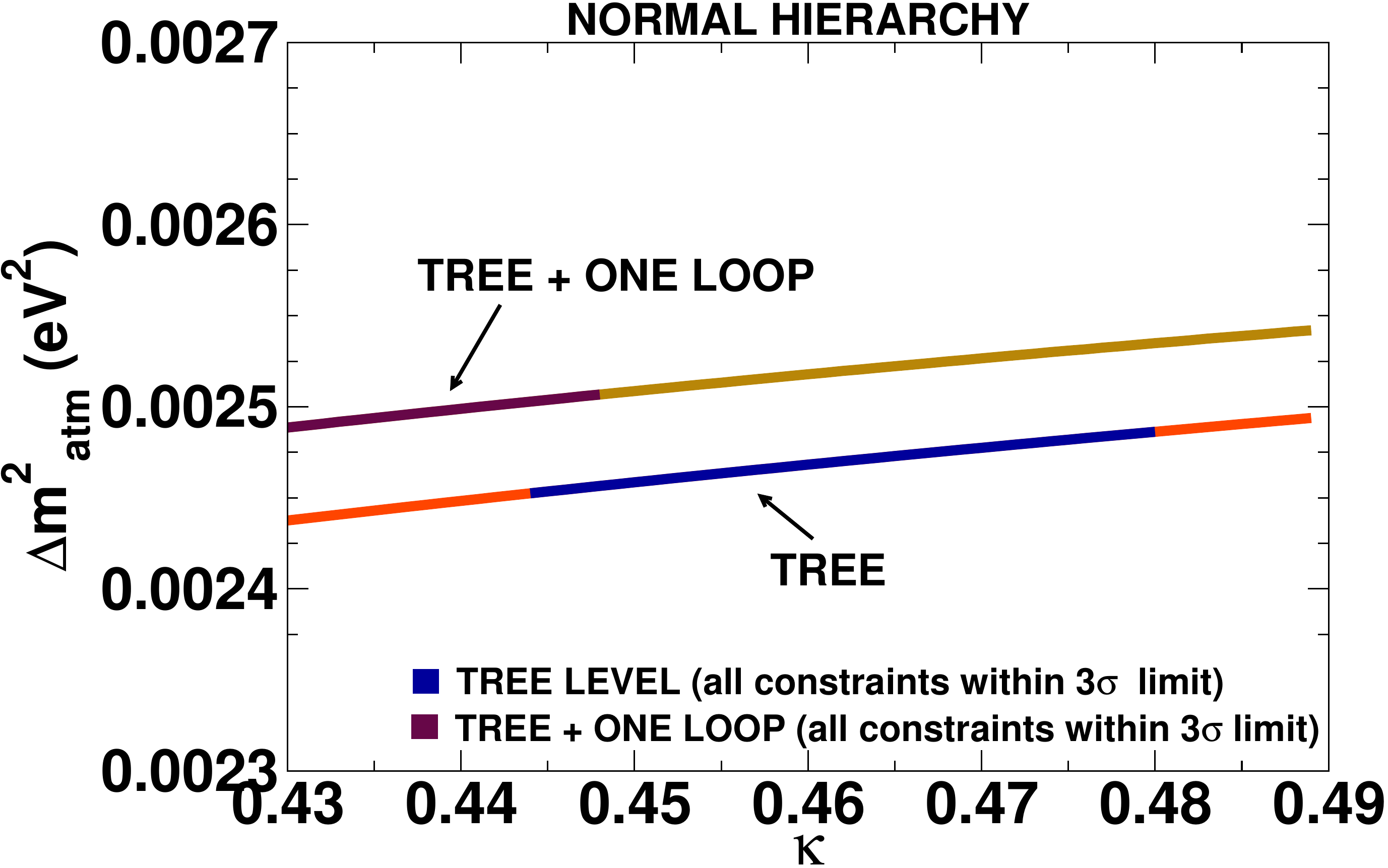}
\includegraphics[width=6.00cm]{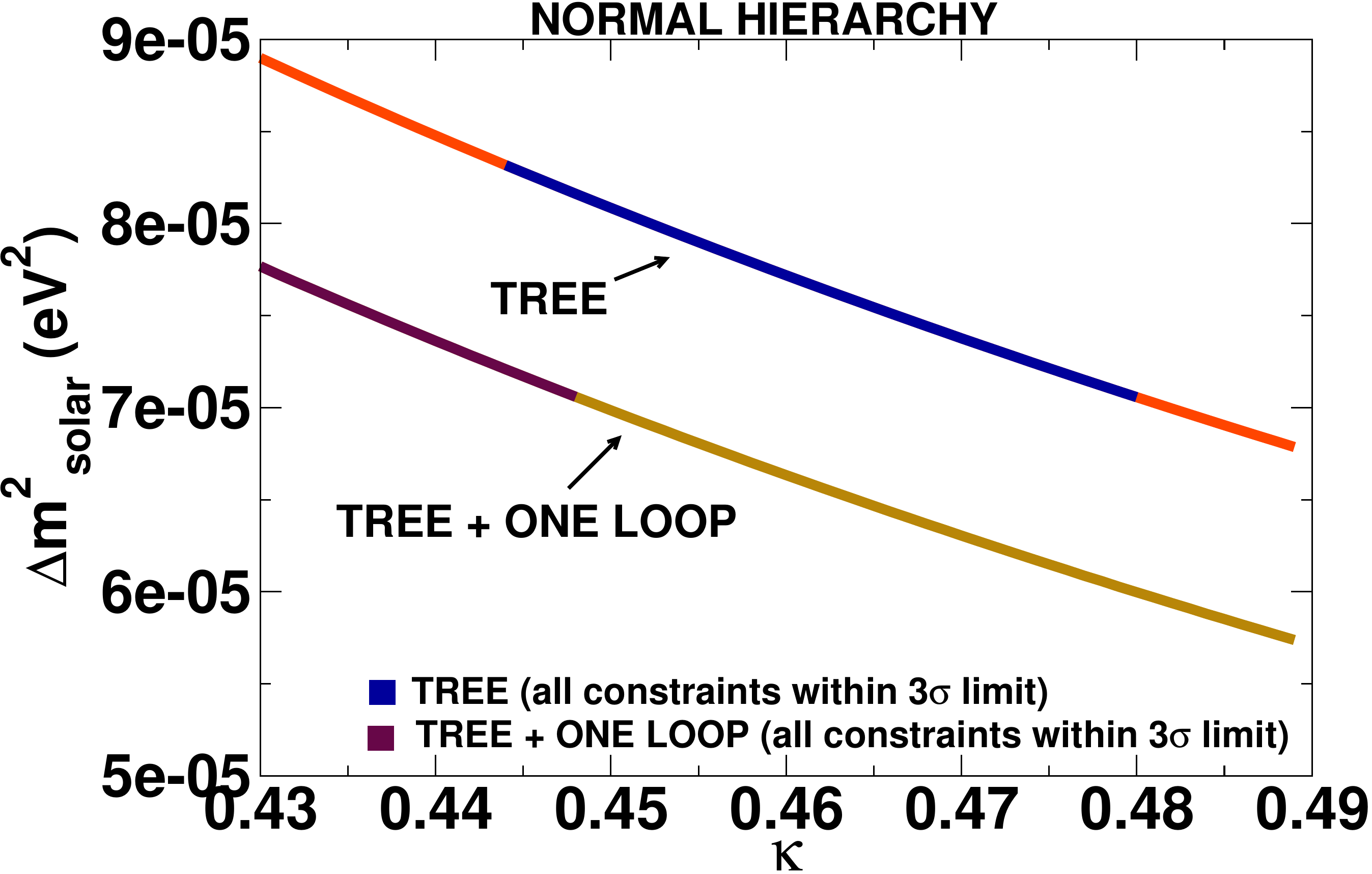}
\vspace{0.2cm}
\includegraphics[width=6.00cm]{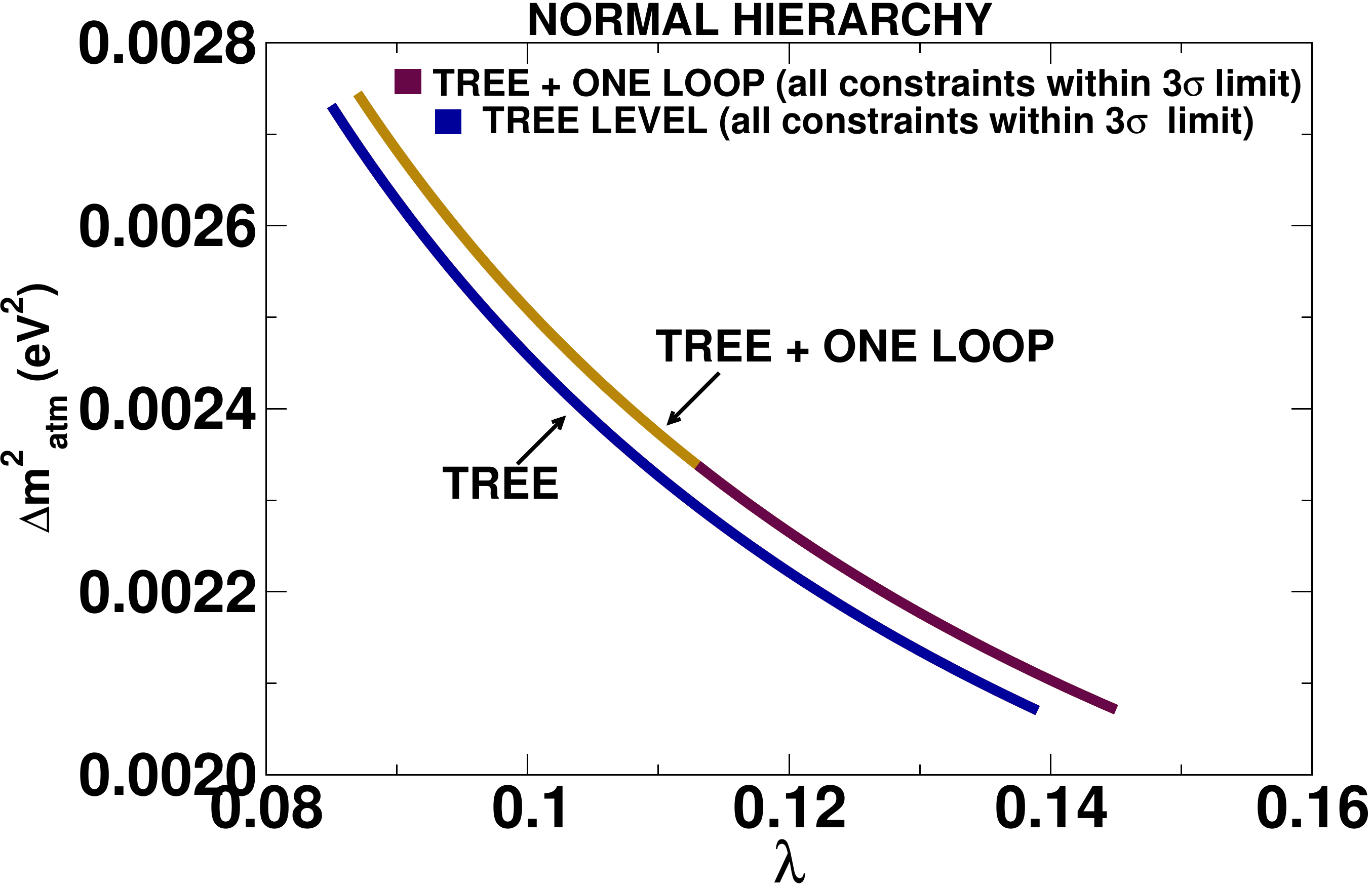}
\includegraphics[width=6.00cm]{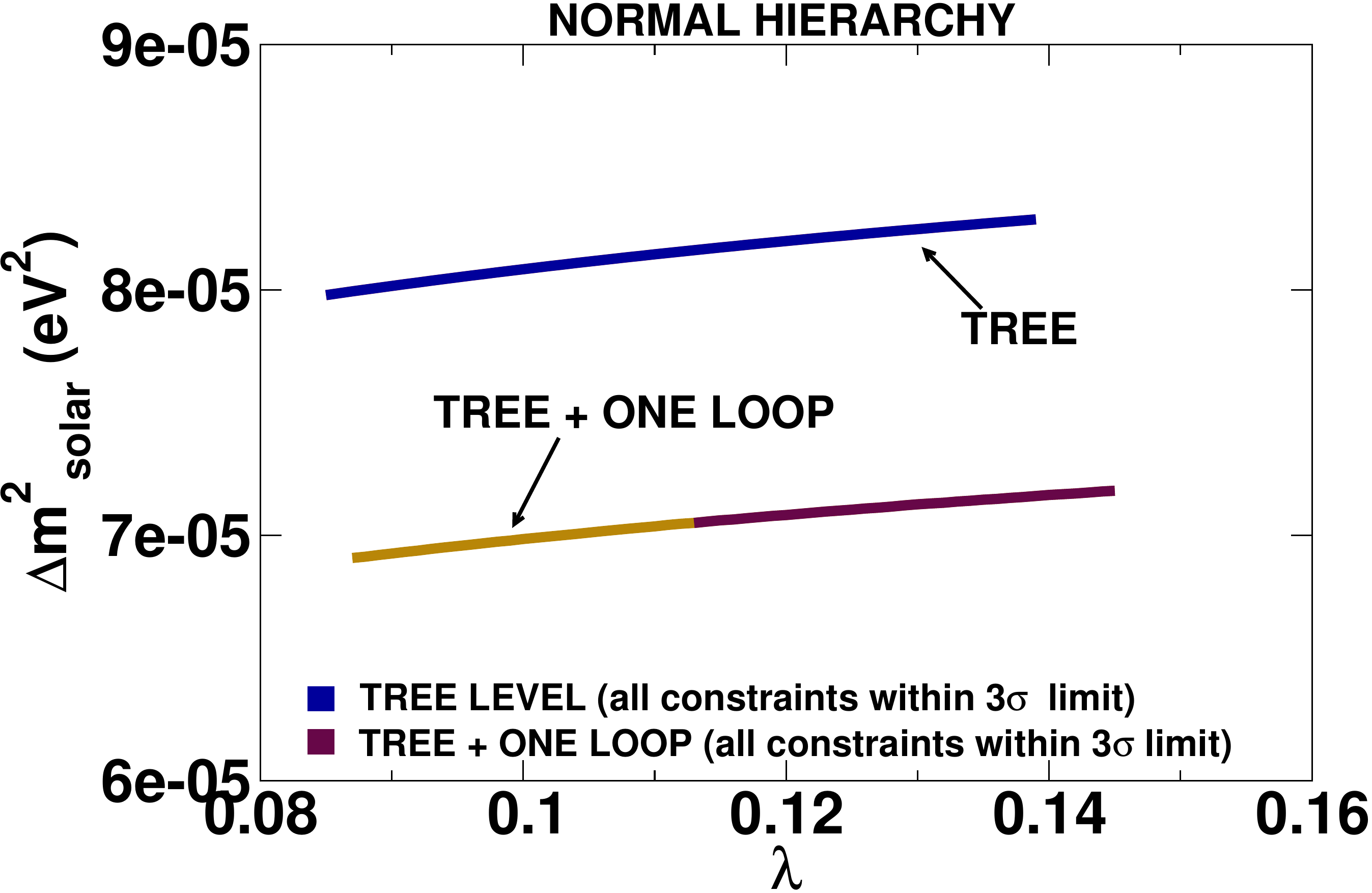}
\caption{Plots showing the variations of $\Delta m^2_{atm},~\Delta
  m^2_{solar}$ with model parameters $\lambda$ and $\kappa$ for
  {\it{normal hierarchy}}. Colour specification is same as 
  described in the context of figure \ref{gsNH}. 
  Parameter choices are shown in tables \ref{loop-param} 
  and \ref{loop-param-2}.}
\label{L-K-NH}
\end{figure}

Next let us discuss the dependence of $\Delta m^2_{atm}$ and 
$\Delta m^2_{solar}$ on two specific model parameters, $\lambda$ and 
$\kappa$, consistent with EWSB conditions. The loop corrections
shift the allowed ranges of $\kappa$ to lower values with some 
amount of overlap with the tree level result. On the other hand,  
the allowed ranges of $\lambda$ shrinks towards higher values when
one-loop corrections are included. These results are shown in 
figure \ref{L-K-NH}. We note in passing that the mass of the  
lightest CP-even scalar decreases with increasing $\lambda$. For example, 
$\lambda = 0.15$ can produce a lightest scalar mass of $40 ~\rm{GeV}$, 
for suitable choices of other parameters. This happens
because with increasing $\lambda$, the lightest scalar state picks up
more and more right-handed sneutrino admixture. This phenomena as
discussed earlier has serious consequence in the mass of the lightest
Higgs boson in $\mu\nu$SSM (see section \ref{munuSSM-scalar} and also 
eqn.(\ref{Higgs-mass-NMSSM-lightest-2})).

\begin{figure}[ht]
\centering
\includegraphics[width=6.00cm]{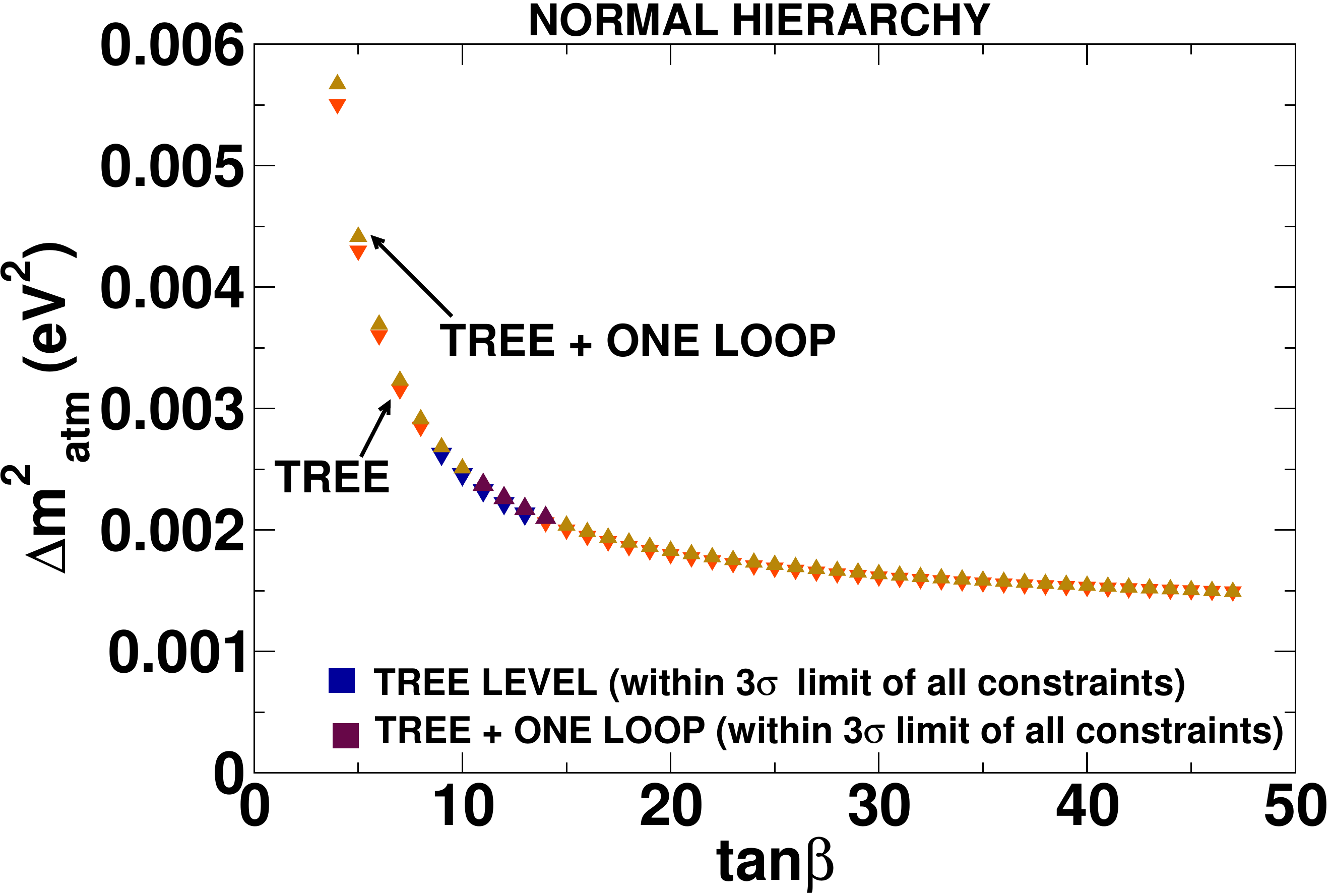}
\includegraphics[width=6.00cm]{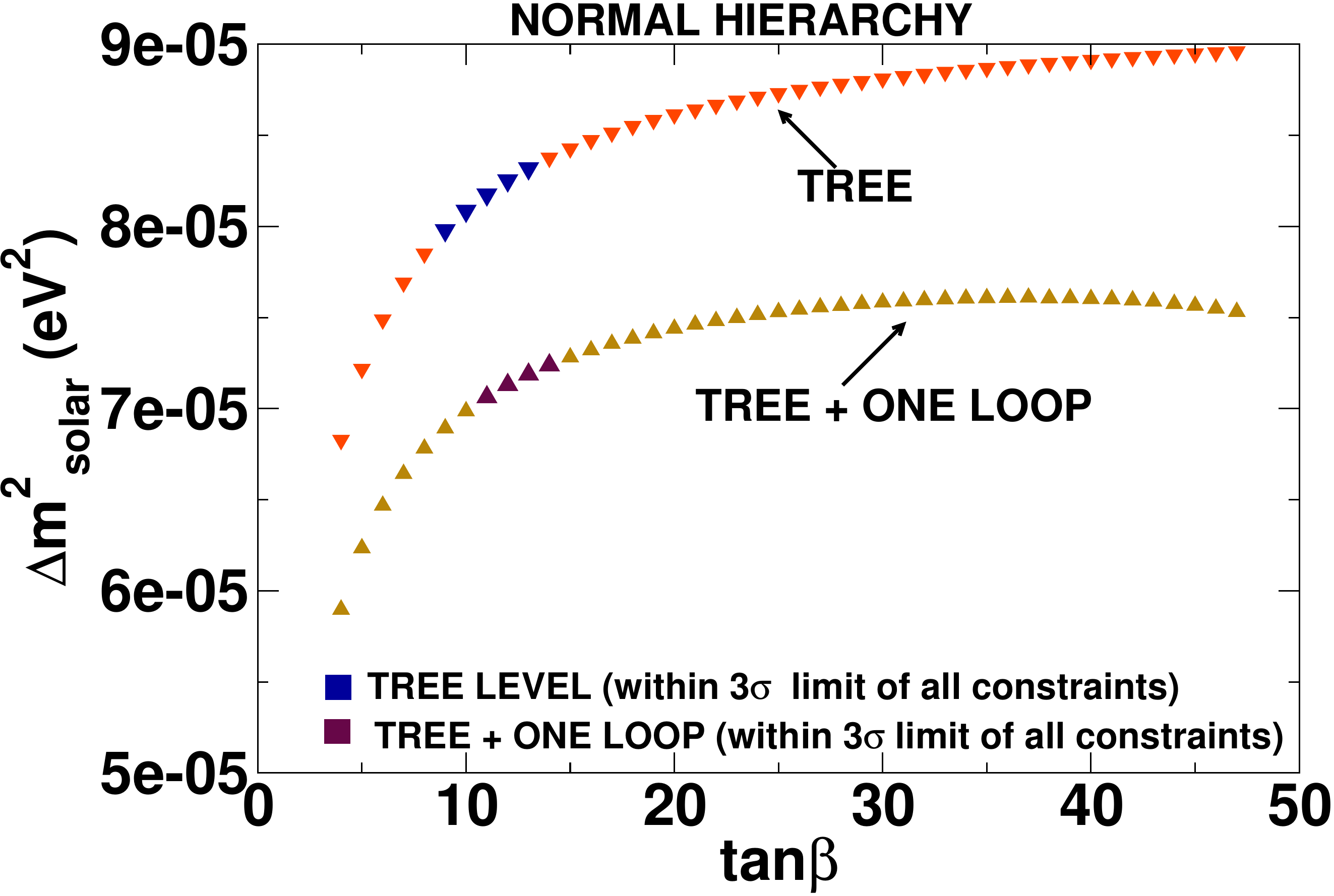}
\caption{$\Delta m^2_{atm},~\Delta m^2_{solar}$ vs $\rm{tan}\beta$
  plots for the {\it{normal hierarchical}} pattern of light neutrino
  masses. The allowed values of $\tan\beta$ are shown by bold points. 
  Other parameter choices are shown in tables \ref{loop-param} 
  and \ref{loop-param-2}.}
\label{tanbetaNH}
\end{figure}

Finally, we will discuss the $\rm{tan}\beta$ dependence of $\Delta
m^2_{atm}$ and $\Delta m^2_{solar}$. These plots are shown in figure 
\ref{tanbetaNH}. The quantity $\Delta m^2_{atm}$ decreases with the
increasing values of $\rm{tan}\beta$ and nearly saturates for larger 
values of $\rm{tan}\beta$. However, the one-loop corrected result 
for $\Delta m^2_{atm}$ is not much different from that at the 
tree level for a particular value of $\tan\beta$. 
On the other hand, the solar mass squared difference initially 
increases with $\tan\beta$ and for higher values of $\tan\beta$ 
the variation slows down and tends to saturate. The one-loop corrections 
result in lower values of $\Delta m^2_{solar}$ for a particular $\tan\beta$. 
The darker and bigger points on both the plots of figure \ref{tanbetaNH} are 
the allowed values of $\tan\beta$, where all the neutrino experimental data 
are satisfied. Note that only a very small range of $\tan\beta$ ($\sim$ 10--14) 
is allowed. This is a very important observation of this analysis. 

Next we will discuss the light neutrino mixing and the effect of 
one-loop corrections on the mixing angles. It was shown in 
ref.\cite{c4Ghosh:2008yh} that for the normal hierarchical pattern of 
neutrino masses, when the parameter $b_i \sim a_i$ (see subsection \ref{tree-neut-mass}), 
the neutrino mixing angles $\theta_{23}$ and $\theta_{13}$ can be 
written as (with the tree level analysis), (see eqns.(\ref{atmos_analytical}),
(\ref{reactor_analytical}))

\bea
\sin^2\theta_{23} \approx \frac{b^2_\mu}{b^2_\mu + b^2_\tau},
\eea
and 
\bea
\sin^2\theta_{13} \approx \frac{b^2_e}{b^2_\mu + b^2_\tau}.
\eea
On the other hand, the mixing angle $\theta_{12}$ is a much more 
complicated function of the parameters $b_i$ and $a_i$ and we do 
not show it here. Now, when $b_i \sim a_i$, we can easily 
see from eqn.(\ref{specifications}), that
\bea
v^\prime_i \sim \frac{Y^{ii}_\nu v_1}{3\lambda}(\tan\beta -1).
\eea 
This implies that for $\tan\beta \gg$ 1 (recall that the allowed 
range of $\tan\beta$ is $\sim$ 10--14), 
\bea
v^\prime_i \gg \frac{Y^{ii}_\nu v_1}{3\lambda}.
\eea 
As we have discussed earlier, for such values of $v^\prime_i$, 
the quantities $b_i \approx c_i$. Hence, the mixing angles 
$\theta_{23}$ and $\theta_{13}$ can be approximately written as
\bea
\sin^2\theta_{23} \approx \frac{c^2_\mu}{c^2_\mu + c^2_\tau},
\label{eq-mixing-23} 
\eea
and
\bea
\sin^2\theta_{13} \approx \frac{c^2_e}{c^2_\mu + c^2_\tau}.
\label{eq-mixing-13} 
\eea

Naively, one would also expect that $\sin^2\theta_{12}$ should
show some correlation with the quantity $c^2_e/c^2_\mu$. However,
as mentioned earlier, this is a very simple minded expectation since
$\sin^2\theta_{12}$ has a more complicated dependence on the model
parameters (see eqn.(\ref{solar_analytical})). 

\begin{figure}[ht]
\centering
\includegraphics[width=7.50cm]{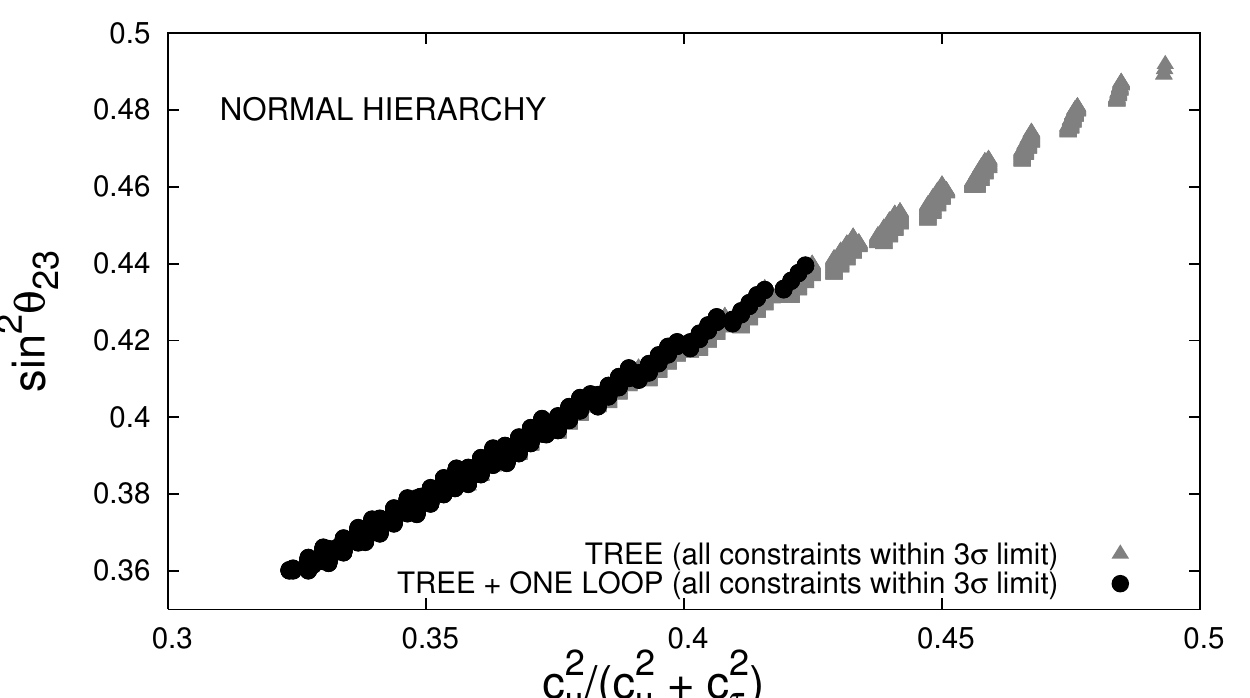}
\vspace{0.2cm}
\includegraphics[width=7.50cm]{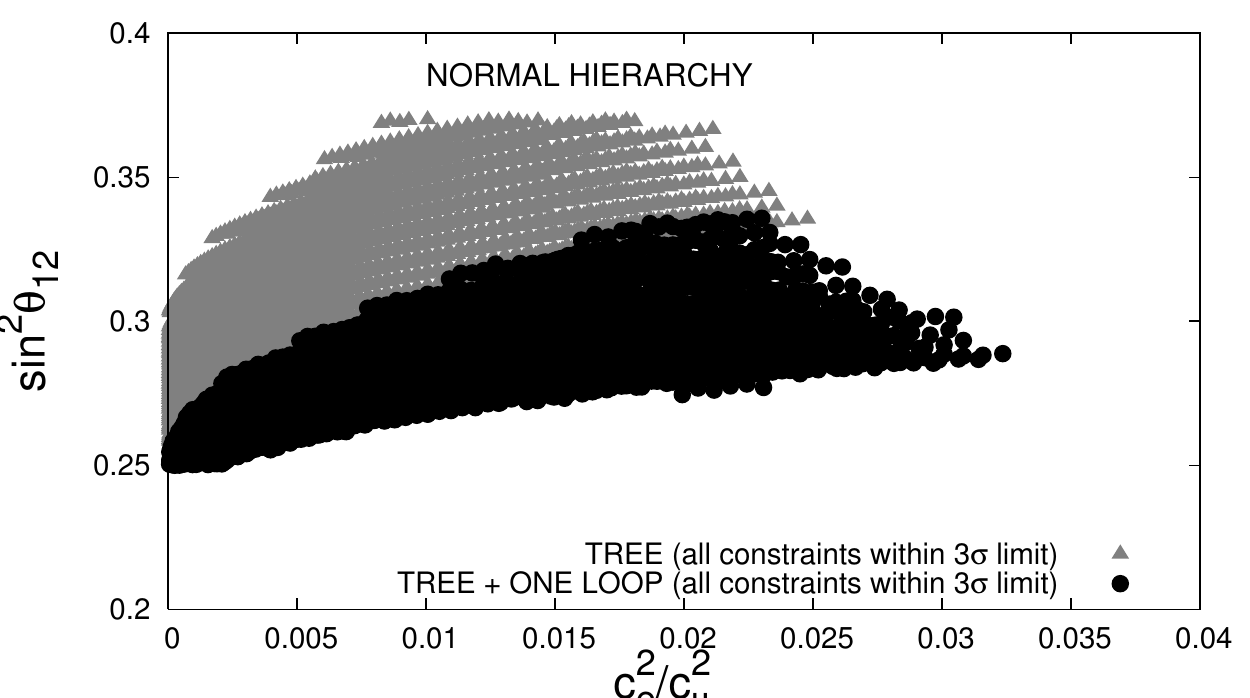}
\vspace{0.2cm}
\includegraphics[width=7.50cm]{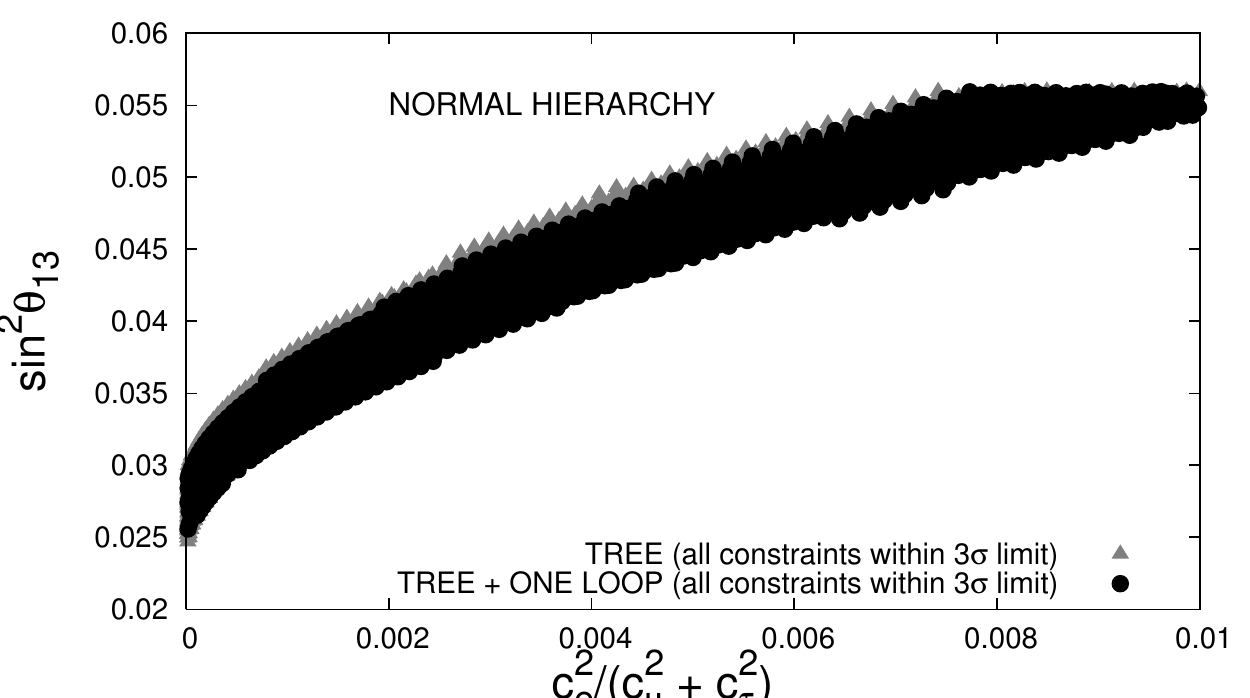}
\caption{Variation of $\rm{sin}^2\theta_{23}$ with
$\frac{c^2_\mu}{(c^2_\mu + c^2_\tau)}$, $\rm{sin}^2\theta_{12}$ with
$\frac{c^2_e}{c^2_\mu}$, $\rm{sin}^2\theta_{13}$ with
$\frac{c^2_e}{(c^2_\mu + c^2_\tau)}$ for normal hierarchy of light
neutrino masses. Parameter choices are shown in tables \ref{loop-param}
and \ref{loop-param-2}.}
\label{mixing-GS-OS}
\end{figure}

The variation of all three mixing angles with the corresponding 
parameters are shown in figure \ref{mixing-GS-OS}. Note that in order 
to generate these plots, we vary only the quantities $c_i$ and 
all the other parameters are fixed at the values given in 
tables \ref{loop-param} and \ref{loop-param-2}. We have chosen the
range of parameters in such a way that the 3-flavour global neutrino 
data are satisfied. The mixing angles have been calculated numerically 
by diagonalizing the neutrino mass matrix in eqn.(\ref{mnuij-compact1}) 
and in eqn.(\ref{one-loop corrected structure of neutralino mass matrix}). 
As expected from our approximate analytical expressions, these plots 
show very nice correlations of the mixing angles $\theta_{23}$ and 
$\theta_{13}$ with the relevant parameters as discussed in 
eqns.(\ref{eq-mixing-23}) and (\ref{eq-mixing-13}). For example, note 
that when $c_\mu \approx c_\tau$, $\sin^2\theta_{23}$ is predicted to 
be $\approx$ 0.5 and that is what we observe in the tree level plot in 
figure \ref{mixing-GS-OS}. However, when one-loop corrections are 
considered, the value of $\sin^2\theta_{23}$ is predicted to be 
somewhat on the lower side of the 3$\sigma$ allowed region. This 
can be understood by looking at the left panel plots of 
figure \ref{gsNH}, where one can see that the one-loop corrected 
results prefer lower values of $c^2_\mu$ and higher values of
$c^2_\tau$. Obviously, this gives smaller $\sin^2\theta_{23}$. On the
other hand, the tree level analysis prefers higher values of $c^2_\mu$
and both lower and higher values of $c^2_\tau$. This gives rise to large 
as well as small values of $\sin^2\theta_{23}$. 

If one looks at the plot of $\sin^2\theta_{13}$ in figure \ref{mixing-GS-OS}, 
then it is evident that the amount of $\nu_e$ flavour in the heaviest state 
$(\nu_3)$ decreases a little bit with the inclusion of one-loop corrections 
for a fixed value of the quantity $\frac{c^2_e}{(c^2_\mu + c^2_\tau)}$. Very 
small $\sin^2\theta_{13}$ demands $c^2_e \ll c^2_\mu, ~c^2_\tau$. This feature
is also consistent with the plots in figure \ref{gsNH}. The correlation of 
$\sin^2\theta_{12}$ with the ratio $c^2_e/c^2_\mu$ is not very sharp as expected 
from the discussion given above. However, a large $\theta_{12}$ mixing angle 
requires a larger value of this ratio. The effect of one-loop correction is
more pronounced in this case and predicts a smaller value of $\sin^2\theta_{12}$ 
compared to the tree level result. There is no specific correlation of the mixing 
angles with the quantities $a^2_i$ and we do not show them here. 

\subsection{{\bf I}nverted hierarchy}\label{loop-inverted}

In this subsection we perform a similar numerical analysis for the
inverted hierarchical scheme of three light neutrino masses. 
Recall that for the inverted hierarchical pattern of light neutrino 
masses, the absolute values of the mass eigenvalues are such that 
$m_2 > m_1\gg m_3$. Thus the solar and the atmospheric mass squared 
differences are defined as $\Delta m^2_{atm} = m^2_1 - m^2_3$ and 
$\Delta m^2_{solar} = m^2_2 - m^2_1$. In order to generate such a mass 
pattern, the choices of neutrino Yukawa couplings $Y_\nu^{ii}$ and the
left-handed sneutrino VEVs $v^\prime_i$ are shown in table \ref{loop-param-2}. 
However, these are just sample choices and other choices also exist 
as we will see during the course of this discussion. The choices of 
other parameters are shown in table \ref{loop-param}. The effect of 
one-loop corrections to the mass eigenvalues are such that the absolute 
values of masses $m_3$ and $m_1$ become smaller whereas $m_2$ grows in 
magnitude. This effect of increasing the absolute value of $m_2$ while 
decreasing that of $m_1$ makes it extremely difficult to account for the 
present 3$\sigma$ limits on $\Delta m^2_{solar}$. 

\begin{figure}[ht]
\centering
\includegraphics[width=6.00cm]{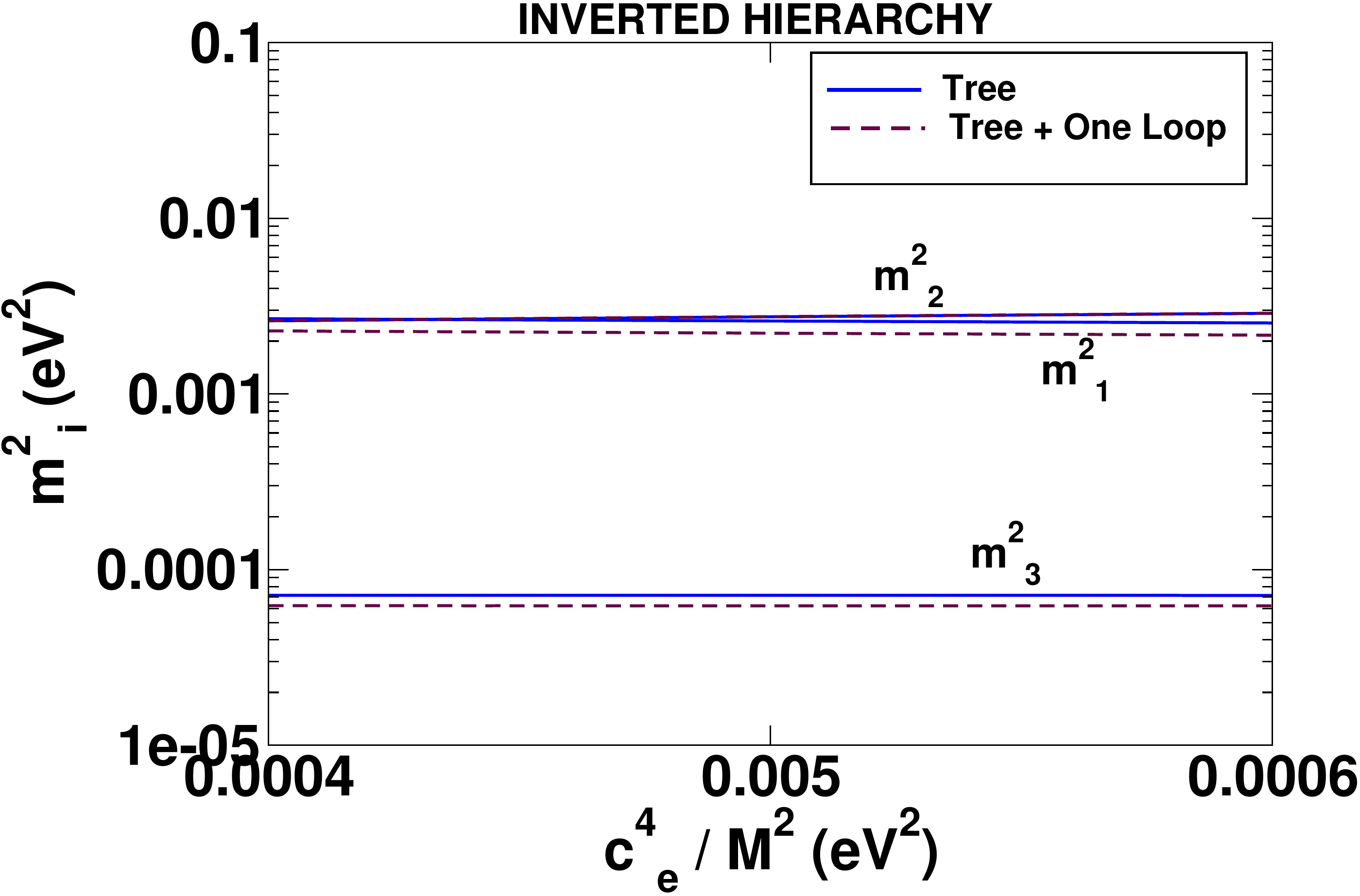}
\includegraphics[width=6.00cm]{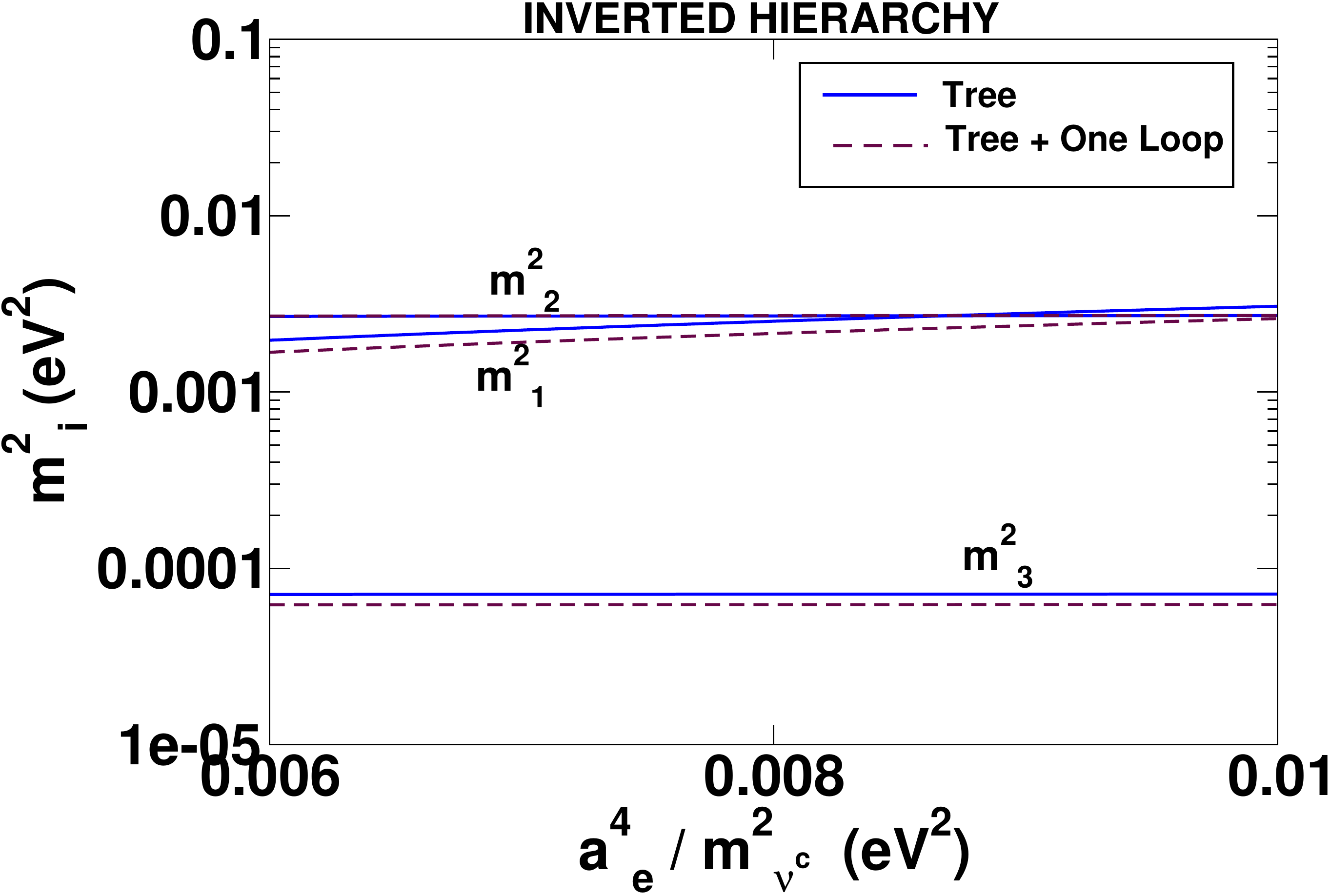}
\vspace{0.2cm}
\includegraphics[width=6.00cm]{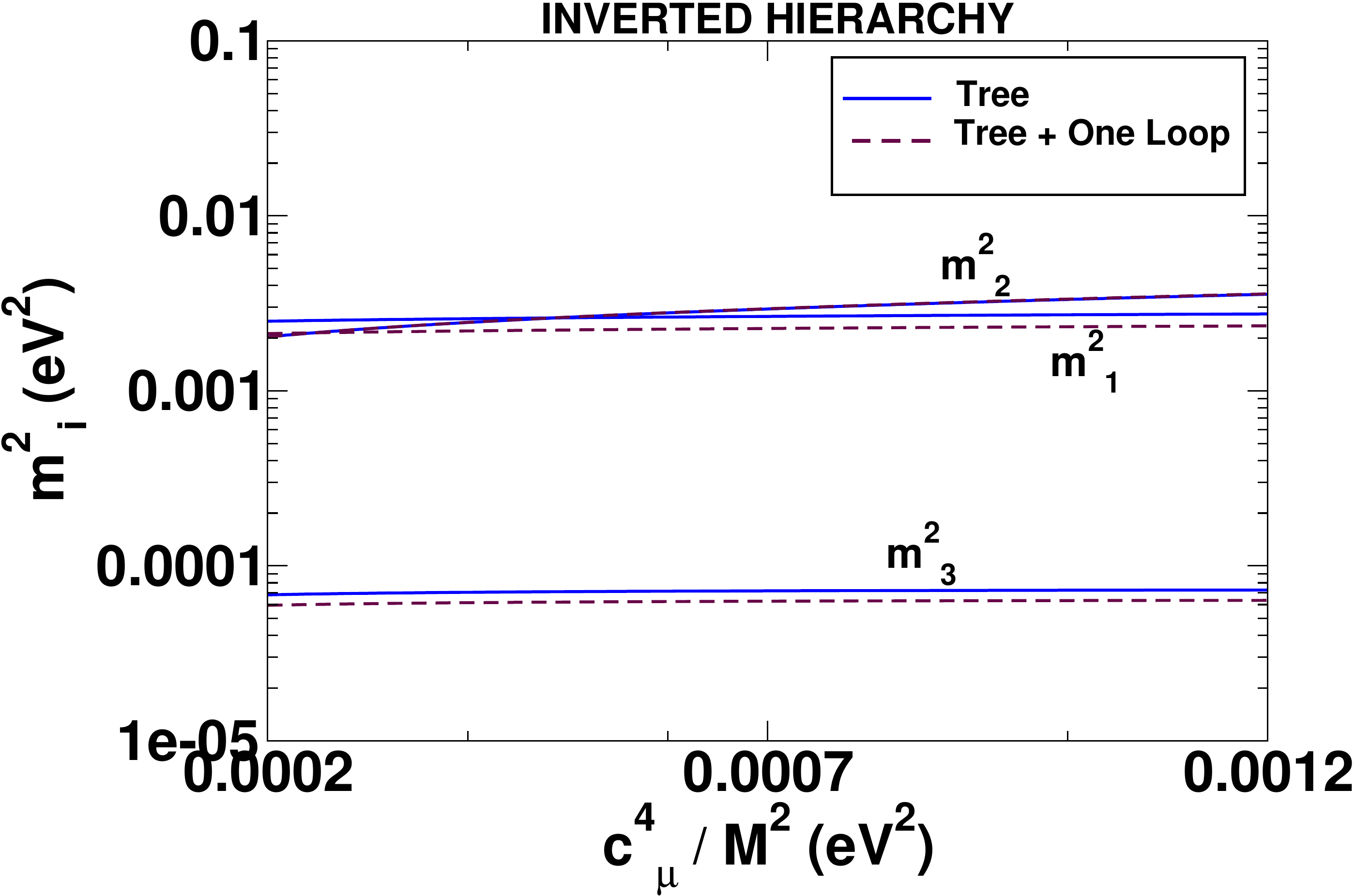}
\includegraphics[width=6.00cm]{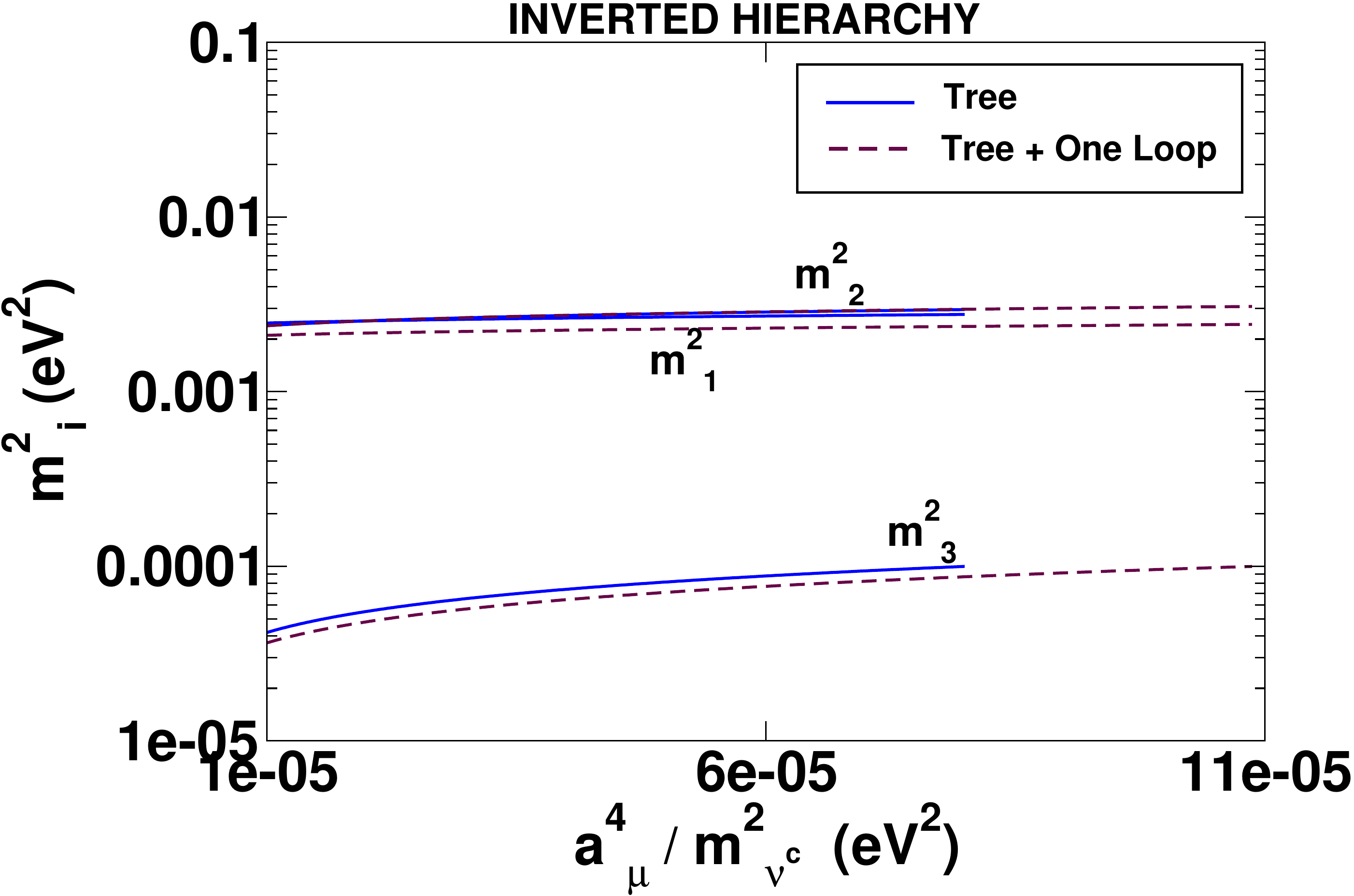}
\vspace{0.2cm}
\includegraphics[width=6.00cm]{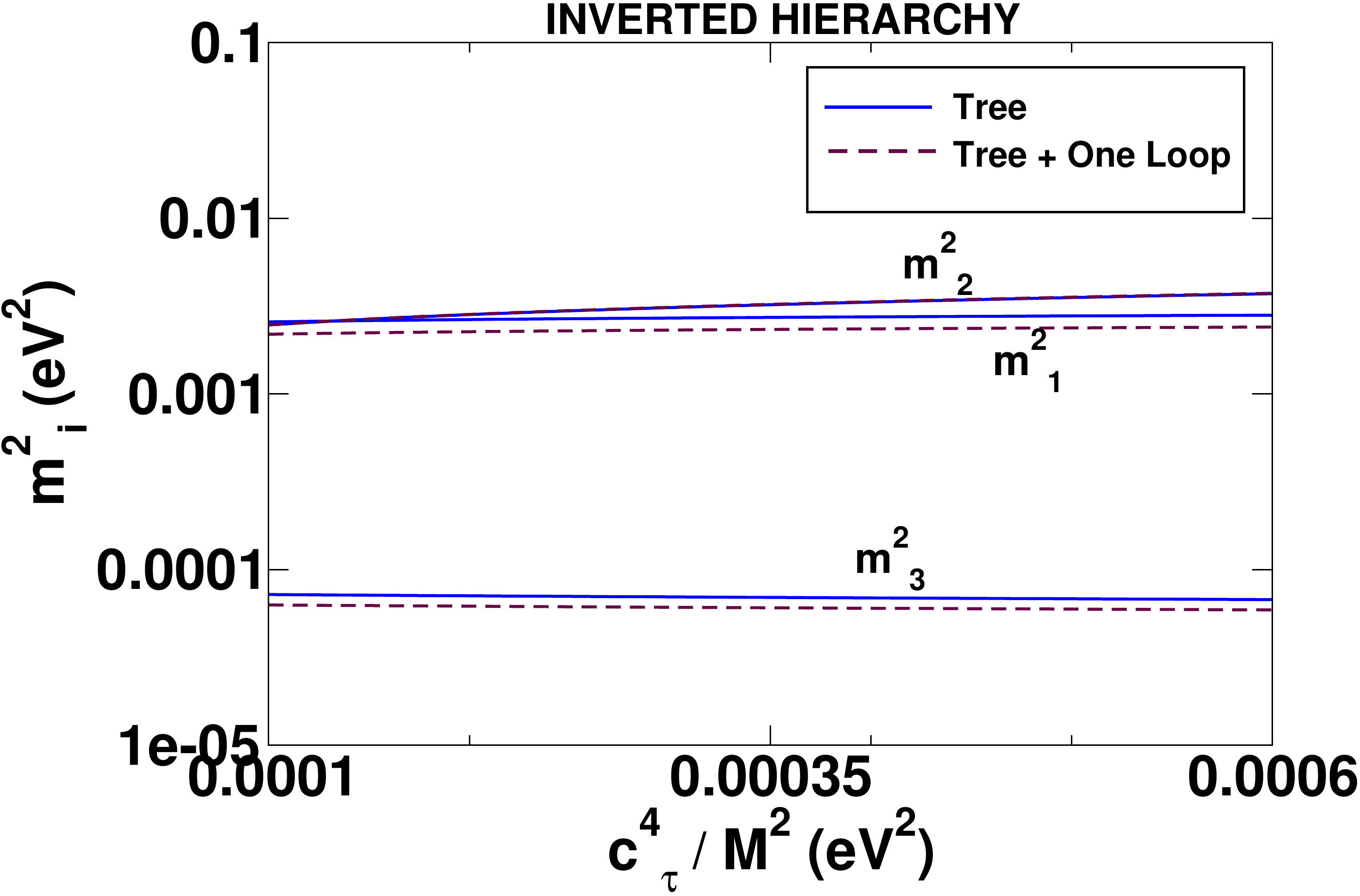}
\includegraphics[width=6.00cm]{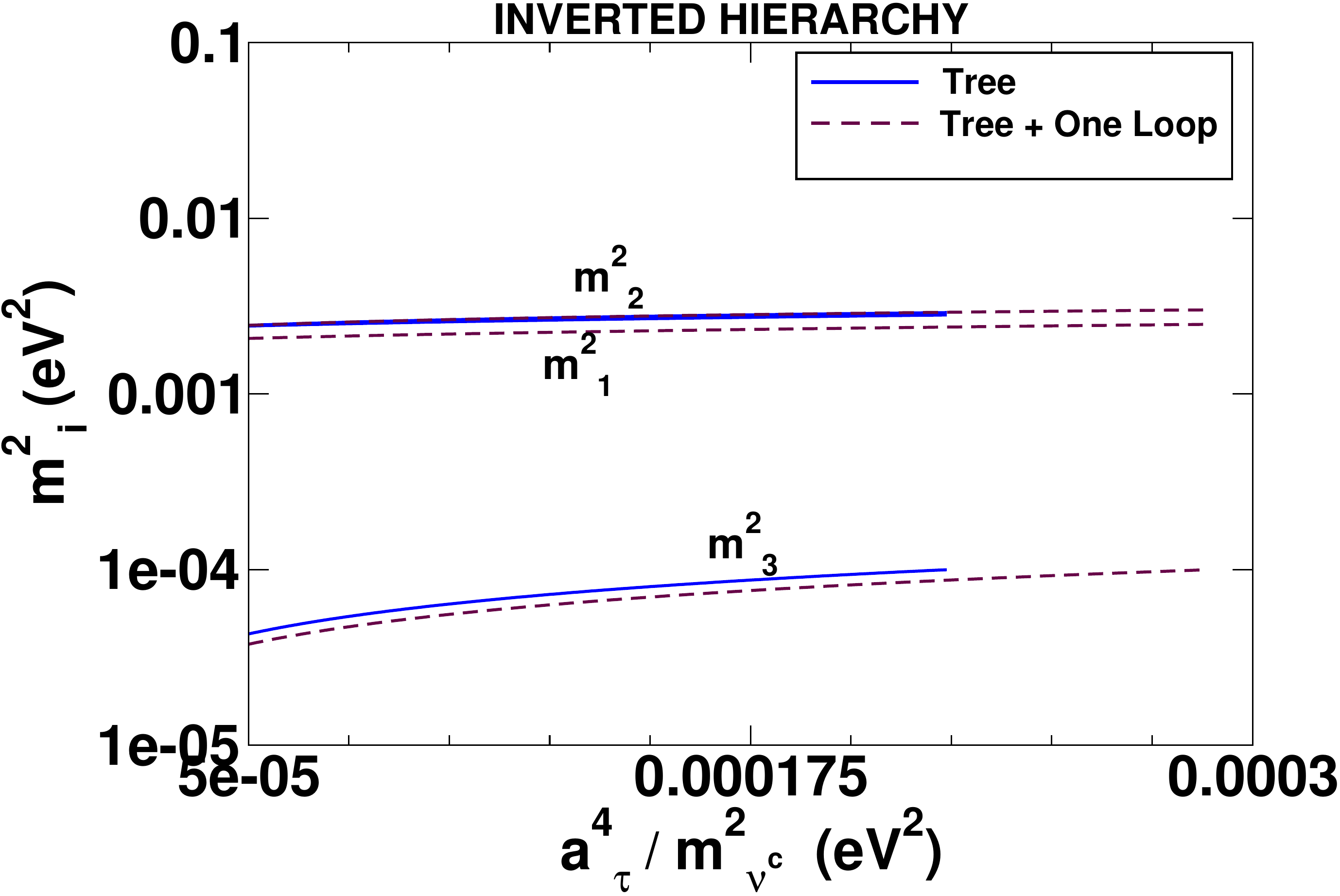}
\caption{Neutrino mass squared values ($m^2_i$) vs $\frac{c^4_i}{M^2}$ 
(left panel) and vs $\frac{a^4_i}{m^2_{\nu^c}}$ (right panel) plots for 
the {\it{inverted hierarchical}} pattern of light neutrino masses, $i = e,
 \mu, \tau$. Parameter choices are shown in tables \ref{loop-param} 
  and \ref{loop-param-2}.}
\label{numsqIH}
\end{figure}

Typical mass spectra are shown in figure \ref{numsqIH}. Once again 
note that a particular model parameter has been varied while the others 
are fixed at values given in tables \ref{loop-param} and \ref{loop-param-2}. As it 
is evident from these plots, the masses $m_1$ and $m_3$ are controlled 
mainly by the parameters $a^2_i/m_\nu^c$, whereas the mass $m_2$ is 
controlled by the seesaw parameters $c^2_i/M$ though there is a small 
contribution coming from $a^2_i/m_\nu^c$ as well. 

\begin{figure}[ht]
\centering
\includegraphics[width=6.00cm]{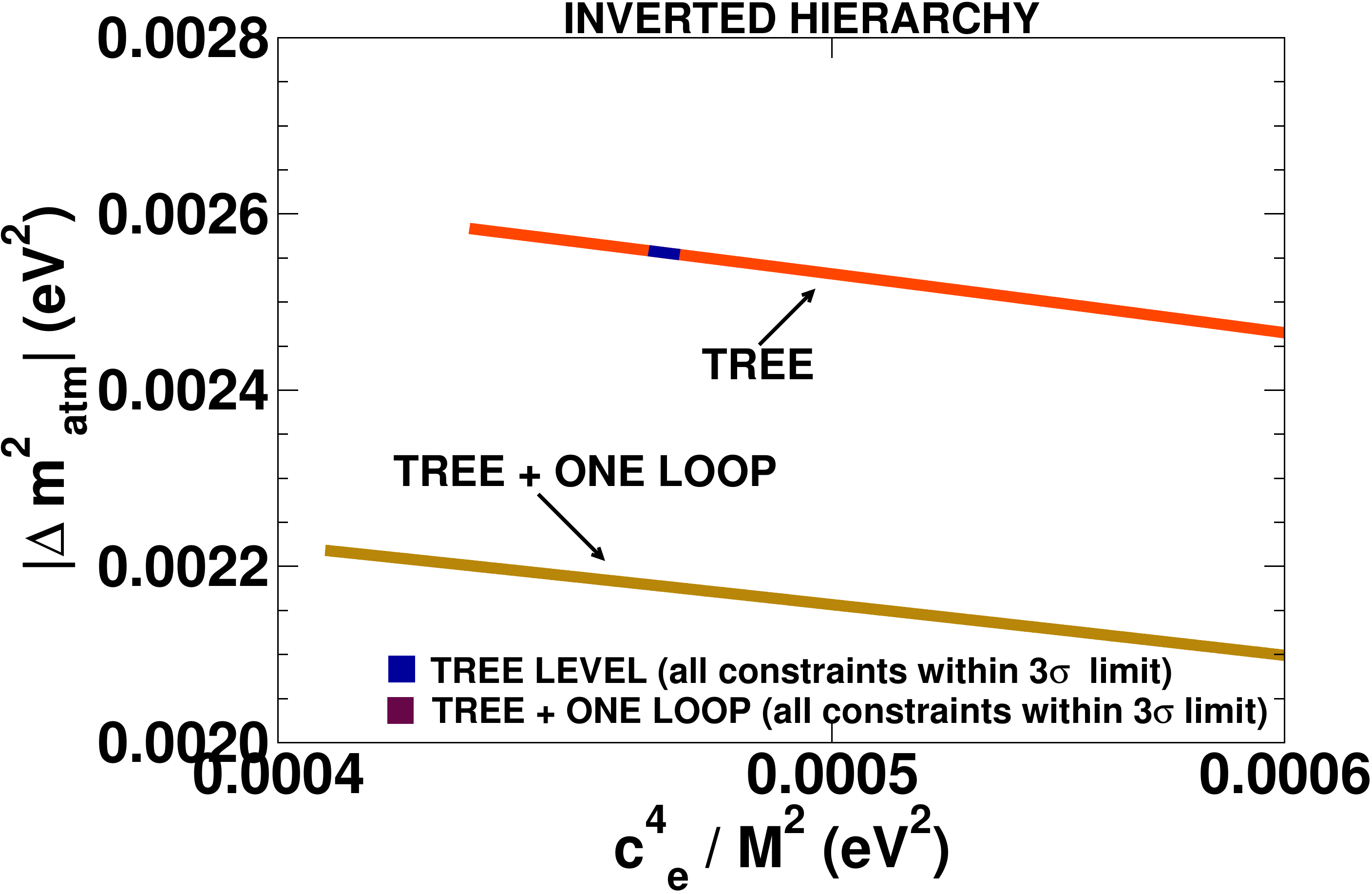}
\includegraphics[width=6.00cm]{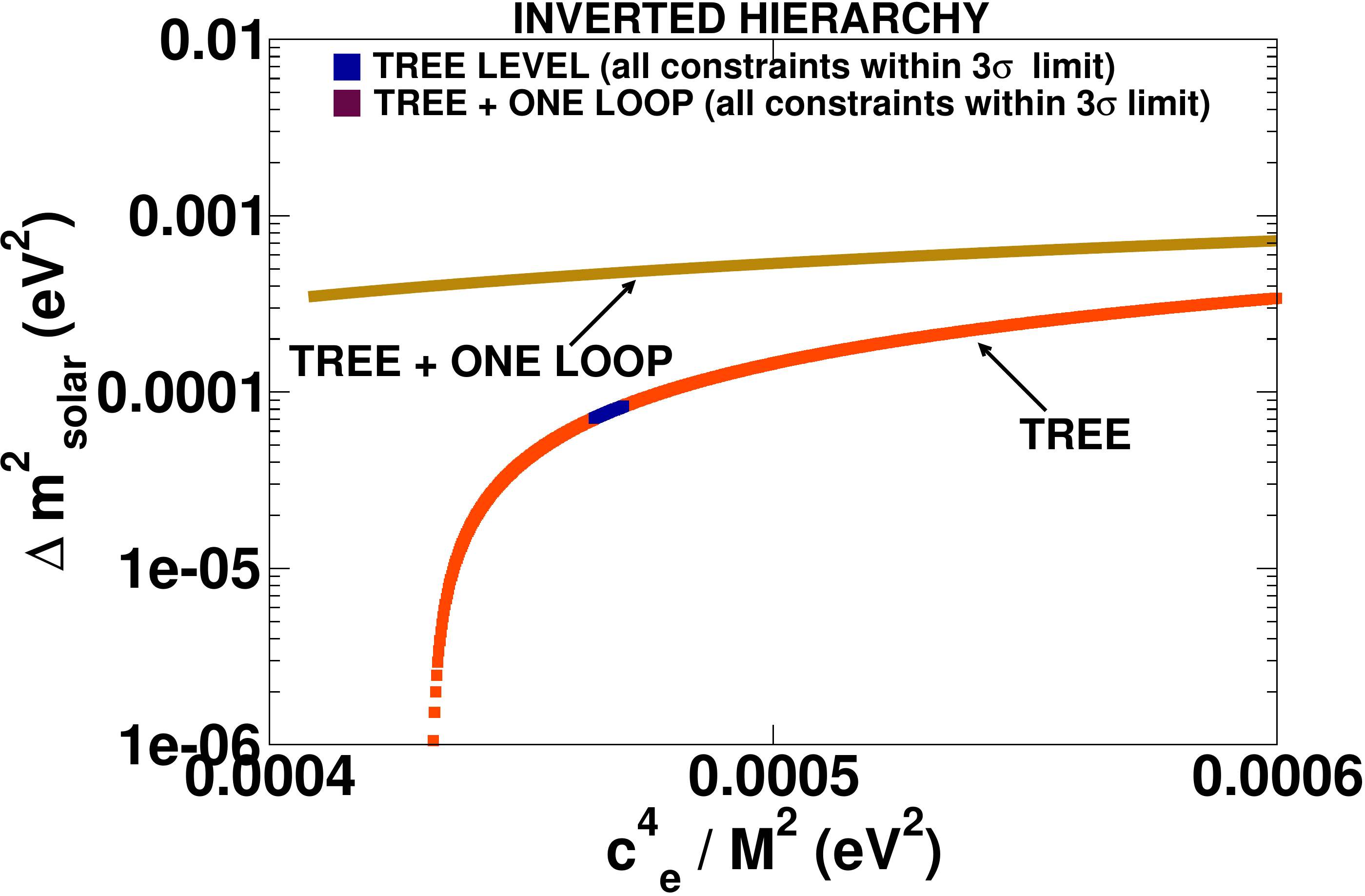}
\vspace{0.2cm}
\includegraphics[width=6.00cm]{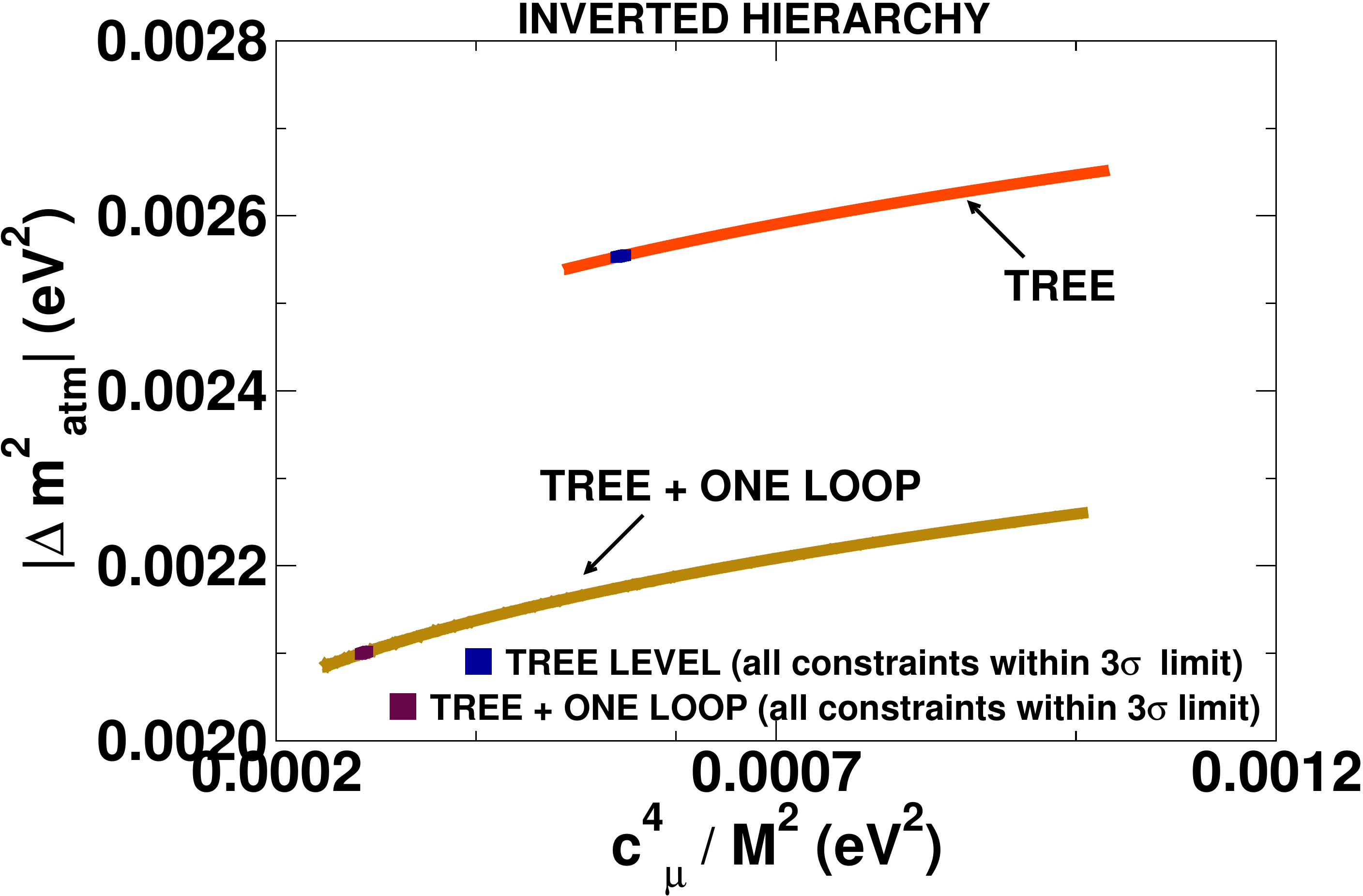}
\includegraphics[width=6.00cm]{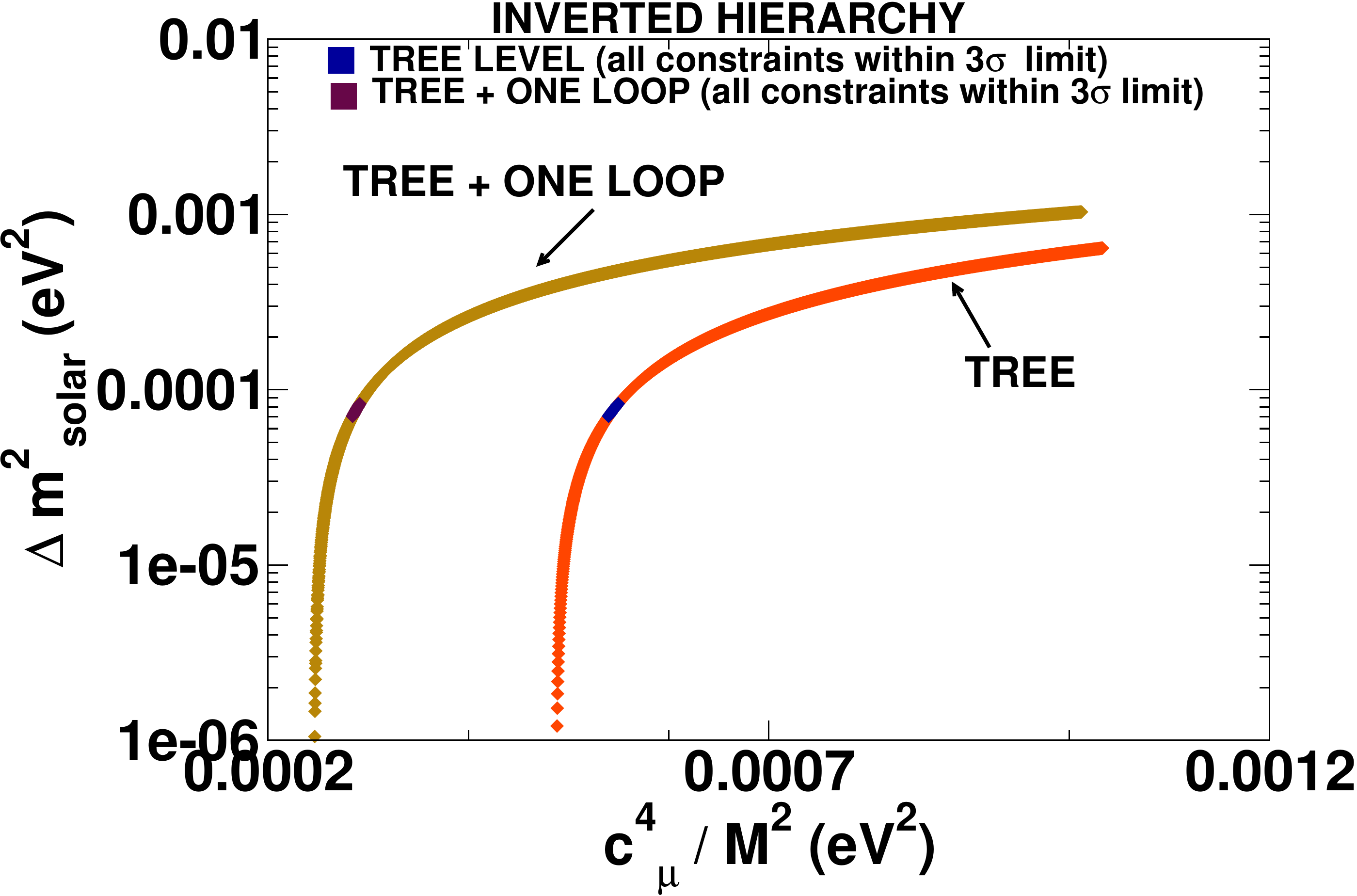}
\vspace{0.2cm}
\includegraphics[width=6.00cm]{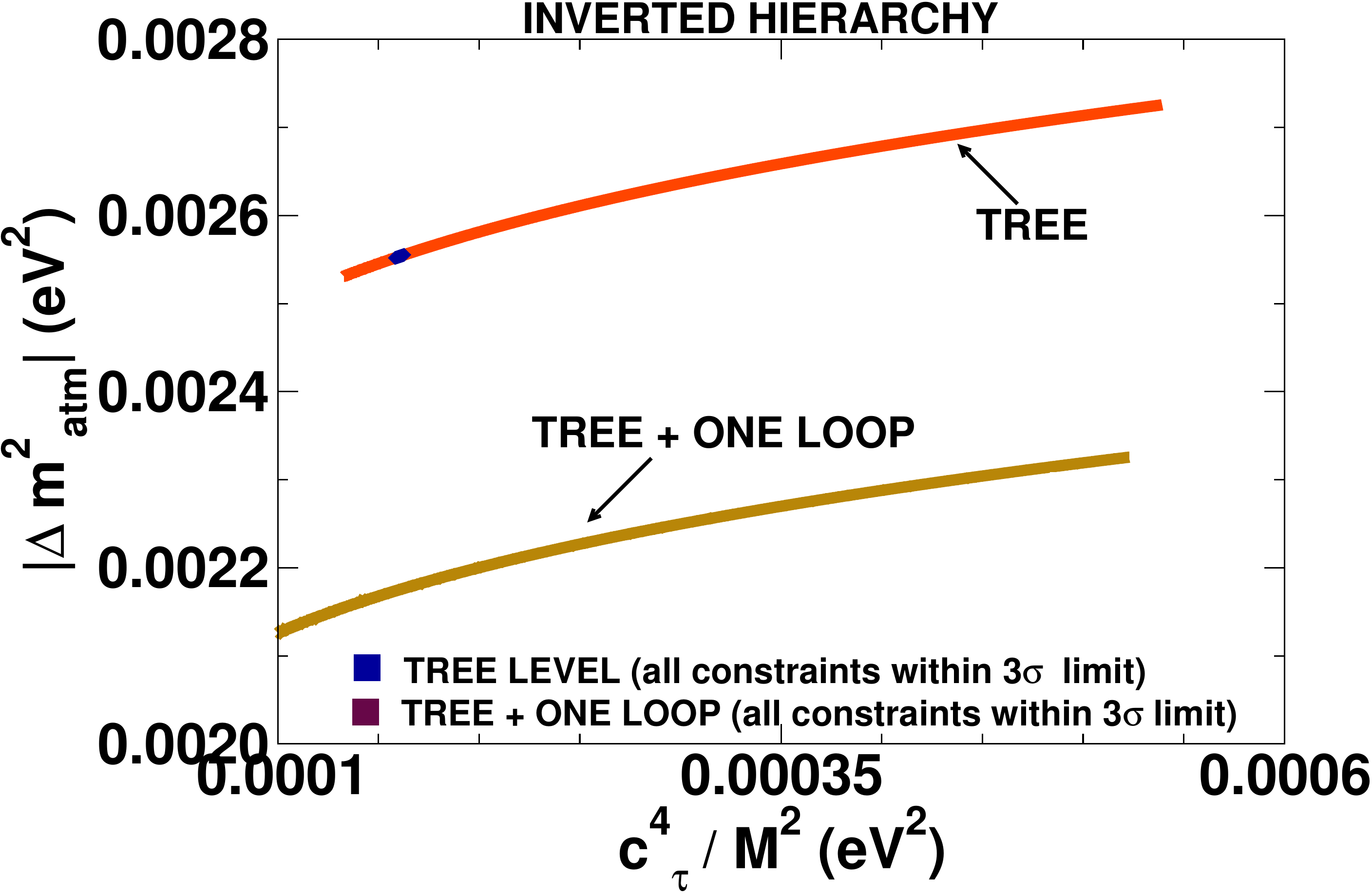}
\includegraphics[width=6.00cm]{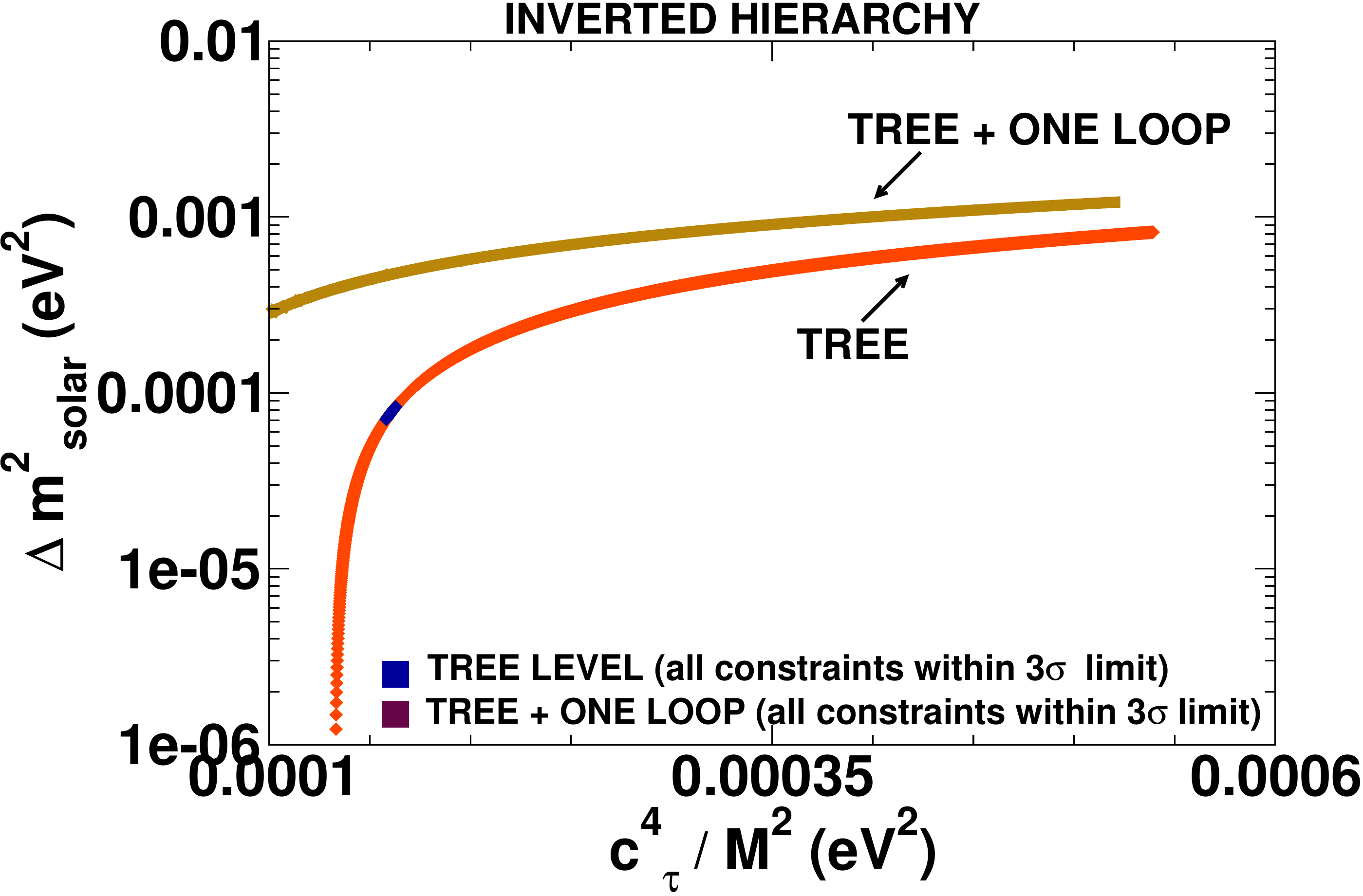}
\caption{Atmospheric and solar mass squared differences $(|\Delta
  m^2_{atm}|,~\Delta m^2_{solar})$ vs $\frac{c_i^4}{M^2}$ plots for
  the {\it{inverted hierarchical}} pattern of light neutrino masses
  with $i = e, \mu, \tau$. Colour specification is same as described 
  in the context of figure \ref{gsNH}. Parameter choices are shown in 
  tables \ref{loop-param} and \ref{loop-param-2}.}
\label{gsIH}
\end{figure}

\begin{figure}[ht]
\centering
\includegraphics[width=6.00cm]{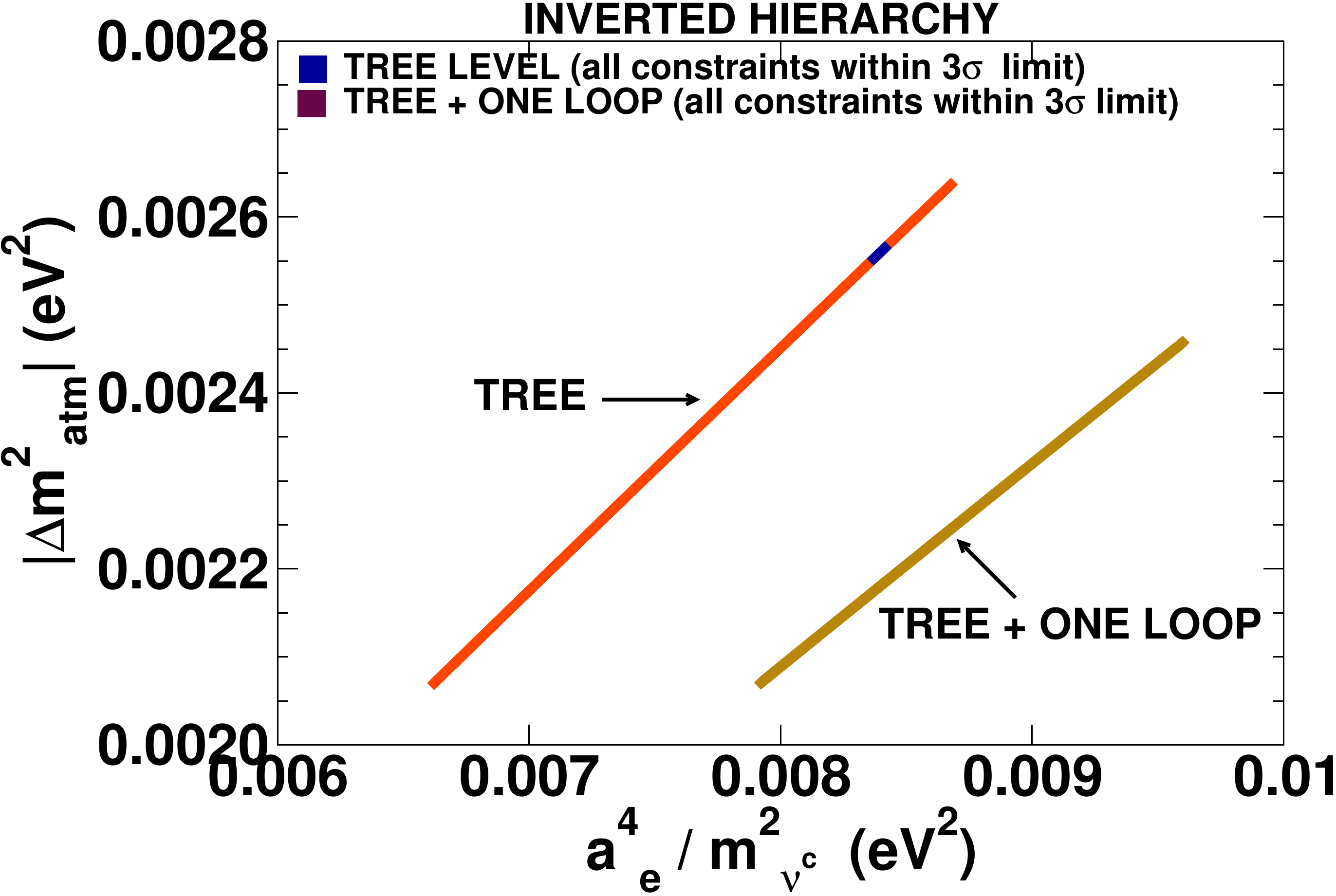}
\includegraphics[width=6.00cm]{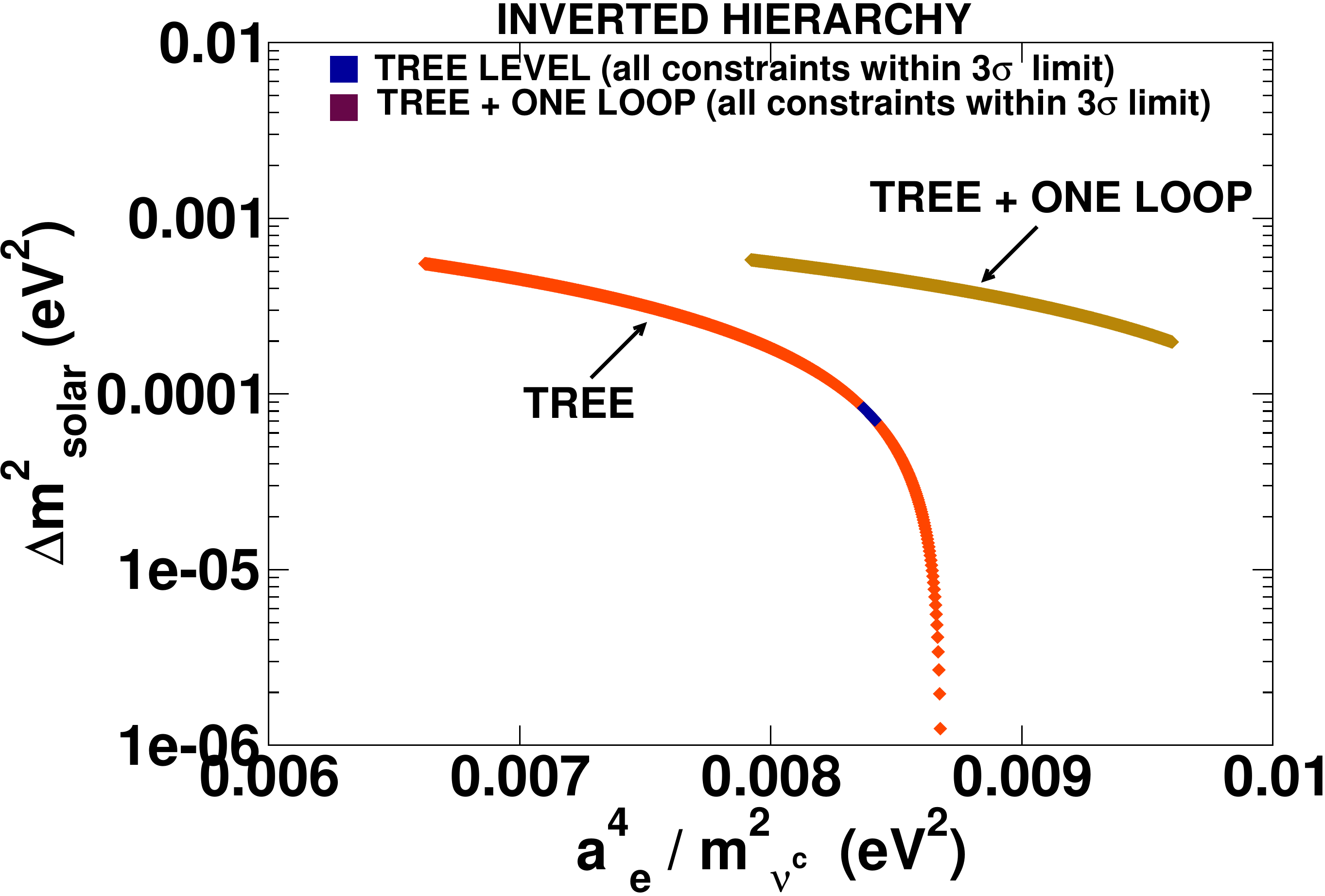}
\vspace{0.2cm}
\includegraphics[width=6.00cm]{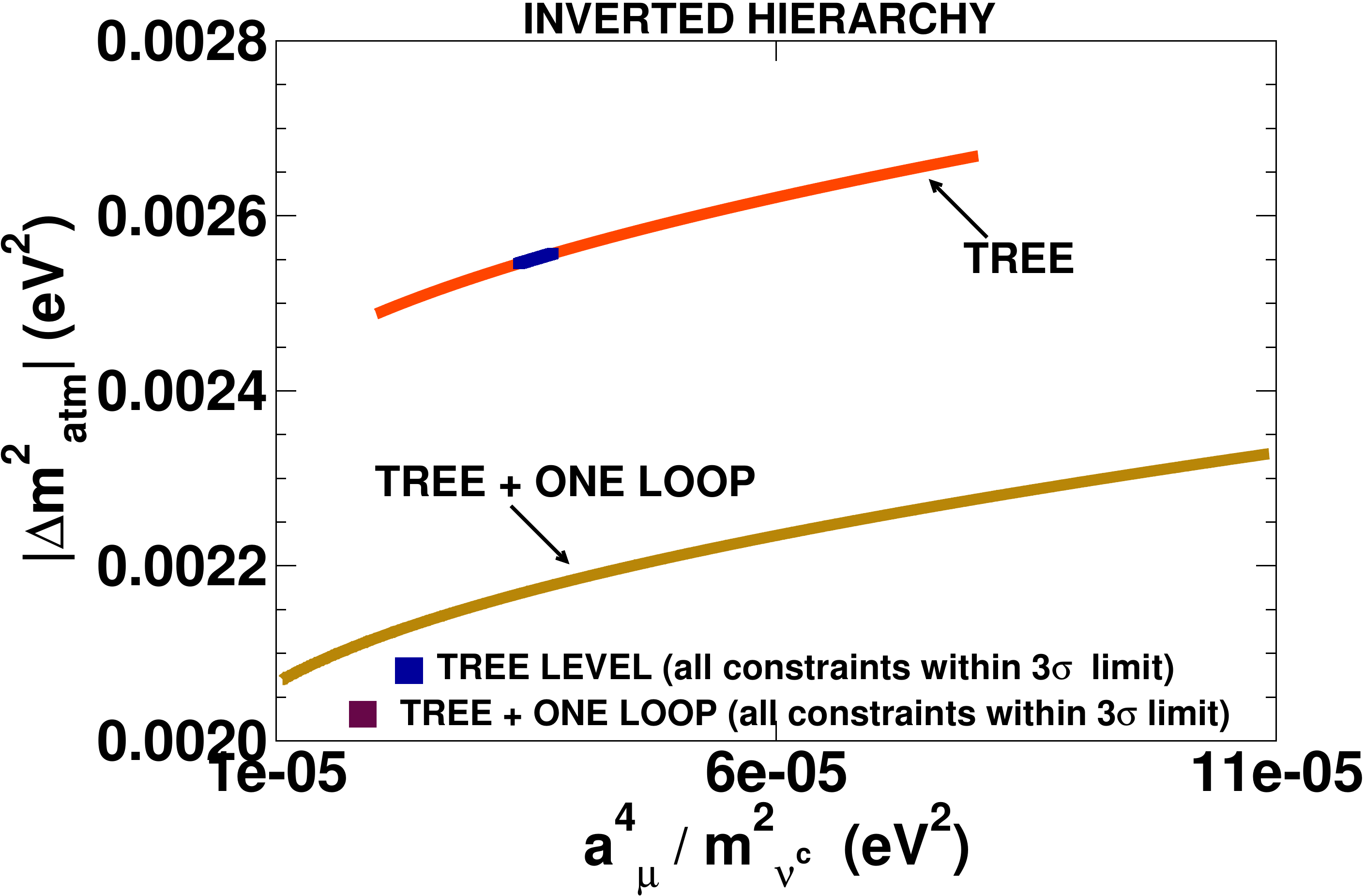}
\includegraphics[width=6.00cm]{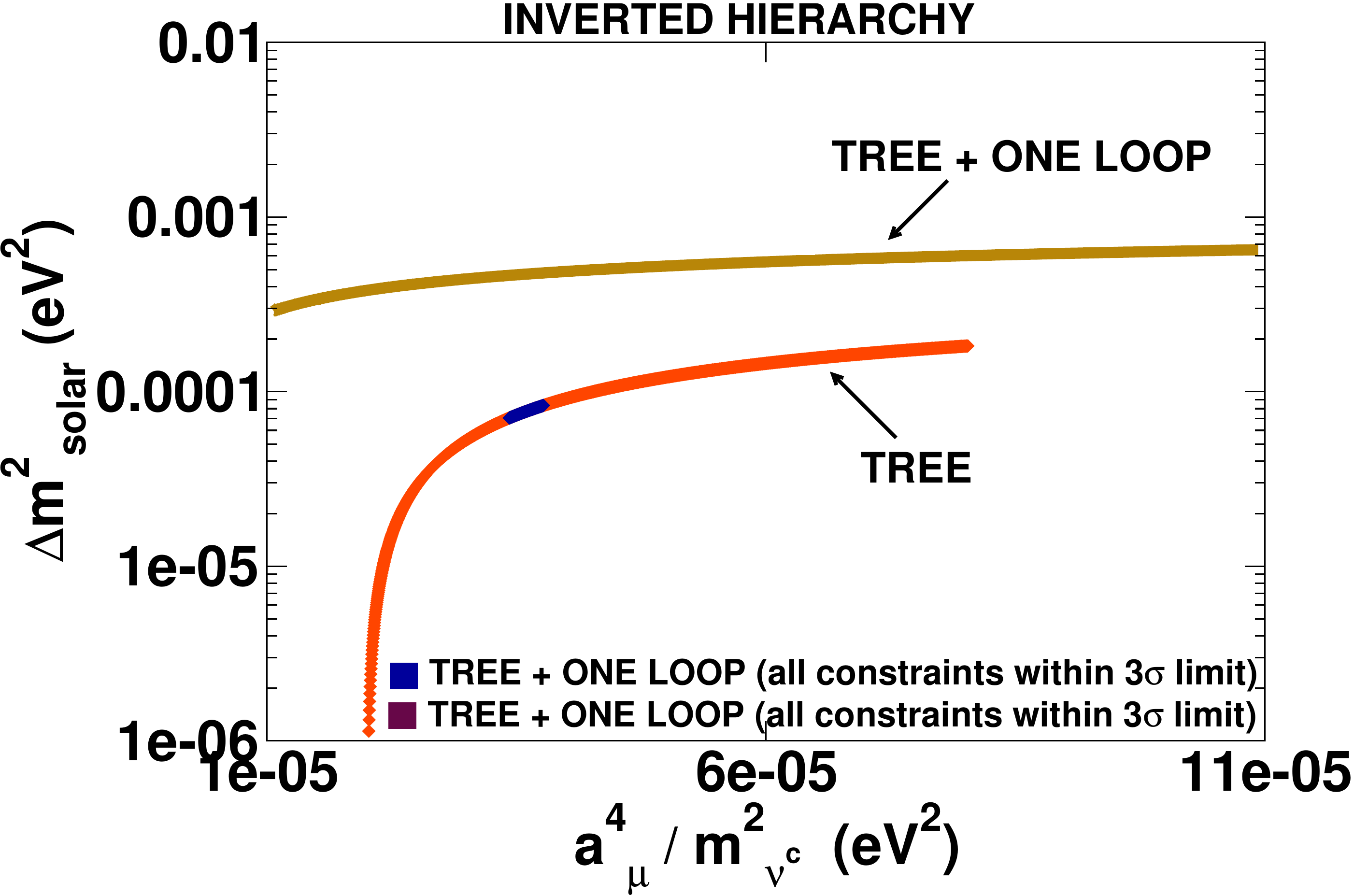}
\vspace{0.2cm}
\includegraphics[width=6.00cm]{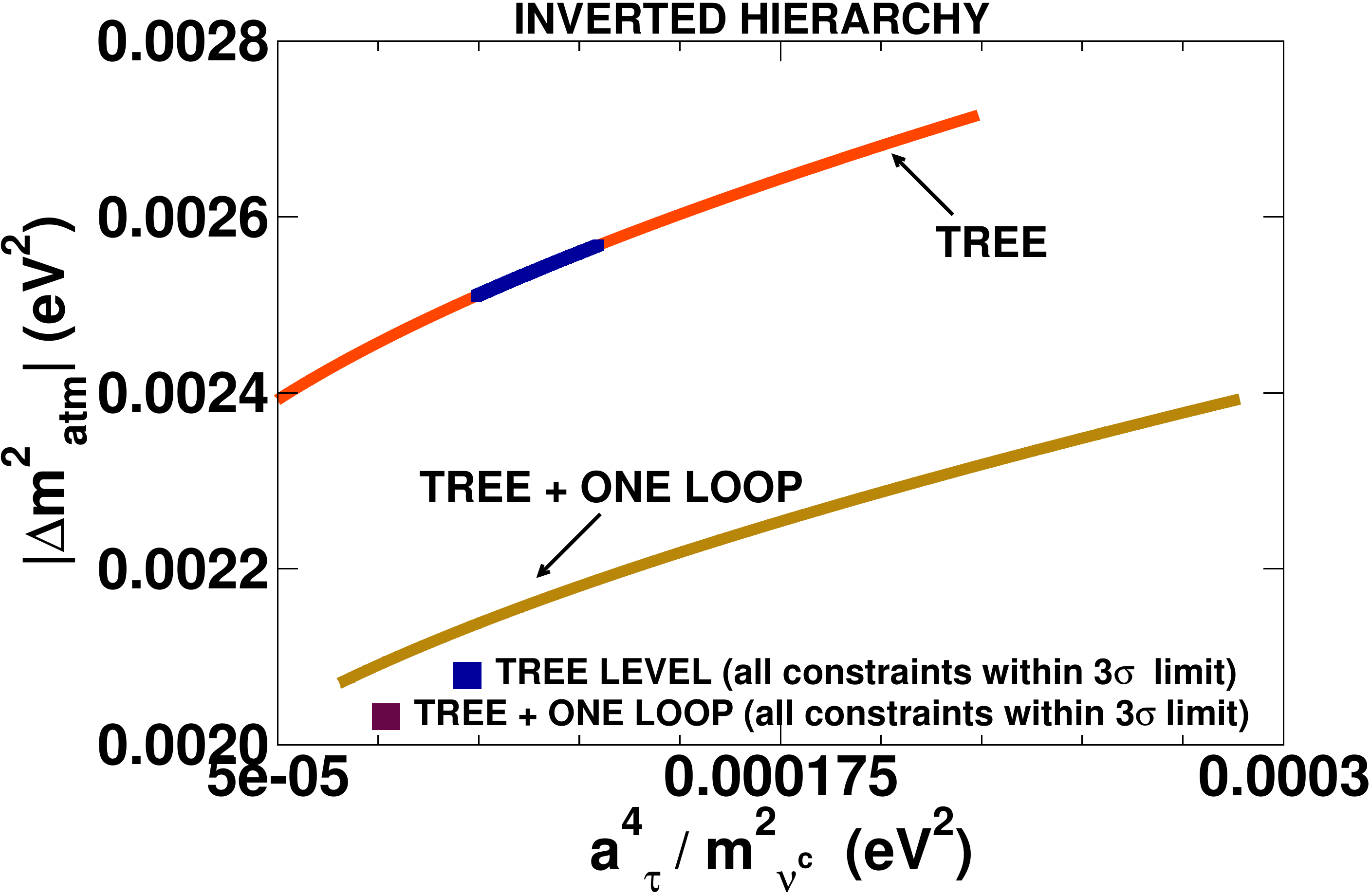}
\includegraphics[width=6.00cm]{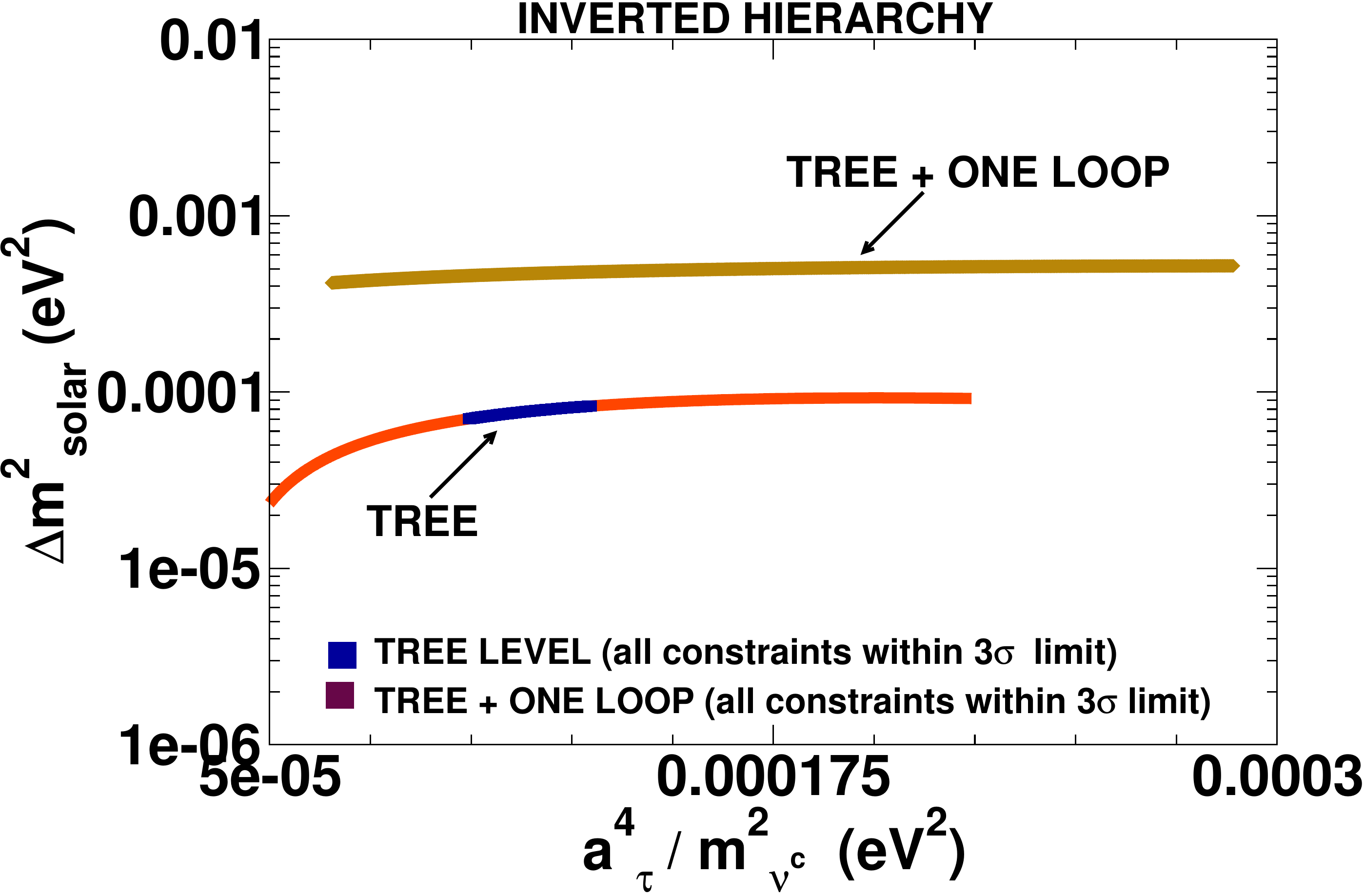}
\caption{Atmospheric and solar mass squared differences $(|\Delta
  m^2_{atm}|,~\Delta m^2_{solar})$ vs $a^4_i/m^2_{\nu^c}$ plots for
  the {\it{inverted hierarchical}} pattern of light neutrino masses
  with $i = e, \mu, \tau$. Colour specification is same as described 
  in the context of figure \ref{gsNH}. Parameter choices are shown in 
  tables \ref{loop-param} and \ref{loop-param-2}.}
\label{osIH}
\end{figure}

Let us now turn our attention to the variation of $|\Delta m^2_{atm}|$ 
and $\Delta m^2_{solar}$ with $c^4_i/M^2$ and $a^4_i/{m^2_{\nu^c}}$ 
shown in figure \ref{gsIH} and figure \ref{osIH}. For our numerical analysis, 
we have set the scale of $m_3$ as  $|m_3|_{max}<0.011~\rm{eV}$. The left panel 
in figure \ref{gsIH} shows that $|\Delta m^2_{atm}|$ increases 
with $c^4_{\mu,\tau}/M^2$ and decreases with $c^4_e/M^2$. This is 
essentially the behaviour shown by $m^2_1$ with the variation of 
$c^4_i/M^2$. Similar behaviour is obtained for the one-loop corrected 
$\Delta m^2_{atm}$. The decrease in the one-loop corrected result 
compared to the tree level one is due to the splitting in $m^2_1$ 
value as shown in figure \ref{numsqIH}. 

The variation of $\Delta m^2_{solar}$ with $c^4_i/M^2$ can be understood 
in a similar manner by looking at figure \ref{numsqIH}. As explained earlier, 
in the case of $\Delta m^2_{solar}$, the one-loop corrected result is larger 
compared to the tree level one. The range of parameters satisfying all the 
three flavour global neutrino data are shown by the fewer dark points on 
the plots. Note that the increase of $\Delta m^2_{solar}$ at the one-loop 
level is such that we do not even see any allowed range of parameters when 
looking at the variation with respect to $c^4_{e,\tau}/M^2$. Once again, the 
behaviour of $\Delta m^2_{atm}$ and $\Delta m^2_{solar}$ with the change 
in the parameters $a^4_i/m^2_{\nu^c}$ (shown in figure \ref{osIH}) can be 
explained by looking at the right panel plots of figure \ref{numsqIH}. 

We have also investigated the nature of variation of $|\Delta m^2_{atm}|$
and $\Delta m^2_{solar}$ with $\varepsilon^2_i$, the squared effective 
bilinear $R_P$-violating parameters. $|\Delta m^2_{atm}|$ was found to 
increase with $\varepsilon^2_i$ (the increase is sharper for $\varepsilon^2_1$), 
whereas $\Delta m^2_{solar}$ initially increases very sharply with $\varepsilon^2_i$ 
(particularly for $\varepsilon^2_1$ and $\varepsilon^2_2$) and then becomes 
flat. In the one-loop corrected results we do not find any range of values for
parameters where the neutrino data are satisfied. These plots are not shown here.

The variation of mass squared differences with $\lambda$ and 
$\kappa$ have also been analyzed. The variation of
$|\Delta m^2_{atm}|$ and $\Delta m^2_{solar}$ with $\lambda$ 
and $\kappa$ are found to be opposite to those of normal hierarchical
scenario. The one-loop corrected results do not show any allowed ranges 
of $\lambda$ and $\kappa$ (for the chosen values of other parameters) 
where the neutrino data can be satisfied.

\begin{figure}[ht]
\centering
\includegraphics[width=6.00cm]{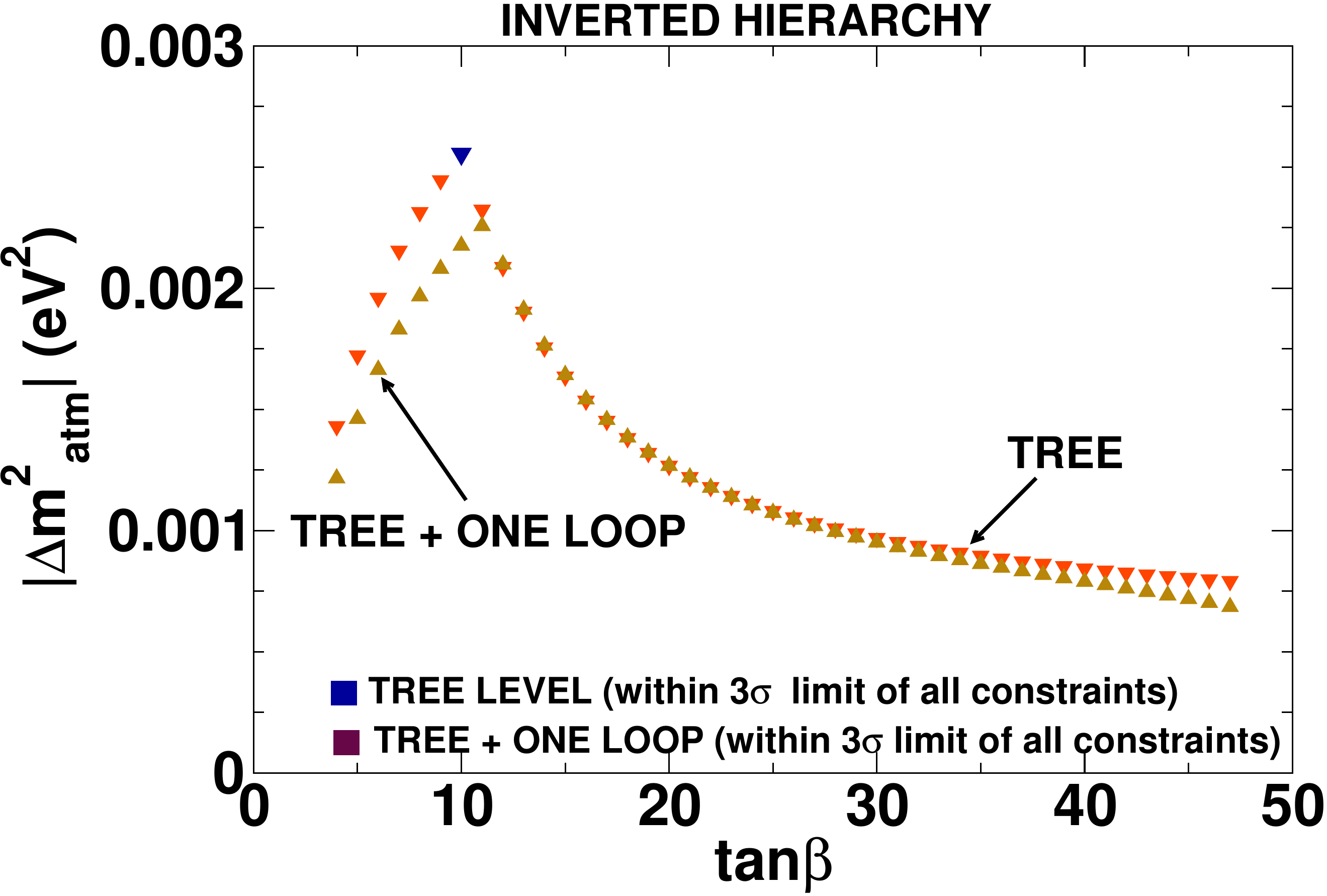}
\includegraphics[width=6.00cm]{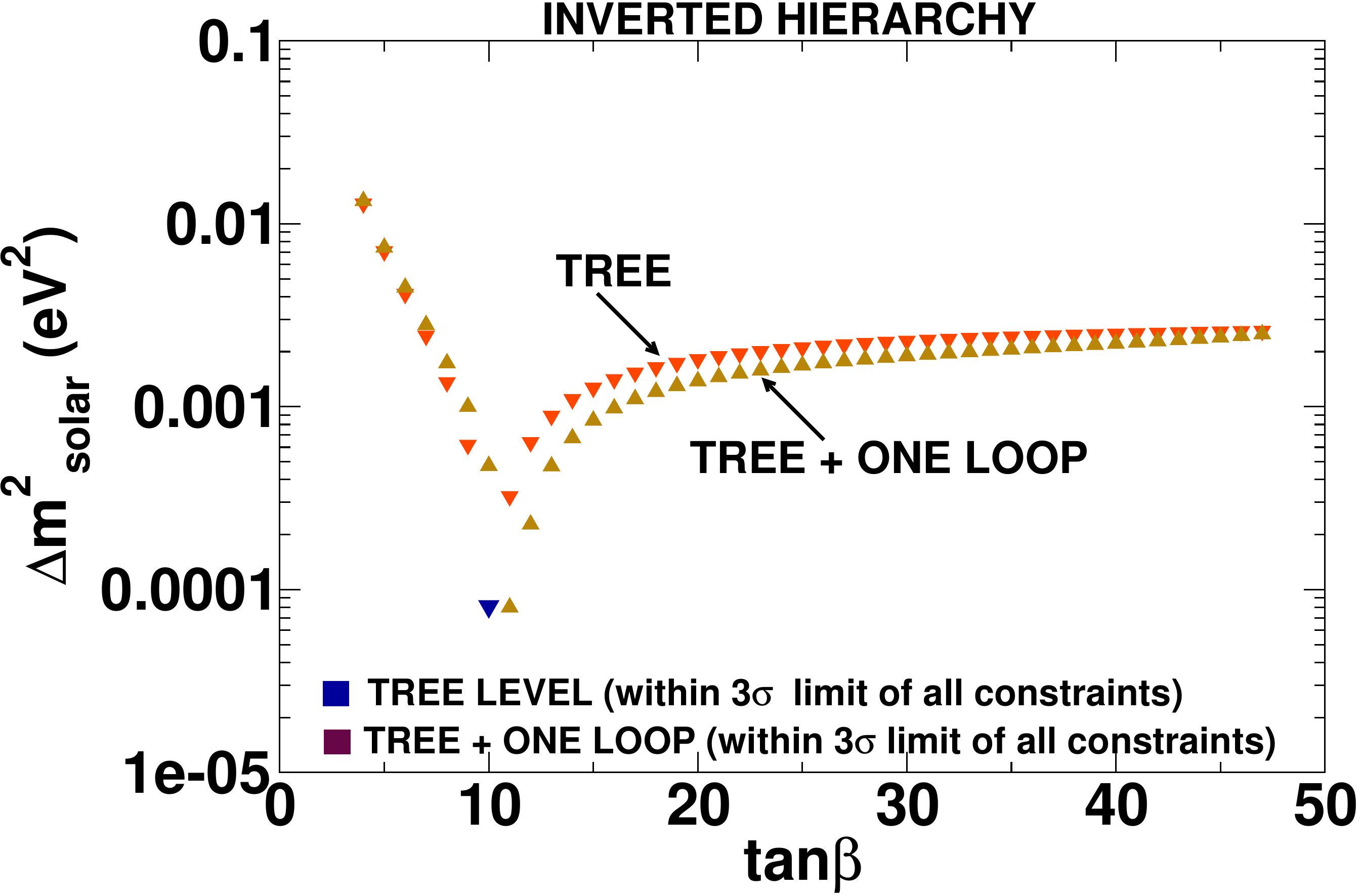}
\caption{$|\Delta m^2_{atm}|,~\Delta m^2_{solar}$ vs $\rm{tan}\beta$
  plots for the {\it{inverted hierarchical}} pattern of light neutrino
  masses. Colour specification is same as described 
  in the context of figure \ref{gsNH}. 
Parameter choices are shown in tables \ref{loop-param} 
  and \ref{loop-param-2}.}
\label{tanbetaIH}
\end{figure}

The $\tan\beta$ dependence of $|\Delta m^2_{atm}|$ and $\Delta m^2_{solar}$ 
is shown in figure \ref{tanbetaIH}. One can see from these two figures that
$|\Delta m^2_{atm}|$ initially increases and then start decreasing at a 
value of $\tan\beta$ around $10$. On the other hand, $\Delta m^2_{solar}$ 
initially decreases and then start increasing around the same value of 
$\tan\beta$. Note that the one-loop corrected result for $|\Delta m^2_{atm}|$
is lower than the corresponding tree level result for $\tan\beta < 10$ whereas
the one-loop corrected result for $\Delta m^2_{solar}$ is lower than the 
corresponding tree level result for $\tan\beta > 10$. For the chosen values 
of other parameters we see that the one-loop corrected analysis does not 
provide any value of $\tan\beta$ where the neutrino data can be satisfied.

\begin{figure}[ht]
\centering
\includegraphics[width=6.00cm]{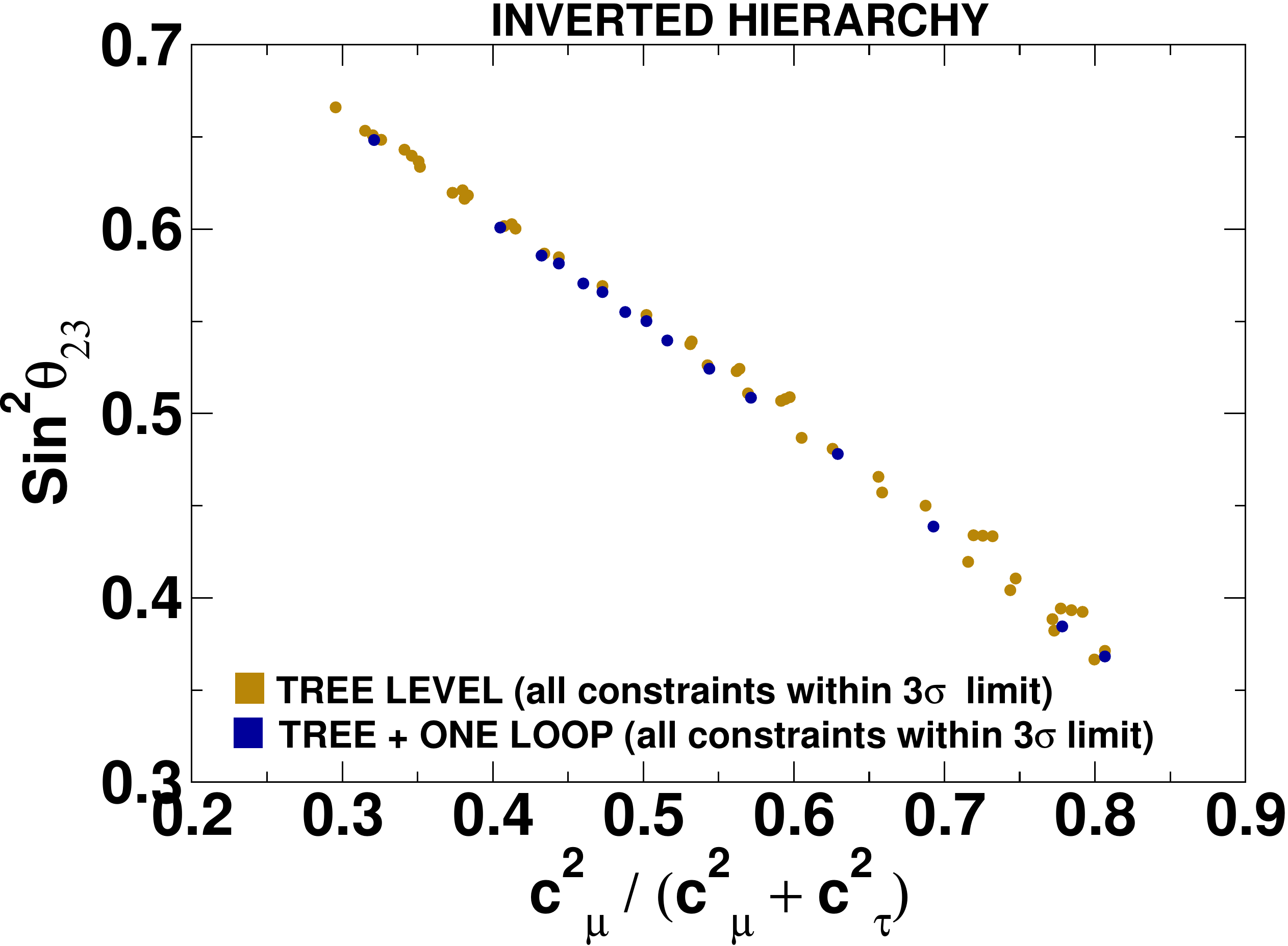}
\includegraphics[width=6.00cm]{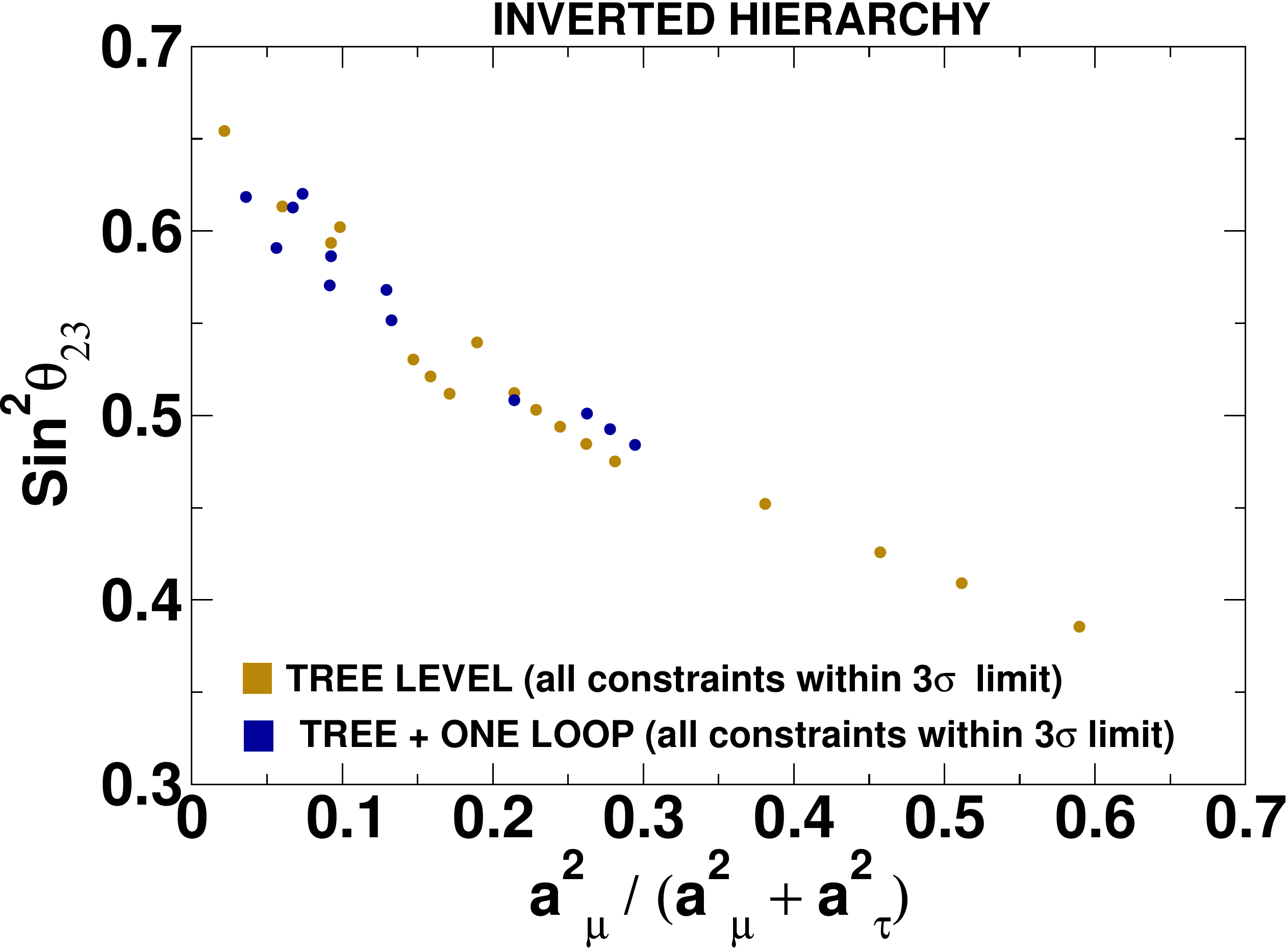}
\caption{Variation of $\sin^2\theta_{23}$ with
  $\frac{c^2_\mu}{(c^2_\mu + c^2_\tau)}$ and 
  $\frac{a^2_\mu}{(a^2_\mu + a^2_\tau)}$ 
  for inverted hierarchy of light neutrino masses. 
Parameter choices are shown in tables \ref{loop-param} 
  and \ref{loop-param-2}.}
\label{mixing-GS-OS-IH-B}
\end{figure}

We conclude the discussion on inverted hierarchy by addressing
the dependence of neutrino mixing angles with the relevant parameters. 
In figure \ref{mixing-GS-OS-IH-B} we show the variation of the neutrino 
mixing angles with the same set of parameters as chosen for the normal
hierarchical scenario. We notice that for inverted hierarchy 
the quantity $\sin^2\theta_{23}$ decreases with increasing 
$\frac{c^2_\mu}{(c^2_\mu + c^2_\tau)}$ which is just opposite to that 
of the normal hierarchy (see, figure \ref{mixing-GS-OS}). Nevertheless, the
correlation of $\sin^2\theta_{23}$ with $\frac{c^2_\mu}{(c^2_\mu + c^2_\tau)}$
is as sharp as in the case of normal hierarchy. A similar feature is obtained
for the variation with $\frac{a^2_\mu}{(a^2_\mu + a^2_\tau)}$.

On the other hand, the correlations of $\sin^2\theta_{12}$ with 
$\frac{c^2_e}{c^2_\mu}$ and $\frac{a^2_e}{a^2_\mu}$ and the 
correlations of $\sin^2\theta_{13}$ with 
$\frac{c^2_e}{(c^2_\mu + c^2_\tau)}$ and 
$\frac{a^2_e}{(a^2_\mu + a^2_\tau)}$ are not very sharp and
we do not show them here. There are allowed values of relevant parameters
where all neutrino data can be satisfied. Remember that, for 
the plots with $c_i$s, we varied all the $c_i$s simultaneously, keeping 
the values of $a_i$s fixed at the ones determined by the parameters in 
table \ref{loop-param-2}. Similarly, for the variation of $a_i$s, the quantities 
$c_i$s were kept fixed. The inclusion of one-loop corrections restrict the 
allowed values of parameter points significantly compared to the tree level 
results. 

\subsection{{\bf Q}uasi-degenerate spectra}
\label{Quasi-degenerate-spectra}

The discussion on the light neutrino mass spectrum remains incomplete
without a note on the so-called ``quasi-degenerate'' scenario. A truly 
degenerate scenario of three light neutrino masses is, however, inconsistent 
with the oscillation data. 
Hence, the quasi-degenerate scenario of light neutrino masses is defined 
in such a way that in this case all the three individual neutrino masses are 
much larger compared to the atmospheric neutrino mass scale. Mathematically,
one writes $m_1 \approx m_2 \approx m_3 \gg \sqrt{|\Delta m^2_{atm}|}$. 
Obviously, the oscillation data suggest that even in such a situation there 
must be a mild hierarchy among the degenerate neutrinos. It is important to note that
unlike the normal or inverted hierarchical scheme of light neutrino masses, in
the case of quasi-degenerate neutrinos all three neutrinos must be massive
in order to satisfy oscillation data (see table \ref{osc-para}). In the case of
normal or inverted hierarchical neutrino masses it is possible to accommodate
the three flavour neutrino data even with two massive neutrinos.

In this subsection we have shown that the huge parameter space of $\mu\nu$SSM 
always leaves us with enough room to accommodate quasi-degenerate spectrum.
For our numerical analysis, we called a set of light neutrino masses to be
quasi-degenerate if the lightest among them is greater than 0.1 eV. We choose
two sets of sample parameter points which are given below in tabular form
(values of other parameters are same as in table \ref{loop-param}). For these 
two sets of neutrino Yukawa couplings ($Y_\nu^{ii}$) and the left-handed 
sneutrino VEVs ($v^\prime_i$) we observe the following patterns of light
neutrino masses at the tree level   

\noindent (i) Quasi-degenerate-I: $m_3 \gtrsim m_2 \gtrsim m_1 \gg 
\sqrt{|\Delta m^2_{atm}|}$
\\
(ii) Quasi-degenerate-II: $m_2 \gtrsim m_1 \gtrsim m_3 \gg \sqrt{|\Delta m^2_{atm}|}.$ 

\begin{figure}[ht]
\centering
\includegraphics[width=6.00cm]{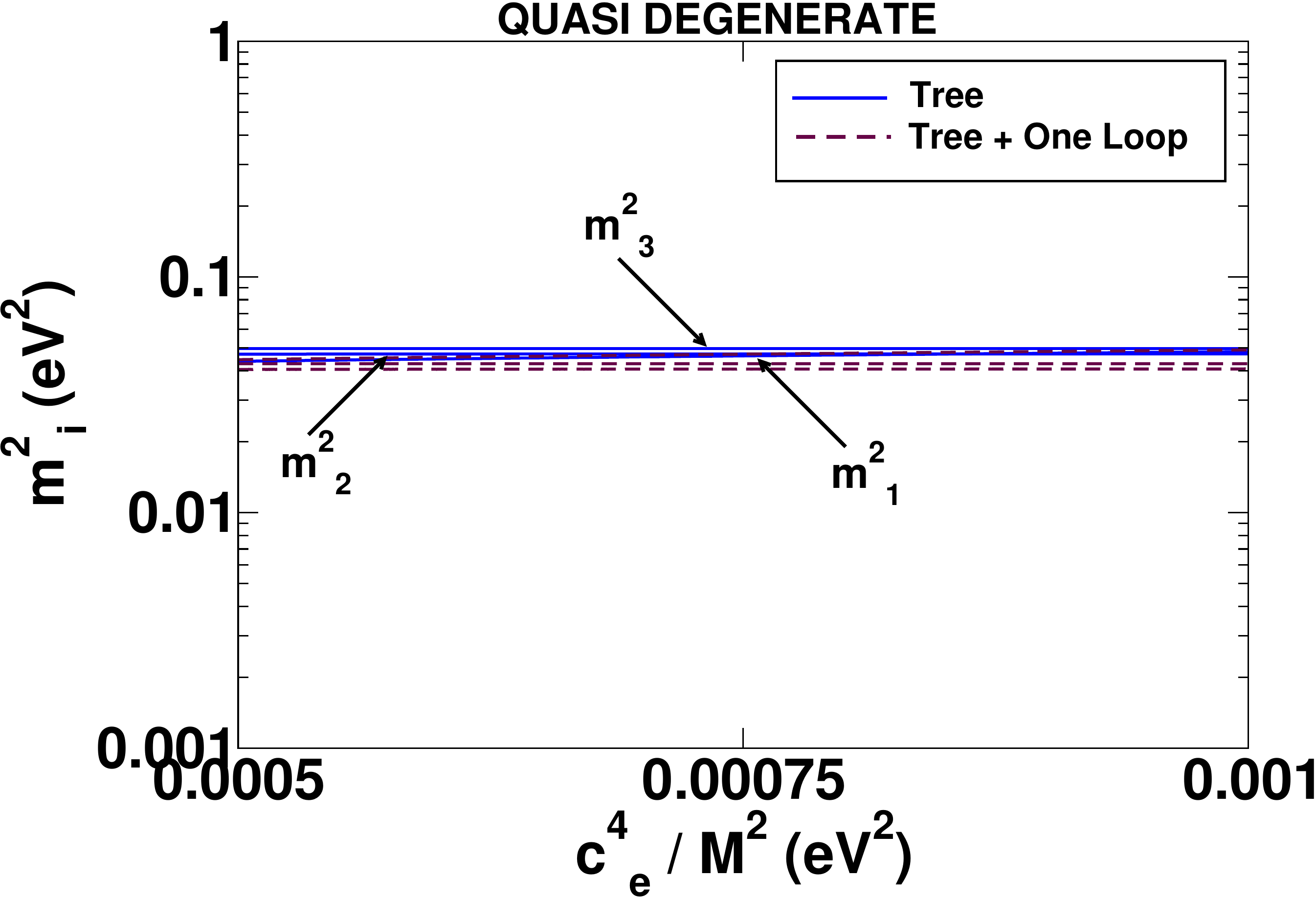}
\includegraphics[width=6.00cm]{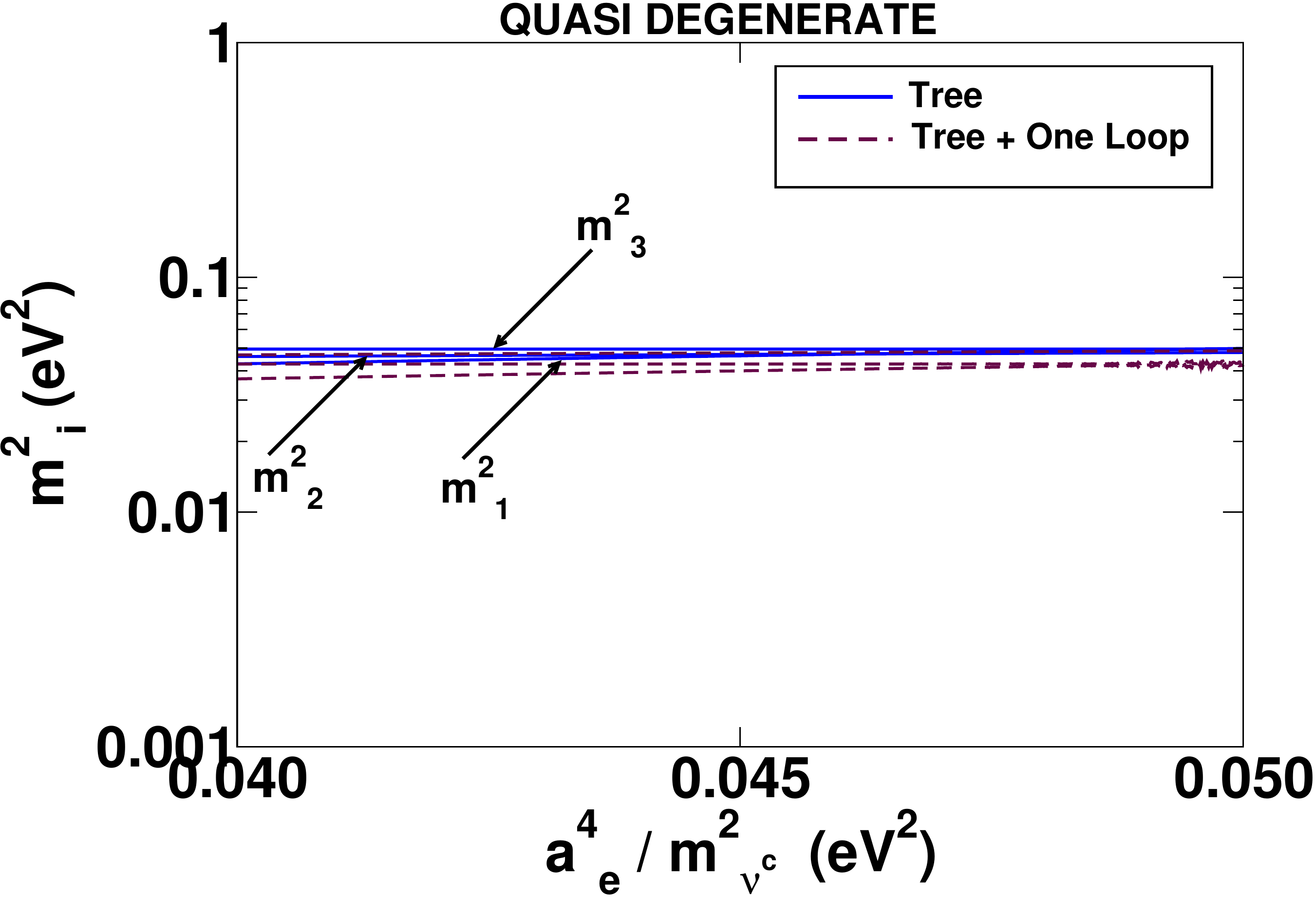}
\caption{Neutrino mass squared values ($m^2_i$) vs $\frac{c^4_e}{M^2}$ 
(left panel) and vs $\frac{a^4_e}{m^2_{\nu^c}}$ (right panel) plots for 
the {\it{quasi-degenerate}} pattern of light neutrino masses.  
Parameter choices are shown in tables \ref{loop-param} 
  and \ref{loop-param-2}.}
\label{numsqQD}
\end{figure}

\noindent For case (i), we have varied the parameters around the values in 
table \ref{loop-param-2} and identified a few extremely fine-tuned points in the 
parameter space where either the tree level or the one-loop corrected result 
is consistent with the three flavour global neutrino data. Two representative
spectrum as function of $\frac{c^4_e}{M^2}$ and $\frac{a^4_e}{m^2_{\nu^c}}$ 
are shown in figure \ref{numsqQD}. The mass spectrum for Quasi-degenerate-I case is
analogous to a normal hierarchical scenario whereas that for Quasi-degenerate-II
resembles a inverted spectrum.

As mentioned earlier, one can play with the model parameters and 
obtain a spectrum with a different ordering of masses 
termed as ``Quasi-degenerate-II" in table \ref{loop-param-2}. However, for
such an ordering of masses, we found that it was rather impossible
to find any region of parameter space where the one-loop corrected result
satisfies all the constraints on neutrino masses and mixing. 
Nevertheless, we must emphasize here that it is not a 
completely generic conclusion and for other choices of soft SUSY breaking 
and other parameters it could be possible to have a spectrum like that 
shown in ``Quasi degenerate II" with neutrino constraints satisfied 
even at the one-loop level. On the other hand, there exist regions where 
neutrino data are satisfied at the tree level with this ordering of masses. 

\section{{\bf S}ummary}\label{com-ana-neut}
So in a nutshell in $\mu\nu$SSM it is possible to account for three flavour global
neutrino data itself at the tree level even with the choice of flavour diagonal
neutrino Yukawa couplings. Besides, different hierarchical (normal, inverted, quasi-degenerate)
scheme of light neutrino mass can be accommodated by playing with the hierarchy
in Yukawa couplings. The tree level results of neutrino masses and mixing show appreciable
variation with the inclusion of the one-loop radiative corrections, depending on
the light neutrino mass hierarchy.

\vspace{2cm}

It seems so far that the $\mu\nu$SSM is extremely successful in accommodating massive
neutrinos both with tree level and one-loop combined analysis, consistent with the three
flavour global data (see table \ref{osc-para}). But how to test these neutrino physics
information in a collider experiment, which can give additional checks for the $\mu\nu$SSM
model? Fortunately for us certain ratios of the decay branching ratios of the lightest
neutralino (which is also the LSP for a large region of the parameter space) show 
nice correlations with certain light neutrino mixing angle \cite{c4Ghosh:2008yh,c4Bartl:2009an}.
These correlations could act as excellent probes to the $\mu\nu$SSM model in the ongoing era of
the colliders. These issues will be considered in details in the next chapter.


\chapter{ \sffamily{{\bf $\mu$}$\nu$SSM: decay of the LSP
 }}\label{munuSSM-LSP}

\section{{\bf A} decaying LSP}\label{LSP-dec-1}
We have learned already in section \ref{R-parity} that
the lightest supersymmetric particle (LSP) is absolutely stable so long
as $R_p$ is conserved. Besides, as argued in section
\ref{corre} that the LSP has to be charge and colour neutral \cite{c5Wolfram:1978gp,
c5Dover:1979sn,c5Ellis:1983ew} so long it preserves its stability. Consequently,
only the electrically neutral colourless sparticles remain to be the only possible 
choice for the LSP. Interestingly,
when $R_p$ is broken (figure \ref{Rpc-Rpv}, see also section \ref{R-parity}),
any sparticle (the lightest neutralino, chargino \cite{c5Feng:1999fu}, 
squark, gluino \cite{c5Raby:1997bpa,c5Baer:1998pg,c5Raby:1998xr}, sneutrino 
\cite{c5Hagelin:1984wv}, (see also ref.\cite{c5Ellis:1983ew})) 
can be the LSP. In a supersymmetric model with broken $R_p$
the LSP will decay into further lighter states namely, into the SM particles.
Apart from the neutrinos rest of these decay products are easily detectable in a collider 
experiment and thus can act as a potential probe for the underlying model. 
Since $\mu\nu$SSM is an $R_p$-violating supersymmetric model, the LSP
for this model is also unstable and can yield striking signatures at the collider 
which we aim to discuss in this chapter.
This remarkable feature 
is absent in the conventional $R_p$ conserving supersymmetric models,
where any sparticle decay ends with LSP in the final state and hence
yield large missing energy signatures. 
For example if the lightest neutralino $(\ntrl1)$ is the LSP then the following
two and three body decay modes are kinematically possible
\bea
\ntrl1 &\to& W^{\pm} \ell^{\mp},~Z^0 \nu_k,
~h^0\nu_k,\nn\\
 &\to&b \bar{b}\nu_k,~\ell^+_i \ell^-_j\nu_k,
~q_i\bar{q_i}\nu_k,~q_i\bar{q}^{\prime}_j\ell^{\mp}_k,~\nu_i\bar{\nu}_j\nu_k.
\label{2-3-body-decays}
\eea
The lightest neutralino $(\ntrl1)$ can be the LSP in a large
region of the parameter space. The three body
decay modes become dominant when mass of the LSP $(m_{\ntrl1})$ is less than that of
the $W$-boson $(m_W)$. 
The corresponding Feynman diagrams are given in appendix \ref{appenG}, section \ref{LN-decay} 
(figures \ref{LSP-decay2}, \ref{LSP-decay3}). 
It is also interesting to note that apart
from these tree level two and three body decays the LSP can also decay into a neutrino
and a photon radiatively \cite{c5Hall:1983id,c5Dawson:1985vr,c5Hempfling:1997je,
c5Mukhopadhyaya:1999gy}.

One more important aspect in the decays of the lightest supersymmetric particle
through $R_p$-violating channel is the appearance of the displaced vertices 
\cite{c5Mukhopadhyaya:1998xj,c5Chun:1998ub,c5Choi:1999tq,c5Porod:2000hv,c5DeCampos:2010yu}.
The displaced vertices appear to be macroscopic $(\sim \rm{a~few~mm}$ or larger)
due to the smallness of the associated $R_p$-violating couplings. 
A displaced vertex is defined as the distance traversed by a neutral particle
between the primary and the secondary interaction points. 
The displaced vertices are extremely useful to remove undesired
backgrounds in case of a collider analysis. The length of the displaced vertices
also vary with the nature of the lightest neutralino or the LSP. 
Thus, before proceeding further it is important to discuss about the various LSP 
scenario in $\mu\nu$SSM.
We note in passing that in this chapter we concentrate on the two-body decays
only and in the next chapter we will discuss about the three body decays.

\section{{\bf D}ifferent LSP scenarios in $\mu\nu$SSM}\label{lsp-nat}

In the $\mu\nu$SSM the neutralino sector is highly enriched compared to that of the MSSM 
due to $R_p$-violating mixing of the MSSM neutralinos with the three generations of left-handed and 
right-handed neutrinos. So mathematically in $\mu\nu$SSM with gaugino mass unification
at the GUT scale (that is at the electroweak scale $M_2=2 M_1$), possible LSP natures are described by

\vspace{0.1cm}
\noindent
1.~$\chi^0_1 \approx \bN_{11} \wt B^0,~~|\bN_{11}|^2 \sim 1 \Rrightarrow$ bino like $\ntrl1$.

\vspace{0.1cm}
\noindent
2.~$\chi^0_1 \approx \bN_{13} \wt H^0_d + \bN_{14} \wt H^0_u,
~~|\bN_{13}|^2 + |\bN_{14}|^2 \sim 1 \Rrightarrow$ higgsino like $\ntrl1$.

\vspace{0.1cm}
\noindent
3.~$\chi^0_1 \approx \sum \bN_{i,\al+4} \nu^c_\al,
~~|\bN_{15}|^2 + |\bN_{16}|^2 + |\bN_{17}|^2 \sim 1 
\Rrightarrow$ right-handed neutrino $(\nu^c)$ like $\ntrl1$.

In terms of the model ingredients the LSP nature in $\mu\nu$SSM
depends on the relative dominance of three parameters, (1) the $U(1)$ gaugino
soft mass $M_1$ (see eqn.(\ref{munuSSM-soft})), (2) the higgsino mass parameter or
the $\mu$-term ($=3\lam v^c$) (see eqns.(\ref{Abbrevations}), (\ref{assumption1})) 
and (3) the right-chiral neutrino Majorana mass term, $m_{\nu^c}$ ($=2\kp v^c$) 
(using eqn.(\ref{assumption1}), see eqn.(\ref{specifications-2})) \cite{c5Ghosh:2008yh,
c5Bartl:2009an,c5Bandyopadhyay:2010cu}. Thus we can write

\vspace{0.1cm}
\noindent
I.~$\mu$,~$m_{\nu^c}~>~M_1$ $\Longrightarrow$ LSP bino (gaugino) like.

\vspace{0.1cm}
\noindent
II.~$M_1$,~$m_{\nu^c}~>~\mu$ $\Longrightarrow$ LSP higgsino like.

\vspace{0.1cm}
\noindent
III.~$M_1$,~$\mu~>~m_{\nu^c}$ $\Longrightarrow$ LSP right-handed neutrino 
like. Since right-handed neutrinos are singlet under the SM gauge group,
a right-handed neutrino like LSP is often called a ``{\it{singlino}}'' LSP.

It is important to mention that the right sneutrinos $(\wt \nu^c)$ are also eligible
candidate for the LSP in $\mu\nu$SSM \cite{c5Ghosh:2008yh,c5Chang:2009dh}. Also as a continuation
of the discussion of the last section, the length of the displaced vertices
can vary from a few mm to a few cm for a bino like LSP to a higgsino like LSP 
\cite{c5Ghosh:2008yh,c5Bartl:2009an}.
On the other hand, for a singlino LSP the length of the displaced vertices 
can be as large as a few meters \cite{c5Bartl:2009an,c5Bandyopadhyay:2010cu}. None of
these are unexpected since a bino like LSP, being a gaugino, has gauge interactions
and the gauge couplings are $\sim$ $\cal{O}$ $(1)$ couplings whereas a higgsino
like LSP involves smaller Yukawa couplings which is responsible for a smaller decay width 
and consequently a larger ($\sim$ a few cm) displaced vertices. A singlino LSP on the other hand 
is mostly a gauge singlet fermion by nature and thus couples to other particles via very small 
$R_p$-violating couplings, which finally yield a large displaced vertex.

\section{{\bf D}ecays of the lightest neutralino in $\mu\nu$SSM}\label{lsp-decay-calc}

In this section we aim to calculate a few tree level two-body decays of
the lightest neutralino $\ntrl1$ in $\mu\nu$SSM model \cite{c5Ghosh:2008yh}. As stated earlier
we denote the lightest neutralino as $\ntrl1$ when the seven neutralino
masses (see eqn.(\ref{neutralino_mass_eigenstate_matrixform}))
are arranged in the increasing order of magnitude ($\ntrl1$ being the 
lightest and $\ntrl7$ being the heaviest).
However, for this chapter from now on, we follow the convention of ref. \cite{c5Ghosh:2008yh}
where the eigenvalues are arranged in reverse order so that $\ntrl7$ denotes
the lightest neutralino. The lightest neutralino considered here is
either the LSP or the next-to LSP (NLSP). The lightest neutralino
mass is set to be more than $m_W$ such that two-body decays dominate. 
Two-body and three-body decays of the LSP in $\mu\nu$SSM has been
discussed in a recent ref. \cite{c5Bartl:2009an} with one generation
of right handed neutrino superfield.
Three-body decays of a singlino like lightest neutralino (which is also the 
LSP) for $\mu\nu$SSM also has been addressed in ref. \cite{c5Bandyopadhyay:2010cu}.

In this section we mainly concentrate on the two-body decays like
\bea\label{general_2body-decay}
&&\ntrl7 \longrightarrow W^\pm + \ell^\mp_k\\
&&\ntrl7 \longrightarrow Z + \nu_k,\nn,
\eea
where $k=1,2,3\equiv e,\mu,\tau$. The required Feynman rules 
are given in appendix {\ref{appenD}}. Let us also remark that the lightest neutralino
can also decay to $h^0 + \nu_k$, if it is kinematically allowed, where $h^0$ is
the MSSM-like lightest Higgs boson (this is true if the amount of admixture
of the MSSM Higgses with the right-handed sneutrinos are very small). 
However, for our illustration purposes we
have considered the mass of the lightest neutralino in such a way that this 
decay is either kinematically forbidden or very much suppressed (assuming a 
lower bound on the mass of $h$ to be 114 GeV). Even if this decay branching 
ratio is slightly larger, it is usually smaller than the branching ratios in
the ($\ell^\pm_i + W^\mp$) channel. 
Hence, this will not affect 
our conclusions regarding the ratios of branching ratios in the charged lepton channel 
($\ell_i + W$), to be discussed later. The lightest neutralino decay 
${\wt \chi}^0_7 \rightarrow \nu + {\wt \nu}^c$, where ${\wt \nu}^c$ 
is the scalar partner of the gauge singlet neutrino $\nu^c$, is always very 
suppressed. 
We will discuss more on this when we consider a $\nu^c$ 
dominated lightest neutralino in subsection \ref{rhnu-LN}.

Consider the following decay process
\beq\label{general_decay}
\wt{\chi}_i \longrightarrow \wt{\chi}_j + V,
\eeq
where $\wt{\chi}_{i(j)}$ is either a neutralino\footnote{Remember that the neutrinos are
also a part of the extended neutralino matrix (eqn.(\ref{neutralino_basis})).}
or chargino, with mass $m_{i(j)}$ and $V$ is the gauge boson which is either $W^{\pm}$ 
or $Z$, with mass $m_v$. The masses $m_i$ and $m_j$ are positive.

The decay width for this process in eqn.(\ref{general_decay}) is given 
by \cite{c5Gunion:1987yh,c5Franke:1995tf,c5Franke:1995tc}
\beq\label{general_decay_width_formula}
\Gamma \left(\wt{\chi}_i \longrightarrow \wt{\chi}_j + V\right) = 
\frac{g^2 \mathcal{K}^{1/2}}{32 ~\pi m^3_i m^2_W} \times \left\{\left( G^2_L 
+ G^2_R \right)\mathcal{F} - G^*_L G_R 
~\mathcal{G} \right \},
\eeq
where $\mathcal{F}$, $\mathcal{G}$ are functions of $m_i, m_j, m_v$ and 
given by
\bea\label{general_decay_width_formula_specifications}
& & \mathcal{F}(m_i, m_j, m_v) =  \mathcal{K} + 3 ~m^2_v \left( m^2_i + m^2_j 
- m^2_v \right), \nonumber \\ 
& & \mathcal{G}(m_i, m_j, m_v) =  12 ~\epsilon_i \epsilon_j m_i m_j m^2_v,
\eea
with $\epsilon_i(j)$ carrying the actual signs $(\pm 1)$ of the 
neutralino masses \cite{c5Gunion:1984yn}. The chargino masses must be positive.
The kinematical factor $\mathcal{K}$ is given by
\beq\label{kinematical_factor}
\mathcal{K}(m^2_i, m^2_j, m^2_v) = \left( m^2_i + m^2_j - m^2_v \right)^2 
- 4 ~m^2_i m^2_j.
\eeq

In order to derive eqn.(\ref{general_decay_width_formula}), we have used 
the relation $m^2_W = m^2_Z \cos^2\theta_W$ and since $v'_i << v_1, v_2$, 
some of the MSSM relations still hold good.
The factors $G_L$, $G_R$ are given here for some possible decay modes
\bea\label{some_possible_decay_modes}
& &\text{For} \quad \wt{\chi}^0_i \longrightarrow \wt{\chi}^0_j Z 
\nonumber \\
& & G_L = O^{\prime \prime L}_{ji}, \quad  G_R = O^{\prime \prime R}_{ji},
\nonumber \\
\quad \
& &\text{For} \quad \wt{\chi}^0_i \longrightarrow \wt{\chi}^+_j {W^-} 
\nonumber \\
& & G_L = O^{L}_{ij}, \quad  G_R = O^{R}_{ij},
\eea
where $O^{\prime \prime L(R)}_{ji}$ and $O^{L(R)}_{ij}$ are given by 
(using eqns.(\ref{nnz-L-R-couplings}),(\ref{ncw-L-R-couplings}) without
the sign factors $\epsilon_i,\eta_j$)

\bea\label{O-Op-coups}
O^{{\prime\prime}^L}_{ij} &=& - \frac{1}{2} \bN_{i3} \bN^*_{j3} + \frac{1}{2} 
\bN_{i4} \bN^*_{j4} - \frac{1}{2} \bN_{i,k+7} \bN^*_{j,k+7}, \nonumber \\
O^{{\prime\prime}^R}_{ij} &=& - {O^{{\prime\prime}^L}_{ij}}^*, k = 1,2,3, \nonumber \\
O^L_{ij} &=& \bN_{i2} \bV^*_{j1} - \frac{1}{\sqrt{2}} \bN_{i4} \bV^*_{j2},
\nonumber \\
O^R_{ij} &=& \bN^*_{i2} \bU_{j1} + \frac{1}{\sqrt{2}} \bN^*_{i3} \bU_{j2} + 
\frac{1}{\sqrt{2}} \bN^*_{i,k+7} \bU_{j,k+2}.
\eea

Now consider the decays shown in eqn.(\ref{general_2body-decay}). At this stage let us 
discuss our notation and convention for calculating these decays \cite{c5Ghosh:2008yh}. 
The neutralino mass matrix is  a 10$\times$10 mass matrix which includes three generations of 
the left-handed as well as the gauge-singlet neutrinos (eqns.(\ref{neutralino_7x7}), 
(\ref{neutralino_3x7})). If the mass eigenvalues of this 
matrix are arranged in the descending order then the three lightest eigenvalues of 
this 10$\times$10 neutralino mass matrix would correspond to the three light 
neutrinos. Out of the remaining seven heavy eigenvalues, the lightest one 
is denoted as the lightest neutralino. Thus, as argued earlier in our notation 
$\wt{\chi}^0_7$ is the lightest neutralino (LN) and 
$\wt{\chi}^0_{j+7}, \text{where}$ $j=1,2,3$ correspond to the three light 
neutrinos \cite{c5Ghosh:2008yh}. Similarly, for the chargino masses, $\wt{\chi}^{\pm}_{l+2}$ 
($l=1,2,3$) corresponds to the charged leptons $e, \mu, \tau$. Immediately,
with this choice, we can write down different natures of the lightest neutralino as

\vspace{0.1cm}
\noindent
A.~$\chi^0_7 \approx \bN_{71} \wt B^0,~~|\bN_{71}|^2 \sim 1 \Rrightarrow$ bino like LN.

\vspace{0.1cm}
\noindent
B.~$\chi^0_7 \approx \bN_{73} \wt H^0_d + \bN_{74} \wt H^0_u,
~~|\bN_{73}|^2 + |\bN_{74}|^2 \sim 1 \Rrightarrow$ higgsino like LN.

\vspace{0.1cm}
\noindent
C.~$\chi^0_7 \approx \sum \bN_{7,\al+4} \nu^c_\al,
~~|\bN_{75}|^2 + |\bN_{76}|^2 + |\bN_{77}|^2 \sim 1 
\Rrightarrow$ $\nu^c$ like LN.

\vspace{0.1cm}
So for $\wt{\chi}^0_{LN} \rightarrow Z + \nu_k$, which is also equivalent
to $\wt{\chi}^0_7 \rightarrow Z + \wt{\chi}^0_{j+7}$ ($j = 1,~2,~3 $), 
one gets from eqn.(\ref{some_possible_decay_modes}) and
eqn.(\ref{O-Op-coups})
\bea\label{G_L-&-G_R-for-first-of-two-selected-decays}
& & G_L = - \frac{1}{2} \bN_{j+7,3} \bN^*_{73} + \frac{1}{2} \bN_{j+7,4} \bN^*_{74} 
- \frac{1}{2} \bN_{j+7,k+7} \bN^*_{7,k+7}, \nonumber \\
& & G_R = -G^*_L,
\eea
where $~j,k = 1,~2,~3$ and this in turn modifies 
eqn.(\ref{general_decay_width_formula}) as 
\beq\label{decay-width-for-process-1}
\Gamma \left(\wt{\chi}^0_7 \rightarrow Z + ~\wt{\chi}^0_{j+7}
\right) = \frac{g^2 \mathcal{K}^{1/2}}{32 ~\pi  m^3_{{\wt \chi}^0_7} 
m^2_W} \times \left\{ 2 ~G^2_L \mathcal{F} + G^{*^2}_L ~\mathcal{G} \right \},
\eeq
with $m_i = m_{{\wt \chi}^0_7}$, $m_j = m_{\nu_j} \approx 0$ 
(eqn.(\ref{neutralino_mass_eigenstate_matrixform})) and $m_v = m_Z$.

Let us now consider the other decay which is $\wt{\chi}^0_{LN} \rightarrow 
W^{\pm} + \ell^{\mp}$ or equivalently $\wt{\chi}^0_7 \rightarrow 
W^{\pm} + \wt{\chi}^{\mp}_j (j=3,4,5)$.

For the process $\wt{\chi}^0_7 \rightarrow W^{-} + \wt{\chi}^{+}_j$
\bea\label{decay-width-for-process-2-a}
\Gamma \left(\wt{\chi}^0_7 \rightarrow W^- 
+ ~\wt{\chi}^+_j\right) &=& \frac{g^2 \mathcal{K}^{1/2}}{32 ~\pi 
m^3_{{\wt \chi}^0_7} m^2_W} \times \left\{ \left(G^2_L + G^2_R \right) \mathcal{F} -G^*_L G_R 
~\mathcal{G} \right \},\nonumber \\
G_L &=& \bN_{72} \bV^*_{j1} - \frac{1}{\sqrt{2}} \bN_{74} \bV^*_{j2},\nonumber \\
G_R &=& \bN^*_{72} \bU_{j1} + \frac{1}{\sqrt{2}} \bN^*_{73} \bU_{j2} 
+ \frac{1}{\sqrt{2}} \bN^*_{7,k+7} \bU_{j,k+2},\nonumber \\
(k &=& 1,~2,~3),
\eea
where eqn.(\ref{some_possible_decay_modes}) and
eqn.(\ref{O-Op-coups}) 
have been used. The process $\wt{\chi}^0_7 \longrightarrow W^{+} 
+ \wt{\chi}^{-}_j$ is obtained by charge conjugation of the process
in eqn.(\ref{decay-width-for-process-2-a}). 

Note that the neutralino mixing
matrix $\bN$ contains the expansion parameter $\xi$ (eqn.(\ref{expansion-parameter})) 
which as shown in appendix \ref{appenC} can be expressed as a function
of the quantities $a_i,b_i,c_i$ (eqn.(\ref{specifications})). On the other
hand as shown in eqns.(\ref{decay-width-for-process-1}), (\ref{decay-width-for-process-2-a})
the decay widths (for $\ntrl7 \to Z+ \nu_j$ and $\ntrl7 \to W^\pm+ \ell^\mp_j$)
contain quadratic power of $\bN$, that is, these decay widths are quadratic in
$\xi$ or even more precisely quadratic in $a_i,b_i,c_i$. This information will
be explored further in the next section.

\section{{\bf L}ight neutrino mixing and the neutralino decay}\label{correlation}

In $\mu\nu$SSM, the light neutrino mixing angles are expressible
in terms of the parameters $a_i,b_i,c_i$ (see eqn.(\ref{specifications})).
These relations were also verified numerically, as shown in figures 
\ref{neut2313-mixing-tree}, \ref{neut12-mixing-tree}, \ref{neut23a-mixing-tree}.
Now it has been already argued in the last section that the two-body
decays of the lightest neutralino are also quadratic in $a_i,b_i,c_i$ parameters. Combining 
these two pictures we found that in $\mu\nu$SSM the light neutrino mixing angles are
correlated with the lightest neutralino (or LSP) decays, to be more precise with the ratios of the 
decay branching ratio $(Br)$ \cite{c5Ghosh:2008yh}.

These correlations are well studied in the context of the $R_p$-violating supersymmetric
model of light neutrino mass generation \cite{c5Mukhopadhyaya:1998xj,c5Chun:1998ub,c5Choi:1999tq,
c5Porod:2000hv,c5DeCampos:2010yu}. 
Nevertheless, one should note certain differences in these two cases. In $\mu\nu$SSM 
lepton number is broken explicitly in the superpotential by terms which are trilinear 
as well as linear in singlet neutrino superfields. In addition to that there are 
lepton number conserving terms involving the singlet neutrino superfields with dimensionless neutrino 
Yukawa couplings. After the electroweak symmetry breaking these terms can generate the effective bilinear 
R-parity violating terms as well as the $\Delta L$ =2 Majorana mass terms for the singlet neutrinos in the 
superpotential. In general, there are corresponding soft supersymmetry breaking terms in the scalar potential.
Thus the parameter space of this model is much larger compared to the bilinear $R_p$ violating model. 
Hence, in general, one would not expect a very tight correlation between the neutrino mixing angles and the
ratios of decay branching ratios of the LSP. However, under certain simplifying assumptions  
\cite{c5Ghosh:2008yh}, one can reduce the number of free parameters and in those cases it is possible that the above 
correlations reappear. This issue has been studied in great detail for the two body $\ell^\pm-W^\mp$ 
final states in ref. \cite{c5Ghosh:2008yh} and for all possible two and three body final states 
in ref. \cite{c5Bartl:2009an}. Let us note in passing that such a nice correlation is lost in 
the general scenario of bilinear-plus-trilinear 
R-parity violation \cite{c5Choi:1999tq}. 

Another important difference between $\mu\nu$SSM and the bilinear R-parity violating model in the context of 
the decay of the LSP (assumed to be the lightest neutralino in this case) is that in $\mu\nu$SSM the lightest 
neutralino can have a significant singlet neutrino ($\nu^c$) contribution. In this case, the correlation between 
neutrino mixing angles and decay branching ratios of the LSP is different \cite{c5Ghosh:2008yh,c5Bartl:2009an} compared 
to the cases when the dominant component of the LSP is either a bino, or a higgsino or a Wino. This gives us a
possibility of distinguishing between different R-parity violating models through the observation of the
decay branching ratios of the LSP in collider experiments \cite{c5Ghosh:2008yh,c5Bartl:2009an}. In addition, the decay of 
the lightest neutralino will show displaced vertices in collider experiments and when the lightest neutralino
is predominantly a singlet neutrino, the decay length can be of the order of several meters for a lightest 
neutralino mass in the neighbourhood of 50 GeV \cite{c5Bartl:2009an}. This is very different from the bilinear R-parity 
violating model where for a Bino LSP of similar mass the decay length is less than or of the order of a meter 
or so \cite{c5Porod:2000hv}. 

In references \cite{c5Mukhopadhyaya:1998xj,c5Porod:2000hv,c5Chun:2002rh}
this correlation was studied for a bino like neutralino LSP. However, the correlations
appear for other natures of the lightest supersymmetric particle as well
\cite{c5Hirsch:2002ys,c5Hirsch:2003fe,c5Aristizabal-Sierra:2004cy}.
These inter-relations reflects
the predictive power of a model where the light neutrino mass generation as well as 
the lightest neutralino/LSP decays are governed by a common set of small number of parameters. 
These correlations are also addressed in a recent review \cite{c5Nath:2010zj}.
So in conclusion, with the help of these nice correlations neutrino mixing angles 
can be indirectly measured in colliders by comparing the branching ratios of
the lightest neutralino or the LSP decay modes.

We observe that the correlations between the lightest neutralino decays and neutrino mixing angles
depend on the nature of the lightest neutralino as well as on the 
mass hierarchies of the neutrinos, i.e.  whether we have a normal 
hierarchical pattern of neutrino masses or an inverted one \cite{c5Ghosh:2008yh}. 
In this section we look into these possibilities in details with three 
different natures of the lightest neutralino. We 
consider that the lightest neutralino to be either (1) bino dominated or 
(2) higgsino dominated or (3) right-handed neutrino dominated. For each of these 
cases we consider both the normal and the inverted hierarchical pattern 
of neutrino masses. In the case of a bino or a higgsino like lightest neutralino,
they are also the LSP but for a right-handed neutrino dominated lightest
neutralino it is the NLSP with right handed sneutrino as the LSP \cite{c5Ghosh:2008yh}.  
The possibility for a right-handed neutrino or singlino like lightest neutralino
LSP has also been addressed in references \cite{c5Bartl:2009an,c5Bandyopadhyay:2010cu}.
We show that for the different natures of the lightest neutralino, the 
ratio of branching ratios of certain decays of the lightest neutralino 
correlates with certain neutrino mixing angle. In some cases
the correlation is with the atmospheric angle $(\theta_{23})$ and the reactor angle $(\theta_{13})$ 
and in other cases the ratio of the branching ratios correlates with the solar mixing
angle $(\theta_{12})$. Nevertheless, there also exists scenarios with no correlations at all. Let us 
now study these possibilities case by case \cite{c5Ghosh:2008yh} in three subsequent subsections. 
As already mentioned, that the interesting difference between this study and similar studies with $R_p$ 
violating scenario \cite{c5Mukhopadhyaya:1998xj,c5Chun:1998ub,c5Choi:1999tq,c5Romao:1999up,
c5Porod:2000hv} in the MSSM is the presence of a gauge singlet neutrino dominated lightest 
neutralino. We will see later that in this case the results can be very different from 
a bino or higgsino dominated lightest neutralino. The lightest neutralino
decays in neutrino mass models with spontaneous R-parity violation have been
studied in ref.\cite{c5Hirsch:2008ur}. Our parameter choices for the next three
subsections are consistent with the constraints of the scalar sector 
(section \ref{munuSSM-scalar}).

\subsection{{\bf B}ino dominated lightest neutralino}\label{bino-LN}
\begin{figure}[ht]
\centering
\vspace*{0.4cm}
\includegraphics[height=4.00cm]{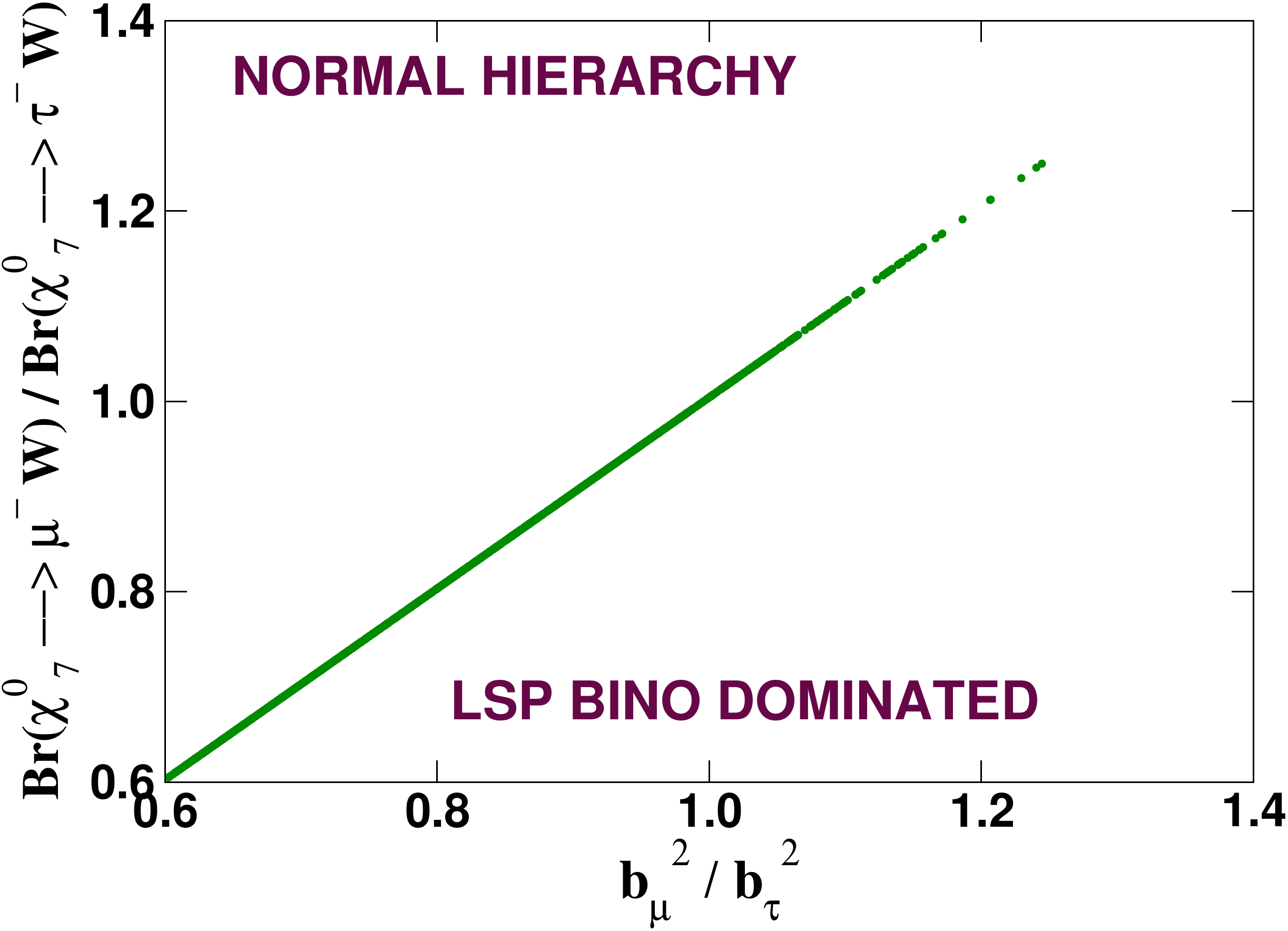}
\includegraphics[height=4.00cm]{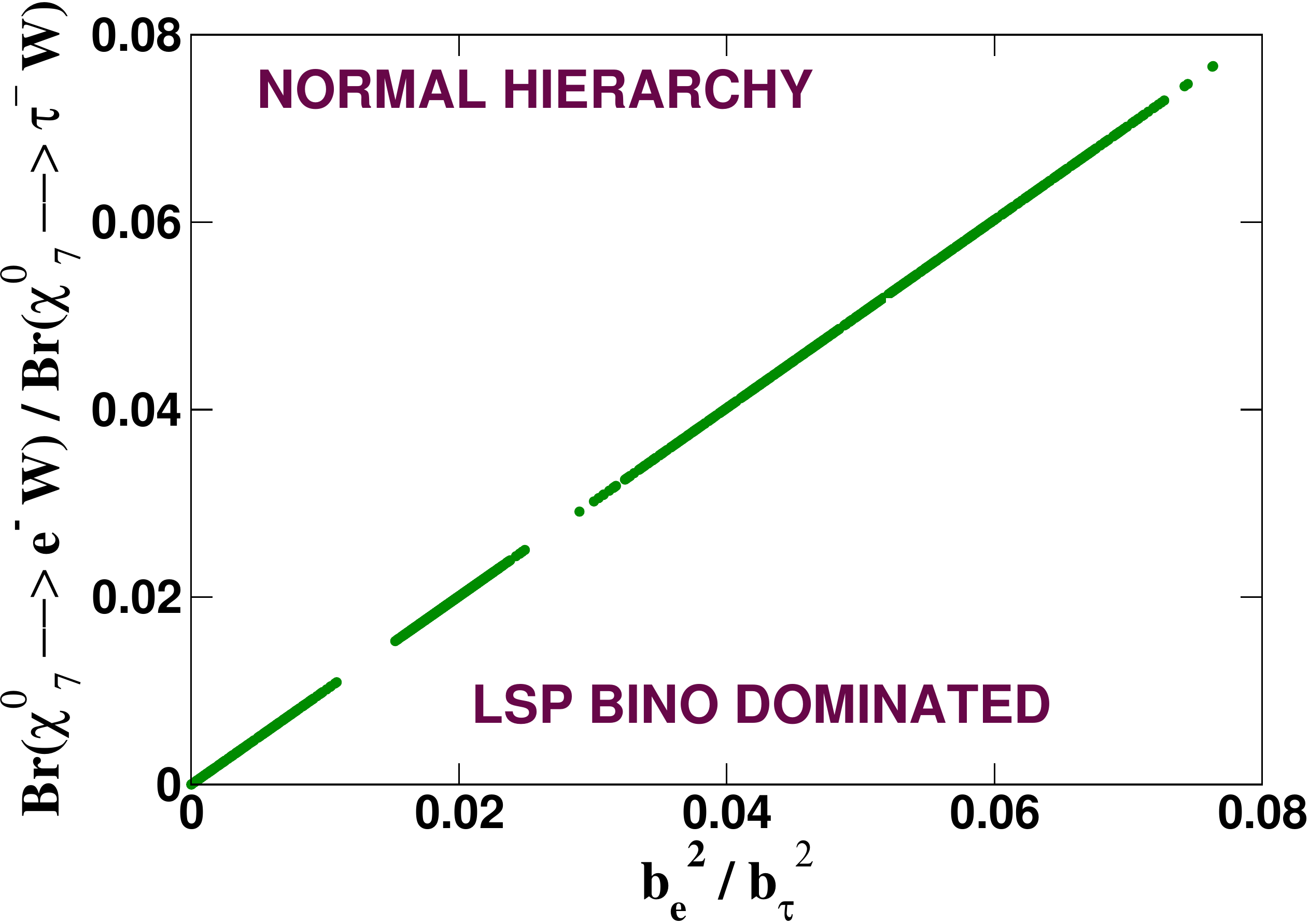}
\includegraphics[height=4.00cm]{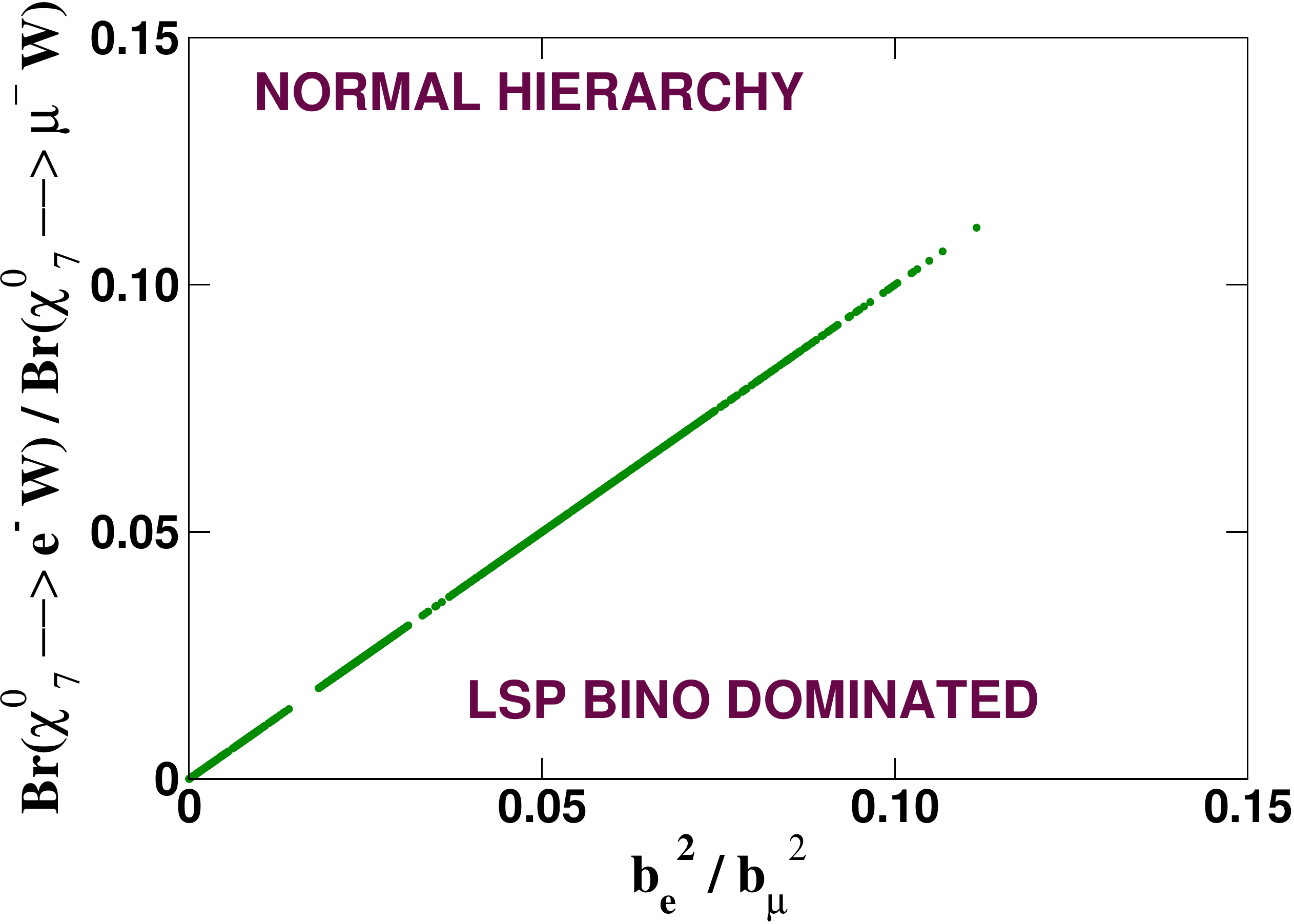}
\caption{Ratio $\frac{Br(\chi^0_7 \longrightarrow \ell_i~W)}{Br(\chi^0_7
\longrightarrow \ell_j~W)}$ versus $\frac{b^2_i}{b^2_j}$ plot for a bino like
lightest neutralino (the LSP) with bino component, $|N_{71}|^2>$~0.92, where 
$i,j,k~=~e,\mu,\tau$.  Neutrino mass pattern is taken to be normal 
hierarchical. Choice of parameters are $M_1=110$ GeV, $\lambda=0.13, 
\kappa=0.65, m_{\tilde{\nu}^c}=300~{\rm GeV} ~{\rm and} ~m_{\tilde{L}}=400 
~{\rm GeV}$. Mass of the LSP is $106.9$ GeV. 
The value of the $\mu$ parameter comes out to be $-228.9$ GeV.}
\label{nor-bino-bmbebt}
\end{figure}

According to our choice, at the EW scale the ratio of the $U(1)$ and $SU(2)$ 
gaugino masses are $M_1:M_2 = 1:2$. If in addition, $M_1<\mu$ and the value of 
$\kappa$ is large (so that the effective gauge singlet neutrino mass  
$2 \kappa v^c$ is large), the lightest neutralino is essentially bino 
dominated and it is the LSP. First we consider the case when the composition 
of the lightest neutralino is such that, the bino-component $|N_{71}|^2>$ 0.92 
and neutrino masses follow the normal hierarchical pattern. We have observed 
that for the bino dominated case, the lightest neutralino 
(${\tilde \chi}^0_7$) couplings to $\ell^{\pm}$--$W^{\mp}$ pair 
(where $\ell = e, \mu$ or $\tau$) depend on the 
quantities $b_i$ along with a factor which is independent of various lepton 
generations. Naturally, we would expect that the ratios of various decay 
branching ratios such as BR(${\tilde \chi}^0_7 \rightarrow e + W$), 
BR(${\tilde \chi}^0_7 \rightarrow \mu + W$), and  
BR(${\tilde \chi}^0_7 \rightarrow \tau + W$) show nice correlations with the
quantities $b^2_i/b^2_j$ with $i,j$ being $e, \mu$ or $\tau$. This feature is
evident from figure \ref{nor-bino-bmbebt}. Here we have scanned the parameter 
space of the three neutrino Yukawa couplings with random values for a 
particular choice of the couplings $\lambda$, $\kappa$ and the associated soft 
SUSY breaking trilinear parameters, as well as other MSSM parameters. The 
trilinear soft parameters $A_\nu$ corresponding to $Y_\nu$s also vary  
randomly in a certain range. In addition we have imposed the condition that 
the lightest neutralino (which is the LSP) is bino dominated and neutrino mass 
pattern is normal hierarchical.  

\begin{figure}[ht]
\centering
\vspace*{0.4cm}
\includegraphics[height=4.00cm]{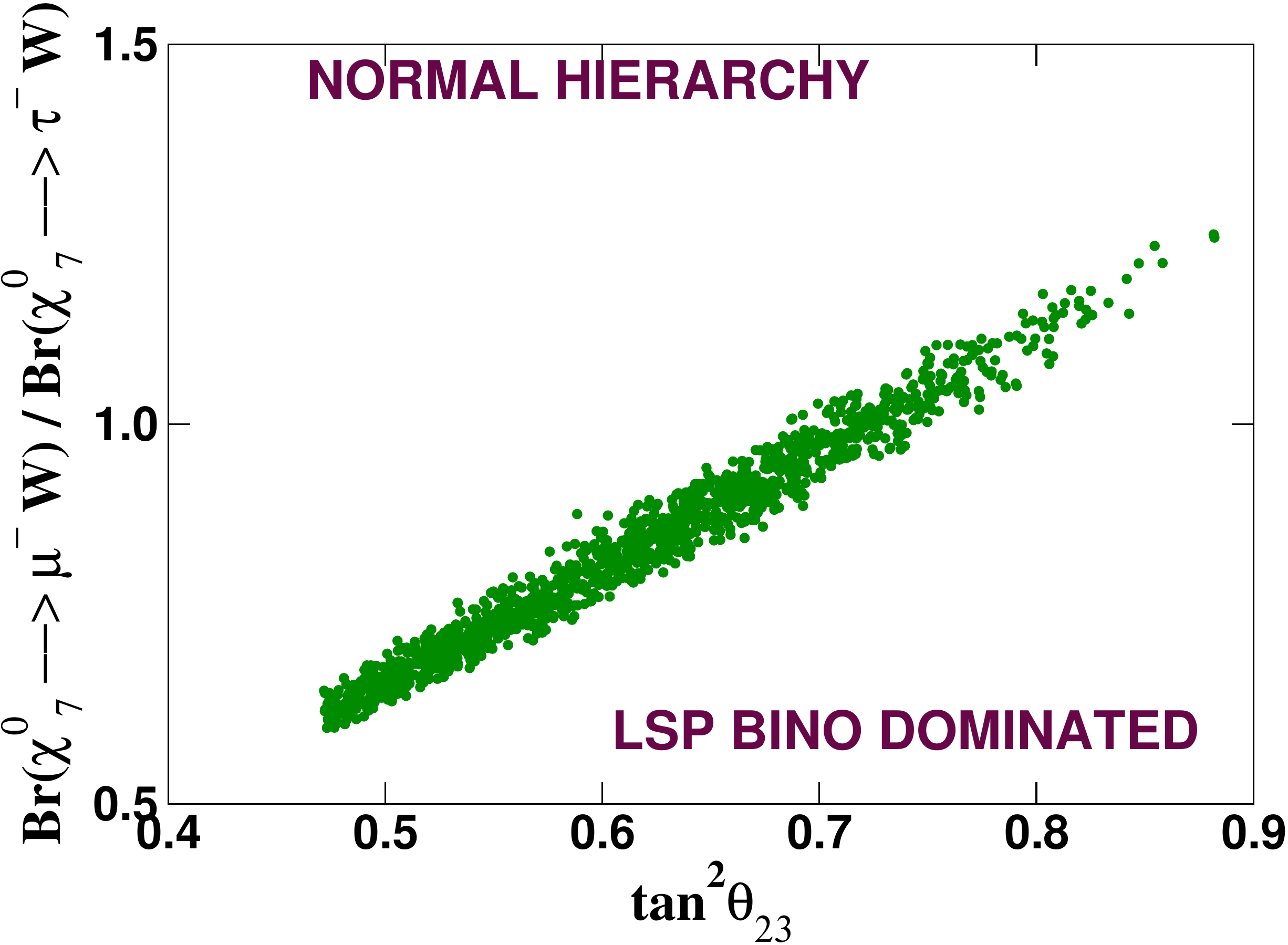}
\includegraphics[height=4.00cm]{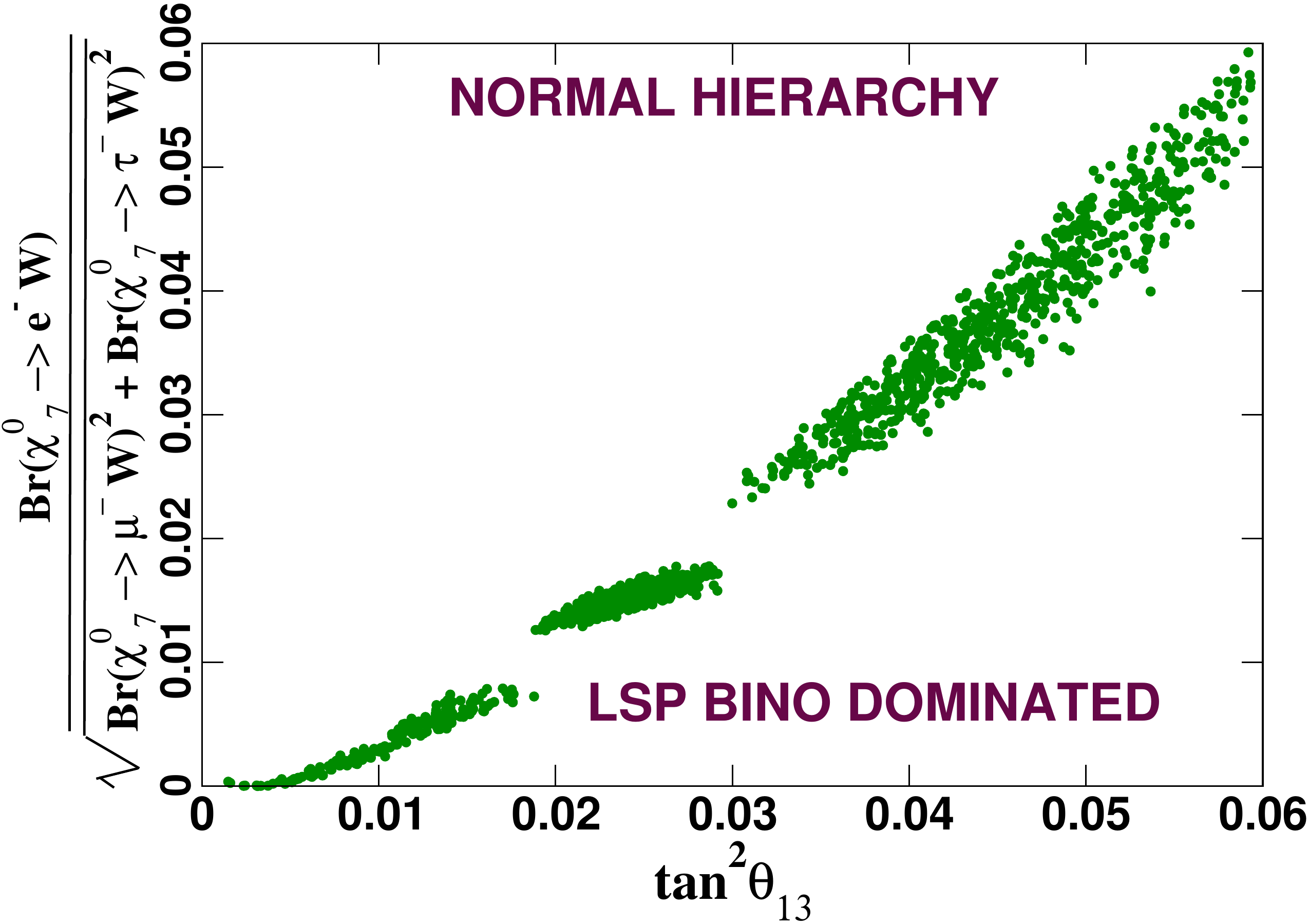}
\caption{Ratio $\frac{Br(\chi^0_7 \longrightarrow \mu~W)}{Br(\chi^0_7
\longrightarrow \tau~W)}$ versus $\tan^2\theta_{23}$~(left),
$\frac{Br(\chi^0_7 \longrightarrow e~W)}{\sqrt{Br(\chi^0_7 \longrightarrow
\mu~W)^2+Br(\chi^0_7 \longrightarrow \tau~W)^2}}$ with $\tan^2\theta_{13}$~(right) 
plot for a bino dominated lightest neutralino (the LSP) with bino component, 
$|N_{71}|^2>$~$0.92$. Neutrino mass pattern is normal hierarchical. Choice of parameters 
are same as that of figure \ref{nor-bino-bmbebt}.}
\label{nor-bino-LSP}
\end{figure}

We have checked that the correlations between the ratios of the lightest 
neutralino decay branching ratios and $b^2_i/b^2_j$ is more prominent 
with increasing bino component of the lightest neutralino. Note that when 
$(b_i/b_j)^2\rightarrow 1$ the ratios of branching ratios shown in 
figure \ref{nor-bino-bmbebt} also tend to $1$. We have seen earlier that the 
neutrino mixing angles $\theta_{23}$ and $\theta_{13}$ also show nice 
correlation with the ratios $b^2_\mu/b^2_\tau$ and $b^2_e/b^2_\tau$, 
respectively (see figure \ref{neut2313-mixing-tree}). Hence we would expect 
that the ratios of the 
branching ratios $\frac{{\rm BR}({\tilde \chi}^0_7 \rightarrow \mu W)}
{{\rm BR}({\tilde \chi}^0_7 \rightarrow \tau W)}$ and 
$\frac{BR({\tilde \chi}^0_7 \longrightarrow e~W)}
{\sqrt{BR({\tilde \chi}^0_7 \longrightarrow \mu~W)^2
+BR({\tilde \chi}^0_7 \longrightarrow \tau~W)^2}}$ show correlations with 
$\tan^2\theta_{23}$ and $\tan^2\theta_{13}$. These correlations are shown in
figure \ref{nor-bino-LSP}. We have seen earlier (see eqn. (\ref{eigenvalues})) that
with the normal hierarchical pattern of the neutrino masses, the atmospheric
mass scale is determined by the quantity $\Omega_b = \sqrt{b^2_e + b^2_\mu 
+ b^2_\tau}$. Naturally one would expect that the atmospheric and the reactor
angles are correlated with the $\ell + W$ final states of the lightest 
neutralino decays and no correlation is expected for the solar angle. This is
what we have observed numerically. Here we have considered the regions of the
parameter space where the neutrino mass-squared differences and mixing angles
are within the 3$\sigma$ allowed range as shown in table \ref{osc-para}.
Figures \ref{nor-bino-LSP} also 
shows the model prediction for the ratios of branching ratios where the 
neutrino experimental data are satisfied. For our sample choice of parameters
in figure \ref{nor-bino-LSP}, one would expect that the ratio 
$\frac{BR({\tilde \chi}^0_7 \longrightarrow \mu~W)}{BR({\tilde \chi}^0_7
\longrightarrow \tau~W)}$ should be in the range 0.45 to 1.25. Similarly, 
the other ratio $\frac{BR({\tilde \chi}^0_7 \longrightarrow e~W)}
{\sqrt{BR({\tilde \chi}^0_7 \longrightarrow \mu~W)^2
+BR({\tilde \chi}^0_7 \longrightarrow \tau~W)^2}}$ is expected in this case 
to be less than 0.07. We can also see from figure \ref{nor-bino-LSP} that the 
ratio of branching ratios in the ($\mu + W$) and ($\tau + W$) channels becomes 
almost equal for the maximal value of the atmospheric mixing angle 
($\theta_{23}~=~45^\circ$). On the other hand, we do not observe any 
correlation with the solar mixing angle $\theta_{12}$ since it is a 
complicated function of $a^2_i$ and $b^2_i$ (see eqn. (\ref{solar_analytical})).

\begin{figure}[ht]
\centering
\vspace*{0.4cm}
\includegraphics[height=4.00cm]{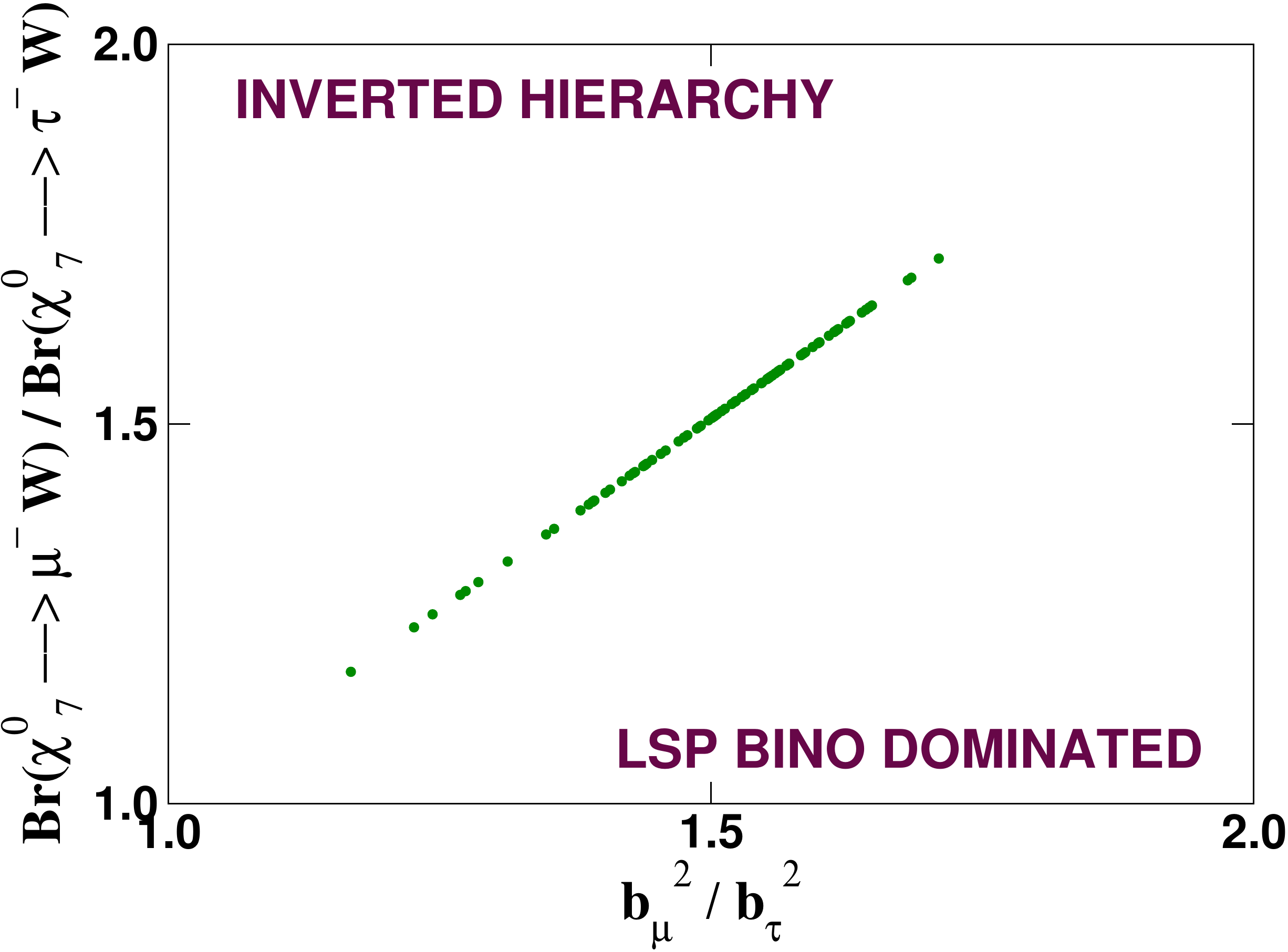}
\includegraphics[height=4.00cm]{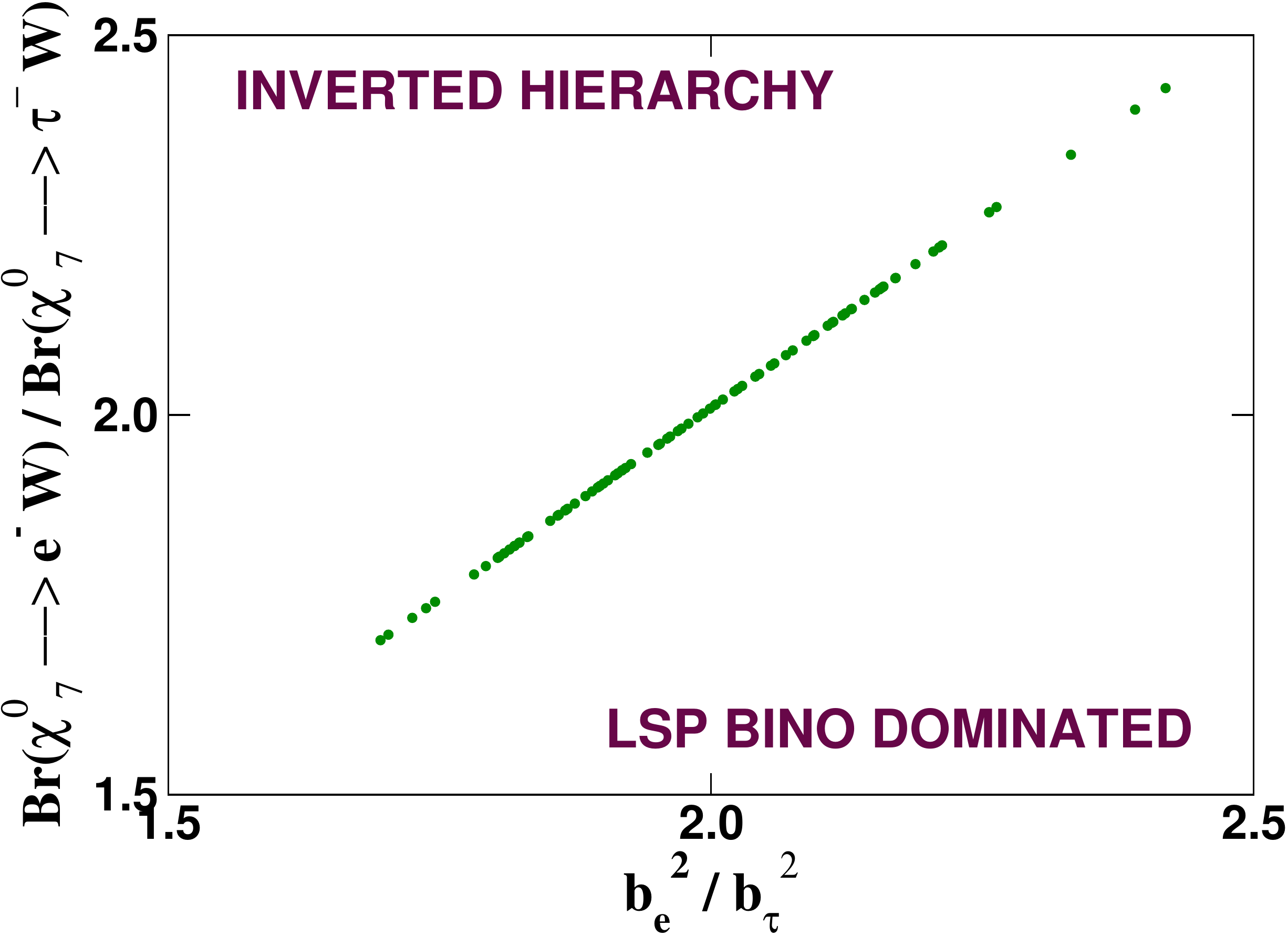}
\includegraphics[height=4.00cm]{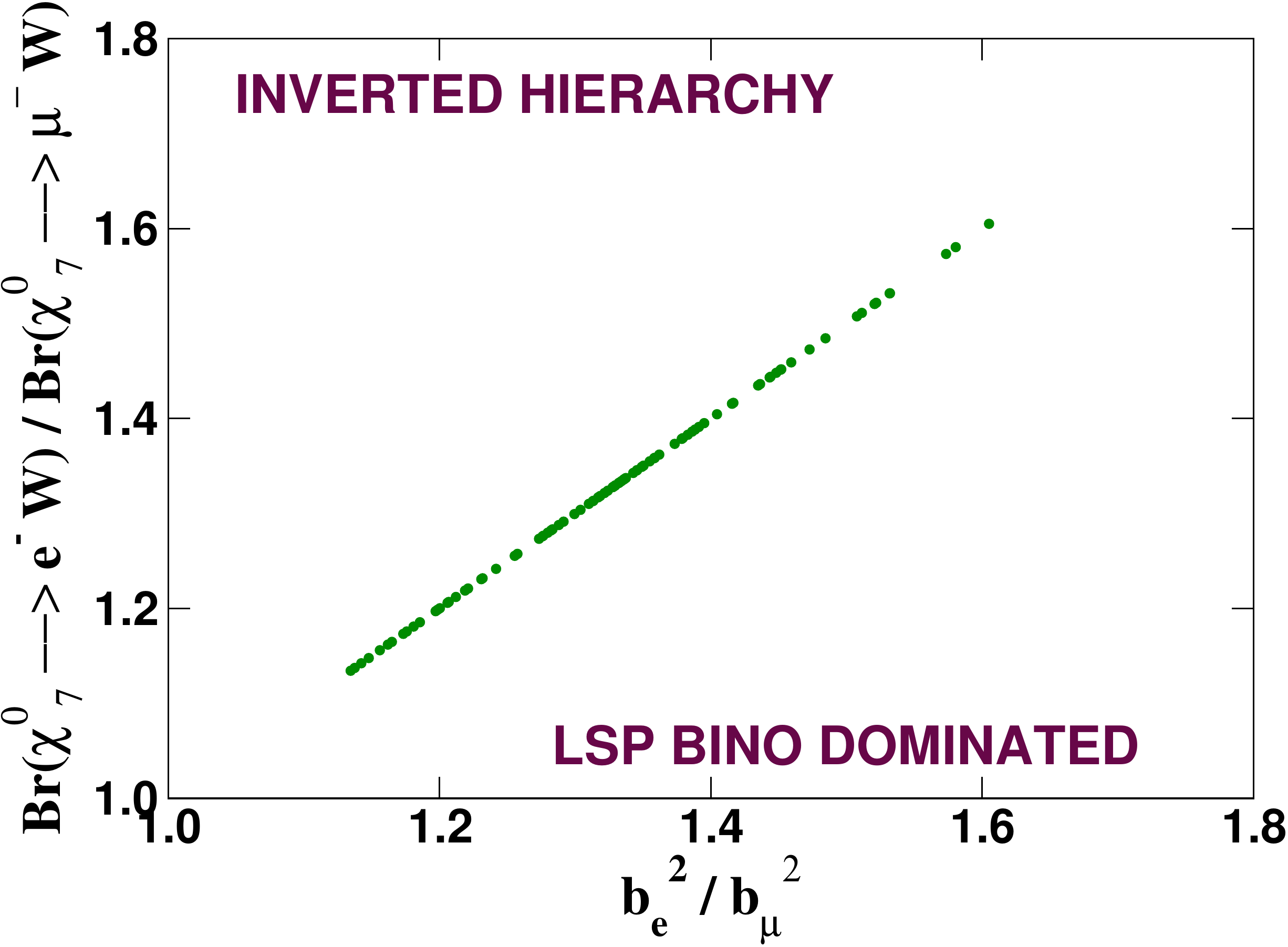}
\caption{Ratio $\frac{BR({\tilde \chi}^0_7 \longrightarrow \ell^-_i~W)}
{BR({\tilde \chi}^0_7 \longrightarrow \ell^-_j~W)}$ versus 
$\frac{b^2_i}{b^2_j}$ plot for a bino like lightest neutralino (the LSP) 
with bino component $|N_{71}|^2>$~0.95, where $i,j,k~=~e,\mu,\tau$. Neutrino 
mass pattern is inverted hierarchical. Choice of parameters are 
$M_1=105$ GeV, $\lambda=0.15, \kappa=0.65, m_{\tilde{\nu}^c}=300~{\rm GeV} 
~{\rm and} ~m_{\tilde{L}}=445~{\rm GeV}$. Mass of the LSP is $103.3$ GeV.
The value of the $\mu$ parameter comes out to be $-263.7$ GeV.}
\label{inv-bino-bmbebt}
\end{figure}

\begin{figure}[ht]
\centering
\vspace*{0.4cm}
\includegraphics[height=4.00cm]{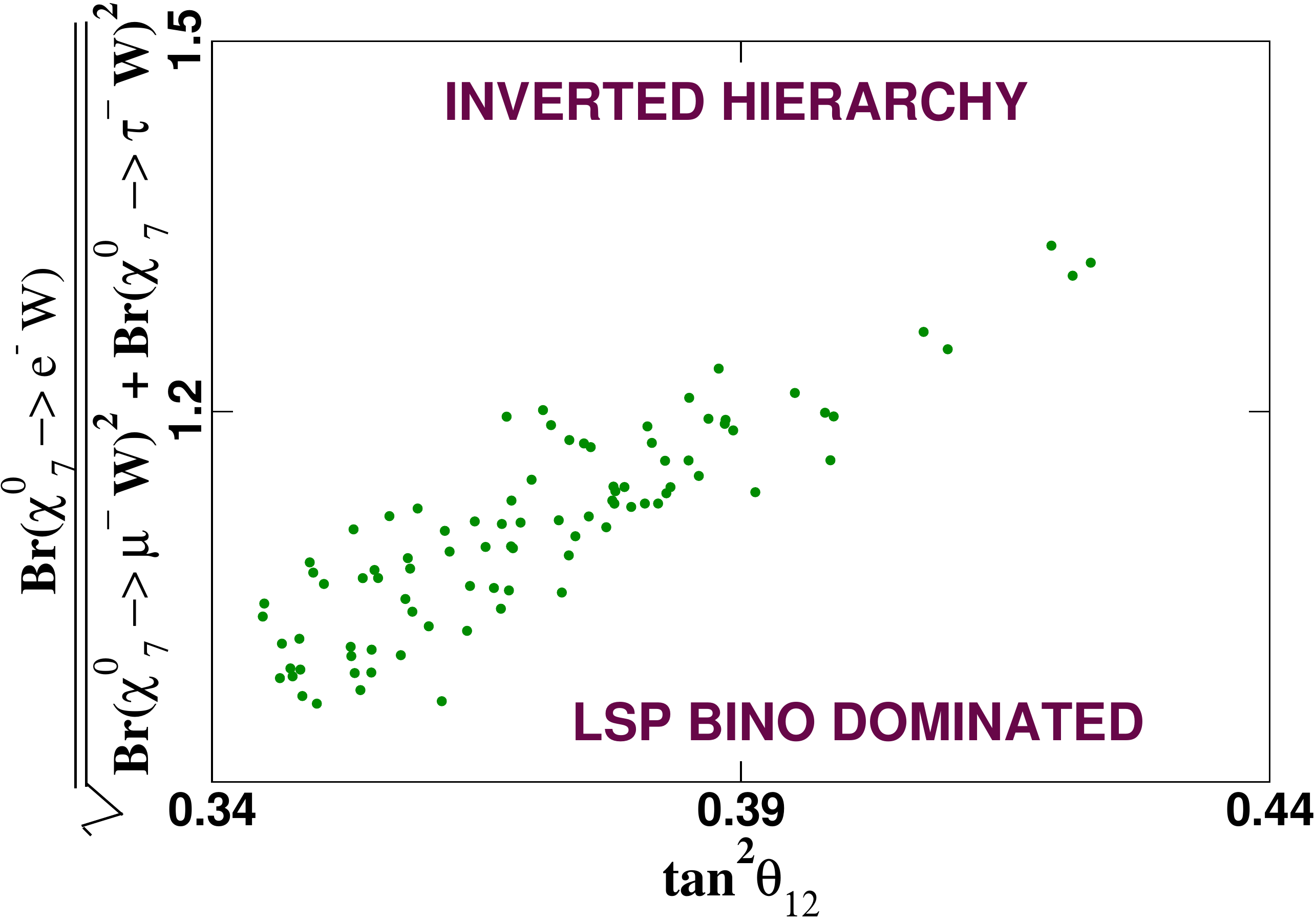}
\includegraphics[height=4.00cm]{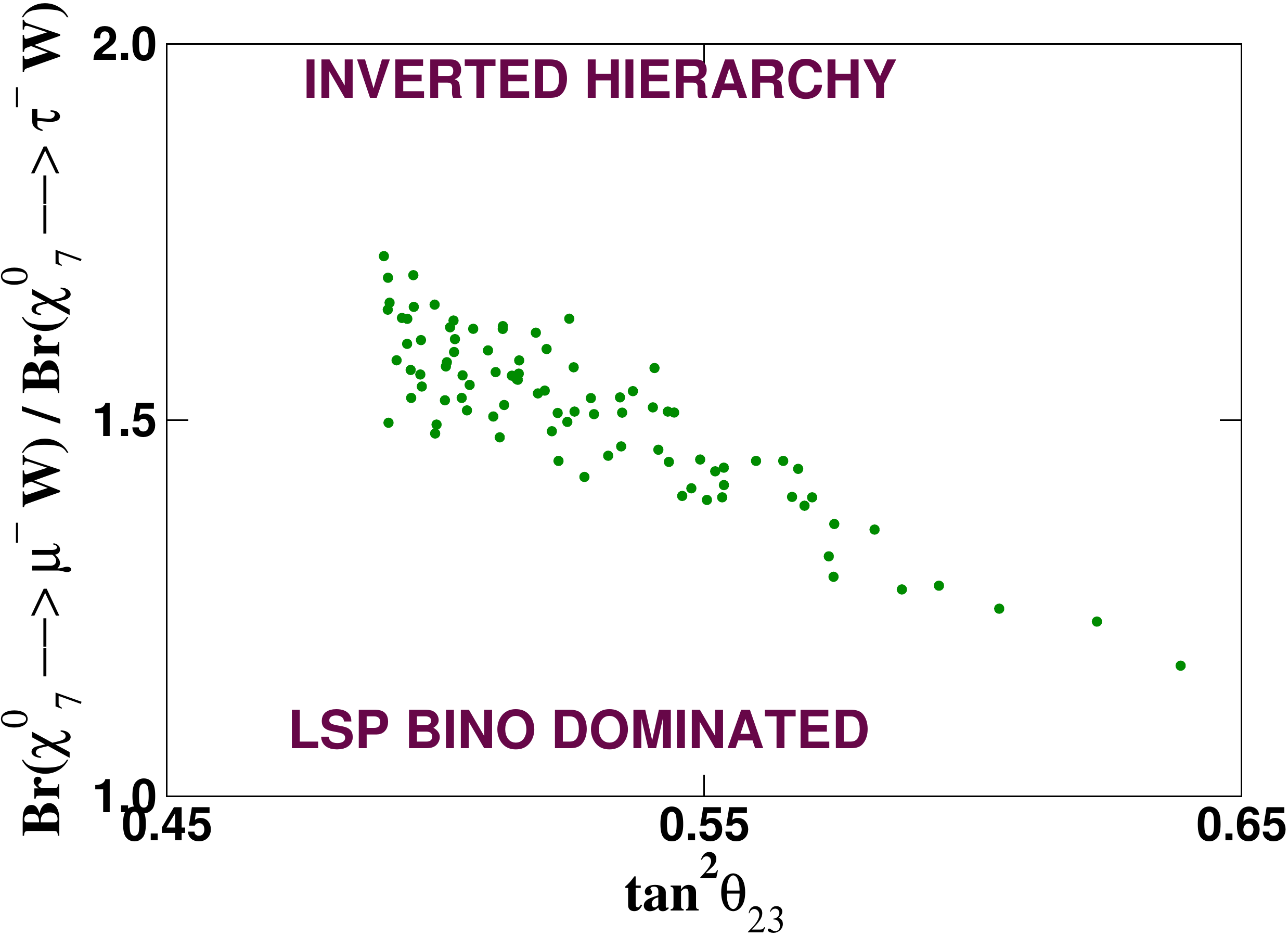}
\caption{Ratio $\frac{BR({\tilde \chi}^0_7 \longrightarrow e~W)}
{\sqrt{BR({\tilde \chi}^0_7 \longrightarrow \mu~W)^2+BR({\tilde \chi}^0_7 
\longrightarrow \tau~W)^2}}$ with $\tan^2\theta_{12}$ (left) plot for a bino 
dominated lightest neutralino (LSP) with bino component $|N_{71}|^2>$~0.95. 
In the right figure the ratio $\frac{BR({\tilde \chi}^0_7 \longrightarrow 
\mu~W)}{BR({\tilde \chi}^0_7 \longrightarrow \tau~W)}$ versus 
$\tan^2\theta_{23}$ is plotted.  Neutrino mass pattern is assumed to be 
inverted hierarchical. Choice of parameters are same as that of 
figure \ref{inv-bino-bmbebt}.}
\label{inv-bino-LSP}
\end{figure}

In the case of inverted hierarchical mass pattern of the light neutrinos, the
${\tilde \chi}^0_7$--$\ell_i$--$W$ coupling is still controlled by the 
quantities $b^2_i$. Hence the ratios of the branching ratios discussed earlier,
show nice correlations with $b^2_i/b^2_j$ (see figure \ref{inv-bino-bmbebt}). 
However, in this case the solar mixing angle shows 
some correlation with the ratio $\frac{BR({\tilde \chi}^0_7 \longrightarrow e~W)}
{\sqrt{\sum BR({\tilde \chi}^0_7 \longrightarrow \ell_i~W)^2}}$ 
with $\ell_i=\mu,\tau$. This is shown in 
figure \ref{inv-bino-LSP}. The correlation is not very sharp and some dispersion 
occurs due to the fact that the two heavier neutrino masses controlling the 
atmospheric mass scale and solar mass-squared difference are not completely 
determined by the quantities $b^2_i$ and there is some contribution of the 
quantities $a^2_i$, particularly for the second heavy neutrino mass eigenstate.

The correlation of the ratio $\frac{BR({\tilde \chi}^0_7 \longrightarrow 
\mu~W)} {BR({\tilde \chi}^0_7 \longrightarrow \tau~W)}$ with 
$\tan^2\theta_{23}$ shows a different behaviour compared to what we have seen
in the case of normal hierarchical scenario. This is because in the case of 
inverted hierarchical mass pattern of the neutrinos, $\tan^2\theta_{23}$ 
decreases with increasing $b^2_\mu/b^2_\tau$. One can observe from 
Figures \ref{nor-bino-LSP} and \ref{inv-bino-LSP} that if the experimental 
value of the ratio $\frac{BR({\tilde \chi}^0_7 \longrightarrow e~W)}
{\sqrt{BR({\tilde \chi}^0_7 \longrightarrow \mu~W)^2+BR({\tilde \chi}^0_7
\longrightarrow \tau~W)^2}}$ is $\ll$ 1 then that indicates a normal 
hierarchical neutrino mass pattern for a bino-dominated lightest neutralino 
LSP whereas a higher value ($\sim$ 1) of this ratio measured in experiments
might indicate that the neutrino mass pattern is inverted hierarchical. 
Similarly a measurement of the ratio $\frac{BR({\tilde \chi}^0_7 
\longrightarrow \mu~W)}{BR({\tilde \chi}^0_7 \longrightarrow \tau~W)}$ can
also give an indication regarding the particular hierarchy of the neutrino
mass pattern in the case of a bino dominated LSP.

\subsection{{\bf H}iggsino dominated lightest neutralino}\label{higgsino-LN}
\begin{figure}[ht]
\centering
\vspace*{0.4cm}
\includegraphics[height=4.00cm]{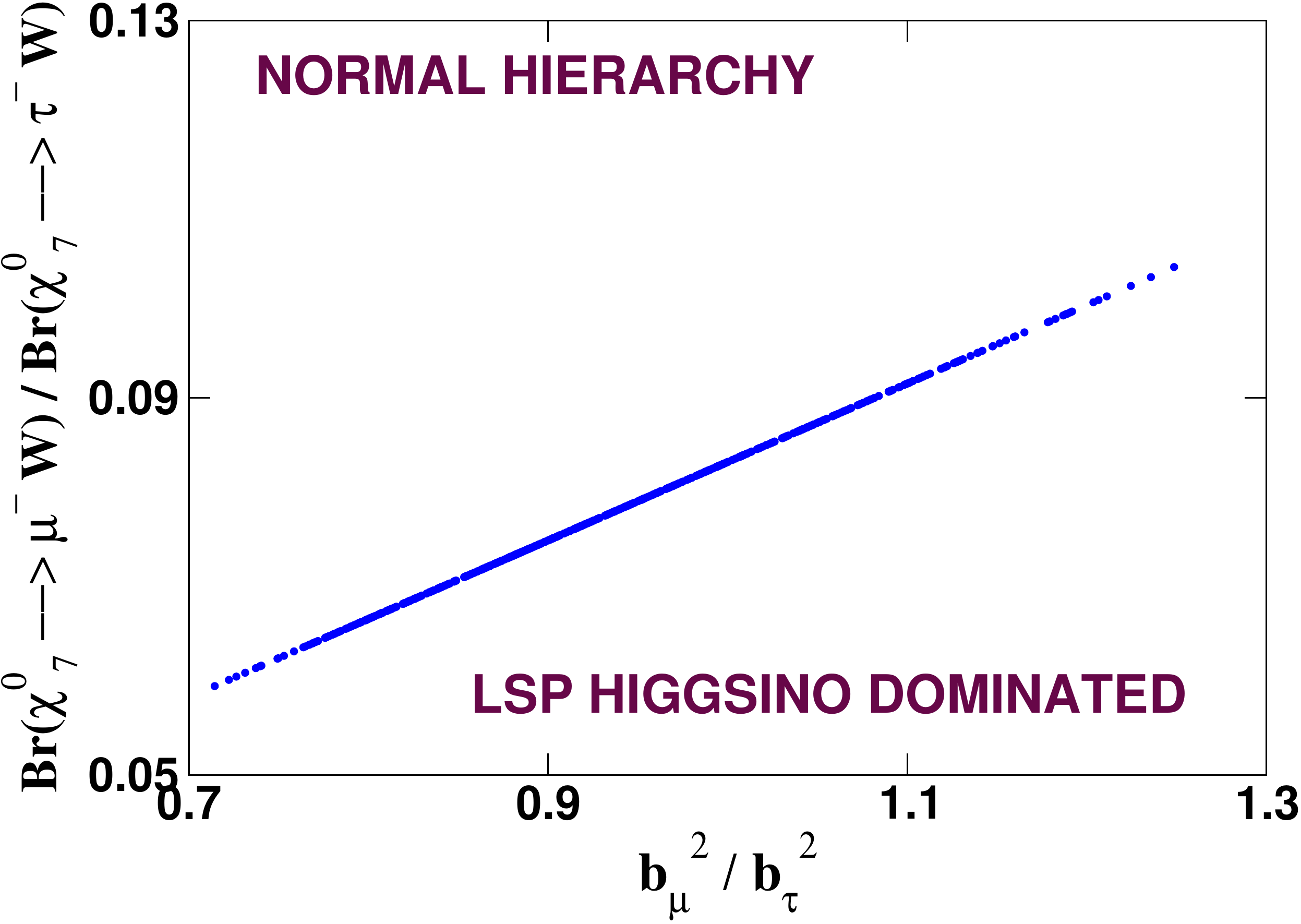}
\includegraphics[height=4.00cm]{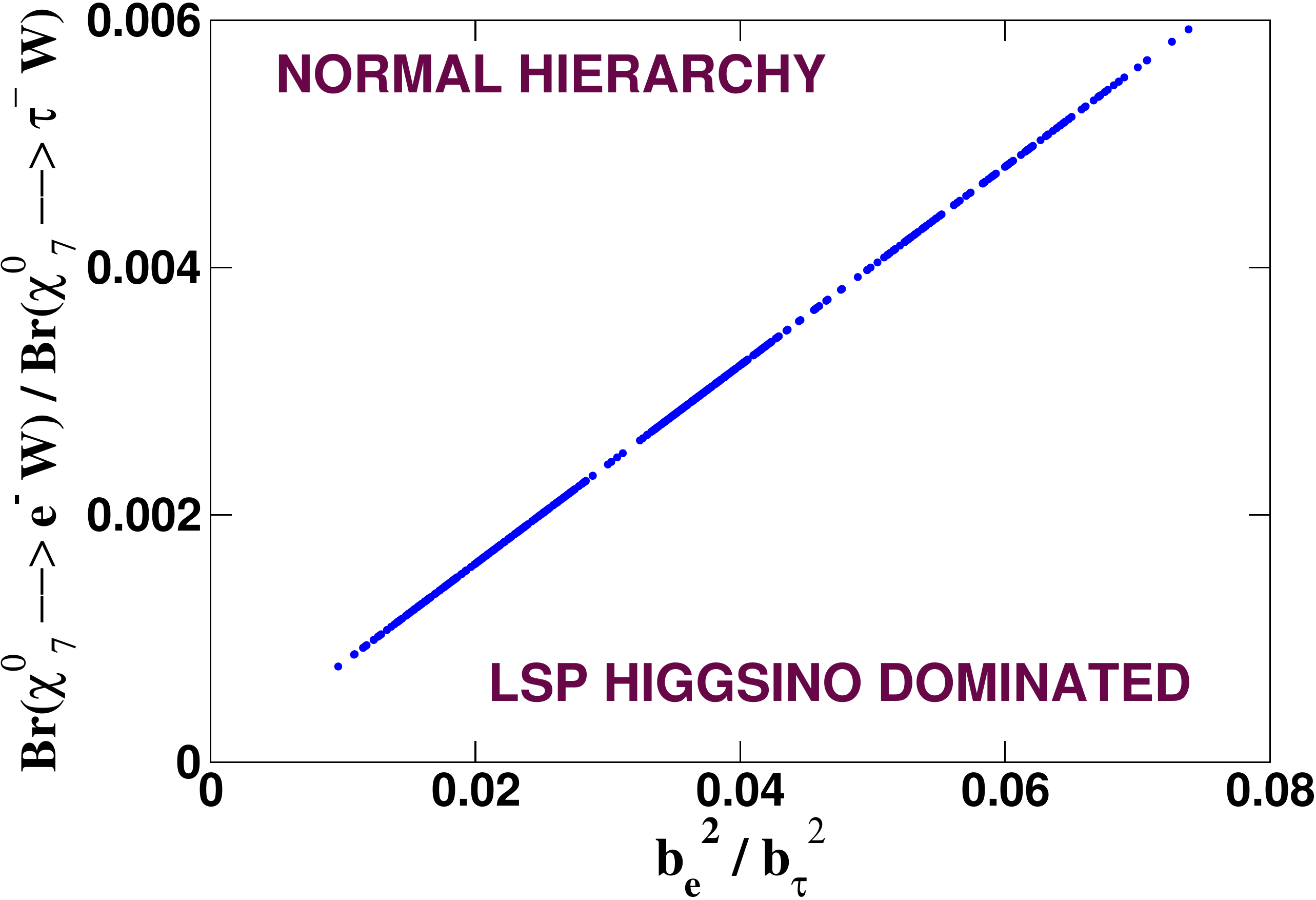}
\includegraphics[height=4.00cm]{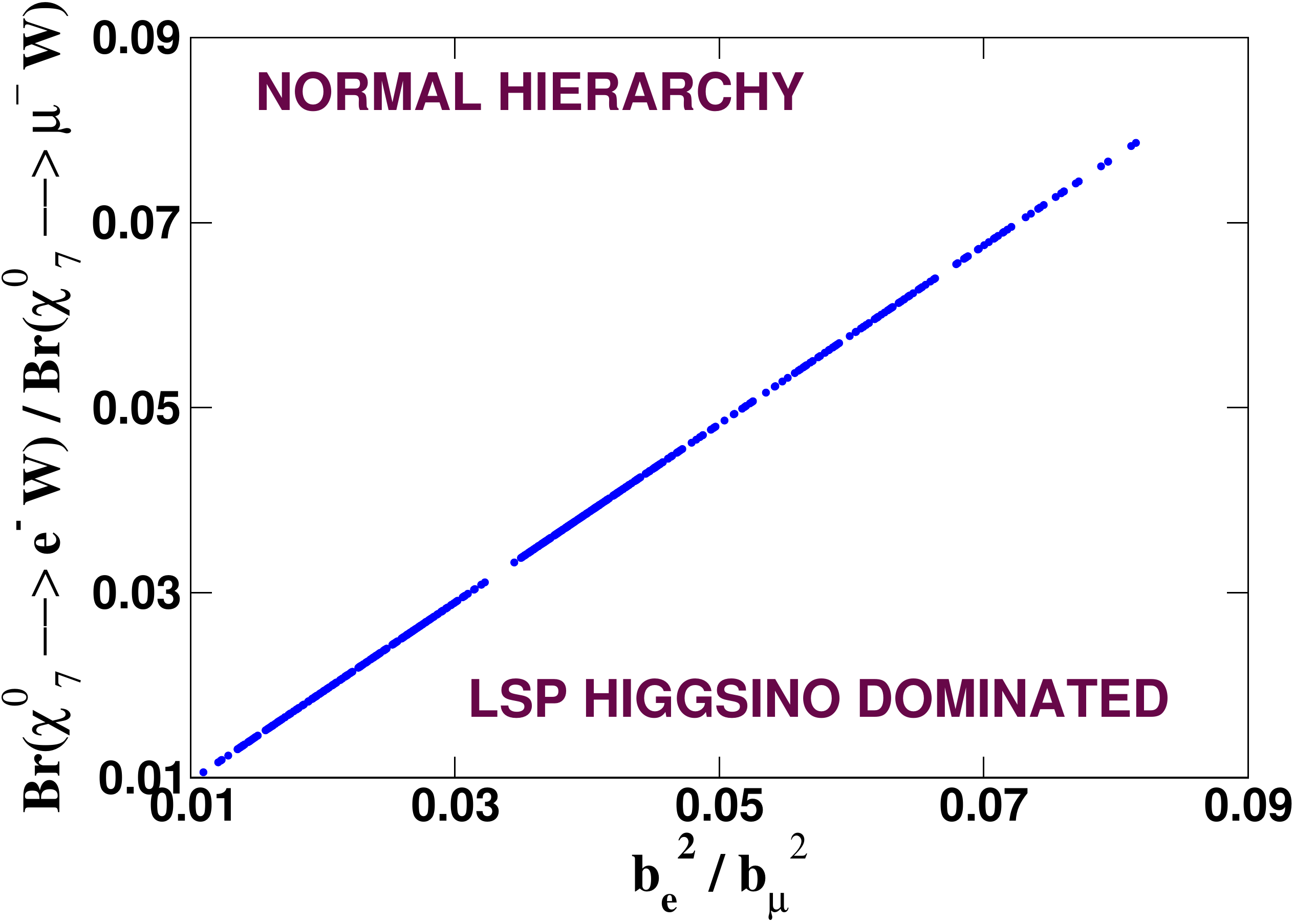}
\caption{Ratio $\frac{BR({\tilde \chi}^0_7 \longrightarrow l_i~W)}
{BR({\tilde \chi}^0_7 \longrightarrow l_j~W)}$ versus $\frac{b^2_i}{b^2_j}$ 
plot for a higgsino like LSP with higgsino component $(|N_{73}|^2+
|N_{74}|^2)>$~0.95, where $i,j,k~=~e,\mu,\tau$. Neutrino mass pattern is 
normal hierarchical. Choice of parameters are $M_1=325$ GeV, $\lambda=0.06, 
\kappa=0.65, m_{\tilde{\nu}^c}=300~{\rm GeV} ~\rm{and} ~m_{\tilde{L}}=
400~{\rm GeV}$. Mass of the LSP is $98.6$ GeV.
The value of the $\mu$ parameter comes out to be $-105.9$ GeV.}
\label{nor-higgsino-bmbebt}
\end{figure}

\begin{figure}[ht]
\centering
\vspace*{0.4cm}
\includegraphics[height=4.00cm]{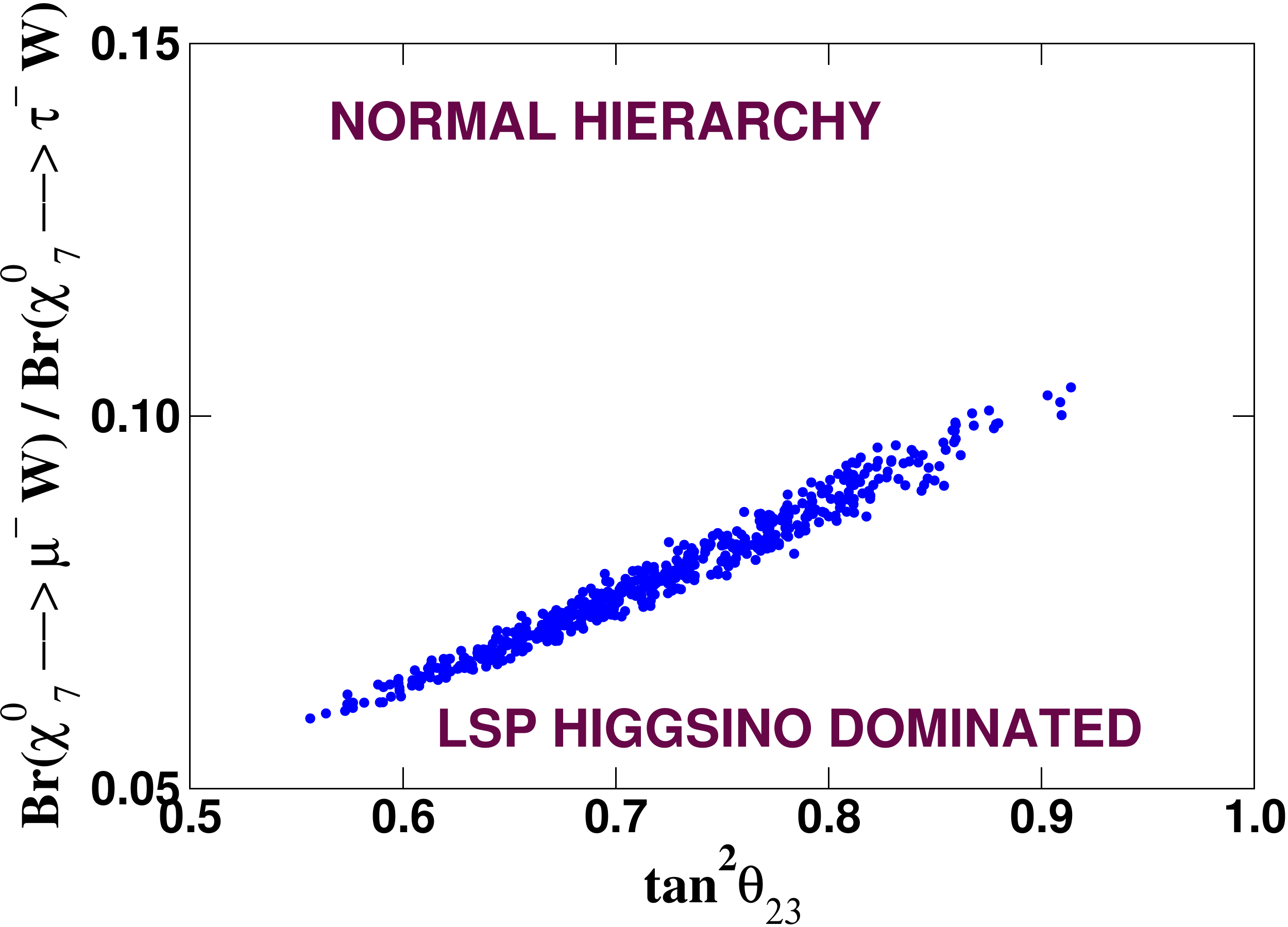}
\includegraphics[height=4.00cm]{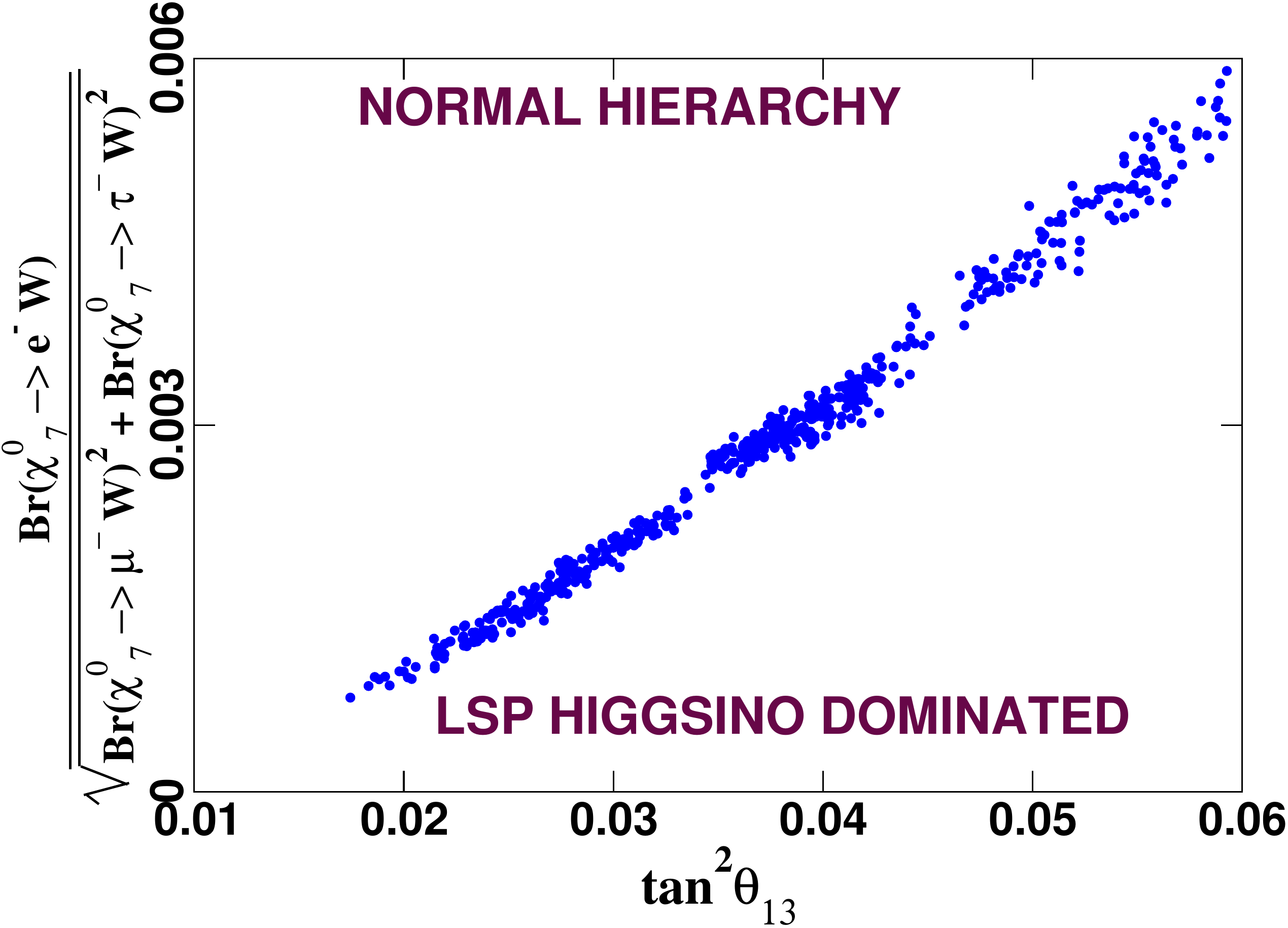}
\caption{Ratio $\frac{BR({\tilde \chi}^0_7 \longrightarrow \mu~W)}
{BR({\tilde \chi}^0_7 \longrightarrow \tau~W)}$ versus $\tan^2\theta_{23}$~
(left), $\frac{BR({\tilde \chi}^0_7 \longrightarrow e~W)}
{\sqrt{BR({\tilde \chi}^0_7 \longrightarrow \mu~W)^2+
BR({\tilde \chi}^0_7 \longrightarrow \tau~W)^2}}$ with $\tan^2\theta_{13}$
~(right) plot for a higgsino LSP with higgsino component  
$(|N_{73}|^2+|N_{74}|^2)>$~$0.95$. Neutrino mass pattern is normal 
hierarchical. Choice of parameters are same as that of 
figure \ref{nor-higgsino-bmbebt}.}
\label{nor-higgsino-LSP}
\end{figure}

When one considers higher values of the $U(1)$ gaugino mass $M_1$, i.e. 
$M_1>\mu$ and large value of $\kappa$ (so that the effective gauge singlet 
neutrino mass $2 \kappa v^c$ is large), the lightest neutralino is  
essentially higgsino dominated and it is the LSP. Naturally one needs to 
consider a small value of the coupling $\lambda$ so that the effective $\mu$ 
parameter ($\mu = 3 \lambda v^c$) is smaller. In order to look at the 
lightest neutralino decay branching ratios in this case, we consider a 
situation where the higgsino component in ${\tilde \chi}^0_7$ is 
$|N_{73}|^2 + |N_{74}|^2 >$ $0.90$. As in the case of a bino dominated LSP,
the generation dependence of the ${\tilde \chi}^0_7$--$\ell_i$--$W$ couplings
comes through the quantities $b^2_i$. However, because of the large value of 
the $\tau$ Yukawa coupling, the higgsino--$\tau$ mixing is larger 
and as a result the partial decay width of ${\tilde \chi}^0_7$ into ($W + \tau$) is 
larger than into ($W + \mu$) and ($W + e$). This feature is shown in 
figure \ref{nor-higgsino-bmbebt}, where the ratios of branching ratios are 
plotted against the quantities $b^2_i/b^2_j$. The domination of
BR(${\tilde \chi}^0_7 \rightarrow \tau + W$) over the other two is clearly
evident. Nevertheless, all the three ratios of branching ratios show 
sharp correlations with the corresponding $b^2_i/b^2_j$. In this figure
the normal hierarchical pattern of the neutrino masses has been considered.
As in the case of a bino LSP, here also the ratios $\frac{BR({\tilde \chi}^0_7 
\longrightarrow \mu~W)} {BR({\tilde \chi}^0_7 \longrightarrow \tau~W)}$ and
$\frac{BR({\tilde \chi}^0_7 \longrightarrow e~W)}
{\sqrt{BR({\tilde \chi}^0_7 \longrightarrow \mu~W)^2+BR({\tilde \chi}^0_7
\longrightarrow \tau~W)^2}}$ show nice correlations with neutrino mixing
angles $\theta_{23}$ and $\theta_{13}$, respectively. This is shown in 
figure \ref{nor-higgsino-LSP}. However, in this case the predictions for these
two ratios are very different from the bino LSP case. The expected value of
the ratio $\frac{BR({\tilde \chi}^0_7 \longrightarrow \mu~W)} 
{BR({\tilde \chi}^0_7 \longrightarrow \tau~W)}$ is approximately between 
0.05 and 0.10 in a region where one can accommodate the experimental 
neutrino data. Similarly, the predicted value of the ratio 
$\frac{BR({\tilde \chi}^0_7 \longrightarrow e~W)}
{\sqrt{BR({\tilde \chi}^0_7 \longrightarrow \mu~W)^2+BR({\tilde \chi}^0_7
\longrightarrow \tau~W)^2}}$ is $\leq$ 0.006. On the other hand, there is 
no such correlations with the solar mixing angle $\theta_{12}$.

\vspace{0.5cm}
\vspace{0.5cm}
\vspace{1.3cm}

\begin{figure}[ht]
\centering
\vspace*{0.4cm}
\includegraphics[height=4.00cm]{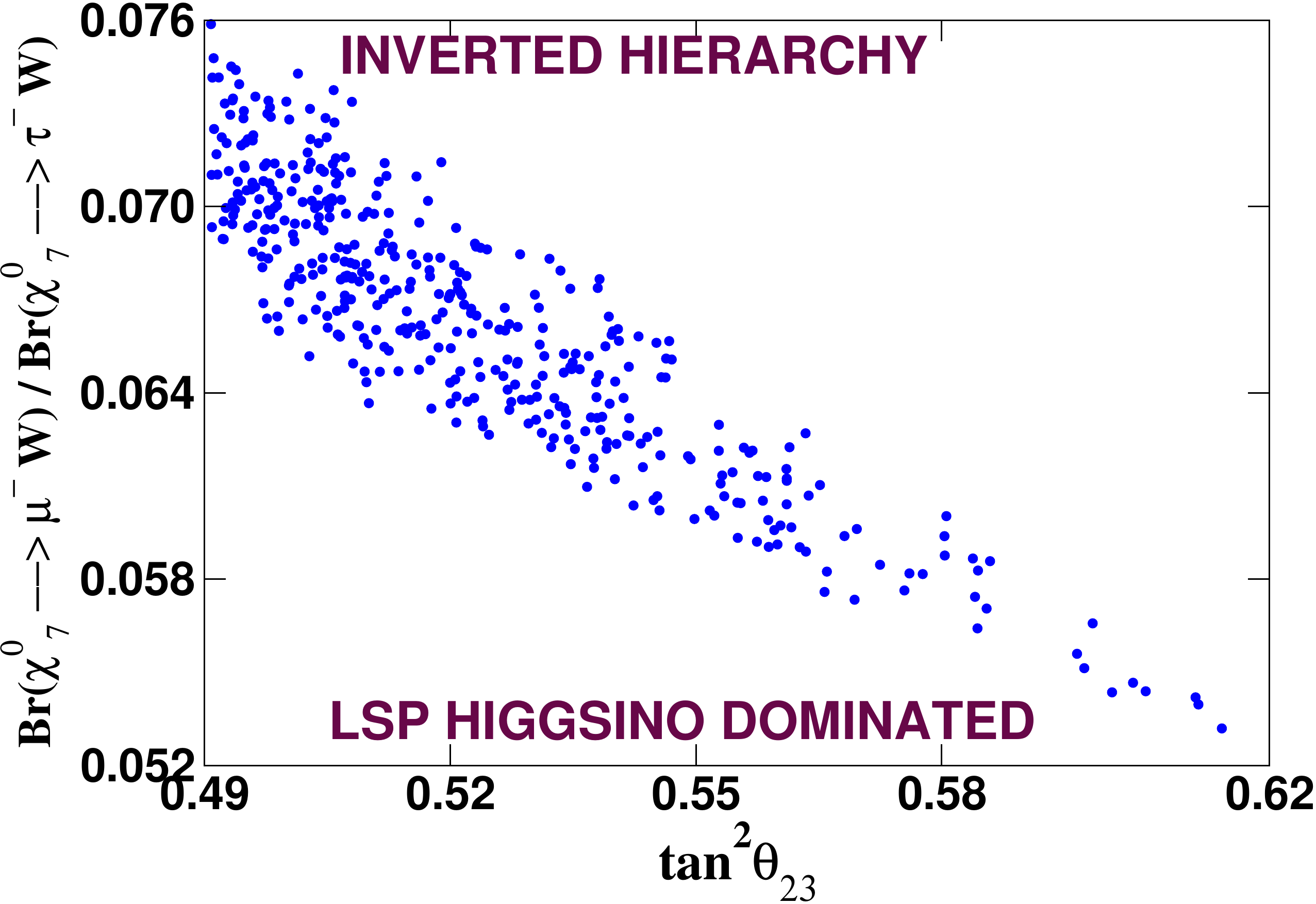}
\includegraphics[height=4.00cm]{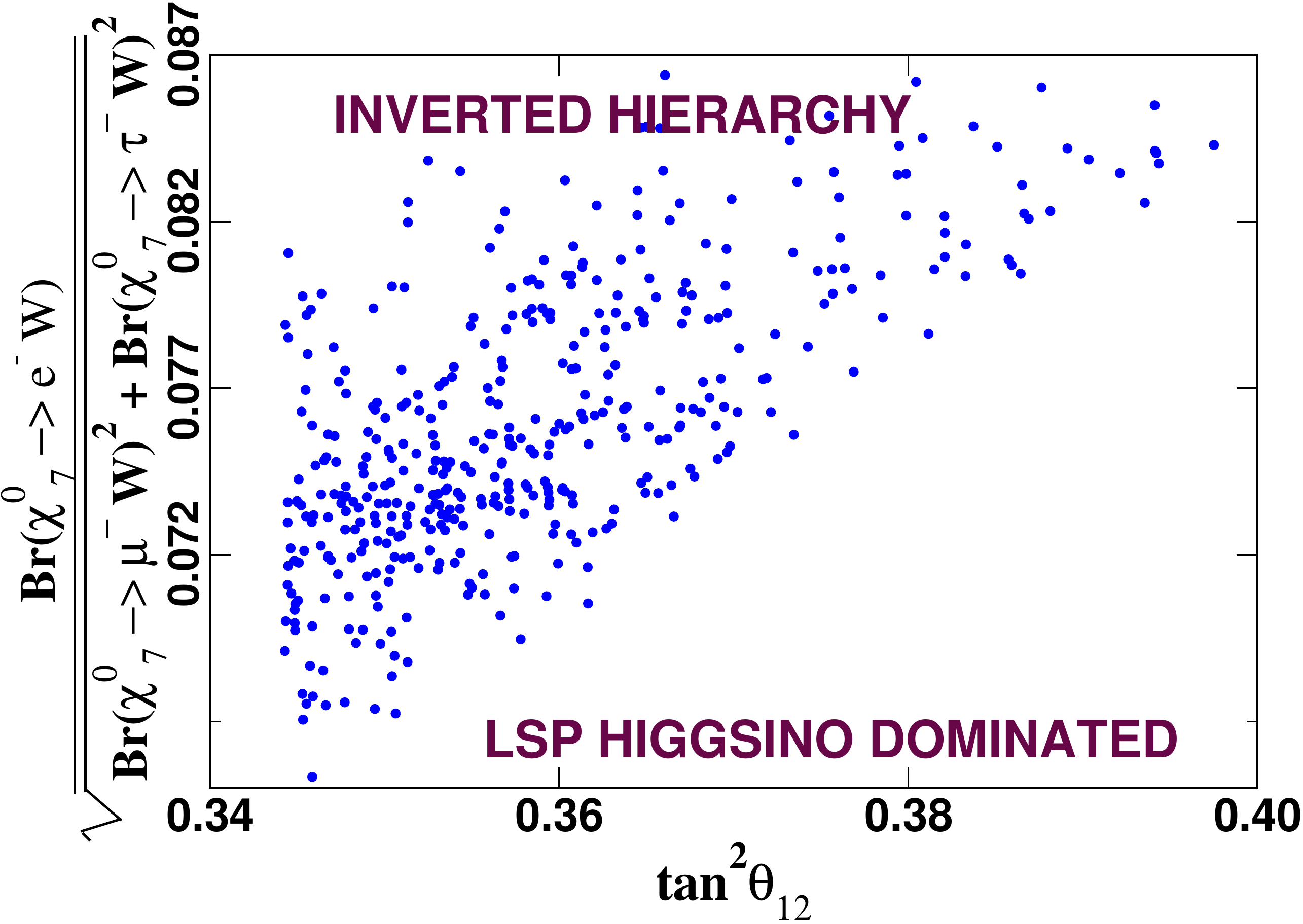}
\caption{Ratio $\frac{BR({\tilde \chi}^0_7 \longrightarrow \mu~W)}
{BR({\tilde \chi}^0_7 \longrightarrow \tau~W)}$ versus $\tan^2\theta_{23}$
~(left), $\frac{BR({\tilde \chi}^0_7 \longrightarrow e~W)}
{\sqrt{BR({\tilde \chi}^0_7 \longrightarrow \mu~W)^2+Br({\tilde \chi}^0_7 
\longrightarrow \tau~W)^2}}$ with $\tan^2\theta_{12}$~(right) plot
for a higgsino LSP with higgsino component $(|N_{73}|^2+|N_{74}|^2)>$~0.95. 
Neutrino mass pattern is inverted hierarchical. Choice of parameters are 
$M_1=490$ GeV, $\lambda=0.07, \kappa=0.65, m_{\tilde{\nu}^c}=320 \rm{GeV} 
~\rm{and} ~m_{\tilde{L}}=430 \rm{GeV}$. Mass of the LSP is $110.8$ GeV.
The value of the $\mu$ parameter comes out to be $-115.3$ GeV.}
\label{inv-higgsino-LSP}
\end{figure}

Similar correlations of the ratios of branching ratios with $b^2_i/b^2_j$ 
are also obtained for a higgsino dominated LSP in the case where the neutrino 
mass pattern is inverted hierarchical. Once again it shows that the 
${\tilde \chi}^0_7$ decays to ($\tau + W$) channel is dominant over the 
channels ($e + W$) and ($\mu + W$) for any values of $b^2_i/b^2_j$ because 
of the larger $\tau$ Yukawa coupling. On the other hand, the correlations with
the neutrino mixing angles show a behaviour similar to that of a bino LSP
with inverted neutrino mass hierarchy though with much smaller values for the
ratios $\frac{BR({\tilde \chi}^0_7 \rightarrow \mu~W)}
{BR({\tilde \chi}^0_7 \rightarrow \tau~W)}$ 
and $\frac{BR({\tilde \chi}^0_7 \rightarrow e~W)}
{\sqrt{BR({\tilde \chi}^0_7 \rightarrow \mu~W)^2+Br({\tilde \chi}^0_7 
\rightarrow \tau~W)^2}}$. These are shown in figure \ref{inv-higgsino-LSP}.
Note that the correlations in this case are not very sharp, especially with 
$\tan^2\theta_{12}$. Thus we see that small values of these ratios 
(for both normal and inverted hierarchy) are characteristic features of a 
higgsino dominated LSP in this model. 
\vspace{1.7cm}

\subsection{{\bf R}ight-handed neutrino dominated lightest neutralino}\label{rhnu-LN}
Because of our choice of parameters i.e., a generation independent coupling 
$\kappa$ of the gauge singlet neutrinos and a common VEV $v^c$ (see eqn.(\ref{assumption1})),
the three neutralino mass eigenstates which are predominantly gauge singlet 
neutrinos are essentially mass degenerate. There is a very small mass 
splitting due to mixing. However, unlike the case of a bino or higgsino 
dominated lightest neutralino, these $\nu^c$ dominated lightest neutralino 
states cannot be considered as the LSP. This is because in this case the 
lightest scalar (which is predominantly a gauge singlet sneutrino 
${\tilde \nu}^c$) is the lightest supersymmetric particle. This is very 
interesting since usually one does not get a ${\tilde \nu}^c$ as an LSP 
in a model where the gauge singlet neutrino superfield has a large Majorana 
mass term in the superpotential. However, in this case the effective Majorana 
mass term is at the EW scale and there is also a contribution from the 
trilinear scalar coupling $A_\kappa \kappa$ which keeps the mass of the 
singlet scalar sneutrino smaller. It is also very interesting to study the 
decay patterns of the lightest neutralino in this case since here one can 
probe the gauge singlet neutrino mass scales at the colliders. 

Before discussing the decay patterns of the lightest neutralino which is 
$\nu^c$ dominated, let us say a few words regarding their production at 
the LHC. The direct production of $\nu^c$ (by $\nu^c$ we mean the $\nu^c$
dominated lightest neutralino in this subsection) is negligible because of
the very small mixing with the MSSM neutralinos. Nevertheless, they can be 
produced at the end of the cascade decay chains of the squarks and gluinos at
the LHC. For example, if the next-to-next-to-lightest SUSY particle (NNLSP) 
is higgsino dominated (this is the state above the three almost degenerate 
lightest neutralinos) and it has a non-negligible mixing with $\nu^c$ (remember 
that the higgsino--$\nu^c$ mixing occurs mainly because of the 
term $\lambda {\hat \nu}^c {\hat H}_d {\hat H}_u$ in the superpotential, 
eqn.(\ref{munuSSM-superpotential})), then 
the branching ratio of the decay ${\tilde H} \rightarrow Z + \nu^c$ can be 
larger than the branching ratios in the $\ell W$ and $\nu Z$ channels. 
This way one can produce $\nu^c$ dominated lightest neutralino. Similarly, 
a higgsino dominated lighter chargino can also produce gauge singlet neutrinos. 
Another way of producing $\nu^c$ is through the decay of an NNLSP ${\tilde \tau}_1$, such as 
${\tilde \tau}_1 \rightarrow \tau + \nu^c$.  
      
\vspace{0.5cm}
\begin{figure}[ht]
\centering
\vspace*{0.4cm}
\includegraphics[height=4.00cm]{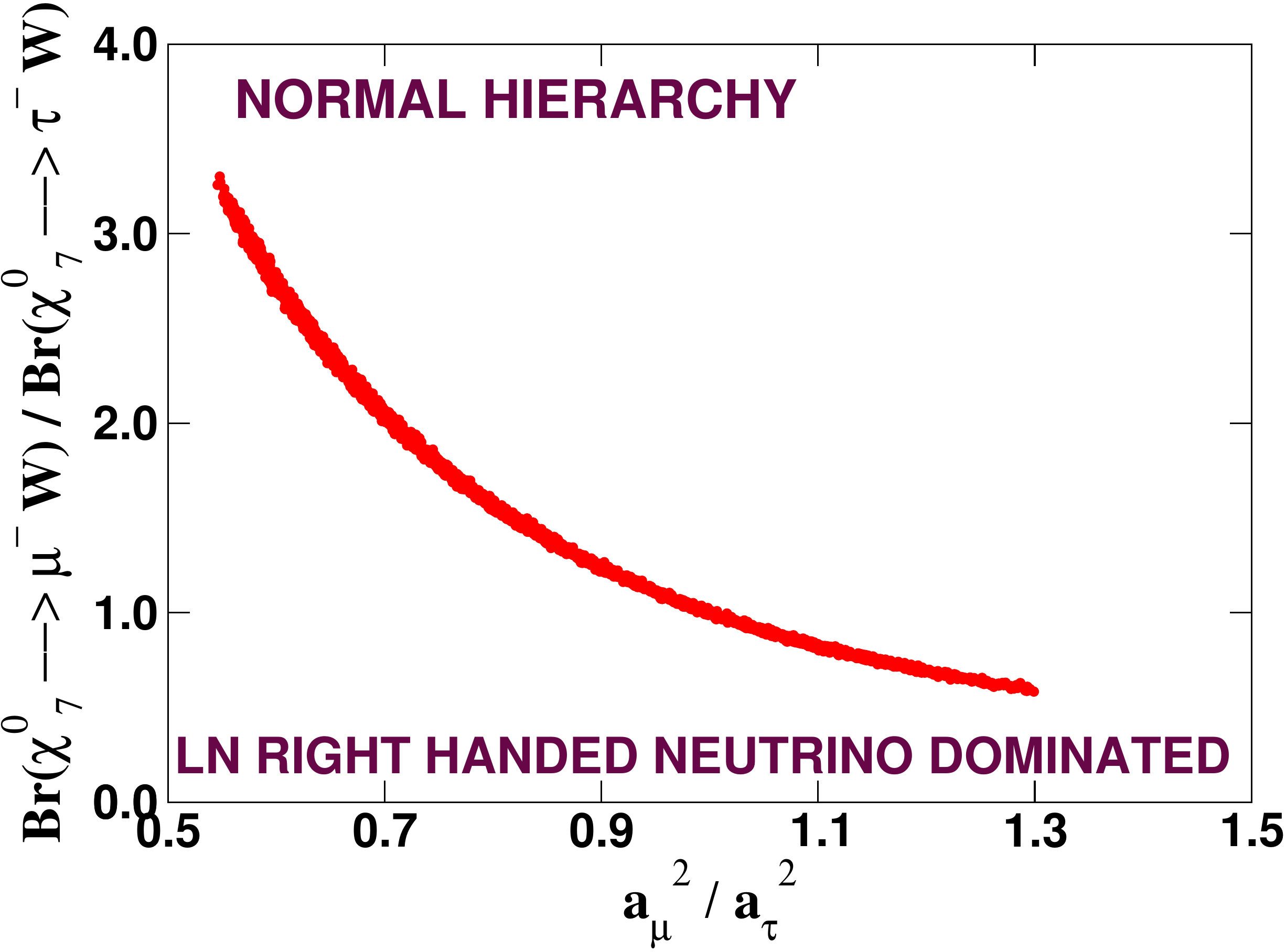}
\includegraphics[height=4.00cm]{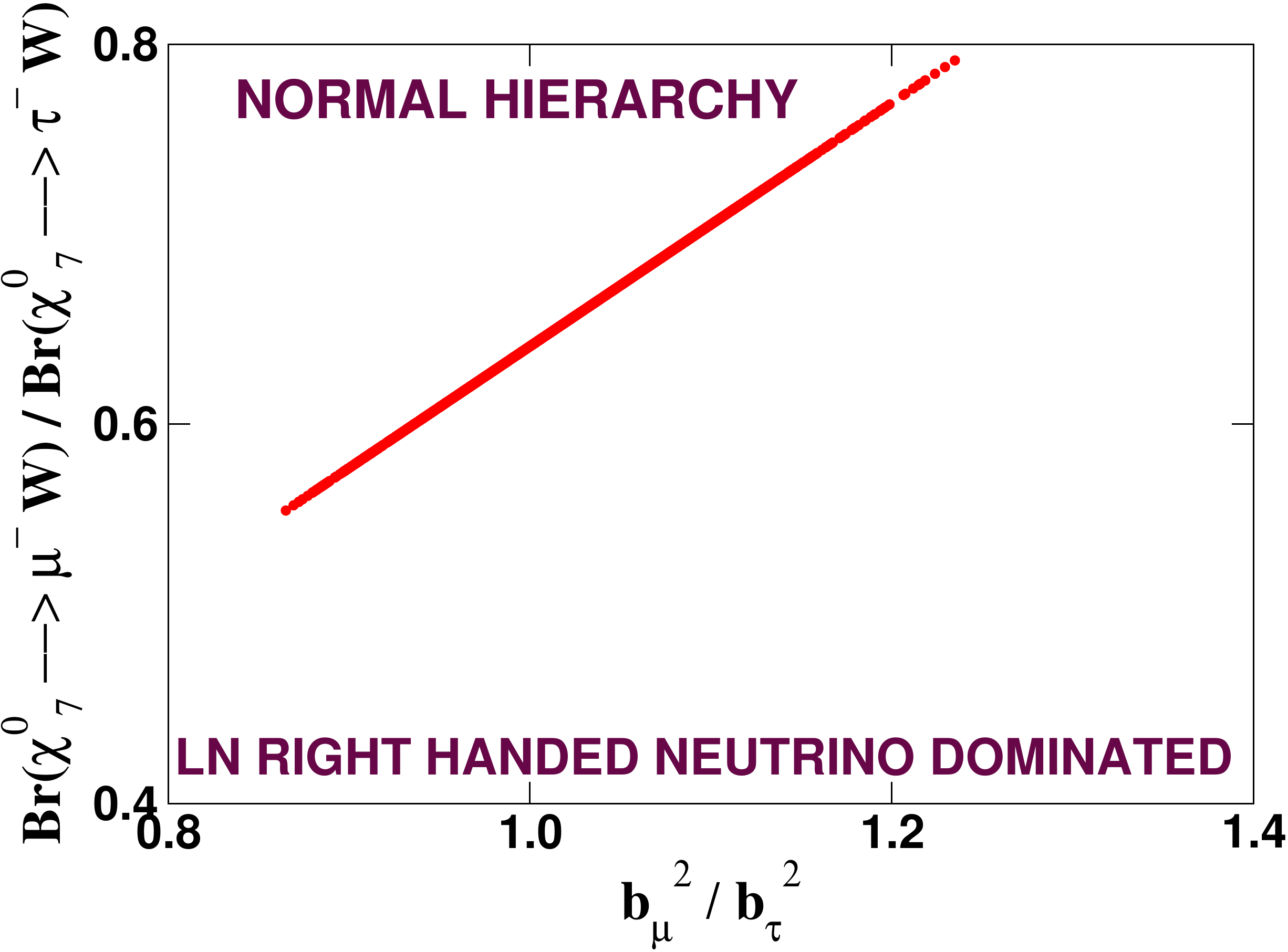}
\caption{Ratio $\frac{BR({\tilde \chi}^0_7 \longrightarrow \mu~W)}
{BR({\tilde \chi}^0_7 \longrightarrow \tau~W)}$ versus 
$\frac{a^2_\mu}{a^2_\tau}$ (left) and versus $\frac{b^2_\mu}{b^2_\tau}$ 
(right) plot for a $\nu^c$ like lightest neutralino (${\tilde \chi}^0_7$) 
with $\nu^c$ component $(|N_{75}|^2+|N_{76}|^2+|N_{77}|^2)>$~0.99, (left) and
$>$0.97 (right). Neutrino mass pattern is normal hierarchical. Choice of 
parameters are for (left) $M_1=405$ GeV, $\lambda=0.29, \kappa=0.07, 
(A_\lambda \lambda)= -8.2~{\rm TeV} \times \lambda,(A_\kappa \kappa)= 165 
~{\rm GeV} \times \kappa,~m_{\tilde{\nu}^c}=50~{\rm GeV} ~\rm{and} 
~m_{\tilde{L}}=825~{\rm GeV}$ and for (right) $M_1=405$ GeV, $\lambda=0.10, 
\kappa=0.07, (A_\lambda \lambda)= -2~{\rm TeV} \times \lambda,(A_\kappa \kappa)
= 165~{\rm GeV} \times \kappa, ~m_{\tilde{\nu}^c}=50~{\rm GeV} ~{\rm and} 
~m_{\tilde{L}}=825~{\rm GeV}$. Mass of the lightest neutralino is $129.4$ GeV 
(left) and $119.8$ GeV (right) respectively.
The values of the $\mu$ parameter are $-803.9$ GeV and $-258.8$ GeV, respectively.}
\label{nor-singlino-bmbt-amat}
\end{figure}

When one considers higher value of the gaugino mass, i.e. $M_1>\mu$ and 
a small value of the coupling $\kappa$ (so that the effective Majorana mass of 
$\nu^c$ is small, i.e. $m_{\nu^c}=2 \kappa v^c < \mu$), the lightest neutralino is 
essentially $\nu^c$ dominated. As we have mentioned earlier, in this case the 
LSP is the scalar partner of $\nu^c$, i.e. ${\tilde \nu}^c$. However, the 
decay of $\nu^c$ into $\nu + {\tilde \nu}^c$ is suppressed compared to the 
decays $\nu^c \rightarrow \ell_i + W$ and $\nu^c \rightarrow \nu_i +Z$ that 
we have considered so far. Because of this, in this section we will neglect the
decay $\nu^c \rightarrow \nu + {\tilde \nu}^c$ while discussing the correlation
of the lightest neutralino (${\tilde \chi}^0_7$) decays with the neutrino 
mixing angles.

In this case the coupling of the lightest neutralino (${\tilde \chi}^0_7$) 
with $\ell_i$--$W$ pair depends on the $\nu^c$ content of ${\tilde \chi}^0_7$. 
Note that the $\nu^c$ has a very small mixing with the MSSM neutralino states. 
However, in some cases the $\nu^c$ dominated lightest neutralino can have a 
non-negligible higgsino component. In such cases the coupling 
${\tilde \chi}^0_7$--$\ell_i$--$W$ depends mainly on the quantities $b_i$. On 
the other hand, if ${\tilde \chi}^0_7$ is very highly dominated by $\nu^c$, 
then the coupling ${\tilde \chi}^0_7$-- $\ell_i$--$W$ has a nice correlation 
with the quantities $a_i$. So in order to study the decay correlations of the 
$\nu^c$ dominated lightest neutralino, we consider two cases (i) $\nu^c$ 
component is $>$ $0.99$, and (ii) $\nu^c$ component is $>0.97$ with some 
non-negligible higgsino admixture.

The correlations of the decay branching ratio 
$\frac{BR({\tilde \chi}^0_7 \longrightarrow \mu~W)}
{BR({\tilde \chi}^0_7 \longrightarrow \tau~W)}$ are shown in 
figure \ref{nor-singlino-bmbt-amat} for the cases (i) and (ii) mentioned above. 
As we have explained already, this figure demonstrates that in case (i) the
ratio of the branching ratio is dependent on the quantity $a^2_\mu/a^2_\tau$
whereas in case (ii) this ratio is correlated with $b^2_\mu/b^2_\tau$ though
there is some suppression due to large $\tau$ Yukawa coupling. 
\vspace{0.5cm}
\begin{figure}[ht]
\centering
\vspace*{0.4cm}
\includegraphics[height=4.00cm]{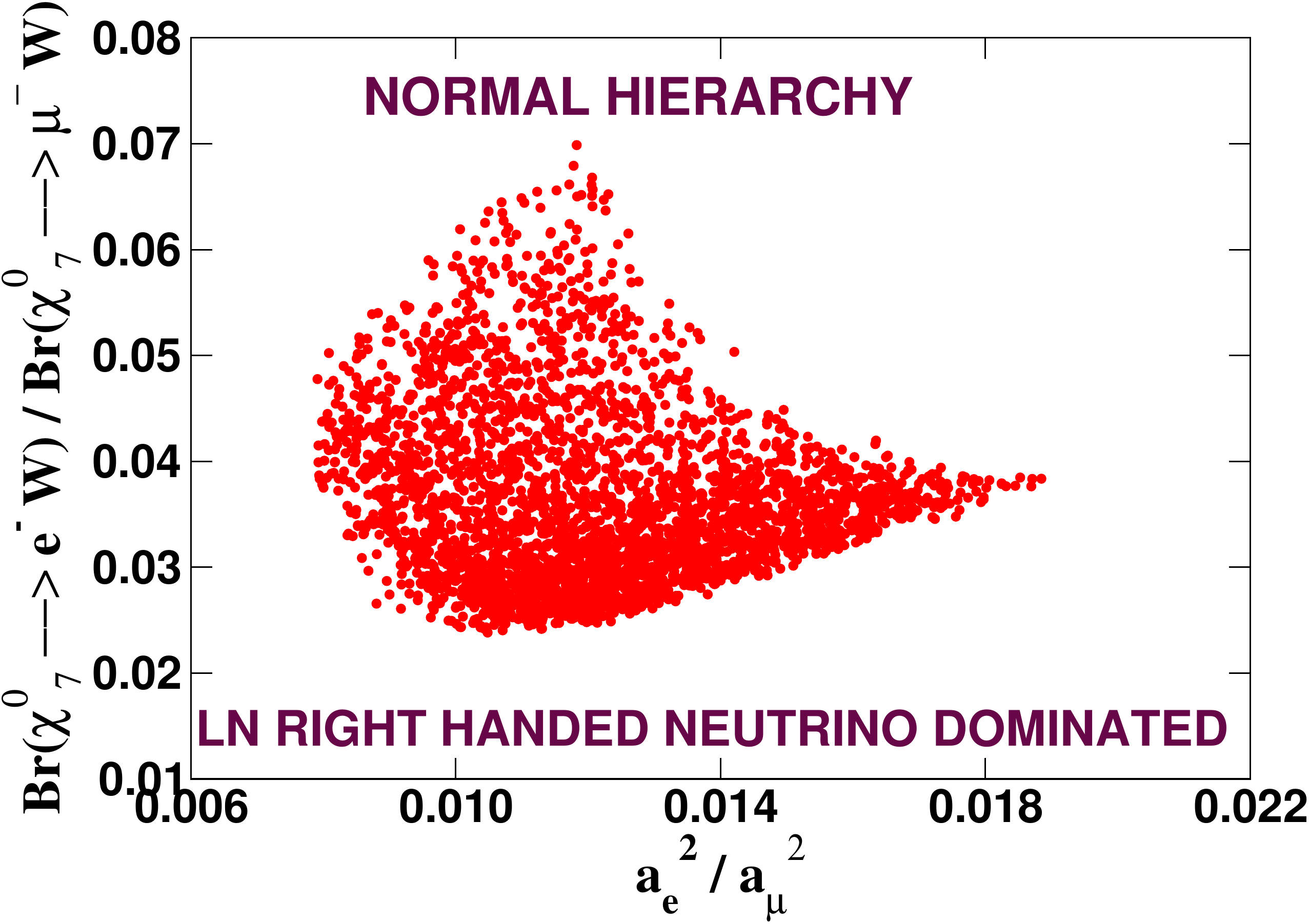}
\includegraphics[height=4.00cm]{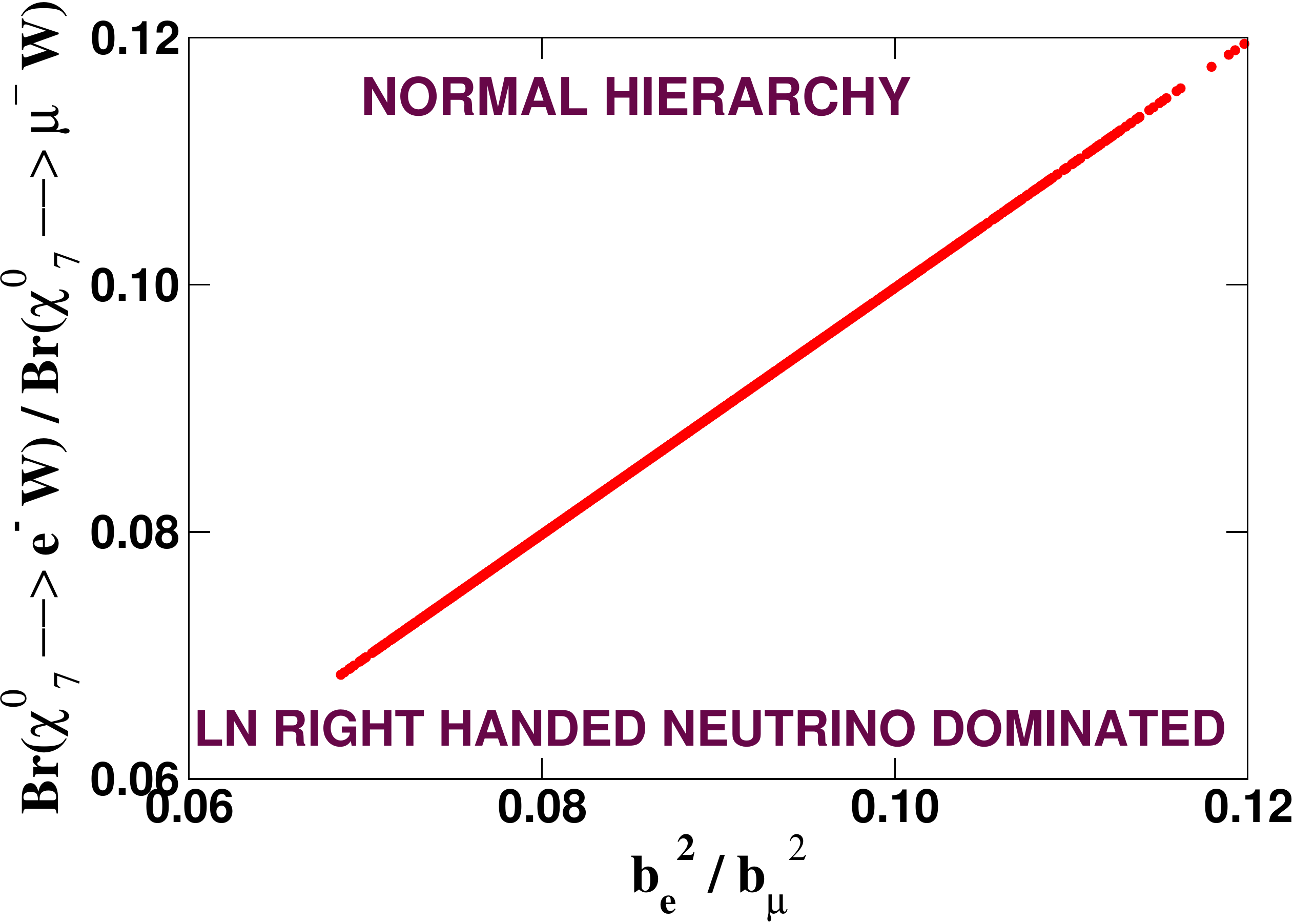}
\caption{Ratio $\frac{BR({\tilde \chi}^0_7 \longrightarrow e~W)}
{BR({\tilde \chi}^0_7 \longrightarrow \mu~W)}$ versus 
$\frac{a^2_e}{a^2_\mu}$ (left) and versus $\frac{b^2_e}{b^2_\mu}$ 
(right) plot for a $\nu^c$ like lightest neutralino (${\tilde \chi}^0_7$) 
with $\nu^c$ component $(|N_{75}|^2+|N_{76}|^2+|N_{77}|^2)>$~0.99 (left), and
$>$0.97 (right). Neutrino mass pattern is normal hierarchical. Choice of 
parameters are same as that of figure \ref{nor-singlino-bmbt-amat}.}
\label{nor-singlino-bebm-aeam}
\end{figure}

Similar calculations were performed also for the other ratios discussed
earlier. For example, in figure \ref{nor-singlino-bebm-aeam} we have shown the 
variations of the ratio $\frac{BR({\tilde \chi}^0_7 \longrightarrow e~W)} 
{BR({\tilde \chi}^0_7 \longrightarrow \mu~W)}$ as functions of 
$\frac{a^2_e}{a^2_\mu}$ and $\frac{b^2_e}{b^2_\mu}$ for the cases (i) and (ii),
respectively. The variation with $\frac{a^2_e}{a^2_\mu}$ is not sharp and 
dispersive in nature whereas the variation with $\frac{b^2_e}{b^2_\mu}$ is 
very sharp and shows that in this case the relevant couplings are proportional 
to $b_e$ and $b_\mu$, respectively. 

\vspace{0.5cm}
\vspace{0.5cm}
\begin{figure}[ht]
\centering
\vspace*{0.4cm}
\includegraphics[height=4.00cm]{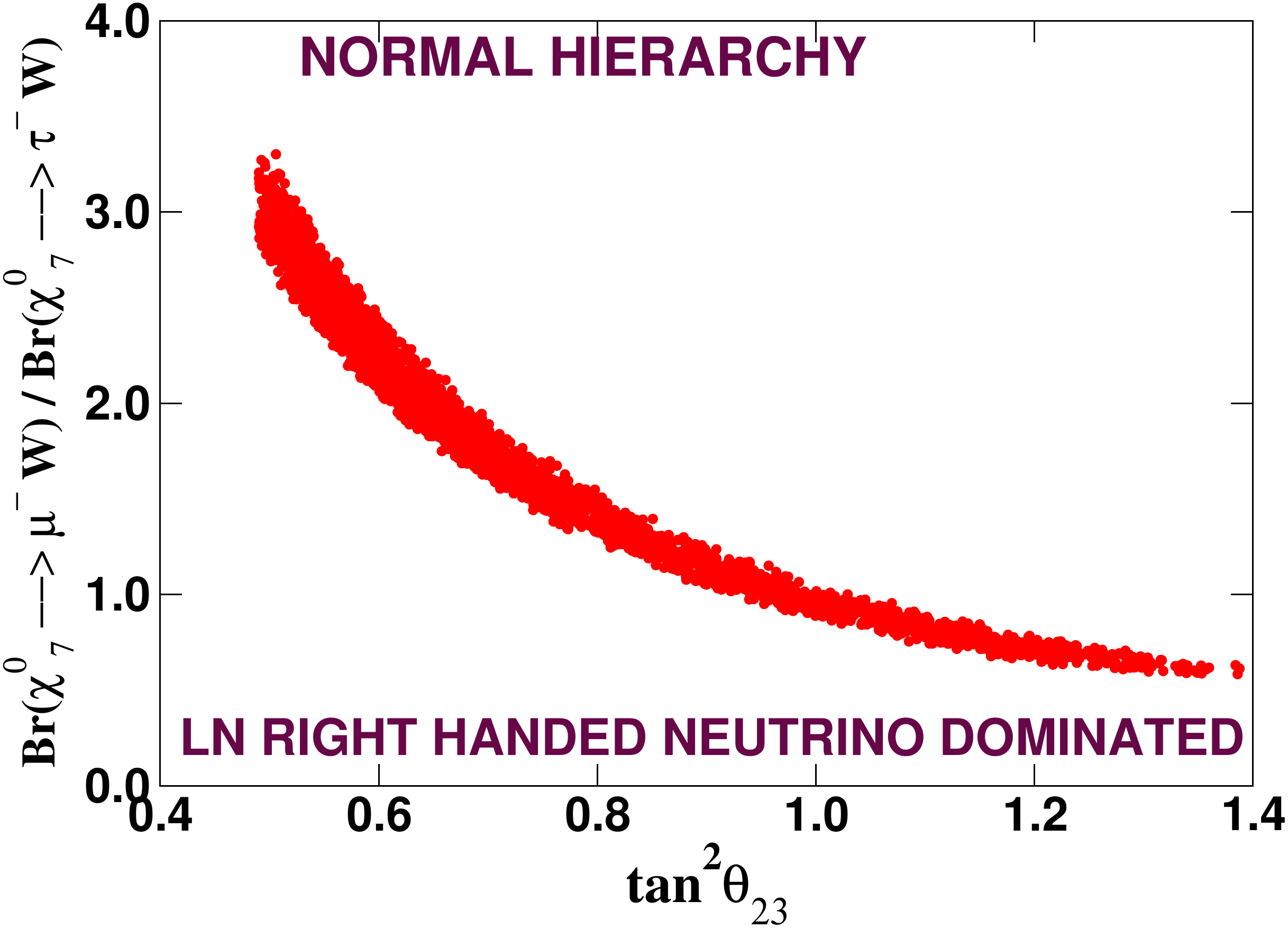}
\includegraphics[height=4.00cm]{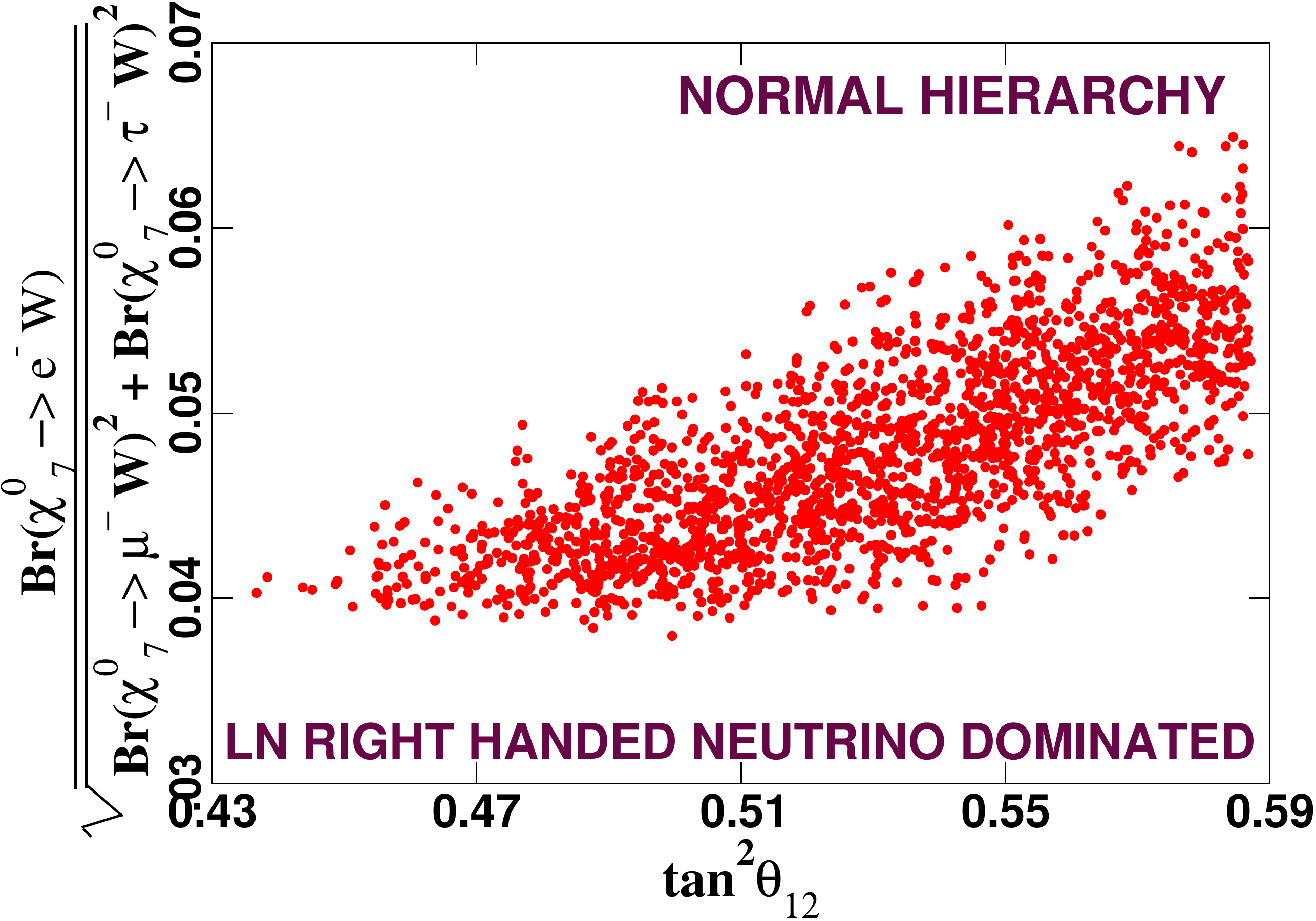}
\caption{Ratio $\frac{BR({\tilde \chi}^0_7 \longrightarrow \mu~W)}
{BR({\tilde \chi}^0_7 \longrightarrow \tau~W)}$ versus 
$\tan^2\theta_{23}$~(left), $\frac{BR({\tilde \chi}^0_7 \longrightarrow e~W)}
{\sqrt{BR({\tilde \chi}^0_7 \longrightarrow \mu~W)^2
+BR({\tilde \chi}^0_7 \longrightarrow \tau~W)^2}}$ with $\tan^2\theta_{12}$
~(right) plot for a $\nu^c$ dominated lightest neutralino with $\nu^c$ component 
$(|N_{75}|^2+|N_{76}|^2+|N_{77}|^2)>$~$0.99$ (left) and $>$ $0.97$ (right). 
Neutrino mass pattern is normal hierarchical. Choice of parameters are same 
as that of figure \ref{nor-singlino-bmbt-amat}.}
\label{nor-singlino-LN}
\end{figure}

On the other hand, in case (i) only $\tan^2\theta_{23}$ shows a nice 
correlation with the ratio $\frac{BR({\tilde \chi}^0_7 \longrightarrow \mu~W)}
{BR({\tilde \chi}^0_7 \longrightarrow \tau~W)}$ (see 
figure \ref{nor-singlino-LN}) and $\tan^2\theta_{12}$ or $\tan^2\theta_{13}$ 
does not show any correlation with the other ratio. The non-linear behaviour 
of the ratios of branching ratios in case(i) is due to the fact that the
parameters $Y_\nu$s (which control the $a_i$) appear both in the neutralino
and chargino mass matrices. The charged lepton Yukawa couplings also play a 
role in determining the ratios. One can also see that the
prediction for this ratio of branching ratio for case (i), as shown in 
figure \ref{nor-singlino-LN}, is in the range $0.5-3.5$, which is larger compared 
to the bino dominated or higgsino dominated cases for both normal and 
inverted hierarchical pattern of neutrino masses. Also, the nature of this 
variation is similar to what we see with the inverted hierarchical pattern of 
neutrino masses in the bino or higgsino dominated cases. 

In case (ii) none of the neutrino mixing angles show very good correlations 
with the ratios of branching ratios that we have been discussing. However, one 
can still observe some kind of correlation between $\tan^2\theta_{12}$ and the 
ratio $\frac{BR({\tilde \chi}^0_7 \longrightarrow e~W)} {\sqrt{BR({\tilde \chi}^0_7 
\longrightarrow \mu~W)^2 +BR({\tilde \chi}^0_7 \longrightarrow \tau~W)^2}}$. 
The prediction for this ratio from the neutrino data
is on the smaller side $(\sim 0.07)$. 

\vspace{0.5cm}
\begin{figure}[ht]
\centering
\vspace*{0.4cm}
\includegraphics[height=4.00cm]{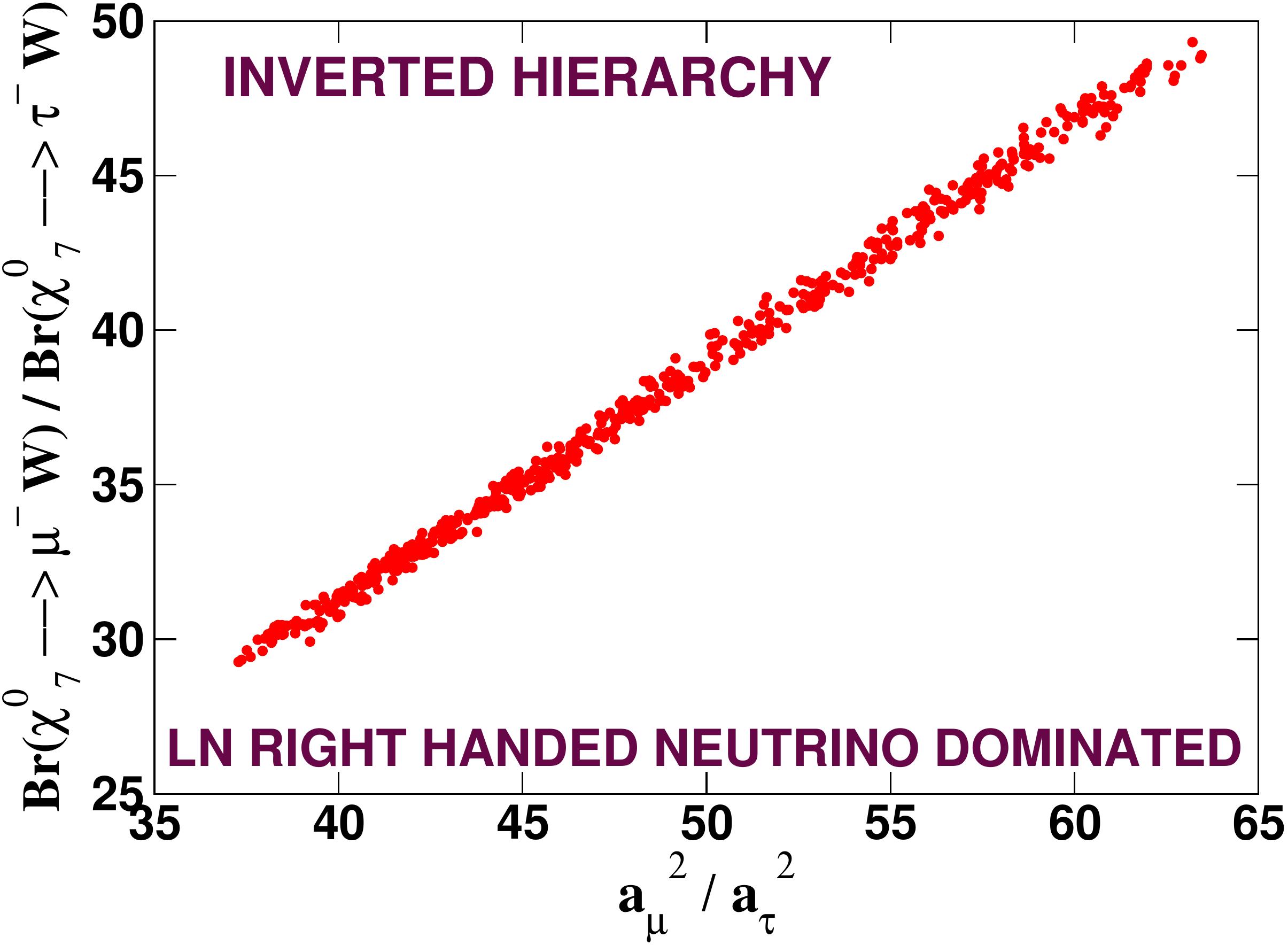}
\includegraphics[height=4.00cm]{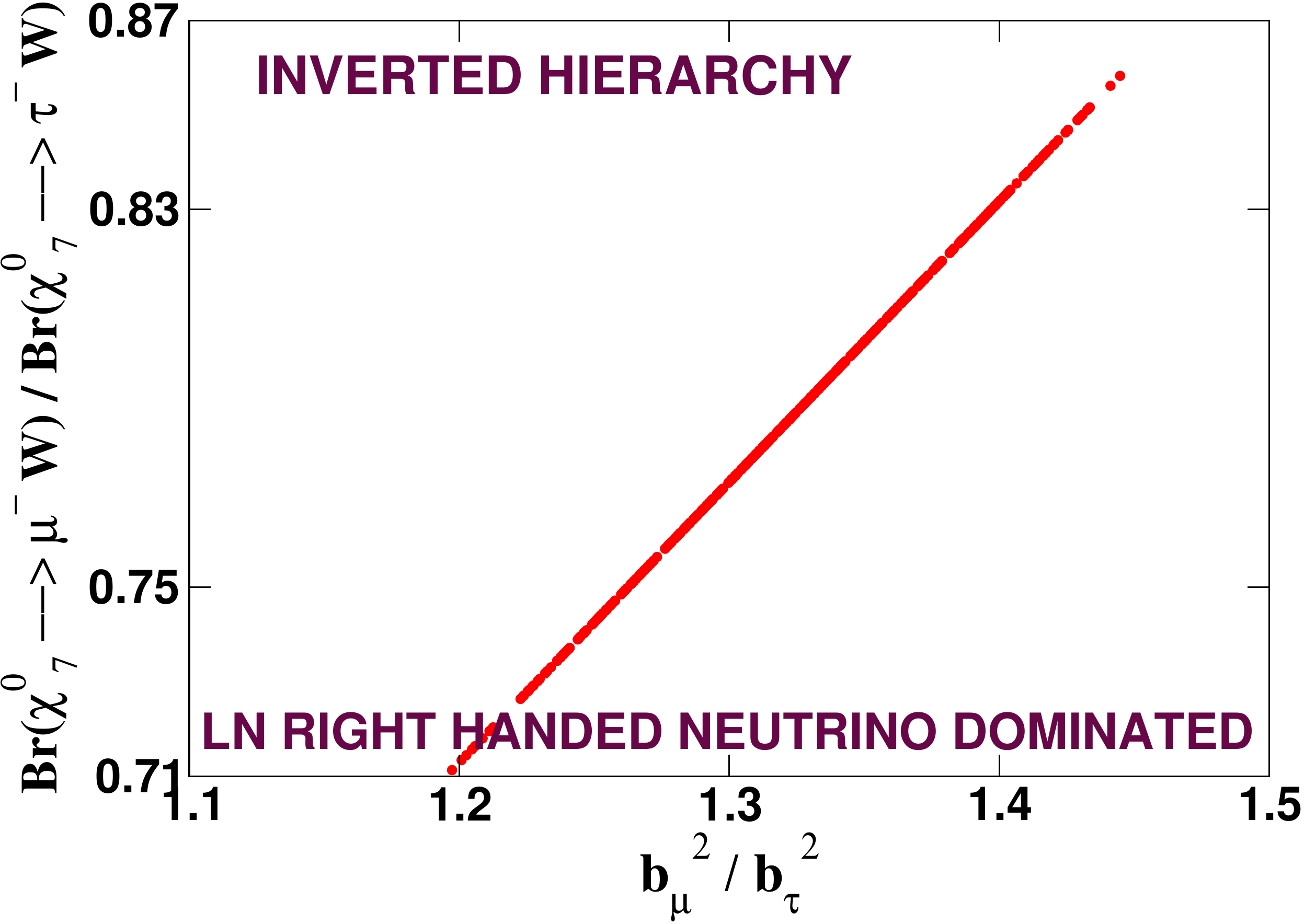}
\caption{Ratio $\frac{BR({\tilde \chi}^0_7 \longrightarrow \mu~W)}
{BR({\tilde \chi}^0_7 \longrightarrow \tau~W)}$ versus 
$\frac{a^2_\mu}{a^2_\tau}$ (left) and versus $\frac{b^2_\mu}{b^2_\tau}$ 
(right) plot for a $\nu^c$ like lightest neutralino (${\tilde \chi}^0_7$) 
with $\nu^c$ component $(|N_{75}|^2+|N_{76}|^2+|N_{77}|^2)>$~$0.99$ (left), and
$>~0.97$ (right). Neutrino mass pattern is inverted hierarchical. Choice of 
parameters are for (left) $M_1=445$ GeV, $\lambda=0.29, \kappa=0.07, 
(A_\lambda \lambda)= -8.2~{\rm TeV} \times \lambda,(A_\kappa \kappa)= 
165~{\rm GeV} \times \kappa,~m_{\tilde{\nu}^c}=50~{\rm GeV} ~{\rm and} 
~m_{\tilde{L}}=835 \rm{GeV}$ and for (right) $M_1=445$ GeV, $\lambda=0.10, 
\kappa=0.07, (A_\lambda \lambda)= -2~{\rm TeV} \times \lambda,(A_\kappa \kappa)
= 165~{\rm GeV} \times \kappa, ~m_{\tilde{\nu}^c}=50~{\rm GeV} ~{\rm and} 
~m_{\tilde{L}}=835~{\rm GeV}$. Mass of the lightest neutralino is $129.4$ GeV 
(left) and $119.8$ GeV (right) respectively.}
\label{inv-singlino-bmbt-amat}
\end{figure}
\vspace{2.3cm}

\begin{figure}[ht]
\centering
\vspace*{0.4cm}
\includegraphics[height=4.00cm]{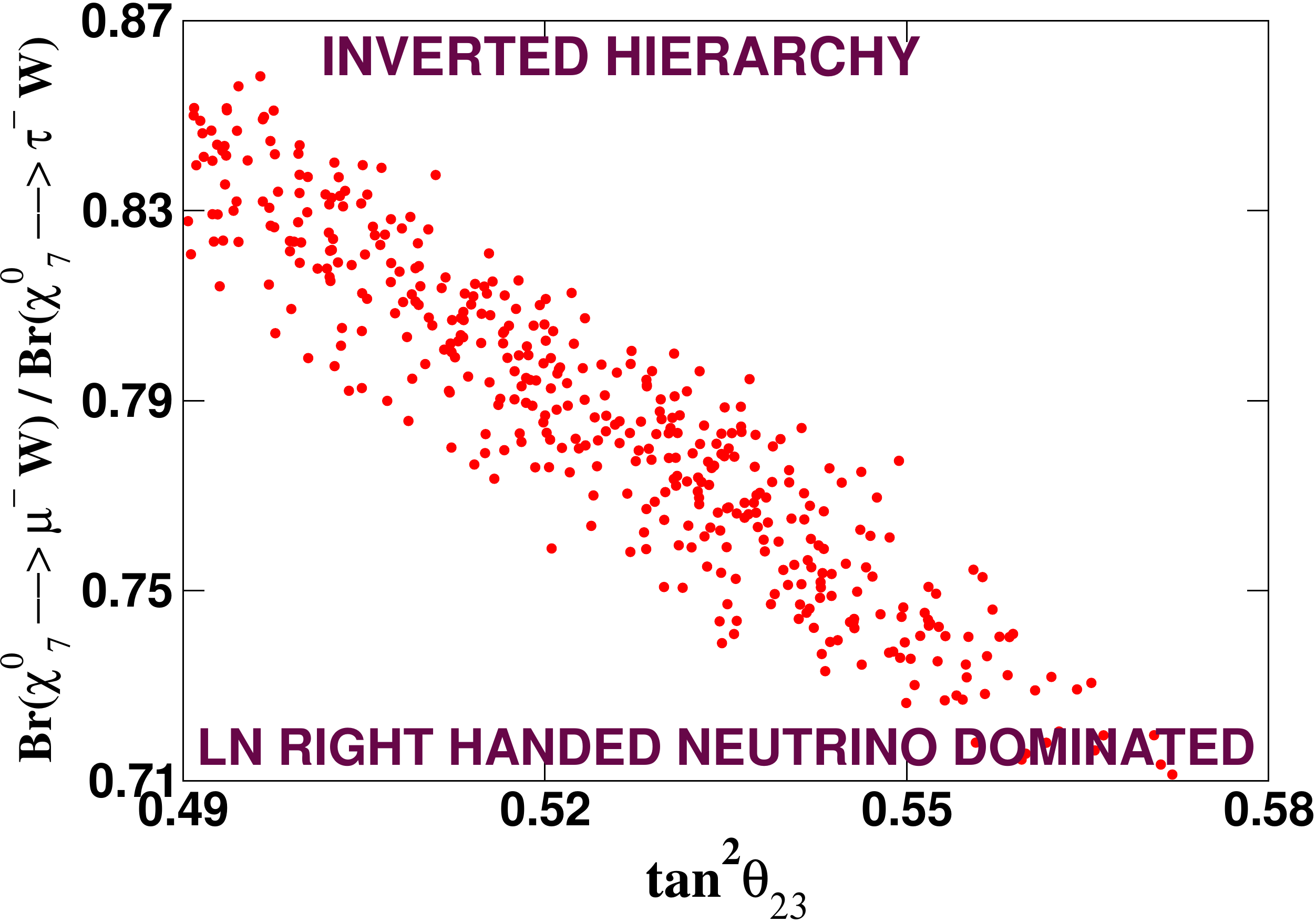}
\includegraphics[height=4.00cm]{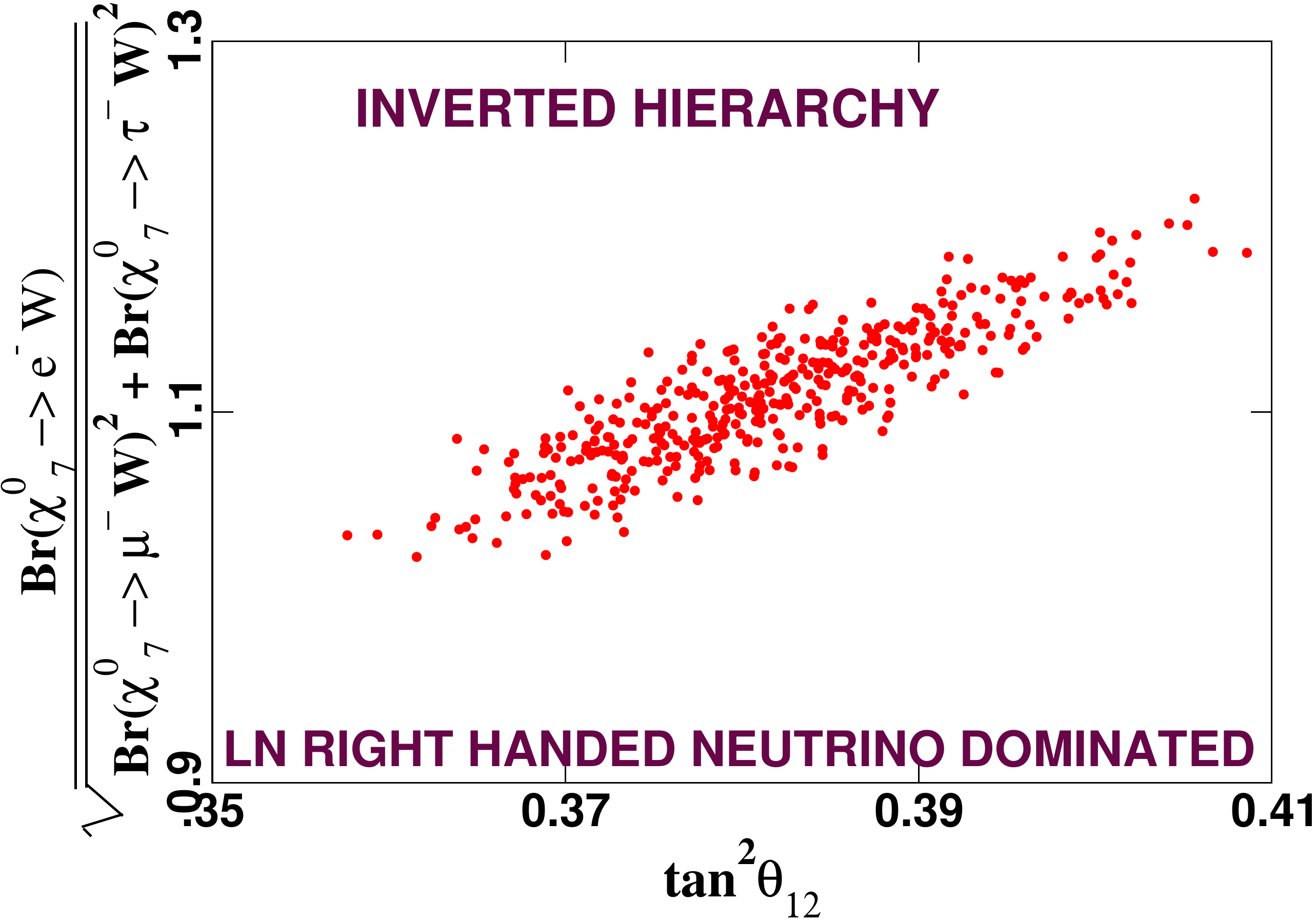}
\caption{Ratio $\frac{BR({\tilde \chi}^0_7 \longrightarrow \mu~W)}
{BR({\tilde \chi}^0_7 \longrightarrow \tau~W)}$ versus $\tan^2\theta_{23}$
~(left), $\frac{BR({\tilde \chi}^0_7 \longrightarrow e~W)}
{\sqrt{BR({\tilde \chi}^0_7 \longrightarrow \mu~W)^2
+BR({\tilde \chi}^0_7 \longrightarrow \tau~W)^2}}$ with $\tan^2\theta_{12}$
~(right) plot for a $\nu^c$ dominated lightest neutralino with $\nu^c$ 
component $(|N_{75}|^2+|N_{76}|^2+|N_{77}|^2)>$~$0.97$. Neutrino mass pattern is 
inverted hierarchical. Choice of parameters are same as that of figure 
\ref{inv-singlino-bmbt-amat}.}
\label{inv-singlino-LN}
\end{figure}

With the inverted hierarchical neutrino mass pattern, in case (i)
one observes a sharp correlation of the ratio 
$\frac{BR({\tilde \chi}^0_7 \longrightarrow \mu~W)}
{BR({\tilde \chi}^0_7 \longrightarrow \tau~W)}$ with 
$\frac{a^2_\mu}{a^2_\tau}$ (see figure \ref{inv-singlino-bmbt-amat}). 
The other two ratios $\frac{BR({\tilde \chi}^0_7 \longrightarrow e~W)}
{BR({\tilde \chi}^0_7 \longrightarrow \mu~W)}$ and 
$\frac{BR({\tilde \chi}^0_7 \longrightarrow e~W)}
{BR({\tilde \chi}^0_7 \longrightarrow \tau~W)}$ do not show very sharp 
correlations with $\frac{a^2_e}{a^2_\mu}$ and $\frac{a^2_e}{a^2_\tau}$,
respectively and we do not plot them here. However, in case (ii) all the
three ratios show nice correlations with the corresponding $b^2_i/b^2_j$. We
have shown this in figure \ref{inv-singlino-bmbt-amat} only for 
$b^2_\mu/b^2_\tau$. In this case the variations of the ratios of branching 
ratios with neutrino mixing angles are shown in figure \ref{inv-singlino-LN}. 

\vspace{0.9cm}

For the case (i), only $\tan^2\theta_{13}$ shows certain correlation with the
ratio of branching ratio shown in figure \ref{inv-singlino-LN} (right), but we
do not show it here.

Finally, we would like to reemphasize that in all these different cases discussed 
above, the lightest neutralino can have a finite decay length which can produce
displaced vertices (also discussed earlier in sections \ref{LSP-dec-1}, \ref{lsp-nat}) 
in the vertex detectors. Depending on the composition of the
lightest neutralino, one can have different decay lengths. For example, a
bino-dominated lightest neutralino can produce a displaced vertex $\sim$ a few
mm. Similarly, for a higgsino dominated lightest neutralino, decay vertices of
the order of a few cms can be observed \cite{c5Ghosh:2008yh,c5Bartl:2009an}. 
On the other hand, if the lightest 
neutralino is $\nu^c$ dominated, then the decay lengths can be of the order 
of a few meters \cite{c5Ghosh:2008yh,c5Bartl:2009an,c5Bandyopadhyay:2010cu}. 
These are very unique predictions of this model which can, 
in principle, be tested at the LHC \cite{c5Bandyopadhyay:2010cu}. 

The advantage of having large displaced vertices for a singlino like 
lightest neutralino makes it easier to kill all of the SM backgrounds
unambiguously. Additionally, it is also difficult to achieve a reasonably large ($\sim$ a 
few meter) displaced vertex in the conventional $R_p$-violating model \cite{c5Mukhopadhyaya:1998xj,
c5Choi:1999tq,c5Porod:2000hv}. As a consequence it is rather difficult for
the $R_p$-violating supersymmetric models to mimic
a specific collider signatures of $\mu\nu$SSM, particularly when a 
gauge singlet LSP is involved in the process. We will use the favour of large displaced
vertex associated with a singlino like LSP to describe an unconventional
signal of the lightest Higgs boson of $\mu\nu$SSM \cite{c5Bandyopadhyay:2010cu} 
in the next chapter.


\chapter{ \sffamily{{\bf $\mu$}$\nu$SSM: Unusual signal of Higgs boson at LHC
 }}\label{munuSSM-Higgs}
\section{{\bf H}iggs boson in $\mu\nu$SSM}

In $\mu\nu$SSM $R_p$ is violated through lepton number violation both in the superpotential and
in the soft terms. In this model neutral Higgs bosons of the MSSM mix with 
three generations of left and right-handed sneutrinos and thus the neutral
scalar and pseudoscalar squared mass matrices are enhanced $(8\times8)$
over their $2\times2$ MSSM structures \cite{c6Escudero:2008jg,c6Ghosh:2008yh}. In a similar fashion 
the charged scalar squared mass matrix is also a $8\times8$ matrix for $\mu\nu$SSM due to
mixing between charged Higgs of the MSSM and charged sleptons \cite{c6Escudero:2008jg,c6Ghosh:2008yh}. 
In general the nature of the lightest neutral scalar state can be very different from
that of the MSSM due to the presence of the gauge singlet right-handed
sneutrino component. It has been already shown that $\mu\nu$SSM is capable
of accommodating neutrino data both from tree level \cite{c6Ghosh:2008yh}
and one loop combined analysis \cite{c6Ghosh:2010zi}. With the initiation of 
the LHC experiment at CERN it is naturally tempting to see
whether this is capable of producing interesting collider signatures apart from 
accommodating the neutrino data.

The issues of Higgs boson discovery have been studied extensively over years
in the literature (see for example \cite{c6Barger:1987nn}).
In this chapter we propose a prodigious signal of Higgs boson in supersymmetry, 
having dilepton and four hadronic jets along with large displaced vertices 
$(\gsim 3 ~m)$ \cite{c6Bandyopadhyay:2010cu}.
Most of the usual signal of Higgs boson are
impaired by undesired backgrounds and one has to remove them somehow for claiming 
a discovery. Often the procedures for background subtraction in turn weaken the 
desired signal significantly. 
On the other hand, it was well known that the advantage of displaced vertices are
always extremely useful to kill all of the SM backgrounds and also some 
of the possible backgrounds arising from the $R_p$ violating MSSM. 
Displaced vertices arising from MSSM with $\rpv$ are usually 
much smaller \cite{c6Mukhopadhyaya:1998xj,c6Choi:1999tq,romao-dl,c6Porod:2000hv}

Now in the last chapter we have learned that in $\mu\nu$SSM, with suitable 
choice of parameters, a right-handed neutrino like lightest neutralino can be a viable candidate
for the LSP. It was also discussed that since a right-handed neutrino is singlet under the 
SM gauge group it can decay only in $R_p$-violating channels through small $R_p$-violating couplings 
and consequently the associated displaced vertices can be very large $(\sim {\rm{meter}})$
\cite{c6Bartl:2009an,c6Bandyopadhyay:2010cu}. Indeed these displaced vertices can
kill all of the SM backgrounds as well as backgrounds 
arising from MSSM with $\rpv$ \cite{c6Mukhopadhyaya:1998xj,
c6Choi:1999tq,romao-dl,c6Porod:2000hv}. Furthermore, imprint of 
this signal is different from that of the cosmic muons which have definite entry and exit 
point in the detector. So this is apparently a clean signal and a discovery, thus is 
definite even with small number of signal events.
In the next section we will discuss how to use the favour of
these large displaced vertices associated with a singlino like LSP for proposing
a new kind of signal of Higgs boson \cite{c6Bandyopadhyay:2010cu}.


\section{{\bf T}he Signal}\label{signal}

There are essentially two key features of our analysis, which collectively can lead
to an unusual signal of the Higgs boson in supersymmetry 

\vspace{0.1cm}
\noindent
1.~The lightest neutralino LSP $(\ntrl1)$ in the $\mu\nu$SSM with the parameter choice 
$M_1,\mu\gg m_{\nu^c}$ (see section \ref{lsp-nat}) can be predominantly composed of 
right-handed neutrinos which, as argued earlier will be called a $\nu^c$-like or
a singlino like LSP \cite{c6Bartl:2009an,c6Bandyopadhyay:2010cu}. For the
analysis of ref. \cite{c6Bandyopadhyay:2010cu} we choose 
$|\bN_{15}|^2 + |\bN_{16}|^2 + |\bN_{17}|^2 \geqslant 0.70$.

\begin{figure}[ht]
\centering
\vspace*{0.5cm}
\includegraphics[height=2.80cm]{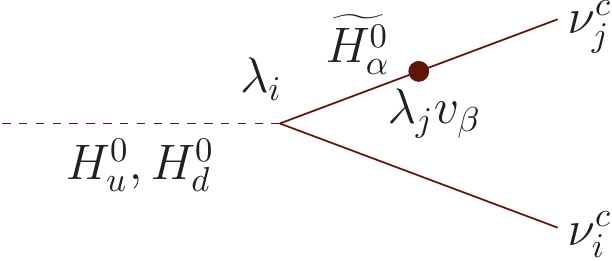}
\caption{Feynman diagram for the singlino singlino Higgs couplings. $\beta=2/1$ for $\al=d/u$.}
\label{lam-sing}
\end{figure}

\vspace{0.1cm}
\noindent
2.~A pair of singlino like LSP can couple to a Higgs boson in $\mu\nu$SSM
mainly through couplings like $\nu^cH_uH_d$ (see fig \ref{lam-sing}). 

The neutralino LSP, $\ntrl1$ in $\mu\nu$SSM can be predominantly 
$(\gsim 70\%)$ $\nu^c$-like (also known as a {\it{singlino}} LSP). 
$\ntrl1$ being singlet, $\ntrl1 \ntrl1 Z$ or $\ntrl1 q \widetilde q$ 
couplings \cite{c6Ghosh:2010zi} are vanishingly small, which in turn results in very small 
cross-section for direct $\ntrl1$ pair production. On the contrary, the coupling 
$\lam\nu^cH_uH_d$ may produce a large $\ntrl1 \ntrl1 S^0_i$ 
\cite{c6Ghosh:2010zi} coupling with $\lambda \sim$ $\cal{O}$ $(1)$, where $S^0_i$ are the scalar 
states. With the chosen set of parameters (see Table \ref{mass-spectrum}) 
we obtained $S^0_4 \equiv h^0$, where $h^0$ is the lightest Higgs boson of MSSM. 
In addition with heavy squark/gluino masses as indicated in Table \ref{mass-spectrum}
for different benchmark points, production of a singlino LSP through
cascade decays is suppressed.
In the backdrop of such a scenario, production of $h^0$ in gluon fusion channel 
followed by the decay process $h^0 \to \ntrl1 \ntrl1$ will be the leading production 
channel for the singlino LSP at the LHC. We want to emphasize here that
for the present analysis we choose to work with the tree level mass of the 
lightest CP-even Higgs boson $(S^0_4 \equiv h^0)$ of the $\mu\nu$SSM. With loop
corrections the Higgs boson mass can be higher \cite{c6Escudero:2008jg,c6Bartl:2009an}.
For loop corrected Higgs boson mass, the process $h^0\to\ntrl1\ntrl1$ will yield
heavy singlino like LSPs with smaller decay lengths \cite{c6Bartl:2009an}.
However, our general conclusions will not change for a singlino LSP in the 
mass range $20-60$ GeV, as long as the decay branching ratio for the 
process $h^0\to\ntrl1\ntrl1$ is substantial.

A set of four benchmark points (BP) used for collider studies
compatible with neutrino data \cite{c6Schwetz:2008er}, upto one-loop level 
analysis \cite{c6Ghosh:2010zi} are given in Table \ref{mass-spectrum}. 
These are sample points and similar spectra can be obtained in a reasonably
large region of the parameter space even after satisfying all the constraints 
from neutrino experiments.


\begin{table}[ht]
\centering
\begin{tabular}{ c || c  c  c  c}
\hline \hline
  & BP-1 & BP-2 & BP-3 & BP-4 \\ \hline \hline
$\mu$ & 177.0 & 196.68 & 153.43 & 149.12 \\ 
tan$\beta$ & 10 & 10 & 30 & 30 \\
$m_{h^0}~(\equiv m_{\ns4})$ & 91.21 & 91.63 & 92.74 & 92.83    \\ 
$m_{\ns1}$ & 48.58 & 49.33 & 47.27 & 49.84   \\
$m_{\nps2}$ & 47.21 & 49.60 & 59.05 & 49.45   \\ 
$m_{\cs2}$ & 187.11 & 187.10 & 187.21 & 187.21   \\ 
$m_{\bsq1}$ & 831.35 & 831.33 & 830.67 & 830.72  \\ 
$m_{\bsq2}$ & 875.03 & 875.05 & 875.72 & 875.67   \\ 
$m_{\tsq1}$ & 763.41 & 763.63 & 761.99 & 761.98   \\ 
$m_{\tsq2}$ & 961.38 & 961.21 & 962.46 & 962.48   \\ 
$m_{\ntrl1}$ & 43.0 & 44.07 & 44.20 & 44.24   \\ 
$m_{\ntrl2}$ & 55.70 & 57.64 & 61.17 & 60.49   \\ 
$m_{{\chpm4}}$ & 151.55 & 166.61 & 133.69 & 130.77 \\ 
\hline \hline
\end{tabular}
\caption{\label{mass-spectrum}
$\mu$-parameter and relevant mass spectrum (GeV) for chosen
benchmark points. $m_{\widetilde{\chi}^{\pm}_{1,2,3}} \equiv ~m_{e,\mu,\tau}$.
Only the relevant masses are shown here. Squark
masses of first two generations are $\sim 800$ GeV, which are not shown here. For our parameter
choices the fourth CP-even scalar state $S^0_4\equiv h^0$ \cite{c6Bandyopadhyay:2010cu}. 
The quantities $S^0,P^0,S^\pm,\wt \chi^0,\wt \chi^\pm$ represent 
physical scalar, pseudoscalar, charged scalar, neutralino and chargino states,
respectively. \cite{c6Escudero:2008jg,c6Ghosh:2008yh,c6Ghosh:2010zi}. The heavy quarks 
namely, bottom, charm and top masses are computed at the $m_Z$ mass scale or 
at the electroweak scale (see ref. \cite{c6Djouadi-Spira} and references therein).}
\end{table}

For the set of specified benchmark points (table \ref{mass-spectrum}), we observe, 
the process $h^0\to \ntrl1 \ntrl1$ to
be one of the dominant decay modes of $h^0$ (branching fraction within $35$-$65\%$), while
the process $h^0 \to b \bar{b}$ remains the main competitor. 
Different Feynman rules concerning Higgs decays are given in appendix \ref{appenH}. With
a suitable choice of benchmark points (table \ref{mass-spectrum}) two body
decays of $h^0$ into lighter scalar or pseudoscalar states were kept kinematically
disfavoured. Squared matrix elements for the processes $h^0\to b \bar{b}$
and $h^0 \to \ntrl1 \ntrl1$ are also given in appendix \ref{appenH}.

The pair produced singlino like $\ntrl1$ will finally decay into standard
model particles as shown in eqn.(\ref{2-3-body-decays}). For a lightest Higgs
boson mass $m_{h^0}$ as shown in table \ref{mass-spectrum}, mass of a singlino
like $\ntrl1$ $(m_{\ntrl1})$ arising from $h^0$ decay (see figure \ref{lam-sing}) 
is $< m_W$, and thus three body decays dominate. Out of the five
possible three body final states we choose to work with the specific decay 
mode $\ntrl1 \to q_i\bar{q}^{\prime}_j\ell^\pm_k$ 
to yield a signal $pp\to 2\ell + 4j + X$ in the final state\footnote{The dilepton have 
same sign on $50\%$ occurrence since $\ntrl1$ is a Majorana particle.}. This particular
final state is free from neutrinos and thus a reliable invariant mass reconstruction is 
very much possible. It has to be emphasized here that as suggested in ref. \cite{c6Barger:1987nn},
a reliable mass reconstruction is possible even for the final states with a single neutrino,
thus apart from the $2\ell + 4j + X$ final state there also exist other equally
interesting final states like $3\ell + 2j + X$ ($\ntrl1 \to q_i\bar{q}^{\prime}_j\ell^\pm_k,
\ntrl1 \to\ell^+_i \ell^-_j\nu_k$), $1\ell + 4j + X$ ($\ntrl1 \to q_i\bar{q}^{\prime}_j\ell^\pm_k
,\ntrl1 \to q_i\bar{q_i}\nu_k$) etc. For the chosen benchmark points, 
$Br(\ntrl1 \to q_i\bar{q}^{\prime}_j\ell^\pm_k)$ lies within $8-10\%$.
Squared matrix elements for all possible three body decays of $\ntrl1$
(see eqn.(\ref{2-3-body-decays})) are given in appendix \ref{appenI}.
At this point the importance of a singlino $\ntrl1$ becomes apparent. Since
all the leptons and jets are originating from the decays of a gauge singlet fermion,
the associated displaced vertices are very large $\sim 3-4$ meter, which
simply wash out any possible backgrounds. Detection of these 
displaced as well as isolated leptons and hadronic jets can lead to 
reliable mass reconstruction for $\ntrl1$ and Higgs boson in the absence of missing
energy in the final state. There is one more merit of this analysis, 
i.e., invariant mass reconstruction for a singlino LSP can give 
us an estimation of the seesaw scale, since the right-handed neutrinos are 
operational in light neutrino mass generation through a TeV scale seesaw 
mechanism \cite{c6LopezFogliani:2005yw,c6Ghosh:2008yh} in $\mu\nu$SSM.

It is important to note that in the real experimental ambience, extra jets can arise from 
initial state radiation (ISR) and final state radiation (FSR). Likewise semi-leptonic 
decays of quarks can accrue extra leptons. Also from the experimental point of view 
one cannot have zero missing $p_T$ in the final state. With this set of information 
we optimize our chosen signal as
%
\bea
(n_{j} \ge 4) + (n_{\ell} \geq 2) + (\ptmiss \le 30 ~{\rm{GeV}}),
\label{signal-choice}
\eea
%
where $n_{j(\ell)}$ represents the number of {\it{jets(leptons}}).

It should also be noted that, similar final states can appear from the decay of 
heavier scalar or pseudoscalar states in the model. Obviously, their production cross 
section will be smaller compared to $h^0$ and the invariant mass distribution 
(some other distributions also) should be different in those cases. So, in a sense
it is possible to discriminate this signal (eqn.(\ref{signal-choice})) from the model
backgrounds. Another possible source of backgrounds can arise from the cosmic muons.
However as discussed earlier, cosmic muons have definite entry and exit points inside a
detector and thus there signatures are different from the proposed signal.

\section{{\bf C}ollider analysis and detection}\label{coll-ana}

{\tt PYTHIA (version 6.4.22) \cite{c6Pythia6.4}} has been used for the purpose of event generation. 
The corresponding mass spectrum and decay branching fractions 
are fed to {\tt PYTHIA} by using the SLHA interface \cite{c6Skands-SLHA}. Subsequent decays of the
produced particles, hadronization and the collider analysis were performed using {\tt PYTHIA}. 
We used {\tt CTEQ5L} parton distribution function (PDF) \cite{c6Lai-CTEQ,c6Pumplin-PDF} for the analysis. 
The renormalization/factorization scale $Q$ was chosen to be the parton level center-of-mass energy, 
$\sqrt{\hat{s}}$. 
We also kept ISR, FSR and multiple interaction on for the analysis. The 
production cross-section of $h^0$ via gluon fusion channel for different benchmark points 
(table \ref{mass-spectrum}) is shown in table \ref{tabcross}.
\begin{table}[ht]
\centering
\begin{tabular}{ c || c  c  c  c}
\hline \hline
$\sqrt{s}$  & BP-1 & BP-2 & BP-3 & BP-4 \\ \hline \hline
$7$ TeV &6837 &7365 &6932 &6948  \\ 
$14$ TeV &23150 &25000 &23580 &23560 \\
\hline \hline
\end{tabular}
\caption{\label{tabcross}
Hard scattering cross-section in fb for the process
$gg \to h^0$ for PDF CTEQ5L with $Q = \sqrt{\hat{s}}$.}
\end{table}

%
We have used {\tt PYCELL}, the toy calorimeter simulation provided in
{\tt PYTHIA}, with the following criteria:

\noindent
I. The calorimeter coverage is $\rm |\eta| < 4.5$ and the segmentation is
given by $\rm\Delta\eta\times\Delta\phi= 0.09 \times 0.09 $ which resembles
a generic LHC detector.

\noindent
II. $\Delta R \equiv \sqrt{(\Delta\eta)^{2}+(\Delta\phi)^{2}} = 0.5$
        has been used in cone algorithm for jet finding.

\noindent
III. $p_{T,min}^{jet} = 10$ GeV.

\noindent
IV. No jet matches with a hard lepton in the event.

In addition, the following set of standard kinematic cuts
were incorporated throughout:

\noindent
1. $p_T^{\ell} \geq 5$ GeV and $\rm |\eta| _{\ell} \le 2.5$,

\noindent
2. $|\eta| _{j}\leq 2.5$, $\Delta R_{\ell j} \geq 0.4$,
$\Delta R_{\ell\ell}\geq 0.2,~$

\noindent
where $\Delta R_{\ell j}$ and $\Delta R_{\ell \ell}$ measure the lepton-jet and 
lepton-lepton isolation, respectively. Events with isolated leptons, having $p_T\ge 5$ 
GeV are taken for the final state analysis.

\begin{figure}[ht]
\centering
\includegraphics[width=11.55cm,keepaspectratio]{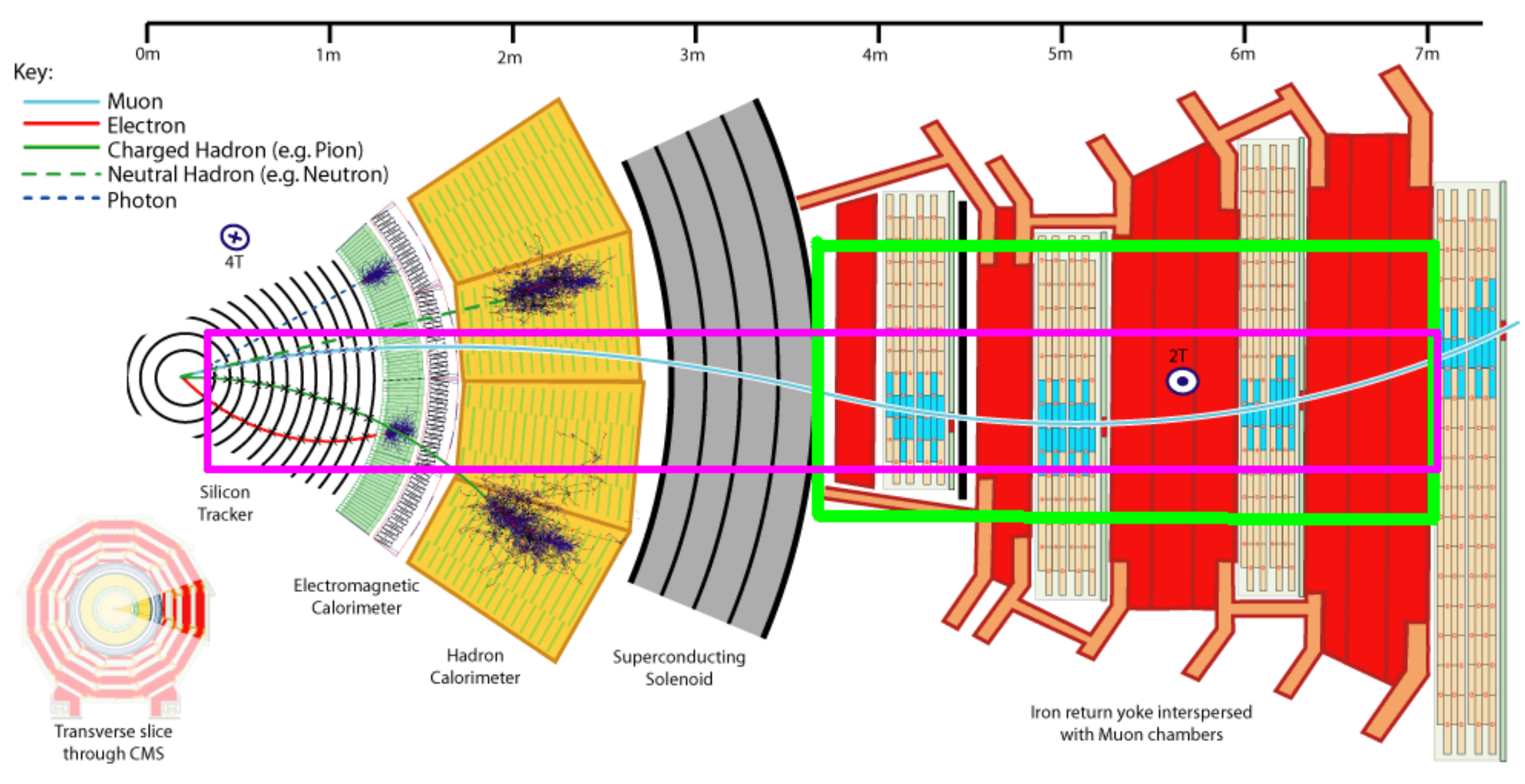}
\caption{Transverse slice from the CMS detector. The maroon square corresponds
to the global muons which travel throughout the detector starting from the interaction
point. The light green square on the other hand corresponds to the stand-alone muons
which leave their imprints only in the muon chamber.}
\label{CMS-detec}
\end{figure}

Now depending on the distribution of the transverse decay length it is 
possible to study the behaviour of this spectacular signal in different 
regions of a generic LHC detector like CMS or ATLAS. For the purpose of
illustration we present a slice like picture of the CMS detector in figure
\ref{CMS-detec} to describe this novel signal in more details.

Let us now analyze this rare signal (see eqn.(\ref{signal-choice})) piece wise
for the CMS detector as shown by figure \ref{CMS-detec}. We choose BP-2 as
the sample benchmark point. To start with we divide the entire detector in 
five different regions on the basis of different transverse decay lengths $(L_{T})$ 
and conduct our analysis. The decay length $(L)$ is given by

\bea\label{decay-length-defn}
L = c \tau (\beta \gamma),
\eea
where $c$ is the speed of light in vacuum ($=1$ in natural unit system),
$\tau$ is the proper life time and the kinematical factor $\beta\gamma=
\frac{|\vec{p}|}{m}$. Here $|\vec{p}|$ is the magnitude of the three momentum
$=\sqrt{|p_x|^2+|p_y|^2+|p_z|^2}$ and $m$ is the mass of the decaying particle. Now it is in general
difficult to measure the longitudinal component of the momentum $(p_z)$ which
lies along the beam axis, thus we choose to work with the transverse decay length
given by
\bea\label{decay-length-defn-T}
L_T = c \tau (\beta \gamma)_T,
\eea
where $(\beta\gamma)_T=\frac{\sqrt{|p_x|^2+|p_y|^2}}{m}$.

\vspace{0.2cm}
\noindent
I.~ $L_{T}\leq1~cm$ $\blacktriangleright$ Roughly $10\%$ to $15\%$ 
of the total number of events appear in this region. These events are close to the 
interaction point and may be mimicked by MSSM models with $\rpv$. Thus we do not consider these 
points in our analysis though these are also free from the SM backgrounds.

\vspace{0.2cm}
\noindent
II.~ $1~cm<L_{T}\leq50~cm$ $\blacktriangleright$  There exist reasonable number of events 
$(\sim 30\%$ of the total events) with decay length in between {\it{$1$ cm {\rm{and}} $50$ cm}}.
For these events the associated electrons and muons\footnote{$\tau$'s are dropped out 
for poor detection efficiency.} will leave charged tracks in the inner silicon tracker
as well as the electrons will deposit their energy at the electronic calorimeter (ECAL).
Associated hadronic jets will also deposit their energy at the hadronic calorimeter (HCAL).
The associated muons are {\it{global}} in nature and leave their marks throughout,
upto the muon chamber starting from few layers on the inner tracker. 
It is easy for the conventional triggers to work for this kind of
signal and a reliable mass reconstruction of these displaced hadronic jets and leptons
can lead to a discovery. The number of signal events in this region are shown in table
\ref{tabevents-cut}.

\vspace{0.2cm}
\noindent
III.~ $50~cm<L_{T}\leq3~m$ $\blacktriangleright$ Almost $40\%$ of the total events
appear in this region. The associated electrons and hadronic jets may or may not get
detected in this situation depending on the length of the displaced vertices. However,
the associated muons will leave tracks either in the muon chamber only or in
the muon chamber along with matching tracks in the inner detector also.
The number of signal events in this region are also given 
in table \ref{tabevents-cut}. 

\vspace{0.2cm}
\noindent
IV.~ $3~m<L_{T}\leq6~m$ $\blacktriangleright$ There exist some number of
events $(\sim 10\%$ of the total number of events) which appear only in the 
territory of the muon chamber. In this case the associated electrons get
absorbed in the thick iron yoke of the muon chamber and thus escape detection.
Besides, it is also difficult to identify the hadronic jets as jets in the muon 
chamber, rather they appear as noise. The muons are, however leave visible tracks
in the muon chamber {\it{only}} indicating their {\it{stand-alone}} natures.
It is indeed difficult for the conventional triggers to work for this specific signal, 
rather this asks for a dedicated special trigger which we believe is a challenging task 
for experimentalists. The corresponding number of events in this region for BP-2
are shown in table \ref{tabevents-cut}.

\vspace{0.2cm}
\noindent
V.~ $L_{T}\geq7~m$ $\blacktriangleright$ There also exist a small number of
events $(\sim 4\%)$ where decays occur outside the detector and yield conventional
missing energy signature.

\begin{table}[ht]
\centering
\begin{tabular}{ c || c || c  c  c}
\hline \hline
&  &  & No. of events & \\
$\sqrt{s}$ & signal & $L_{T_1}$ & $L_{T_2}$ & $L_{T_3}$\\ \hline\hline
& $\ge 4j + \geq 2\ell + \ptmiss \le 30~\rm{GeV}$
&45  &69   &17 \\ 
7 & $\ge 4j + \geq 2\mu + \ptmiss \le 30~\rm{GeV}$ 
&27  &38  &11   \\ 
$\rm{TeV}$ & $\ge 4j + \geq 2e + \ptmiss \le 30~\rm{GeV}$ 
&6  &10  &2  \\ 
& $\ge 4j + 1e +1\mu + \ptmiss \le 30~\rm{GeV}$
&12  &21 &4 \\ \hline\hline
 & $\ge 4j + \geq 2\ell + \ptmiss \le 30~\rm{GeV}$
&234 &373  &98 \\
14 & $\ge 4j + \geq 2\mu + \ptmiss \le 30~\rm{GeV}$ 
&128 &218  &58 \\ 
$\rm{TeV}$ & $\ge 4j + \geq 2e + \ptmiss \le 30~\rm{GeV}$
&37 &45 &16  \\ 
& $\ge 4j + + 1e +1\mu + \ptmiss \le 30~\rm{GeV}$
&69 &113  &24\\
\hline \hline
\end{tabular}
\caption{\label{tabevents-cut}
Number of signal events for $\mathcal{L} =$ $5 ~\rm{fb}^{-1}$ 
for $\sqrt{s}=7$ and $14~\rm{TeV}$ at different ranges of the decay length for BP-2
with $1~cm<L_{T_1}\le50~cm$, $50~cm<L_{T_2}\le3~m$ and $3~m<L_{T_3}\le6~m$.
$L_{T_i}$s are different transverse decay lengths.}
\end{table}

\begin{figure}[ht]
\centering
\includegraphics[width=6.05cm,angle=-90]{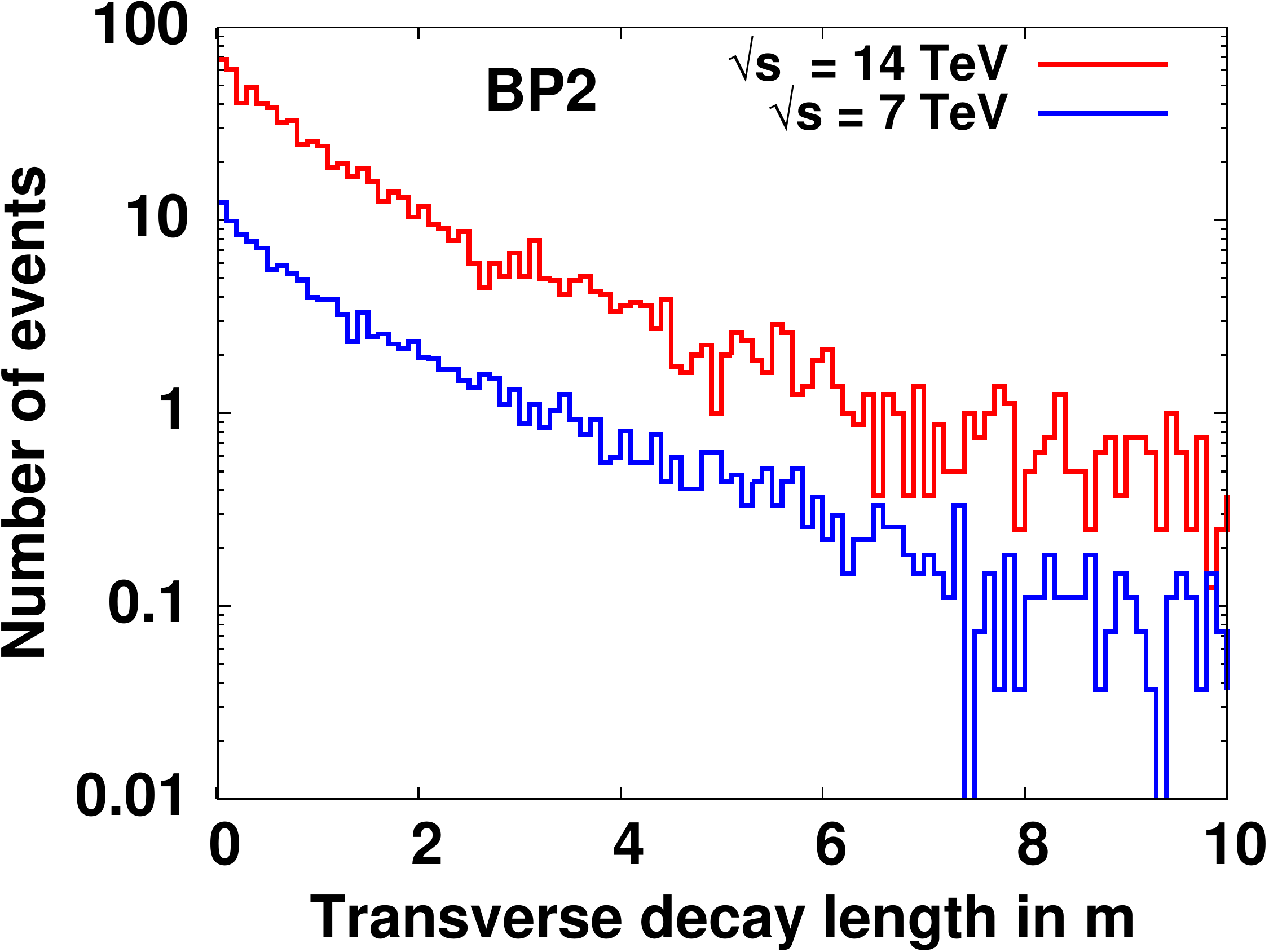}
\caption{Transverse decay length distribution of $\ntrl1$ for $\sqrt{s} =  7$ 
and $14$ TeV with BP-2 for a typical detector size $\sim$ $10~m$ with $\mathcal{L}=$
$5 ~\rm{fb}^{-1}$. Minimum bin size is $10~cm$. 
The signal is given by eqn.(\ref{signal-choice}).}
\label{Decay-Length}
\end{figure}

The number of events for different length of displaced vertices as addressed
earlier are shown in table \ref{tabevents-cut} both for center-of-mass energy
$7$ and $14$ TeV with an integrated luminosity of $5 ~\rm{fb}^{-1}$. Since this
is a background free signal, even with this number of events this spectacular signal can lead to  
discovery at 14 TeV run of the LHC with $\mathcal{L} =$ 5 fb$^{-1}$. At 7 TeV the 
situation looks less promising and higher luminosity might be required for 
discovering such an event. Distribution of the transverse decay length is shown
by figure \ref{Decay-Length}.

In summary, this signal can give rise to non-standard activities 
in the muon chamber with two muons and four hadronic jets. There are, however,
number of events which can leave their imprints not only at the 
muon chamber but also in the inner tracker and calorimeters concurrently. 
Integrating these two signatures can lead to discovery of an unusual signal 
of Higgs boson at the $14$ TeV run of the LHC. 
Though with higher luminosity discovery at $\sqrt{s}=7$ TeV is also possible. Indubitably, 
development of new triggers and event reconstruction tools are essential. 

It is also important to note that the average decay length for a singlino like LSP 
is determined by the LSP mass as well as by a set of parameters
$(\lambda,\kappa,v^c,Y^{ii}_\nu,v'_i)$ so
that the constraints on neutrino masses and mixing are satisfied.
Here $v^c$ and $v'_i$ stand for the vacuum expectation values of
the right and left-handed sneutrino fields.


\section{{\bf C}orrelations with neutrino mixing angles}\label{neut-corr}

\begin{center}
\begin{figure}[ht]
\centering
\includegraphics[width=10.55cm]{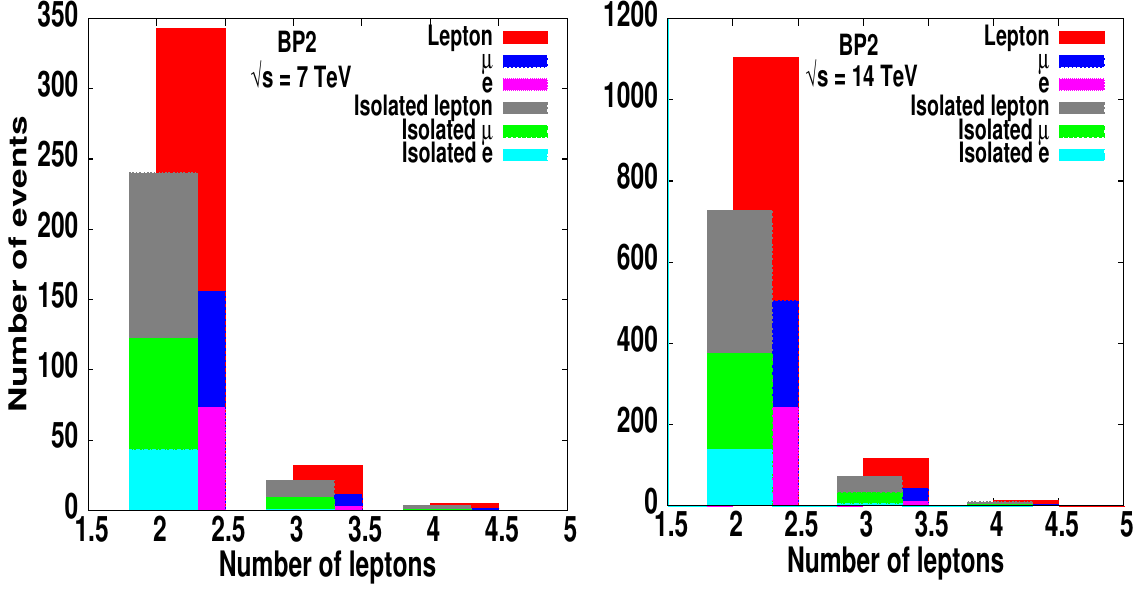}
\caption{Lepton multiplicity  distribution of signal
for $\sqrt{s} = 7$ and $14$ TeV with $1~\rm{fb}^{-1}$
of integrated luminosity.}
\label{lmlt}
\end{figure} 
\end{center}

One of the striking features in $\mu\nu$SSM is that certain ratios of branching fractions of 
the LSP decay modes are correlated with the neutrino mixing angles \cite{c6Ghosh:2008yh,c6Bartl:2009an}. 
These correlations have been explored in details in chapter \ref{munuSSM-LSP}. A
consequence of the correlation with solar mixing angle $\theta_{12}$
implies $n_{\mu} > n_{e}$ in the final state. Figure \ref{lmlt} shows the lepton 
multiplicity distribution for inclusive $\geq 2\ell$ ($\geq 2 \mu~+\geq 2 e+1\mu,1e$)
and exclusive ($\geq 2 \mu$, $\geq 2 e$) for BP-2, without the signal criteria 
(eqn.(\ref{signal-choice})). Muon dominance of the higher histograms 
(without any isolation cuts) continues to the lower ones even after the application 
of $\Delta R_{\ell j},~\Delta R_{\ell \ell}$ cuts. Consequently we observe that the 
correlation between $n_e$ and $n_{\mu}$ also appears in the lower 
histograms (figure \ref{lmlt}) with a ratio $n_e:n_\mu$ $\sim$ $1:3$.

We present number of events for final state signal (eqn.(\ref{signal-choice})) 
in table \ref{tabevents} both for $\sqrt{s}=7$ and $14$ TeV for $\mathcal{L} = 5~ {\rm fb}^{-1}$, 
without a cut on the actual $\ntrl1$ decay position (like table \ref{tabevents-cut}). It is important 
to note from table \ref{tabevents} that the correlation between $n_e$ and $n_{\mu}$ in the final
state is still well maintained, similar to what was shown in the lower histograms of 
figure \ref{lmlt} even with the final state signal topology (eqn.(\ref{signal-choice})). 

\begin{table}[ht]
\centering
\begin{tabular}{ c || c || c  c  c c}
\hline \hline
$\sqrt{s}$ & signal & BP-1 & BP-2 & BP-3 & BP-4 \\ \hline\hline
& $\ge 4j + \geq 2\ell + \ptmiss \le 30~\rm{GeV}$
&181  &153    &170 &173  \\ 
7 & $\ge 4j + \geq 2\mu + \ptmiss \le 30~\rm{GeV}$ 
&100  &85    &97  &100 \\ 
$\rm{TeV}$ & $\ge 4j + \geq 2e + \ptmiss \le 30~\rm{GeV}$ 
&27  &23   &21  & 23\\ 
& $\ge 4j + 1e +1\mu + \ptmiss \le 30~\rm{GeV}$
&54  &46  &52 &50\\ \hline\hline
 & $\ge 4j + \geq 2\ell + \ptmiss \le 30~\rm{GeV}$
&1043 &878  & 951& 929 \\
14 & $\ge 4j + \geq 2\mu + \ptmiss \le 30~\rm{GeV}$ 
&580 &463  & 533 & 513\\ 
$\rm{TeV}$ & $\ge 4j + \geq 2e + \ptmiss \le 30~\rm{GeV}$
&160 &139  & 121&129  \\ 
& $\ge 4j + + 1e +1\mu + \ptmiss \le 30~\rm{GeV}$
&306 &279 &  300&290 \\
\hline \hline
\end{tabular}
\caption{\label{tabevents}
Expected number of events of signals for $\mathcal{L} =$ $5 ~\rm{fb}^{-1}$ 
for $\sqrt{s}=7$ and $14~\rm{TeV}$.}
\end{table}

\section{{\bf I}nvariant mass}\label{inv-mass}
\begin{figure}[ht]
\centering
\includegraphics[width=8.55cm]{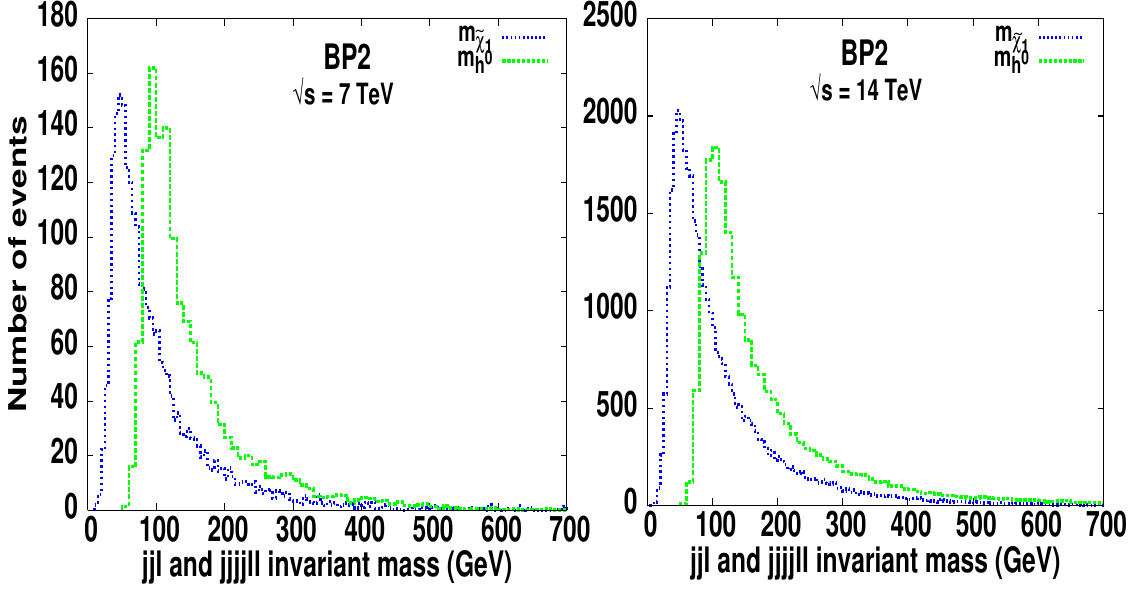}
\caption{Invariant mass distribution for (a) $\ntrl1$ $(jj\ell )$ 
and (b) the Higgs boson $(jjjj\ell\ell )$. Plots are shown for
$\sqrt{s} = 7$ and $14$ TeV with $1~\rm{fb}^{-1}$
of integrated luminosity.
Number of events for reconstructing
$m_{\ntrl1}$ for $\sqrt{s} = 7(14)$ TeV
are scaled by a multiplicative factor $4(7)$.}
\label{invariant-masses}
\end{figure}

It has been already argued in section \ref{coll-ana} that with a trustworthy 
detection of the two isolated and displaced muons and(or) electrons and four associated
hadronic jets a background free signal of this kind can lead to
definite discovery. We have already discussed about the possibility
for invariant mass reconstruction using those leptons and jets, not only
for a singlino like LSP but also for $h^0$. Results of invariant mass reconstruction
for $\ntrl1$ and $h^0$ for BP-2 are shown in figure \ref{invariant-masses}.
We choose $jj\ell$ invariant mass $M(jj\ell)$ for $m_{\ntrl1}$ reconstruction.
Reconstruction of $m_{h^0}$ was achieved through $M(jjjj\ell\ell)$,
invariant mass of $jjjj\ell\ell$ (see eqn.(\ref{signal-choice})).
We take the jets and leptons from the window of 
$35\, \rm{GeV}\, \leq M(jj\ell)\leq \, 45\, \rm{GeV}$ to construct 
$M(jjjj\ell\ell)$. Even a narrow window like this cannot kill all the 
combinatorial backgrounds. As a corollary, effect of combinatorial background 
for $m_{\ntrl1}$ reconstruction ($^4C_2$ for $j$ and $^2C_1$ for $\ell$)
also causes long tail for Higgs mass distribution.

In conclusion, we have studied an unusual but spectacular signal of 
Higgs boson in supersymmetry. This signal can give rise to non-standard activities 
in the muon chamber with two muons and four hadronic jets. There are, however,
number of events which can leave their imprints not only at the 
muon chamber but also in the inner tracker and calorimeters concurrently. 
Integrating these two signatures can lead to discovery of an unusual signal 
of Higgs boson at the 14 TeV run of the LHC. 
Though with higher luminosity discovery at $\sqrt{s}=$7 TeV is also possible. Indubitably, 
development of new triggers and event reconstruction tools are essential. This signal is 
generic to a class of models where gauge-singlet neutrinos and $\rpv$ take part simultaneously 
in generating neutrino masses and mixing. Another interesting feature of this study is that 
the number of muonic events in the final state is larger than the number of electron events 
and the ratio of these two numbers can be predicted from the study of the neutrino mixing angles.



\chapter{ \sffamily{{\bf S}ummary and Conclusion
 }}\label{con-sum}

The standard model (SM) of particle physics has already been firmly 
established as one of the very successful theories in physics
as revealed by a host of experiments. However, there are issues
where the SM is an apparent failure. Perhaps the severe most problem
of the SM is that the scalar mass is not protected by any symmetry arguments.
Thus the Higgs boson mass (only scalar in the SM) can be as large as the
Planck scale with radiative corrections. It appears that in the SM an unnatural
fine-tuning in the Higgs sector is essential for a Higgs boson mass consistent
with the requirements of theory and experiment.
On the other side, non-vanishing
neutrino masses as have been confirmed by experiments, are impossible
to explain with the SM alone. These shortcomings, as discussed in chapter \ref{SM},
ask for some new physics requirement at and beyond the TeV scale.

As a candidate theory to explain new physics beyond the TeV scale together
with solutions to the drawbacks of the SM, supersymmetry has sought
tremendous attention for the last few decades. 
A supersymmetric theory includes new particles
having spin difference of half-integral unit with that of the SM members.
The scalar masses are no longer unprotected and consequently the Higgs boson mass
remains free from quadratic divergences under radiative corrections.
However, missing experimental evidences for sparticles have confirmed
that supersymmetry must be broken in nature so that sparticles remain
heavier compared to their SM partners. It was pointed out in chapter \ref{susy}
that supersymmetry must be broken softly, 
so that only logarithmic divergences appear in the 
Higgs boson mass which requires sparticle masses around
the TeV scale. This is the prime reason why the discovery of TeV scale superpartners are highly
envisaged at the LHC. The definite mechanism
for supersymmetry breaking remains yet a debatable issue and consequently different
mechanisms exist in literature.
Turning our attention to the neutrino sector it appears that it is possible
to accommodate massive neutrinos in supersymmetric theories either through
$R$-parity violation or using seesaw mechanisms with extra particle content.
It must be emphasized here that in spite of being successful in solving some of the
drawbacks of the SM, supersymmetric theories
are also not free from shortcomings, which in turn result in a wide variant of
models. To mention one, as briefly reviewed in chapter \ref{susy}, the non-minimal supersymmetric
standard model was required to propose a solution to the $\mu$-problem of the minimal version.

Issues of the neutrino masses and mixing remain the prime focus of this thesis.
Requirement of massive neutrinos were essential to explain phenomena like
atmospheric and solar neutrino problem as observed in oscillation experiments. 
From experimental constraints, a neutrino mass is expected to be very small.
So it remains to be answered how one can generate tiny neutrino masses consistent with the 
oscillation data. Moreover, it also remains to be answered whether the neutrinos
are Dirac or Majorana particles by nature. We review these issues in
chapter \ref{neut} along with different mechanism of light neutrino mass generation 
both in supersymmetric and non-supersymmetric theories. The seesaw mechanisms
turn out to be the simple most ways to generate small neutrino masses both in
supersymmetric and non-supersymmetric theories at the cost of enhanced particle
content. But there also exists models of neutrino mass generation through radiative
effects. On the contrary, neutrino mass generation through the violation of $R$-parity
is a pure supersymmetric phenomena without any SM analogue. 
Sources of $R$-parity violation can be either spontaneous or explicit.
In the conventional
$R$-parity violating (bilinear and trilinear) models loop corrections are unavoidable 
to accommodate neutrino data. Bilinear $R$-parity violating models of neutrino mass generation
have one more striking feature, that is the existence of nice correlations
between the neutrino mixing angles and the lightest supersymmetric particle decay modes.
In addition decays of the lightest supersymmetric particle for these class of models
produce measurable displaced vertices which together with the fore stated correlations
can act as a very promising probe for these models at the colliders. 
All of these spectacular features of the $R$-parity
violating models have made them perhaps the most well studied models in the context
of supersymmetry.
 
Apart from inevitable loop corrections to satisfy three flavour
neutrino data, models with bilinear $R$-parity violation suffer from the naturalness 
problem similar to the $\mu$-problem, which is better known as the 
$\epsilon$-problem. A new kind of supersymmetric model of neutrino mass generation with 
a simultaneous solution to the $\mu$-problem has been introduced in chapter \ref{munuSSM-neut}.
This model is known as the $\mu\nu$SSM which by virtue of an imposed $Z_3$ symmetry 
is completely free from naturalness
problem like $\mu$ or $\epsilon$-problem. 
$\mu\nu$SSM introduces the gauge singlet right-handed neutrino 
superfields (${\hat\nu}^c_i$) to solve 
the $\mu$ problem in a way similar to that of NMSSM. These right-handed neutrinos
are also instrumental for light neutrino mass generation in $\mu\nu$SSM.
The terms in the superpotential involving the ${\hat\nu}^c_i$ include the neutrino 
Yukawa couplings, the trilinear interaction terms among the singlet neutrino 
superfields as well as a term which couples the Higgs superfields to the 
$\hat\nu^c_i$. In addition, there are corresponding soft SUSY breaking terms in 
the scalar potential. When the scalar components of ${\hat\nu}^c_i$ get VEVs 
through the minimization conditions of the scalar potential, an effective 
$\mu$ term with an EW scale magnitude is generated.
Explicit $\rpv$ in $\mu\nu$SSM through lepton number violation
both in the superpotential and in the soft supersymmetry
breaking Lagrangian result in enlarged $(8\times8)$ scalar,
pseudoscalar and charged scalar squared mass matrix. Also
the neutralino and chargino mass matrices are enhanced.
Small Majorana masses of the active neutrinos are generated due 
to the mixing with the neutralinos as well as due to the seesaw mechanism 
involving the gauge singlet neutrinos. In such a scenario, 
we show that it is possible to provide a theory of neutrino masses and mixing 
explaining the experimental data, even with a flavour diagonal neutrino Yukawa 
coupling matrix, without resort to an arbitrary flavour structure in the 
neutrino sector. This essentially happens because of the mixing involved in 
the neutralino-neutrino (both doublet and singlet) system mentioned above.
Light neutrino mass generation in $\mu\nu$SSM
is a combined effect of $R$-parity violation and a TeV scale seesaw mechanism using 
right handed neutrinos. Alternatively, as shown in chapter \ref{munuSSM-neut}
a combined effect of Type-I and Type-III seesaw mechanisms are instrumental for neutrino
mass generation in the $\mu\nu$SSM. 
The TeV scale seesaw mechanism itself is very interesting since it may 
provide a direct way to probe the gauge singlet neutrino mass scale at the
LHC and does not need to introduce a very high energy scale in the theory, 
as in the case of GUT seesaw. 
We present a detailed analytical and numerical work and show 
that the three flavour neutrino data can be accommodated in such a scenario. 
In addition, we observe that in this model different neutrino mass hierarchies 
can also be obtained with correct mixing pattern, at the tree level. 

Though all three neutrinos acquire masses at the tree
level, it is always important to judge the stability of tree level analysis
in the exposure of radiative corrections. In chapter \ref{munuSSM-neut} effect of the complete
set of one-loop corrections to the light neutrino masses and mixing are considered. The 
tree level and the one-loop corrected neutrino mass matrix are observed to posses similar
structure but with different coefficients arising from the loop corrections. 
The effects of one-loop corrections are found
to be capable of altering the tree level analysis 
in an appreciable manner depending on the concerned
mass hierarchy. We also explore different regions in the SUSY parameter
space, which can accommodate the three patterns in turn.

In conclusion, $\mu\nu$SSM can accommodate neutrino masses and mixing consistent
with the three flavour global neutrino data for different mass hierarchies at the 
tree level itself even with the choice of flavour diagonal neutrino Yukawa
couplings. Inclusion of one-loop radiative corrections to light neutrino masses
and mixing can alter the results of tree level analysis in a significant manner,
depending on the concerned mass orderings.

Correlations between the light neutrino mixing angles with the 
ratios of certain decay branching ratios of the lightest supersymmetric
particle (usually the lightest neutralino for a large region of
the parameter space) 
in $\mu\nu$SSM have been explored in chapter \ref{munuSSM-LSP}.
These correlations are very similar to the bilinear $\rpv$ models and
have drawn immense
attention as a test of neutrino mixing at a collider experiment. 
However, there exist certain differences between these two scenarios.
In $\mu\nu$SSM 
lepton number is broken explicitly in the superpotential by terms which are trilinear 
as well as linear in singlet neutrino superfields. In addition to that there are 
lepton number conserving terms involving the singlet neutrino superfields with dimensionless neutrino 
Yukawa couplings. After the electroweak symmetry breaking these terms can generate the effective bilinear 
R-parity violating terms as well as the $\Delta L$ =2 Majorana mass terms for the singlet neutrinos in the 
superpotential. In general, there are corresponding soft supersymmetry breaking terms in the scalar potential.
Thus the parameter space of this model is much larger compared to the bilinear R-parity violating model. 
Hence, in general, one would not expect a very tight correlation between the neutrino mixing angles and the
ratios of decay branching ratios of the LSP. However, under certain simplifying assumptions 
one can reduce the number of free parameters and in those cases it is possible that the above 
correlations reappear. As mentioned earlier, we have studied these correlations 
in great detail for the two body $\ell^\pm W^\mp$ 
final states. These nice correlations are lost in the general scenario of bilinear-plus-trilinear 
R-parity violation.

Another important difference between $\mu\nu$SSM and the bilinear R-parity violating model in the context of 
the decay of the LSP (assumed to be the lightest neutralino in this case) is that in $\mu\nu$SSM the lightest 
neutralino can have a significant singlet neutrino ($\nu^c$) contribution. In this case, the correlation between 
neutrino mixing angles and decay branching ratios of the LSP is different compared 
to the cases when the dominant component of the LSP is either a bino, or a higgsino or a Wino. This gives us a
possibility of distinguishing between different R-parity violating models through the observation of the
decay branching ratios of the LSP in collider experiments. In addition, the decay of 
the lightest neutralino will show displaced vertices in collider experiments and when the lightest neutralino
is predominantly a singlet neutrino, the decay length can be of the order of a few meters for a lightest 
neutralino mass in the neighbourhood of $50$ GeV. This is very different from the bilinear R-parity 
violating model where for a Bino LSP of similar mass the decay length is less than or of the order of a meter 
or so. 

In a nutshell we have studied the correlations among the ratio of the branching
ratios of the lightest supersymmetric particle decays into $W$-boson and a charged lepton
with different relevant parameters. These correlations are analysed for different natures
of the lightest neutralino which is usually the lightest supersymmetric particle for
a novel region of the parameter space. Besides, effect of different light neutrino mass hierarchies 
in the correlation study are
also taken into account. These spectacular and nice correlations together with a measurement of the 
displaced vertices can act as an important experimental signature for the $\mu\nu$SSM.

We shift our attention to a different aspect of the $\mu\nu$SSM in chapter \ref{munuSSM-Higgs}, 
where a new kind of unconventional signal for the Higgs boson in supersymmetry has
been advocated. The basic idea lies in the fact that with suitable choice of model parameters
a right handed neutrino like lightest supersymmetric particle is possible in the $\mu\nu$SSM
and a pair of such gauge singlet fermions can couple to a MSSM like Higgs boson. We show that
with heavy squark and gluino masses,
pair production of the right handed neutrino like lightest supersymmetric particles from the 
decay of a MSSM like Higgs boson, produced in the gluon fusion channel at the LHC can 
be the dominant source for singlino pair production.

We analyze a specific final state signal with two isolated
and displaced leptons (electron and(or) muon) and four isolated and displaced hadronic jets
arising from the three body decay of a pair of right handed neutrino like 
lightest supersymmetric particles.
This particular final state has the advantage of zero missing energy since no neutrinos
appear in the process and thus a reliable Higgs boson mass reconstruction as well as
the same for a right handed neutrino are highly envisaged.
Appearance of reasonably large displaced vertices associated with the gauge singlet nature
of a right handed neutrino are extremely useful to abate any SM backgrounds for this proposed
signal. Besides, presence
of the definite entry and the exit points for the cosmic muons also helps to discriminate this
signal from the cosmic muon signature. Depending on the length of the associated displaced
vertices this rare signal can either leave its imprints in the entire detector, starting
from the tracker to the muon chamber with conventional global muon signature or can leave
visible tracks in the muon chamber only from stand alone muons. The latter case also
requires development of special kind of triggers. Combining the two fore mentioned scenarios
a discovery with this signal criteria is expected with the $14$ TeV run of the LHC. 
This unusual signal is also
testable in the $7$ TeV LHC run but requires much higher luminosity 
compared to the $14$ TeV scenario. Ratio of the number
of electrons to that of the muons in the final state signal is again observed to show
correlation with the concerned neutrino mixing angle. We present a set of four benchmark points
where the neutrino data are satisfied up to one-loop level. Apart from the Higgs discovery,
a signal of this kind with a faithful mass reconstruction for right handed neutrino like
lightest supersymmetric particle offers a possibility to probe the seesaw scale which is
one more appealing feature of the $\mu\nu$SSM. It must be emphasized here that though
we performed this analysis with tree level Higgs boson mass in $\mu\nu$SSM, but
even for loop corrected Higgs boson mass our general
conclusions will not change for a singlino LSP in the
mass range $20-60$ GeV.

To conclude, $\mu\nu$SSM is a potential candidate for explaining physics beyond the standard
model. Not only this model can accommodate massive neutrinos consistent with the three
flavour global data but at the same time also offers a solution to the $\mu$-problem
of supersymmetry with the use of same set of right handed neutrino superfields. This model 
is also phenomenologically very rich and can yield new kind of signatures at collider experiments.
Diverse interesting aspects of the $\mu\nu$SSM have been addressed in this thesis 
and more studies are expected to reveal more phenomenological wonders in the near future. There
are a host of areas yet to be explored for this model like the effect of complete one-loop corrections 
in the scalar sector, more detailed analysis of new kind of Higgs signal at the colliders, a
comparative study of different lightest supersymmetric particle scenarios in the context
of an accelerator experiment and many more. In a nutshell,
with the LHC running at the corner we expect to explore more wonders of the $\mu\nu$SSM.

\appendix
\chapter{ \sffamily{{\bf }
 }}\label{appenA}

\section{Scalar mass squared matrices in MSSM}\label{MSSM-matsq-scalar}

\begin{flushleft}
{\it{$\maltese$ Neutral scalar, 
$(\mathcal{M}^2_{MSSM-scalar})_{2\times2}$}}
in the basis $(\Re H^0_d, \Re H^0_u) \Rrightarrow$
\end{flushleft}
%
\bea
\left(\begin{array}{c c}
B_\mu tan\beta -\frac{\mu}{v_1} \vp^\al v'_\al + 2\gamma_g v^2_1
 & -2 \gamma_g v_1 v_2 + B_\mu \\
-2 \gamma_g v_1 v_2 + B_\mu &
B_\mu cot\beta + B_{\vp_\al} \frac{v'_\al}{v_2} + 2\gamma_g v^2_2
\end{array}\right).
\label{MSSM-cp-even-scalar}
\eea
%
\begin{flushleft}
{\it{$\maltese$ Neutral pseudoscalar 
$(\mathcal{M}^2_{MSSM-pseudoscalar})_{2\times2}$}}
in the basis $(\Im H^0_d, \Im H^0_u) \Rrightarrow$
\end{flushleft}
%
\bea
\left(\begin{array}{c c}
B_\mu tan\beta -\mu \vp^\al \frac{v'_\al}{v_1} 
 &  B_\mu \\
 B_\mu &
B_\mu cot\beta + B_{\vp_\al} \frac{v'_\al}{v_2}
\end{array}\right).
\label{MSSM-cp-odd-scalar}
\eea
%
\begin{flushleft}
{\it{$\maltese$ Charged scalar 
$(\mathcal{M}^2_{MSSM-charged})_{2\times2}$}}
in the basis $(H^+_d, H^+_u) \Rrightarrow$
\end{flushleft}
%
\bea
(\mathcal{M}^2_{MSSM-charged})_{2\times2} = \left(\begin{array}{c c}
C^2_{11} & C^2_{12}\\
C^2_{21} & C^2_{22}
\end{array}\right),
\label{MSSM-charged-scalar1}
\eea
where
\bea
C^2_{11} &=& B_\mu tan\beta -\mu\vp^\al \frac{v'_\al}{v_1} 
+ Y^{\al\rho}_e Y^{\beta\rho}_e v'_\al v'_\beta 
-\frac{g^2_2}{2}\{v'^2_\al-v^2_2\}, \nn \\
C^2_{12} &=& B_\mu + \frac{g^2_2}{2}v_1v_2, ~C^2_{21} = C^2_{12}, \nn \\
C^2_{22} &=& B_\mu cot\beta + B_{\vp_\al} \frac{v'_\al}{v_2}  
+ \frac{g^2_2}{2}\{v'^2_\al+v^2_1\}.
\label{MSSM-charged-scalar2}
\eea
In these derivations minimization equations for $H_u,H_d$ has been
used, which are given by
\bea
 {\vp^2_\al} v_2- B_{\vp_\al}{v'_\al}-B_\mu v_1 + ({m^2_{H_u}}+{\mu^2}) v_2
-{\gamma_g}{\xi_{\upsilon}} {v_2} 
= 0,\nn \\
\mu \vp^\al v'_\al -B_\mu v_2 + ({m^2_{H_d}}+\mu^2)v_1 
+{\gamma_g}{\xi_{\upsilon}} {v_1} = 0,
\label{MSSM-Higgs-minima}
\eea
with $\gamma_{g} = \frac{1}{4}({g_1^2 + g_2^2})$ and 
$\xi_{\upsilon} ={\sum {v'^2_\al} + v_1^2 -v_2^2}$.

\section{Fermionic mass matrices in MSSM}
\label{MSSM-matsq-fermion}

\begin{flushleft}
{\it{$\maltese$ Chargino mass matrix 
$(M^{chargino}_{MSSM})_{2\times2}$}}
in the basis $-i \wt {\lambda}^{+}_{2}, \wt{H}^+_u$ (column)
and $-i \wt {\lambda}^{-}_{2}, \wt{H}^-_d$ (row)
$\Rrightarrow$
\end{flushleft}
\bea
(M^{chargino}_{MSSM})_{2\times2} =
\left(\begin{array}{c c}
M_2 & g_2v_2\\
g_2v_1& \mu
\end{array}\right).
\label{MSSM-chargino}
\eea

\begin{flushleft}
{\it{$\maltese$ Neutralino mass matrix 
$(M^{neutralino}_{MSSM})_{4\times4}$}}
in the basis $-i\wt B^0, -i\wt W_3^0, \wt H_d^0,
\wt H_u^0\Rrightarrow$
\end{flushleft}
\bea
(M^{neutralino}_{MSSM})_{4\times4} =
\left(\begin{array}{c c c c}
M_1 & 0 & -\frac{g_1}{\sqrt{2}}v_1 & \frac{g_1}{\sqrt{2}}v_2  \\
0 & M_2 & \frac{g_2}{\sqrt{2}}v_1 & -\frac{g_2}{\sqrt{2}}v_2 \\
-\frac{g_1}{\sqrt{2}}v_1 & \frac{g_2}{\sqrt{2}}v_1 & 0 & -{\mu} \\
\frac{g_1}{\sqrt{2}}v_2 & -\frac{g_2}{\sqrt{2}}v_2 & -{\mu} & 0 
\end{array}\right).
\label{MSSM-neutralino}
\eea

\chapter{ \sffamily{{\bf }
 }}\label{appenB}

\section{Scalar mass squared matrices in $\mu\nu$SSM}\label{munuSSM-matsq-scalar}

Decomposition of various neutral scalar fields of $\mu\nu$SSM in 
real ($\Re$) and imaginary (${\Im}$) parts are as follows
\bea
{H^0_d}&=&\Re{H^0_d}+\Im{H^0_d}=
{H^0_{d{\mathcal R}}}+i{H^0_{d{\mathcal I}}},\nonumber \\
{H^0_u}&=&\Re{H^0_u}+\Im{H^0_u}=
{H^0_{u{\mathcal R}}}+i{H^0_{u{\mathcal I}}}, \nonumber \\
{\wt \nu}^c_k&=&\Re{\wt \nu^c_k}+\Im{\wt \nu^c_k}=
{\wt \nu}^c_{k{\mathcal R}} 
+i{\wt \nu}^c_{k{\mathcal I}},\nonumber \\
{\wt \nu}_k&=&\Re{\wt \nu_k}+\Im{\wt \nu_k}=
{\wt \nu}_{k{\mathcal R}}+i{\wt \nu}_{k{\mathcal I}}.
\label{decomposition}
\eea
Only the real components get VEVs as indicated in eqn.(\ref{munuSSM-vevs}).  

The entries of the scalar and pseudoscalar mass-squared matrices are defined as
\beq
({M^2_S})^{\alpha \beta}=\langle\frac{1}{2}\frac{\partial^2{V_{neutral}}}
{\partial{\phi^{\alpha}_{\mathcal R}}\partial{\phi^{\beta}_{\mathcal R}}}
\rangle,
~~({M^2_P})^{\alpha \beta}=\langle\frac{1}{2}\frac{\partial^2{V_{neutral}}}
{\partial{\phi^{\alpha}_{\mathcal I}}\partial{\phi^{\beta}_{\mathcal I}}}
\rangle, 
\label{mass_square_matrix_working_formula}
\eeq
where 
\bea
{\phi^{\alpha}_{\mathcal R}}&=&{H^0_{d{\mathcal R}}}, {H^0_{u{\mathcal R}}}, 
{\wt \nu}^c_{k{\mathcal R}}, {\wt \nu}_{k{\mathcal R}},\nonumber \\
{\phi^{\alpha}_{\mathcal I}}&=&{H^0_{d{\mathcal I}}}, {H^0_{u{\mathcal I}}}, 
{\wt \nu}^c_{k{\mathcal I}}, {\wt \nu}_{k{\mathcal I}}. 
\eea
Note that the Greek indices $\alpha, \beta$ are used to refer 
various scalar and pseudoscalar Higgs and both $\rm{SU(2)_L}$ doublet and singlet 
sneutrinos, that is $H^0_d, H^0_u, {\wt \nu}^c_k, {\wt \nu}_k$, 
whereas k is used as a subscript to specify various flavours of doublet and 
singlet sneutrinos i.e., $k=e, {\mu}, {\tau}$ in the flavour 
(weak interaction) basis.

\begin{flushleft}
{\it{$\maltese$ Neutral scalar}} $\blacktriangleright$ 
\end{flushleft}
In the flavour basis or weak interaction basis $\Phi^T_{S}=({H^0_{d{\mathcal R}}},{H^0_{u{\mathcal R}}},
{{\widetilde{\nu}}^c_{n{\mathcal R}}},{{\widetilde{\nu}_{n{\mathcal R}}}})$,\footnote{In refs.
\cite{AppGhosh:2008yh,AppGhosh:2010zi} ${H^0_{1{\mathcal R}}},{H^0_{2{\mathcal R}}}$ was used
in lieu of ${H^0_{d{\mathcal R}}},{H^0_{u{\mathcal R}}}$.} the scalar mass term in 
the Lagrangian is of the form
\beq\label{scalar_Lagrangian}
{\mathcal{L}_{scalar}^{mass}} = {\Phi_{S}^T} {M}^2_{S} {\Phi_{S}},
\eeq
where ${M}^2_{S}$ is an $8\times8$ symmetric matrix. The mass eigenvectors are 
\beq
S^0_\alpha = \bRs_{\alpha \beta} \Phi_{S_\beta},
\label{scalar-mass-basis}
\eeq
with the diagonal mass matrix
\beq
(\mathcal{M}^{diag}_{S})^2_{\alpha \beta} = \bRs_{\alpha \gamma} {M}^2_{S_{\gamma \delta}} 
\bRs_{\beta \delta}.
\label{scalar-diagonal-mass-matrix}
\eeq
%
\begin{flushleft}
{\it{$\maltese$ Neutral pseudoscalar}} $\blacktriangleright$
\end{flushleft}
In the weak interaction basis $\Phi^T_{P}=(H^0_{d{\mathcal I}},H^0_{u{\mathcal I}}, 
{\widetilde \nu}^c_{n{\mathcal I}},{\widetilde \nu}_{n{\mathcal I}})$, the pseudoscalar 
mass term in the Lagrangian is of the form
\beq\label{pseudoscalar_Lagrangian}
{\mathcal{L}_{pseudoscalar}^{mass}} = {\Phi_{P}^T} {M}^2_{P} {\Phi_{P}},
\eeq
where ${M}^2_{P}$ is an $8\times8$ symmetric matrix. The mass eigenvectors are defined as
\bea
P^0_\alpha = \bRp_{\alpha \beta} \Phi_{P_\beta},
\label{pseudoscalar-mass-basis}
\eea
with the diagonal mass matrix
\bea
(\mathcal{M}^{diag}_{P})^2_{\alpha \beta} = \bRp_{\alpha \gamma}
{M}^2_{P_{\gamma \delta}} \bRp_{\beta \delta}.
\label{pseudoscalar-diagonal-mass-matrix}
\eea
\begin{flushleft}
{\it{$\maltese$ Charged scalar}} $\blacktriangleright$ 
\end{flushleft}
In the weak basis $\Phi^{+^T}_{C}=({H^+_{d}}, {H^+_{u}}, 
{{\widetilde{e}}^+_{Rn}} ,{{\widetilde{e}}^+_{Ln}})$\footnote{In refs.\cite{AppGhosh:2008yh,AppGhosh:2010zi} 
$\Phi^{+^T}_{C}=({H^+_{1}}, {H^+_{2}}, {{\widetilde{e}}^+_{Rn}} ,{{\widetilde{e}}^+_{Ln}})$ 
basis was used.} the charged scalar mass term in 
the Lagrangian is of the form
\beq\label{charged-scalar_Lagrangian}
{\mathcal{L}_{charged~scalar}^{mass}} = {\Phi_{C}^{-^T}} {M}^2_{C^{\pm}} 
{\Phi_{C}^{+}},
\eeq
where ${M}^2_{C^{\pm}}$ is an $8\times8$ symmetric matrix. The mass eigenvectors are 
\bea
S^{\pm}_\alpha = \bRc_{\alpha \beta} \Phi^{\pm}_{C_\beta},
\label{charged-scalar-mass-basis}
\eea
with the diagonal mass matrix
\bea
(\mathcal{M}^{diag}_{C^{\pm}})^2_{\alpha \beta} = \bRc_{\alpha \gamma}
{M}^2_{C^{\pm}_{\gamma \delta}} \bRc_{\beta \delta}.
\label{charged-scalar-diagonal-mass-matrix}
\eea

The independent entries of the $8\times8$ symmetric matrix ${M}^2_{S}$ 
(eqn. (\ref{scalar_Lagrangian})) using 
eqn. (\ref{Minim-munuSSM}) and eqn. (\ref{Abbrevations}) \footnote{A typo in
$(M^2_S)^{H^0_{d{\mathcal R}} H^0_{u{\mathcal R}}}$ in ref.\cite{AppGhosh:2008yh} 
has been corrected.} 
are given by
\bea
(M^2_S)^{H^0_{d{\mathcal R}} H^0_{d{\mathcal R}}}&=&{\frac{1}{v_1}}\left[{\sum_j}{\lambda^j}{v_2}
\left({\sum_{ik}} {\kappa^{ijk}}{v^c_i}{v^c_k}\right)+{\sum_j}{\lambda^j}
{r^j}{v^2_2}+{\mu} {\sum_j}{r^j_c}{v'_j}\right. \nonumber\\
&+& \left.
{\sum_i}(A_{\lambda}{\lambda})^{i}{v^c_i}{v_2}\right] +2{\gamma_g}{v^2_1}, \nonumber \\
(M^2_S)^{H^0_{d{\mathcal R}} H^0_{u{\mathcal R}}}&=&-2{\sum_j}{\lambda^j}{\rho^j}{v_2}-{\sum_{i,j,k}}{\lambda^j}
{\kappa^{ijk}}{v^c_i}{v^c_k}-2{\gamma_g}{v_1}{v_2}-{\sum_i}{({A_{\lambda}}
{\lambda})^i}{v^c_i}, \nonumber\\
(M^2_S)^{H^0_{u{\mathcal R}} H^0_{u{\mathcal R}}}&=&{\frac{1}{v_2}}\left[-{\sum_{j}}{\rho^{j}}\left({\sum_{l,k}}
\kappa^{ljk}{v^c_l}{v^c_k}\right)-{\sum_{i,j}}(A_{\nu}{Y_{\nu}})^{ij}
{v'_i}{v^c_j}\right. \nonumber\\
&+& \left.
{\sum_i}(A_\lambda {\lambda})^{i} {v^c_i}{v_1}\right]+2{\gamma_g}{v^2_2},\nonumber \\
(M^2_S)^{H^0_{d{\mathcal R}} {\wt{\nu}^c_{m{\mathcal R}}}}&=&-2{\sum_j}{\lambda^j}{u^{mj}_c}{v_2}+2{\mu}
{v_1}{\lambda^m}-{\lambda^m}{\sum_i}{r^i_c}{v'_i}-{\mu}{r^m}-
{({A_{\lambda}}{\lambda})^m}{v_2}, \nonumber\\
(M^2_S)^{H^0_{d{\mathcal R}} {\wt{\nu}_{m{\mathcal R}}}}&=&-{\sum_j}{\lambda^j}{Y^{mj}_{\nu}}{v^2_2}-
{\mu}{r^m_c}+2{\gamma_g}{v'_m}{v_1},\nonumber \\
(M^2_S)^{H^0_{u{\mathcal R}} {\wt{\nu}^c_{m{\mathcal R}}}}&=&2{\sum_j}{u^{mj}_c}{\rho^j}+2{\lambda^m}
{\mu}{v_2}+2{\sum_i}{Y^{im}_{\nu}}{r^i_c}{v_2} + {\sum_i}{(A_{\nu}{Y_{\nu}})^{im}}{v'_i}\nn\\
&-&{(A_{\lambda}{\lambda})^{m}}{v_1},\nonumber \\
(M^2_S)^{H^0_{u{\mathcal R}} {\wt{\nu}_{m{\mathcal R}}}}&=&2{\sum_j}{Y^{mj}_{\nu}}{\rho^j}{v_2}
+{\sum_{i,j,k}}{Y^{mj}_{\nu}}{\kappa^{ijk}}{v^c_i}{v^c_k}-2{\gamma_g}{v'_m}{v_2}
+{\sum_j}{(A_{\nu}{Y_{\nu}})^{mj}}{v^c_j},\nonumber \\
(M^2_S)^{{\wt{\nu}^c_{n{\mathcal R}}} {\wt{\nu}^c_{m{\mathcal R}}}}&=&2{\sum_j}{\kappa^{jnm}}{\zeta^j}
+4{\sum_j}{u^{mj}_c}{u^{nj}_c}+{\rho^m}{\rho^n}+{h^{nm}}{v^2_2}\nonumber \\
&+&{(m^2_{\wt{\nu}^c})^{mn}}+2{\sum_i}{(A_{\kappa}{\kappa})^{imn}}{v^c_i},
\nonumber \\
(M^2_S)^{{\wt{\nu}^c_{n{\mathcal R}}} {\wt{\nu}_{m{\mathcal R}}}}&=&2{\sum_j}{Y^{nj}_{\nu}}{u^{mj}_c}{v_2}
+{Y^{nm}_{\nu}}{\sum_i}{r^i_c}{v'_i}+{r^n_c}{r^m}-{\mu}{v_1}{Y^{nm}_{\nu}}\nonumber \\
&-&{\lambda^m}{r^n_c}{v_1}+{(A_{\nu}{Y_{\nu}})^{nm}}{v_2},\nonumber \\
(M^2_S)^{{\wt{\nu}_{n{\mathcal R}}} {\wt{\nu}_{m{\mathcal R}}}}&=&{\sum_j}{Y^{nj}_{\nu}}{Y^{mj}_{\nu}}{v^2_2}
+{r^m_c}{r^n_c}+{\gamma_g}{\xi_{\upsilon}}{\delta_{nm}}+2{\gamma_g}{v'_n}
{v'_m}+{(m^2_{\wt{L}})^{mn}}.
\label{element_of_scalar_mass_matrix}
\eea

%
Similarly independent elements of $8\times8$ symmetric matrix $\mathcal{M}^2_{P}$ 
(eqn. (\ref{pseudoscalar_Lagrangian}))
using eqn. (\ref{Minim-munuSSM}) and eqn. (\ref{Abbrevations}) are given by
\bea
(M^2_P)^{H^0_{d{\mathcal I}} H^0_{d{\mathcal I}}}&=&{\frac{1}{v_1}}
\left[{\sum_j}{\lambda^j}{v_2} \left({\sum_{ik}} {\kappa^{ijk}}{v^c_i}
{v^c_k}\right)+{\sum_j} {\lambda^j}{r^j}{v^2_2}+ {\mu}{\sum_j}{r^j_c}{v'_j}
\right.\nn\\
&+&\left.
{\sum_i} (A_{\lambda}{\lambda})^{i}{v^c_i}{v_2} \right], \nonumber \\
(M^2_P)^{H^0_{d{\mathcal I}} H^0_{u{\mathcal I}}}&=&{\sum_{i,j,k}}{\lambda^j}
{\kappa^{ijk}} {v^c_i}{v^c_k} + \sum_i (A_{\lambda} {\lambda})^{i}{v^c_i},
\nonumber \\
(M^2_P)^{H^0_{u{\mathcal I}} H^0_{u{\mathcal I}}}&=&{\frac{1}{v_2}}
\left[-{\sum_{j}}{\rho^{j}} \left({\sum_{l,k}} \kappa^{ljk}{v^c_l}
{\nu^c_k}\right)-{\sum_{i,j}} (A_{\nu}{Y_{\nu}})^{ij} {v'_i}{v^c_j}\right.\nn\\
&+&\left.
{\sum_i}(A_\lambda {\lambda})^{i} {v^c_i}{v_1}\right], 
\nonumber \\
(M^2_P)^{H^0_{d{\mathcal I}} {\wt \nu}^c_{m{\mathcal I}}}&=&-2 {\sum_j}
{\lambda^j}{u^{mj}_c}{v_2} - {\mu}{r^m} + {\lambda^m}{\sum_i}{r^i_c}{v'_i} 
+ (A_{\lambda}{\lambda})^m {v_2}, \nonumber \\
(M^2_P)^{H^0_{d{\mathcal I}} {\wt \nu}_{m{\mathcal I}}}&=&-{\sum_j}
{\lambda^j}{Y^{mj}_{\nu}} {v^2_2} - {\mu}{r^m_c},\nonumber \\
(M^2_P)^{H^0_{u{\mathcal I}} {\wt \nu}^c_{m{\mathcal I}}}&=&2{\sum_j}
{u^{mj}_c}{\rho^j} - {\sum_i}(A_{\nu} Y_{\nu})^{im}{v'_i} + 
(A_{\lambda}{\lambda})^m {v_1}, \nonumber \\
(M^2_P)^{H^0_{u{\mathcal I}} {\wt \nu}_{m{\mathcal I}}}&=&-{\sum_{i,j,k}}
{Y^{mj}_{\nu}} {\kappa^{ijk}} {v^c_i}{v^c_k}-{\sum_j}
({A_{\nu}}{Y_{\nu}})^{mj}{v^c_j}, 
\nonumber \\
(M^2_P)^{{\wt \nu}^c_{n{\mathcal I}} {\wt \nu}^c_{m{\mathcal I}}}&=&
-2{\sum_j}{\kappa^{jnm}} {\zeta^j}+4{\sum_j} {u^{mj}_c}{u^{nj}_c}+{\rho^m}
{\rho^n}+{h^{nm}}{v^2_2} \nn\\
&+&(m^2_{\wt{\nu}^c})^{nm} -2{\sum_{i}} ({A_{\kappa}}{\kappa})^{inm} {v^c_i},\nonumber \\
(M^2_P)^{{\wt \nu}^c_{n{\mathcal I}} {\wt \nu}_{m{\mathcal I}}}&=&
2{\sum_j}{u^{mj}_c} {Y^{nj}_{\nu}}{v_2}-{Y^{nm}_{\nu}}{\sum_i}{r^i_c}{v'_i}
+{r^n_c}{r^m} +{\mu}{v_1}{Y^{nm}_{\nu}}\nn\\
&-&{\lambda^m}{r^n_c}{v_1} - {({A_{\nu}}{Y_{\nu}})^{nm}} {v_2},\nonumber \\
(M^2_P)^{{\wt \nu}_{n{\mathcal I}} {\wt \nu}_{m{\mathcal I}}}&=&
{\sum_j}{Y^{mj}_{\nu}} {Y^{nj}_{\nu}}{v^2_2} +{r^m_c}{r^n_c}
+(m^2_{\wt{L}})^{nm}+{\gamma_g} {\xi_{\upsilon}} {\delta_{mn}}.
\label{element_of_pseudoscalar_mass_matrix}
\eea
In eqns.(\ref{element_of_scalar_mass_matrix}), (\ref{element_of_pseudoscalar_mass_matrix})
$h^{nm}={\lambda^n}{\lambda^m}+{\sum}{Y_{\nu}^{in}} {Y_{\nu}^{im}}$ has been used. 
One eigenvalue of $\mathcal{M}^2_{P}$ matrix is zero which corresponds
to the neutral Goldstone boson.

Finally, the independent entries of $\mathcal{M}^2_{C}$ 
using eqn. (\ref{Minim-munuSSM}) and eqn. (\ref{Abbrevations}) are given by

\bea
(M^2_C)^{H_{d} H_{d}}&=&{\frac{1}{v_1}}\left[{\sum_j} \lambda^j \zeta^j v_2 
+ \mu {\sum_j} r^j_c v'_j + {\sum_i} ( A_{\lambda} \lambda )^i v^c_i v_2 
\right] \nn\\
&+&  {\sum_{i,j,k}} Y^{ij}_e Y^{kj}_e v'_i v'_k - \frac{{g_2}^2}{2} 
({\sum_i} v'^2_i - v^2_2 ), \nonumber \\
(M^2_C)^{H_{d} H_{u}}&=& -{\sum_j} \lambda^{j^2} v_1 v_2 + {\sum_j} \lambda^j r^j v_2 
+ {\sum_j} \lambda^j u^{ij}_c v^c_i + \frac{{g_2}^2}{2} v_1 v_2 \nonumber\\
&+& {\sum_i} (A_{\lambda} \lambda )^i v^c_i, \nonumber\\
(M^2_C)^{H_{u} H_{u}}&=&{\frac{1}{v_2}}\left[ - {\sum_j} \rho^j  \zeta^j  
- {\sum_{i,j}} ( A_\nu Y_\nu )^{ij} v'_i v^c_j + {\sum_i} ( A_\lambda 
\lambda )^i v^c_i v_1\right] \nn\\
&+& \frac{{g_2}^2}{2} ( {\sum_i} v'^2_i + v^2_1 ),\nonumber \\
(M^2_C)^{H_{d} {{\wt{e}}_{Rm}}}&=& - {\sum_i} r^i_c Y^{im}_e v_2 - {\sum_i} 
( A_e Y_e )^{im} v'_i, \nonumber\\
(M^2_C)^{H_{d} {{\wt{e}}_{Lm}}}&=& - \mu r^m_c - {\sum_{i,j}} Y^{mj}_e Y^{ij}_e 
v'_i v_1 + \frac{g_2^2}{2} v'_m v_1,\nonumber \\
(M^2_C)^{H_{u} {{\wt{e}}_{Rm}}}&=& - \mu {\sum_i} Y^{mi}_e v'_i - {\sum_i} 
Y^{im}_e r^i_c v_1,\nonumber \\
(M^2_C)^{H_{u} {{\wt{e}}_{Lm}}}&=& - {\sum_j} Y^{mj}_{\nu} \zeta^j  + 
\frac{g_2^2}{2} v'_m v_2 - {\sum_i} (A_{\nu} Y_{\nu})^{mi} v^c_i,\nonumber \\
(M^2_C)^{{{\wt{e}}_{Rn}} {{\wt{e}}_{Rm}}}&=& {\sum_{i,j}} Y^{im}_e Y^{jn}_e v'_i v'_j 
+ {\sum_i} Y^{im}_e Y^{in}_e v^2_1 + (m^2_{\wt{e}^c})^{mn} - 
\frac{g_1^2}{2} {\xi_\upsilon} \delta_{mn},\nonumber \\
(M^2_C)^{{{\wt{e}}_{Rn}} {{\wt{e}}_{Lm}}}&=& - \mu Y^{mn}_e v_2 + (A_e Y_e)^{nm} v_1 ,\nonumber \\
(M^2_C)^{{{\wt{e}}_{Ln}} {{\wt{e}}_{Lm}}}&=& r^m_c r^n_c + {\sum_j} Y^{mj}_e Y^{nj}_e 
v^2_1 + \gamma_g {\xi_\upsilon} \delta_{mn} -\frac{g_2^2}{2}{\xi_\upsilon} 
\delta_{mn} \nonumber \\
&+& \frac{g_2^2}{2} v'_m v'_n + (m^2_{\wt{L}})^{mn}.
\label{element_of_charged-scalar_mass_matrix}
\eea
For the charged scalar mass-squared matrix, seven out of eight eigenvalues are 
positive and the remaining one is a massless charged Goldstone boson. 

Note that in eqns. (\ref{element_of_scalar_mass_matrix}), 
(\ref{element_of_pseudoscalar_mass_matrix}), (\ref{element_of_charged-scalar_mass_matrix})
we have used $v^c_i$ and $v'_i$ to represent VEV of $i$-th right handed and left handed sneutrino,
respectively. In ref. \cite{AppGhosh:2008yh} these were represented by $\nu^c_i$ and $\nu_i$,
respectively.

\begin{flushleft}
{\it{$\maltese$ Squark mass matrices}} $\blacktriangleright$
\end{flushleft}

In the weak basis,
$\widetilde{u'}_i=(\widetilde{u}_{L_i},\widetilde{u}^*_{R_i})$ and 
$\widetilde{d'}_i=(\widetilde{d}_{L_i},\widetilde{d}^*_{R_i} )$, we get
\beq
\mathcal{L}^{mass}_{squark}=
     \frac{1}{2}\widetilde{u'_i}^\dag M_{\widetilde{u_{ij}}}^2\ \widetilde{u'_j}
     +\frac{1}{2}\widetilde{d'_i}^{\dag} M_{\widetilde{d_{ij}}}^2\ \widetilde{d'_j}\ ,
     \label{squark-mass-lagrangian}
\eeq
where $\widetilde{q}=(\widetilde{u'},\widetilde{d'})$. Explicitly for up and down type 
squarks $(\widetilde{u},\widetilde{d})$, using eqn.(\ref{Abbrevations}) the entries are
\bea
(M^2_{\widetilde{u}})^{L_iL_j}&=&
    (m^2_{\widetilde{Q}})^{ij} + \frac{1}{6}(\frac{3g^2_2}{2}  
    - \frac{g_1^2}{2}){\xi_{\upsilon}} \delta^{ij}+ 
    \sum_n {Y_u^{in} Y_u^{jn} v_2^2}\ ,  \nonumber\\ 
(M^2_{\widetilde{u}})^{R_iR_j}&=&
    (m^2_{\widetilde{u}^c})^{ij}+ \frac{g^2_1}{3}{\xi_{\upsilon}}\delta^{ij}
    + \sum_n {Y_u^{ni} Y_u^{nj}v_2^2}\ ,  \nonumber\\
(M^2_{\widetilde{u}})^{L_iR_j}&=& 
    (A_u Y_u)^{ij} v_2 -Y_u^{ij} v_1 \mu  
    + Y_u^{ij} \sum_l{r^l_c v'_l}\ ,\nonumber\\
(M^2_{\widetilde{u}})^{R_iL_j}&=&
    (M^2_{\widetilde{u}})^{L_jR_i}\ ,
\label{entries-of-usquark-mass-matrix}
\eea 
and
\bea
(M^2_{\widetilde{d}})^{L_iL_j}&=&
    (m^2_{\widetilde{Q}})^{ij} - \frac{1}{6}(\frac{3g^2_2}{2} 
    + \frac{g_1^2}{2}){\xi_{\upsilon}}\delta^{ij}+ 
    \sum_n {Y_d^{in} Y_d^{jn} v_1^2}\ ,  \nonumber\\ 
(M^2_{\widetilde{d}})^{R_iR_j}&=&
    (m^2_{\widetilde{d}^c})^{ij} - \frac{g^2_1}{6}{\xi_{\upsilon}}\delta^{ij}
    + \sum_n {Y_d^{ni} Y_d^{nj}v_1^2}\ ,  \nonumber\\
(M^2_{\widetilde{d}})^{L_iR_j}&=& 
    (A_d Y_d)^{ij} v_1 -Y_d^{ij} v_2 \mu \ ,\nonumber\\
(M^2_{\widetilde{d}})^{R_iL_j}&=&
    (M^2_{\widetilde{d}})^{L_jR_i}\ .
\label{entries-of-dsquark-mass-matrix}
\eea 
For the mass eigenstate $\widetilde{\mathbf{q}}_i$ we have
\beq
\widetilde{\mathbf{q}}_i = \Rsq_{ij} \widetilde{q}_j\ , 
\label{squark-mass-basis}
\eeq
with the diagonal mass matrix
\beq
 (\mathcal{M}^{\text{diag}}_{\widetilde{q}})^2_{ij} 
   = \Rsq_{il}  M^2_{\widetilde{q}_{lk}} \Rsq_{jk} .
\label{squark-diagonal-mass-matrix}
\eeq


\section{Quark mass matrices in $\mu\nu$SSM}
\label{munuSSM-matsq-fermion}

The mass matrices for up and down quarks are $3\times3$ and they are diagonalized using bi-unitary 
transformation. Entries of up and down quark mass matrices $m^u_{3\times3}~\rm{and}~m^d_{3\times3}$ are 
same as the MSSM and are given below
\bea
(m^u_{3\times3})_{ij} &=& Y_u^{ij} v_2, \nonumber \\
(m^d_{3\times3})_{ij} &=& Y_d^{ij} v_1. 
\label{quark-mass-matrix}
\eea
The quark mass matrices are diagonalized as follows
\bea\label{quark_mass_eigenstate_matrixform}
{\Ru_L}^* m^u_{3\times3} {\Ru_R}^{-1} &=& \mathcal{M}^{diag}_U, \nonumber \\
{\Rd_L}^* m^d_{3\times3} {\Rd_R}^{-1} &=& \mathcal{M}^{diag}_D.
\eea

\chapter{ \sffamily{{\bf }
 }}\label{appenC}

\section{Details of expansion matrix $\xi$}\label{Details-xi}
In this appendix the entries of the expansion matrix $\xi$ are given in details
\bea\label{expansion-parameter-terms}
& &\xi_{i1} \approx \frac{\sqrt{2} g_1 \mu m^2_{\nu^c} M_2 A}{12 D}b_i, \nonumber \\
& &\xi_{i2} \approx -\frac{\sqrt{2} g_2 \mu m^2_{\nu^c} M_1 A}{12 D}b_i, \nonumber \\
& &\xi_{i3} \approx -\frac{m^2_{\nu^c} M^{\prime}}{2 D} \left\{ 
{\left(\lambda v_2 v^2 - 4 \mu A {\frac{M}{v_2}}\right)}a_i + {m_{\nu^c} v_2 v^c}b_i\right.\nn\\
& &-\left.
{3\lam \left(\lam v_1 v^2-2 m_{\nu^c} v^c v_2\right) }c_i\right\}, \nonumber \\
& &\xi_{i4} \approx -\frac{m^2_{\nu^c} M^{\prime}}{2 D} \left\{ {\lam v_1 v^2}a_i 
+ {m_{\nu^c} v_1 v^c}b_i+{3\lam^2 v_2 v^2}c_i\right\}, \nonumber \\
& &\xi_{i,4+i} \approx \frac{m_{\nu^c} M^{\prime}}{2 D}\left\{ 2 \lam \left( \lam v^4 (1-\frac{1}{2}
{\rm{sin}^2{2\beta}}) + \frac{m_{\nu^c}}{2} v^c v^2 {\rm{sin}{2\beta}} 
\right.\right.\nn\\
& &+\left.\left.
 A v^2 {\rm{sin}{2\beta}} - 4 \mu M A\right) a_i 
- {\mu m_{\nu^c} v^2 {\rm{cos}{2\beta}}}b_i\right\}, \nonumber \\
& &\xi_{16} \approx \xi_{17} \approx -\frac{m_{\nu^c}  M^{\prime}}{2 D}\left\{ {\lam \left
( \lam {v^4} - 4 \mu M A\right)}a_1 + \frac{\mu m_{\nu^c} v^2}{3}b_1 
- {2 \lam \mu m_{\nu^c} v^2_2}c_1\right\}, \nonumber \\
& &\xi_{25} \approx \xi_{27} \approx -\frac{m_{\nu^c}  M^{\prime}}{2 D}\left\{ {\lam \left( \lam {v^4} 
- 4 \mu M A\right)}a_2 + \frac{\mu m_{\nu^c} v^2}{3}b_2 - {2 \lam \mu m_{\nu^c} v^2_2}c_2\right\}, \nonumber \\
& &\xi_{35} \approx \xi_{36} \approx -\frac{m_{\nu^c}  M^{\prime}}{2 D}\left\{ {\lam \left( \lam {v^4} 
- 4 \mu M A\right)}a_3 + \frac{\mu m_{\nu^c} v^2}{3}b_3 - {2 \lam \mu m_{\nu^c} v^2}c_3\right\}, \nn\\
\eea
where using eqn.(\ref{assumption1})
\bea
& &a_i = Y_{\nu}^{ii} v_2, ~b_i = (Y_{\nu}^{ii} v_1 + 3 {\lambda} {v'_i}),~c_i = {v'_i}, \nonumber \\ 
& &m_{\nu^c} = 2 \kp v^c, ~\mu = 3 \lambda v^c,~A = ({\kappa}{v^c}^2 + {\lambda} v_1 v_2),\nonumber \\
& &v_2 = v {\rm{sin}{\beta}},~v_1 = v {\rm{cos}{\beta}}, ~D = Det\left[M_{7\times7}\right],\nonumber \\
& &\frac{1}{M} = \frac{g^2_1}{M_1} +\frac{g^2_2}{M_2}, ~M^{\prime} = \frac{M_1 M_2}{M}, 
\label{specifications-2}
\eea
with ${i} = {e,\mu,\tau} ~\equiv{1,2,3}$.

\section{Tree level analysis with perturbative calculation}\label{tree-perturbed}

In the unperturbed basis ${\mathcal{B}} b_ib_j$ with ${\mathcal{B}}
={\frac{2}{3}}{\frac{A {v^c}}{\Delta}}$ the eigenvalues and eigenvectors are
given by
\bea
& & 0, 0, {\mathcal{B}}{(b^2_e + b^2_\mu + b^2_\tau)}, \nn\\
& & \left(\begin{array}{ccc}
-\frac{b_{\tau}}{b_e} & 0 & 1
\end{array}\right)^{T}, 
\left(\begin{array}{ccc}
-\frac{b_{\mu}}{b_e} & 1 & 0
\end{array}\right)^{T},  
\left(\begin{array}{ccc}
\frac{b_e}{b_{\tau}} & \frac{b_{\mu}}{b_{\tau}} & 1
\end{array}\right)^{T},
\label{unperturbed}
\eea
where $b_i$s are given by eqn.(\ref{specifications}). We choose the co-efficient of
$a_ia_j$ term to be ${\mathcal{A}} (= {\frac{1}{6 {\kappa}{v^c}}})$.
The set of orthonormal eigenvectors are obtained using Gram-Schmidt orthonormalization
procedure. The set of orthonormal eigenvectors obtained in this case are
\bea
& &\text{y}_1 =
{\frac{b_e}{\sqrt{b^2_e + b^2_{\tau}}}}\left(\begin{array}{c}
-\frac{b_{\tau}}{b_e} \\
0 \\
1
\end{array}\right),\nonumber \\
& &\text{y}_2 =
{\frac{\sqrt{b^2_e + b^2_{\tau}}}{{{\Omega}_b}}}
\left(\begin{array}{c}
-\frac{b_e b_{\mu}}{b^2_e + b^2_{\tau}} \\
1 \\
-\frac{b_{\mu} b_{\tau}}{b^2_e + b^2_{\tau}}
\end{array}\right),\nonumber \\
& &\text{y}_3 =
{\frac{b_{\tau}}{{{\Omega}_b}}}\left(\begin{array}{c}
\frac{b_e}{b_{\tau}} \\
\frac{b_{\mu}}{b_{\tau}} \\
1
\end{array}\right),
\label{eigenvectors}
\eea
where
\beq\label{omegab}
{\Omega}_b={\sqrt{b^2_e + b^2_{\mu} + b^2_{\tau}}}.
\eeq
Using degenerate perturbation theory for this set of orthonormal eigenvectors, 
the modified eigenvalues $m^\prime_{\pm}$ and $m^\prime_3$ are obtained as 
\bea
m^\prime_{\pm} &=& -{\frac{\mathcal{A}}{{\Omega}_b^2}} \left\{{\Pi_{ab}} 
\pm {\sqrt{\left[-3 {{\Omega}^2_b}{\left({\Sigma_{ab}}\right)^2}  +
\left({\Pi_{ab}}\right)^2 \right]}} \right\},
\nonumber \\
m^\prime_3 &=& {\mathcal{B}}{{\Omega}^2_b} - 
{\frac{2{\mathcal{A}}}{{\Omega}_b^2}} \left\{(\sum_{i}{a_i}{b_i})^2 - 
3 {\Lambda}_{ab}\right\},
\label{eigenvalues}
\eea
where
\bea
{\Lambda}_{ab}={{\sum_{i < j}} {a_i} {a_j} {b_i} {b_j}},~~
{\Pi_{ab}}={\sum_{i < j}}(a_i b_j+ a_j b_i)^2-{{\Lambda}_{ab}},
~~{\Sigma_{ab}}&=&{\sum_{i \neq j \neq k}}{a_i}{a_j}{b_k}.\nn\\
\label{eigenvalues-details}
\eea
As one can see from eqn.(\ref{eigenvalues}), the correction to the eigenvalues
are proportional to the coefficient $\mathcal{A}$ appearing in 
ordinary seesaw (eqn.(\ref{ordinaryseesaw})). This is 
a well expected result since we treat the ordinary seesaw terms
as the perturbation. Let us note in passing that this effect is absent if only one 
generation of left chiral neutrino is considered, whereas for two and three 
generations of left chiral neutrino the ordinary seesaw effect exists. 
This can be understood from the most general calculation involving 
n-generations of left chiral neutrinos, where the coefficients of 
$\mathcal{A}$ pick up an extra factor $(n-1)$ (see section \ref{n-gen}).

With the set of orthonormal eigenvectors in 
eqn. (\ref{eigenvectors}) and the eigenvalues in eqn.(\ref{eigenvalues}), it is 
possible to write down the eigenvectors of matrix given by eqn.(\ref{mnuij-compact1}) 
in the following form
\bea
(\mathcal{Y}_1)_{3\times1} = {\alpha_1}{y_1} + {\alpha_2}{y_2},
~~(\mathcal{Y}_2)_{3\times1} = {{\alpha}^\prime}_{1}{y_1} + {{\alpha}^\prime}_{2}
{y_2},
~~(\mathcal{Y}_3)_{3\times1} = {y_3},
\label{evec_anal}
\eea
where ${\alpha_1}$, ${\alpha_2}$, $\alpha^\prime_1$, 
$\alpha^\prime_2$, are calculated using degenerate perturbation 
theory and their analytical expressions are given by
\bea
&&\alpha_1 = \pm \left( \frac {h_{12}} {\sqrt{h^2_{12} + (h_{11} - 
m^\prime_+)^2}}\right),
~~\alpha_2 = \mp \left( \frac {h_{11} - m^\prime_+} {\sqrt{h^2_{12} + (h_{11} 
- m^\prime_+)^2}}\right),\nn\\
&&\alpha^\prime_1 = \pm \left( \frac {h_{12}} {\sqrt{h^2_{12} + (h_{11} 
- m^\prime_-)^2}}\right),
~~\alpha^\prime_2 = \mp \left( \frac {h_{11} - m^\prime_-} 
{\sqrt{h^2_{12} + (h_{11} - m^\prime_-)^2}}\right).\nn\\
\label{alphas}
\eea
Here $m^\prime_+$, $m^\prime_-$ are given by eqn.(\ref{eigenvalues}) and 
$h_{11}$, $h_{12}$ are given by
\bea
&&h_{11} = -\frac {2{\mathcal{A}} \left({a^2_{\tau}} {b^2_e} + 
{a_e}{a_{\tau}}{b_e}{b_{\tau}} + {a^2_e} {b^2_{\tau}}\right)}{b^2_+},\nn\\
&&h_{12} = \frac{{\mathcal{A}}\left[a_{\mu}(a_{\tau} b_e - a_e b_{\tau}){b^2_+} 
- {b_{\mu}}\left(2 b_e b_{\tau}{a^2_-}+ a_e 
a_{\tau}{b^2_-}\right)\right]}{{\Omega_b}{b^2_+}},
\label{h1112}
\eea
where
\bea\label{h12_specifications}
b^2_{\pm} = (b^2_e \pm b^2_{\tau}),~~a^2_- = (a^2_e - a^2_{\tau}),
\eea
and $\Omega_b$ has been defined in eqn.({\ref{omegab}}).

The light neutrino mixing matrix or PMNS matrix $U$ (eqn.(\ref{PMNS1})) can be constructed using 
the eigenvectors given in eqn.(\ref{evec_anal}) and it looks like
\bea
{U}&=&
\left(\begin{array}{ccc}
{\mathcal{Y}_1} & {\mathcal{Y}_2} & {\mathcal{Y}_3}
\end{array}\right)_{3\times3}.
\label{mneutrino_mixing_numerical}
\eea

\section{See-saw masses with n generations}\label{n-gen}

For the sake of completeness we mention the neutrino mass generation
in $\mu\nu$SSM with $n$ generations of lepton family. 
The most general form of effective neutrino mass matrix is given by
\bea
({M^{seesaw}_{\nu}})_{ij} &=& {\frac{1}{2 n \kappa {v^c}}} {a_{i}} {a_{j}} 
(1-n\delta_{ij}) + {\frac{2 A {v^c}}{n \Delta}} {b_{i}} {b_{j}}.
\nonumber \\
\label{mnuij-compact-general}
\eea
In this situation eqn.(\ref{specifications}), eqn.(\ref{omegab}) and eqn.(\ref{eigenvalues})  
are modified as follows
\bea\label{omegab-general}
{\Omega}_b&=&{\sum_m} b^2_m \nonumber \\
\text{where} ~~b_m~&=&~(Y^{mm}_\nu v_1 + n \lambda {\nu}_m)~~m=1,..,n,
\eea
\bea
m^\prime_r &=& -{\frac{(n-1)\mathcal{A}}{2{\Omega}^2_b}} \left\{{\Pi_{ab}} -~(-1)^{n-r} 
{\sqrt{\left[-3 {{\Omega}^2_b}{\left({\Sigma_{ab}}\right)^2}  +\left({\Pi_{ab}}\right)^2 \right]}} 
\right\},
\nonumber \\
m^\prime_n &=&{\mathcal{B}}{{\Omega}^2_b} - {\frac{(n-1){\mathcal{A}}}{{\Omega}^2_b}} 
\left\{(\sum_{i}{a^2_i}{b^2_i})^2 - 3(n-2) {\Lambda}_{ab}\right\},
\label{eigenvalues-general}
\eea 
where ${\mathcal{A}} = {\frac{1}{2~n {\kappa}{v^c}}} $, ${\mathcal{B}}
={\frac{2}{n}}{\frac{A {v^c}}{\Delta}}$, $\mu ~= n \lambda v^c$, $r~=~1,...,(n-1)$ and
\bea
{\Lambda}_{ab}&=&{{\sum_{i < j}} {a_i} {a_j} {b_i} {b_j}},\nonumber \\
{\Pi_{ab}}&=&{\sum_{i < j}}{(a_i b_j + a_j b_i)}^2-(n-2){{\Lambda}_{ab}},\nonumber \\
{\Sigma_{ab}}&=&{\sum_{i \neq j \neq k}}{a_i}{a_j}{b_k}\nonumber \\
\text{where}~ i,~j,~k ~&=&~1,.....,n.
\label{notation-general}
\eea


\chapter{ \sffamily{{\bf }
 }}\label{appenD}

 \section{Feynman rules}\label{Feynman rules-1}
The relevant Feynman rules required for the calculation of the 
one-loop contributions to the neutralino masses (see figure \ref{one-loop-diagrams}, 
section \ref{loop-neut-calc}) are shown here \cite{AppGhosh:2010zi}. 
Some of these Feynman rules have been derived also in ref.\cite{AppGhosh:2008yh} 
for calculating two body decays of the lightest neutralino, $\ntrl1$. 
Feynman rules for MSSM are given in references \cite{AppHaber:1984rc,AppRosiek:1989rs,AppRosiek:1995kg} 
and in references \cite{AppGunion:1984yn,AppGunion:1986nh,AppFranke:1995tf,AppFranke:1995tc} for MSSM 
with singlet superfields. Feynman rules for $R_p$-violating MSSM are studied in 
references \cite{AppHempfling:1995wj,AppHirsch:2000ef,AppDiaz:2003as}. 
The required Feynman rules are (using relations of 
form {\it{neutralino-fermion-scalar/gauge boson}} and they are listed below.


\subsection*{$\bigstar$~Neutralino-neutralino-neutral scalar}\label{nnh}

The Lagrangian using four component spinor notation can be written as
\beq
\mathcal{L}^{nnh}= - \frac{\widetilde g}{\sqrt{2}} \ovl{{\widetilde \chi}^0_i} 
(O^{nnh}_{Lijk} P_L + O^{nnh}_{Rijk} P_R) {{\widetilde \chi}^0_j} S^0_k,
\label{Neutralino-neutralino-neutral-scalar}
\eeq
where ${\widetilde g} O^{nnh}_{Lijk}$ is given by
\bea
&& \eta_{j} \frac{1}{2} \left[\bRs_{k1} 
\left( \frac{g_2}{\sqrt{2}} {\bN}^*_{i2} \bN^*_{j3} - \frac{g_1}{\sqrt{2}} \bN^*_{i1} \bN^*_{j3} 
- \lam^m \bN^*_{i4} \bN^*_{j,m+4}\right)
\right. \nonumber \\
&&- \bRs_{k2} \left(\frac{g_2}{\sqrt{2}} {\bN}^*_{i2} \bN^*_{j4} 
-\frac{g_1}{\sqrt{2}} \bN^*_{i1} \bN^*_{j4} + \lam^m \bN^*_{i3} \bN^*_{j,m+4} 
- Y^{mn}_{\nu} \bN^*_{i,n+4} \bN^*_{j,m+7}\right)\nonumber \\
&&+ \bRs_{k,m+2} \left( Y^{mn}_{\nu} \bN^*_{i4} \bN^*_{j,n+7} 
- \lam^m \bN^*_{i3} \bN^*_{j4} + \kp^{mnp} \bN^*_{i,n+4} \bN^*_{j,p+4}\right)\nonumber \\
&&+  \left.
\bRs_{k,m+5} \left( \frac{g_2}{\sqrt{2}} {\bN}^*_{i2} \bN^*_{j,m+7} 
- \frac{g_1}{\sqrt{2}} \bN^*_{i1} \bN^*_{j,m+7} + Y^{mn}_{\nu} \bN^*_{i4} \bN^*_{j,n+4}\right)\right] \nonumber\\
&&+ 
(i\leftrightarrow j),\nonumber\\
\label{nnh-L-couplings}
\eea
and\footnote{A typos \cite{AppGhosh:2010zi} in the expression of $O^{nnh}_{Rijk}$ has been corrected.}
\beq
O^{nnh}_{Rijk} =(O^{nnh}_{Ljik})^*.
\label{nnh-R-couplings}
\eeq
\subsection*{$\bigstar$~Neutralino-neutralino-neutral pseudoscalar}\label{nna}

The Lagrangian using four component spinor notation can be written as
\beq
\mathcal{L}^{nna}= - i\frac{\widetilde g}{\sqrt{2}} \ovl{{\widetilde \chi}^0_i} 
(O^{nna}_{Lijk} P_L + O^{nna}_{Rijk} P_R) {{\widetilde \chi}^0_j} P^0_k,
\label{Neutralino-neutralino-neutral-pseudoscalar}
\eeq
where ${\widetilde g} O^{nna}_{Lijk}$ is given as
\bea
&&\eta_{j} \frac{1}{2} \left[\bRp_{k1} 
\left( -\frac{g_2}{\sqrt{2}} {\bN}^*_{i2} \bN^*_{j3} + \frac{g_1}{\sqrt{2}} \bN^*_{i1} \bN^*_{j3} 
- \lam^m \bN^*_{i4} \bN^*_{j,m+4}\right)
\right. \nonumber \\
&&+ \bRp_{k2} \left( \frac{g_2}{\sqrt{2}} {\bN}^*_{i2} \bN^*_{j4} 
- \frac{g_1}{\sqrt{2}} \bN^*_{i1} \bN^*_{j4} - \lam^m \bN^*_{i3} \bN^*_{j,m+4} 
+ Y^{mn}_{\nu} \bN^*_{i,n+4} \bN^*_{j,m+7}\right)\nonumber \\
&&+ \bRp_{k,m+2} \left( Y^{mn}_{\nu} \bN^*_{i4} \bN^*_{j,n+7} 
- \lam^m \bN^*_{i3} \bN^*_{j4} + \kp^{mnp} \bN^*_{i,n+4} \bN^*_{j,p+4}\right)\nonumber \\
&&+  \left.
\bRp_{k,m+5} \left( -\frac{g_2}{\sqrt{2}} {\bN}^*_{i2} \bN^*_{j,m+7} 
+ \frac{g_1}{\sqrt{2}} \bN^*_{i1} \bN^*_{j,m+7} + Y^{mn}_{\nu} \bN^*_{i4} \bN^*_{j,n+4}\right)\right] \nonumber \\
&&+(i\leftrightarrow j),\nonumber\\
\label{nna-L-couplings}
\eea
and\footnote{A typos \cite{AppGhosh:2010zi} in the expression of $O^{nna}_{Rijk}$ has been corrected.}
\beq
O^{nna}_{Rijk} =-(O^{nna}_{Ljik})^*.
\label{nna-R-couplings}
\eeq
\subsection*{$\bigstar$~Neutralino-neutralino-$Z^0$}\label{nnz}

The Lagrangian using four component spinor notation can be written as
\beq
\mathcal{L}^{nnz}= - \frac{g_2}{2} \ovl{{\widetilde \chi}^0_i} \gamma^{\mu} 
(O^{nnz}_{Lij} P_L + O^{nnz}_{Rij} P_R) {{\widetilde \chi}^0_j} Z^0_{\mu},
\label{Neutralino-neutralino-neutral-Z-boson}
\eeq
where
\bea
O^{nnz}_{Lij} = & & \eta_{i} \eta_{j} \frac{1}{2 \cos_{\theta_W}} \left(\bN_{i3} \bN^*_{j3} 
- \bN_{i4} \bN^*_{j4} + \bN_{i,m+7} \bN^*_{j,m+7}\right), \nonumber \\
O^{nnz}_{Rij} = & & \frac{1}{2 \cos_{\theta_W}} \left(-\bN^*_{i3} \bN_{j3} + \bN^*_{i4} \bN_{j4} 
- \bN^*_{i,m+7} \bN_{j,m+7}\right). 
\label{nnz-L-R-couplings}
\eea
\subsection*{$\bigstar$~Neutralino-chargino-charged scalar}\label{ncs}
The Lagrangian using four component spinor notation can be written as
\beq
\mathcal{L}^{ncs}= - {\widetilde g} \ovl{{\widetilde \chi}_i} (O^{cns}_{Lijk} P_L 
+ O^{cns}_{Rijk} P_R) {{\widetilde \chi}^0_j} S^+_k 
- {\widetilde g} \ovl{{\widetilde \chi}^0_i} (O^{ncs}_{Lijk} P_L 
+ O^{ncs}_{Rijk} P_R) {{\widetilde \chi}_j} S^-_k,
\label{Neutralino-chargino-chargedscalar}
\eeq
where
\bea
{\widetilde g} O^{cns}_{Lijk} = & &\eta_{j} \left[\bRc_{k1} 
\left( -\frac{g_2}{\sqrt{2}} {\bU}^*_{i2} \bN^*_{j2} - \frac{g_1}{\sqrt{2}} \bU^*_{i2} \bN^*_{j1} 
+ {g_2} \bU^*_{i1} \bN^*_{j3}\right) 
\right. \nonumber \\
& &+ \bRc_{k2} \left( \lam^m \bU^*_{i2} \bN^*_{j,m+4} - Y^{mn}_{\nu} \bU^*_{i,m+2} \bN^*_{j,n+4}\right) \nonumber \\
& &+ \bRc_{k,m+2} \left(Y^{mn}_e \bU^*_{i,n+2} \bN^*_{j3} - Y^{mn}_e \bU^*_{i2} \bN^*_{j,n+7} \right)\nonumber \\
& &+ \left.
\bRc_{k,m+5} \left( {g_2} {\bU}^*_{i1} \bN^*_{j,m+7} - \frac{g_2}{\sqrt{2}} \bU^*_{i,m+2} \bN^*_{j2} 
- \frac{g_1}{\sqrt{2}} \bU^*_{i,m+2} \bN^*_{j1}\right)\right],\nonumber \\
\nonumber\\
\label{cns-L-couplings}
{\widetilde g} O^{cns}_{Rijk} = & &\epsilon_{i} \left[\bRc_{k1} 
\left( \lam^m \bV_{i2} \bN_{j,m+4} -Y^{mn}_e \bV_{i,n+2} \bN_{j,m+7}\right) 
\right. \nonumber \\
& &+ \bRc_{k2} \left( \frac{g_2}{\sqrt{2}} {\bV}_{i2} \bN_{j2} 
+ \frac{g_1}{\sqrt{2}} \bV_{i2} \bN_{j1} + {g_2} \bV_{i1} \bN_{j4}\right) \nonumber \\
& &+ \sqrt{2} {g_1} \bRc_{k,m+2} \bV_{i,m+2} \bN_{j1}\nonumber \\
& &+ \left.
\bRc_{k,m+5} \left(Y^{mn}_e \bV_{i,n+2} \bN_{j3} 
- Y^{mn}_{\nu} \bV_{i2} \bN_{j,n+4} \right)\right],\nonumber \\
\label{cns-R-couplings}
\eea
and
\beq
O^{ncs}_{Lijk} =(O^{cns}_{Rjik})^*, \quad O^{ncs}_{Rijk} =(O^{cns}_{Ljik})^*.
\label{ncs-L-R-couplings}
\eeq
\subsection*{$\bigstar$~Neutralino-chargino-$W$}\label{ncw}
The Lagrangian using four component spinor notation can be written as
\beq
\mathcal{L}^{ncw}=  -  {g_2} \ovl{{\widetilde \chi}_i} \gamma^{\mu} 
(O^{cnw}_{Lij} P_L + O^{cnw}_{Rij} P_R) {{\widetilde \chi}^0_j} 
W^+_{\mu} -  {g_2} \ovl{{\widetilde \chi}^0_i} \gamma^{\mu} (O^{ncw}_{Lij} P_L 
+ O^{ncw}_{Rij} P_R) {{\widetilde \chi}_j} W^-_{\mu}.
\label{Neutralino-chargino-W-boson}
\eeq
where
\bea{\label{ncw-L-R-couplings}}
O^{cnw}_{Lij} = & &- \epsilon_{i} \eta_{j} \left(\bV_{i1} \bN^*_{j2} 
- \frac{1}{\sqrt{2}}  \bV_{i2} \bN^*_{j4}\right), \nonumber \\
O^{cnw}_{Rij} = & &- \bU^*_{i1} \bN_{j2} - \frac{1}{\sqrt{2}}  \bU^*_{i2} \bN_{j3}
- \frac{1}{\sqrt{2}} \bU^*_{i,n+2} \bN_{j,n+7}, 
\eea
and
\beq
O^{ncw}_{Lij} =(O^{cnw}_{Lji})^*, \quad O^{ncw}_{Rij} =(O^{cnw}_{Rji})^*.
\label{cnw-L-R-couplings}
\eeq
The factors $\eta_{j}$ and $\epsilon_{i}$ are the proper signs of neutralino 
and chargino masses \cite{AppGunion:1984yn}. They have values $\pm{1}$. 
\subsection*{$\bigstar$~Neutralino-quark-squark}\label{nqsq}
The Lagrangian using four component spinor notation can be written as
\beq
\mathcal{L}^{nq\q}=  -  {\widetilde g} \ovl{q_i} (O^{qn\q}_{Lijk} P_L 
+ O^{qn\q}_{Rijk} P_R) {{\widetilde \chi}^0_j} \q_k 
-  {\widetilde g} \ovl{{\widetilde \chi}^0_i} (O^{nq\q}_{Lijk} P_L 
+ O^{nq\q}_{Rijk} P_R) q_j \q^*_k.
\label{Neutralino-quark-squark}
\eeq
where
\beq
O^{qn\q}_{Lijk} =(O^{nq\q}_{Rjik})^*, \quad O^{qn\q}_{Rijk} =(O^{nq\q}_{Ljik})^*,
\label{q-n-squark-L-R-couplings}
\eeq
and
\bea
{\widetilde g} O^{nu\u}_{Lijk} = & & \Rsu_{km} \left(\frac{g_2}{\sqrt{2}} \bN^*_{i2} \Ru_{L_{jm}} 
+ \frac{g_1}{3\sqrt{2}} \bN^*_{i1} \Ru_{L_{jm}}\right) 
+ Y^{nm}_u \Rsu_{k,m+3} \bN^*_{i4} \Ru_{L_{jn}} , \nonumber \\
{\widetilde g} O^{nu\u}_{Rijk} = & & Y^{{mn}^*}_u \Rsu_{km} \bN_{i4} R^{u^*}_{R_{jn}} 
- \frac{4{g_1}}{3 \sqrt{2}} \Rsu_{k,m+3} \bN_{i1} R^{u^*}_{R_{jm}},\nonumber \\
{\widetilde g} O^{nd\d}_{Lijk} = & & \Rsd_{km} \left(-\frac{g_2}{\sqrt{2}} \bN^*_{i2} \Rd_{L_{jm}} 
+ \frac{g_1}{3\sqrt{2}} \bN^*_{i1} \Rd_{L_{jm}}\right) 
+ Y^{nm}_d \Rsd_{k,m+3} \bN^*_{i3} \Rd_{L_{jn}} , \nonumber \\
{\widetilde g} O^{nd\d}_{Rijk} = & & Y^{{mn}^*}_d \Rsd_{km} \bN_{i3} R^{d^*}_{R_{jn}} 
+ \frac{2{g_1}}{3 \sqrt{2}} \Rsd_{k,m+3} \bN_{i1} R^{d^*}_{R_{jm}}.
\label{n-q-squark-L-R-couplings}
\eea

Note that for couplings of the type $\wt \chi^0 \wt \chi^0 \B$,
with $\B$ as either a scalar (CP-even, CP-odd) or a vector boson ($Z$)
the associated Feynman rules must be multiplied by a $2$ factor in calculations. This feature
is a special property of a Majorana fermion since a Majorana field, being self conjugate 
(eqn.(\ref{neut-mass-a2})) contains both creation and annihilation operators \cite{AppHaber:1984rc}.

We have extensively used a set of relations between weak or flavour eigenbasis and mass eigenbasis,
both for the scalars and fermions to derive all these Feynman rules.
For the scalars (CP-even scalar, CP-odd scalar, charged 
scalar and scalar quarks) these relations are given by eqns.(\ref{scalar-mass-basis}), 
(\ref{pseudoscalar-mass-basis}), (\ref{charged-scalar-mass-basis}) and (\ref{squark-mass-basis}).
Similar relations for the four component neutralinos and charginos (eqn.(\ref{neutralino-chargino})) 
are given below.

\vspace{0.1cm}
\noindent
$\blacksquare$~{\it{Neutralinos}}
\bea\label{neutralino_mass-basis_weak-basis_relations}
& &{P_L}{\widetilde{B}}^0 = P_L N^*_{i1} \widetilde{\chi}^0_i, 
\quad\ {P_L}{\widetilde{W}}^0_3 = P_L N^*_{i2} \widetilde{\chi}^0_i, 
\quad\ {P_L}{\widetilde{H}}_j = P_L N^*_{i,j+2} \widetilde{\chi}^0_i,\nonumber \\
& &{P_L}{\nu}_k = P_L N^*_{i,k+7} \widetilde{\chi}^0_i, 
\quad\ {P_L}{\nu}^{c}_k = P_L N^*_{i,k+4} \widetilde{\chi}^0_i,\nonumber \\
& &{P_R}{\widetilde{B}}^0 = P_R N_{i1} \widetilde{\chi}^0_i, 
\quad\ {P_R}{\widetilde{W}}^0_3 = P_R N_{i2} \widetilde{\chi}^0_i, 
\quad\ {P_R}{\widetilde{H}}_j = P_R N_{i,j+2},\quad \nonumber \\
& &{P_R}{\nu}_k = P_R N_{i,k+7} \widetilde{\chi}^0_i, 
\quad\ {P_R}{\nu}^{c}_k = P_R N_{i,k+4} \widetilde{\chi}^0_i, \nonumber \\
& & \text{where} \quad j = 1,2 \quad k = 1,2,3, \quad i = 1,2,...,10
\eea
and 
\beq\label{P-L-P-R}
P_{L}=\left(\frac{1-{\gamma^5}}{2}\right), \quad 
P_{R}=\left(\frac{1+{\gamma^5}}{2}\right).
\eeq
\vspace{0.1cm}
\noindent
$\blacksquare$~{\it{Charginos}}
\bea\label{chargino_mass-basis_weak-basis_relations}
& &{P_L}{\widetilde{W}} = P_L V^*_{i1} \widetilde{\chi}_i, 
\quad\ {P_L}{\widetilde{H}} = P_L V^*_{i2} \widetilde{\chi}_i, 
\quad\ {P_L}{l_k} = P_L U^*_{i,k+2} \widetilde{\chi}^c_i,\nonumber \\
& &{P_R}{\widetilde{W}} = P_R U_{i1} \widetilde{\chi}_i, 
\quad\ {P_R}{\widetilde{H}} = P_R U_{i2} \widetilde{\chi}_i, 
\quad\ {P_R}{l_k} = P_R V_{i,k+2} \widetilde{\chi}^c_i,\nonumber \\
& &{P_L}{\widetilde{W}^c} = P_L U^*_{i1} \widetilde{\chi}^c_i, 
\quad\ {P_L}{\widetilde{H}^c} = P_L U^*_{i2} \widetilde{\chi}^c_i, 
\quad\ {P_L}{l^c_k} = P_L V^*_{i,k+2} \widetilde{\chi}_i,\nonumber \\
& &{P_R}{\widetilde{W}^c} = P_R V_{i1} \widetilde{\chi}^c_i, 
\quad\ {P_R}{\widetilde{H}^c} = P_R V_{i2} \widetilde{\chi}^c_i, 
\quad\ {P_R}{l^c_k} = P_R U_{i,k+2} \widetilde{\chi}_i,\nonumber \\
\eea
where $k = 1,2,3$, and $i$ varies from $1$ to $5$.

\chapter{ \sffamily{{\bf }
 }}\label{appenE}

In this appendix we give the detail expressions for the renormalized self energy functions 
${\widetilde \Sigma}^V_{ij}$ and ${\widetilde \Pi}^V_{ij}$. Different $(O^{ff^{\prime}b})$\footnote{$f$ 
is a neutralino, $f'$ is either a neutralino or a chargino or a quark and $b$ is either a scalar 
(CP-even or CP-odd or charged or squark) or a vector boson $(W^\pm,Z)$.} couplings are given in
appendix \ref{appenD}.
\section{The ${\widetilde \Sigma}^V_{ij}$ function}\label{self-energy-sigma}
The regularized function ${\widetilde \Sigma}^V_{ij}$ is given as

\bea\label{Sigma-part}
 -\frac{1}{16 \pi^2}&&\left[ \frac{{\widetilde g}^2}{2} 
\sum_{r=1}^{8} \sum_{k=1}^{10}  \left(O^{nnh}_{Lkir} O^{nnh}_{Rjkr} 
+ O^{nnh}_{Ljkr} O^{nnh}_{Rkir}\right) B_1(p^2,m^2_{\n_k},m^2_{S^0_r}) 
\right. \nonumber\\
- && \frac{{\widetilde g}^2}{2} \sum_{r=1}^{7} \sum_{k=1}^{10}  \left(O^{nna}_{Lkir} O^{nna}_{Rjkr} 
+ O^{nna}_{Ljkr} O^{nna}_{Rkir}\right) B_1(p^2,m^2_{\n_k},m^2_{P^0_r})\nonumber\\
+ && {g^2_2} \sum_{k=1}^{10}  \left( O^{nnz}_{Lki} O^{nnz}_{Ljk} 
+ O^{nnz}_{Rki} O^{nnz}_{Rjk} \right) {B_1(p^2,m^2_{\n_k},m^2_{Z^0_{\mu}})}\nonumber\\
+ && {2 g^2_2} \sum_{k=1}^{5}  \left( O^{cnw}_{Lki} O^{ncw}_{Ljk} 
+ O^{cnw}_{Rki} O^{ncw}_{Rjk} \right) {B_1(p^2,m^2_{{\widetilde\chi}^{\mp}_k},m^2_{W^{\pm}_{\mu}})}\nonumber\\
+ && {{\widetilde g}^2} \sum_{r=1}^{7} \sum_{k=1}^{5} \left(O^{cns}_{Lkir} O^{ncs}_{Rjkr} 
+ O^{ncs}_{Ljkr} O^{cns}_{Rkir}\right) {B_1(p^2,m^2_{{\widetilde\chi}^{\mp}_k},m^2_{S^{\pm}_r})}\nonumber\\
+ && 3 {{\widetilde g}^2} \sum_{r=1}^{6} \sum_{k=1}^{3} \left(O^{un\u}_{Lkir} O^{nu\u}_{Rjkr} 
+ O^{nu\u}_{Ljkr} O^{un\u}_{Rkir}\right) {B_1(p^2,m^2_{u_k},m^2_{\u_r})}\nonumber\\
+ && \left.
3{{\widetilde g}^2} \sum_{r=1}^{6} \sum_{k=1}^{3} \left(O^{dn\d}_{Lkir} O^{nd\d}_{Rjkr} 
+ O^{nd\d}_{Ljkr} O^{dn\d}_{Rkir}\right) {B_1(p^2,m^2_{d_k},m^2_{\d_r})} \right].\nonumber\\
\eea
\section{The ${\widetilde \Pi}^V_{ij}$ function}\label{self-energy-pi}
In similar fashion ${\widetilde \Pi}^V_{ij}$ looks like
\bea\label{Pi-part}
\nonumber \\
 -\frac{1}{16 \pi^2}&&\left[ {{\widetilde g}^2} \sum_{r=1}^{8} 
\sum_{k=1}^{10} \frac{m_{\n_k}}{2} \left(O^{nnh}_{Lkir} O^{nnh}_{Ljkr} 
+ O^{nnh}_{Rkir} O^{nnh}_{Rjkr}\right) B_0(p^2,m^2_{\n_k},m^2_{S^0_r}) 
\right. \nonumber\\
- &&{{\widetilde g}^2 } \sum_{r=1}^{7} \sum_{k=1}^{10} \frac{m_{\n_k}}{2} \left(O^{nna}_{Lkir} O^{nna}_{Ljkr} 
+ O^{nna}_{Rkir} O^{nna}_{Rjkr}\right) B_0(p^2,m^2_{\n_k},m^2_{P^0_r})\nonumber\\
- && {2 g^2_2} \sum_{k=1}^{10} {m_{\n_k}} \left(O^{nnz}_{Lki} O^{nnz}_{Rjk} 
+ O^{nnz}_{Ljk} O^{nnz}_{Rki}\right) {B_0(p^2,m^2_{\n_k},m^2_{Z^0_{\mu}})}\nonumber\\
- && {4 g^2_2} \sum_{k=1}^{5} {m_{\widetilde{\chi}^{\pm}_k}} \left(O^{cnw}_{Lki} O^{ncw}_{Rjk} 
+ O^{cnw}_{Rki} O^{ncw}_{Ljk}\right) {B_0(p^2,m^2_{{\widetilde\chi}^{\mp}_k},m^2_{W^{\pm}_{\mu}})}\nonumber\\
+ && {{\widetilde g}^2} \sum_{r=1}^{7} \sum_{k=1}^{5} {m_{\widetilde{\chi}^{\pm}_k}} 
\left(O^{cns}_{Lkir} O^{ncs}_{Ljkr} + O^{ncs}_{Rjkr} O^{cns}_{Rkir}\right) 
{B_0(p^2,m^2_{{\widetilde\chi}^{\mp}_k},m^2_{S^{\pm}_r})}\nonumber\\
+ && 3{{\widetilde g}^2} \sum_{r=1}^{6} \sum_{k=1}^{3} {m_{u_k}} \left(O^{un\u}_{Lkir} O^{nu\u}_{Ljkr} 
+ O^{un\u}_{Rkir} O^{nu\u}_{Rjkr}\right) {B_0(p^2,m^2_{u_k},m^2_{\u_r})}\nonumber\\
+ && \left.
3{{\widetilde g}^2} \sum_{r=1}^{6} \sum_{k=1}^{3} {m_{d_k}} \left(O^{dn\d}_{Lkir} O^{nd\d}_{Ljkr} 
+ O^{dn\d}_{Rkir} O^{nd\d}_{Rjkr}\right) {B_0(p^2,m^2_{d_k},m^2_{\d_r})} \right].\nonumber\\
\eea

Note that the quark - squark loops (second row, right most diagram of figure \ref{one-loop-diagrams})
receive an extra enhancement by a factor of $3$ from {\it{three}} different quark colours.
The Passarino-Veltman functions $(B_0,B_1)$ are given in appendix \ref{appenF}.

\chapter{ \sffamily{{\bf }
 }}\label{appenF}

\section{The ${\mathbf{B_0,~B_1}}$ functions}\label{b0-b1}

The $B_0$ and $B_1$ functions are Passarino-Veltman \cite{App'tHooft:1978xw,AppPassarino:1978jh} 
functions defined in the notation of \cite{AppHahn:1998yk} as

\bea{\label{Passarino-Veltman-Functions}}
\frac{i}{16 {\pi}^2} {B_0(p^2,m^2_{f^\prime_k},m^2_{b_r})} &=& 
{(\mu^2)^{4-D}}\int \frac{d^Dq}{(2\pi)^D} \frac{1}{(q^2-m^2_{f^\prime_k})((q+p)^2-m^2_{b_r})},\nonumber \\
\frac{i}{16 {\pi}^2} {B_{\mu}(p^2,m^2_{f^\prime_k},m^2_{b_r})} &=& 
{(\mu^2)^{4-D}}\int \frac{d^Dq}{(2\pi)^D} \frac{q_{\mu}}{(q^2-m^2_{f^\prime_k})((q+p)^2-m^2_{b_r})},\nonumber \\
{B_{\mu}(p^2,m^2_{f^\prime_k},m^2_{b_r})} &=& p_{\mu} {B_1(p^2,m^2_{f^\prime_k},m^2_{b_r})}.\nonumber \\
\eea
$D$ is the dimension of the integral. In the $D$ dimension mass dimension $[M]$ for a fermion
is $[M]^{\frac{D-1}{2}}$ and that of a scalar is $[M]^{\frac{D-2}{2}}$. Consequently, the
4-dimensional couplings are scaled by a factor $(\mu^2)^{4-D}$, where $[\mu]=[M]$.


\chapter{ \sffamily{{\bf }
 }}\label{appenG}

\section{Feynman diagrams for the tree level $\ntrl1$ decay}\label{LN-decay}

Possible two-body and three-body final states (at the tree level) arising from the $R_p$-violating
decays of a lightest neutralino, $\ntrl1$ are shown here
\begin{figure}[ht]
\centering
\vspace*{0.5cm}
\includegraphics[height=2.00cm]{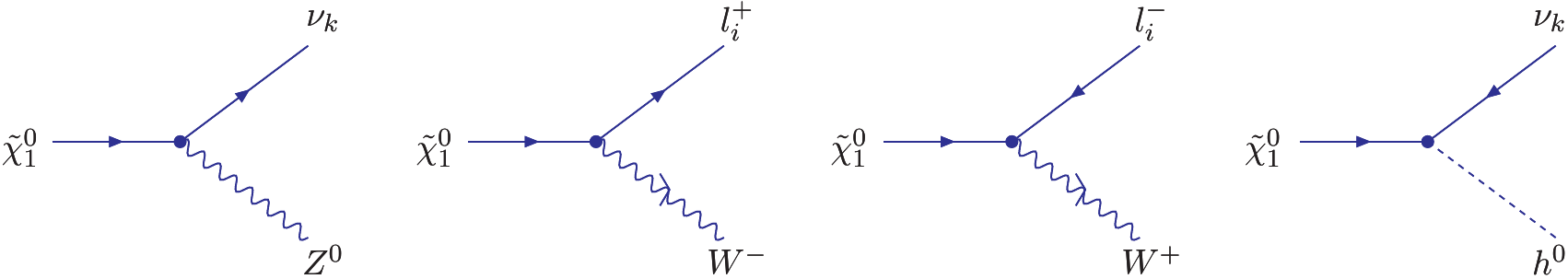}
\caption{Feynman diagrams for the possible two body decays of the lightest neutralino. 
$h^0$ is the lightest Higgs boson of the MSSM which is similar to the SM Higgs boson.}
\label{LSP-decay2}
\end{figure}
\begin{figure}[ht]
\centering
\vspace*{0.5cm}
\includegraphics[height=15.80cm,keepaspectratio]{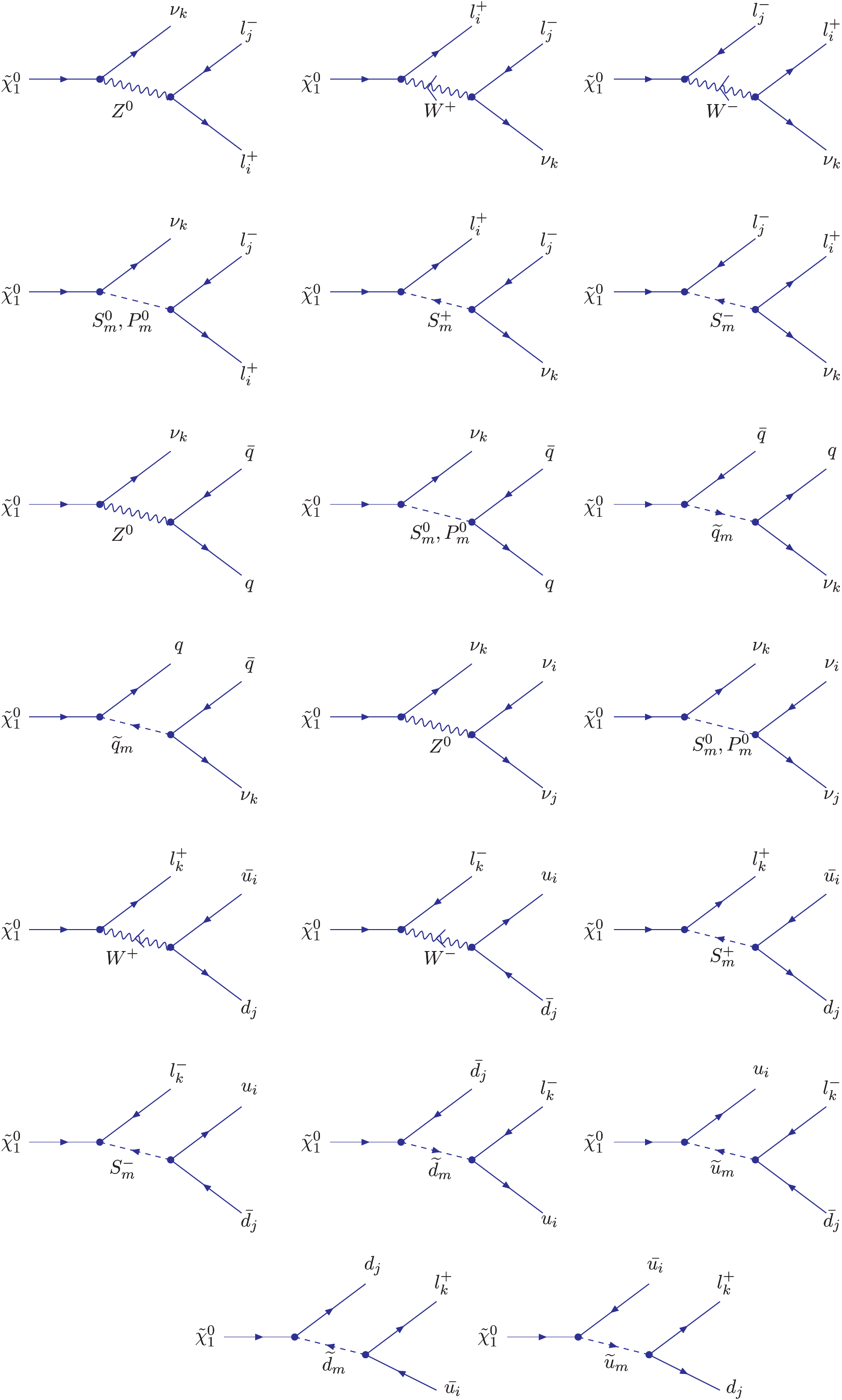}
\caption{Feynman diagrams for the possible three body decays of the lightest neutralino.
$S^0,P^0,S^\pm$ represent the scalar, the pseudoscalar and the charged scalar states, respectively.}
\label{LSP-decay3}
\end{figure}


\chapter{ \sffamily{{\bf }
 }}\label{appenH}

 \section{Feynman rules}\label{Feynman rules-2}
The relevant Feynman rules required for the calculation of the 
possible two-decays of the scalar and pseudoscalar states are shown in this
appendix. The factors $\eta_{i}$,$\eta_{j}$ and 
$\epsilon_{i}$, $\epsilon_{j}$ are the 
proper signs of neutralino and chargino masses \cite{AppGunion:1984yn}.

\subsection*{$\bigstar$~Chargino-chargino-neutral scalar}\label{cch}

The Lagrangian using four component spinor notation can be written as
\beq
\mathcal{L}^{cch}= - \frac{\widetilde g}{\rt2} \ovl{{\widetilde \chi}_i} 
(O^{cch}_{Lijk} P_L + O^{cch}_{Rijk} P_R) {{\widetilde \chi}_j} S^0_k,
\label{chargino-chargino-neutral scalar}
\eeq
where
\bea
{\widetilde g}O^{cch}_{Lijk} = & & \epsilon_j \left[\bRs_{k1} \left(
Y^{mn}_e \bU^*_{i,m+2} \bV^*_{j,n+2} + g_2 \bU^*_{i2} \bV^*_{j1} \right) 
\right. \nonumber\\
& &+ g_2 \bRs_{k2} \bU^*_{i1} \bV^*_{j2} \nonumber\\
& &+ \bRs_{k,m+2} \left(\lam^m \bU^*_{i2} \bV^*_{j2} 
-Y^{mn}_\nu  \bU^*_{i,n+2} \bV^*_{j2} \right) \nonumber\\
& &+ \left. 
\bRs_{k,m+5} \left(g_2 \bU^*_{i,m+2} \bV^*_{j1} 
-Y^{mn}_e  \bU^*_{i2} \bV^*_{j,n+2} \right) \right],
\label{cch-couplings}
\eea
and
\beq
\tg O^{cch}_{Rijk} =(\tg O^{cch}_{Ljik})^*.
\label{cch-R-couplings}
\eeq
\subsection*{$\bigstar$~Chargino-chargino-neutral pseudoscalar}\label{cca}

The Lagrangian using four component spinor notation can be written as
\beq
\mathcal{L}^{cch}= - i\frac{\widetilde g}{\rt2} \ovl{{\widetilde \chi}_i} 
(O^{cca}_{Lijk} P_L + O^{cca}_{Rijk} P_R) {{\widetilde \chi}_j} P^0_k,
\label{chargino-chargino-neutral pseudoscalar}
\eeq
where
\bea
{\widetilde g}O^{cca}_{Lijk} = & & \epsilon_j \left[\bRp_{k1} \left(
Y^{mn}_e \bU^*_{i,m+2} \bV^*_{j,n+2} - g_2 \bU^*_{i2} \bV^*_{j1} \right) 
\right. \nonumber\\
& &- g_2 \bRp_{k2} \bU^*_{i1} \bV^*_{j2} \nonumber\\
& &+ \bRp_{k,m+2} \left(\lam^m \bU^*_{i2} \bV^*_{j2} 
-Y^{mn}_\nu  \bU^*_{i,n+2} \bV^*_{j2} \right) \nonumber\\
& &- \left. 
\bRp_{k,m+5} \left(g_2 \bU^*_{i,m+2} \bV^*_{j1} 
+ Y^{mn}_e  \bU^*_{i2} \bV^*_{j,n+2} \right) \right],
\label{cca-couplings}
\eea
and
\beq
\tg O^{cca}_{Rijk} =-(\tg O^{cca}_{Ljik})^*.
\label{cca-R-couplings}
\eeq
\subsection*{$\bigstar$~Down-quark-down-quark-neutral scalar}\label{ddh}

The Lagrangian using four component spinor notation can be written as
\beq
\mathcal{L}^{ddh}= - {\widetilde g} \ovl{d_i} 
(O^{ddh}_{Lijk} P_L + O^{ddh}_{Rijk} P_R) {d_j} S^0_k ,
\label{down-down-neutral scalar}
\eeq
where
\bea
{\widetilde g}O^{ddh}_{Lijk} = & & \frac{1}{\rt2} 
Y^{mn}_d \bRs_{k1} \Rd_{L_{im}} \Rd_{L_{jn}},\nonumber\\
{\widetilde g}O^{ddh}_{Rijk} = & & ({\widetilde g} O^{ddh}_{Ljik})^*.
\label{ddh-couplings}
\eea
\subsection*{$\bigstar$~Down-quark-down-quark-neutral pseudoscalar}\label{dda}

The Lagrangian using four component spinor notation can be written as
\beq
\mathcal{L}^{dda}= - i{\widetilde g} \ovl{d_i} 
(O^{dda}_{Lijk} P_L + O^{dda}_{Rijk} P_R) {d_j} P^0_k ,
\label{down-down-neutral pseudoscalar}
\eeq
where
\bea
{\widetilde g}O^{dda}_{Lijk} = & & \frac{1}{\rt2} 
Y^{mn}_d \bRp_{k1} \Rd_{L_{im}} \Rd_{L_{jn}},\nonumber\\
{\widetilde g}O^{dda}_{Rijk} = & & -({\widetilde g} O^{dda}_{Ljik})^*.
\label{dda-couplings}
\eea
\subsection*{$\bigstar$~Up-quark-up-quark-neutral scalar}\label{uuh}

The Lagrangian using four component spinor notation can be written as
\beq
\mathcal{L}^{uuh}= - {\widetilde g} \ovl{u_i} 
(O^{uuh}_{Lijk} P_L + O^{uuh}_{Rijk} P_R) {u_j} S^0_k ,
\label{up-up-neutral scalar}
\eeq
where
\bea
{\widetilde g}O^{uuh}_{Lijk} = & & \frac{1}{\rt2} 
Y^{mn}_u \bRs_{k2} \Ru_{L_{im}} \Ru_{L_{jn}},\nonumber\\
{\widetilde g}O^{uuh}_{Rijk} = & & ({\widetilde g} O^{uuh}_{Ljik})^*.
\label{uuh-couplings}
\eea
\subsection*{$\bigstar$~Up-quark-up-quark-neutral pseudoscalar}\label{uua}

The Lagrangian using four component spinor notation can be written as
\beq
\mathcal{L}^{uua}= - i{\widetilde g} \ovl{u_i} 
(O^{uua}_{Lijk} P_L + O^{uua}_{Rijk} P_R) {u_j} P^0_k ,
\label{up-up-neutral pseudoscalar}
\eeq
where
\bea
{\widetilde g}O^{uua}_{Lijk} = & & \frac{1}{\rt2} 
Y^{mn}_u \bRp_{k2} \Ru_{L_{im}} \Ru_{L_{jn}},\nonumber\\
{\widetilde g}O^{uua}_{Rijk} = & & -({\widetilde g} O^{uua}_{Ljik})^*.
\label{uua-couplings}
\eea
\subsection*{$\bigstar$~Quark-squark-chargino}\label{qsqc}

The Lagrangian using four component spinor notation can be written as
\beq
\mathcal{L}^{q\q c}= -\wt{g} \ovl{\wt {\chi}^c_i}
\left(O^{cdu}_{Lijk} P_L + O^{cdu}_{Rijk} P_R\right) d_j \u^*_k
-\wt{g} \ovl{u_i} \left(O^{ucd}_{Lijk} P_L + O^{ucd}_{Rijk} P_R\right)
\wt{\chi}_j \d_k + h.c,
\label{q-sq-chargino}
\eeq
where
\bea
\wt{g} O^{cdu}_{Lijk} = & & -Y^{mn}_u \bV^*_{i2} \Rd_{L_{jm}} \Rsu_{k,n+3}
+g_2 \bV^*_{i1} \Rd_{L_{jm}} \Rsu_{km},\nn\\
\wt{g} O^{cdu}_{Rijk} = & & -Y^{{mn}^*}_d \bU_{i2} {\Rd}^*_{R_{jn}} \Rsu_{km},\nn\\
\wt{g} O^{ucd}_{Lijk} = & & -Y^{mn}_u \bV^*_{j2} \Ru_{R_{in}} {\Rsd}^*_{km},\nn\\
\wt{g} O^{ucd}_{Rijk} = & & -Y^{{mn}^*}_d \bU_{j2} {\Ru}^*_{L_{im}} {\Rsd}^*_{k,n+3}
+g_2 \bU_{j1} {\Ru}^*_{L_{im}} {\Rsd}^*_{km}.
\label{qsqc-couplings}
\eea
The charge conjugated chargino spinor $\wt{\chi}^c$ is defined
by eqn.(\ref{neutralino-chargino}).
\subsection*{$\bigstar$~Quark-quark-charged scalar}\label{qqcs}

The Lagrangian is written as
\beq
\mathcal{L}^{qqs}= -\wt{g} \ovl{u_i}
\left(O^{uds}_{Lijk} P_L + O^{uds}_{Rijk} P_R\right) d_j S^+_k
+ h.c,
\label{q-q-charged-scalar}
\eeq
where
\bea
\wt{g} O^{uds}_{Lijk} = & & -Y^{mn}_u \Ru_{R_{in}} \Rd_{L_{jm}} \bRc_{k2},\nn\\
\wt{g} O^{uds}_{Rijk} = & & -Y^{{mn}^*}_d {\Ru}^*_{L_{im}} {\Rd}^*_{R_{jn}} \bRc_{k1}.
\label{qqs-couplings}
\eea
 \section{Squared matrix elements for $h^0\to \ntrli \ntrlj, b \bar{b}
$}\label{h-mat-ele}
\bea
\bullet~ |M|^2 (h^0\to \ntrli \ntrlj) &&=
2\wt{g}^2(m^2_{h^0}-(m^2_{\ntrli}+m^2_{\ntrlj}))\left(
O^{{nnh}^*}_{Lij4}O^{nnh}_{Lij4} + O^{{nnh}^*}_{Rij4}O^{nnh}_{Rij4}\right)\nn\\
&&-4 \wt{g}^4
m_{\ntrli} m_{\ntrlj}\left(
O^{{nnh}^*}_{Rij4}O^{nnh}_{Lij4} + O^{{nnh}^*}_{Lij4}O^{nnh}_{Rij4}\right),
\label{math-n1n1}
\eea
where we have used the favour of an extra $2$ factor for $\ntrli-\ntrlj-h^0$
vertex \cite{AppHaber:1984rc} (also see appendix \ref{appenD}).
\bea
\bullet~ |M|^2 (h^0\to b \bar{b}) &&=
3\wt{g}^2(m^2_{h^0}-2m^2_b)\left(
O^{{ddh}^*}_{L334}O^{ddh}_{L334} + O^{{ddh}^*}_{R334}O^{ddh}_{R334}\right)\nn\\
&&-6 \wt{g}^4
m^2_{b} \left(
O^{{ddh}^*}_{R334}O^{ddh}_{L334} + O^{{ddh}^*}_{L334}O^{ddh}_{R334}\right),
\label{math-bb}
\eea
where we have used relations from appendix \ref{appenD}, section \ref{Feynman rules-2} 
and put $3$ for the colour factor.
\chapter{ \sffamily{{\bf }
 }}\label{appenI}

\section{Three body decays of the $\ntrl1$ LSP}\label{LSP-3body-decay}

In this appendix we write down the spin-averaged (sum over spins
of the final state particles and average over the spin of initial particle) matrix
element square ($\ovl{|\mathbb{M}|^2}$) for possible three body decays of a neutralino LSP 
$\ntrl1$. The possible decays are given by eqn. (\ref{2-3-body-decays}). Since neutralinos 
are fermion by nature, an average over the initial particle spin will yield a factor
of $\frac{1}{2}$, that is, mathematically,
\beq
\ovl{|\mathbb{M}|^2}=\frac{N_c X_1 X_2}{(2.{\frac{1}{2}}+1)} 
\left[\sum_i M^\dagger_i M_i + 2 \Re \left(\sum_{i \neq j}  M^\dagger_i M_j\right)\right],
\label{mat-ele}
\eeq
where we put spin of the neutralino, $S_{\ntrli}=\frac{1}{2}$ in the factor
$\frac{1}{(2.S_{\ntrli}+1)}$. The second terms of eqn.(\ref{mat-ele}) represent
interference terms in case multiple Feynman diagrams exist for a given process.
$M_i$ represents amplitude of the $i$-th Feynman Diagram of a given process.
$N_c$ is the colour factor which is $3(1)$ for processes involving quarks(leptons).
The quantities $X_{1,2}$ are associated with two vertices of a three body decay 
process (see figure \ref{LSP-qqnu} for example). These factors are $2$ for a $\ntrl1-\nu 
-{\rm{neutral~boson}}$ vertex\footnote{Also true for $\nu-\nu -{\rm{neutral~boson}}$ vertex,
appears in $\ntrl1 \to \nu \nu \bar{\nu}$ process.} since $\ntrl1,\nu$ are Majorana particles 
\cite{AppHaber:1984rc} and equal to $1$ for all other vertices. All processes are calculated 
using 't-Hooft-Feynman gauge.
\section{Process $\ntrl1\to q \bar{q} \nu$}\label{LSP-qqnu-decay}
\begin{figure}[ht]
\centering
\vspace*{0.5cm}
\includegraphics[height=6.00cm]{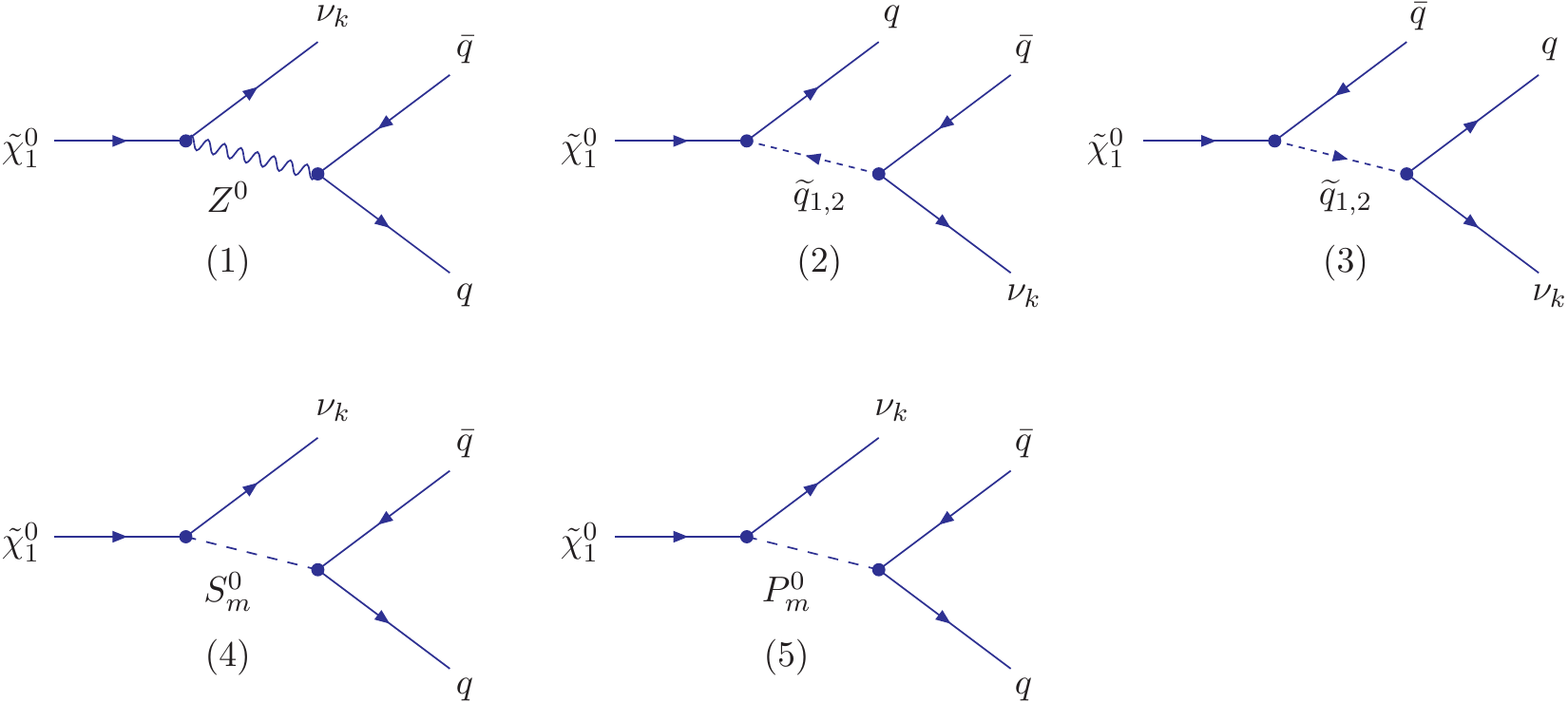}
\caption{Feynman diagrams for the possible three body decays of the lightest supersymmetric
particle into $q\bar{q}\nu$ final states, with $q\neq t$. $\wt{q}_{1,2}$ are the squarks
in the mass eigenbasis (see eqn.(\ref{squark-mass-basis})).
$S^0_m,P^0_m$ are the neutral scalar and pseudoscalar states of the $\mu\nu$SSM as
shown by eqns.(\ref{scalar-mass-basis}), (\ref{pseudoscalar-mass-basis}).}
\label{LSP-qqnu}
\end{figure}

We start with the processes involving down type quarks $(q=d,s,b)$ first and later
for $q=u,c$. We represent different down and up type quarks generically by $d$ and
$u$, respectively. We write down all possible $M^\dagger_iM_j$ for the five diagrams
shown in figure \ref{LSP-qqnu}. The four-momentum assignments are as follows
\beq
\ntrl1(P) \to q(k)+\bar{q}(k')+\nu_i(p),
\label{momentum-qqnu}
\eeq
where $i$ stands for $i$-th neutrino flavour. $i=1,2,3$ or $e,\mu,\tau$.
$\ntrl1$ is the lightest of the seven heavy neutralino states (see 
eqn.(\ref{neutralino_mass_eigenstate_matrixform})).
\bea
\bullet~ M^\dagger_1M_1 (\ntrl1\to q \bar{q}\sum \nu_i) &=& 
\frac{4 g^4_2}{cos^2_{\theta_W}\left[((k+k')^2-m^2_Z)^2+m^2_Z\Gamma^2_Z\right]} \nn\\
&&\sum_i
\left[(P.k)(p.k')A^{qq\nu_i}_{11}+(P.k')(p.k)B^{qq\nu_i}_{11}+m^2_q(P.p)C^{qq\nu_i}_{11}\right],\nn\\
\label{mat11-qqnu}
\eea
where $q=d(u)$, $\Gamma_Z$ is the $Z$-boson decay width and
\bea
A^{qq\nu_i}_{11} &=& \left(O^{{1iZ}^*}_{Li1} O^{{1iZ}}_{Li1}
+ O^{{1iZ}^*}_{Ri1} O^{{1iZ}}_{Ri1}\right)\left(\frac{1(4)}{9}sin^4{\theta_W}
-\frac{1(2)}{6}sin^2{\theta_W}+\frac{1}{8}\right)\nn\\
&+& \left(O^{{1iZ}^*}_{Li1} O^{{1iZ}}_{Li1}
- O^{{1iZ}^*}_{Ri1} O^{{1iZ}}_{Ri1}\right)
\left(\frac{1(2)}{6}sin^2{\theta_W}-\frac{1}{8}\right),\nn\\
B^{qq\nu_i}_{11} &=& \left(O^{{1iZ}^*}_{Li1} O^{{1iZ}}_{Li1}
+ O^{{1iZ}^*}_{Ri1} O^{{1iZ}}_{Ri1}\right)\left(\frac{1(4)}{9}sin^4{\theta_W}
-\frac{1(2)}{6}sin^2{\theta_W}+\frac{1}{8}\right)\nn\\
&-& \left(O^{{1iZ}^*}_{Li1} O^{{1iZ}}_{Li1}
- O^{{1iZ}^*}_{Ri1} O^{{1iZ}}_{Ri1}\right)
\left(\frac{1}{6}sin^2{\theta_W}-\frac{1}{8}\right),\nn\\
C^{qq\nu_i}_{11} &=& \left(O^{{1iZ}^*}_{Li1} O^{{1iZ}}_{Li1}
+ O^{{1iZ}^*}_{Ri1} O^{{1iZ}}_{Ri1}\right)\left(\frac{1(4)}{9}sin^4{\theta_W}
-\frac{1(2)}{6}sin^2{\theta_W}\right).
\label{mat11-qqnu-det}
\eea
\bea
\bullet~ M^\dagger_2M_2 (\ntrl1\to q \bar{q}\sum \nu_i) &=& \sum^2_{r,s=1}
\frac{4 \wt {g}^4}{\left[((k'+p)^2-m^2_{\wt {q}_r})
((k'+p)^2-m^2_{\wt {q}_s})\right]} \nn\\
&&\sum_i
\left[(P.k)(p.k')A^{qq\nu_i}_{22}+ m_q m_{\ntrl1} (p.k')B^{qq\nu_i}_{22}\right],
\label{mat22-qqnu}
\eea
where
\bea
A^{qq\nu_i}_{22} &=& \left(O^{q1\q}_{Lq1\q_s}O^{{q1\q}^*}_{Lq1\q_r}
+ O^{q1\q}_{Rq1\q_s}O^{{q1\q}^*}_{Rq1\q_r}\right)
\left(O^{iq\q}_{Liq\q_s}O^{{iq\q}^*}_{Liq\q_r}
+ O^{iq\q}_{Riq\q_s}O^{{iq\q}^*}_{Riq\q_r}\right),\nn\\
B^{qq\nu_i}_{22} &=& \left(O^{q1\q}_{Lq1\q_s}O^{{q1\q}^*}_{Rq1\q_r}
+ O^{q1\q}_{Rq1\q_s}O^{{q1\q}^*}_{Lq1\q_r}\right)
\left(O^{iq\q}_{Liq\q_s}O^{{iq\q}^*}_{Liq\q_r}
+ O^{iq\q}_{Riq\q_s}O^{{iq\q}^*}_{Riq\q_r}\right).
\label{mat22-qqnu-det}
\eea
\bea
\bullet~ M^\dagger_3M_3 (\ntrl1\to q \bar{q}\sum \nu_i) &=& \sum^2_{r,s=1}
\frac{4 \wt {g}^4}{\left[((k+p)^2-m^2_{\wt {q}_r})
((k+p)^2-m^2_{\wt {q}_s})\right]} \nn\\
&&\sum_i
\left[(P.k')(p.k)A^{qq\nu_i}_{33}+ m_q m_{\ntrl1} (p.k)B^{qq\nu_i}_{33}\right],
\label{mat33-qqnu}
\eea
where
\bea
A^{qq\nu_i}_{33} &=& \left(O^{1q\q}_{L1q\q_s}O^{{1q\q}^*}_{L1q\q_r}
+ O^{1q\q}_{R1q\q_s}O^{{1q\q}^*}_{R1q\q_r}\right)
\left(O^{qi\q}_{Lqi\q_s}O^{{qi\q}^*}_{Lqi\q_r}
+ O^{qi\q}_{Rqi\q_s}O^{{qi\q}^*}_{Rqi\q_r}\right),\nn\\
B^{qq\nu_i}_{33} &=& \left(O^{1q\q}_{L1q\q_s}O^{{1q\q}^*}_{R1q\q_r}
+ O^{1q\q}_{R1q\q_s}O^{{1q\q}^*}_{L1q\q_r}\right)
\left(O^{qi\q}_{Lqi\q_s}O^{{qi\q}^*}_{Lqi\q_r}
+ O^{qi\q}_{Rqi\q_s}O^{{qi\q}^*}_{Rqi\q_r}\right).
\label{mat33-qqnu-det}
\eea
\bea
\bullet~ M^\dagger_4M_4 (\ntrl1\to q \bar{q}\sum \nu_i) &=& \sum^8_{k,l=1}
\frac{2 \wt {g}^4}{\left[((k+k')^2-m^2_{S^0_k})
((k+k')^2-m^2_{S^0_l})\right]} \nn\\
&&\sum_i
\left[(P.p)(k.k')A^{qq\nu_i}_{44}- m^2_q (P.p)B^{qq\nu_i}_{44}\right],
\label{mat44-qqnu}
\eea
where
\bea
A^{qq\nu_i}_{44} &=& \left(O^{{qqh}^*}_{Lqqk} O^{qqh}_{Lqql}
+ O^{{qqh}^*}_{Rqqk} O^{qqh}_{Rqql} \right)
\left(O^{{i1h}^*}_{Li1k} O^{i1h}_{Li1l}
+ O^{{i1h}^*}_{Ri1k} O^{i1h}_{Ri1l} \right),\nn\\
B^{qq\nu_i}_{44} &=& \left(O^{{qqh}^*}_{Lqqk} O^{qqh}_{Rqql}
+ O^{{qqh}^*}_{Rqqk} O^{qqh}_{Lqql} \right)
\left(O^{{i1h}^*}_{Li1k} O^{i1h}_{Li1l}
+ O^{{i1h}^*}_{Ri1k} O^{i1h}_{Ri1l} \right).
\label{mat44-qqnu-det}
\eea
\bea
\bullet~ M^\dagger_5M_5 (\ntrl1\to q \bar{q}\sum \nu_i) &=& \sum^8_{k,l=1}
\frac{2 \wt {g}^4}{\left[((k+k')^2-m^2_{P^0_k})
((k+k')^2-m^2_{P^0_l})\right]} \nn\\
&&\sum_i
\left[(P.p)(k.k')A^{qq\nu_i}_{55}- m^2_q (P.p)B^{qq\nu_i}_{55}\right],
\label{mat55-qqnu}
\eea
where
\bea
A^{qq\nu_i}_{55} &=& \left(O^{{qqa}^*}_{Lqqk} O^{qqa}_{Lqql}
+ O^{{qqa}^*}_{Rqqk} O^{qqa}_{Rqql} \right)
\left(O^{{i1a}^*}_{Li1k} O^{i1a}_{Li1l}
+ O^{{i1a}^*}_{Ri1k} O^{i1a}_{Ri1l} \right),\nn\\
B^{qq\nu_i}_{55} &=& \left(O^{{qqa}^*}_{Lqqk} O^{qqa}_{Rqql}
+ O^{{qqa}^*}_{Rqqk} O^{qqa}_{Lqql} \right)
\left(O^{{i1a}^*}_{Li1k} O^{i1a}_{Li1l}
+ O^{{i1a}^*}_{Ri1k} O^{i1a}_{Ri1l} \right).
\label{mat55-qqnu-det}
\eea
\bea
&&\bullet~ M^\dagger_1M_2 (\ntrl1\to q \bar{q}\sum \nu_i) = -(+)\sum^2_{r=1}
\frac{2 g^2_2 \wt {g}^2 sec\theta_W}{\left[((k+k')^2-m^2_Z-im_Z\Gamma_Z)
((p+k')^2-m^2_{\q_r})\right]} \nn\\
&&\sum_i
\left[2(p.k')(P.k)(A^{qq\nu_i}_{12}C^{qq\nu_i}_{12}+B^{qq\nu_i}_{12}D^{qq\nu_i}_{12})
+ m_q m_{\ntrl1}(p.k)(A^{qq\nu_i}_{12}D^{qq\nu_i}_{12}+B^{qq\nu_i}_{12}C^{qq\nu_i}_{12})
\right.\nn\\
&&+ \left.
2m_q m_{\ntrl1}(p.k')(B^{qq\nu_i}_{12}E^{qq\nu_i}_{12}+A^{qq\nu_i}_{12}F^{qq\nu_i}_{12})
+m^2_q(P.p)(A^{qq\nu_i}_{12}E^{qq\nu_i}_{12}+B^{qq\nu_i}_{12}F^{qq\nu_i}_{12})
\right],\nn\\
\label{mat12-qqnu}
\eea
where
\bea
A^{qq\nu_i}_{12} &=& O^{{1iZ}^*}_{Li1} O^{iq\q}_{Riq\q_r}, 
~~B^{qq\nu_i}_{12} = O^{{1iZ}^*}_{Ri1} O^{iq\q}_{Liq\q_r},
~~C^{qq\nu_i}_{12} = \frac{1(2)}{3}sin^2\theta_W O^{q1\q}_{Lq1\q_r},\nn\\
D^{qq\nu_i}_{12} &=& \frac{1(2)}{3}sin^2\theta_W O^{q1\q}_{Rq1\q_r}
-\frac{1}{2} O^{q1\q}_{Rq1\q_r},
~~E^{qq\nu_i}_{12} = \frac{1(2)}{3}sin^2\theta_W O^{q1\q}_{Lq1\q_r}
-\frac{1}{2} O^{q1\q}_{Lq1\q_r},\nn\\
F^{qq\nu_i}_{12} &=& \frac{1(2)}{3}sin^2\theta_W O^{q1\q}_{Rq1\q_r}.
\label{mat12-qqnu-det}
\eea
\bea
&&\bullet~ M^\dagger_1M_3 (\ntrl1\to q \bar{q}\sum \nu_i) = (-)\sum^2_{r=1}
\frac{2 g^2_2 \wt {g}^2 sec\theta_W}{\left[((k+k')^2-m^2_Z-im_Z\Gamma_Z)
((p+k)^2-m^2_{\q_r})\right]} \nn\\
&&\sum_i
\left[2(p.k)(P.k')(A^{qq\nu_i}_{13}C^{qq\nu_i}_{13}+B^{qq\nu_i}_{13}D^{qq\nu_i}_{13})
+ m_q m_{\ntrl1}(p.k')(A^{qq\nu_i}_{13}D^{qq\nu_i}_{13}+B^{qq\nu_i}_{13}C^{qq\nu_i}_{13})
\right.\nn\\
&&+ \left.
2m_q m_{\ntrl1}(p.k)(B^{qq\nu_i}_{13}E^{qq\nu_i}_{13}+A^{qq\nu_i}_{13}F^{qq\nu_i}_{13})
+m^2_q(P.p)(A^{qq\nu_i}_{13}E^{qq\nu_i}_{13}+B^{qq\nu_i}_{13}F^{qq\nu_i}_{13})
\right],\nn\\
\label{mat13-qqnu}
\eea
where
\bea
A^{qq\nu_i}_{13} &=& O^{{1iZ}^*}_{Ri1} O^{qi\q}_{Lqi\q_r}, 
~~B^{qq\nu_i}_{13} = O^{{1iZ}^*}_{Li1} O^{qi\q}_{Rqi\q_r},
~~C^{qq\nu_i}_{13} = \frac{1(2)}{3}sin^2\theta_W O^{1q\q}_{R1q\q_r},\nn\\
D^{qq\nu_i}_{13} &=& \frac{1(2)}{3}sin^2\theta_W O^{1q\q}_{L1q\q_r}
-\frac{1}{2} O^{1q\q}_{L1q\q_r},
~~E^{qq\nu_i}_{13} = \frac{1(2)}{3}sin^2\theta_W O^{1q\q}_{R1q\q_r}
-\frac{1}{2} O^{1q\q}_{R1q\q_r},\nn\\
F^{qq\nu_i}_{13} &=& \frac{1(2)}{3}sin^2\theta_W O^{1q\q}_{L1q\q_r}.
\label{mat13-qqnu-det}
\eea
\bea
\bullet~ M^\dagger_1M_4 (\ntrl1\to q \bar{q}\sum \nu_i) &=& -(+)\sum^8_{k=1}
\frac{\rt2 g^2_2 \wt {g}^2 sec\theta_W}{\left[((k+k')^2-m^2_Z-im_Z\Gamma_Z)
((k+k')^2-m^2_{S^0_k})\right]} \nn\\
&&\sum_i
\left[m_q m_{\ntrl1} (p.k') A^{qq\nu_i}_{14} - m_q m_{\ntrl1} (p.k) B^{qq\nu_i}_{14}
\right],
\label{mat14-qqnu}
\eea
where
\bea
A^{qq\nu_i}_{14} &=& \left(O^{{1iZ}^*}_{Ri1} O^{i1h}_{Li1k}
+ O^{{1iZ}^*}_{Li1} O^{i1h}_{Ri1k}\right)\nn\\
&\times&\left\{\left(\frac{1(2)}{3}sin^2\theta_W-\frac{1}{2}\right)O^{qqh}_{Lbbk}
+ \frac{1(2)}{3}sin^2\theta_W O^{qqh}_{Rbbk}\right\},\nn\\
B^{qq\nu_i}_{14} &=& \left(O^{{1iZ}^*}_{Ri1} O^{i1h}_{Li1k}
+ O^{{1iZ}^*}_{Li1} O^{i1h}_{Ri1k}\right)\nn\\
&\times&\left\{\left(\frac{1(2)}{3}sin^2\theta_W-\frac{1}{2}\right)O^{qqh}_{Rbbk}
+ \frac{1(2)}{3}sin^2\theta_W O^{qqh}_{Lbbk}\right\}.
\label{mat14-qqnu-det}
\eea
\bea
\bullet~ M^\dagger_1M_5 (\ntrl1\to q \bar{q}\sum \nu_i) &=& (-)\sum^8_{k=1}
\frac{\rt2 g^2_2 \wt {g}^2 sec\theta_W}{\left[((k+k')^2-m^2_Z-im_Z\Gamma_Z)
((k+k')^2-m^2_{P^0_k})\right]} \nn\\
&&\sum_i
\left[m_q m_{\ntrl1} (p.k') A^{qq\nu_i}_{15} - m_q m_{\ntrl1} (p.k) B^{qq\nu_i}_{15}
\right],
\label{mat15-qqnu}
\eea
where
\bea
A^{qq\nu_i}_{15} &=& \left(O^{{1iZ}^*}_{Ri1} O^{i1a}_{Li1k}
+ O^{{1iZ}^*}_{Li1} O^{i1a}_{Ri1k}\right)\nn\\
&\times&\left\{\left(\frac{1(2)}{3}sin^2\theta_W-\frac{1}{2}\right)O^{qqa}_{Lbbk}
+ \frac{1(2)}{3}sin^2\theta_W O^{qqa}_{Rbbk}\right\},\nn\\
B^{qq\nu_i}_{15} &=& \left(O^{{1iZ}^*}_{Ri1} O^{i1a}_{Li1k}
+ O^{{1iZ}^*}_{Li1} O^{i1a}_{Ri1k}\right)\nn\\
&\times&\left\{\left(\frac{1(2)}{3}sin^2\theta_W-\frac{1}{2}\right)O^{qqa}_{Rbbk}
+ \frac{1(2)}{3}sin^2\theta_W O^{qqa}_{Lbbk}\right\}.
\label{mat15-qqnu-det}
\eea
\bea
&&\bullet~ M^\dagger_2M_3 (\ntrl1\to q \bar{q}\sum \nu_i) = -\sum^2_{r,s=1}
\frac{2 \wt {g}^4}{\left[((p+k')^2-m^2_{\q_r})
((p+k)^2-m^2_{\q_s})\right]} \nn\\
&&\sum_i
\left[\{(P.k)(p.k')-(P.p)(k.k')+(P.k')(p.k)\} (A^{qq\nu_i}_{23} O^{{q1\q}^*}_{Lq1\q_r}
+B^{qq\nu_i}_{23} O^{{q1\q}^*}_{Rq1\q_r})
\right.\nn\\
&+& m_q m_{\ntrl1} (p.k') (A^{qq\nu_i}_{23} O^{{q1\q}^*}_{Rq1\q_r}
+B^{qq\nu_i}_{23} O^{{q1\q}^*}_{Lq1\q_r})
+ m_q m_{\ntrl1} (p.k) (C^{qq\nu_i}_{23} O^{{q1\q}^*}_{Rq1\q_r}
+D^{qq\nu_i}_{23} O^{{q1\q}^*}_{Lq1\q_r})\nn\\
&+& \left.
m^2_q(P.p) (C^{qq\nu_i}_{23} O^{{q1\q}^*}_{Lq1\q_r}
+D^{qq\nu_i}_{23} O^{{q1\q}^*}_{Rq1\q_r})
\right],
\label{mat23-qqnu}
\eea
where
\bea
A^{qq\nu_i}_{23} &=& O^{qi\q}_{Lqi\q_s} O^{{iq\q}^*}_{Liq\q_r} O^{1q\q}_{L1q\q_s},
~~B^{qq\nu_i}_{23} = O^{qi\q}_{Rqi\q_s} O^{{iq\q}^*}_{Riq\q_r} O^{1q\q}_{R1q\q_s},\nn\\
C^{qq\nu_i}_{23} &=& O^{qi\q}_{Rqi\q_s} O^{{iq\q}^*}_{Riq\q_r} O^{1q\q}_{L1q\q_s},
~~D^{qq\nu_i}_{23} = O^{qi\q}_{Lqi\q_s} O^{{iq\q}^*}_{Liq\q_r} O^{1q\q}_{R1q\q_s}.
\label{mat23-qqnu-det}
\eea
\bea
&&\bullet~ M^\dagger_2M_4 (\ntrl1\to q \bar{q}\sum \nu_i) = \sum^2_{r=1}\sum^8_{k=1}
\frac{\rt2 \wt {g}^4}{\left[((p+k')^2-m^2_{\q_r})
((k+k')^2-m^2_{S^0_k})\right]} \nn\\
&&\sum_i
\left[\{(P.p)(k.k')-(P.k')(p.k)+(p.k')(P.k)\} (A^{qq\nu_i}_{24} O^{{q1\q}^*}_{Rq1\q_r}
+B^{qq\nu_i}_{24} O^{{q1\q}^*}_{Lq1\q_r})
\right.\nn\\
&-& i \epsilon_{\mu\nu\rho\sigma} p^\mu P^\nu k^\rho {k'}^\sigma
(A^{qq\nu_i}_{24} O^{{q1\q}^*}_{Rq1\q_r}
-B^{qq\nu_i}_{24} O^{{q1\q}^*}_{Lq1\q_r})\nn\\
&+& m_q m_{\ntrl1} (p.k') (A^{qq\nu_i}_{24} O^{{q1\q}^*}_{Lq1\q_r}
+B^{qq\nu_i}_{24} O^{{q1\q}^*}_{Rq1\q_r})
- m_q m_{\ntrl1} (p.k) (C^{qq\nu_i}_{24} O^{{q1\q}^*}_{Lq1\q_r}
+D^{qq\nu_i}_{24} O^{{q1\q}^*}_{Rq1\q_r})\nn\\
&-& \left.
m^2_q(P.p) (C^{qq\nu_i}_{24} O^{{q1\q}^*}_{Rq1\q_r}
+D^{qq\nu_i}_{24} O^{{q1\q}^*}_{Lq1\q_r})
\right],
\label{mat24-qqnu}
\eea
where
\bea
A^{qq\nu_i}_{24} &=& O^{qqh}_{Rqqk} O^{{iq\q}^*}_{Riq\q_r} O^{i1h}_{Ri1k},
~~B^{qq\nu_i}_{24} = O^{qqh}_{Lqqk} O^{{iq\q}^*}_{Liq\q_r} O^{i1h}_{Li1k},\nn\\
C^{qq\nu_i}_{24} &=& O^{qqh}_{Lqqk} O^{{iq\q}^*}_{Riq\q_r} O^{i1h}_{Ri1k},
~~D^{qq\nu_i}_{24} = O^{qqh}_{Rqqk} O^{{iq\q}^*}_{Liq\q_r} O^{i1h}_{Li1k}.
\label{mat24-qqnu-det}
\eea
\bea
&&\bullet~ M^\dagger_2M_5 (\ntrl1\to q \bar{q}\sum \nu_i) = -\sum^2_{r=1}\sum^8_{k=1}
\frac{\rt2 \wt {g}^4}{\left[((p+k')^2-m^2_{\q_r})
((k+k')^2-m^2_{P^0_k})\right]} \nn\\
&&\sum_i
\left[\{(P.p)(k.k')-(P.k')(p.k)+(p.k')(P.k)\} (A^{qq\nu_i}_{25} O^{{q1\q}^*}_{Rq1\q_r}
+B^{qq\nu_i}_{25} O^{{q1\q}^*}_{Lq1\q_r})
\right.\nn\\
&-& i \epsilon_{\mu\nu\rho\sigma} p^\mu P^\nu k^\rho {k'}^\sigma
(A^{qq\nu_i}_{25} O^{{q1\q}^*}_{Rq1\q_r}
-B^{qq\nu_i}_{25} O^{{q1\q}^*}_{Lq1\q_r})\nn\\
&+& m_q m_{\ntrl1} (p.k') (A^{qq\nu_i}_{25} O^{{q1\q}^*}_{Lq1\q_r}
+B^{qq\nu_i}_{25} O^{{q1\q}^*}_{Rq1\q_r})
- m_q m_{\ntrl1} (p.k) (C^{qq\nu_i}_{25} O^{{q1\q}^*}_{Lq1\q_r}
+D^{qq\nu_i}_{25} O^{{q1\q}^*}_{Rq1\q_r})\nn\\
&-& \left.
m^2_q(P.p) (C^{qq\nu_i}_{25} O^{{q1\q}^*}_{Rq1\q_r}
+D^{qq\nu_i}_{25} O^{{q1\q}^*}_{Lq1\q_r})
\right],
\label{mat25-qqnu}
\eea
where
\bea
A^{qq\nu_i}_{25} &=& O^{qqa}_{Rqqk} O^{{iq\q}^*}_{Riq\q_r} O^{i1a}_{Ri1k},
~~B^{qq\nu_i}_{25} = O^{qqa}_{Lqqk} O^{{iq\q}^*}_{Liq\q_r} O^{i1a}_{Li1k},\nn\\
C^{qq\nu_i}_{25} &=& O^{qqa}_{Lqqk} O^{{iq\q}^*}_{Riq\q_r} O^{i1a}_{Ri1k},
~~D^{qq\nu_i}_{25} = O^{qqa}_{Rqqk} O^{{iq\q}^*}_{Liq\q_r} O^{i1a}_{Li1k}.
\label{mat25-qqnu-det}
\eea
\bea
&&\bullet~ M^\dagger_3M_4 (\ntrl1\to q \bar{q}\sum \nu_i) = \sum^2_{r=1}\sum^8_{k=1}
\frac{\rt2 \wt {g}^4}{\left[((p+k)^2-m^2_{\q_r})
((k+k')^2-m^2_{S^0_k})\right]} \nn\\
&&\sum_i
\left[\{(P.k')(p.k)-(P.k)(p.k')+(P.p)(k.k')\} (A^{qq\nu_i}_{34} O^{{1q\q}^*}_{L1q\q_r}
+B^{qq\nu_i}_{34} O^{{1q\q}^*}_{R1q\q_r})
\right.\nn\\
&+& m_q m_{\ntrl1} (p.k) (A^{qq\nu_i}_{34} O^{{1q\q}^*}_{R1q\q_r}
+B^{qq\nu_i}_{34} O^{{1q\q}^*}_{L1q\q_r})
- m_q m_{\ntrl1} (p.k') (C^{qq\nu_i}_{34} O^{{1q\q}^*}_{R1q\q_r}
+D^{qq\nu_i}_{34} O^{{1q\q}^*}_{L1q\q_r})\nn\\
&-& \left.
m^2_q(P.p) (C^{qq\nu_i}_{34} O^{{1q\q}^*}_{L1q\q_r}
+D^{qq\nu_i}_{34} O^{{1q\q}^*}_{R1q\q_r})
\right],
\label{mat34-qqnu}
\eea
where
\bea
A^{qq\nu_i}_{34} &=& O^{qqh}_{Lqqk} O^{{qi\q}^*}_{Lqi\q_r} O^{i1h}_{Li1k},
~~B^{qq\nu_i}_{34} = O^{qqh}_{Rqqk} O^{{qi\q}^*}_{Rqi\q_r} O^{i1h}_{Ri1k},\nn\\
C^{qq\nu_i}_{34} &=& O^{qqh}_{Rqqk} O^{{qi\q}^*}_{Lqi\q_r} O^{i1h}_{Li1k},
~~D^{qq\nu_i}_{34} = O^{qqh}_{Lqqk} O^{{qi\q}^*}_{Rqi\q_r} O^{i1h}_{Ri1k}.
\label{mat34-qqnu-det}
\eea
\bea
&&\bullet~ M^\dagger_3M_5 (\ntrl1\to q \bar{q}\sum \nu_i) = -\sum^2_{r=1}\sum^8_{k=1}
\frac{\rt2 \wt {g}^4}{\left[((p+k)^2-m^2_{\q_r})
((k+k')^2-m^2_{P^0_k})\right]} \nn\\
&&\sum_i
\left[\{(P.k')(p.k)-(P.k)(p.k')+(P.p)(k.k')\} (A^{qq\nu_i}_{35} O^{{1q\q}^*}_{L1q\q_r}
+B^{qq\nu_i}_{35} O^{{1q\q}^*}_{R1q\q_r})
\right.\nn\\
&+& m_q m_{\ntrl1} (p.k) (A^{qq\nu_i}_{35} O^{{1q\q}^*}_{R1q\q_r}
+B^{qq\nu_i}_{35} O^{{1q\q}^*}_{L1q\q_r})
- m_q m_{\ntrl1} (p.k') (C^{qq\nu_i}_{35} O^{{1q\q}^*}_{R1q\q_r}
+D^{qq\nu_i}_{35} O^{{1q\q}^*}_{L1q\q_r})\nn\\
&-& \left.
m^2_q(P.p) (C^{qq\nu_i}_{35} O^{{1q\q}^*}_{L1q\q_r}
+D^{qq\nu_i}_{35} O^{{1q\q}^*}_{R1q\q_r})
\right],
\label{mat35-qqnu}
\eea
where
\bea
A^{qq\nu_i}_{35} &=& O^{qqa}_{Lqqk} O^{{qi\q}^*}_{Lqi\q_r} O^{i1a}_{Li1k},
~~B^{qq\nu_i}_{35} = O^{qqa}_{Rqqk} O^{{qi\q}^*}_{Rqi\q_r} O^{i1a}_{Ri1k},\nn\\
C^{qq\nu_i}_{35} &=& O^{qqa}_{Rqqk} O^{{qi\q}^*}_{Lqi\q_r} O^{i1a}_{Li1k},
~~D^{qq\nu_i}_{35} = O^{qqa}_{Lqqk} O^{{qi\q}^*}_{Rqi\q_r} O^{i1a}_{Ri1k}.
\label{mat35-qqnu-det}
\eea
\bea
&&\bullet~ M^\dagger_4M_5 (\ntrl1\to q \bar{q}\sum \nu_i) = -\sum^8_{k,l=1}
\frac{2 \wt {g}^4}{\left[((k+k')^2-m^2_{S^0_k})
((k+k')^2-m^2_{P^0_l})\right]} \nn\\
&&\sum_i
\left[(P.p)(k.k')\left(O^{{i1h}^*}_{Li1k} O^{i1a}_{Li1l} + O^{{i1h}^*}_{Ri1k} O^{i1a}_{Ri1l} \right)
\left(O^{{qqh}^*}_{Lqqk} O^{qqa}_{Lqql} + O^{{qqh}^*}_{Rqqk} O^{qqa}_{Rqql} \right)
\right.\nn\\
&&-\left.
m^2_q(P.p)\left(O^{{i1h}^*}_{Li1k} O^{i1a}_{Li1l} + O^{{i1h}^*}_{Ri1k} O^{i1a}_{Ri1l} \right)
\left(O^{{qqh}^*}_{Rqqk} O^{qqa}_{Lqql} + O^{{qqh}^*}_{Lqqk} O^{qqa}_{Rqql} \right)
\right].
\label{mat45-qqnu}
\eea
Values for Weinberg angle $\theta_W$ and $\Gamma_Z$ are given in ref. \cite{AppNakamura-c2}.
Quark masses are also taken from ref. \cite{AppNakamura-c2}. All the relevant couplings
are given in appendices \ref{appenD} and \ref{appenH}.

\section{Process $\ntrl1\to \ell^+_i \ell^-_j \nu_k$}\label{LSP-llnu-decay}
\begin{figure}[ht]
\centering
\vspace*{0.5cm}
\includegraphics[height=9.00cm]{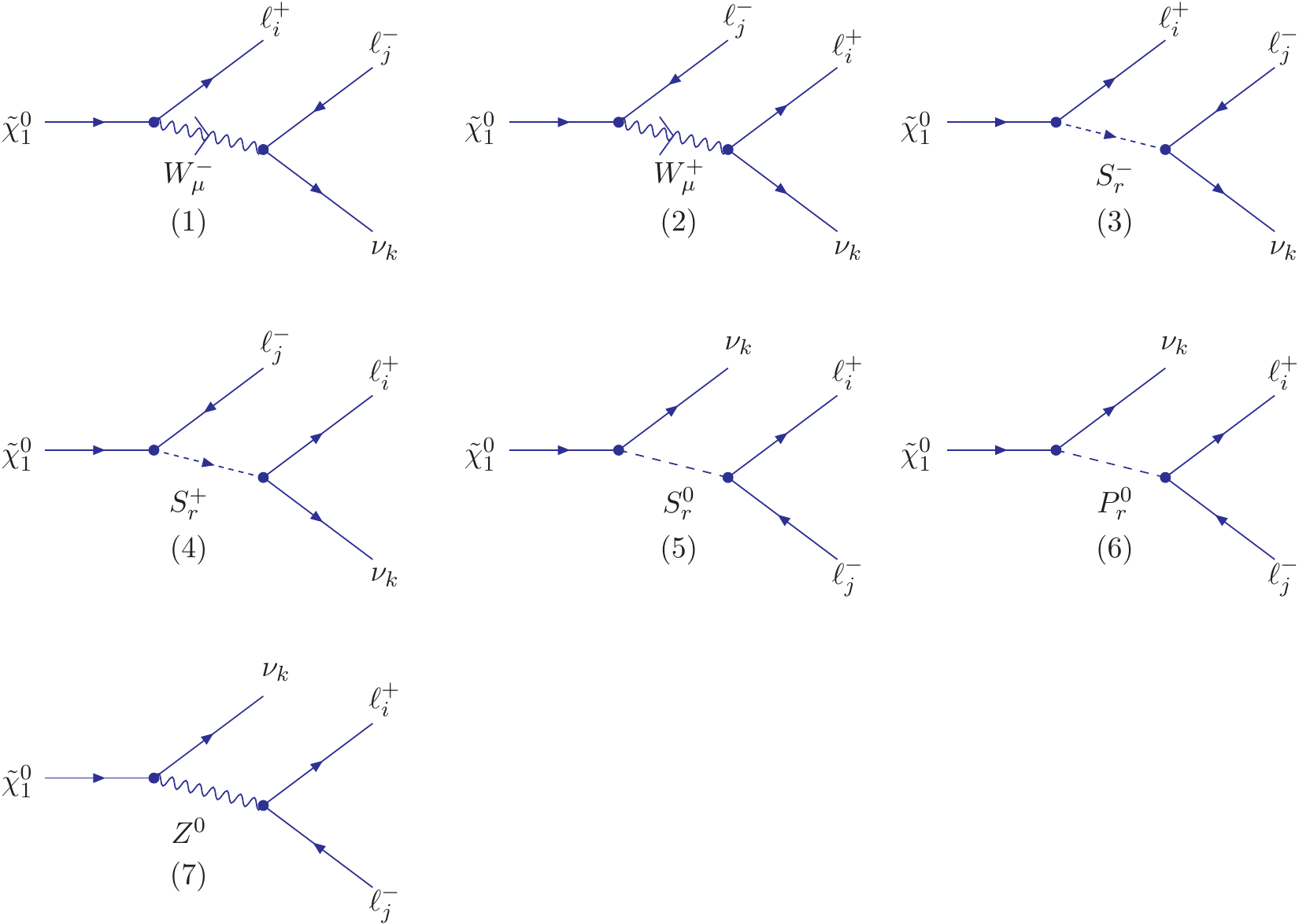}
\caption{Feynman diagrams for the possible three body decays of the lightest supersymmetric
particle into $\ell^+_i+\ell^-_j+\nu_k$ final states. 
$S^\pm_r,S^0_r,P^0_r$ are the charged scalar, neutral scalar and pseudoscalar states of 
the $\mu\nu$SSM as shown by eqns.(\ref{charged-scalar-mass-basis}), (\ref{scalar-mass-basis}), 
(\ref{pseudoscalar-mass-basis}).}
\label{LSP-llnu}
\end{figure}

We represent different leptons $(e,\mu,\tau)$ generically by $\ell$. 
We write down all possible $M^\dagger_iM_j$ for the seven diagrams
shown in figure \ref{LSP-llnu}. We treat the charge conjugate leptons 
as charginos (see eqn.(\ref{neutralino-chargino}))
as shown in eqn.(\ref{chargino_mass-basis_weak-basis_relations}).
The four-momentum assignments are as follows
\beq
\ntrl1(P) \to \ell^+_i(k)+\ell^-_j(k')+\nu_k(p),
\label{momentum-llnu}
\eeq
where $i,j,k$ stand for various lepton flavours. $i,j,k=1,2,3$ or $e,\mu,\tau$.
\bea
&&\bullet~ M^\dagger_1M_1 (\ntrl1\to \sum \ell^+_i \ell^-_j \nu_k) = 
\frac{8{g}^4_2}{\left[((p+k')^2-m^2_{W})^2+ m^2_W \Gamma^2_W\right]} \nn\\
&&\sum_{i,j,k}
\left[2(P.k')(p.k) A^{\ell^+_i \ell^-_j \nu_k}_{11} 
+ 2(P.p)(k.k') B^{\ell^+_i \ell^-_j \nu_k}_{11}
-m_{\ell_i}m_{\ntrl1}(p.k') C^{\ell^+_i \ell^-_j \nu_k}_{11}
\right],\nn\\
\label{mat11-llnu}
\eea
where
\bea
A^{\ell^+_i \ell^-_j \nu_k}_{11} &=& \left(O^{{cnw}*}_{Li1} O^{cnw}_{Li1}
O^{{ncw}*}_{Lkj} O^{ncw}_{Lkj} + O^{{cnw}*}_{Ri1} O^{cnw}_{Ri1}
O^{{ncw}*}_{Rkj} O^{ncw}_{Rkj}\right),\nn\\
B^{\ell^+_i \ell^-_j \nu_k}_{11} &=& \left(O^{{cnw}*}_{Li1} O^{cnw}_{Li1}
O^{{ncw}*}_{Rkj} O^{ncw}_{Rkj} + O^{{cnw}*}_{Ri1} O^{cnw}_{Ri1}
O^{{ncw}*}_{Lkj} O^{ncw}_{Lkj}\right),\nn\\
C^{\ell^+_i \ell^-_j \nu_k}_{11} &=& \left(O^{{cnw}*}_{Ri1} O^{cnw}_{Li1}
+ O^{{cnw}*}_{Li1} O^{cnw}_{Ri1}\right)
\left(O^{{ncw}*}_{Lkj} O^{ncw}_{Lkj}+ O^{{ncw}*}_{Rkj} O^{ncw}_{Rkj}\right).
\label{mat11-llnu-det}
\eea
\bea
&&\bullet~ M^\dagger_2M_2 (\ntrl1\to \sum \ell^+_i \ell^-_j \nu_k) = 
\frac{8{g}^4_2}{\left[((p+k)^2-m^2_{W})^2+ m^2_W \Gamma^2_W\right]} \nn\\
&&\sum_{i,j,k}
\left[2(P.k)(p.k') A^{\ell^+_i \ell^-_j \nu_k}_{22} 
+ 2(P.p)(k.k') B^{\ell^+_i \ell^-_j \nu_k}_{22}
-m_{\ell_j}m_{\ntrl1}(p.k) C^{\ell^+_i \ell^-_j \nu_k}_{22}
\right],\nn\\
\label{mat22-llnu}
\eea
where
\bea
A^{\ell^+_i \ell^-_j \nu_k}_{22} &=& \left(O^{{ncw}*}_{L1j} O^{ncw}_{L1j}
O^{{cnw}*}_{Lik} O^{cnw}_{Lik} + O^{{ncw}*}_{R1j} O^{ncw}_{R1j}
O^{{cnw}*}_{Rik} O^{cnw}_{Rik}\right),\nn\\
B^{\ell^+_i \ell^-_j \nu_k}_{22} &=& \left(O^{{ncw}*}_{L1j} O^{ncw}_{L1j}
O^{{cnw}*}_{Rik} O^{cnw}_{Rik} + O^{{ncw}*}_{R1j} O^{ncw}_{R1j}
O^{{cnw}*}_{Lik} O^{cnw}_{Lik}\right),\nn\\
C^{\ell^+_i \ell^-_j \nu_k}_{22} &=& \left(O^{{ncw}*}_{R1j} O^{ncw}_{L1j}
+ O^{{ncw}*}_{L1j} O^{ncw}_{R1j}\right)
\left(O^{{cnw}*}_{Lik} O^{cnw}_{Lik}+ O^{{cnw}*}_{Rik} O^{cnw}_{Rik}\right).
\label{mat22-llnu-det}
\eea
\bea
\bullet~ M^\dagger_3M_3 (\ntrl1\to \sum \ell^+_i \ell^-_j \nu_k) &=& \sum^8_{r,l=1}
\frac{4\wt{g}^4}{\left[((p+k')^2-m^2_{S^\pm_r})((p+k')^2-m^2_{S^\pm_l})\right]} \nn\\
&&\sum_{i,j,k}
\left[(P.k)(p.k') A^{\ell^+_i \ell^-_j \nu_k}_{33} 
+m_{\ell_i}m_{\ntrl1}(p.k') B^{\ell^+_i \ell^-_j \nu_k}_{33}
\right],\nn\\
\label{mat33-llnu}
\eea
where
\bea
A^{\ell^+_i \ell^-_j \nu_k}_{33} &=& \left(O^{{cns}*}_{Li1r} O^{cns}_{Li1l}
+ O^{{cns}*}_{Ri1r} O^{cns}_{Ri1l}\right)
\left(O^{{ncs}*}_{Lkjr} O^{ncs}_{Lkjl}+ O^{{ncs}*}_{Rkjr} O^{ncs}_{Rkjl}\right),\nn\\
B^{\ell^+_i \ell^-_j \nu_k}_{33} &=& \left(O^{{cns}*}_{Ri1r} O^{cns}_{Li1l}
+ O^{{cns}*}_{Li1r} O^{cns}_{Ri1l}\right)
\left(O^{{ncs}*}_{Lkjr} O^{ncs}_{Lkjl}+ O^{{ncs}*}_{Rkjr} O^{ncs}_{Rkjl}\right).
\label{mat33-llnu-det}
\eea
\bea
\bullet~ M^\dagger_4M_4 (\ntrl1\to \sum \ell^+_i \ell^-_j \nu_k) &=& \sum^8_{r,l=1}
\frac{4\wt{g}^4}{\left[((p+k)^2-m^2_{S^\pm_r})((p+k)^2-m^2_{S^\pm_l})\right]} \nn\\
&&\sum_{i,j,k}
\left[(P.k')(p.k) A^{\ell^+_i \ell^-_j \nu_k}_{44} 
+m_{\ell_j}m_{\ntrl1}(p.k) B^{\ell^+_i \ell^-_j \nu_k}_{44}
\right],\nn\\
\label{mat44-llnu}
\eea
where
\bea
A^{\ell^+_i \ell^-_j \nu_k}_{44} &=& \left(O^{{ncs}*}_{L1jr} O^{ncs}_{L1jl}
+ O^{{ncs}*}_{R1jr} O^{ncs}_{R1jl}\right)
\left(O^{{cns}*}_{Likr} O^{cns}_{Likl}+ O^{{cns}*}_{Rikr} O^{cns}_{Rikl}\right),\nn\\
B^{\ell^+_i \ell^-_j \nu_k}_{44} &=& \left(O^{{ncs}*}_{R1jr} O^{ncs}_{L1jl}
+ O^{{ncs}*}_{L1jr} O^{ncs}_{R1jl}\right)
\left(O^{{cns}*}_{Likr} O^{cns}_{Likl}+ O^{{cns}*}_{Rikr} O^{cns}_{Rikl}\right).
\label{mat44-llnu-det}
\eea
\bea
\bullet~ M^\dagger_5M_5 (\ntrl1\to \sum \ell^+_i \ell^-_j \nu_k) &=& \sum^8_{r,l=1}
\frac{\wt{g}^4}{\left[((k+k')^2-m^2_{S^0_r})((k+k')^2-m^2_{S^0_l})\right]} \nn\\
&&\sum_{i,j,k}
\left[(P.p)(k.k') A^{\ell^+_i \ell^-_j \nu_k}_{55} 
-m_{\ell_i}m_{\ell_j}(P.p) B^{\ell^+_i \ell^-_j \nu_k}_{55}
\right],\nn\\
\label{mat55-llnu}
\eea
where
\bea
A^{\ell^+_i \ell^-_j \nu_k}_{55} &=& \left(O^{{nnh}*}_{Lk1r} O^{nnh}_{Lk1l}
+ O^{{nnh}*}_{Rk1r} O^{nnh}_{Rk1l}\right)
\left(O^{{cch}*}_{Lijr} O^{cch}_{Lijl}+ O^{{cch}*}_{Rijr} O^{cch}_{Rijl}\right),\nn\\
B^{\ell^+_i \ell^-_j \nu_k}_{55} &=& \left(O^{{nnh}*}_{Lk1r} O^{nnh}_{Lk1l}
+ O^{{nnh}*}_{Rk1r} O^{nnh}_{Rk1l}\right)
\left(O^{{cch}*}_{Rijr} O^{cch}_{Lijl}+ O^{{cch}*}_{Lijr} O^{cch}_{Rijl}\right).
\label{mat55-llnu-det}
\eea
\bea
\bullet~ M^\dagger_6M_6 (\ntrl1\to \sum \ell^+_i \ell^-_j \nu_k) &=& \sum^8_{r,l=1}
\frac{\wt{g}^4}{\left[((k+k')^2-m^2_{P^0_r})((k+k')^2-m^2_{P^0_l})\right]} \nn\\
&&\sum_{i,j,k}
\left[(P.p)(k.k') A^{\ell^+_i \ell^-_j \nu_k}_{66} 
-m_{\ell_i}m_{\ell_j}(P.p) B^{\ell^+_i \ell^-_j \nu_k}_{66}
\right],\nn\\
\label{mat66-llnu}
\eea
where
\bea
A^{\ell^+_i \ell^-_j \nu_k}_{66} &=& \left(O^{{nna}*}_{Lk1r} O^{nna}_{Lk1l}
+ O^{{nna}*}_{Rk1r} O^{nna}_{Rk1l}\right)
\left(O^{{cca}*}_{Lijr} O^{cca}_{Lijl}+ O^{{cca}*}_{Rijr} O^{cca}_{Rijl}\right),\nn\\
B^{\ell^+_i \ell^-_j \nu_k}_{66} &=& \left(O^{{nna}*}_{Lk1r} O^{nna}_{Lk1l}
+ O^{{nna}*}_{Rk1r} O^{nna}_{Rk1l}\right)
\left(O^{{cca}*}_{Rijr} O^{cca}_{Lijl}+ O^{{cca}*}_{Lijr} O^{cca}_{Rijl}\right).
\label{mat66-llnu-det}
\eea
\bea
&&\bullet~ M^\dagger_7M_7 (\ntrl1\to \sum \ell^+_i \ell^-_j \nu_k) = 
\frac{2{g}^4_2}{cos^2\theta_W\left[((k+k')^2-m^2_{Z})^2 + m^2_Z \Gamma^2_Z\right]} \nn\\
&&\sum_{i,j,k}
\left[2(P.k')(p.k) A^{\ell^+_i \ell^-_j \nu_k}_{77} 
+ 2(P.k)(p.k')B^{\ell^+_i \ell^-_j \nu_k}_{77}
-m_{\ell_i}m_{\ell_j}(P.p) C^{\ell^+_i \ell^-_j \nu_k}_{77}
\right],\nn\\
\label{mat77-llnu}
\eea
where
\bea
A^{\ell^+_i \ell^-_j \nu_k}_{77} &=& \left(O^{{nnz}^*}_{Lk1} O^{nnz}_{Lk1}
O^{{ccz}^*}_{Lij} O^{ccz}_{Lij} + O^{{nnz}^*}_{Rk1} O^{nnz}_{Rk1}
O^{{ccz}^*}_{Rij} O^{ccz}_{Rij}\right),\nn\\
B^{\ell^+_i \ell^-_j \nu_k}_{77} &=& \left(O^{{nnz}^*}_{Lk1} O^{nnz}_{Lk1}
O^{{ccz}^*}_{Rij} O^{ccz}_{Rij} + O^{{nnz}^*}_{Rk1} O^{nnz}_{Rk1}
O^{{ccz}^*}_{Lij} O^{ccz}_{Lij}\right),\nn\\
C^{\ell^+_i \ell^-_j \nu_k}_{77} &=& \left(O^{{nnz}^*}_{Lk1} O^{nnz}_{Lk1}
+ O^{{nnz}^*}_{Rk1} O^{nnz}_{Rk1}\right)\left(
O^{{ccz}^*}_{Rij} O^{ccz}_{Lij} + O^{{ccz}^*}_{Lij} O^{ccz}_{Rij}\right).
\label{mat77-llnu-det}
\eea
\bea
&&\bullet~ M^\dagger_1M_2 (\ntrl1\to \sum \ell^+_i \ell^-_j \nu_k) = \nn\\
&&\frac{8{g}^4_2}{\left[((p+k')^2-m^2_{W}) -i m_W \Gamma_W\right]
\left[((p+k)^2-m^2_{W})^2 +i m_W \Gamma_W\right]} 
\nn\\
&&\sum_{i,j,k}
\left[-2(P.p)(k.k') (A^{\ell^+_i \ell^-_j \nu_k}_{12} O^{{cnw}^*}_{Li1}
+ B^{\ell^+_i \ell^-_j \nu_k}_{12} O^{{cnw}^*}_{Ri1})
\right.\nn\\
&&+ m_{\ell_i} m_{\ntrl1} (p.k') (A^{\ell^+_i \ell^-_j \nu_k}_{12} O^{{cnw}^*}_{Ri1}
+ B^{\ell^+_i \ell^-_j \nu_k}_{12} O^{{cnw}^*}_{Li1}) \nn\\
&&+m_{\ell_j} m_{\ntrl1} (p.k) (C^{\ell^+_i \ell^-_j \nu_k}_{12} O^{{cnw}^*}_{Ri1}
+ D^{\ell^+_i \ell^-_j \nu_k}_{12} O^{{cnw}^*}_{Li1})\nn\\
&&+\left.
m_{\ell_i} m_{\ell_j} (P.p) (C^{\ell^+_i \ell^-_j \nu_k}_{12} O^{{cnw}^*}_{Li1}
+ D^{\ell^+_i \ell^-_j \nu_k}_{12} O^{{cnw}^*}_{Ri1})
\right],\nn\\
\label{mat12-llnu}
\eea
where
\bea
A^{\ell^+_i \ell^-_j \nu_k}_{12} &=& O^{{ncw}^*}_{Rkj} O^{ncw}_{L1j} O^{cnw}_{Lik},
~~B^{\ell^+_i \ell^-_j \nu_k}_{12} = O^{{ncw}^*}_{Lkj} O^{ncw}_{R1j} O^{cnw}_{Rik},\nn\\
C^{\ell^+_i \ell^-_j \nu_k}_{12} &=& O^{{ncw}^*}_{Lkj} O^{ncw}_{L1j} O^{cnw}_{Rik},
~~D^{\ell^+_i \ell^-_j \nu_k}_{12} = O^{{ncw}^*}_{Rkj} O^{ncw}_{R1j} O^{cnw}_{Lik},\nn\\
\label{mat12-llnu-det}
\eea
\bea
\bullet~ M^\dagger_1M_3 (\ntrl1\to \sum \ell^+_i \ell^-_j \nu_k) &=& \sum^8_{l=1}
\frac{4g^2_2\wt{g}^2}{\left[(((p+k')^2-m^2_{W}) -i m_W \Gamma_W) 
((p+k')^2-m^2_{S^\pm_l})\right]} 
\nn\\
&&\sum_{i,j,k}
\left[ m_{\ell_j}m_{\ntrl1} (p.k) A^{\ell^+_i \ell^-_j \nu_k}_{13}
+ m_{\ell_i} m_{\ell_j} (P.p) B^{\ell^+_i \ell^-_j \nu_k}_{13} \right],\nn\\
\label{mat13-llnu}
\eea
where
\bea
A^{\ell^+_i \ell^-_j \nu_k}_{13} &=& \left(O^{{cnw}^*}_{Ri1} O^{cns}_{Li1l}
+ O^{{cnw}^*}_{Li1} O^{cns}_{Ri1l}\right)
\left(O^{{ncw}^*}_{Rkj} O^{ncs}_{Lkjl}
+ O^{{ncw}^*}_{Lkj} O^{ncs}_{Rkjl}\right),\nn\\
B^{\ell^+_i \ell^-_j \nu_k}_{13} &=& \left(O^{{cnw}^*}_{Li1} O^{cns}_{Li1l}
+ O^{{cnw}^*}_{Ri1} O^{cns}_{Ri1l}\right)
\left(O^{{ncw}^*}_{Rkj} O^{ncs}_{Lkjl}
+ O^{{ncw}^*}_{Lkj} O^{ncs}_{Rkjl}\right).
\label{mat13-llnu-det}
\eea
\bea
\bullet~ M^\dagger_1M_4 (\ntrl1\to \sum \ell^+_i \ell^-_j \nu_k) &=& - \sum^8_{r=1}
\frac{4g^2_2\wt{g}^2}{\left[(((p+k')^2-m^2_{W}) -i m_W \Gamma_W) 
((p+k)^2-m^2_{S^\pm_r})\right]} 
\nn\\
&&\sum_{i,j,k}
\left[2(P.k')(p.k) (A^{\ell^+_i \ell^-_j \nu_k}_{14} O^{{cnw}^*}_{Li1}
+ B^{\ell^+_i \ell^-_j \nu_k}_{14} O^{{cnw}^*}_{Ri1})
\right.\nn\\
&&-m_{\ell_i} m_{\ntrl1} (p.k') (A^{\ell^+_i \ell^-_j \nu_k}_{14} O^{{cnw}^*}_{Ri1}
+ B^{\ell^+_i \ell^-_j \nu_k}_{14} O^{{cnw}^*}_{Li1})\nn\\
&&+2 m_{\ell_j} m_{\ntrl1} (p.k) (C^{\ell^+_i \ell^-_j \nu_k}_{14} O^{{cnw}^*}_{Ri1}
+ D^{\ell^+_i \ell^-_j \nu_k}_{14} O^{{cnw}^*}_{Li1})\nn\\
&&-\left.
m_{\ell_i}m_{\ell_j} (P.p) (C^{\ell^+_i \ell^-_j \nu_k}_{14} O^{{cnw}^*}_{Li1}
+ D^{\ell^+_i \ell^-_j \nu_k}_{14} O^{{cnw}^*}_{Ri1})\right],
\label{mat14-llnu}
\eea
where
\bea
A^{\ell^+_i \ell^-_j \nu_k}_{14} &=& O^{{ncw}^*}_{Lkj} O^{ncs}_{L1jl} O^{cns}_{Rikl},
~~B^{\ell^+_i \ell^-_j \nu_k}_{14} = O^{{ncw}^*}_{Rkj} O^{ncs}_{R1jl} O^{cns}_{Likl},\nn\\
C^{\ell^+_i \ell^-_j \nu_k}_{14} &=& O^{{ncw}^*}_{Rkj} O^{ncs}_{L1jl} O^{cns}_{Likl},
~~D^{\ell^+_i \ell^-_j \nu_k}_{14} = O^{{ncw}^*}_{Lkj} O^{ncs}_{R1jl} O^{cns}_{Rikl}.
\label{mat14-llnu-det}
\eea
\bea
\bullet~ M^\dagger_1M_5 (\ntrl1\to \sum \ell^+_i \ell^-_j \nu_k) &=& - \sum^8_{r=1}
\frac{2g^2_2\wt{g}^2}{\left[(((p+k')^2-m^2_{W}) -i m_W \Gamma_W) 
((k+k')^2-m^2_{S^0_r})\right]} 
\nn\\
&&\sum_{i,j,k}
\left[2(P.p)(k.k') (A^{\ell^+_i \ell^-_j \nu_k}_{15} O^{{cnw}^*}_{Li1}
+ B^{\ell^+_i \ell^-_j \nu_k}_{15} O^{{cnw}^*}_{Ri1})
\right.\nn\\
&&-m_{\ell_i} m_{\ntrl1} (p.k') (A^{\ell^+_i \ell^-_j \nu_k}_{15} O^{{cnw}^*}_{Ri1}
+ B^{\ell^+_i \ell^-_j \nu_k}_{15} O^{{cnw}^*}_{Li1})\nn\\
&&+m_{\ell_j} m_{\ntrl1} (p.k) (C^{\ell^+_i \ell^-_j \nu_k}_{15} O^{{cnw}^*}_{Ri1}
+ D^{\ell^+_i \ell^-_j \nu_k}_{15} O^{{cnw}^*}_{Li1})\nn\\
&&-\left.
2m_{\ell_i}m_{\ell_j} (P.p) (C^{\ell^+_i \ell^-_j \nu_k}_{15} O^{{cnw}^*}_{Li1}
+ D^{\ell^+_i \ell^-_j \nu_k}_{15} O^{{cnw}^*}_{Ri1})\right],
\label{mat15-llnu}
\eea
where
\bea
A^{\ell^+_i \ell^-_j \nu_k}_{15} &=& O^{{ncw}^*}_{Rkj} O^{nnh}_{Lk1l} O^{cch}_{Rijl},
~~B^{\ell^+_i \ell^-_j \nu_k}_{15} = O^{{ncw}^*}_{Lkj} O^{nnh}_{Rk1l} O^{cch}_{Lijl},\nn\\
C^{\ell^+_i \ell^-_j \nu_k}_{15} &=& O^{{ncw}^*}_{Rkj} O^{nnh}_{Lk1l} O^{cch}_{Lijl},
~~D^{\ell^+_i \ell^-_j \nu_k}_{15} = O^{{ncw}^*}_{Lkj} O^{nnh}_{Rk1l} O^{cch}_{Rijl}.
\label{mat15-llnu-det}
\eea
\bea
\bullet~ M^\dagger_1M_6 (\ntrl1\to \sum \ell^+_i \ell^-_j \nu_k) &=&  \sum^8_{r=1}
\frac{2g^2_2\wt{g}^2}{\left[(((p+k')^2-m^2_{W}) -i m_W \Gamma_W) 
((k+k')^2-m^2_{P^0_r})\right]} 
\nn\\
&&\sum_{i,j,k}
\left[2(P.p)(k.k') (A^{\ell^+_i \ell^-_j \nu_k}_{16} O^{{cnw}^*}_{Li1}
+ B^{\ell^+_i \ell^-_j \nu_k}_{16} O^{{cnw}^*}_{Ri1})
\right.\nn\\
&&-m_{\ell_i} m_{\ntrl1} (p.k') (A^{\ell^+_i \ell^-_j \nu_k}_{16} O^{{cnw}^*}_{Ri1}
+ B^{\ell^+_i \ell^-_j \nu_k}_{16} O^{{cnw}^*}_{Li1})\nn\\
&&+m_{\ell_j} m_{\ntrl1} (p.k) (C^{\ell^+_i \ell^-_j \nu_k}_{16} O^{{cnw}^*}_{Ri1}
+ D^{\ell^+_i \ell^-_j \nu_k}_{16} O^{{cnw}^*}_{Li1})\nn\\
&&-\left.
2m_{\ell_i}m_{\ell_j} (P.p) (C^{\ell^+_i \ell^-_j \nu_k}_{16} O^{{cnw}^*}_{Li1}
+ D^{\ell^+_i \ell^-_j \nu_k}_{16} O^{{cnw}^*}_{Ri1})\right],\nn\\
\label{mat16-llnu}
\eea
where
\bea
A^{\ell^+_i \ell^-_j \nu_k}_{16} &=& O^{{ncw}^*}_{Rkj} O^{nna}_{Lk1l} O^{cca}_{Rijl},
~~B^{\ell^+_i \ell^-_j \nu_k}_{16} = O^{{ncw}^*}_{Lkj} O^{nna}_{Rk1l} O^{cca}_{Lijl},\nn\\
C^{\ell^+_i \ell^-_j \nu_k}_{16} &=& O^{{ncw}^*}_{Rkj} O^{nna}_{Lk1l} O^{cca}_{Lijl},
~~D^{\ell^+_i \ell^-_j \nu_k}_{16} = O^{{ncw}^*}_{Lkj} O^{nna}_{Rk1l} O^{cca}_{Rijl}.
\label{mat16-llnu-det}
\eea
\bea
&&\bullet~ M^\dagger_1M_7 (\ntrl1\to \sum \ell^+_i \ell^-_j \nu_k) =\nn\\  
&&-\frac{4g^4_2see\theta_W}{\left[(((p+k')^2-m^2_{W}) -i m_W \Gamma_W)\right]
\left[(((k+k')^2-m^2_{Z}) + i m_Z \Gamma_Z)\right]} 
\nn\\
&&\sum_{i,j,k}
\left[2(P.k')(p.k) (A^{\ell^+_i \ell^-_j \nu_k}_{17} O^{{cnw}^*}_{Li1}
+ B^{\ell^+_i \ell^-_j \nu_k}_{17} O^{{cnw}^*}_{Ri1})
\right.\nn\\
&&-m_{\ell_i} m_{\ntrl1} (p.k') (A^{\ell^+_i \ell^-_j \nu_k}_{17} O^{{cnw}^*}_{Ri1}
+ B^{\ell^+_i \ell^-_j \nu_k}_{17} O^{{cnw}^*}_{Li1})\nn\\
&&+m_{\ell_j} m_{\ntrl1} (p.k) (C^{\ell^+_i \ell^-_j \nu_k}_{17} O^{{cnw}^*}_{Ri1}
+ D^{\ell^+_i \ell^-_j \nu_k}_{17} O^{{cnw}^*}_{Li1})\nn\\
&&+\left.
m_{\ell_i}m_{\ell_j} (P.p) (C^{\ell^+_i \ell^-_j \nu_k}_{17} O^{{cnw}^*}_{Li1}
+ D^{\ell^+_i \ell^-_j \nu_k}_{17} O^{{cnw}^*}_{Ri1})\right],\nn\\
\label{mat17-llnu}
\eea
where
\bea
A^{\ell^+_i \ell^-_j \nu_k}_{17} &=& O^{{ncw}^*}_{Lkj} O^{nnz}_{Lk1} O^{ccz}_{Lij},
~~B^{\ell^+_i \ell^-_j \nu_k}_{17} = O^{{ncw}^*}_{Rkj} O^{nnz}_{Rk1} O^{ccz}_{Rij},\nn\\
C^{\ell^+_i \ell^-_j \nu_k}_{17} &=& O^{{ncw}^*}_{Lkj} O^{nnz}_{Lk1} O^{ccz}_{Rij},
~~D^{\ell^+_i \ell^-_j \nu_k}_{17} = O^{{ncw}^*}_{Rkj} O^{nnz}_{Rk1} O^{ccz}_{Lij}.
\label{mat17-llnu-det}
\eea
\bea
\bullet~ M^\dagger_2M_3 (\ntrl1\to \sum \ell^+_i \ell^-_j \nu_k) &=& \sum^8_{r=1}
\frac{4g^2_2 \wt{g}^2}{\left[(((p+k)^2-m^2_{W}) -i m_W \Gamma_W)
((p+k')^2-m^2_{S^\pm_r})\right]} 
\nn\\
&&\sum_{i,j,k}
\left[2(P.k)(p.k') (A^{\ell^+_i \ell^-_j \nu_k}_{23} O^{{ncw}^*}_{L1j}
+ B^{\ell^+_i \ell^-_j \nu_k}_{23} O^{{ncw}^*}_{R1j})
\right.\nn\\
&&-m_{\ell_j} m_{\ntrl1} (p.k) (A^{\ell^+_i \ell^-_j \nu_k}_{23} O^{{ncw}^*}_{R1j}
+ B^{\ell^+_i \ell^-_j \nu_k}_{23} O^{{ncw}^*}_{L1j})\nn\\
&&+2m_{\ell_i} m_{\ntrl1} (p.k') (C^{\ell^+_i \ell^-_j \nu_k}_{23} O^{{ncw}^*}_{R1j}
+ D^{\ell^+_i \ell^-_j \nu_k}_{23} O^{{ncw}^*}_{L1j})\nn\\
&&-\left.
m_{\ell_i}m_{\ell_j} (P.p) (C^{\ell^+_i \ell^-_j \nu_k}_{23} O^{{ncw}^*}_{L1j}
+ D^{\ell^+_i \ell^-_j \nu_k}_{23} O^{{ncw}^*}_{R1j})\right],\nn\\
\label{mat23-llnu}
\eea
where
\bea
A^{\ell^+_i \ell^-_j \nu_k}_{23} &=& O^{{cnw}^*}_{Rik} O^{cns}_{Li1r} O^{ncs}_{Rkjr},
~~B^{\ell^+_i \ell^-_j \nu_k}_{23} = O^{{cnw}^*}_{Lik} O^{cns}_{Ri1r} O^{ncs}_{Lkjr},\nn\\
C^{\ell^+_i \ell^-_j \nu_k}_{23} &=& O^{{cnw}^*}_{Lik} O^{cns}_{Li1r} O^{ncs}_{Lkjr},
~~D^{\ell^+_i \ell^-_j \nu_k}_{23} = O^{{cnw}^*}_{Rik} O^{cns}_{Ri1r} O^{ncs}_{Rkjr}.
\label{mat23-llnu-det}
\eea
\bea
\bullet~ M^\dagger_2M_4 (\ntrl1\to \sum \ell^+_i \ell^-_j \nu_k) &=& -\sum^8_{r=1}
\frac{4g^2_2 \wt{g}^2}{\left[(((p+k)^2-m^2_{W}) -i m_W \Gamma_W)
((p+k)^2-m^2_{S^\pm_r})\right]} 
\nn\\
&&\sum_{i,j,k}
\left[m_{\ell_i} m_{\ntrl1} (p.k') A^{\ell^+_i \ell^-_j \nu_k}_{24} 
+ m_{\ell_i} m_{\ell_j} (P.p) B^{\ell^+_i \ell^-_j \nu_k}_{24} \right],\nn\\
\label{mat24-llnu}
\eea
where
\bea
A^{\ell^+_i \ell^-_j \nu_k}_{24} &=& \left(O^{{ncw}^*}_{R1j} O^{ncs}_{L1jr}
+ O^{{ncw}^*}_{L1j} O^{ncs}_{R1jr}\right)
\left(O^{{cnw}^*}_{Lik} O^{cns}_{Likr}
+ O^{{cnw}^*}_{Rik} O^{cns}_{Rikr}\right),\nn\\
B^{\ell^+_i \ell^-_j \nu_k}_{24} &=& \left(O^{{ncw}^*}_{L1j} O^{ncs}_{L1jr}
+ O^{{ncw}^*}_{R1j} O^{ncs}_{R1jr}\right)
\left(O^{{cnw}^*}_{Lik} O^{cns}_{Likr}
+ O^{{cnw}^*}_{Rik} O^{cns}_{Rikr}\right).
\label{mat24-llnu-det}
\eea
\bea
\bullet~ M^\dagger_2M_5 (\ntrl1\to \sum \ell^+_i \ell^-_j \nu_k) &=& \sum^8_{r=1}
\frac{2g^2_2 \wt{g}^2}{\left[(((p+k)^2-m^2_{W}) -i m_W \Gamma_W)
((k+k')^2-m^2_{S^0_r})\right]} 
\nn\\
&&\sum_{i,j,k}
\left[2(P.p)(k.k') (A^{\ell^+_i \ell^-_j \nu_k}_{25} O^{{ncw}^*}_{L1j}
+ B^{\ell^+_i \ell^-_j \nu_k}_{25} O^{{ncw}^*}_{R1j})
\right.\nn\\
&&-m_{\ell_j} m_{\ntrl1} (p.k) (A^{\ell^+_i \ell^-_j \nu_k}_{25} O^{{ncw}^*}_{R1j}
+ B^{\ell^+_i \ell^-_j \nu_k}_{25} O^{{ncw}^*}_{L1j})\nn\\
&&+m_{\ell_i} m_{\ntrl1} (p.k') (C^{\ell^+_i \ell^-_j \nu_k}_{25} O^{{ncw}^*}_{R1j}
+ D^{\ell^+_i \ell^-_j \nu_k}_{25} O^{{ncw}^*}_{L1j})\nn\\
&&-\left.
2m_{\ell_i}m_{\ell_j} (P.p) (C^{\ell^+_i \ell^-_j \nu_k}_{25} O^{{ncw}^*}_{L1j}
+ D^{\ell^+_i \ell^-_j \nu_k}_{25} O^{{ncw}^*}_{R1j})\right],\nn\\
\label{mat25-llnu}
\eea
where
\bea
A^{\ell^+_i \ell^-_j \nu_k}_{25} &=& O^{{cnw}^*}_{Lik} O^{nnh}_{Lk1r} O^{cch}_{Rijr},
~~B^{\ell^+_i \ell^-_j \nu_k}_{25} = O^{{cnw}^*}_{Rik} O^{nnh}_{Rk1r} O^{cch}_{Lijr},\nn\\
C^{\ell^+_i \ell^-_j \nu_k}_{25} &=& O^{{cnw}^*}_{Lik} O^{nnh}_{Lk1r} O^{cch}_{Lijr},
~~D^{\ell^+_i \ell^-_j \nu_k}_{25} = O^{{cnw}^*}_{Rik} O^{nnh}_{Rk1r} O^{cch}_{Rijr}.
\label{mat25-llnu-det}
\eea
\bea
\bullet~ M^\dagger_2M_6 (\ntrl1\to \sum \ell^+_i \ell^-_j \nu_k) &=& -\sum^8_{r=1}
\frac{2g^2_2 \wt{g}^2}{\left[(((p+k)^2-m^2_{W}) -i m_W \Gamma_W)
((k+k')^2-m^2_{P^0_r})\right]} 
\nn\\
&&\sum_{i,j,k}
\left[2(P.p)(k.k') (A^{\ell^+_i \ell^-_j \nu_k}_{26} O^{{ncw}^*}_{L1j}
+ B^{\ell^+_i \ell^-_j \nu_k}_{26} O^{{ncw}^*}_{R1j})
\right.\nn\\
&&-m_{\ell_j} m_{\ntrl1} (p.k) (A^{\ell^+_i \ell^-_j \nu_k}_{26} O^{{ncw}^*}_{R1j}
+ B^{\ell^+_i \ell^-_j \nu_k}_{26} O^{{ncw}^*}_{L1j})\nn\\
&&+m_{\ell_i} m_{\ntrl1} (p.k') (C^{\ell^+_i \ell^-_j \nu_k}_{26} O^{{ncw}^*}_{R1j}
+ D^{\ell^+_i \ell^-_j \nu_k}_{26} O^{{ncw}^*}_{L1j})\nn\\
&&-\left.
2m_{\ell_i}m_{\ell_j} (P.p) (C^{\ell^+_i \ell^-_j \nu_k}_{26} O^{{ncw}^*}_{L1j}
+ D^{\ell^+_i \ell^-_j \nu_k}_{26} O^{{ncw}^*}_{R1j})\right],\nn\\
\label{mat26-llnu}
\eea
where
\bea
A^{\ell^+_i \ell^-_j \nu_k}_{26} &=& O^{{cnw}^*}_{Lik} O^{nna}_{Lk1r} O^{cca}_{Rijr},
~~B^{\ell^+_i \ell^-_j \nu_k}_{26} = O^{{cnw}^*}_{Rik} O^{nna}_{Rk1r} O^{cca}_{Lijr},\nn\\
C^{\ell^+_i \ell^-_j \nu_k}_{26} &=& O^{{cnw}^*}_{Lik} O^{nna}_{Lk1r} O^{cca}_{Lijr},
~~D^{\ell^+_i \ell^-_j \nu_k}_{26} = O^{{cnw}^*}_{Rik} O^{nna}_{Rk1r} O^{cca}_{Rijr}.
\label{mat26-llnu-det}
\eea
\bea
&&\bullet~ M^\dagger_2M_7 (\ntrl1\to \sum \ell^+_i \ell^-_j \nu_k) = \nn\\
&&-\frac{4g^4_2sec\theta_W}{\left[(((p+k)^2-m^2_{W}) -i m_W \Gamma_W)\right]
\left[(((k+k')^2-m^2_{Z}) + i m_Z \Gamma_Z)\right]} 
\nn\\
&&\sum_{i,j,k}
\left[2(P.k)(p.k') (A^{\ell^+_i \ell^-_j \nu_k}_{27} O^{{ncw}^*}_{L1j}
+ B^{\ell^+_i \ell^-_j \nu_k}_{27} O^{{ncw}^*}_{R1j})
\right.\nn\\
&&-m_{\ell_j} m_{\ntrl1} (p.k) (A^{\ell^+_i \ell^-_j \nu_k}_{27} O^{{ncw}^*}_{R1j}
+ B^{\ell^+_i \ell^-_j \nu_k}_{27} O^{{ncw}^*}_{L1j})\nn\\
&&+m_{\ell_i} m_{\ntrl1} (p.k') (C^{\ell^+_i \ell^-_j \nu_k}_{27} O^{{ncw}^*}_{R1j}
+ D^{\ell^+_i \ell^-_j \nu_k}_{27} O^{{ncw}^*}_{L1j})\nn\\
&&+\left.
m_{\ell_i}m_{\ell_j} (P.p) (C^{\ell^+_i \ell^-_j \nu_k}_{27} O^{{ncw}^*}_{L1j}
+ D^{\ell^+_i \ell^-_j \nu_k}_{27} O^{{ncw}^*}_{R1j})\right],\nn\\
\label{mat27-llnu}
\eea
where
\bea
A^{\ell^+_i \ell^-_j \nu_k}_{27} &=& O^{{cnw}^*}_{Rik} O^{nnz}_{Lk1} O^{ccz}_{Rij},
~~B^{\ell^+_i \ell^-_j \nu_k}_{27} = O^{{cnw}^*}_{Lik} O^{nnz}_{Rk1} O^{ccz}_{Lij},\nn\\
C^{\ell^+_i \ell^-_j \nu_k}_{27} &=& O^{{cnw}^*}_{Rik} O^{nnz}_{Lk1} O^{ccz}_{Lij},
~~D^{\ell^+_i \ell^-_j \nu_k}_{27} = O^{{cnw}^*}_{Lik} O^{nnz}_{Rk1} O^{ccz}_{Rij}.
\label{mat27-llnu-det}
\eea
\bea
&&\bullet~ M^\dagger_3M_4 (\ntrl1\to \sum \ell^+_i \ell^-_j \nu_k) = -\sum^8_{r,l=1}
\frac{2\wt{g}^4}{\left[((p+k')^2-m^2_{S^\pm_r})((p+k)^2-m^2_{S^\pm_l})\right]} 
\nn\\
&&\sum_{i,j,k}
\left[\left\{(P.k)(p.k')-(P.p)(k.k')+(P.k')(p.k)\right\} 
(A^{\ell^+_i \ell^-_j \nu_k}_{34} O^{{cns}^*}_{Li1r}
+ B^{\ell^+_i \ell^-_j \nu_k}_{34} O^{{cns}^*}_{Ri1r})
\right.\nn\\
&&+m_{\ell_i} m_{\ntrl1} (p.k) (A^{\ell^+_i \ell^-_j \nu_k}_{34} O^{{cns}^*}_{Ri1r}
+ B^{\ell^+_i \ell^-_j \nu_k}_{34} O^{{cns}^*}_{Li1r})\nn\\
&&+m_{\ell_j} m_{\ntrl1} (p.k') (C^{\ell^+_i \ell^-_j \nu_k}_{34} O^{{cns}^*}_{Ri1r}
+ D^{\ell^+_i \ell^-_j \nu_k}_{34} O^{{cns}^*}_{Li1r})\nn\\
&&+\left.
m_{\ell_i}m_{\ell_j} (P.p) (C^{\ell^+_i \ell^-_j \nu_k}_{34} O^{{cns}^*}_{Li1r}
+ D^{\ell^+_i \ell^-_j \nu_k}_{34} O^{{cns}^*}_{Ri1r})\right],
\label{mat34-llnu}
\eea
where
\bea
A^{\ell^+_i \ell^-_j \nu_k}_{34} &=& O^{{ncs}^*}_{Lkjr} O^{ncs}_{L1jl} O^{cns}_{Likl},
~~B^{\ell^+_i \ell^-_j \nu_k}_{34} = O^{{ncs}^*}_{Rkjr} O^{ncs}_{R1jl} O^{cns}_{Rikl},\nn\\
C^{\ell^+_i \ell^-_j \nu_k}_{34} &=& O^{{ncs}^*}_{Rkjr} O^{ncs}_{L1jl} O^{cns}_{Rikl},
~~D^{\ell^+_i \ell^-_j \nu_k}_{34} = O^{{ncs}^*}_{Lkjr} O^{ncs}_{R1jl} O^{cns}_{Likl}.
\label{mat34-llnu-det}
\eea
\bea
&&\bullet~ M^\dagger_3M_5 (\ntrl1\to \sum \ell^+_i \ell^-_j \nu_k) = \sum^8_{r,l=1}
\frac{\wt{g}^4}{\left[((p+k')^2-m^2_{S^\pm_r})((k+k')^2-m^2_{S^0_l})\right]} 
\nn\\
&&\sum_{i,j,k}
\left[\left\{(P.k)(p.k')-(P.k')(p.k)+(P.p)(k.k')\right\} 
(A^{\ell^+_i \ell^-_j \nu_k}_{35} O^{{cns}^*}_{Li1r}
+ B^{\ell^+_i \ell^-_j \nu_k}_{35} O^{{cns}^*}_{Ri1r})
\right.\nn\\
&&+m_{\ell_i} m_{\ntrl1} (p.k') (A^{\ell^+_i \ell^-_j \nu_k}_{35} O^{{cns}^*}_{Ri1r}
+ B^{\ell^+_i \ell^-_j \nu_k}_{35} O^{{cns}^*}_{Li1r})\nn\\
&&-m_{\ell_j} m_{\ntrl1} (p.k) (C^{\ell^+_i \ell^-_j \nu_k}_{35} O^{{cns}^*}_{Ri1r}
+ D^{\ell^+_i \ell^-_j \nu_k}_{35} O^{{cns}^*}_{Li1r})\nn\\
&&-\left.
m_{\ell_i}m_{\ell_j} (P.p) (C^{\ell^+_i \ell^-_j \nu_k}_{35} O^{{cns}^*}_{Li1r}
+ D^{\ell^+_i \ell^-_j \nu_k}_{35} O^{{cns}^*}_{Ri1r})\right],
\label{mat35-llnu}
\eea
where
\bea
A^{\ell^+_i \ell^-_j \nu_k}_{35} &=& O^{{ncs}^*}_{Lkjr} O^{nnh}_{Lk1l} O^{cch}_{Lijl},
~~B^{\ell^+_i \ell^-_j \nu_k}_{35} = O^{{ncs}^*}_{Rkjr} O^{nnh}_{Rk1l} O^{cch}_{Rijl},\nn\\
C^{\ell^+_i \ell^-_j \nu_k}_{35} &=& O^{{ncs}^*}_{Lkjr} O^{nnh}_{Lk1l} O^{cch}_{Rijl},
~~D^{\ell^+_i \ell^-_j \nu_k}_{35} = O^{{ncs}^*}_{Rkjr} O^{nnh}_{Rk1l} O^{cch}_{Lijl}.
\label{mat35-llnu-det}
\eea
\bea
&&\bullet~ M^\dagger_3M_6 (\ntrl1\to \sum \ell^+_i \ell^-_j \nu_k) = -\sum^8_{r,l=1}
\frac{\wt{g}^4}{\left[((p+k')^2-m^2_{S^\pm_r})((k+k')^2-m^2_{P^0_l})\right]} 
\nn\\
&&\sum_{i,j,k}
\left[\left\{(P.k)(p.k')-(P.k')(p.k)+(P.p)(k.k')\right\} 
(A^{\ell^+_i \ell^-_j \nu_k}_{36} O^{{cns}^*}_{Li1r}
+ B^{\ell^+_i \ell^-_j \nu_k}_{36} O^{{cns}^*}_{Ri1r})
\right.\nn\\
&&+m_{\ell_i} m_{\ntrl1} (p.k') (A^{\ell^+_i \ell^-_j \nu_k}_{36} O^{{cns}^*}_{Ri1r}
+ B^{\ell^+_i \ell^-_j \nu_k}_{36} O^{{cns}^*}_{Li1r})\nn\\
&&-m_{\ell_j} m_{\ntrl1} (p.k) (C^{\ell^+_i \ell^-_j \nu_k}_{36} O^{{cns}^*}_{Ri1r}
+ D^{\ell^+_i \ell^-_j \nu_k}_{36} O^{{cns}^*}_{Li1r})\nn\\
&&-\left.
m_{\ell_i}m_{\ell_j} (P.p) (C^{\ell^+_i \ell^-_j \nu_k}_{36} O^{{cns}^*}_{Li1r}
+ D^{\ell^+_i \ell^-_j \nu_k}_{36} O^{{cns}^*}_{Ri1r})\right],
\label{mat36-llnu}
\eea
where
\bea
A^{\ell^+_i \ell^-_j \nu_k}_{36} &=& O^{{ncs}^*}_{Lkjr} O^{nna}_{Lk1l} O^{cca}_{Lijl},
~~B^{\ell^+_i \ell^-_j \nu_k}_{36} = O^{{ncs}^*}_{Rkjr} O^{nna}_{Rk1l} O^{cca}_{Rijl},\nn\\
C^{\ell^+_i \ell^-_j \nu_k}_{36} &=& O^{{ncs}^*}_{Lkjr} O^{nna}_{Lk1l} O^{cca}_{Rijl},
~~D^{\ell^+_i \ell^-_j \nu_k}_{36} = O^{{ncs}^*}_{Rkjr} O^{nna}_{Rk1l} O^{cca}_{Lijl}.
\label{mat36-llnu-det}
\eea
\bea
\bullet~ M^\dagger_3M_7 (\ntrl1\to \sum \ell^+_i \ell^-_j \nu_k) &=& -\sum^8_{r=1}
\frac{2g^2_2\wt{g}^2}{\left[((p+k')^2-m^2_{S^\pm_r})((k+k')^2-m^2_{Z}+im_Z\Gamma_Z)\right]} 
\nn\\
&&\sum_{i,j,k}
\left[2(P.k)(p.k') (A^{\ell^+_i \ell^-_j \nu_k}_{37} O^{{ccz}}_{Lij}
+ B^{\ell^+_i \ell^-_j \nu_k}_{37} O^{{ccz}}_{Rij})
\right.\nn\\
&&+m_{\ell_i} m_{\ell_j} (P.p) (A^{\ell^+_i \ell^-_j \nu_k}_{37} O^{{ccz}}_{Rij}
+ B^{\ell^+_i \ell^-_j \nu_k}_{37} O^{{ccz}}_{Lij})\nn\\
&&+2m_{\ell_i} m_{\ntrl1} (p.k') (C^{\ell^+_i \ell^-_j \nu_k}_{37} O^{{ccz}}_{Rij}
+ D^{\ell^+_i \ell^-_j \nu_k}_{37} O^{{ccz}}_{Lij})\nn\\
&&+\left.
m_{\ell_j}m_{\ntrl1} (p.k) (C^{\ell^+_i \ell^-_j \nu_k}_{37} O^{{ccz}}_{Lij}
+ D^{\ell^+_i \ell^-_j \nu_k}_{37} O^{{ccz}}_{Rij})\right],
\label{mat37-llnu}
\eea
where
\bea
A^{\ell^+_i \ell^-_j \nu_k}_{37} &=& O^{{ncs}^*}_{Lkjr} O^{nnz}_{Rk1} O^{{cns}^*}_{Ri1r},
~~B^{\ell^+_i \ell^-_j \nu_k}_{37} = O^{{ncs}^*}_{Rkjr} O^{nnz}_{Lk1} O^{{cns}^*}_{Li1r},\nn\\
C^{\ell^+_i \ell^-_j \nu_k}_{37} &=& O^{{ncs}^*}_{Rkjr} O^{nnz}_{Lk1} O^{{cns}^*}_{Ri1r},
~~D^{\ell^+_i \ell^-_j \nu_k}_{37} = O^{{ncs}^*}_{Lkjr} O^{nnz}_{Rk1} O^{{cns}^*}_{Li1r}.
\label{mat37-llnu-det}
\eea
\bea
&&\bullet~ M^\dagger_4M_5 (\ntrl1\to \sum \ell^+_i \ell^-_j \nu_k) = \sum^8_{r,l=1}
\frac{\wt{g}^4}{\left[((p+k)^2-m^2_{S^\pm_r})((k+k')^2-m^2_{S^0_l})\right]} 
\nn\\
&&\sum_{i,j,k}
\left[\left\{(P.k')(p.k)-(P.k)(p.k')+(P.p)(k.k')\right\} 
(A^{\ell^+_i \ell^-_j \nu_k}_{45} O^{{ncs}^*}_{L1jr}
+ B^{\ell^+_i \ell^-_j \nu_k}_{45} O^{{ncs}^*}_{R1jr})
\right.\nn\\
&&+m_{\ell_j} m_{\ntrl1} (p.k) (A^{\ell^+_i \ell^-_j \nu_k}_{45} O^{{ncs}^*}_{R1jr}
+ B^{\ell^+_i \ell^-_j \nu_k}_{45} O^{{ncs}^*}_{L1jr})\nn\\
&&-m_{\ell_i} m_{\ntrl1} (p.k') (C^{\ell^+_i \ell^-_j \nu_k}_{45} O^{{ncs}^*}_{R1jr}
+ D^{\ell^+_i \ell^-_j \nu_k}_{45} O^{{ncs}^*}_{L1jr})\nn\\
&&-\left.
m_{\ell_i}m_{\ell_j} (P.p) (C^{\ell^+_i \ell^-_j \nu_k}_{45} O^{{ncs}^*}_{L1jr}
+ D^{\ell^+_i \ell^-_j \nu_k}_{45} O^{{ncs}^*}_{R1jr})\right],
\label{mat45-llnu}
\eea
where
\bea
A^{\ell^+_i \ell^-_j \nu_k}_{45} &=& O^{{cns}^*}_{Likr} O^{nnh}_{Lk1l} O^{cch}_{Lijl},
~~B^{\ell^+_i \ell^-_j \nu_k}_{45} = O^{{cns}^*}_{Rikr} O^{nnh}_{Rk1l} O^{cch}_{Rijl},\nn\\
C^{\ell^+_i \ell^-_j \nu_k}_{45} &=& O^{{cns}^*}_{Likr} O^{nnh}_{Lk1l} O^{cch}_{Rijl},
~~D^{\ell^+_i \ell^-_j \nu_k}_{45} = O^{{cns}^*}_{Rikr} O^{nnh}_{Rk1l} O^{cch}_{Lijl}.
\label{mat45-llnu-det}
\eea
\bea
&&\bullet~ M^\dagger_4M_6 (\ntrl1\to \sum \ell^+_i \ell^-_j \nu_k) = -\sum^8_{r,l=1}
\frac{\wt{g}^4}{\left[((p+k)^2-m^2_{S^\pm_r})((k+k')^2-m^2_{P^0_l})\right]} 
\nn\\
&&\sum_{i,j,k}
\left[\left\{(P.k')(p.k)-(P.k)(p.k')+(P.p)(k.k')\right\} 
(A^{\ell^+_i \ell^-_j \nu_k}_{46} O^{{ncs}^*}_{L1jr}
+ B^{\ell^+_i \ell^-_j \nu_k}_{46} O^{{ncs}^*}_{R1jr})
\right.\nn\\
&&+m_{\ell_j} m_{\ntrl1} (p.k) (A^{\ell^+_i \ell^-_j \nu_k}_{46} O^{{ncs}^*}_{R1jr}
+ B^{\ell^+_i \ell^-_j \nu_k}_{46} O^{{ncs}^*}_{L1jr})\nn\\
&&-m_{\ell_i} m_{\ntrl1} (p.k') (C^{\ell^+_i \ell^-_j \nu_k}_{46} O^{{ncs}^*}_{R1jr}
+ D^{\ell^+_i \ell^-_j \nu_k}_{46} O^{{ncs}^*}_{L1jr})\nn\\
&&-\left.
m_{\ell_i}m_{\ell_j} (P.p) (C^{\ell^+_i \ell^-_j \nu_k}_{46} O^{{ncs}^*}_{L1jr}
+ D^{\ell^+_i \ell^-_j \nu_k}_{46} O^{{ncs}^*}_{R1jr})\right],
\label{mat46-llnu}
\eea
where
\bea
A^{\ell^+_i \ell^-_j \nu_k}_{46} &=& O^{{cns}^*}_{Likr} O^{nna}_{Lk1l} O^{cca}_{Lijl},
~~B^{\ell^+_i \ell^-_j \nu_k}_{46} = O^{{cns}^*}_{Rikr} O^{nna}_{Rk1l} O^{cca}_{Rijl},\nn\\
C^{\ell^+_i \ell^-_j \nu_k}_{46} &=& O^{{cns}^*}_{Likr} O^{nna}_{Lk1l} O^{cca}_{Rijl},
~~D^{\ell^+_i \ell^-_j \nu_k}_{46} = O^{{cns}^*}_{Rikr} O^{nna}_{Rk1l} O^{cca}_{Lijl}.
\label{mat46-llnu-det}
\eea
\bea
&&\bullet~ M^\dagger_4M_7 (\ntrl1\to \sum \ell^+_i \ell^-_j \nu_k) = \sum^8_{r=1}
\frac{2g^2_2\wt{g}^2}{\left[((p+k)^2-m^2_{S^\pm_r})((k+k')^2-m^2_{Z}+im_Z\Gamma_Z)\right]} 
\nn\\
&&\sum_{i,j,k}
\left[2(P.k')(p.k)(A^{\ell^+_i \ell^-_j \nu_k}_{47} O^{ccz}_{Rij}
+ B^{\ell^+_i \ell^-_j \nu_k}_{47} O^{ccz}_{Lij})
\right.\nn\\
&&+m_{\ell_i} m_{\ell_j} (P.p) (A^{\ell^+_i \ell^-_j \nu_k}_{47} O^{ccz}_{Lij}
+ B^{\ell^+_i \ell^-_j \nu_k}_{47} O^{ccz}_{Rij})\nn\\
&&+2m_{\ell_j} m_{\ntrl1} (p.k) (C^{\ell^+_i \ell^-_j \nu_k}_{47} O^{ccz}_{Lij}
+ D^{\ell^+_i \ell^-_j \nu_k}_{47} O^{ccz}_{Rij})\nn\\
&&+\left.
m_{\ell_i}m_{\ntrl1} (p.k') (C^{\ell^+_i \ell^-_j \nu_k}_{47} O^{ccz}_{Rij}
+ D^{\ell^+_i \ell^-_j \nu_k}_{47} O^{ccz}_{Lij})\right],
\label{mat47-llnu}
\eea
where
\bea
A^{\ell^+_i \ell^-_j \nu_k}_{47} &=& O^{{cns}^*}_{Likr} O^{nnz}_{Rk1} O^{{ncs}^*}_{R1jr},
~~B^{\ell^+_i \ell^-_j \nu_k}_{47} = O^{{cns}^*}_{Rikr} O^{nnz}_{Lk1} O^{{ncs}^*}_{L1jr},\nn\\
C^{\ell^+_i \ell^-_j \nu_k}_{47} &=& O^{{cns}^*}_{Rikr} O^{nnz}_{Lk1} O^{{ncs}^*}_{R1jr},
~~D^{\ell^+_i \ell^-_j \nu_k}_{47} = O^{{cns}^*}_{Likr} O^{nnz}_{Rk1} O^{{ncs}^*}_{L1jr}.
\label{mat47-llnu-det}
\eea
\bea
\bullet~ M^\dagger_5M_6 (\ntrl1\to \sum \ell^+_i \ell^-_j \nu_k) &=& -\sum^8_{r,l=1}
\frac{\wt{g}^4}{\left[((k+k')^2-m^2_{S^0_r})((k+k')^2-m^2_{P^0_l})\right]} 
\nn\\
&&\sum_{i,j,k}
\left[(P.p)(k.k')A^{\ell^+_i \ell^-_j \nu_k}_{56}
-m_{\ell_i}m_{\ell_j}(P.p)B^{\ell^+_i \ell^-_j \nu_k}_{56}\right],\nn\\
\label{mat56-llnu}
\eea
where
\bea
A^{\ell^+_i \ell^-_j \nu_k}_{56} &=& \left(O^{{cch}^*}_{Lijr} O^{cca}_{Lijl}
+O^{{cch}^*}_{Rijr} O^{cca}_{Rijl}\right)
\left(O^{{nnh}^*}_{Lk1r} O^{nna}_{Lk1l}
+O^{{nnh}^*}_{Rk1r} O^{nna}_{Rk1l}\right),\nn\\
B^{\ell^+_i \ell^-_j \nu_k}_{56} &=& \left(O^{{cch}^*}_{Rijr} O^{cca}_{Lijl}
+O^{{cch}^*}_{Lijr} O^{cca}_{Rijl}\right)
\left(O^{{nnh}^*}_{Lk1r} O^{nna}_{Lk1l}
+O^{{nnh}^*}_{Rk1r} O^{nna}_{Rk1l}\right).
\label{mat56-llnu-det}
\eea
\bea
\bullet~ M^\dagger_5M_7 (\ntrl1\to \sum \ell^+_i \ell^-_j \nu_k) &=& -\sum^8_{r=1}
\frac{g^2_2\wt{g}^2see\theta_W}{\left[((k+k')^2-m^2_{S^0_r})((k+k')^2-m^2_{Z}+im_Z\Gamma_Z)
\right]} \nn\\
&&\sum_{i,j,k}
\left[m_{\ell_i}m_{\ntrl1}(p.k')A^{\ell^+_i \ell^-_j \nu_k}_{57}
-m_{\ell_j}m_{\ntrl1}(p.k)B^{\ell^+_i \ell^-_j \nu_k}_{57}\right],\nn\\
\label{mat57-llnu}
\eea
where
\bea
A^{\ell^+_i \ell^-_j \nu_k}_{57} &=& \left(O^{{cch}^*}_{Rijr} O^{ccz}_{Rij}
+O^{{cch}^*}_{Lijr} O^{ccz}_{Lij}\right)
\left(O^{{nnh}^*}_{Rk1r} O^{nnz}_{Lk1}
+O^{{nnh}^*}_{Lk1r} O^{nnz}_{Rk1}\right),\nn\\
B^{\ell^+_i \ell^-_j \nu_k}_{57} &=& \left(O^{{cch}^*}_{Rijr} O^{ccz}_{Lij}
+O^{{cch}^*}_{Lijr} O^{ccz}_{Rij}\right)
\left(O^{{nnh}^*}_{Rk1r} O^{nnz}_{Lk1}
+O^{{nnh}^*}_{Lk1r} O^{nnz}_{Rk1}\right).
\label{mat57-llnu-det}
\eea
\bea
\bullet~ M^\dagger_6M_7 (\ntrl1\to \sum \ell^+_i \ell^-_j \nu_k) &=& \sum^8_{r=1}
\frac{g^2_2\wt{g}^2see\theta_W}{\left[((k+k')^2-m^2_{P^0_r})((k+k')^2-m^2_{Z}+im_Z\Gamma_Z)
\right]} \nn\\
&&\sum_{i,j,k}
\left[m_{\ell_i}m_{\ntrl1}(p.k')A^{\ell^+_i \ell^-_j \nu_k}_{67}
-m_{\ell_j}m_{\ntrl1}(p.k)B^{\ell^+_i \ell^-_j \nu_k}_{67}\right],\nn\\
\label{mat67-llnu}
\eea
where
\bea
A^{\ell^+_i \ell^-_j \nu_k}_{67} &=& \left(O^{{cca}^*}_{Rijr} O^{ccz}_{Rij}
+O^{{cca}^*}_{Lijr} O^{ccz}_{Lij}\right)
\left(O^{{nna}^*}_{Rk1r} O^{nnz}_{Lk1}
+O^{{nna}^*}_{Lk1r} O^{nnz}_{Rk1}\right),\nn\\
B^{\ell^+_i \ell^-_j \nu_k}_{67} &=& \left(O^{{cca}^*}_{Rijr} O^{ccz}_{Lij}
+O^{{cca}^*}_{Lijr} O^{ccz}_{Rij}\right)
\left(O^{{nna}^*}_{Rk1r} O^{nnz}_{Lk1}
+O^{{nna}^*}_{Lk1r} O^{nnz}_{Rk1}\right).
\label{mat67-llnu-det}
\eea
$\Gamma_W$ and $\Gamma_Z$ are the decay width for $W$ and $Z$-boson, respectively
and their values are given in ref. \cite{AppNakamura-c2}. All the lepton
masses are also taken from ref. \cite{AppNakamura-c2}.

\section{Process $\ntrl1\to \nu_i \ovl{\nu}_j \nu_k$}\label{LSP-3nu-decay}
\begin{figure}[ht]
\centering
\vspace*{0.5cm}
\includegraphics[height=3.00cm]{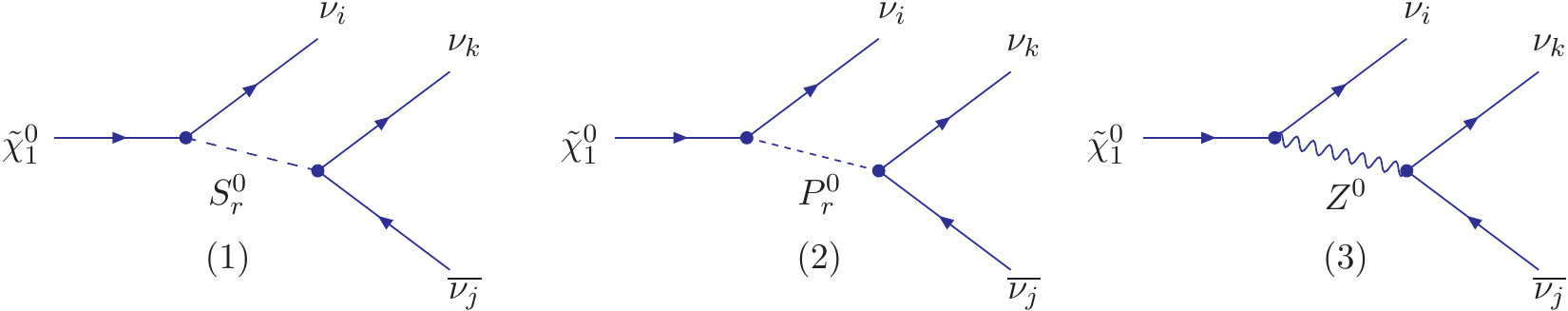}
\caption{Feynman diagrams for the possible three body decays of the lightest supersymmetric
particle into $\nu_i\ovl{\nu_j}\nu_k$ final states.
$S^0_r,P^0_r$ are the neutral scalar and pseudoscalar states of the $\mu\nu$SSM as
shown by eqns.(\ref{scalar-mass-basis}), (\ref{pseudoscalar-mass-basis}).}
\label{LSP-3nu}
\end{figure}

We represent different lepton flavours $(e,\mu,\tau)$ by $i,j,k$. 
We write down all possible $M^\dagger_iM_j$ for the three diagrams
shown in figure \ref{LSP-3nu}. The four-momentum assignments are as follows
\beq
\ntrl1(P) \to \nu_i(p)+\ovl{\nu_j}(k)+\nu_k(k').
\label{momentum-3nu}
\eeq
\bea
&&\bullet~ M^\dagger_1M_1 (\ntrl1\to \sum \nu_i \ovl{\nu_j} \nu_k) = \sum^8_{r,l=1}
\frac{\wt {g}^4}{\left[((k+k')^2-m^2_{S^0_r})
((k+k')^2-m^2_{S^0_l})\right]} \nn\\
&&\times\sum_{i,j,k}(P.p)(k.k')
\left(O^{{nnh}^*}_{Li1r}O^{nnh}_{Li1l}+O^{{nnh}^*}_{Ri1r}O^{nnh}_{Ri1l}\right)
\left(O^{{nnh}^*}_{Lkjr}O^{nnh}_{Lkjl}+O^{{nnh}^*}_{Rkjr}O^{nnh}_{Rkjl}\right).\nn\\
\label{mat11-3nu}
\eea
\bea
&&\bullet~ M^\dagger_2M_2 (\ntrl1\to \sum \nu_i \ovl{\nu_j} \nu_k) = \sum^8_{r,l=1}
\frac{\wt {g}^4}{\left[((k+k')^2-m^2_{P^0_r})
((k+k')^2-m^2_{P^0_l})\right]} \nn\\
&&\times\sum_{i,j,k}(P.p)(k.k')
\left(O^{{nna}^*}_{Li1r}O^{nna}_{Li1l}+O^{{nna}^*}_{Ri1r}O^{nna}_{Ri1l}\right)
\left(O^{{nna}^*}_{Lkjr}O^{nna}_{Lkjl}+O^{{nna}^*}_{Rkjr}O^{nna}_{Rkjl}\right).\nn\\
\label{mat22-3nu}
\eea
\bea
&&\bullet~ M^\dagger_3M_3 (\ntrl1\to \sum \nu_i \ovl{\nu_j} \nu_k) = 
\frac{{g}^4}{\left[((k+k')^2-m^2_{Z})^2 + m^2_Z \Gamma^2_Z\right]} \nn\\
&&\times\sum_{i,j,k}
\left[(P.k)(p.k')\left(O^{{nnz}^*}_{Li1}O^{nnz}_{Li1}O^{{nnz}^*}_{Lkj}O^{nnz}_{Lkj}
+O^{{nnz}^*}_{Ri1}O^{nnz}_{Ri1}O^{{nnz}^*}_{Rkj}O^{nnz}_{Rkj}\right)
\right.\nn\\
&&+\left.
(P.k')(p.k)\left(O^{{nnz}^*}_{Li1}O^{nnz}_{Li1}O^{{nnz}^*}_{Rkj}O^{nnz}_{Rkj}
+O^{{nnz}^*}_{Ri1}O^{nnz}_{Ri1}O^{{nnz}^*}_{Lkj}O^{nnz}_{Lkj}\right)\right].
\label{mat33-3nu}
\eea
\bea
&&\bullet~ M^\dagger_1M_2 (\ntrl1\to \sum \nu_i \ovl{\nu_j} \nu_k) = -\sum^8_{r,l=1}
\frac{\wt {g}^4}{\left[((k+k')^2-m^2_{S^0_r})
((k+k')^2-m^2_{P^0_l})\right]} \nn\\
&&\times\sum_{i,j,k}(P.p)(k.k')
\left(O^{{nnh}^*}_{Li1r}O^{nna}_{Li1l}+O^{{nnh}^*}_{Ri1r}O^{nna}_{Ri1l}\right)
\left(O^{{nnh}^*}_{Lkjr}O^{nna}_{Lkjl}+O^{{nnh}^*}_{Rkjr}O^{nna}_{Rkjl}\right).\nn\\
\label{mat12-3nu}
\eea
\bea
&&\bullet~ M^\dagger_1M_3 (\ntrl1\to \sum \nu_i \ovl{\nu_j} \nu_k) = 0.
\label{mat13-3nu}
\eea
\bea
&&\bullet~ M^\dagger_2M_3 (\ntrl1\to \sum \nu_i \ovl{\nu_j} \nu_k) = 0.
\label{mat23-3nu}
\eea

\section{Process $\ntrl1\to \bar{u}_i d_j \ell^+_k$}\label{LSP-qql1-decay}
We represent different lepton flavours $(e,\mu,\tau)$ by $k$. $u_i(d_j)$ stands
for different up-type and down-type quarks $(u,c(d,s,b))$, except the top. 
We write down all possible $M^\dagger_iM_j$ for the four diagrams
shown in figure \ref{LSP-qql1}. 
Required couplings are given in appendices \ref{appenD} and \ref{appenH}.
The four-momentum assignments are as follows
\beq
\ntrl1(P) \to \ell^+_k(p)+\bar{u}_i(k)+d_j(k').
\label{momentum-qql1}
\eeq
\begin{figure}[ht]
\centering
\vspace*{0.5cm}
\includegraphics[height=6.00cm]{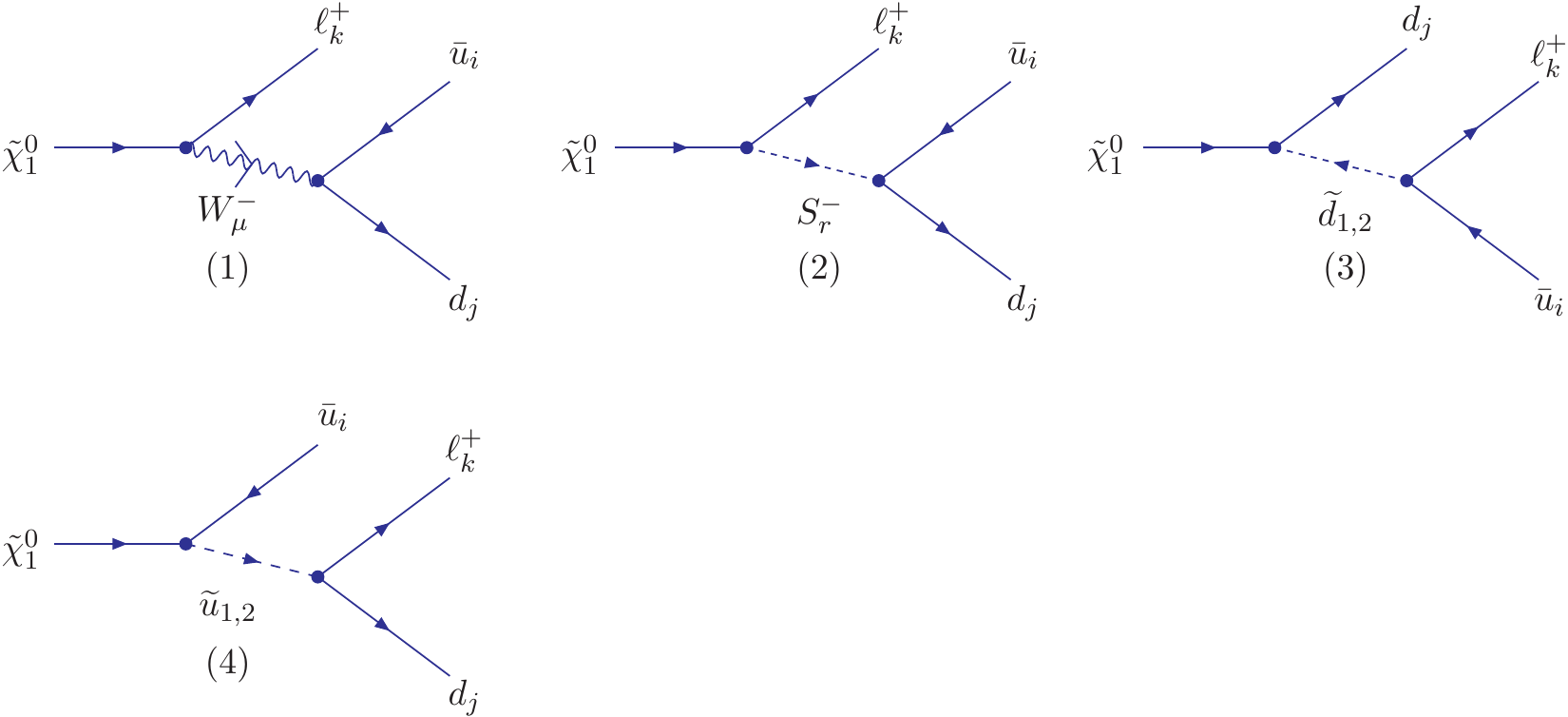}
\caption{Feynman diagrams for the possible three body decays of the lightest supersymmetric
particle into $\bar{u}_i d_j \ell^+_k$ final states.
$S^-_r$ are the charged scalar states of the $\mu\nu$SSM as
shown by eqn.(\ref{charged-scalar-mass-basis}). $\u(\d)$ are the up and down-type
squarks as shown by eqn.(\ref{squark-mass-basis}) corresponding
to $\bar{u}_i$ and $d_j$.}
\label{LSP-qql1}
\end{figure}
\bea
&&\bullet~ M^\dagger_1M_1 (\ntrl1\to \sum \bar{u}_i d_j \ell^+_k) = \sum_{i,j,k}
\frac{4 {g}^4 |V^{CKM}_{ij}|^2}{\left[(((k+k')^2-m^2_{W})^2+m^2_W\Gamma^2_W)\right]} \nn\\
&&\left[2(P.k)(p.k')O^{{cnw}^*}_{Lk1} O^{cnw}_{Lk1}
+ 2(P.k')(p.k) O^{{cnw}^*}_{Rk1} O^{cnw}_{Rk1}
\right.\nn\\
&&-\left.
m_{\ell_k} m_{\ntrl1} (k.k') \left(O^{{cnw}^*}_{Rk1} O^{cnw}_{Lk1}
+ O^{{cnw}^*}_{Lk1} O^{cnw}_{Rk1}\right)\right].
\label{mat11-qql1}
\eea
\bea
&&\bullet~ M^\dagger_2M_2 (\ntrl1\to \sum \bar{u}_i d_j \ell^+_k) = \sum_{i,j,k}
\sum^8_{r,l=1}\frac{4 \wt{g}^4}{\left[((k+k')^2-m^2_{S^\pm_r})((k+k')^2-m^2_{S^\pm_l})
\right]} \nn\\
&&\left[(P.p)\left(O^{{cns}^*}_{Lk1r} O^{cns}_{Lk1l} 
+ O^{{cns}^*}_{Rk1r} O^{cns}_{Rk1l}\right)+ 
m_{\ell_k} m_{\ntrl1} \left(O^{{cns}^*}_{Lk1r} O^{cns}_{Rk1l} 
+ O^{{cns}^*}_{Rk1r} O^{cns}_{Lk1l}\right)\right]\nn\\
&&\times \left[(k.k')\left(O^{{uds}^*}_{Lijl} O^{uds}_{Lijr} 
+ O^{{uds}^*}_{Rijl} O^{uds}_{Rijr}\right)-
m_{u_i}m_{d_j} \left(O^{{uds}^*}_{Rijl} O^{uds}_{Lijr} 
+ O^{{uds}^*}_{Lijl} O^{uds}_{Rijr}\right)\right].\nn\\
\label{mat22-qql1}
\eea
\bea
&&\bullet~ M^\dagger_3M_3 (\ntrl1\to \sum \bar{u}_i d_j \ell^+_k) = \sum_{i,j,k}
\sum^2_{r,l=1}\frac{4 \wt{g}^4}{\left[((p+k)^2-m^2_{\d_r})((p+k)^2-m^2_{\d_l})
\right]} \nn\\
&&\left[(P.k')\left(O^{{dn\d}^*}_{Lj1r} O^{dn\d}_{Lj1l} 
+ O^{{dn\d}^*}_{Rj1r} O^{dn\d}_{Rj1l}\right)+ 
m_{d_j} m_{\ntrl1} \left(O^{{dn\d}^*}_{Rj1r} O^{dn\d}_{Lj1l} 
+ O^{{dn\d}^*}_{Lj1r} O^{dn\d}_{Rj1l}\right)\right]\nn\\
&&\times \left[(p.k)\left(O^{{ucd}^*}_{Likl} O^{ucd}_{Likr} 
+ O^{{ucd}^*}_{Rikl} O^{ucd}_{Rikr}\right)-
m_{u_i}m_{\ell_k} \left(O^{{ucd}^*}_{Rikl} O^{ucd}_{Likr} 
+ O^{{ucd}^*}_{Likl} O^{ucd}_{Rikr}\right)\right].\nn\\
\label{mat33-qql1}
\eea
\bea
&&\bullet~ M^\dagger_4M_4 (\ntrl1\to \sum \bar{u}_i d_j \ell^+_k) = \sum_{i,j,k}
\sum^2_{r,l=1}\frac{4 \wt{g}^4}{\left[((p+k')^2-m^2_{\u_r})((p+k')^2-m^2_{\u_l})
\right]} \nn\\
&&\left[(P.k)\left(O^{{nu\u}^*}_{L1ir} O^{nu\u}_{L1il} 
+ O^{{nu\u}^*}_{R1ir} O^{nu\u}_{R1il}\right)+ 
m_{u_i} m_{\ntrl1} \left(O^{{nu\u}^*}_{R1ir} O^{nu\u}_{L1il} 
+ O^{{nu\u}^*}_{L1ir} O^{nu\u}_{R1il}\right)\right]\nn\\
&&\times \left[(p.k')\left(O^{{cdu}^*}_{Lkjl} O^{cdu}_{Lkjr} 
+ O^{{cdu}^*}_{Rkjl} O^{cdu}_{Rkjr}\right)-
m_{d_j}m_{\ell_k} \left(O^{{cdu}^*}_{Rkjl} O^{cdu}_{Lkjr} 
+ O^{{cdu}^*}_{Lkjl} O^{cdu}_{Rkjr}\right)\right].\nn\\
\label{mat44-qql1}
\eea
\bea
&&\bullet~ M^\dagger_1M_2 (\ntrl1\to \sum \bar{u}_i d_j \ell^+_k) = \nn\\
&&\sum_{i,j,k}
\sum^8_{r=1}\frac{2\rt2 g^2_2\wt{g}^2 V^{CKM}_{ij}}
{\left[((k+k')^2-m^2_{W}-im_W\Gamma_W)((k+k')^2-m^2_{S^\pm_r})
\right]} \nn\\
&&\left[m_{u_i}m_{\ntrl1}(p.k')O^{{uds}^*}_{Lijr}
A^{\bar{u}_i d_j \ell^+_k}_{12} 
+m_{\ell_k} m_{u_i}(P.k') O^{{uds}^*}_{Lijr}
B^{\bar{u}_i d_j \ell^+_k}_{12}
\right.\nn\\
&&-\left.
m_{d_j}m_{\ntrl1}(p.k) O^{{uds}^*}_{Rijr}
A^{\bar{u}_i d_j \ell^+_k}_{12}
-m_{d_j}m_{\ell_k}(P.k) O^{{uds}^*}_{Rijr}
B^{\bar{u}_i d_j \ell^+_k}_{12}\right].\nn\\
\label{mat12-qql1}
\eea
where
\bea
A^{\bar{u}_i d_j \ell^+_k}_{12} &=& \left(O^{{cnw}^*}_{Lk1} O^{cns}_{Rk1r}
+O^{{cnw}^*}_{Rk1} O^{cns}_{Lk1r}\right),
~~B^{\bar{u}_i d_j \ell^+_k}_{12} = \left(O^{{cnw}^*}_{Lk1} O^{cns}_{Lk1r}
+O^{{cnw}^*}_{Rk1} O^{cns}_{Rk1r}\right).\nn\\
\label{mat12-qql1-det}
\eea
\bea
&&\bullet~ M^\dagger_1M_3 (\ntrl1\to \sum \bar{u}_i d_j \ell^+_k) = \nn\\
&&-\sum_{i,j,k}
\sum^2_{r=1}\frac{2\rt2 g^2_2\wt{g}^2 V^{CKM}_{ij}}
{\left[((k+k')^2-m^2_{W}-im_W\Gamma_W)((p+k)^2-m^2_{\d_r})
\right]} \nn\\
&&\left[2(P.k')(p.k)A^{\bar{u}_i d_j \ell^+_k}_{13}
-m_{\ell_k}m_{\ntrl1} (k.k') B^{\bar{u}_i d_j \ell^+_k}_{13}
+m_{u_i}m_{\ntrl1}(p.k')C^{\bar{u}_i d_j \ell^+_k}_{13}
\right.\nn\\
&&-2m_{u_i}m_{\ell_k} (P.k') D^{\bar{u}_i d_j \ell^+_k}_{13}
+2m_{d_j}m_{\ntrl1}(p.k)E^{\bar{u}_i d_j \ell^+_k}_{13}
-m_{d_j}m_{\ell_k}(P.k)F^{\bar{u}_i d_j \ell^+_k}_{13}\nn\\
&&+\left.
m_{u_i}m_{d_j}(P.p)G^{\bar{u}_i d_j \ell^+_k}_{13}
-2m_{u_i}m_{d_j}m_{\ell_k}m_{\ntrl1}H^{\bar{u}_i d_j \ell^+_k}_{13}\right],
\label{mat13-qql1}
\eea
where
\bea
A^{\bar{u}_i d_j \ell^+_k}_{13} &=& O^{{cnw}^*}_{Rk1} O^{{ucd}^*}_{Rikr}
O^{dn\d}_{Rj1r},
~~B^{\bar{u}_i d_j \ell^+_k}_{13} = O^{{cnw}^*}_{Lk1} O^{{ucd}^*}_{Rikr}
O^{dn\d}_{Rj1r},\nn\\
C^{\bar{u}_i d_j \ell^+_k}_{13} &=& O^{{cnw}^*}_{Lk1} O^{{ucd}^*}_{Likr}
O^{dn\d}_{Rj1r},
~~D^{\bar{u}_i d_j \ell^+_k}_{13} = O^{{cnw}^*}_{Rk1} O^{{ucd}^*}_{Likr}
O^{dn\d}_{Rj1r},\nn\\
E^{\bar{u}_i d_j \ell^+_k}_{13} &=& O^{{cnw}^*}_{Rk1} O^{{ucd}^*}_{Rikr}
O^{dn\d}_{Lj1r},
~~F^{\bar{u}_i d_j \ell^+_k}_{13} = O^{{cnw}^*}_{Lk1} O^{{ucd}^*}_{Rikr}
O^{dn\d}_{Lj1r},\nn\\
G^{\bar{u}_i d_j \ell^+_k}_{13} &=& O^{{cnw}^*}_{Lk1} O^{{ucd}^*}_{Likr}
O^{dn\d}_{Lj1r},
~~H^{\bar{u}_i d_j \ell^+_k}_{13} = O^{{cnw}^*}_{Rk1} O^{{ucd}^*}_{Likr}
O^{dn\d}_{Lj1r}.
\label{mat13-qql1-det}
\eea
\bea
&&\bullet~ M^\dagger_1M_4 (\ntrl1\to \sum \bar{u}_i d_j \ell^+_k) = \nn\\
&&\sum_{i,j,k}
\sum^2_{r=1}\frac{2\rt2 g^2_2\wt{g}^2 V^{CKM}_{ij}}
{\left[((k+k')^2-m^2_{W}-im_W\Gamma_W)((p+k')^2-m^2_{\u_r})
\right]} \nn\\
&&\left[2(P.k)(p.k')A^{\bar{u}_i d_j \ell^+_k}_{14}
-m_{\ell_k}m_{\ntrl1} (k.k') B^{\bar{u}_i d_j \ell^+_k}_{14}
+m_{d_j}m_{\ntrl1}(p.k)C^{\bar{u}_i d_j \ell^+_k}_{14}
\right.\nn\\
&&-2m_{d_j}m_{\ell_k} (P.k) D^{\bar{u}_i d_j \ell^+_k}_{14}
+2m_{u_i}m_{\ntrl1}(p.k')E^{\bar{u}_i d_j \ell^+_k}_{14}
-m_{u_i}m_{\ell_k}(P.k')F^{\bar{u}_i d_j \ell^+_k}_{14}\nn\\
&&+\left.
m_{u_i}m_{d_j}(P.p)G^{\bar{u}_i d_j \ell^+_k}_{14}
-2m_{u_i}m_{d_j}m_{\ell_k}m_{\ntrl1}H^{\bar{u}_i d_j \ell^+_k}_{14}\right],
\label{mat14-qql1}
\eea
where
\bea
A^{\bar{u}_i d_j \ell^+_k}_{14} &=& O^{{cnw}^*}_{Lk1} O^{{cdu}^*}_{Lkjr}
O^{nu\u}_{L1ir},
~~B^{\bar{u}_i d_j \ell^+_k}_{14} = O^{{cnw}^*}_{Rk1} O^{{cdu}^*}_{Lkjr}
O^{nu\u}_{L1ir},\nn\\
C^{\bar{u}_i d_j \ell^+_k}_{14} &=& O^{{cnw}^*}_{Rk1} O^{{cdu}^*}_{Rkjr}
O^{nu\u}_{L1ir},
~~D^{\bar{u}_i d_j \ell^+_k}_{14} = O^{{cnw}^*}_{Lk1} O^{{cdu}^*}_{Rkjr}
O^{nu\u}_{L1ir},\nn\\
E^{\bar{u}_i d_j \ell^+_k}_{14} &=& O^{{cnw}^*}_{Lk1} O^{{cdu}^*}_{Lkjr}
O^{nu\u}_{R1ir},
~~F^{\bar{u}_i d_j \ell^+_k}_{14} = O^{{cnw}^*}_{Rk1} O^{{cdu}^*}_{Lkjr}
O^{nu\u}_{R1ir},\nn\\
G^{\bar{u}_i d_j \ell^+_k}_{14} &=& O^{{cnw}^*}_{Rk1} O^{{cdu}^*}_{Rkjr}
O^{nu\u}_{R1ir},
~~H^{\bar{u}_i d_j \ell^+_k}_{14} = O^{{cnw}^*}_{Lk1} O^{{cdu}^*}_{Rkjr}
O^{nu\u}_{R1ir}.
\label{mat14-qql1-det}
\eea
\bea
&&\bullet~ M^\dagger_2M_3 (\ntrl1\to \sum \bar{u}_i d_j \ell^+_k) = \sum_{i,j,k}
\sum^8_{r=1}\sum^2_{l=1}\frac{2\wt{g}^4}
{\left[((k+k')^2-m^2_{S^\pm_r})((p+k)^2-m^2_{\d_l})
\right]} \nn\\
&&\left[\{(P.p)(k.k')-(P.k)(p.k')+(P.k')(p.k)\}
\left(A^{\bar{u}_i d_j \ell^+_k}_{23}O^{{cns}^*}_{Lk1r}
+B^{\bar{u}_i d_j \ell^+_k}_{23}O^{{cns}^*}_{Rk1r}\right)
\right.\nn\\
&&+m_{\ell_k}m_{\ntrl1}(k.k')\left(A^{\bar{u}_i d_j \ell^+_k}_{23}O^{{cns}^*}_{Rk1r}
+B^{\bar{u}_i d_j \ell^+_k}_{23}O^{{cns}^*}_{Lk1r}\right)\nn\\
&&-m_{u_i}m_{\ntrl1}(p.k')\left(C^{\bar{u}_i d_j \ell^+_k}_{23}O^{{cns}^*}_{Rk1r}
+D^{\bar{u}_i d_j \ell^+_k}_{23}O^{{cns}^*}_{Lk1r}\right)\nn\\
&&-m_{u_i}m_{\ell_k}(P.k')\left(C^{\bar{u}_i d_j \ell^+_k}_{23}O^{{cns}^*}_{Lk1r}
+D^{\bar{u}_i d_j \ell^+_k}_{23}O^{{cns}^*}_{Rk1r}\right)\nn\\
&&+m_{d_j}m_{\ntrl1}(p.k)\left(E^{\bar{u}_i d_j \ell^+_k}_{23}O^{{cns}^*}_{Rk1r}
+F^{\bar{u}_i d_j \ell^+_k}_{23}O^{{cns}^*}_{Lk1r}\right)\nn\\
&&+m_{d_j}m_{\ell_k}(P.k)\left(E^{\bar{u}_i d_j \ell^+_k}_{23}O^{{cns}^*}_{Lk1r}
+F^{\bar{u}_i d_j \ell^+_k}_{23}O^{{cns}^*}_{Rk1r}\right)\nn\\
&&-m_{u_i}m_{d_j}(P.p)\left(G^{\bar{u}_i d_j \ell^+_k}_{23}O^{{cns}^*}_{Lk1r}
+H^{\bar{u}_i d_j \ell^+_k}_{23}O^{{cns}^*}_{Rk1r}\right)\nn\\
&&-\left.
m_{u_i}m_{d_j}m_{\ell_k}m_{\ntrl1}\left(G^{\bar{u}_i d_j \ell^+_k}_{23}O^{{cns}^*}_{Rk1r}
+H^{\bar{u}_i d_j \ell^+_k}_{23}O^{{cns}^*}_{Lk1r}\right)\right],
\label{mat23-qql1}
\eea
where
\bea
A^{\bar{u}_i d_j \ell^+_k}_{23} &=& O^{{ucd}^*}_{Rikl} O^{uds}_{Rijr}
O^{dn\d}_{Lj1l},
~~B^{\bar{u}_i d_j \ell^+_k}_{23} = O^{{ucd}^*}_{Likl} O^{uds}_{Lijr}
O^{dn\d}_{Rj1l},\nn\\
C^{\bar{u}_i d_j \ell^+_k}_{23} &=& O^{{ucd}^*}_{Likl} O^{uds}_{Rijr}
O^{dn\d}_{Lj1l},
~~D^{\bar{u}_i d_j \ell^+_k}_{23} = O^{{ucd}^*}_{Rikl} O^{uds}_{Lijr}
O^{dn\d}_{Rj1l},\nn\\
E^{\bar{u}_i d_j \ell^+_k}_{23} &=& O^{{ucd}^*}_{Likl} O^{uds}_{Lijr}
O^{dn\d}_{Lj1l},
~~F^{\bar{u}_i d_j \ell^+_k}_{23} = O^{{ucd}^*}_{Rikl} O^{uds}_{Rijr}
O^{dn\d}_{Rj1l},\nn\\
G^{\bar{u}_i d_j \ell^+_k}_{23} &=& O^{{ucd}^*}_{Rikl} O^{uds}_{Lijr}
O^{dn\d}_{Lj1l},
~~H^{\bar{u}_i d_j \ell^+_k}_{23} = O^{{ucd}^*}_{Likl} O^{uds}_{Rijr}
O^{dn\d}_{Rj1l}.
\label{mat23-qql1-det}
\eea
\bea
&&\bullet~ M^\dagger_2M_4 (\ntrl1\to \sum \bar{u}_i d_j \ell^+_k) = \sum_{i,j,k}
\sum^8_{r=1}\sum^2_{l=1}\frac{2\wt{g}^4}
{\left[((k+k')^2-m^2_{S^\pm_r})((p+k')^2-m^2_{\u_l})
\right]} \nn\\
&&\left[\{(P.p)(k.k')-(P.k')(p.k)+(P.k)(p.k')\}
\left(A^{\bar{u}_i d_j \ell^+_k}_{24}O^{{cns}^*}_{Lk1r}
+B^{\bar{u}_i d_j \ell^+_k}_{24}O^{{cns}^*}_{Rk1r}\right)
\right.\nn\\
&&+m_{\ell_k}m_{\ntrl1}(k.k')\left(A^{\bar{u}_i d_j \ell^+_k}_{24}O^{{cns}^*}_{Rk1r}
+B^{\bar{u}_i d_j \ell^+_k}_{24}O^{{cns}^*}_{Lk1r}\right)\nn\\
&&-m_{d_j}m_{\ntrl1}(p.k)\left(C^{\bar{u}_i d_j \ell^+_k}_{24}O^{{cns}^*}_{Rk1r}
+D^{\bar{u}_i d_j \ell^+_k}_{24}O^{{cns}^*}_{Lk1r}\right)\nn\\
&&-m_{d_j}m_{\ell_k}(P.k)\left(C^{\bar{u}_i d_j \ell^+_k}_{24}O^{{cns}^*}_{Lk1r}
+D^{\bar{u}_i d_j \ell^+_k}_{24}O^{{cns}^*}_{Rk1r}\right)\nn\\
&&+m_{u_i}m_{\ntrl1}(p.k')\left(E^{\bar{u}_i d_j \ell^+_k}_{24}O^{{cns}^*}_{Rk1r}
+F^{\bar{u}_i d_j \ell^+_k}_{24}O^{{cns}^*}_{Lk1r}\right)\nn\\
&&+m_{u_i}m_{\ell_k}(P.k')\left(E^{\bar{u}_i d_j \ell^+_k}_{24}O^{{cns}^*}_{Lk1r}
+F^{\bar{u}_i d_j \ell^+_k}_{24}O^{{cns}^*}_{Rk1r}\right)\nn\\
&&-m_{u_i}m_{d_j}(P.p)\left(G^{\bar{u}_i d_j \ell^+_k}_{24}O^{{cns}^*}_{Lk1r}
+H^{\bar{u}_i d_j \ell^+_k}_{24}O^{{cns}^*}_{Rk1r}\right)\nn\\
&&-\left.
m_{u_i}m_{d_j}m_{\ell_k}m_{\ntrl1}\left(G^{\bar{u}_i d_j \ell^+_k}_{24}O^{{cns}^*}_{Rk1r}
+H^{\bar{u}_i d_j \ell^+_k}_{24}O^{{cns}^*}_{Lk1r}\right)\right],
\label{mat24-qql1}
\eea
where
\bea
A^{\bar{u}_i d_j \ell^+_k}_{24} &=& O^{{cdu}^*}_{Rkjl} O^{uds}_{Rijr}
O^{nu\u}_{L1il},
~~B^{\bar{u}_i d_j \ell^+_k}_{24} = O^{{cdu}^*}_{Lkjl} O^{uds}_{Lijr}
O^{nu\u}_{R1il},\nn\\
C^{\bar{u}_i d_j \ell^+_k}_{24} &=& O^{{cdu}^*}_{Lkjl} O^{uds}_{Rijr}
O^{nu\u}_{L1il},
~~D^{\bar{u}_i d_j \ell^+_k}_{24} = O^{{cdu}^*}_{Rkjl} O^{uds}_{Lijr}
O^{nu\u}_{R1il},\nn\\
E^{\bar{u}_i d_j \ell^+_k}_{24} &=& O^{{cdu}^*}_{Lkjl} O^{uds}_{Lijr}
O^{nu\u}_{L1il},
~~F^{\bar{u}_i d_j \ell^+_k}_{24} = O^{{cdu}^*}_{Rkjl} O^{uds}_{Rijr}
O^{nu\u}_{R1il},\nn\\
G^{\bar{u}_i d_j \ell^+_k}_{24} &=& O^{{cdu}^*}_{Rkjl} O^{uds}_{Lijr}
O^{nu\u}_{L1il},
~~H^{\bar{u}_i d_j \ell^+_k}_{24} = O^{{cdu}^*}_{Lkjl} O^{uds}_{Rijr}
O^{nu\u}_{R1il}.
\label{mat24-qql1-det}
\eea
\bea
&&\bullet~ M^\dagger_3M_4 (\ntrl1\to \sum \bar{u}_i d_j \ell^+_k) = -\sum_{i,j,k}
\sum^2_{r,l=1}\frac{2\wt{g}^4}
{\left[((p+k)^2-m^2_{\d_r})((p+k')^2-m^2_{\u_l})
\right]} \nn\\
&&\left[\{(P.k')(p.k)-(P.p)(k.k')+(P.k)(p.k')\}
\left(A^{\bar{u}_i d_j \ell^+_k}_{34}O^{{dn\d}^*}_{Lj1r}
+B^{\bar{u}_i d_j \ell^+_k}_{34}O^{{dn\d}^*}_{Rj1r}\right)
\right.\nn\\
&&+m_{d_j}m_{\ntrl1}(p.k)\left(A^{\bar{u}_i d_j \ell^+_k}_{34}O^{{dn\d}^*}_{Rj1r}
+B^{\bar{u}_i d_j \ell^+_k}_{34}O^{{dn\d}^*}_{Lj1r}\right)\nn\\
&&-m_{\ell_k}m_{\ntrl1}(k.k')\left(C^{\bar{u}_i d_j \ell^+_k}_{34}O^{{dn\d}^*}_{Rj1r}
+D^{\bar{u}_i d_j \ell^+_k}_{34}O^{{dn\d}^*}_{Lj1r}\right)\nn\\
&&-m_{d_j}m_{\ell_k}(P.k)\left(C^{\bar{u}_i d_j \ell^+_k}_{34}O^{{dn\d}^*}_{Lj1r}
+D^{\bar{u}_i d_j \ell^+_k}_{34}O^{{dn\d}^*}_{Rj1r}\right)\nn\\
&&+m_{u_i}m_{\ntrl1}(p.k')\left(E^{\bar{u}_i d_j \ell^+_k}_{34}O^{{dn\d}^*}_{Rj1r}
+F^{\bar{u}_i d_j \ell^+_k}_{34}O^{{dn\d}^*}_{Lj1r}\right)\nn\\
&&+m_{u_i}m_{d_j}(P.p)\left(E^{\bar{u}_i d_j \ell^+_k}_{34}O^{{dn\d}^*}_{Lj1r}
+F^{\bar{u}_i d_j \ell^+_k}_{34}O^{{dn\d}^*}_{Rj1r}\right)\nn\\
&&-m_{u_i}m_{\ell_k}(P.k')\left(G^{\bar{u}_i d_j \ell^+_k}_{34}O^{{dn\d}^*}_{Lj1r}
+H^{\bar{u}_i d_j \ell^+_k}_{34}O^{{dn\d}^*}_{Rj1r}\right)\nn\\
&&-\left.
m_{u_i}m_{d_j}m_{\ell_k}m_{\ntrl1}\left(G^{\bar{u}_i d_j \ell^+_k}_{34}O^{{dn\d}^*}_{Rj1r}
+H^{\bar{u}_i d_j \ell^+_k}_{34}O^{{dn\d}^*}_{Lj1r}\right)\right],
\label{mat34-qql1}
\eea
where
\bea
A^{\bar{u}_i d_j \ell^+_k}_{34} &=& O^{{cdu}^*}_{Rkjl} O^{ucd}_{Rikr}
O^{nu\u}_{L1il},
~~B^{\bar{u}_i d_j \ell^+_k}_{34} = O^{{cdu}^*}_{Lkjl} O^{ucd}_{Likr}
O^{nu\u}_{R1il},\nn\\
C^{\bar{u}_i d_j \ell^+_k}_{34} &=& O^{{cdu}^*}_{Lkjl} O^{ucd}_{Rikr}
O^{nu\u}_{L1il},
~~D^{\bar{u}_i d_j \ell^+_k}_{34} = O^{{cdu}^*}_{Rkjl} O^{ucd}_{Likr}
O^{nu\u}_{R1il},\nn\\
E^{\bar{u}_i d_j \ell^+_k}_{34} &=& O^{{cdu}^*}_{Lkjl} O^{ucd}_{Likr}
O^{nu\u}_{L1il},
~~F^{\bar{u}_i d_j \ell^+_k}_{34} = O^{{cdu}^*}_{Rkjl} O^{ucd}_{Rikr}
O^{nu\u}_{R1il},\nn\\
G^{\bar{u}_i d_j \ell^+_k}_{34} &=& O^{{cdu}^*}_{Rkjl} O^{ucd}_{Likr}
O^{nu\u}_{L1il},
~~H^{\bar{u}_i d_j \ell^+_k}_{34} = O^{{cdu}^*}_{Lkjl} O^{ucd}_{Rikr}
O^{nu\u}_{R1il}.
\label{mat34-qql1-det}
\eea
\section{Process $\ntrl1\to u_i \bar{d}_j \ell^-_k$}\label{LSP-qql2-decay}
We represent different lepton flavours $(e,\mu,\tau)$ by $k$. $u_i(d_j)$ stands
for different up-type and down-type quarks $(u,c(d,s,b))$, except the top. 
We write down all possible $M^\dagger_iM_j$ for the four diagrams
shown in figure \ref{LSP-qql2}. 
Required couplings are given in appendices \ref{appenD} and \ref{appenH}.
The four-momentum assignments are as follows
\beq
\ntrl1(P) \to \ell^-_k(p)+u_i(k)+\bar{d}_j(k').
\label{momentum-qql2}
\eeq
\begin{figure}[ht]
\centering
\vspace*{0.5cm}
\includegraphics[height=6.00cm]{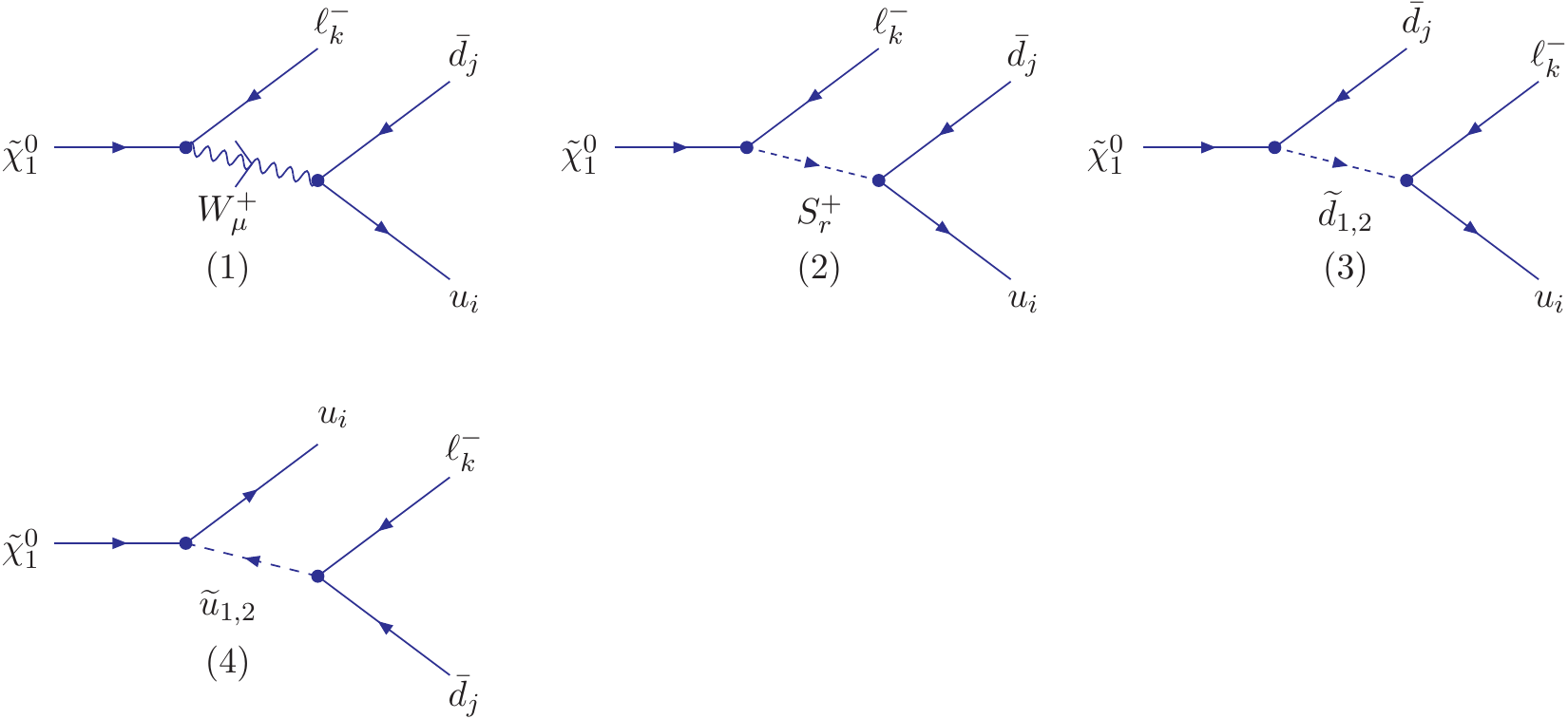}
\caption{Feynman diagrams for the possible three body decays of the lightest supersymmetric
particle into $u_i \bar{d}_j \ell^-_k$ final states.
$S^+_r$ are the charged scalar states of the $\mu\nu$SSM as
shown by eqn.(\ref{charged-scalar-mass-basis}). $\u(\d)$ are the up and down-type
squarks as shown by eqn.(\ref{squark-mass-basis}) corresponding
to $\bar{u}_i$ and $d_j$.}
\label{LSP-qql2}
\end{figure}
\bea
&&\bullet~ M^\dagger_1M_1 (\ntrl1\to \sum u_i \bar{d}_j  \ell^-_k) = 
M^\dagger_1M_1 (\ntrl1\to \sum \bar{u}_i d_j \ell^+_k)
\label{mat11-qql2}
\eea
\bea
&&\bullet~ M^\dagger_2M_2 (\ntrl1\to \sum u_i \bar{d}_j  \ell^-_k) = 
M^\dagger_2M_2 (\ntrl1\to \sum \bar{u}_i d_j \ell^+_k)
\label{mat22-qql2}
\eea
\bea
&&\bullet~ M^\dagger_3M_3 (\ntrl1\to \sum u_i \bar{d}_j  \ell^-_k) = 
M^\dagger_3M_3 (\ntrl1\to \sum \bar{u}_i d_j \ell^+_k)
\label{mat33-qql2}
\eea
\bea
&&\bullet~ M^\dagger_4M_4 (\ntrl1\to \sum u_i \bar{d}_j  \ell^-_k) = 
M^\dagger_4M_4 (\ntrl1\to \sum \bar{u}_i d_j \ell^+_k)
\label{mat44-qql2}
\eea
\bea
&&\bullet~ M^\dagger_1M_2 (\ntrl1\to \sum u_i \bar{d}_j \ell^-_k) = \nn\\
&&\sum_{i,j,k}
\sum^8_{r=1}\frac{2\rt2 g^2_2\wt{g}^2 V^{{CKM}^*}_{ij}}
{\left[((k+k')^2-m^2_{W}-im_W\Gamma_W)((k+k')^2-m^2_{S^\pm_r})
\right]} \nn\\
&&\left[m_{d_j}m_{\ntrl1}(p.k)O^{{uds}}_{Rijr}
A^{u_i \bar{d}_j \ell^-_k}_{12} 
+m_{\ell_k} m_{d_j}(P.k) O^{{uds}}_{Rijr}
B^{u_i \bar{d}_j \ell^-_k}_{12}
\right.\nn\\
&&-\left.
m_{u_i}m_{\ntrl1}(p.k') O^{{uds}}_{Lijr}
A^{u_i \bar{d}_j \ell^-_k}_{12}
-m_{u_i}m_{\ell_k}(P.k') O^{{uds}}_{Lijr}
B^{u_i \bar{d}_j \ell^-_k}_{12}\right],\nn\\
\label{mat12-qql2}
\eea
where
\bea
A^{u_i \bar{d}_j \ell^-_k}_{12} &=& \left(O^{{cnw}}_{Rk1} O^{{cns}^*}_{Rk1r}
+O^{{cnw}}_{Lk1} O^{{cns}^*}_{Lk1r}\right),
~~B^{u_i \bar{d}_j \ell^-_k}_{12} = \left(O^{{cnw}}_{Lk1} O^{{cns}^*}_{Rk1r}
+O^{{cnw}}_{Rk1} O^{{cns}^*}_{Lk1r}\right).\nn\\
\label{mat12-qql2-det}
\eea
\bea
&&\bullet~ M^\dagger_1M_3 (\ntrl1\to \sum u_i \bar{d}_j \ell^-_k) = \nn\\
&&\sum_{i,j,k}
\sum^2_{r=1}\frac{2\rt2 g^2_2\wt{g}^2 V^{{CKM}^*}_{ij}}
{\left[((k+k')^2-m^2_{W}-im_W\Gamma_W)((p+k)^2-m^2_{\d_r})
\right]} \nn\\
&&\left[2(P.k)(p.k')A^{u_i \bar{d}_j \ell^-_k}_{13}
-m_{\ell_k}m_{\ntrl1} (k.k') B^{u_i \bar{d}_j \ell^-_k}_{13}
+m_{u_i}m_{\ntrl1}(p.k')C^{u_i \bar{d}_j \ell^-_k}_{13}
\right.\nn\\
&&-2m_{u_i}m_{\ell_k} (P.k') D^{u_i \bar{d}_j \ell^-_k}_{13}
+2m_{d_j}m_{\ntrl1}(p.k)E^{u_i \bar{d}_j \ell^-_k}_{13}
-m_{d_j}m_{\ell_k}(P.k)F^{u_i \bar{d}_j \ell^-_k}_{13}\nn\\
&&+\left.
m_{u_i}m_{d_j}(P.p)G^{u_i \bar{d}_j \ell^-_k}_{13}
-2m_{u_i}m_{d_j}m_{\ell_k}m_{\ntrl1}H^{u_i \bar{d}_j \ell^-_k}_{13}\right],
\label{mat13-qql2}
\eea
where
\bea
A^{u_i \bar{d}_j \ell^-_k}_{13} &=& O^{{cnw}}_{Lk1} O^{{ucd}}_{Rikr}
O^{{dn\d}^*}_{Rj1r},
~~B^{u_i \bar{d}_j \ell^-_k}_{13} = O^{{cnw}}_{Rk1} O^{{ucd}}_{Rikr}
O^{{dn\d}^*}_{Rj1r},\nn\\
C^{u_i \bar{d}_j \ell^-_k}_{13} &=& O^{{cnw}}_{Rk1} O^{{ucd}}_{Likr}
O^{{dn\d}^*}_{Rj1r},
~~D^{u_i \bar{d}_j \ell^-_k}_{13} = O^{{cnw}}_{Lk1} O^{{ucd}}_{Likr}
O^{{dn\d}^*}_{Rj1r},\nn\\
E^{u_i \bar{d}_j \ell^-_k}_{13} &=& O^{{cnw}}_{Lk1} O^{{ucd}}_{Rikr}
O^{{dn\d}^*}_{Lj1r},
~~F^{u_i \bar{d}_j \ell^-_k}_{13} = O^{{cnw}}_{Rk1} O^{{ucd}}_{Rikr}
O^{{dn\d}^*}_{Lj1r},\nn\\
G^{u_i \bar{d}_j \ell^-_k}_{13} &=& O^{{cnw}}_{Rk1} O^{{ucd}}_{Likr}
O^{{dn\d}^*}_{Lj1r},
~~H^{u_i \bar{d}_j \ell^-_k}_{13} = O^{{cnw}}_{Lk1} O^{{ucd}}_{Likr}
O^{{dn\d}^*}_{Lj1r}.
\label{mat13-qql2-det}
\eea
\bea
&&\bullet~ M^\dagger_1M_4 (\ntrl1\to \sum u_i \bar{d}_j \ell^-_k) = \nn\\
&&-\sum_{i,j,k}
\sum^2_{r=1}\frac{2\rt2 g^2_2\wt{g}^2 V^{{CKM}^*}_{ij}}
{\left[((k+k')^2-m^2_{W}-im_W\Gamma_W)((p+k')^2-m^2_{\u_r})
\right]} \nn\\
&&\left[2(P.k)(p.k')A^{u_i \bar{d}_j \ell^-_k}_{14}
-m_{\ell_k}m_{\ntrl1} (k.k') B^{u_i \bar{d}_j \ell^-_k}_{14}
+m_{d_j}m_{\ntrl1}(p.k)C^{u_i \bar{d}_j \ell^-_k}_{14}
\right.\nn\\
&&-2m_{d_j}m_{\ell_k} (P.k) D^{u_i \bar{d}_j \ell^-_k}_{14}
+2m_{u_i}m_{\ntrl1}(p.k')E^{u_i \bar{d}_j \ell^-_k}_{14}
-m_{u_i}m_{\ell_k}(P.k')F^{u_i \bar{d}_j \ell^-_k}_{14}\nn\\
&&+\left.
m_{u_i}m_{d_j}(P.p)G^{u_i \bar{d}_j \ell^-_k}_{14}
-2m_{u_i}m_{d_j}m_{\ell_k}m_{\ntrl1}H^{u_i \bar{d}_j \ell^-_k}_{14}\right],
\label{mat14-qql2}
\eea
where
\bea
A^{u_i \bar{d}_j \ell^-_k}_{14} &=& O^{{cnw}}_{Rk1} O^{{cdu}}_{Lkjr}
O^{{nu\u}^*}_{L1ir},
~~B^{u_i \bar{d}_j \ell^-_k}_{14} = O^{{cnw}}_{Lk1} O^{{cdu}}_{Lkjr}
O^{{nu\u}^*}_{L1ir},\nn\\
C^{u_i \bar{d}_j \ell^-_k}_{14} &=& O^{{cnw}}_{Lk1} O^{{cdu}}_{Rkjr}
O^{{nu\u}^*}_{L1ir},
~~D^{u_i \bar{d}_j \ell^-_k}_{14} = O^{{cnw}}_{Rk1} O^{{cdu}}_{Rkjr}
O^{{nu\u}^*}_{L1ir},\nn\\
E^{u_i \bar{d}_j \ell^-_k}_{14} &=& O^{{cnw}}_{Rk1} O^{{cdu}}_{Lkjr}
O^{{nu\u}^*}_{R1ir},
~~F^{u_i \bar{d}_j \ell^-_k}_{14} = O^{{cnw}}_{Lk1} O^{{cdu}}_{Lkjr}
O^{{nu\u}^*}_{R1ir},\nn\\
G^{u_i \bar{d}_j \ell^-_k}_{14} &=& O^{{cnw}}_{Lk1} O^{{cdu}}_{Rkjr}
O^{{nu\u}^*}_{R1ir},
~~H^{u_i \bar{d}_j \ell^-_k}_{14} = O^{{cnw}}_{Rk1} O^{{cdu}}_{Rkjr}
O^{{nu\u}^*}_{R1ir}.
\label{mat14-qql2-det}
\eea
\bea
&&\bullet~ M^\dagger_2M_3 (\ntrl1\to \sum u_i \bar{d}_j  \ell^-_k) 
\underbrace{=}_{L\Longleftrightarrow R} 
M^\dagger_2M_3 (\ntrl1\to \sum \bar{u}_i d_j \ell^+_k)^*.
\label{mat23-qql2}
\eea
\bea
&&\bullet~ M^\dagger_2M_4 (\ntrl1\to \sum u_i \bar{d}_j  \ell^-_k) 
\underbrace{=}_{L\Longleftrightarrow R} 
M^\dagger_2M_4 (\ntrl1\to \sum \bar{u}_i d_j \ell^+_k)^*.
\label{mat24-qql2}
\eea
\bea
&&\bullet~ M^\dagger_3M_4 (\ntrl1\to \sum u_i \bar{d}_j  \ell^-_k) 
\underbrace{=}_{L\Longleftrightarrow R} 
M^\dagger_3M_4 (\ntrl1\to \sum \bar{u}_i d_j \ell^+_k)^*.
\label{mat34-qql2}
\eea
$V^{CKM}_{ij}$ are the entries of the CKM matrix and their values are given
in ref.\cite{AppNakamura-c2}.



\end{document}